\documentclass{memo-l}

\usepackage[scaled]{helvet} 
\usepackage{courier} 
\usepackage{eulervm} 
\normalfont
\usepackage[T1]{fontenc}

\usepackage{mathrsfs}
\usepackage{hyperref}
\usepackage{amssymb,amsthm}
\usepackage{amsmidx}
\usepackage{imakeidx}
\usepackage{amsmath}
\usepackage{accents}
\usepackage{pict2e}
\usepackage{bbold}
\usepackage{color}
\usepackage{mathtools}
\usepackage{hyperref}
\usepackage{amsmath}

\DeclareSymbolFont{greek}{OML}{cmr}{m}{n}  
\DeclareMathSymbol{\varrho}{0}{greek}{"25}

\newtheorem{theorem}{Theorem}[section]
\newtheorem{corollary}[theorem]{Corollary}
\newtheorem{proposition}[theorem]{Proposition}
\newtheorem{lemma}[theorem]{Lemma}

\theoremstyle{definition}
\newtheorem{definition}[theorem]{Definition} 
\newtheorem{example}[theorem]{Example}

\theoremstyle{remark}
\newtheorem{remark}[theorem]{Remark}
\newtheorem{remarks}[theorem]{Remarks}
\newtheorem{examples}[theorem]{Examples}

\usepackage[cal=boondox]{mathalfa}

\newcommand{\bbH}{\mathcal{H} \kern -.33cm \raisebox{ 0.047cm}{\texttt{\small | }} \kern -.05cm}
\newcommand{\bbh}{ \footnotesize \mathcal{H} \kern -.26cm \raisebox{ 0.037cm}{\rm {\texttt{\tiny | }}}}

\newcommand{\Kgeo}{ \mathfrak{k} }                     % Generator of rotations on dS
\newcommand{\Lgeo}{ \mathfrak{l} }                     % Generator of boosts on dS
\newcommand{\Cgeo}{ \mathfrak{c} }                     % Casimir on dS
\newcommand{\Jgeo}{ \mathfrak{j} }                     % complex boosts on dS

\newcommand{\KV}{ \mathcal{k} }                     % Generator of rotations on Hilbert space on V^+
\newcommand{\LV}{ \mathcal{l} }                     % Generator of boosts on Hilbert space on V^+
\newcommand{\CV}{ \mathcal{c} }                     % Casimir on Hilbert space on V^+

\newcommand{\KH}{ \mathcal{k} }                     % Generator of rotations on Hilbert space on masshells on V^+
\newcommand{\LH}{ \mathcal{l} }                     % Generator of boosts on Hilbert space on masshells on V^+
\newcommand{\CH}{ \mathcal{c} }                     % Casimir on Hilbert space on masshells on V^+

\newcommand{\KS}{ \widehat{k} }                     % Generator of rotations on Hilbert space on S^1
\newcommand{\KStwo}{\mathbb{k}}                     % Generator of rotations on Hilbert space on S^2
\newcommand{\LS}{\widehat{\ell}}                     % Generator of boosts on Hilbert space on S^1
\newcommand{\CS}{ \widehat{c}}                     % Casimir on Hilbert space on Hilbert space on S^1

\newcommand{\KFockdS}{ {K} }                     % Generator of rotations on Fock space on dS
\newcommand{\LFockdS}{ {L} }                     % Generator of boosts on Fock space on dS
\newcommand{\KFock}{\widehat{K}}                     % Generator of rotations on Fock space on S^1
\newcommand{\LFock}{\widehat{L}}                     % Generator of boosts on Fock space on S^1
\newcommand{\HFock}{ \widehat{H} }                     % Casimir on Hilbert space on Fock space on S^1

\newcommand{\KLS}{\widetilde{K}}                     % Generator of rotations on modified Fock space on S^1
\newcommand{\LLS}{\widetilde{L}}                     % Generator of boosts on modified Fock space on S^1
\DeclareMathOperator\arcsinh{arcsinh}

\numberwithin{section}{chapter}
\numberwithin{equation}{section}

\makeindex

\frontmatter

\title{The $\mathscr{P}(\varphi)_2$ Model on 
de Sitter Space}

\author[J.~Barata]{Jo\~{a}o C.A.~Barata}
\address{
Dept. de Fisica Matematica \\
Universidade de S\~ao Paulo (USP), Brasil}
\email{jbarata@if.usp.br}

\author[C.\ J\"akel]{Christian D.\ J\"akel}
\address{
Dept. Matematica Aplicada\\
Universidade de S\~ao Paulo (USP), Brasil}
\email{jaekel@ime.usp.br}

\author[J. Mund]{Jens Mund}
\address{Departamento de Fisica\\
Universidade de Juiz de Fora, Brasil}
\email{mund@fisica.ufjf.br}

\begin{document}



\thanks{The second author is grateful to IHES, Bures-sur-Yvette, France, for generous 
hospitality offered  during a visit in late 2014 and CERN, Geneva, were final revisions of the 
manuscript were undertaken during a visit in June 2016. The second and the third author were 
partially supported by Conselho Nacional de Desenvolvimento Cientifico e Tecnológico (CNPq), 
in particular by Projeto Universal 438435/2018-4
and by the São Paulo Research Foundation (FAPESP);  in particular, 
the second author was supported by a visiting professorship in 2012-2013 and
by Projeto Regular 2018/09613-9 and 
the third author was supported by FAPESP grant number 2014/24522-9.}

\dedicatory{Dedication text (use \\[2pt] for line break if necessary)}

\begin{abstract}
In 1975 Figari, H\o egh-Krohn and Nappi~\cite{FHN} constructed 
the ${\mathscr P}(\varphi)_2$ model on the de Sitter space. Here 
we complement their work with new results, which connect 
this model to various areas of mathematics. In particular, 

\begin{itemize}
\item[$i.)$] we discuss the causal structure of de Sitter space and  
the induces representations of the Lorentz group. We show that the 
UIRs of $SO_0(1,2)$ for both the principal and the complementary 
series can be formulated on Hilbert spaces whose functions 
are supported on a Cauchy surface. We describe the free classical 
dynamical system in both its covariant and canonical form, 
and present the associated quantum one-particle KMS structures 
in the sense of Kay \cite{Kay1}. Furthermore, we discuss the localisation 
properties of one-particle wave functions and how these properties are 
inherited by the algebras of local observables. 
\item[$ii.)$] we describe the relations between the modular objects 
(in the sense of Tomita-Takesaki theory) associated to wedge algebras 
and the representations of the Lorentz group. We connect the 
representations of SO(1,2) to unitary representations
of $SO(3)$ on the Euclidean sphere, and discuss how 
the ${\mathscr P}(\varphi)_2$ interaction can be represented by a 
rotation invariant vector in the Euclidean Fock space. 
We present a novel Osterwalder-Schrader reconstruction theorem, 
which shows that physical \emph{infrared problems}\footnote{As 
shown in ~\cite{FHN}, the ultraviolet problems are resolved 
just like on flat Minkowski space.} are absent on de Sitter 
space. We state the Haag--Kastler axioms for the ${\mathscr P}(\varphi)_2$ 
model and we explain how the \emph{generators} of the boosts and the 
rotations for the interacting quantum field theory arise from 
the \emph{stress-energy tensor}. Finally, we show that the interacting 
quantum fields satisfy the \emph{equations of motion} in their covariant form. 
\end{itemize}
In summary, we argue that the de Sitter  ${\mathscr P}(\varphi)_2$ model 
is the simplest and most explicit relativistic quantum field theory, which 
satisfies basic expectations, like covariance, particle creation, 
stability and finite speed of propagation. 
\end{abstract}

\maketitle

\tableofcontents
\addcontentsline{toc}{chapter}{Abstract\hfill}

\chapter*{List of Symbols}

\quad

\smallskip

{\bf Space-Time}

\bigskip

$(\mathbb{R}^{1+2}, \mathbb{g})$ \hfill  Minkowski space-time in 1+2 dimensions \quad \pageref{dSpage}

$(dS, g)$ \hfill  de Sitter space-time $dS \subset \mathbb{R}^{1+2}$ \quad \pageref{dSpage}

$\mathbb{g}$ \hfill  metric  on Minkowski space $\mathbb{R}^{1+2}$   \quad \pageref{metricpage}

$g = \mathbb{g}_{\upharpoonright dS} $ \hfill  metric  restricted to $dS$     \quad \pageref{metricpage}

${\mathcal C}$ \hfill  a Cauchy surface   \quad \pageref{ccCpage}

$S^1$ \hfill  time-zero circle   \quad \pageref{ccCpage}

$g_{\upharpoonright S^1} $ \hfill  metric  restricted to $S^1$   \quad \pageref{seinspage}

${\rm d} l (\psi)$  \hfill  induced surface element on $S^1$ \quad \pageref{dlpage}		

$I_+$ \hfill  the half-circle $W _1 \cap S^1$      \quad \pageref{halfcirclepage}

$V^+$  \hfill  forward light-cone in $\mathbb{R}^{1+2}$ \quad \pageref{vpluspage}

$\Gamma^\pm (x)$  \hfill  future  and past of a space-time point $x \in dS$
 \quad \pageref{vpluspage}

${\mathcal O}$ \hfill  a open, bounded space-time region 
 \quad \pageref{cOpage}

${\mathcal O}'$ \hfill  space-like complement of  ${\mathcal O} \subset dS$
 \quad \pageref{cOpage}
 
$W$ \hfill   a wedge in $dS$ \quad \pageref{tWpage}

$W _1$ \hfill  the wedge $\bigl\{ x \in dS \mid x_2 > \vert x_0 \vert \bigr\} $  
\quad \pageref{tWpage}

$W^{(\alpha)}$ \hfill  the wedge $R_0(\alpha)W_1$ 
\quad \pageref{w-alpha-page}

${\mathbb{W}}$ \hfill  the double wedge $W \cup W'$ 
\quad \pageref{dWpage}

${\mathcal O}_I$ \hfill the double-cone $I''$, with basis $I\subset S^1$   
 \quad \pageref{doubleconepage}

$H^\pm_m$ \hfill  mass hyperboloid in $\mathbb{R}^{1+2}$ 
\quad \pageref{masshyperboloid}

$P_\tau$  \hfill  horosphere \quad \pageref{horosphere}
	
\bigskip

{\bf Complex Space-Time}

\smallskip

$dS_{\mathbb{C}}$ \hfill complex de Sitter space \quad \pageref{dSccpage}

${\mathcal T}_\pm $ \hfill forward (backward) tuboid 
\quad \pageref{tuboidspage}

$S^2$ \hfill Euclidean sphere  \quad \pageref{euclidspherepage}

\bigskip

{\bf Euclidean space-time}

\smallskip

$S^2$  \hfill Euclidean space-time \quad \pageref{eulidspherepage}

$S_\pm$  \hfill upper (resp.~lower) hemisphere \quad \pageref{spherepluspage}

$S^1$ \hfill time-zero circle \quad \pageref{equatoreuclidpage}

$I_\pm$ \hfill half-circle formed by $ W_1 \cap S^1 $ or  $ W_1' \cap S^1 $ 
\quad \pageref{eulidspherepage}

$I_\alpha$ \hfill  the half-circle $I_\alpha = {\tt R}_{0} (\alpha) I_+$ 
\quad \pageref{halfcirclealphapage}

\bigskip

\goodbreak
{\bf De Sitter Group}

\smallskip

$O(1,2)$  \hfill de Sitter group, \emph{i.e.}, Lorentz group in 1+2 dimensions 
\quad \pageref{metricpage}

$SO_0(1,2)$  \hfill  proper, orthochronous de Sitter group 
\quad \pageref{metricpage}

$R_0$  \hfill  a rotation around the $x_0$-axis  \quad \pageref{rropage}

$\Lambda$ \hfill  an arbitrary element in $SO_0(1,2)$  \quad \pageref{lambdapage}

$\Lambda_1 (t)$ \hfill   the boost which leaves $W_1$ invariant  
\quad \pageref{Lambdaalpha}

$\Lambda^{(\alpha)} (t)$ \hfill   the boost which leaves $W^{(\alpha)}$ invariant  
\quad \pageref{Lambdaalpha}

$\Kgeo_0$, $\Lgeo_1$, $\Lgeo_2$ \hfill generators of Lorentz transformations 
\quad \pageref{Lambdaalpha}

$\Lgeo^{(\alpha)}$ \hfill  generator of the boost $t \mapsto \Lambda^{(\alpha)} (t)$  
\quad \pageref{Lambdaalpha}

$T \colon  \mathbb{R}^3 \to \mathbb{R}^3$ \hfill time reflection 
\quad \pageref{timerefl-page}

$P_1 \colon \mathbb{R}^3 \to \mathbb{R}^3$ \hfill parity reflection 
\quad \pageref{timerefl-page}

$\Theta_{W}$ \hfill reflection at the edge of the wedge $W$ 
\quad \pageref{PTwedgepage}

\bigskip
\goodbreak
{\bf Test-Functions on $dS$}

\smallskip

${\mathcal D}_{\mathbb{R}} (dS) $ \hfill  real $C^\infty$-functions with 
compact support on  $ dS $ \quad \pageref{ccCpage}

$f$, $g$ \hfill  elements of ${\mathcal D}_{\mathbb{R}} (dS) $ 
\quad \pageref{ccCpage}

\bigskip

{\bf Unitary irreducible representations of $SO_0(1,2)$}

\smallskip

${\gamma}$ \hfill a path on the forward light cone~$\partial V^+$ 
\quad \pageref{Gammacpluspage}

${\rm d} \mu_{\gamma}$  \hfill  
restriction of~${\rm d} \mu_{\partial V^+}$  to a path ${\gamma} \subset \partial V^+$ 
\quad \pageref{Gammacpluspage}

${\mathfrak h}_{\nu,0}$  \hfill $f \in C (SO_0(1,2)) $ satisfying 
$f(gh)= \pi_{\nu} (h^{-1}) f(g) $ for $h \in AN$
\quad \pageref{h-nu-0-page}

$\Pi_\nu$ \hfill induced representation of $SO_0(1,2)$ on ${\mathfrak h}_{\nu,0}$
\quad \pageref{h-nu-0-page}

$\widetilde u (\Lambda)$ \hfill  UIR of $SO_0(1,2)$ 
on $\widetilde {{\mathfrak h}}  (\partial V^+)$ \quad \pageref{umLambdapage}

$\KV_0$, $\LV_1$, $\LV_2$ \hfill generators of $SO_0(1,2)$
on $L^2 (\partial V^+, \frac{{\rm d} \alpha}{2 \pi} {\rm d} p_0)$
\quad \pageref{bargmangeneratorpage}

$\CV^2$ \hfill  the Casimir operator on the light cone
\quad \pageref{casimir}

$p = (p_0, p_1, p_2) \in \partial V^+$ \hfill coordinates on the light-cone
\quad \pageref{bargmangeneratorpage}

$S (S+1) \, \Phi= \mu^2 \, \Phi $ \hfill KG equation on the light-cone $\partial V^+$
\quad \pageref{lightconecoordinateKGpage}

\bigskip

{\bf Fourier Transformation}

\smallskip

$  ( x_\pm \cdot p \, )^s$ \hfill the Harish-Chandra plane-wave  
\quad \pageref{planewavepage}

$ \widetilde {f}_\pm ( p, s)  $ \hfill Fourier transform 
\quad \pageref{Fouriertransformpage}

${\mathcal F}_{+ \upharpoonright \nu} $ \hfill FH-transformation restricted to the mass shell
\quad \pageref{mass-shell-ft-page}

$\widetilde f_\nu (p )$ \hfill  restriction of the Fourier transformation  to the mass shell 
\quad \pageref{mass-shell-ft-page}

${\mathfrak h}  (dS)$ \hfill completion of 
${\mathcal D}_{\mathbb{R}} (dS)/{\ker ( {\mathbb E}_\mu  {\mathcal F}_+ )} $ 
\quad \pageref{ophs1page}

$\langle  \, . \, , \, . \, \rangle_{{\mathfrak h} (dS)}$ \hfill scalar product on 
${\mathfrak h}  (dS)$  \quad \pageref{ophs1page}

\bigskip

\goodbreak
{\bf Sobolev spaces}

\smallskip

$\mathbb{H}^{\pm 1} (S^2)$ \hfill Sobolev spaces   \quad \pageref{sobolevpage}

$e ({S^1}), e \left(S_\pm\right)$ \hfill orthogonal projections 
\quad \pageref{sobolevpage}

$\widehat{\mathfrak h} (S^1)$ \hfill a subspace of $\mathbb{H}^{- 1} (S^2)$ 
\quad \pageref{hs1-page}

\bigskip

{\bf (Pseudo-)Differential Operators}

\smallskip

$\square_{dS}+\mu^2$ \hfill Klein--Gordon operator 
\quad \pageref{squarepage}

$n$ \hfill future
pointing normal vector field ${\tt  n}(t,\psi) = \,  \mathbb{cos}_\psi  ^{-1} \,\partial_t$ 
\quad \pageref{vnpage}

$\varepsilon$ \hfill generator of the boosts $t \mapsto \Lambda_1(t)$ 
\quad \pageref{epsilonpage}

$ \mathbb{cos}_\psi $ \hfill 
multiplication operator by  $\cos\psi$ \quad 
\pageref{cosinepsipage}
 
\bigskip

\goodbreak
{\bf Covariant Dynamical System}

\smallskip

${\mathfrak k}(dS)$ \hfill  space of solutions of the Klein--Gordon equation   
\quad \pageref{kldypage-a}

$\sigma$ \hfill  symplectic form associated to  ${\mathcal E}$  
\quad \pageref{sigmapage}

$\mathscr{E}$ \hfill  the commutator function for the Klein--Gordon equation  
\quad \pageref{ccCfpage}

${\mathfrak z} (\Lambda)$ \hfill  representation of $O(1,2)$  on  $({\mathfrak k}(dS), \sigma)$  
\quad \pageref{ccTpage}

$\mathbb{\Phi}$  \hfill a solution of the Klein--Gordon equation (an element  
in ${\mathfrak k}(dS)$)  \quad \pageref{kldypage}

${\mathbb P}$ \hfill  projection from ${\mathcal D}_{\mathbb{R}} (dS) $ 
to $ {\mathfrak k}(dS)$  \quad \pageref{kldypage}

$\mathbb{f}$ \hfill  solution of the KG equation for $f  \in {\mathcal D}_{\mathbb{R}} (dS)$
\quad \pageref{kldypage}

\bigskip

\goodbreak
{\bf Canonical Dynamical System}

\smallskip
 
$\widehat {\mathfrak k} (S^1)$ \hfill the space of Cauchy data for the 
Klein--Gordon equation
\quad \pageref{hatckpage} 

$(\mathbb{\phi}, \mathbb{\pi})$ \hfill  Cauchy data (an element of 
$\widehat {\mathfrak k} (S^1)$) \quad \pageref{hatckpage} 

$\widehat\sigma$   \hfill the canonical symplectic form on $\widehat {\mathfrak k} (S^1)$
\quad \pageref{hatckpage} 

$\widehat {\mathfrak z} (\Lambda)$ \hfill  representation of $O(1,2)$  on   
$(\widehat {\mathfrak k} (S^1), \widehat\sigma)$  \quad \pageref{widehatccTpage}

$\widehat {\mathbb P}$  \hfill  a map from ${\mathcal D}_{\mathbb{R}} (dS) $ 
to $ \widehat {\mathfrak k} (S^1)$  \quad \pageref{widehatcxC}

\bigskip

\goodbreak
{\bf Covariant One-Particle Structure}

\smallskip
 
${\mathfrak h}  (dS)$ \hfill Hilbert space \quad \pageref{umLambdapage2a}

$ u (\Lambda)$ \hfill  unitary irreducible representation of $SO_0(1,2)$ 
on ${\mathfrak h}  (dS)$ \quad \pageref{umLambdapage2b}

$K$ \hfill  maps ${\mathfrak k} (S^1) $ into ${\mathfrak h}  (dS)$ 
\quad \pageref{umLambdapage2}

$\left( K, {\mathfrak h}  (dS), u  \right)$ \hfill   one-particle
structure for $( {\mathfrak k} (dS) ,\sigma, {\mathfrak z} )$ 
\quad \pageref{covonepartpage}

\bigskip

{\bf Canonical One-Particle Structure}

\smallskip
$\widehat{{\mathfrak h}}  (S^1)$ \hfill  time-zero Hilbert space  
\quad \pageref{chfdpage}

$\langle  \, . \, , \, . \, \rangle_{\widehat{{\mathfrak h}}  (S^1)}$ \hfill  scalar product 
on $\widehat{{\mathfrak h}}  (S^1)$ \quad  \pageref{chfdpage}

$\widehat{u}(\Lambda)$ \hfill  unitary irreducible representation of $SO_0(1,2)$ 
on $\widehat{{\mathfrak h}}  (S^1)$ \quad \pageref{Umhatpage}

$\KS$  \hfill	generator of the rotations 
on $\widehat{{\mathfrak h}}  (S^1)$ \quad \pageref{KShatpage}

$\LS_1$, $\LS_2$  \hfill generators of the boosts 
on $\widehat{{\mathfrak h}}  (S^1)$ \quad \pageref{LShatpage}

$\CS^{\, 2}$  \hfill  Casimir operator 
on $\widehat{{\mathfrak h}}  (S^1)$ \quad \pageref{Chatpage}

$\widehat{K}$ \hfill  maps $\widehat {\mathfrak k} (S^1) $ 
into $\widehat{{\mathfrak h}}  (S^1)$ \quad \pageref{Kmhatpage}

$\bigl(\widehat{K} , \widehat{{\mathfrak h}}  (S^1), \widehat{u} \bigr) $ \hfill   one-particle
structure for $(\widehat {\mathfrak k} (S^1) ,\widehat \sigma, \widehat   {\mathfrak z} )$ 
\quad \pageref{Kmhatpage}

\bigskip 
\goodbreak
{\bf Operator Algebras and States}

\smallskip

${\mathfrak W} ({\mathfrak k} , \sigma)$ \hfill  Weyl algebra 
\quad \pageref{weylalgebrapage}

$\alpha_\Lambda $ \hfill automorphic representation of  $SO_0(1,2)$ on 
${\mathfrak W}({\mathfrak k}(dS),\sigma)$ \quad \pageref{alphapage}

$\bigl({\mathfrak W}(dS), \alpha_\Lambda^\circ \bigr)$ \hfill covariant quantum 
dynamical system  \quad \pageref{cqds-page}
 
$\bigl(\widehat {\mathfrak W}(dS), \widehat \alpha_\Lambda^\circ \bigr)$ \hfill 
canonical quantum dynamical system  \quad \pageref{cqds-page}

$\widehat {\alpha}_\Lambda $ \hfill automorphic representation of  $SO_0(1,2)$ 
on  $\widehat {\mathfrak W}({\mathfrak k}(S^1),
\widehat \sigma)$ \quad \pageref{alphapage}

$\omega^\circ$ \hfill free de Sitter vacuum state  \quad \pageref{freevacuumstatepage}

$\widehat \omega^\circ$ \hfill free de Sitter vacuum state  
\quad  \pageref{freevacuumstatepage}

${\mathscr A}_\circ  ( {\mathcal O}) $ \hfill v.~N.~algebra for  the free fields 
in a double cone ${\mathcal O}\subset dS$ \quad \pageref{AO-page}

${\mathcal R} ( I) $ \hfill v.~N.~algebra for  the free fields in the interval 
$I \subset S^1$ \quad \pageref{RI-page}

\bigskip 
\goodbreak

{\bf Euclidean Fock Space}

\smallskip

$C (f,g)$ \hfill covariance \quad \pageref{dualitypage} 
  
$L^{p}( \mathscr{U}(S^2) , \Omega_\circ )$ \hfill $L^p$ spaces 
\quad \pageref{dualitypage}
 
$C_{| s |} (h_{1}, h_{2})$ \hfill  time-zero covariance \quad  \pageref{stcpage} 

$\Phi (\theta, h)$ \hfill  sharp-time field \quad  \pageref{stfpage}

$\Gamma( \mathbb{H}^{-1}(S^2))$ \hfill Fock space over the Sobolev 
space $\mathbb{H}^{-1}(S^2)$  \quad \pageref{fockpage}

\bigskip

\goodbreak
{\bf Interaction}

\smallskip

${:} \Phi(f)^{n} {:}_c$ \hfill   Wick ordering  \quad \pageref{wickpage}

$ V \left(S_+ \right)$ \hfill interaction on the upper hemisphere 
\quad \pageref{interactionspherepage}

${\rm d}\mu_{V}$  \hfill perturbed measure  on the sphere 
\quad \pageref{interactionspherepage}

${V_0} (\operatorname{\mathbb{cos}}) $ \hfill the interaction on 
the half-circle $I_+$\quad \pageref{vcospage}

$\mathbb{\Omega}$  
\hfill interacting Euclidean vacuum vector 
in $\bbH_+$ \quad \pageref{intvacuumpage}

$\widehat{\Omega}$  
\hfill interacting vacuum vector 
in $\widehat{\mathcal H} (S^1)$ \quad \pageref{intvacuumpage}

$\Omega$  
\hfill interacting vacuum vector 
in ${\mathcal H} (dS)$ \quad \pageref{covintvacuumpage}

\bigskip

{\bf Unitary Groups on Covariant Fock Space}

\smallskip

${\mathcal H} (dS)$ \hfill  Fock space over the one-particle space  
${\mathfrak h} (dS)$\quad\pageref{HdS-page}

$\KFockdS$  \hfill	generator of the rotations on ${\mathcal H} (dS)$ 
\quad \pageref{LFockdSpage}

$\LFockdS_1^\circ$, $\LFockdS_2^\circ$  \hfill  generators of the free boosts 
on ${\mathcal H} (dS)$ \quad \pageref{LFockdSpage}

\bigskip
\goodbreak

{\bf Unitary Groups on Canonical Fock Space}

\smallskip

$\widehat{\mathcal H} (S^1)$ \hfill  Fock space over the one-particle space  
$\widehat{\mathfrak h} (S^1)$\quad\pageref{widehat-Fock-space-page}

$\KFock$  \hfill generator of the rotations on $\widehat{\mathcal H} (S^1)$ 
\quad \pageref{KFockhatpage}

$\LFock_1^\circ$, $\LFock_2^\circ$    \hfill   generators of the free boosts 
on $\widehat{\mathcal H} (S^1)$ \quad \pageref{LFockhatpage}

$\widehat{L}^{(0)}$  \hfill  generators of the interacting boosts 
on $\widehat{\mathcal H} (S^1)$ \quad \pageref{interacting-boost-page}

$\widehat{U}$  \hfill a unitary representation of $SO_0(1,2)$ 
on~$\widehat{\mathcal H} (S^1)$ \quad\pageref{umospage}	

$\HFock^{(0)}$       \hfill           perturbed generators of the boosts 
on $\widehat{\mathcal H} (I_+)$ \quad \pageref{HFockhatpage}

\bigskip
\goodbreak

{\bf Unitary Groups on Fock Space over $L^2(S^1, {\rm d} \psi)$}

\smallskip

$\LLS_1^\circ$, $\LLS_2^\circ$   \hfill   generators of the free boosts on 
Fock Space over $L^2(S^1, {\rm d} \psi)$ \quad \pageref{LLSpage}

$\LLS_1$, $\LLS_2$  \hfill generators of the interacting 
boosts on Fock Space over $L^2(S^1, {\rm d} \psi)$ \quad \pageref{KLSpage}

$\KLS$  \hfill	generator of the rotations on on Fock Space 
over $L^2(S^1, {\rm d} \psi)$ \quad \pageref{KLSpage}

\bigskip
\goodbreak

{\bf Symbols Appendices}

\smallskip

$(K, \sigma, T_t)$ \hfill classical dynamical system \quad  \pageref{page48}

$\bigl({\mathfrak h}_{\scriptscriptstyle \rm AW},K_{\scriptscriptstyle \rm AW},
U_{\scriptscriptstyle \rm AW} (t) \bigr)$ \hfill Araki-Woods one-particle structure
\quad  \pageref{page49}

\chapter*{Preface}

Quantum field theory, which originated in 1926 with the work of Born, 
Heisenberg, and Jordan~\cite{BHJ}, underlies (in form of the \emph{standard 
model}) our understanding of physics (with the notable 
exception of gravitational phenomena). There is widespread faith in 
quantum field theory and its validity, even on curved space-times. It should 
predict and explain many of the exciting astrophysical and cosmological phenomena 
currently discovered  in one of the most thriving branches of experimental physics.  
Unfortunately, many physically relevant questions 
are beyond the scope of validity of (re\-normalised) perturbation theory\footnote{See,  
\emph{e.g.},~\cite{BFr, BFK, BrFr,  HW2, HW3, HW1, H1, H2} and references therein.
However, note that even on flat space-time, the ${\mathscr P}(\varphi)_2$-models, 
do \emph{not} allow a \emph{Borel summation} \index{Borel summation} of the 
perturbation series, unless the order of the polynomial is less or equal four, as the 
number of Feynman diagrams \index{Feynman diagram} grows too rapidly for 
polynomials of higher order. Although in each order of perturbation theory there 
are no divergences, the Green's functions are not analytic in the coupling 
constant, neither are the proper self-energy and the two-particle scattering 
amplitude \cite{Jaffe}. For the $\varphi^4_2$-model on Minkowski space, 
perturbation theory yields a  Borel summable 
asymptotic series \index{asymptotic series} for the Schwinger functions.} on 
curved space-time. Non-perturbative methods, which can be used on 
curved space-times\footnote{The so called {\em static} 
space-times allow analytic continuations to Riemannian manifolds, and Ritter 
and Jaffe \cite{JR1, JR2, JR3} 
pioneered a non-perturba\-tive, constructive approach to interacting fields defined on them. 
They have shown that one can reconstruct a unitary representation of the isometry group 
of the static space-time under consideration, starting from the corresponding Euclidean field 
theory~\cite{JR2}. Some progress has also been made in case the space-time is
asymptotically flat,  see, \emph{e.g.},~\cite{D84, GeP, GM}.}, will 
have to be developed, before explicit 
calculations addressing specific phenomena can be carried out. 
From the perspective of general relativity, 
the case of a maximal symmetric space-time is rather exceptional. 
But if the curvature of space-time is small, de Sitter space 
may be a good approximation (at least locally).
More importantly, de Sitter space should provide a valuable test ground for new techniques 
which intend to treat the quantum field theory on a general curved space-time.

Quantum field theory has also been a fruitful 
source of mathematical challenges,  inspiring the development of entire branches of 
mathematics\footnote{Many of the results in differential geometry~\cite{Frankel, KN}, 
harmonic analysis~\cite{DyMcK, Folland, He, Mo1, Vara}, complex analysis in several 
variables~\cite{Fa, Hoer1, Vlad}, operator algebras \cite{BR, KR,Takesaki}, 
the representation theory of semi-simple Lie groups~\cite{Ba, BaF, Knapp, Mack, T, Vil} and 
the theory special functions \cite{Lebedev, snow} that we will use, were originally 
inspired by questions 
posed within quantum field theory.}.  But as a mathematical 
subject by \emph{itself}, it has not yet been casted in 
an axiomatic form (at least on four dimensional 
Minkwoski space), which is  appealing to mathematicians\footnote{Unfortunately,  
the vast knowledge accumulated in axiomatic quantum field theory~\cite{SW, Jost}, 
local quantum field theory~\cite{A0, H}, constructive quantum field 
theory~\cite{GJ, GJcp, S} and quantum 
statistical mechanics \cite{BR, Ruelle} has not --- from the viewpoint 
of the authors' --- found the recognition 
it deserves, both in the physics and the mathematics community.} and 
which allows to derive its consequences. 
However, we believe that there is hope that this fact may change 
once the importance of Tomita-Takesaki  modular theory 
in quantum field theory is revealed: the analyticity properties (which in case 
of the de Sitter space guarantee the stability of the vacuum state) provide 
an intimate connection between the representation theory of semi-simple 
Lie groups and the Tomita-Takesaki  modular theory 
(see~\cite{Bo} for a visionary perspective on this subject). This fact 
is most evident on the de Sitter space, where the Lorentz boosts (which 
generate the whole space-time symmetry group) are implemented by 
modular groups associated to von Neumann algebras for wedges
and the (free or interacting) vacuum vector. 

In order to illustrate this general fact, we present a detailed and very explicit, 
\emph{non-perturbative} description of the ${\mathscr P}(\varphi)_2$ 
model on de Sitter space. This model, originally constructed in 1975 
by Figari, H\o egh-Krohn and Nappi~\cite{FHN}, is not only the 
first \emph{interacting} quantum field theory on a curved space-time but, to 
the best of our knowledge, the only one established so far. In this work 
we reconsider the original formulation of this model~\cite{FHN} in the light 
of more recent work by Birke and Fr\"ohlich~\cite{BF}, Dimock~\cite{D} 
and Fr\"ohlich, Osterwalder and Seiler~\cite{FOS}. The peculiar role of 
the ${\mathscr P}(\varphi)_2$ model is best illustrated by comparing it with 
the role the Ising model plays in (quantum) statistical mechanics or the role  
$SL(2, \mathbb{R})$ plays in harmonic analysis. In many ways, the 
${\mathscr P}(\varphi)_2$ model is the \emph{simplest example} 
of an interacting relativistic quantum theory one can imagine, 
as it is free of both (serious) ultraviolet and (physical) infrared problems 
while still satisfying all the basic properties (like finite speed 
of propagation, stability of the vacuum,  etc.), 
one would usually require in an axiomatic approach. 

Models with polynomial interactions (like the ${\mathscr P}(\varphi)_2$ model) 
were the first interacting quantum field theories (on Min\-kowski space), which 
gained a precise mathematical meaning and up till now they remain the most 
thoroughly studied models in the axiomatic framework. The original construction 
of these models (without cutoffs) is due to Glimm and 
Jaffe~\cite{GJ1, GJ2,GJ3,GJ4,GJ5,GJ6}.  Following their pioneering works,  
an enormous amount of work has been invested to understand 
the scattering theory, the bound states, the low energy particle structure and the 
properties of the correlation functions of these models (see the books by Glimm 
and Jaffe~\cite{GJ,GJcp} and Simon \cite{S}, and the references therein).  So far, 
the ${\mathscr P}(\varphi)_2$ models are the only interacting quantum field theories, 
for which the non-relativistic limit (including bound states) has been analysed in 
detail, demonstrating that the low energy regime of these models can be equally 
well  described by non-relativistic bosons interacting with $\delta$-potentials \cite{D57,SZ}. 
In addition, Hepp  demonstrated  that (on Minkowski space) the classical field equations for the 
${\mathscr P}(\varphi)_2$ models can be recovered by taking the classical limit \cite{Hepp}.
On de Sitter space, that same question still poses an interesting challenge. 

\vskip .5cm

\aufm{Jo\~{a}o C.A.~Barata  ,
Christian D.\ J\"akel \&
Jens Mund}

\mainmatter

\part{De Sitter Space}

\chapter{De Sitter Space as a Lorentzian Manifold}
\label{geometry}

A \emph{Lorentzian manifold}\index{Lorentzian manifold} is a $(1+n)$-dimensional 
manifold $M$ together with a pseudo-Riemannian \emph{metric} $g$ of signature 
	\[
		(+, \underbrace{-, \ldots , -}_{\text{$n$-times}}) \; . 
	\]
For $n=3$, such Lorentzian manifolds appear as the solutions of the Einstein equations,
and are interpreted as space-times. The tangent space of \emph{any} $(1+n)$-dimensional 
Lorentzian manifold is the $(1+n)$-dimensional Minkowski space. The latter is also the 
simplest example of a Lorentzian manifold: it consists of the manifold~$ \mathbb{R}^{1+n}$ 
together with the metric
\label{metricpage}
	\begin{equation}
		\label{metrik}
		\mathbb{g} = {\rm d} x_0 \otimes {\rm d} x_0 
		- {\rm d} x_1 \otimes {\rm d} x_1 \;  \ldots \; - {\rm d} x_n \otimes {\rm d} x_n  \; ,  
 	\end{equation}
which does not depend on $x \in \mathbb{R}^{1+n}$. For $n=2$, we denote the points 
of~$\mathbb{R}^{1+2}$  either as triples $(x_0, x_1, x_2)$ or as column vectors
$\left(\begin{smallmatrix} x_0 \\ x_1 \\ x_2\end{smallmatrix}
\right)$, which ever is more convenient. The Minkowski  product 
	\[
		x \cdot y \doteq x_0 y_0 - x_1 y_1 - x_2 y_2 
	\]
of two vectors $x, y$ in~$\mathbb{R}^{1+2}$ is indicated by a dot. 

Next, we turn to Lorentzian manifolds of constant (non-zero) \emph{curvature}. There are 
only two such Lorentzian manifolds, namely \emph{de Sitter space}\index{de Sitter space} 
$(dS, g)$ with constant \emph{positive} curvature \index{positive curvature} and 
\emph{anti-de Sitter space} with constant \emph{negative} curvature. The latter plays 
a prominent role in the context of string theory, but will be of no importance for us. 
On the other hand, de Sitter space is our main focus. We will show 
that it provides an infrared cut-off for the quantum theory, without destroying 
the space-time symmetries. The latter get merely deformed, and in the limit of 
curvature to zero the theory on flat Minkowski space is recovered.  

As the aim of this work is to describe interacting quantum fields defined 
on de Sitter space, we will \emph{neither} consider more general Lorentzian 
manifolds \emph{nor} present the mathematical framework for Einstein's general 
relativity. But we can not resist the temptation to  add a few remarks on the 
peculiar role de Sitter space played in the historical development of the theory 
of gravitation. The main text will start with Section \ref{sec:1.2}. 

\goodbreak
\section{The Einstein equations}
\label{sec:1.1}

Albert Einstein's theory of gravitation, published in 1915, 
relates the metric tensor\index{metric tensor} $g_{\mu \nu}$ 
and the stress-energy tensor $T_{\mu \nu}$. In particular, 
the Einstein equations\index{Einstein equations},
	\begin{equation}
	\label{eineq}
		\underbrace{R_{\mu \nu} - \frac{1}{2} R \, g_{\mu \nu} }_{\doteq 	G_{\mu \nu}}
		= 8 \pi \, T_{\mu \nu} \; ,  \qquad \mu, \nu = 0, 1, \ldots, n,
	\end{equation}
describe the curvature\index{curvature}\footnote{The Ricci tensor\index{Ricci tensor} 
$R_{\mu \nu}$ and the scalar curvature\index{scalar curvature} $R$ both depend only 
on the metric tensor $g_{\mu \nu}$.}  of space-time resulting from the distribution of 
classical matter fields. 

\goodbreak
Although today there is ample evidence that for $n=3$ the equations \eqref{eineq} 
correctly describe gravitation, the situation was less clear in 1915. 
Just like Isaac Newton, Einstein was convinced that there should be some 
\emph{repulsive} mechanism, which would ensure stability against gravitational 
collapse. Hence, in 1916, Einstein (re-) introduced a positive (\emph{i.e.}, repulsive) 
cosmological constant $\Lambda>0$ in the Einstein equations, 
requesting\footnote{In space-time dimension two, the Einstein 
tensor $G_{\mu \nu}$  is always zero. Nevertheless, $R$ may be non-zero, as  
there is no Bianchi identity in space-time dimension two.} 
	\begin{equation}
	\label{Einstein-Lambda}
		R_{\mu \nu} - \tfrac{1}{2} R \, g_{\mu \nu} + \Lambda \, g_{\mu \nu} 
		= 8 \pi \, T_{\mu \nu} \; , 
	\end{equation}
in an attempt to ensure the existence of \emph{static} solutions. 

However, only a few months later Willem de Sitter showed that for 
$T_{\mu \nu}=0$ (\emph{i.e.}, the empty space), the new constant 
$\Lambda >0$ does not resolve the stability issues:  the 
solution\footnote{The equations simplify substantially, if one assumes 
that the solutions are highly symmetric. Among the 
most symmetric solutions are the de Sitter spaces.} 
of the equation \eqref{Einstein-Lambda} describes a 
universe, which undergoes \emph{accelerated expansion} \cite{deS1,deS2}. Einstein   
discarded de Sitter's solution as physically irrelevant~\cite{Schu}, but
experimental evidence nowadays suggests that on a large scale our universe is indeed 
\emph{isotropic} (\emph{i.e.}, there is no preferred direction), 
\emph{homogeneous} (\emph{i.e.}, there is no preferred location)
and undergoing accelerated expansion. The latter is compatible with the 
existence of a positive cosmological constant \cite{Riess}. 

\section{De Sitter space}
\label{sec:1.2}

Every $n$-dimensional Lorentzian manifold can be embedded as a sub-manifold 
in $\mathbb{R}^{1+ 2n}$. Any Lorentzian manifold, which is maximally 
symmetric\footnote{Vector fields that preserve the metric are called 
\index{Killing vector field} \emph{Killing vector fields}. 
A $d$-dimensional space-time is called \index{maximally symmetric space-time}
\emph{maximally symmetric}, if it has $d(d+1)/2$ linearly  
independent Killing vector fields.}, has constant curvature, and can thus be embedded in 
$\mathbb{R}^{1+ n}$, whereby its metric coincides with the restriction of a 
pseudo-Riemannian metric on~$\mathbb{R}^{1+n}$.  De Sitter space is the 
maximally symmetric Lorentzian manifold with constant \emph{positive} curvature. 
In more than two space-time dimensions, it is simply-con\-nected. 
The two-dimensional de Sitter space $dS$ can be viewed as a one-sheeted 
\emph{hyperboloid}\index{hyperboloid}, embedded in $(1+2)$-dimensional Minkowski 
space~$\mathbb{R}^{1+2}$: following \cite{Schr}, we may identify de Sitter space 
with the submanifold
	\begin{equation}
		\label{eqdSMin}
		dS \doteq \left\{  
		x \equiv (x_0, x_1, x_2) \in \mathbb{R}^{1+2} 
		\mid 
		x_{0}^{2} - x_{1}^{2} - x_{2}^{2} = - r^2 \right\} \; , 
		\quad r >0 \; ,
	\end{equation}
\label{dSpage}
of $\mathbb{R}^{1+2}$. 
The point $o \equiv (0, 0, r )\in dS$ is called the 
\emph{origin}\index{origin of de Sitter space} of~$dS$.  

\goodbreak

\begin{lemma}
\label{lm:1.2.1}
The embedding $(dS, g) \hookrightarrow (\mathbb{R}^{1+2}, \mathbb{g})$ is 
compatible with 
\begin{itemize}
\item [$i.)$] the metric structure, \emph{i.e.}, the \emph{metric}\index{metric} 
$g$ on $dS$ equals the induced metric $\mathbb{g}_{\upharpoonright dS }$,
\emph{i.e.}, $g= \mathbb{g}_{\upharpoonright dS } $, 
with  $\mathbb{g}$ the metric of the ambient space $(\mathbb{R}^{1+2},  \mathbb{g})$;
see \eqref{metrik}. 
\item [$ii.)$] the intrinsic \emph{causal structure}\index{causal structure} of $dS$ 
coincides with the one inherited from the ambient Minkowski space.  
\end{itemize}
\end{lemma}

We will now define a number of charts, which together provide an atlas for the de 
Sitter space $dS$. \index{global coordinates}\index{geographical 
coordinates}\index{coordinates for $dS$}

\subsection{Geographical coordinates}
\label{geo-chart}
The chart\footnote{\, Where necessary, we will restrict this map 
to $- \pi < \varrho <  \pi$, so that it provides a proper chart in the sense of 
differential geometry.}
	\[
			\begin{pmatrix}
					        {\tt x}_0 \\
						{\tt x}_1 \\
						{\tt x}_2 
			\end{pmatrix} 
				= \begin{pmatrix}
						r \sinh t  \\
 						r \cosh t \sin \varrho  \\
 						r \cosh t \cos \varrho  \\
				\end{pmatrix} \; , \qquad t \in \mathbb{R} \;, 
				\quad  - \pi \le \varrho <  \pi \; , 
	\]
covers the whole hyperboloid. We refer to $(t, \varrho )$ as {\em geographical 
coordinates}\index{geographical coordinates}. 
The restriction of the Lorentz metric in $\mathbb{R}^{1+2}$ to this chart  is
	\[
		g = r^2 {\rm d} t \otimes {\rm d} t - r^2  \cosh^2 t \; 
		({\rm d} \varrho \otimes {\rm d} \varrho )
	\]
and the \emph{Laplace-Beltrami operator} \index{Laplace-Beltrami operator}
	\[
		\square_{dS} \doteq |g|^{-1/2} \partial_\mu 
		\Bigl( | g|^{1/2} g^{\mu \nu} \partial_\nu \Bigr)
	\]
on $dS$ takes the form
	\begin{align}
		\label{L1a}
		\square_{dS}
			& =  \frac{1}{r^2 \cosh^2 t} 
				\left( \Bigl( \cosh t \frac{\partial}{\partial t} \Bigr)^2
				- \frac{\partial^2}{\partial \varrho^2}  \right) 
					\nonumber \\
			& = \frac{\partial^2}{\partial t^2} + \tanh t \frac{\partial}{\partial t}
				- \frac{1}{r^2 \cosh^2 t} \frac{\partial^2}{\partial \varrho^2}   
					\; .
	\end{align}
Note that $g^{\mu \nu}$ denotes the inverse matrix to~$g_{\mu \nu}$; 
thus $g^{\varrho \varrho}= (r^{2}  \cosh^{2} t)^{-1}$. 
In the limit of $t \to \pm \infty$,  the second term converges to~$\pm \partial_t$ 
while the third term in~\eqref{L1a} vanishes.
Changing the variable $t$ to the \index{conformal time} \emph{conformal 
time}~$\eta$ via 
	\[
		\tan  \eta   =  \sinh  t  \; , 
	\]
we obtain
	\[
		\begin{pmatrix} 
			x_0 	\\
			x_1	\\
			x_2
		\end{pmatrix}  
		= 
			\begin{pmatrix} 
			 r  \tan \eta  \\ 
			 r  \, \frac{\sin \varrho}{\cos \eta}  \\ 
			 r  \, \frac{\cos \varrho}{\cos \eta} 
		\end{pmatrix}   \; , \qquad \eta \in \left(- \tfrac{\pi}{2} , 
		\tfrac{\pi}{2} \right) \; , \; \; 
		\varrho \in [0, 2 \pi) \; . 
	\] 
\goodbreak
The metric $g$ takes the form
	\[
		g= \frac{r^2}{\cos^2 \eta} \left( {\rm d} \eta \otimes {\rm d}  \eta 
		-  {\rm d} \varrho \otimes {\rm d} \varrho \right) \; ,  
	\]
and, because $| g|^{1/2} g^{\mu \nu} = 1$, the Laplace-Beltrami operator is
	\[
		\square_{dS} 
		=   
		 \frac{\cos^2 \eta }{r^2} 
		  \left( \partial^2_\eta - \partial^2_\varrho  \right) \; .  
	\]
In order to analyse the peculiar structure of de Sitter space, it will be 
convenient to specify alternative coordinate systems. The latter will be  
introduced in Section~\ref{SS:2.6.3} using one-parameter groups of 
space-time symmetries.

\bigskip
In the meantime, we turn our attention to the causal structure of de Sitter space. 
As a consequence of Lemma \ref{lm:1.2.1}
$ii.)$, we may use the same criteria as on Minkowski space to describe it: 

\subsection{The future and the past}
A point $y \in dS$ is called {\em causal, time-like, light-like}\index{time-like 
point} \index{causal point} \index{light-like point}
and {\em space-like separated}\index{space-like point} to $x\in dS$, if 
$(y-x) \cdot (y-x) $ is larger or equal than, larger than, 
equal to or smaller than zero, respectively. Since $x \cdot x = y \cdot y = -r^2$, 
these notions are equivalent to
	\[
		x\cdot y \le - r^2, \quad x\cdot y < -r^2, \quad x\cdot y = -r^2, 
		\quad -r^2 < x\cdot y \; , 
	\]
respectively. The boundaries of the \emph{future}\index{future} (and 
the \emph{past}\index{past})
	\begin{equation} 
		\label{lklkl}
			\Gamma^\pm ( x ) \doteq \bigl\{  y \in dS \mid \pm ( y-  x ) 
			\in \overline{V^+} \bigr\} 
	\end{equation} 
of a  point $ x \in dS$ 
are\footnote{In particular, $\Gamma^\pm (0,0,r) = \{ y \in dS \mid \pm y_0 > 0, y_2 \ge r \}$.} 
given by two \emph{light rays}\index{light ray}, which form the intersection of $ dS $ with a 
Minkowski space future (respectively, 
past) \emph{light cone}\index{light cone}
	\begin{equation*}  
		\label{cone}
				C^\pm ( x ) 
				= \bigl\{  y \in \mathbb{R}^{1+2} 
				 \mid ( y -  x ) \cdot  ( y -  x ) = 0, \;
				\pm ( y_0 - x_0) > 0 \bigr\}  
	\end{equation*}
with apex at $x$. The future light cone $C^+ ( (0,0,0) ) $ with apex at the origin coincides 
with the boundary set $\partial V^+$ of the \emph{forward cone}\index{forward 
cone}\label{vpluspage}
	\[
		V^+ \doteq \bigl\{   y \in \mathbb{R}^{1+2} \mid  y \cdot  y >0,  
		y_0 > 0 \bigr\} \, .  
	\]
The bar in \eqref{lklkl} denotes the closure of $V^+$ in the ambient space 
$(\mathbb{R}^{1+2}, \mathbb{g}  )$.  

\begin{remark}
\label{R-1.2.2}
As it turns out, the two light rays mentioned are also given\footnote{For the 
origin $o$, the light rays $\{ o + \lambda (\pm 1, 0 , 1) \mid \lambda \in \mathbb{R} \}$ 
are given by the intersection of $dS$ with the plane $\{ x \in \mathbb{R}^{1+2} \mid x_2= r \}$.}
by the intersection of $dS$ with the tangent plane at $x \in dS$. They  
separate the future, the past and the space-like regions relative to the point $x$. 
\end{remark}

\subsection{Space-like complements and causal completions}
\label{sec:1.2.3}
The complement of the union $\Gamma^+ ( x) \cup \Gamma^- ( x )$  
consists of the points which are space-like to $x$. The \emph{space-like 
complement}\index{space-like complement} of a simply 
connected set $\mathcal{O} \subset dS$ is the set 
\label{cOpage}
	\begin{equation*} 
	\label{2.12}
		 \mathcal{O}' \doteq \left\{  y \in dS 
		 \mid  y \notin  \Gamma^+ ( x)  \cup  \Gamma^- ( x)  \;
		\; \forall   x \in \overline{\mathcal{O} }\right\} \; . 
	\end{equation*} 
The \emph{causal completion}\index{causal completion} $\mathcal{O}''$ 
of $\mathcal{O}$ is defined as the space-like complement of~$\mathcal{O}'$. 
Note that one always has $\mathcal{O} \subset \mathcal{O} ''$. 
In case $\mathcal{O}'' = \mathcal{O} $ holds, the  subset  $\mathcal{O} 
\subset dS$ is called \emph{causally complete}\index{causally complete subset}. 

\subsection{Cauchy surfaces}
\label{sec:1.2.4}
De Sitter space is \emph{globally hyperbolic}\index{globally hyperbolic 
space-time}, \emph{i.e.}, it has no time-like closed curves 
and for every pair of points $ x,  y \in dS$ the set
	\begin{equation*} 
		\label{globalhyper}
 			\Gamma^- ( x ) \cap  \Gamma^+ ( y)
	\end{equation*} 
is compact (eventually empty). These two properties imply 
that $dS$ is diffeomorphic to $\mathcal{C} \times  \mathbb{R} $, 
with $\mathcal{C}$  a \emph{Cauchy surface}\index{Cauchy surface} 
for $dS$ (see, \emph{e.g.},~\cite{BS}). It is convenient 
to choose $\mathcal{C} = S^1$, with  
\label{ccCpage}
	\begin{equation} 
		\label{SeinsC}
		S^1 \doteq  \bigl\{ ( 0, r \sin \psi, r \cos \psi) \in 
		\mathbb{R}^{1+2} \mid - \tfrac{\pi}{2} \le \psi 
		< \tfrac{3\pi}{2}  \bigr\} \; .   
	\end{equation} 
We frequently refer to $S^1$ as the \emph{time-zero circle}.  \index{time-zero 
circle} One may arrive at this choice by first choosing an arbitrary point $ x \in dS$ 
and a space-like geodesic $\mathcal{C}$ (see Section~\ref{geodesics}) passing 
through~$ x$, and then introducing coordinates in \eqref{eqdSMin} such 
that~$\mathcal{C}$ equals~\eqref{SeinsC}. 

\subsection{Geodesics}
\label{geodesics}

In the presence of a metric, a \emph{geodesic}\index{geodesic} can be defined 
as the curve joining $  x$ and $ y $ with maximum possible length in time --- 
for a time-like curve\footnote{A smooth curve $t \mapsto \gamma (t)$ on $dS$  
(with nowhere vanishing tangent vector $\dot \gamma$) is called {\em causal, 
time-like, light-like} and {\em space-like}, according to whether the tangent vector  
satisfies $0 \le \dot \gamma \cdot \dot \gamma $, $0 < \dot \gamma \cdot 
\dot \gamma $, $0 = \dot \gamma \cdot \dot \gamma  $, 
or $\dot \gamma \cdot \dot \gamma < 0 $, 
everywhere along the curve.} --- or the minimum 
possible length in space --- for a space-like curve. 
The null-geodesics on the de Sitter space are 
light rays, \emph{i.e.}, straight lines. 

De Sitter space is \emph{geodesically complete}\index{geodesically complete}, 
\emph{i.e.}, the affine parameter $t$ of any geodesic $ t \mapsto \gamma(t)$ passing 
through an arbitrary point $x \in dS$ can be extended to reach arbitrary values. However, 
given \emph{two} points $  x, y \in dS$, one may ask whether there exist geodesics 
passing through \emph{both} $x$ and $y$: 
\begin{itemize}
\item[$ i.)$] if $y$ is  time- or light-like to the antipode~$-x$ of $x$, then 
there is \emph{no} geodesic passing through both points $x$ and $y$.  
There exists a plane in $\mathbb{R}^{1+2}$ passing through $ x, y $ 
and~$0$, but $x$ and $y$ lie on disconnected hyperbolas which arise as 
the intersection of this plane with $dS$; 
\item[$ ii.)$] the  case $ y = -x$ is degenerated, as \emph{every} space-like 
geodesics passing through $x$ also passes through $- x$;
\item[$ iii.)$] in all the other cases, there exists\footnote{For a proof, 
we refer the reader to 
\cite[Bemerkung (4.3.14)]{Thirring}.} 
(in complete analogy with the great circles of a sphere) a \emph{unique} 
geodesic passing through $ x$ and~$ y $. It is a connected component of 
the intersection of~$dS$ with the plane in $\mathbb{R}^{1+2}$ passing 
through $ x, y $ and~$0$~\cite{oneil}. 
\end{itemize}

\begin{remark} If a \emph{time-like} curve is contained in the intersection 
of $dS$ with a plane in $\mathbb{R}^{1+2}$ \emph{not} passing through 
the origin, then it describes the trajectory of a \emph{uniformly accelerated} 
observer\index{uniformly accelerated observer}.
\end{remark}

\subsection{Geodesic distance}
\label{goedigst}
If two points are connected by a geodesic, the \emph{geodesic 
distance}\index{geodesic distance} can be defined:
\begin{itemize}
\item[$i.)$] if $x$ and $y$ are space-like to each other \emph{and}  
$ |  x \cdot  y  |  \le  r^2 $, a \emph{spatial distance}\index{spatial distance}  
	\begin{equation} 
		\label{dlength}
			d( x ,  y ) \doteq 
			r \arccos \left(-   \tfrac{x \cdot  y}{r^2} \right) 
	\end{equation} 
is defined as the length of the arc on the ellipse connecting $x$ and~$y$. 
Note\footnote{Recall that the principle values of the function $[-1,1] \ni z \to 
\arccos (z)$ are monotonically decreasing between $\arccos (-1) = \pi$ 
and $\arccos (1) = 0$.} that $d( x ,  x )=0$, iff $x \cdot  x = - r^2 $;

\goodbreak
\item[$ii.)$] if $y$ is  time- or light-like to 
the \index{antipode} \emph{antipode}~$-x$ (\emph{i.e.}, $x \cdot  y >   r^2 $), 
then $d(x,y)$ is not defined. In case $y \in \Gamma^+(-x)$, 
the light rays passing through $y$ will intersect the 
light rays passing through $x$ only in the past, while if $y \in \Gamma^-(-x)$, 
the light rays will intersect only in the future (see Remark~\ref{R-1.2.2}). Thus 
if two observers are located at $x$ and $y \in \Gamma^+(-x)$, there is no 
possibility that a third observer will receive an information emerging from 
those two in the future, but both may encountered information which emerged 
from a common source in the past;

\item[$iii.)$] if $x$ and $y$ are time-like to each other,
the \emph{proper time-difference}\index{proper time-difference} 
	\begin{equation} 
		\label{dlength2}
			d( x ,  y ) \doteq 
			r \; {\rm arcosh} \left(-   \tfrac{x \cdot  y}{r^2} \right) 
			= r \ln \left(-   \tfrac{x \cdot  y}{r^2} 
			+ \sqrt{ \tfrac{|x \cdot  y|^2}{r^4} -1} \, \right)   
	\end{equation} 
is defined as the length\footnote{In particular, if $x=(0, 0, r)$ and $y 
= \Lambda_1(t) x$, then $d(x,y) = rt$.}  of the arc on the hyperbola 
connecting $x$ and~$y$.
\end{itemize}

\section{The Lorentz group}
\label{subsec:1.3}

The two-dimensional de Sitter space $(dS,g)$ has three linearly
independent Killing vector fields 
which give rise to three one-parame\-ter sub\-groups 
leaving one of the coordinate axes in $\mathbb{R}^{1+2}$ invariant and generating 
the space-time symmetry group $SO_0(1,2)$:  the \emph{rotation} \index{rotations} 
subgroup $\{R_0(\alpha) \mid \alpha\in [0, 2 \pi)\}$, with
\label{rropage}
	\[
		R_0(\alpha) \; \doteq \; \begin{pmatrix}
					1 &  0 &0 \\
					0 &  \cos \alpha &  - \sin \alpha  \\
					0  & \sin \alpha & \cos \alpha   
				\end{pmatrix} \;,
	\]
and the two subgroups of
\emph{Lorentz boosts} \index{Lorentz boost} \index{boost} $\{\Lambda_1(t) 
\mid t\in\mathbb{R} \}$ and
$ \{\Lambda_2(s) \mid s\in\mathbb{R} \}$, with
\label{lambdapage}
	\[
		\Lambda_1 (t)   \; \doteq \;   \begin{pmatrix}
				 \cosh t  &  0 &\sinh t \\
     					  0  &  1 & 0  \\
 				  \sinh t &  0 & \cosh t   
				\end{pmatrix} 
	\; \mbox{ and } \;
		\Lambda_2(s) \doteq \begin{pmatrix}
				\cosh s &  \sinh s &0 \\
				\sinh s &  \cosh s &0  \\
					0 & 0 & 1  
				\end{pmatrix} \;.
	\]
We will also need the rotated boosts  
		\begin{equation} 
			\label{Lambdaalpha}
				\Lambda^{(\alpha)} (t)= 
				R_{0}(\alpha) \Lambda_{1}(t) R_{0}(-\alpha) \; , 
				\qquad t \in \mathbb{R} \; . 
		\end{equation} 
According to our convention, the boosts $\Lambda_1 (s)$ (respectively,  $\Lambda_2(t)$) keep 
the $x_1$-axis (respectively,  the $x_2$-axis) invariant, and therefore correspond  to  boosts
in the $x_2$-direction  (respectively, in the $x_1$-direction). The matrices $R_0(\alpha)$ 
are \emph{orthogonal},   
while the matrices $\Lambda_1 (t)$ and $\Lambda_2 (t)$ are \emph{symmetric}. 

\section{Hyperbolicity}
\label{sec:1.4}
\index{hyperbolicity}

The notions introduced in Section \ref{sec:1.2.3}
apply as well to subsets of lower dimension, 
\emph{e.g.},  line-segments  in $dS$. For example, one can easily compute the 
causal completion of an open interval $I\subset S^1$: let
	\[
		x(\psi_\mp) = ( 0, r \sin \psi_\mp, r \cos \psi_\mp) \; , \qquad \text{with} 
		\quad 0\le \psi_- < \psi_+ <\pi \; . 
	\]
The two intersecting (half-) light rays passing through $x(\psi_\mp)$ are 
	\begin{equation}
	\label{intersectinglightrays}
				\mathbb{R}^+ \ni \lambda \mapsto 
				R_0 ( \psi_\mp )
									\left[ 
				\begin{pmatrix}
					0\\
					0\\
					r
				\end{pmatrix}
	+ \lambda 	\begin{pmatrix} 
					 1 \\
					\mp 1\\
					0  
				\end{pmatrix} \right] = 
								\begin{pmatrix}
					\lambda \\
					- r
					\sin\psi_\mp \mp \lambda \cos \psi_\mp \\
					r
					\cos \psi_\mp \mp \lambda \sin \psi_\mp
				\end{pmatrix} \; . 
	\end{equation}
They intersect at a point in $dS$ given by inserting  
	\begin{equation}
		\label{height}
		\lambda =  	r \tan \bigl( \tfrac{\psi_+ - \psi_-}{2}\bigr)  . 
	\end{equation}
in \eqref{intersectinglightrays}. Now, any space-like geodesic can be identified 
with $S^1$ by applying a coordinate transformation. Therefore the causal 
completion of an open interval $I$ on an \emph{arbitrary} space-like geodesic  
is a bounded space-time region in $dS$, if the length $| I |$ (measured by 
inserting the endpoints of the interval $I$ in~\eqref{dlength}, see below) 
is less than $\pi \, r $. 

Half-circles\index{half-circle}, \emph{i.e.}, intervals $I \subset S^1$ with length
$\pi \, r $, will play a special role in the sequel. We denote by $I_+$ (respectively, 
by $I_-$) the open subset of $S^1$ with strictly 
positive (respectively, strictly negative) $x_1$ coordinate: 
\label{halfcirclepage}
	\begin{equation} 
		\label{halfC}
		I_+\doteq \big\{(0, r \sin \psi, r \cos \psi ) \in 
		\mathbb{R}^{1+2} \mid - \tfrac{\pi}{2} < \psi < \tfrac{\pi}{2} \big\}
	\end{equation} 
and $I_-\doteq \big\{(0, r \sin \psi, r \cos \psi ) \in 
		\mathbb{R}^{1+2} \mid  \tfrac{\pi}{2} < \psi < \tfrac{3\pi}{2}\big\}$.
The half-circle $ R_{0} (\alpha) I_+$ is denoted by $I_\alpha$.
Unless the radius $r$ of the time-zero circle plays a significant
role, we will suppress the dependence on $r$. 

\goodbreak

The support of Cauchy data that can
influence events at some point $x\equiv (x_0, x_1, x_2)\in dS$ with $x_0>0$ is given by the
intersection $\Gamma^-(x)\cap S^1$ of the past $\Gamma^-(x) $ of $x$ with
the Cauchy surface $S^1$. It will be of particular importance\footnote{In Section \ref{sec:fsol3}
we will  show that the ${\mathscr P}(\varphi)_2$ models
respect finite speed of propagation.} to describe the evolution of this
set as the point $x$ is subject to a Lorentz boost. We start by considering a special case.

\begin{lemma} 
\label{FSoL}
Consider a point $x (\psi) =(0, r \sin \psi, r \cos \psi ) \in I_+$. For $\tau>0$ the intersection 
	\[ 
		\Gamma^- (\Lambda_1 \left( \tau \right) x ) \cap S^1 
		= \bigl\{ x ( \psi'  ) \in S^1 \mid \psi_- 
		\le  \psi'  \le \psi_+ \bigr\}  \; , 
	\]
of the past $\Gamma^- (\Lambda_1 \left( \tau \right) x )$ of the point 
	\[
	 	\Lambda_1 \left( \tau \right) x  =   	\left( \begin{smallmatrix}
										 \cosh \tau &  0 &\sinh \tau \\
										0 &  1&0  \\
										\sinh \tau & 0 & \cosh \tau   
 									\end{smallmatrix} \right)
									\left( \begin{smallmatrix}
										0 \\
									 	r \sin \psi \\
									 	r \cos \psi 
									 \end{smallmatrix} \right)
	\]
with the time-zero circle $S^1$ is an interval of length 
	\begin{equation}
	\label{length}
		r \cdot | \psi_+ - \psi_- | = 2 r \arctan  (\sinh  \tau   \cos \psi)  
	\end{equation}
centred at 
	\[
		x \bigl(\tfrac{\psi_+  + \psi_- }{2} \bigr) 
		= \bigl( 0, r \sin \tfrac{\psi_+  + \psi_- }{2} , 
		r \cos \tfrac{\psi_+  + \psi_- }{2} \bigr) \; , 
	\]
where the angle $\frac{\psi_+  + \psi_- }{2} 
= \arcsin \frac{\sin \psi}{\sqrt{1+ (\sinh   \tau  \cos \psi)^2}} \, $. 
\end{lemma}

\goodbreak
\begin{proof} We compute:
	\[
		\Lambda_1 \left( \tau \right) x 
		= \left( \begin{smallmatrix} r \sinh  \tau  \cos \psi \\
									r \sin \psi \\
									r \cosh  \tau \cos \psi
						\end{smallmatrix} \right) \; . 
	\]
Eq.~\eqref{length} now follows 
directly from \eqref{height}, and the localisation of the interval follows from 
	\[
		\sin \left( \tfrac{\psi_+  + \psi_- }{2} \right) = \frac{\sin \psi}{\sqrt{\sin^2 \psi 
		+ \cosh^2 \tau \cos^2 \psi}} \; , 
			\qquad \psi_\pm \equiv \psi_\pm (\tau, \psi) \; , 
	\]
using $\sin^2 \psi = 1 - \cos^2 \psi$ and $\cosh^2 \tau= 1+ \sinh^2 \tau$.
\end{proof}

It is not difficult to extend this result to intervals lying on space-like geodesics:

\begin{proposition}
\label{ialpha}
Let $I $ be an arbitrary interval in $S^1$. 
Consider a boost
	\[
		\tau \mapsto \Lambda^{(\alpha)} \left(\tau\right) \; , 
		\qquad \alpha \in [0, 2 \pi ) \; ,  
	\]
as defined in \eqref{Lambdaalpha}. It follows that the set  
	\begin{align}
		\label{Ialphat}
		 I (\alpha , \tau)  \doteq S^1 \cap 
		 \Bigl( \bigcup_{y \in \Lambda^{(\alpha)}  \left(\tau\right) I }   
		\Gamma^- (y )\cup  \Gamma^+ (y ) \Bigr) \; , 
				\nonumber
	\end{align}
which describes the localisation region for the Cauchy data that 
can influence space-time points in the set $\Lambda^{(\alpha)} (\tau) I$, equals
	\[
		 I (\alpha , \tau) 
		 =  \bigcup_{(0, r \sin \psi, r \cos \psi ) \in I } 
		\bigl\{ x( \psi' ) \mid   \psi_- (\pm \tau, \psi+\alpha) 
		\le  \psi' + \alpha \le \psi_+ (\pm \tau, \psi+\alpha) \bigr\} .
	\]
As before, $x ( \psi ) =(0, r \sin \psi, r \cos \psi )$. 
Explicit formulas for the angles $\psi_\pm = \psi_\pm(\tau, \psi)$ are 
provided in Lemma \ref{FSoL}. 
\end{proposition}

\goodbreak
\begin{remarks}
\quad
\begin{itemize}
\item [$i.)$] The \emph{speed of propagation}\index{speed of 
propagation}\footnote{Note that speed refers to proper time and 
spatial geodesic distances as defined in \eqref{dlength}.}
	\[
		{\rm v}_\mp = r \; \frac{d \psi_\mp \bigl( \tfrac{\tau}{r}, 
		\psi + \alpha \bigr) }{d\tau} 
	\]
(to the left and to the right) along the circle $S^1$ goes to zero as $x$ approaches 
the \emph{fixed  points} $R_0 (\alpha)x$, with $x=(0,\pm r,0)$, for the boost 
$\tau \mapsto \Lambda^{(\alpha)} \left( \tau \right)$.
\item [$ii.)$] For $\tau$ small, the interval $I \left( \alpha , \tfrac{\tau}{r} \right)$ grows 
at most\footnote{This is the case for small $\tau$, if a small  
interval is centred at $R_0 (\alpha)x$, with $x=(0,\pm r,0)$.} with the speed
of light (on both sides), while for~$\tau$ large, the growth rate decreases to zero. In fact, 
for any interval $I \subset I_\alpha$, $I_\alpha = R_0 (\alpha) I_+$, we have
	\[
		I (\alpha , \tau) \subset I_\alpha
		\qquad \forall \tau \ge 0 \; . 
	\]
Recall that $ \bigcup_{\tau \in \mathbb{R}} \Lambda^{(\alpha)}  
(\tau) I_\alpha = I_\alpha''  $ and $I_\alpha'' \cap S^1= I_\alpha$.  
\item [$iii.)$] Let $I \subset I_\alpha $ be an open interval. It follows that 
	\[
		\lim_{\tau \to \infty} I (\alpha , \tau) = I_\alpha \; . 
	\]
In fact, for every point $x \in  I_\alpha'' $ one has 
	\[
		\lim_{\tau \to \pm \infty}  \Gamma^\mp  
		\bigr(\Lambda^{(\alpha)} ( \tau ) x\bigr) \cap S^1 = I_\alpha \; .
	\]
\end{itemize}
\end{remarks}

Another question one may ask concerns the spatial distance of two observers
which are both free falling, \emph{e.g.}, 
	\[
		\Lambda_1(t) o \quad \text{and} \quad \Lambda^{(\alpha)} (t) R_0 (\alpha) 
		o \; , \qquad o 
		= \left( \begin{smallmatrix} 0 \\ 0 \\ r \end{smallmatrix} \right)\; . 
	\]
As we have seen in Section \ref{goedigst}, the spatial distance is 
the length of the arc of the 
ellipse given by the intersection of $dS$ with the plane spanned by the 
vectors $\Lambda_1(t) o$ and $\Lambda^{(\alpha)} (t) R_0 (\alpha)$.

What one finds is that, even if initially the spatial distance $|\alpha r|$ is very small, 
the spatial distance between these points will increase rapidly,  so that eventually the 
two observers will no longer be able to make contact. This happens as the second 
observer exits the past of the trajectory of the first, \emph{i.e.}, the region
	\[
		\bigcup_{t \in \mathbb{R}} \Gamma^- \bigl(  \Lambda_1(t) o \bigr) \; . 
	\]
At this time, the future of the second observer no longer intersects the orbit of the first, 
so that signals (light signals or signals propagating with a velocity slower than the 
speed of light) send from the second observer can no longer reach the first observer. 
He simply disappears behind the \emph{horizon}, \emph{i.e.}, 
its orbit enters in the region of $dS$ which is time- or light-like to the 
antipode $-\Lambda_1(t) o$ of $\Lambda_1(t) o$; 
see Section~\ref{geodesics}.

\section{Causally complete regions}
\label{sec:1.5}

If $x \in dS$, the point $-x$, called the \emph{antipode}\index{antipode}, is in $dS$ as well. 
The light rays going through $x$ and $-x$ lie in the tangent planes at $x$ and $-x$, respectively. 
These tangent planes are parallel to each other. It follows that  a point $x \in dS$ determines 
four closed regions, namely $\Gamma^\pm (x)$ and $\Gamma^\pm (-x)$. Since  
	\[
		\Gamma^{+}(  x) \cap \Gamma^{-} ( x) = \{  x \} \; ,
		\qquad 
		\Gamma^{+}( - x) \cap \Gamma^{-} (- x) = \{ - x \} \; ,
	\]
their union consists of two disjoint, connected components. 
The complement of the union of these two sets consists of two open and disjoint sets, which 
we call \emph{wedges}. 

\subsection{Wedges}
The points $(0, \pm r, 0) \in dS$ are the \emph{edges}\index{edge of wedge} of the wedges
	\[ 
		W _1\doteq \bigl\{  x \in dS \mid x_2 > |x_0 | \bigr\} \quad \text{and} \quad
		W _1' \doteq \bigl\{  x \in dS \mid  - x_2 >  |x_0 | \bigr\}  \; .
	\]
Since the proper, orthochronous Lorentz group $SO_0(1,2)$ is transitive\footnote{In fact, 
the orbit $\{g x \mid g \in SO_0(1,2)\}$ of \emph{any} point $x \in dS$ is all of $dS$.} 
on the de Sitter space $dS$, an \emph{arbitrary} wedge~$W$ is of the form 
\label{wedgegeo}
\label{lambdawpage}
\label{tWpage}
	\[  
		W \doteq  \Lambda W _1 \; , \qquad \Lambda \in SO_0(1,2)  \; .
	\]
The wedges
\label{w-alpha-page} 
	\[
		W^{(\alpha)} \doteq R_0(\alpha)W_1 \; , \qquad \alpha \in [0 , 2\pi) \; , 
	\]
whose edges lie on the circle $S^1$,
will frequently appear in the sequel; we note that 
	\[
		W^{(\alpha)} = I_{(\alpha)}'' \; , \qquad \alpha \in [0 , 2\pi) \; .
	\]

\goodbreak

A one-to-one correspondence \cite[p.~1203]{GL} between points $x \in dS$ and wedges 
is established by requiring that 
\begin{itemize}
\item [---] $x$ is an edge of the wedge $W_x$; 
\item [---] for any point $y$ in the interior of $W_x$ the 
triple $\{ (1,0,0), x, y \}$ has positive orientation. 
\end{itemize}
For example, the origin $o$ lies in the wedge $ W_1 = W_{(0,  r  , 0)} $, 
but not in $W_1' = W_{(0,  -r  , 0)}$.  

\goodbreak
\begin{remarks}
\quad
\begin{itemize}
\item [$i.)$] Any point $x \in dS$ is of the form 
	\[	
		x =  \Lambda^{\alpha+\pi/2}(t) R_0 (\alpha) o 
	\]
for some $\alpha \in [0, 2\pi)$ and some $t \in \mathbb{R}$. Hence, 
any wedge $W_x$ is of the form
	\[	
		W_x =  \Lambda^{\alpha+\pi/2}(t) R_0 (\alpha) 
		W_o \; ,  \qquad W_o = R_0 \bigl( \tfrac{\pi}{2} \bigr) W_1 \; , 
	\]
for some $\alpha \in [0, 2\pi)$ and some $t \in \mathbb{R}$.  
\item [$ii.)$] 
Two wedges $W_x$ and $W_y$  have empty intersection, if the 
edge $y$ of the wedge $W_y$ lies in the future or the past of the 
antipode $-x$ of $x$, \emph{i.e.}, iff \cite[Lemma 5.1]{GL}
	\[
		y \in \Gamma^+(-x) \cup \Gamma^- (-x) \; . 
	\]
\item [$iii.)$] 
Given a wedge $W$, there is exactly one time-like geodesic  ${\mathscr G}$, 
which lies entirely within $W$. Indeed, the wedge $W$ itself 
is the causal completion of~${\mathscr G}$, \emph{i.e.}, 
	\begin{equation}
	\label{G-W}
		{\mathscr G}'' = W \; .
	\end{equation} 
For example, $W_1$ is the causal completion of the worldline 
$t \mapsto \Lambda_1(t)  o$. 
\item [$iv.)$] 
To each freely falling observer on $dS$ one can associate 
a (unique) Killing vector field, generated by the one-parameter 
group of Lorentz boosts $t \mapsto \Lambda_W(t)$ (with $W$ 
satisfying \eqref{G-W}), which describes the 
time evolution of this specific observer. No Killing vector fields 
is \emph{globally} time-like, but the vector field given by $t \mapsto \Lambda_W(t)$ 
\emph{is} time-like in the wedge~$W$. Despite many claims in the literature 
that the boosts $t \mapsto \Lambda_W(t)$ 
provide a `time-evolution' in the whole wedge $W$, we do not subscribe to this 
choice of words. Even though an observer may 
\emph{enforce}\footnote{Material points which are (initially) at a spatial distances 
strictly less then $r \pi/ 2$
from the observer can be kept (with constant force) at constant distance in the 
observers laboratory~\cite{BMS02}. However, 
the force necessary to keep this spatial distance diverges as the (initial) spatial 
distance from the observer approaches $r \pi/ 2$.}
$t \mapsto \Lambda_W(t)$ as a `time-evolution' in the open wedge, 
the free falling motion of a point in the neighbourhood of ${\mathscr G}$
is not described by $t \mapsto \Lambda_W(t)$, $t \in \mathbb{R}$. 
\item[$v.)$] 
The union of $ \Gamma^{+}( W )$ with $ \Gamma^{-}( W ')$ 
covers the de Sitter space $dS$; the intersection of $\Gamma^{+}( W )$ 
and $ \Gamma^{-}( W')$ are two light-rays. 
\item[$vi.)$] The space-like complement $W'$ of a wedge $W$ is a wedge, called
the \emph{opposite wedge}\index{opposite wedge}. The \emph{double 
wedge}\index{double wedge}
	\label{dWpage}
	\begin{equation} 
			\label{double-wedge}
			\mathbb{W} \doteq W \cup W'  
	\end{equation} 
is uniquely specified by fixing  (one of) its edges (the other one is just the antipode). 
\end{itemize}
\end{remarks}

\goodbreak 
\subsection{Double cones}
Open, bounded, connected, causally complete space-time regions in $dS$ are 
called \emph{double cones}\index{double cone}. Such regions play an important role in 
local quantum physics; thus we provide various characterisations. 

\begin{proposition} Let ${\mathcal O}$ be a double cone. Then there exist
\begin{itemize}
\item [$i.)$] two\footnote{Note that \emph{every} bounded non-empty region ${\mathcal O}$
given as the intersection of wedges, is an intersection of \emph{two} (canonically determined)
wedges \cite[Lemma~5.2]{GL}.} wedges such that ${\mathcal O}$ is equal to their intersection; 
\item [$ii.)$] a time-like geodesic ${\mathscr G}$ and an open bounded 
interval  $J \subset {\mathscr G}$ such that the causal completion $J''$ 
(which lies entirely within the wedge~${\mathscr G}''$) equals ${\mathcal O}$;
\item [$iii.)$] two points\footnote{Both $x$ and $y$ can be identified as boundary 
points of the segment $J$ appearing in $ii.)$} 
$x, y \in dS$ such that ${\mathcal O}$ is the interior of the intersection of the future 
of $x$ and the past of~$y \in \Gamma^+ (x)$;
\item [$iv.)$] an interval  $I$ of length $| I | < \pi \, r  $ on a space-like geodesic such 
that  the causal completion $I''$
equals ${\mathcal O}$.
\end{itemize}
\end{proposition} 

For double cones with \emph{base}\index{base of double cone} $I$ on 
$S^1$, we introduce the following notation:
\label{doubleconepage}
	\begin{equation*} 
		\mathcal{O}_{ I } \doteq I'' \subset dS \; , \qquad | I | < \pi \, r \; , 
		\quad I \in S^1 \; . 	
	\end{equation*} 
Note that any double cone is of the form 
	\begin{equation*} 
		\Lambda \mathcal{O}_{ I }   \; , \qquad \Lambda \in SO_0(1,2) \; , 
		\quad I \subset S^1 \; , 
		\quad | I | < \pi \, r  \; . 	
	\end{equation*} 
As $| I | \to \pi \,  r $, the light rays in \eqref{intersectinglightrays}
become parallel, and $I''$ itself becomes a wedge~$W$.

\begin{remarks}
\quad 
\begin{itemize}
\item [$i.)$] Wedges and double cones are  causally
complete. 
\item [$ii.)$] Wedges are also \emph{geodesically closed}\index{geodesically 
closed}, in the sense that if $x, y \in W$, then there is an 
interval $I$ on some geodesic connecting these two points, which lies entirely 
in~$W$. In fact, the causal completion~$I''$ of $I$ automatically lies in $W$ 
as well. A similar statement holds for double cones. 
\item [$iii.)$] The converse holds as well: 
if two points $x, y \in dS$ are connected by a geodesic, 
then there is a wedge whose closure contains both of them. 
\end{itemize}
\end{remarks}

\subsection{Boosts associated to wedges}
The boosts $t \mapsto \Lambda_1(t)$ leave the wedge~$W_1$ invariant. 
For an arbitrary  wedge $W = \Lambda W_1$, $\Lambda \in SO_0(1,2)$, 
	\begin{equation}
	\label{L-W-t}
	\Lambda_{\scriptscriptstyle W}(t) \doteq 
	\Lambda \Lambda_1(t) \Lambda^{-1} \; ,  \qquad t \in \mathbb{R} \; , 
	\end{equation}
defines a future directed boost leaving $W$ invariant, \emph{i.e.},  
	\[ 
		\Lambda_{\scriptscriptstyle W}(t) W =W \; , \qquad t \in \mathbb{R} \; .  
	\] 
In particular,  $\Lambda_1(t) = \Lambda_{W_1} (t)$ for all $t \in \mathbb{R}$. 

\goodbreak
The Killing vector field\footnote{As mentioned before, 
Killing fields are the infinitesimal generators of isometries; that is, flows generated 
by Killing fields are continuous isometries of the manifold.} induced by~$\Lambda_{ W } (t)$ 
leaves the opposite wedge~$W'$ invariant  too. It is, however, past directed in~$W'$. 
One may fix the scaling factor  by normalising the Killing vector field  
on the time-like geodesic~${\mathscr G}$ satisfying~${\mathscr G}'' = W$. 
Uniqueness then implies 
	\[
		\Lambda_{ W } (t) = \Lambda_{ W' } (-t) \; , \qquad t \in \mathbb{R} \; .  
	\]
The double-wedge $\mathbb{W}$ introduced in \eqref{double-wedge} is  invariant 
under  both $\Lambda_{ W } (t)$ and $\Lambda_{ W' } (t)$, $t \in \mathbb{R}$. 

Another interesting property of the boosts $t \mapsto \Lambda_1(t)$ is that they 
leave the points $(0, \pm r, 0)$ invariant. In fact, they form the {\em stabilizer} --- within 
the group $SO_0(1,2)$ --- of the point $(0,r, 0) \in dS$ (and, at the same time, 
the antipode $-(0,  r, 0)$). Similarly, the origin $ o \,$ and its antipode $-o\, $ are 
invariant under the boosts $\Lambda_{2}(s)$, $s \in \mathbb{R}$. More generally, 
the group 
	\[ 
		t \mapsto \Lambda_{ W_x } (t) \; , \qquad t \in \mathbb{R} \; , 
	\]
is the unique --- up to rescaling --- one-parameter subgroup of $SO_0(1,2)$,  
which leaves the edges $\pm x$ of the wedge $W_x$ 
invariant and induces a future directed Killing vector field in the wedge $W_x$. 
Clearly, it also leaves  the light rays passing through the edges $\pm x$ invariant. 

\begin{remark}
A free falling observer passing through the origin $o$ interprets the  
boost $\upsilon \mapsto \Lambda_{2} ( \upsilon) =\exp  (  \upsilon  \Lgeo_2 )$  
as a Lorentz transformation, the boosts (re-scaled to proper time) 
	\[ 
		\tau \mapsto \Lambda_{1} \left( \tfrac{\tau}{r} \right) 
		= {\rm e}^{  \frac{\tau}{r} \, \Lgeo_1 }
	\]
as his geodesic time evolution and the 
rotation~$a \mapsto R_{0} \left( \tfrac{a}{r} \right) 
= \exp \left( \tfrac{a}{r} \, \Kgeo_0 \right)$, $a \in [0, 2 \pi \, r)$,  
as a spatial translation\footnote{Alternatively, one may also view 
a motion along a horosphere, given by the 
map $q \mapsto D(q/r)$ as a translation; see \eqref{1.5.1} below.}. 
Unless $x =o$, the path 
	\[
		\tau \mapsto \Lambda_1 \left( \tfrac{\tau}{r} \right)  x \; , 
		\qquad x \in I_+ \; , 
	\]
describes a uniformly accelerated observer. Note that such a path 
lies on the intersection of $dS$ with a plane parallel to the $(x_2= 0)$-plane, 
passing through~$x$. 
\end{remark}

\subsection{Analytic continuation of boosts}
In the complex  de Sitter group 
$O_{\mathbb{C}}  (1,2)$ (see Section~\ref{sec:2.8.4} for further details) 
the reflections 
		\[
		P_1 T \doteq \begin{pmatrix}
								-1 &  0 &0 \\
								0 &  1 & 0  \\
								0  & 0 & -1   
 						\end{pmatrix},  \; 
		P_2 T \doteq \begin{pmatrix}
								-1 &  0 &0 \\
								0 & -1 & 0  \\
								0  & 0 & 1   
						\end{pmatrix},  \; 
							P\doteq  \begin{pmatrix}
								1 &  0 &0 \\
								0 &  -1 & 0  \\
								0  & 0 & -1   
						\end{pmatrix}  , 
		\]
are topologically connected to the identity $\mathbb{1}$.  
In fact, the matrix-valued 
function $t\mapsto \Lambda_{1}(t)$  extends to an entire analytic function\footnote{Note that, 
for $x \in dS$ and  $0 <  \theta < \pi$,  
we have $ \Lambda_{1}(t+i\theta) x = x'+ i x'' \in \mathcal{T}_+$.}
		\begin{equation} 
		\label{eqBooW} 
			\Lambda_{1}(t+i\theta) 
				= \Lambda_{1}(t) 
					\left[ 
					\underbrace{\begin{pmatrix}
								\cos \theta &  0&0  \\
								0&  1 &0  \\
								0 & 0 & \cos \theta  
 						\end{pmatrix}}_{  \doteq \Jgeo_1(\theta)}
					+ i  \sin \theta \underbrace{\begin{pmatrix}
								0 &  0&1  \\
								0&  0 &0  \\
								1 & 0 & 0  
							\end{pmatrix}}_{ \doteq \Lgeo_1} \right] \, ;
		\end{equation} 
see also \eqref{elle-1}.  
The first matrix in the square brackets 
continuously deforms the unit $\mathbb{1}$ to 
$P_1 T$, as~$\theta$ takes values starting at~$\theta=0$ and ending at $\theta =\pm \pi$. The 
second matrix in the square brackets projects the wedge~$W_1$ 
continuously into the $x_1=0$ section of the forward light cone, \emph{cf.}~\cite{H}. 

\subsection{Coordinates for the wedge $W_1$}
The chart 
	\begin{equation} 
	 	\label{w1psi}
		 x (t, \psi) \doteq \Lambda_{1} \left( t \right) \, R_0 \left( - \psi \right) 
			\begin{pmatrix}
					0 \\
					0  \\
					r  
			\end{pmatrix} \; , \qquad \, t \in \mathbb{R} \; , 
		\quad  {\textstyle  -\frac{\pi}{2}< \psi < \frac{ \pi}{2}}   \; ,
	\end{equation} 
provides coordinates for the wedge $W _1$.  
Allowing $\psi \in \;  [-  \tfrac{\pi}{2} ,  \tfrac{3\pi}{2} )$,  these coordinates 
extend\footnote{In the sequel, we will 
always take care of the fact that these coordinates are degenerated at the 
fixed-points $(0, r, 0), (0, -r, 0) \in dS$ for the boosts $t \mapsto \Lambda_1(t)$, 
$t \in \mathbb{R}$.} to the space-time region   
	\begin{equation} 
		\label{3.32}
		\mathbb{W}_1  \cup \{ (0, r, 0), (0, -r, 0)\} 
		= \bigcup_{ t \in \mathbb{R} } \Lambda_{1}(t) S^1  \; . 
	\end{equation} 
The r.h.s.~is the union of the boosted time-zero 
circles $\Lambda_{1}(t) S^1$, $t \in \mathbb{R}$.  

The restriction of the metric~$g$ to $\mathbb{W}_1$ is 
	\[ 
		g_{\upharpoonright {\mathbb{W}}_1} 
		=    r^2 \cos^2 \psi \, {\rm d} t \otimes {\rm d} t -  r^2 \, 
		{\rm d} \psi \otimes {\rm d} \psi \;   . 
	\]
The restriction of the \index{Lorentz invariant  measure}
Lorentz invariant  measure ${\rm d} \mu_{dS} $ to $\mathbb{W}_1$ is
\label{dlpage}
	\begin{equation}
		\label{new-surface} 
		{\rm d} \mu_{\mathbb{W}_1 } (t,\psi) = r {\rm d} t\, {\rm d} l (\psi) \;  , 
		\qquad \text{with} \quad 
		{\rm d} l (\psi) =  |\cos \psi | \; r {\rm d} \psi  \; . 
	\end{equation}
The line element \index{line element} on the circle $S^1$ is
\label{seinspage}
\begin{equation}\label{new-eqVolInd}
 | g_{\upharpoonright S^1} |^{1/2} \, {\rm d}\psi  =r {\rm d}\psi \; .  
\end{equation}
Let us recall from \eqref{varepsilon} that, restricted to the \index{double wedge} 
double wedge $\mathbb{W}_1 \, $, the Klein--Gordon operator takes the form 
\label{epsilonpage}
	\begin{equation}
		\label{varepsilon}
			\square_{\mathbb{W}_1}+\mu^2
				=  \frac{1}{r^2 \cos^2 \psi}\,(\partial_t^2+  \varepsilon^2) \; , 
	\end{equation}
with 
	\[
		\varepsilon^2  \doteq  - (\cos \psi  \, \partial_\psi)^2 + (\cos \psi )^2 \, \mu^2 r^2 \; . 
	\]

\begin{remark}
For $\psi \in ( -\pi/2,\pi/2 )$ define a new spatial coordinate $\chi=\chi(\psi)$  by
	\begin{equation*}
		\label{eqphichi}
		\frac{{\rm d}\chi}{{\rm d}\psi} = \frac{1}{\cos \psi} \; ,
		\qquad \chi(0)=0 \; . 
\end{equation*}
Find
	$
		\chi(\psi)= \ln \tan (\psi/2+\pi/4) 
	$
and
	\[
			\cos\psi=(\cosh\chi)^{-1},\quad \sin\psi=\tanh\chi \; .
	\]
$\chi$ is a diffeomorphism \index{diffeomorphism} from $(-\pi/2,\pi/2) $ 
onto $\mathbb{R}$, and 
	\[
		g_{\upharpoonright W_1}
		=   \tfrac{r^2}{\cosh^2 \chi} (  {\rm d} t \otimes {\rm d} t - 
		{\rm d} \chi \otimes {\rm d} \chi) \; .
	\]
Thus $W_1$ is conformally equivalent \index{conformal equivalence} to Minkowski 
space $\mathbb{R}^{1+1}$ \cite{HaEl}. In these coordinates 
	\[
			\square_{\mathbb{W}_1}+\mu^2
				=  \frac{\cosh^2 \chi}{r^2} \,(\partial_t^2+  \epsilon^2) \; , 
	\]
with $\epsilon^2  \doteq  - \partial_\chi^2 + (\cosh \chi )^{-2} \, \mu^2 r^2 $. 
\end{remark}

\begin{lemma}
\label{Asadj}
Identify $S^1 \cong [-\frac{\pi}{2} ,\frac{3\pi}{2}) $,  $I_+  \cong (- \frac{\pi}{2}, \frac{\pi}{2} )$ and 
$I_- \cong (\frac{\pi}{2} ,\frac{3\pi}{2})$. It follows that $\varepsilon^2$ is positive and symmetric on  
	\[
		{\textstyle{\mathcal D} \left(S^1 \setminus \{ - \frac{\pi}{2}, \frac{\pi}{2}\} \right) }
		\subset L^2(S^1, \, | \cos \psi |^{-1}  \, r {\rm d} \psi) \; . 
	\]
Denote its Friedrich extension \index{Friedrich extension} by the same symbol.  Then  ${\rm Sp}
(\varepsilon^2) = [ \, 0, \infty ) $. 
\end{lemma}

\begin{proof} 
Clearly, $\varepsilon^2$ is positive and symmetric on 
	\[
		{\mathcal D} \left(S^1 
		\setminus \{ - \tfrac{\pi}{2}, \tfrac{\pi}{2}\} \right) 
		={\mathcal D} (I_+) \oplus{\mathcal D}(I_-) \; . 
	\]
We next show that ${\mathcal D} \left(S^1 \setminus \{ - \frac{\pi}{2}, \frac{\pi}{2}\} \right)$ is 
dense in $L^2(S^1, \, | \cos \psi |^{-1}  \, r {\rm d}\psi)$.
First note that ${\mathcal D} \left(S^1 \setminus \{ - \frac{\pi}{2}, \frac{\pi}{2}\} \right)$ 
is dense in~$L^2(S^1,  \, r {\rm d}\psi)$. Moreover, a function 
\label{cosinepsipage}
	\begin{equation}
		\label{4.9}
		 \,  \mathbb{cos}_\psi^{1/2}h \in L^2(S^1, \, | \cos \psi |^{-1}  \, 
		 r {\rm d}\psi) \qquad \text{iff}  \quad 
		 h \in L^2(S^1, \,  r {\rm d}\psi) \; . 
	\end{equation}
Here $\mathbb{cos}_\psi$ denotes the multiplication 
operator \index{multiplication operator}
	$
		(\mathbb{cos}_\psi h)(\psi) \doteq \cos \psi  \, h(\psi)
	$
acting on functions of $\psi$. It follows that
	\[
	 	\textstyle { \,  \mathbb{cos}_\psi^{1/2}{\mathcal D} 
	 	\left(S^1 \setminus \{ - \frac{\pi}{2}, \frac{\pi}{2}\} \right) =
		{\mathcal D} \left(S^1 \setminus \{ - \frac{\pi}{2}, \frac{\pi}{2}\} \right) }
	\]
is dense in $L^2(S^1, \, | \cos \psi |^{-1}  \, r {\rm d}\psi)$. 
Thus \cite[Theorem X.23, p.177]{RS} applies and defines 
a positive self-adjoint operator, 
known as the Friedrichs extension. 
\end{proof}

\begin{remark}
Since the spectrum ${\rm Sp} (\varepsilon^2) =  [0, \infty) $ of $\varepsilon^2$ has no gap at zero, 
the choice of coordinates \eqref{w1psi} may  lead to artificial infrared problems \index{infrared 
problems} if one adds an interaction, similar to the ones encountered in~\cite{FHN}. 
We will avoid this problem later on by working with functions in the Hilbert 
space $\widehat{{\mathfrak h}} (S^1)$, whose scalar product is rotation-invariant;
see Section~\ref{Sect: canon-HS}.
\end{remark}

$\varepsilon^2 $ is a differential operator, thus $\varepsilon^2$ acts locally 
and maps the subspaces 
	\[
 		{\mathscr D}^\pm \doteq{\mathscr D} (\varepsilon^2) 
		\cap L^2 \bigl(I_\pm,|\cos\psi|^{-1} r {\rm d} \psi \bigr)
	\] 
into $L^2 \left(I_\pm,|\cos\psi|^{-1} r {\rm d} \psi \right)$, respectively. 
It therefore is consistent to define 
	\begin{equation}
		\label{vaepsdef}
		\varepsilon (h_+ + h_-) 
			\doteq \sqrt{{\varepsilon^2}_{\upharpoonright 	 {\mathscr D}^+}} \; h_+   
			- \sqrt{{\varepsilon^2}_{\upharpoonright 	{\mathscr D}^-}} \; h_-\; , 
				\qquad h_\pm\in {\mathscr D}^\pm\;  . 
	\end{equation}
$\varepsilon$ is densely defined by \eqref{vaepsdef}, 
as ${\mathscr D}^+ \oplus {\mathscr D}^- 
={\mathscr D}(\varepsilon^2)$. The pseudo-differential 
operator~$\varepsilon$ is non-local, but  does not mix 
functions supported on the half-circles $I_+$ and~$I_-$. Denote the restrictions by 
$\varepsilon_{\upharpoonright I_+}$ and $\varepsilon_{\upharpoonright I_-}$.

\begin{lemma}
\label{Lm3.5}
There exits a self-adjoint  operator $\varepsilon$ on $L^2(S^1, 
\, | \cos \psi |^{-1}  \, r {\rm d}\psi)$  
such that~\eqref{vaepsdef} holds.  ${\rm Sp} (\varepsilon) = \mathbb{R} $, 
${\rm Sp} (\varepsilon_{\upharpoonright I_+}) =  [0, \infty) $ and 
${\rm Sp} (\varepsilon_{\upharpoonright I_-}) =  (-\infty, 0] $. 
Moreover, zero is not an eigenvalue of $\varepsilon$.
\end{lemma}

\begin{proof} Use the spectral theorem to define the square roots in 
\eqref{vaepsdef} as self-adjoint operators. One has
${\mathscr D}^+ \cap {\mathscr D}^- =\{0\}$, in fact ${\mathscr D}^+
$ and ${\mathscr D}^-$ are orthogonal to each other in  $L^2(S^1, 
\, | \cos \psi |^{-1}  \, r {\rm d}\psi)$.  
Thus  the sum of the square roots is self-adjoint on the direct sum of their domains 
(see \cite[Theorem VIII.6]{RS}) and ${\rm Sp} (\varepsilon) = \mathbb{R} $.
\end{proof}

\section{Complexified de Sitter space}
\label{tuboidsds}

Consider  the complex de Sitter space
\label{dSccpage}
	\[  
		dS_{\mathbb{C}} \doteq \bigl\{  z \in \mathbb{C}^{1+2} 
		\mid z_0^2 - z_1^2 - z_{2}^2 = - r^2 \bigr\} \;  .
	\] 
A \emph{tuboid}\index{tuboid} for the de Sitter space is a subset 
of $dS_{\mathbb{C}}$ which  is $i.)$~bordered by  real de Sitter 
space $dS$ (and allows boundary values on $dS$ of functions 
holomorphic in the tuboid to be controlled by methods of complex analysis) 
and $ii.)$ whose {\em shape} (called its {\em profile})
near each point $x$ of $dS$ can be mapped to
a  cone ${\mathcal P}_{ x}$ in the tangent space $T_{ x }dS$. 
The exact definition needs some preparation, so we 
proceed in several steps, following closely \cite{BM}. 
The first step is to provide a precise definition of a profile.

\goodbreak
\begin{definition}
A \emph{profile}\index{profile of a tuboid}~${\mathcal P}$ is an open 
subset of the tangent bundle $TdS$ of the form 
	\[
		{\mathcal P} = \bigcup_{ x \in dS} (  x , {\mathcal P}_{ x}) \; , 
	\]
where each fibre\index{fibre} ${\mathcal P}_{ x}$ is a non-empty cone 
with apex at the origin in $T_{ x}dS$.
\end{definition}

Next we describe a class of 
local maps from  $TdS$ to $dS_{\mathbb{C}}$
\cite[Definition~A.1] {BM}:

\goodbreak
\begin{definition}
Let ${\mathcal N}_{\scriptscriptstyle TdS} ( x_\circ,  0 ) \subset TdS$ be 
a neighbourhood of $(  x_\circ ,  0 )$. 
A diffeomorphism $\Xi \colon {\mathcal N}_{\scriptscriptstyle TdS} ( x_\circ,  0 )  
\to dS_{\mathbb{C}}$ 
is called an \emph{admissible local diffeomorphism}\index{admissible 
local diffeomorphism} at a point $ x_\circ \in dS$, if 
\begin{itemize}
\item[$i.)$] the image of ${\mathcal N}_{\scriptscriptstyle TdS} ( x_\circ,  0 )$,  
	\[
		{\mathcal N}_{\mathbb{C}}( x_\circ)
		\doteq \Xi \left({\mathcal N}_{\scriptscriptstyle TdS} ( x_\circ,  0 )\right) \; , 
	\]
is a neighbourhood of $x_\circ $ in $dS_{\mathbb{C}}$, 
considered as a 4-dimensional $C^\infty$-manifold; 
\item[$ii.)$] locally it acts as the identity map from the base $dS$ of 
the tangent bundle $TdS$ to the (real) de Sitter space $dS$
considered as a submanifold of $dS_{\mathbb{C}}$, 
\emph{i.e.}, for $( x,  0) 
\in {\mathcal N}_{\scriptscriptstyle TdS} ( x_\circ,  0 )$
	\[
		\Xi ( x,  0 ) =  x \; , 	
	\]
with $x \in {\mathcal N}_{\mathbb{C}}( x_\circ) $; 
\item[$iii.)$] for all $( x ,  y) \in {\mathcal N}_{\scriptscriptstyle TdS} ( x_\circ, 0 )  $ 
with $ y \ne  0$ the differentiable function 
	\[
		t \mapsto f(t) \doteq \Xi ( x,  t  y ) \in dS_\mathbb{C}
	\]
is such that 
	$
	\tfrac{1}{i} \left(\tfrac{{\rm d}f}{{\rm d}t} \right)_{\upharpoonright t= 0} = \alpha   y $ 
	for some $\alpha > 0 $. 
\end{itemize}
\end{definition}

The causal structure in the tangent bundle 
can most conveniently be rephrased using a \emph{projective 
representation} of the tangent bundle:

\begin{definition} Let $\dot T_{ x} dS$ denote the directions of 
vectors in $T_{ x} dS$, \emph{i.e.}, 
	\[
		\dot T_{ x} dS \doteq   \bigl( T_{ x} dS\setminus \{  0 \} \bigr) / \mathbb{R}^+ \; . 
	\] 
Similarly, let $\dot {\mathcal P}_{ x} \doteq   \bigl( {{\mathcal P}}_{ x} 
\setminus \{  0 \} \bigr) / \mathbb{R}^+ $.
The image of each point $ y \in T_{ x} dS$ in ${\dot T}_{ x} dS$ is 
$\dot { y}= \{ \lambda  y \mid \lambda >0 \}$.
Then the projective tangent bundle $\dot T dS$ is
	\[
		\dot T dS \doteq \bigcup_{ x \in dS} \left(  x, \dot T_{ x} dS \right) \;. 
	\]
Similarly, let $\dot {\mathcal P} \doteq   \bigcup_{ x \in dS} ( x, \dot {\mathcal P}_{ x} ) $. 
The complement $\dot {\mathcal P}'$ of $\dot {\mathcal P} $ in $\dot T dS$ is the open set 
	\[
		\dot {\mathcal P}^c \doteq \dot T dS \setminus \overline{\dot {\mathcal P}}  \;  .
	\]
This set equals $\bigcup_{ x \in dS} ( x, \dot {\mathcal P}_{ x}^c ) $, where 
$\dot {\mathcal P}_{ x}^c = \dot T_{ x} dS \setminus \overline{\dot {\mathcal P}_{ x}} $. 
\end{definition}

Taking advantage of these notions, one can now define tuboids for the de Sitter space
\cite[Bros and Moschella]{BM}:

\begin{definition}
\label{def-1.6.4}
A connected open subset (\emph{i.e.}, a domain) ${\mathcal T} \subset dS_{\mathbb{C}}$ 
is called a {\em tuboid} with profile ${\mathcal P}$ above $dS$ if, for 
every point $ x_\circ \in dS$, there exists an admissible local diffeomorphism $\Xi$ at 
$ x_\circ$, which \emph{respects the causal structure}, \emph{i.e.},
\begin{itemize}
\item[$i.)$] for every point $( x_\circ, \dot { y}_1) \in \dot {\mathcal P}$ there exists a compact 
neighbourhood 
	\[
		{\mathcal K}( x_\circ, \dot { y}_1) \subset \dot {\mathcal P}
	\]
and, in the sequel, a sufficiently small neighbourhood ${\mathcal N}_{\scriptscriptstyle TdS} 
( x_\circ,  0 ) \subset TdS$ of~$( x_\circ,  0 )$ such that 
	\[
		\Xi \left(   \bigl\{ ( x,   y) \in {\mathcal N}_{\scriptscriptstyle TdS} ( x_\circ,  0) 
		\mid ( x, \dot  y) \in {\mathcal K}( x_\circ, \dot  {y}_1)   \, \bigr\} \right) 
		\subset {\mathcal T} \; ;
	\]
\item[$ii.)$] for every point $( x_\circ, \dot { y}_2) \in {\dot {\mathcal P}}^c$ there exists 
a \emph{compact} neighbourhood 
	\[
		{\mathcal K}^{c}( x_\circ, \dot { y}_2) \subset \dot {{\mathcal P}}^c
	\]
and, in the sequel, a sufficiently small neighbourhood ${\mathcal N}_{\scriptscriptstyle TdS}^c 
( x_\circ,  0 ) \subset TdS$ of~$( x_\circ,  0 )$ such that 
	\[
		\Xi \left( \bigl\{ ( x, { y}) \in {\mathcal N}_{\scriptscriptstyle TdS}^{c} ( x_\circ,  0 )
		\mid ( x, \dot { y}) \in {\mathcal K}^{c}( x_\circ, \dot { y}_2) \bigr\} \right) 
		\cap {\mathcal T} = \emptyset \; .
	\]
\end{itemize}
Note that in $i.)$ and $ii.)$ the neighbourhoods ${\mathcal N}_{\scriptscriptstyle TdS}  ( x_\circ,  0 )$ 
and  ${\mathcal N}_{\scriptscriptstyle TdS}^{c} ( x_\circ,  0 )$  may depend on $\dot { y}_1$ and 
$\dot { y}_2$, respectively.
\end{definition}

\goodbreak
Complex de Sitter space $dS_{\mathbb{C}}$ is equipped \cite{BM2} with four 
distinguished tuboids, which are invariant under the action of $SO_0(1,2)$:
	\begin{align} 
		{\mathcal T}_\pm &\doteq  \bigl\{  \Lambda \; ( i r \sin \theta , 0, r \cos \theta) 
		\in dS_{\mathbb{C}} \mid  0 <  \mp \theta < \pi ,  \;  \Lambda \in SO_0 (1,2) \bigr\} \; , 
		 \nonumber \\  
				{\mathcal T}_{\hskip -.1cm {\quad \atop \rightarrow} \atop  
				\hskip -.1cm \leftarrow} \hskip -.2cm
		&\doteq \bigl\{ \Lambda  ( 0, i r \sinh t , r \cosh t) 
		\in dS_{\mathbb{C}} \mid  \mp t >0 ,  \;  
		\Lambda \in SO_0 (1,2) \bigr\} \; . 
		\label{chiral-tuboids}  
	\end{align}
The {\em chiral} tuboids ${\mathcal T}_{\leftarrow}$ and~${\mathcal T}_{\rightarrow}$ are not 
simply-connected. Their profiles at the origin~$ o = (0,0, r)$ of $dS$ are  the cones
	\[
		   T_{ o} dS \cap \bigl\{  y \in \mathbb{R}^{1+2} \mid  \mp y_1 > |y_0|   \bigr\} \; , 
		   \qquad  T_{ o} dS = \bigl\{ \mathbb{R}^1+2 \mid x_2 = 0 \bigr\} \; .
	\]
The chiral tuboids play a key role for quantum fields on anti-de Sitter space \cite{BuSu, AdS}, 
but are of no relevance for this work. 
\label{tuboidspage}

\goodbreak
The {\em Lorentzian  tuboids}  ${\mathcal T}_\pm$ are  similar in many respects to the 
tubes\footnote{Of course, $V^+$ here denotes the future light cone in $\mathbb{R}^{1+2}$. 
Note that our sign convention follows~\cite{SW}, 
in contrast to the less common sign convention chosen in~\cite{BM2}.} 
	\begin{equation*}
		{\mathfrak T}_\pm \doteq \mathbb{R}^{1+2} \mp i V^+ 
	\end{equation*}
defined in complex Minkowski space. In fact~\cite[Proposition 2]{BM2}, 
	\begin{equation*} 
 		{\mathcal T}_\pm = {\mathfrak T}_\pm \cap dS_{\mathbb{C}} \; .  
	\end{equation*} 

\begin{proposition}[Proposition 1, \cite{BM2}]
The tuboids ${\mathcal T}_\pm$ consists of all points $z \in dS_{\mathbb{C}}$ for which 
the inequality $\mp \Im z \cdot p >0$ holds for all $p \in \partial V^+$, \emph{i.e.}, 
	\begin{equation} 
		\label{tube-0}
 		{\mathcal T}_\pm = \bigl\{ z \in dS_{\mathbb{C}} 
		\mid  \mp \Im z \cdot p >0 \; \forall p \in \partial V^+ 
		\setminus \{(0,0,0)\}
		 \bigr\} \; .  
	\end{equation} 
\end{proposition}

\begin{proof}
Consider a point $x+ i y  \in dS_{\mathbb{C}}$, with $x, y \in 
\mathbb{R}^{1+2}$. This point is in the tuboid ${\mathcal T}_\pm$ if 
$y \in \mp i V^+$. Thus we only have to show that $y \in \mp i V^+$ is equivalent 
to $\mp y \cdot p >0 \; \forall p \in \partial V^+$. In order to do so, 
consider the vectors $ p_\circ =(1,0,-1)$ and $q= (0,r,0)$. Since $ p_\circ  
\cdot p_\circ  =0$ and $p_\circ \cdot q = 0$, we find 
	\begin{equation}
	\label{zero-plane}
	( \lambda  p_\circ  + \mu q) \cdot 
	p_\circ  = 0   \qquad \forall \lambda , \mu \in \mathbb{R} \; . 
	\end{equation}
The plane\footnote{This plane contains the light rays
$(0, \pm r,0) + \lambda (1, 0, -1)$, $\lambda \in \mathbb{R}$, forming 
the horosphere $P_{- \infty }$, see \eqref{horosphere}.}
spanned by $p_\circ$ and $q$ separates the regions in $\mathbb{R}^3 \ni x$ for 
which $x \cdot p_\circ  <0$ and 
$x \cdot p_\circ >0$, respectively. The latter half-space includes the positive $x_0$-axis. 
Rotating the vector, setting $p = R_0(\alpha) p_\circ $, 
$\alpha \in [0, 2\pi)$ and taking the intersection of the 
resulting regions  for which the scalar product $x \cdot p$
is positive, yields the forward light cone $\partial V^+ \setminus \{(0,0,0)\}$.   
\end{proof}

\goodbreak 
The  profile of the {\em forward tuboid}  ${\mathcal T}_+$ near each 
point $ x$ of $dS$ (in the space 
of $\Im  z$ and for $\Im  z \searrow 0$) is the cone
	\begin{equation}  
		\label{px}
		{\mathcal P}^+_{ x}=  T_{ x } dS \cap (-V^+)  
	\end{equation} 
in the tangent space $T_{ x } dS$  at the point $ x \in dS$. 
(Note that in \eqref{px} the tangent space $T_{ x } dS \cong \mathbb{R}^2$ at $ x \in dS$ 
is viewed as a subspace of $T_{ x } \mathbb{R}^3 \cong \mathbb{R}^3$.) For the origin $o \in dS$
this yields ${\mathcal P}^+_{o}=  \bigl\{ y \in \mathbb{R}^3 \mid - y_0 > |y_1|, y_2 = r \bigr\}$.

\subsection{A first glance at the Euclidean sphere}
\label{SS:1.6.1}
Applying\footnote{Applying the boosts $\Lambda_1 (t)$, $t \in \mathbb{R}$, to 
the half-circles \eqref{half-circle-S1}, followed by  
the rotations $R_{0} (\alpha)$, $\alpha \in [0, 2 \pi)$, yields the interior 
of the one-sheeted hyperboloid in $( i \mathbb{R}) \times 
\mathbb{R}^2$ with the (closed) past light cone and the interior of the
future mass shell for the value $m=r$ removed.}
the rotations $R_{0} (\alpha)$, $\alpha \in [0, 2 \pi)$, to the half-circles 
	\begin{equation}
	\label{half-circle-S1}
	 \bigl\{  ( i r \sin \theta , 0, r \cos \theta) \in dS_{\mathbb{C}} 
	 \mid  0 <  \mp \theta < \pi \bigr\} 
	 \subset {\mathcal T}_+ \; 
	\end{equation}
we find  the (open) \emph{lower}\index{lower hemisphere} and  
\emph{upper hemispheres}\index{upper hemisphere} 
	\begin{equation} 	
		\label{spl}
		S_\mp = \bigl\{ ( i\lambda_0,x_1, x_2) \in  (i\mathbb{R}) 
		\times \mathbb{R}^2 \mid \lambda_0^2 +  x_1^2 +  x_2^2 = r^2, 
		\mp \lambda_0 > 0 \bigr\}
	\end{equation} 
of the \emph{Euclidean sphere}\index{Euclidean sphere}\footnote{Note that 
the definition of $S^2$ in \eqref{euclidsphere} refers to the Lorentz 
metric~\eqref{metrik}. }
\label{euclidspherepage}
	\begin{equation} 
		\label{euclidsphere}
		S^2 = \Bigl\{ \left( \begin{smallmatrix} i r \sin \theta \cos \psi \\ 
		r \sin \psi \\ r \cos \theta \cos \psi
		\end{smallmatrix} \right)  \in \mathbb{C}^3 
		\mid \theta \in (- {\textstyle  \frac{\pi}{2}, \frac{\pi}{2}}  ], 
		\psi \in ( - {\textstyle  \frac{\pi}{2}, \frac{3\pi}{2}} ]   \Bigr\} \;  .
	\end{equation} 
Thus ${\mathcal T}_\pm =  \{  \Lambda \; S_\mp \mid    \Lambda \in SO_0 (1,2)\} $  and, consequently,
	$
		S^2 \subset \overline{{\mathcal T}_+ \cup {\mathcal T}_-} 
		\subset dS_{\mathbb{C}} $ . 

\goodbreak
\begin{remark}
Clearly, the decomposition of the Euclidean sphere into a lower and an upper hemisphere in
\eqref{spl} distinguishes a Cauchy surface $S^1= \partial S_\mp$. However, 
as ${\mathcal T}_+$ is invariant under the action of $SO_0 (1,2)$, one might as well 
consider the Lorentz transformed Cauchy surface $\Lambda S^1\subset dS$ together with a 
Lorentz transformed sphere 
	\[
		\Lambda S^2 \subset \overline{{\mathcal T}_+ 
		\cup {\mathcal T}_-} \subset dS_{\mathbb{C}} \; , 
		\qquad \Lambda \in SO_0(1,2) \; . 
	\]
\end{remark}

Next, consider a point $x$ on the Cauchy surface $S^1$, subject to future oriented boosts
passing through $x$. The trajectories of these boosts span the future of the point $x$. Moreover, 
they allow analytic  continuations, which lie inside ${\mathcal T}_+$: 

\begin{lemma} 
\label{lemma1.2}
Let $x = R_{0} (\psi) o \in S^1$, $\psi \in [0, 2 \pi)$. Then 
	\begin{equation} 
		\label{waw}
		\Gamma^\pm ( x  )
		= 	\overline{ \bigl\{ 
				{\Lambda^{(\alpha)}}(t)  x  \in dS  
				\mid t  \in \mathbb{R}^\pm \; , \; 
				\alpha \in  ( {\textstyle  \psi - \frac{\pi}{2}, 
				\psi + \frac{\pi}{2}  })  \bigr\} } \; . 
	\end{equation} 
Moreover,  the map 
	\begin{equation} 
		\label{waw1}
		\tau \mapsto{\Lambda^{(\alpha)}} (\tau)   x \; , 
		\qquad \alpha \in (  {\textstyle  \psi-\frac{\pi}{2}, \psi+\frac{\pi}{2} } )  \; , 
	\end{equation} 
is entire,  and for $\pm \tau \in {\mathbb S} \doteq \mathbb{R} - i ( 0,\pi ) $ 
the map \eqref{waw1} takes values in $ {\mathcal T}_\pm$. 
\end{lemma}

\begin{proof} Let $x = R_{0} (\psi) o \in S^1$, $\psi \in [0, 2 \pi)$, be fixed.
Let $y$ be a point in the interior of $\Gamma^+ ( x )$. Then an explicit computation 
shows that there exist unique $t>0$ and 
$\alpha \in (  {\textstyle  \psi-\frac{\pi}{2}, \psi+\frac{\pi}{2} } )$ such that 
	\[ 
		y = \Lambda^{(\alpha)}(t)  x \; . 
	\]
Analyticity of the map \eqref{waw1} follows from \eqref{Lambdaalpha} and 
the explicit form of $\Lambda_1(t)$ 
provided at the beginning of Section \ref{subsec:1.3}; see also \eqref{eqBooW} below. 
The explicit form of $\Lambda_1(- i \tau)$, $\tau \in (0, \pi)$, also shows that 
$\Lambda^{(\alpha)}(-i \tau) x$ takes values in $S_-$, and consequently
$\Lambda^{(\alpha)}(t) \Lambda^{(\alpha)}(-i \tau) x \in \{  \Lambda \; S_- 
\mid    \Lambda \in SO_0 (1,2)\}
= {\mathcal T}_+ $. Similar argument holds for the other 
choices of signs. 
\end{proof}

\begin{remark} 
Given an arbitrary point $ x \in dS$, formulas analogous to \eqref{waw} 
and~\eqref{waw1} hold true for all possible choices of  space-like geodesics passing through
the point~$ x$. Note that a space-like geodesic is used to define ${\Lambda^{(\alpha)}}$. 
\end{remark}

\begin{lemma} Every point on the hemisphere $S_+$ 
can be reached by applying a boost $\Lambda^{(\alpha)}(i \theta)$, 
$\theta  \in (0, \pi)$, $\alpha \in  ( - \frac{\pi}{2}, \frac{\pi}{2})$, 
to the origin $o$, \emph{i.e.}, 
	\[
		S_+ = \bigl\{ 
				\Lambda^{(\alpha)}(i \theta)  o  \in dS   
				\mid \theta  \in (0, \pi) \; , \; 
				\alpha \in  ( - \tfrac{\pi}{2}, \tfrac{\pi}{2})  \bigr\}  \; .
	\]
Moreover, 
	\[
		\underbrace{ \Gamma^+ ( o  ) \cup \Gamma^- ( o  ) }_{ \doteq \Gamma (o) }
		= 	\overline{ \bigl\{ 
				{\Lambda^{(\alpha)}}(t)  o  \in dS  
				\mid t  \in \mathbb{R} \; , \; 
				\alpha \in  ( - \tfrac{\pi}{2},  \tfrac{\pi}{2} )  \bigr\}}  \; 
	\]  
and $dS \setminus \Gamma ( o  ) 
		=  \bigl\{ 
				{\Lambda^{(\alpha)}}(t+i \pi )  o  \in dS  
				\mid t  \in \mathbb{R} \; , \; 
				\alpha \in  ( - \tfrac{\pi}{2},  \tfrac{\pi}{2} )  \bigr\}$.   
A similar result holds for $S_-$. 
\end{lemma}

\begin{proof}
Let $( i\lambda_0,x_1, x_2) \in S_+$ be an arbitrary point in $S_+$, 
then $|x_2|<r$ and 
the vector 
	\[ 
		\lambda	\left[ \begin{pmatrix}
		0 \\
		x_1 \\ 
		x_2 
		\end{pmatrix} - 
		\begin{pmatrix}
		0 \\
		0 \\ 
		r
		\end{pmatrix}
		\right]
		= \begin{pmatrix}
		0 \\
		r \sin \alpha \\ 
		r \cos \alpha 
		\end{pmatrix}
	\]
fixes an angle 
	$
	 \alpha = \arctan \frac{x_1}{x_2-r} $, $
	\alpha \in  ( - \tfrac{\pi}{2},  \tfrac{\pi}{2} ) $, 
such that 
	\[ 
		\begin{pmatrix}
		 i\lambda_0 \\
		 x_1 \\
		 x_2
		 \end{pmatrix}
		 = \Lambda^{(\alpha)}(i \theta) \begin{pmatrix}
		 0 \\
		 0 \\
		 r
		 \end{pmatrix}
	\]
with 
	\[ 	
		 \theta = \arcsin \frac{2\lambda}{\sqrt{x_1^2 + (x_2 -r)^2}} \; , 
		\qquad \theta \in 
		\begin{cases}
		(0, \tfrac{\pi}{2} ] & x_2 \ge 0 \; , \\
		(\tfrac{\pi}{2}, \pi ) & x_2 <0 \; . 
		\end{cases}		
	\] 
Finally, the statements concerning the boundary of the tube  
${\mathcal T}_+$ are consequences 
of the fact that for $\alpha \in  ( - \frac{\pi}{2},  \frac{\pi}{2} )$ fixed,
the trajectories of $\bigl\{  \Lambda^{(\alpha)}(t ) o  
\mid t  \in \mathbb{R}  \bigr\}$ and $\bigl\{\Lambda^{(\alpha)}(t+i \pi ) o  
\mid t  \in \mathbb{R}  \bigr\}$ form the intersection of a plane orthogonal 
to the $x_1-x_2$-plane whose intersection with the $x_1-x_2$-plane passed
through the points $(0, 0, r)$ and $(0, r \sin \alpha, r \cos \alpha)$. 
\end{proof}

\begin{remarks}
\quad
\begin{itemize}
\item [$i.)$]
For $\alpha \ne 0$ the map $\mathbb{R} \ni t \mapsto {\Lambda^{(\alpha)}}(t) o$ 
no longer describes the geodesic motion of a free falling observer. 
As $\alpha \to \pm \pi /2$, the observer following the path 
$\bigl\{ {\Lambda^{(\alpha)}} (\frac{t}{r}) o \mid t \in \mathbb{R} \bigr\}$ is exposed 
to a {\em uniformly accelerated motion}, namely a boost, and will  observe 
a temperature~$\bigl( (2 \pi r ) \cos \alpha \bigr)^{-1}$;   
see also~\eqref{eqBooW}.  
This result follows by parameterising the path~\eqref{waw} in the proper 
time, see~\eqref{Lalpha} and 
also~\cite{NPT}. In other words,  the result 
of {\em Bisognano-Wichmann}~\cite{BiWia, BiWib} and 
Unruh~\cite{Unruh} remains valid on $dS$ (see also~\cite{DeBeM}). 
\item [$ii.)$]
Alternatively, one may consider an observer following the path 
	\begin{equation}
		t \mapsto {\Lambda^{(0)}} (t) R_0(\alpha) o \; ,
		\qquad 
		t \in \mathbb{R} \; , 
	\end{equation}
with $\alpha \in  ( - \frac{\pi}{2}, \frac{\pi}{2})$ fixed. The trajectory lies in the 
$(x_1= r \sin \alpha)$-plane, and it is given by projecting (in the $x_1$-direction)
the trajectory 
	\[
		t \mapsto \Lambda^{(0)} (t) \left( \begin{smallmatrix}
		 0 \\ 0 \\ r \cos \alpha  \end{smallmatrix} \right)  \; ,
		\qquad 
		t \in \mathbb{R} \; , 
	\]
onto the $(x_1= r \sin \alpha)$-plane. In particular, if $t$ is analytically continued into the 
strip~$\mathbb{S}$ the \emph{imaginary part} lies entirely in the $x_0$-direction, and its 
maximum value $r \cos \alpha$ is reached at $\Im t = \frac{\pi}{2}$.
\end{itemize}
\end{remarks}

Finally, we may consider trajectories of the type
	\[	
		t \mapsto \Lambda_2 (s) \Lambda_1 (t) \Lambda_2 (-s)
		o = \Lambda_2 (s) \Lambda_1 (t) o
		\; , \qquad s, t \in \mathbb{R} \; . 
	\] 
This is, of course, the time-like geodesic associated to the 
wedge $\Lambda_2 (s) W_1$, $s \in \mathbb{R}$. Note that, for all $s \in \mathbb{R}$,
the edges of the wedge $\Lambda_2 (s) W_1$ are space-like to $o$. 
For $s \in \mathbb{R}$ fixed and $\theta \in  [0, \pi] $, we find
	\begin{align}	
		\Lambda_2 (s) \Lambda_1 (t - i \theta) 
		o 
		& =  \begin{pmatrix}
				\cosh s &  \sinh s &0 \\
				\sinh s &  \cosh s &0  \\
					0 & 0 & 1  
				\end{pmatrix}
				\left[ \begin{pmatrix}
				r \cos \theta \sinh t  \\
     					  0  \\
 				 r \cos \theta \cosh t   
				\end{pmatrix}
		- i \begin{pmatrix}
				 r \sin \theta \cosh t    \\
     				0  \\
 				 r \sin \theta \sinh t    
				\end{pmatrix} \right] 
		\nonumber \\
				& = r \cos \theta
				\begin{pmatrix}
				 \sinh t  \cosh s \\
     				\sinh t  \sinh s  \\
 				\cosh t   
				\end{pmatrix}
		- i r \sin \theta \begin{pmatrix}
				 \cosh t \cosh s \\
				 \cosh t \sinh s \\
				 \sinh t    
				\end{pmatrix}	\; . 
		\label{eq-1.6.11} 
	\end{align} 
For $\theta = 0$, we have $\sin \theta = 0$ and $\cos \theta =1$. Thus, 
the imaginary part vanishes and
	\[ 
		\bigl\{ 	\Lambda_2 (s) \Lambda_1 (t)  o\mid s, t \in \mathbb{R} \bigr\}
		= \Gamma^+ (o) \cup \Gamma^- (o) \; . 
	 \] 
On the other hand, for $t=0$ we have 
	\begin{align}	
		\Lambda_2 (s) \Lambda_1 ( - i \theta) 
		o 
		 = 	
		 \begin{pmatrix} 0  \\ 0 \\ r \cos \theta 
		\end{pmatrix}
		- i \Lambda_2(s) 
		\begin{pmatrix}
				r \sin \theta \\
				0 \\
				0    
				\end{pmatrix}		\; , 
		\qquad s \in \mathbb{R} \; , \quad \theta \in [0, \pi] \; .   
	\label{boosted-ms}
	\end{align} 
Hence, the real part depends only on $\cos \theta$ and the imaginary
part arises from the mass shell\footnote{In physics,  
a set of the form $\bigl\{ \Lambda 
\left( \begin{smallmatrix} m \\ 0 \\ 0\end{smallmatrix} \right) \mid
\Lambda \in S_0(1,2) \bigr\}$, $m > 0$, is called a mass shell.}
which forms the imaginary 
part in \eqref{boosted-ms}. 

\bigskip
In Lemma \ref{lemma1.2} we have associated a subset 
of ${\mathcal T}_+$ to a fixed point $x \in S^1$. 
We will now show that the union of the sets associated to 
all points $x$ in $S^1$ is, in fact, the whole 
tuboid ${\mathcal T}_+$. Somewhat surprisingly, the double 
twisted M\"obius strip plays a role in the statement of the result.

\begin{lemma}
\label{t+2} 
Let ${\mathbb S}$ be the strip introduced in Lemma~\ref{lemma1.2}, 
let $T^1$ be the unit circle, and let
	\[
		M \doteq \bigl\{ (\psi, \alpha) \in T^1 \times T^1
		\mid   | \alpha -\psi  | < \pi /2 \bigr\}
	\]
be the double twisted M\"obius strip. Here $| \alpha -\psi  | \, $ denotes 
the minimal distance on~$S^1$. The map
	\begin{align} 
		\label{ganzetube}
				 {\mathbb S}  \times  M 
				& \to  {\mathcal T}_+ \nonumber \\ 
				 (\tau, \psi, \alpha) 
				 & \mapsto  {\Lambda^{(\alpha)}} (\tau) R_{0} (  \psi)  o   
	\end{align} 
is surjective. 
\end{lemma}

\goodbreak
\begin{proof} 
Let $\tau=t +i\theta$, with $- \pi/2 < \theta <0$. Then (see \eqref{eqBooW} below)
	\begin{align} 
		\label{3.16}
		{\Lambda^{(\alpha)}} (\tau) R_{0} ( \psi)  o 
		& =  u +i y
	\end{align} 
 with
	\begin{equation} 
		\label{3.17}
 		u =  r \begin{pmatrix}
			\cos (\psi- \alpha) \cos \theta \sinh t \\
			- \cos \alpha \sin (\psi- \alpha)- \sin \alpha 
			\cos (\psi- \alpha) \cos \theta \cosh t \\
			- \sin \alpha \sin (\psi- \alpha)- \cos \alpha 
			\cos (\psi- \alpha) \cos \theta \cosh t  \\
		\end{pmatrix}
	\end{equation} 
and
	\begin{equation} 
		\label{3.18}
		    y =  r \sin \theta \, \cos (\psi - \alpha) 
		   	\begin{pmatrix}
					\cosh t \\
					- \sin \alpha \sinh t \\
					\cos \alpha \sinh t \\
			\end{pmatrix} \; . 
	\end{equation} 
The vector $y$ is time-like, \emph{i.e.},  $0 \le  y \cdot y   \le r^2 $, and 
	\[
		 x =\frac{1}{\sqrt{1- \frac{y \cdot y}{r^2} }} \;  u \in dS \; . 
	\]
Moreover, $  u  \cdot  y = 0$. The equality $  u \cdot  y = 0$ implies 
that $ u + i  y \in dS_{\mathbb{C}}$, as 
	\begin{equation} 
		\label{rmp3.20}
		dS_{\mathbb{C}} = \left\{ ( u,  y) \in \mathbb{R}^6 \mid  
		u \cdot u -  y \cdot y = - r^2 ,   \; u \cdot  y =  0 \right\} \; .
	\end{equation} 
In fact, choosing $\psi = \alpha$ and keeping $\theta$ fixed, 
	\begin{align*}
		\bigl\{ {\Lambda^{(\alpha)}} (t + i \theta) & R_{0} ( \alpha)  o \mid
		\alpha \in [0, 2 \pi) \, , \; t \in \mathbb{R} \bigr\} 
		\\
		& = \bigl\{ R_{0} ( \alpha) \Lambda_1 (t) \left( \begin{smallmatrix} 0 \\ 0 \\ 
		r \cos \theta  \end{smallmatrix}\right)  \mid
		\alpha \in [0, 2 \pi) \, , \; t \in \mathbb{R} \bigr\} 
		\\
		& \qquad + i \bigl\{ R_{0} ( \alpha) \Lambda_1 (t) 
		\left( \begin{smallmatrix} r \sin \theta 
		\\ 0 \\ 
		0  \end{smallmatrix}\right)  \mid
		\alpha \in [0, 2 \pi) \, , \; t \in \mathbb{R} \bigr\} 
		\\
		& = \bigl\{ v \in \mathbb{R}^{1+2} \mid  v \cdot v = - r^2 \cos^2 \theta \bigr\}
		+ i \bigl\{ w \in \mathbb{R}^{1+2} \mid  w \cdot w =  r^2 \sin^2 \theta \bigr\} \; . 
	\end{align*}
Hence, the image of the map \eqref{ganzetube} equals
	\begin{align*}
		\bigcup_{0 <  - \theta < \frac{\pi}{2} } 
		& \Biggl[ \bigl\{ x \in \mathbb{R}^{1+2} \mid  x \cdot x 
		= - r^2 \cos^2 \theta \bigr\} 
		+ i \bigl\{  \Lambda \; \left( 
		\begin{smallmatrix} r \sin \theta \\ 0 \\ 0 \end{smallmatrix}\right) 
		\mid \Lambda \in SO_0 (1,2) \bigr\} \Biggr] 
		\\
		& = \bigl\{  \Lambda \; \left( \begin{smallmatrix} i r \sin \theta \\
		0 
		\\ r \cos \theta \end{smallmatrix}\right) 
		\in dS_{\mathbb{C}} \mid  0 <  - \theta < \tfrac{\pi}{2} \, ,  \;  
		\Lambda \in SO_0 (1,2) \bigr\}  
		\nonumber
		\\
		& =	{\mathcal T}_+ \setminus  \bigl\{  z \in dS_\mathbb{C} 
		\mid \Re  z = 0  \bigr\} \; . 
	\end{align*}
This shows that the map \eqref{ganzetube} is surjective. 
\end{proof}

\begin{lemma}
\label{zWsx} 
For every point $z$ in the tuboid ${\mathcal T}_+$ one can find a wedge $W$, 
a point $x\in W$, as well as an angle $\theta \in (0,\pi/2]$ such that     
	\begin{equation} 
		\label{eqzWthetax} 
			z=\Lambda_{W}( - i \theta )x \; .  
	\end{equation}
Furthermore, $x$ can be chosen on the (unique) time-like geodesic
whose causal completion coincides with $W$. Then there is a de Sitter
transformation $\Lambda$ such that $W=\Lambda W_1$ and $x=\Lambda o$, 
\emph{i.e.}, 
	\begin{equation} 
		\label{eqzLambda}
		z= \Lambda \Lambda_1( - i \theta )\,o 
				\equiv \Lambda\bigl(\cos\theta \cdot o 
				 - i  \sin\theta \cdot \Lgeo_1 o \bigr) \; . 
	\end{equation}
Here, $\Lgeo_1$ is the generator --- see \eqref{elle-1} --- of 
the $\Lambda_1$-boosts, which acts as the
Pauli matrix $\sigma_1$ on the $x_0$-$x_2$ 
plane, and as zero on $x_1$.   
\end{lemma}

\begin{proof}
This is just~\cite[Lemma A.2]{MSY}, the last equation being implicit
in the proof. 
\end{proof}

\begin{remark}
It is noteworthy that $\theta$, $x$ and $W$ can
be directly characterized in a coordinate-independent manner: 
let $z=u+iy$. The real part $u$ satisfies 
$\tfrac{u \cdot u}{r^2} \in (-1,0]$ and is orthogonal to $y$.
Then
	\[
		\theta =  \arccos  \sqrt{-\frac{ u\cdot u}{r^2} } \; , \qquad   x = 
		\frac{ u } {\cos \theta}   \; ,
	\]
and $W$ is the causal completion of the unique time-like geodesic in
$dS$ starting at $x$ with initial velocity
$y$. (Note that $y$ is orthogonal to $x$ and can therefore be
identified with a tangential vector at $x$.) 
\end{remark} 

\begin{lemma} \label{lm:1.16.12}
Let $x \in dS$ be the intersection point of 
two (non-identical) time-like geodesics. Then there is a de Sitter
transformation $\Lambda$ and a real number $s\neq 0$ such that $x=\Lambda
o$, and the geodesics are $t\mapsto \Lambda_{W}(t) x$ and $ \hat{t} \mapsto 
\Lambda_{\widehat{W}}(\hat{t}) x$, where $W\doteq \Lambda W_1$ and
$\widehat W\doteq \Lambda \Lambda_2(s) W_1$. 
\end{lemma}
\begin{proof}
Every time-like geodesic is of the form $t\mapsto \Lambda
\Lambda_{W_1}(t)o$
for some de Sitter transformation $\Lambda$. This can be written
$\Lambda_W(t)x$ with $W\doteq \Lambda W_1$ and $x\doteq \Lambda o$.  
This geodesic has inicial velocity $\Lambda\Lgeo_1 o$.
Every other time-like geodesic through $x\equiv \Lambda o$ has initial velocity
$\Lambda \Lambda_2(s)\Lgeo_1 o$ for some 
$s\neq 0$. The unique corresponding geodesic is
$t\mapsto \Lambda \Lambda_2(s)\Lambda_1(t)o$, which can be written
$\Lambda_{\widehat W}(t)x$ with $\widehat{W}\doteq \Lambda
\Lambda_2(s) W_1$.   
\end{proof}

\goodbreak

\begin{lemma} \label{L'Lx}
Let $x$, $W$ and $\hat W$ 
be as in Lemma~\ref{lm:1.16.12}, and let
$\tau,\tau', \widehat{\tau},\widehat{\tau}'
\in{\mathbb C}$ with $\Im \tau, \Im \tau' \in   
(-  \pi , \pi )$ and 
$\Im \widehat{\tau} , \Im \widehat{\tau}' \in 
\bigl(- \frac{\pi}{2}, \frac{\pi}{2} \bigr)$. 
Then the equality     
	\begin{equation} 
		\label{eqL'Lx}
			\Lambda_W (\tau) 
			\Lambda_{\widehat W}(\widehat{\tau}) x 
		= \Lambda_W (\tau')
                \Lambda_{\widehat W}( \widehat{\tau}') x
	\end{equation}
holds if, and only if, $\tau'=\tau$
and $\widehat{\tau}'=\widehat{\tau}$.
\end{lemma}

\begin{proof}
As in the proof of
Lemma~\ref{lm:1.16.12}, it suffices to consider the case $x= o$,
$W=W_1$ and $\widehat{W} \doteq \Lambda_2(s) W_1$. 
We compute
	\begin{align*}
	\Lambda_1 (\tau) \Lambda_{\widehat W}( \widehat{\tau}) o
	& = 	 \begin{pmatrix}
				 \cosh \tau  &  0 &\sinh \tau \\
     					  0  &  1 & 0  \\
 				  \sinh \tau &  0 & \cosh \tau   
				\end{pmatrix} 
			\begin{pmatrix}
				r \cosh s \sinh \widehat{\tau}   \\
				r \sinh s \sinh \widehat{\tau}   \\
				r \cosh \widehat{\tau}   
				\end{pmatrix}
	\\
	& = \begin{pmatrix}
				r  \cosh \tau \cosh s \sinh \widehat{\tau} 
				+ r \sinh \tau  \cosh \widehat{\tau}  \\
				r \sinh s \sinh \widehat{\tau}     \\
				r  \sinh \tau \cosh s \sinh \widehat{\tau} 
				+ r \cosh \tau  \cosh \widehat{\tau}  
				\end{pmatrix}  .   
		\end{align*}
We now assume that 
	\begin{equation}
	\label{identity-1}
		\Lambda_1 (\tau) \Lambda_{\widehat W}( \widehat{\tau}) o
		= \Lambda_1 (\tau') \Lambda_{\widehat W}( \widehat{\tau}') o\; , 
		\qquad \tau, \tau' , \widehat{\tau} , \widehat{\tau}' \in \mathbb{C} \; .   	
	\end{equation}
As $\sinh s \ne 0$, the \emph{second vector component} of this identity yields
$\sinh \widehat{\tau} = \sinh \widehat{\tau}'$, which is 
equivalent to   
	\[
		u - u^{-1} = 2 \sinh \widehat{\tau}',  
	\] 
where we have set $u \doteq {\rm e}^{\widehat{\tau}}$. 
This equation has the two solutions, namely
	\begin{align*}
		u & = \sinh \widehat{\tau}' \pm \cosh \widehat{\tau}' 
		= \begin{cases} {\rm e}^{\widehat{\tau}'} & \text{ for
 			 ``+''} \\  
			{\rm e}^{-\widehat{\tau}'+i\pi} & \text{ for ``-''}. 
		\end{cases}
	\end{align*}
Hence, 
	\[
		\widehat{\tau} = \widehat{\tau}' + i 2 \pi \widehat{n}
		\quad
		\text{or}
		\quad
		\widehat{\tau} = - \widehat{\tau}' + i \pi (2 \widehat{n} +1)
	\]
for some  $\widehat{n} \in \mathbb{Z}$.  The second solution is excluded, 
since $\Im \widehat{\tau}, 
\Im \widehat{\tau}' \in \bigl(- \frac{\pi}{2}, \frac{\pi}{2} \bigr)$. 
The first implies $\widehat{\tau} = \widehat{\tau}'$. 
Now, \eqref{identity-1} is equivalent to 
	\[ 
		\Lambda_{\widehat W}( \widehat{\tau}) o
		= \Lambda_1 (\tau'- \tau ) \Lambda_{\widehat W}( \widehat{\tau}) o
		= \Lambda_1 (\tau'- \tau ) \Lambda_2 (s) \Lambda_1( \widehat{\tau}) o
		\; , 
	\]
for $\tau, \tau' , \widehat{\tau}  \in \mathbb{C}$. 
Thus 
	\begin{equation}
	\label{new-identity-for-tau-2}
		 		 \begin{pmatrix}
		 r \cosh s \sinh \widehat{\tau} 
		 \\
		 r \sinh s \sinh \widehat{\tau} \\
		 r \cosh \widehat{\tau} 
		 \end{pmatrix} 
		= \Lambda_1 \bigl( \underbrace{\tau'- \tau}_{ = \widetilde{\tau}}  \bigr) 
		 \begin{pmatrix}
		 r \cosh s \sinh \widehat{\tau} 
		 \\
		 r \sinh s \sinh \widehat{\tau} \\
		 r \cosh \widehat{\tau} 
		 \end{pmatrix} , 
		 \qquad \tau, \tau' , \widehat{\tau}  \in \mathbb{C}
		\; . 
	\end{equation}
The one-parameter group of complex boosts 
$\mathbb{C} \ni z \mapsto \Lambda_1(z)$ leaves 
the $z_1$-compo\-nent of the vector \eqref{new-identity-for-tau-2} 
invariant. Thus we can concentrate on the 
$z_0$ and $z_1$ components in $dS_\mathbb{C}$. 
For the plus sign, Equ.~\eqref{new-identity-for-tau-2} is of the from
	\[
		\begin{pmatrix}
				 \cosh \widetilde{\tau}  &  \sinh \widetilde{\tau} \\
   				  \sinh \widetilde{\tau} &   \cosh \widetilde{\tau}   
				\end{pmatrix}
		\begin{pmatrix}
				 z_0 \\
   				  z_2   
				\end{pmatrix}= \begin{pmatrix}
				 z_0 \\
   				  z_2   
				\end{pmatrix} \; . 
	\]
This is equal to 
	\[
		\begin{pmatrix}
				 \cosh\widetilde{\tau}  - 1&  \sinh \widetilde{\tau} \\
   				  \sinh \widetilde{\tau} &   \cosh \widetilde{\tau}  -1 
				\end{pmatrix}
		\begin{pmatrix}
				 z_0 \\
   				  z_2   
				\end{pmatrix}= \begin{pmatrix}
				0 \\
   				0   
				\end{pmatrix} \; . 
	\]
The determinant $\cosh^2\widetilde{\tau}  - 2 \cosh \widetilde{\tau} 
		+ 1 - \sinh^2 \widetilde{\tau}
		= 2 (1- \cosh \widetilde{\tau})$ 
is only zero for $\cosh \widetilde{\tau} = 1$, \emph{i.e.}, for 
	\[
		\tau = \tau' + i 2 \pi n \; , 
		\qquad
		n \in \mathbb{Z} \; . 
	\]
The restrictions $\Im \tau, \Im \tau' \in  (- \pi , \pi) $ 
imply $\tau = \tau' $. 
\end{proof}

\begin{lemma} 
\label{Chart}
Let $x$, $W$ and $\widehat{W}$ be as in Lemma~\ref{lm:1.16.12}, and
let $U$ be the strip ${\mathbb R}+ i (- \pi, \pi)$
and $\widehat{U}\doteq {\mathbb R}+ i (- \pi/2, \pi/2)$. Then
the map  
	\begin{equation} 
		\label{eqHoloChart}
		(\tau , \widehat{\tau}) \mapsto 
                \Lambda_{W}(\tau) \Lambda_{\widehat{W}}(\widehat{\tau}\,) x  
	\end{equation}
is a holomorphic chart from $U \times \widehat{U}$ into complexified
de Sitter space. Further, let $I_{x,W,\widehat W}$
be the open interval $\bigl\{t \in{\mathbb R} \mid | \sinh s \sinh t | < 1
\bigr\}$, and let 
	\[
		T\doteq \sup I_{x,W,\widehat W}\equiv
			\arcsinh(1/|\sinh s|) \; . 
	\]
Then, for all $t \in {\mathbb R}$, $\hat{t} \in
I_{x,W,\widehat{W}} \, $, $\widehat{\theta} \in [-\pi/2,0]$ and $\theta \in
[-\pi/2,0]$ with $| \theta |$ small enough such that 
	\begin{equation} 
		\label{eqCosTheta}
		\cos^2 \theta > \frac{\sinh^2s}{\cosh^2s-\tanh^2T}
	\end{equation}
with $\theta \cdot \widehat{\theta} \neq 0$, 
the point 
	\begin{equation} 
		\label{eqzT+}
		 \Lambda_{W}(t+i\theta) \Lambda_{\widehat{W}} (\hat{t}+i
                 \widehat{\theta} ) x
	\end{equation}
is in the tuboid $\mathcal{T}_+$.
(Note that the set of $\theta$ satisfying the
        condition~\eqref{eqCosTheta} contains an open interval
        centered around the origin, since the
        right hand side is strictly smaller than one.) 
\end{lemma}

Note that the image of the map~\eqref{eqHoloChart}
contains the open set of real points of the form
	\[
		\Lambda_{W}(t) 
		\Lambda_{\widehat{W}}(\hat{t}) x  \in dS \; , 
		\qquad \hat t, t\in \mathbb{R} \; , 
	\]
as well as the points of the form 
	\[ 
		\Lambda_{W} (- i\theta)  x\in {\mathcal T}_+ \; , 
		\qquad \theta\in (0,\pi) \; . 
	\]         
Thus, every $z\in{\mathcal T}_+$ has a 
\emph{complex} neighbourhood in $dS_{\mathbb{C}}$ 
which extends to real de Sitter space and on which the 
map \eqref{eqHoloChart} is a holomorphic chart.   

\begin{proof}
Lemma~\ref{L'Lx} implies that the map~\eqref{eqHoloChart}
is a diffeomorphism from $U \times \widehat{U}$ 
onto some open set in complexified de Sitter space.     
Since the boosts $\Lambda_W(\tau)$ are entire analytic transformations
in ambient Minkowski space, this proves   
the statement about the holomorphic chart. 

We now prove the last statement. As in the proof of
Lemma~\ref{lm:1.16.12}, it suffices to consider the case $x= o$,
$W=W_1$ and $\widehat{W} \doteq \Lambda_2(s) W_1$.
For simplicity, denote 
$\Lambda (\cdot) \doteq \Lambda_{W_1}(\cdot) $, 
$\widehat\Lambda (\cdot) \doteq \Lambda_{\widehat W}(\cdot) $ etc. 
Inspecting \eqref{eqBooW}, we have 
	\begin{align}
  		\Lambda(t + i \theta) & = \Lambda(t)\big( \Jgeo(\theta) +
    		i \sin (\theta) \Lgeo \big) \; , \\
		\widehat \Lambda(\hat{t} + i\widehat{\theta}) 
		&= \widehat{\Lambda}(\hat{t})\big( \,\hat{\Jgeo}(\widehat{\theta}) 
		+  i \sin (\widehat{\theta}) \hat{\Lgeo} \, \big) \;  . 
	\end{align}
The point \eqref{eqzT+} is in the tuboid iff its imaginary part is in
the backward light cone (of ambient Minkowski space).
But its imaginary part is, according to the above equations,
	\begin{equation} 
		\label{eqzT+'}
		 \Lambda(t) \big(\sin( \widehat{\theta})   \, \Jgeo(\theta) \, \hat{\Lgeo}
		+  \sin(\theta) \,
			\Lgeo \, \hat{\Jgeo}(\widehat{\theta}) \big)  
			\widehat{\Lambda}( \hat{t} ) o \; .
	\end{equation}
The real Lorentz transformation $\Lambda(t)$ preserves the forward light cone;  
it is therefore sufficient to consider the case $t=0$, \emph{i.e.}, 
$\Lambda (0) = \mathbb{1}$. 

For the first term, we find
	\begin{align*}
	\Jgeo_1(\theta) \hat{\Lgeo} \, \widehat{\Lambda}( \hat{t} )o
	& = 	 \begin{pmatrix} 
				\cos \theta & 0 & 0 \\
				0 & 1 & 0\\
				0 & 0 &  \cos \theta
			\end{pmatrix} \begin{pmatrix}
				 0
				 &  
				 0 
				  & \cosh s  \\
  				0 & 0 &  \sinh s   \\
 				 \cosh s  &  
				  -  \sinh s & 0   
				\end{pmatrix} 
			\begin{pmatrix}
				r \cosh s \sinh \hat{t}   \\
				r \sinh s \sinh \hat{t}   \\
				r \cosh \hat{t}   
				\end{pmatrix}
	\\
	& 
	= \begin{pmatrix}
				r  \cos \theta \cosh s \cosh  \hat{t}   \\
				r \sinh s \cosh  \hat{t}    \\
				r  \cos \theta \sinh  \hat{t}  
				\end{pmatrix}  .   
		\end{align*}
This vector is time-like if, and only if, the variables
$s,\theta$ and $\hat t$ satisfy the inequality \eqref{eqCosTheta},
with $T$ replaced by $\hat t$. For $|\hat t|<T$, this yields
condition~\eqref{eqCosTheta}.   

We next show that the second term in \eqref{eqzT+'} is always in the forward
light cone given that $\hat t\in I_{x,W,\widehat W}$: By that hypothesis,
the vector $\widehat{\Lambda}( \hat{t} ) o$ is in the wedge $W$, and
the map $\Jgeo(\widehat{\theta})$ just acts on it as multiplication by
$\cos\widehat \theta$ (see the explicit formulas below), hence still
is in the wedge. Thus, $l$ maps this vector into the forward light
cone.  In explicit formulas,  
	\begin{align*}
			\Lgeo \, \hat{\Jgeo} (\widehat{\theta}) \big)  \widehat{\Lambda}( \hat{t} ) o
			& = \Lgeo \, \Lambda_2(s)
			\Jgeo_1 (\widehat{\theta}) \Lambda_1( \hat{t} ) 
			\begin{pmatrix} 
				0 \\
				0 \\
				r	
			\end{pmatrix}
			= l \begin{pmatrix}
				\cosh s &  \sinh s &0 \\
				\sinh s &  \cosh s &0  \\
					0 & 0 & 1  
				\end{pmatrix}
			\begin{pmatrix} 
  			r \cos \widehat{\theta} \sinh \hat{t}  \\
			0 \\
			r \cos \widehat{\theta} \cosh \hat{t} \\
			\end{pmatrix}   
			\\
			&   
			=    r \cos \widehat{\theta}  
			\begin{pmatrix} 
  			\cosh \hat{t} \\
			0  \\
			 \cosh s \sinh \hat{t}  
			\end{pmatrix}   .
	\end{align*}
Since $\cosh^2\hat{t} - \cosh^2 s \sinh^2 \hat{t}\equiv 1-\sinh^2 s
\sinh^2 \hat{t}>0$ by hypothesis, this vector is indeed time-like.

Thus, taking care of the minus sign, the two terms in
Eq.~\eqref{eqzT+'} are past-directed time-like vectors, and so is
their sum. This completes the proof.  
\end{proof}

Next we provide a  {\em flat tube theorem}  (see, e.g., \cite{BB,
BEGS}; an early result of this type  
is due to Malgrange and Zerner) for the de Sitter space. 

\begin{theorem}
\label{flattube}
Let $f$ be a continuous function on $dS$ with the
following property: for any wedge $W \subset dS$ and any $x\in W $, the map 
	\begin{equation}
  		\label{eqStripHolo}
		t \mapsto f \bigl( \Lambda_{W}(t)x \bigr)
	\end{equation}
can be extended to a function defined  and  analytic in the strip ${\mathbb S}  
= \mathbb{R}  - i ( 0,\pi )  $, whose boundary values 
(as the imaginary part goes to zero)  
are described 
by~\eqref{eqStripHolo}. Then $f$ is the boundary value of a unique continuous 
function~$F$, which is analytic in the tuboid~${\mathcal T}_+$. 
\end{theorem}

\begin{remarks} 
\quad
\begin{itemize}
\item [$i.)$] 
In case $f \in L^2 (dS, {\rm d} \mu_{dS})$, Theorem 1 
in \cite{BM2} (see Theorem~\ref{hardy} below) provides an explicit formula, 
namely formula \eqref{ck}, for the analytic continuation into the tube. 
\item [$ii.)$] 
A generalisation to distributions should follow from an appropriate generalisation 
of the Bros-Epstein-Glaser Lemma \cite[Vol.~II, Theorem~IX.15]{RS}.
For an example of a vector-valued distribution, which satisfies the 
assumptions of the theorem, see Proposition~\ref{prop:2.2}.
\end{itemize}
\end{remarks} 

\begin{proof} 
\emph{(Part I: An application of the local flat tube theorem).}
Let~$\widehat\Lambda \equiv
\Lambda_{\widehat{W}}$ be defined  
as in the preceding lemmas, and denote as before 
by $I_{W_1,\widehat{W}}$ the finite real interval 
$\bigl\{ \hat{t} \in{\mathbb R} \mid \Lambda_{\widehat{W}}
( \hat{t} ) o \subset W_1 \bigr\}$. By assumption, we know: 
\goodbreak
\begin{itemize}
\item [$i.)$] For any $\hat{t}$ fixed in $I_{W_1, {\widehat W}}$, 
the map 
		\begin{equation}
  		\label{eqStripHolo-3}
			 t \mapsto f \bigl( 
			\Lambda_{1}(t) \widehat{\Lambda} (\hat{t}) o \bigr) \; , 
			\qquad t \in \mathbb{R} \; , 
	\end{equation}
can be extended to a function
	\begin{equation}
  		\label{eqStripHolo-4}
		\tau \mapsto F_{W_1, {\widehat W} } ( \tau, \hat{t}  ) \; , 
		\qquad \tau \in {\mathbb S}   \; ,  
	\end{equation}
defined  and  analytic in the strip ${\mathbb S} 
\doteq \mathbb{R} - i ( 0,\pi )$, 
whose boundary values  (as the imaginary part goes to zero)  
are described by~\eqref{eqStripHolo-3}. 
\item [$ii.)$] For any $t \in \mathbb{R}$ fixed, 
the map 
		\begin{equation}
  		\label{eqStripHolo-3a}
			 \hat{t} \mapsto f \bigl( 
			\Lambda_{1}(t) \widehat{\Lambda} ( \hat{t} )o \bigr) \; , 
			\qquad \hat{t} \in \mathbb{R} \; , 
	\end{equation}
can be extended to a function
	\begin{equation}
  		\label{eqStripHolo-4a}
		\widehat{\tau} \mapsto F_{W_1, {\widehat W} } ( t, \widehat{\tau}  ) \; , 
		\qquad \widehat{\tau} \in {\mathbb S}   \; ,  
	\end{equation}
defined  and  analytic in the strip ${\mathbb S} 
\doteq \mathbb{R} - i ( 0,\pi )$, 
whose boundary values  (as the imaginary part goes to zero)  
are described by~\eqref{eqStripHolo-3a}.  
\end{itemize}
However, in order to get a jointly analytic function we
can exploit this fact only for $\widehat{\tau}$ in the rectangle
$ I_{W_1,\widehat W}  - i ( 0,\pi ) \subset {\mathbb S}$. 
In fact, we will consider an even smaller region. 
The relevant theorem about joint analyticity (the \emph{local flat tube
theorem} due to D'Antoni and Zsido~\cite{DA-Z-2001a}),
Theorem~\ref{local-flat-tube}, is tailored for  
disc segments: for a real interval $I = (a,b)$, let
$\mathbb{G}_{-\pi, I}$  
be the \emph{disk segment} bounded by $I$ and by a
circular arc through the three points $a, - i \pi, b$  
in the complex
plane. Since it is contained in the strip ${\mathbb S}$, we have: 
\begin{itemize}
\item [$ii'.)$]
For any $t \in \mathbb{R}$ fixed, 
the map 
		\begin{equation}
  		\label{eqStripHolo-3b}
			 \hat{t} \mapsto f \bigl( 
			\Lambda_{1}(t) \widehat{\Lambda} ( \hat{t} )o
                        \bigr) \; ,  \qquad \hat{t} \in 
                        I_{W_1,\widehat W}  \equiv I \; , 
	\end{equation}
can be extended to a function
	\begin{equation}
  		\label{eqStripHolo-4b}
		\widehat{\tau} \mapsto 
		F_{W_1, {\widehat W} } ( t, \widehat{\tau}  ) \; , 
		\qquad \widehat{\tau} \in 
		{\mathbb G}_{-\pi, I} \; ,  
	\end{equation}
defined  and  analytic in the disc segment 
${\mathbb G}_{-\pi, I}$,  
whose boundary values  (as the imaginary part goes to zero)  
are described by~\eqref{eqStripHolo-3b}. 
\end{itemize}
By Theorem~\ref{local-flat-tube}, the functions appearing in 
\eqref{eqStripHolo-4} and \eqref{eqStripHolo-4b} are the boundary value of a
holomorphic function $F_{W_1, {\widehat W} } (\tau, \widehat{\tau} )$ defined 
on the set 
	\begin{equation}
	\label{local-tube} 
		V_{W_1,\widehat W}\doteq \bigl\{ (\tau, \widehat{\tau}) \in  
		{\mathbb S}\times {\mathbb G}_{-\pi, I}
		\mid    | \Im \tau| + |\Im  \Phi_{  I }  
		(\widehat{\tau}) | < \pi
		\bigr\} \; , 
	\end{equation}
which is analytic in the interior. Here $\Phi_{I }$ is the
biholomorphic function from the disc segment 
${\mathbb G}_{-\pi, I} $ onto the strip ${\mathbb S}$ from D'
Antoni and Zsid\'o, see Eq.~\eqref{DAZ-phi} in Appendix A. 
Note that $I_{W_1,\widehat W}=(-T,T)$ with $T=
\arcsinh(1/|\sinh s|)$, and that the set of $\widehat \tau \in  
{\mathbb G}_{-\pi, I} $ with  $\Im
\Phi(\widehat{\tau})= - \theta
=$ const.\ is just the circular arc through the
three points $-T, - i \theta, T$ in the disc segment 
${\mathbb G}_{-\pi, I} $.  
This result can be improved further, by applying the
same techniques to rectangles of the form 
	\[
		[ a, b ] - i (0, \pi) \subset [ -T, T ] - i (0, \pi) 
	\]
and taking the union over the regions of analyticity ${\mathbb G}_{- \pi, 
(a, b) }$, $ -T < a < b < T$. 	 

\bigskip
\noindent
\emph{(Part II: A tuboid over $W_1$).}
Let $z(\cdot,\cdot)$ be the map from
Eq.~\eqref{eqHoloChart},  
	\begin{equation} 
		\label{eqztau}
		z ( \tau, \widehat{\tau} ) \doteq \Lambda_1 ( \tau) 
		\widehat{\Lambda} (\widehat{\tau}) o \; , 
		\qquad  \widehat{\tau}, \tau  \in \mathbb{C} \; . 
\end{equation}
By Lemma~\ref{Chart}, the map $z(\cdot,\cdot)$ is bijective (and
biholomorphic) on the set $\big({\mathbb R}+i(-\pi,\pi)\big)\times
\big({\mathbb R}+i(-\tfrac{\pi}{2},\tfrac{\pi}{2})\big)$. 
Let, for $ 0 < \epsilon \ll \pi$, 
	\[
		V (\epsilon) \doteq \bigcup_{-T < a < b < T} \bigl\{ (\tau, \widehat{\tau}) \in  
		{\mathbb S}\times  {\mathbb G}_{- \pi, (a, b) }  
		\mid    | \Im \tau| + |\Im \Phi_{(a, b)} (\widehat{\tau}) | < \pi, 
		| \Im \tau| < \epsilon , |\Im \widehat\tau|< \tfrac{\pi}{2}
		\bigr\} \; .  
	\]  
On the set $V (\epsilon)$, the map $z(\cdot,\cdot)$ is biholomorphic. 
We set 
	\begin{equation}
	\label{f-extension}
		f_{0}\big(z(\tau,\widehat \tau)\big)
		\doteq F_{W_1,{\widehat W} } (\tau, \widehat{\tau} ) \; ,
		\qquad  (\tau,\widehat{\tau}) \in  V (\epsilon)  \;  .  
	\end{equation}
This is a well-defined and analytic function on the \emph{image} of
$V (\epsilon) $ under the map $z(\cdot,\cdot)$,
namely the set
	\[
		U  (\epsilon) \doteq \bigl\{ \Lambda_1( \tau) 
		 \widehat{\Lambda}(\widehat \tau \, ) o\,\mid\,
		(\tau,\widehat \tau)\in V (\epsilon) \bigr\} \; .
	\]
Note that $U(\epsilon)$ is contained in the tuboid
${\mathcal T}_+$ if $\epsilon$ is small enough\footnote{ 
How small $\epsilon$ should be will be determined in Lemma~\ref{path-in-tube}.}
 such that the bound
\eqref{eqCosTheta} holds for all $|\theta|< \epsilon$.
For reasons we will explain in the sequel, we prefer to 
work with a smaller set, namely the set
	\[
		U_0  (\epsilon)   
		\doteq \bigl\{ z \in U  (\epsilon) 
		 \mid  | \Re z_0 | < \Re z_2 \bigr\} \; .
	\]
This set is the image of the set 
	\[
		V_0 (\epsilon)  
		\doteq \bigl\{ (\tau,\widehat{\tau}) 
		\in V (\epsilon)  
		\mid z ( \tau, \widehat{\tau} )
		\in  U_0 (\epsilon)  \bigr\} \; . 
	\]
Note that the real boundary of the set $ U_0 (\epsilon)$ 
is the wedge 
	\[
		\bigl\{  \Lambda_1(t) \widehat{\Lambda}(\hat{t}) o 
		\mid t \in {\mathbb R}
		  \, , \; \hat{t} \in I_{W_1, {\widehat W} }  \bigr\} =  W_1  \; .   
	\] 
In the sequel, we will consider 
the extension of $f|_{W_1}$ to an
analytic function $f_{0}$ on the 
set $U_0 (\epsilon)$.

\bigskip
\noindent
\emph{(Part III: A countable covering of $\mathcal{T}_+  
		\setminus \{  z \in dS_\mathbb{C} \mid \Re  z = 0 \}
		$).}
In the sequel, we assume that $s>0$,  hence  $\sinh s >0$. 
We first consider the real part of $U_0 (\epsilon)$. 
Since $U_0 (\epsilon)$ is contained in the
tuboid, the real part of $U_0 (\epsilon)$ consists of 
space-like vectors $x \equiv \Re z$ with 
$|x\cdot x| \in (0,r^2)$. Further, $x$ is in the wedge
$\widetilde{W_1}$ of ambient Minkowski space, \emph{i.e.}, the wedge
	\[
		\widetilde{W_1} 
		\doteq \bigl\{ x \in \mathbb{R}^{1+2} \mid |x_0| < x_2 \bigr\} \; .  
	\]
Thus, $\Re U_0 (\epsilon) $ is the disjoint union 
	\[
		\Re U_0 (\epsilon)  =  
		\bigcup_{0< r' < r} W_0  (r'),\quad  
	\]
where $W_0 (r')$ is the intersection of 
$U_0(\epsilon)$ with the set of space-like
vectors in ambient Minkowski space with $x\cdot x = -(r')^2$, which we
denote here by $dS(r')$. Now recall that the real part of $z(t+i \theta,\hat t + i \hat  \theta)$ is 
	\[
		x (\tau,\widehat \tau) 
		= \Lambda_1(t) \bigl( \Jgeo(\theta) \cos \widehat{\theta} 
		- \sin \theta \sin \widehat{\theta} \, \Lgeo \, \hat{\Lgeo} \, \bigr) 
		\widehat{\Lambda} (\hat{t} )\, o \; . 
	\]
In particular, for $\theta = 0$ we have $x(t,\widehat \tau) = \cos
\widehat \theta \cdot \Lambda_1(t) \hat \Lambda(\hat t) o$.
Since $\cos \widehat{\theta}$ runs through the open 
interval $(0,1)$ and $\Lambda_1(t)  
\widehat{\Lambda}(\hat{t}) o$ runs through the entire wedge~$W_1$ (in $dS$),
it follows that for each $r'$ in $(0,r)$ the set of
$x(\tau,\widehat\tau)$ with $\theta =0$, intersected with $dS(r')$, 
coincides with $\widetilde{W_1}\cap dS(r')$. On the other hand, this set is
contained in the closure of $W_0(r')$. We therefore have 
	\[
		W_0(r')
		\subset \widetilde{W_1}\cap dS(r') \subset 
		\overline{W_0 (r')} \; .
	\] 
Thus, $W_0(r')$ is an open\footnote{The 
set $W_0(r')$ is open in the topology 
induced on $dS(r')$ from ambient Minkowski space.} subset of the wedge
$\widetilde{W_1}\cap dS(r')$ whose closure 
coincides with the closure of the wedge, and which is invariant
under the $\Lambda_1$-boosts. The only such set is the wedge itself, and we
conclude that 
	\[
		W_0 (r')= \widetilde{W_1}\cap dS(r') \; . 
	\]
If we include the rotations, setting $W_\alpha (r') \doteq R_0(\alpha) W_0 (r') $, 
we find the \emph{real part of the entire tuboid} $\mathcal{T}_+$: 
	\begin{equation}
	\label{tub-Re}
		\Re \mathcal{T}_+
		= \bigcup_{\alpha \in [0, 2 \pi) }
		\bigcup_{0 \le r' < r} W_\alpha (r') \; .
	\end{equation}

Next, we consider the imaginary part of $U_0 (\epsilon)$: 
according to \eqref{eqzT+'}, the imaginary part of a point in 
$U_0 (\epsilon)$ is of the form
	\begin{align*}
		y (\tau,\widehat{\tau})
		& \doteq 
		\Im \Lambda_1 (t + i \theta)  \Lambda_2 (s) 
		\Lambda_1( \hat{t} + i \widehat{\theta}) o
		\\
		& =  
		r \cdot \Lambda_1 (t) \left[ 
		 \sin \widehat{\theta}    
		 \begin{pmatrix} 
		 \cos \theta \cosh s \cosh \hat{t} 
		\\  \sinh s \cosh \hat{t}  \\  \cos \theta \sinh \hat{t} 
		\end{pmatrix} +  \sin \theta
		\cos \widehat{\theta} \begin{pmatrix}
				\cosh \hat{t}  \\
				0  \\
				\cosh s \sinh \hat{t}  
				\end{pmatrix} \right]\; . 
	\end{align*}
For $(t + i \theta, \hat{t} + i \widehat{\theta}) \in V_0 (\epsilon)$ the 
coefficients $\sin \widehat{\theta}$ and 
$\sin \theta \cos \widehat{\theta}$ of the vectors in 
the square brackets both take values in the interval $(-1, 0)$.
In particular, for $ \theta =0$, 
	\begin{align*}
		y (t,\widehat{\tau})
		& =  
		\Lambda_1 (t)  \Lambda_2 (s)
		\begin{pmatrix} r  \sin \widehat{\theta} 
		\cosh \hat{t}  \\ 0 \\  r  \sin \widehat{\theta} \sinh \hat{t} 
		\end{pmatrix} \; . 
	\end{align*}
As $\hat{t}$ goes from $-T$ to~$T$, the vector   
	\[
		\hat{t} \mapsto \Lambda_2 (s) \begin{pmatrix} r  \sin \widehat{\theta} 
		\cosh \hat{t}  \\ 0 \\  r  \sin \widehat{\theta} \sinh \hat{t} 
		\end{pmatrix} \; ,  \qquad 0< \widehat{\theta} < \frac{\pi}{2} \; , 
		\qquad   | \hat{t} \,  | < T \; ,   
	\]
yields a (bounded, connected) segment  of a hyperbola 
on the mass shell with ``mass'' $r  \sin \widehat{\theta}$, 
which is tilted 
by the boosts $\Lambda_2 (s)$, with $s \in \mathbb{R}$ fixed, 
but remains on the same mass shell. 
As $\sin \widehat{\theta} < 0 $ for $\widehat{\theta} \in (-\pi, 0)$, 
the $y_1$-component of this vector 
is always negative. As $- T <  \hat{t}  < T$, it is also bounded.
Next, we apply the boosts $t \mapsto \Lambda_1 (t)$, which also 
preserve the mass shell with ``mass'' $r  \sin \widehat{\theta}$. 
The resulting regions are unbounded, but the  
$y_1$-component satisfies
	\[
		r \sinh s \sin \widehat{\theta}
		<  y_1 (t , \widehat{\tau})  = 
		 y_1 (0 , \widehat{\tau}) < r \sinh s
		\cosh T \sin \widehat{\theta}
		\qquad \forall t \in \mathbb{R} \; . 
	\]
The vectors of 
the form $y(t , \hat{t}+ i \widehat{\theta})$,  with $\widehat{\theta}$ fixed, 
span the region
	\begin{align*}
		Z_0 (\widehat{\theta}) & \doteq \bigl\{ y \in \mathbb{R}^{1+2}  
		\mid  r \sinh s \sin \widehat{\theta} < y_1 < r \sinh s
		\cosh T \sin \widehat{\theta} \bigr\}
		\\
		& 
		\qquad \cap 
		\bigl\{ 
		y \in \mathbb{R}^{1+2}
		\mid y \cdot y =  r^2  \sin^2 \widehat{\theta} , y_0 > 0	\bigr\}   \; .  
	\end{align*}
Taking the union over 
$\widehat{\theta} \in (- \pi/2 , 0]$ still yields a subset   
of $\Im U_0 (\epsilon) $, \emph{i.e.}, 
	\[
		\Im U_0 (\epsilon)  \supseteq 
		\bigcup_{- \pi/2< \widehat{\theta} < 0} Z_0 (\widehat{\theta}) \; . 
	\]
If we include the rotations, setting $Z_\alpha (\widehat{\theta}) 
		\doteq R_0(\alpha) Z_0 (\widehat{\theta}) $, 
we find the \emph{imaginary part of the entire 
tuboid} $\mathcal{T}_+$: 
	\begin{equation}
	\label{tub-Im}
		\Im \mathcal{T}_+
		= \bigcup_{\alpha \in [0, 2 \pi) }
		\bigcup_{- \pi/2< \widehat{\theta} < 0} Z_\alpha (\widehat{\theta}) \; .
	\end{equation}
We note that from \eqref{tub-Im} and \eqref{tub-Im} one can not conclude that 
	\begin{equation}
	\label{tub-Re-Im}
		\bigcup_{\alpha \in [0, 2 \pi) } R_0 (\alpha)  
		U_0 (\epsilon)  
 		= 
		\mathcal{T}_+ 
		\setminus \bigl\{  z \in dS_\mathbb{C} \mid \Re  z = 0 \bigr\} 
		\; ;  
	\end{equation}
however, \eqref{tub-Re-Im} follows directly from \eqref{cartan-2} and
\eqref{chiral-tuboids}. 
Inspecting the restrictions on~$\widehat{\tau}$ in 
$V_0 (\epsilon)$, we note that  $U_0 (\epsilon)$
actually contains \emph{open 
complex neighbourhoods} of the subsets $W_\alpha (r')$
and $Z_\alpha (\widehat{\theta}) $. 
Hence, it is sufficient to take the union over \emph{rational} rotations: 
	\begin{equation}
	\label{U-union}
		\bigcup_{\alpha \in \mathbb{Q} \cap [0, 2 \pi) } R_0 (\alpha) 
		U_0 (\epsilon) = 
		\mathcal{T}_+ 
		\setminus \bigl\{  z \in dS_\mathbb{C} \mid \Re  z = 0 \bigr\}
		\; .  
	\end{equation}
This union yields a countable\footnote{A countable
covering allows us to iteratively extend the domain of the analytic function.}
covering of the tuboid $\mathcal{T}_+$ 
(with the exception of the set $\bigl\{  z \in dS_\mathbb{C} \mid \Re  z = 0 \bigr\}$)
with regions of the form  
$R_0 (\alpha) U_0 (\epsilon)  \; $, $\alpha \in [0, 2 \pi)$. 	
The next step is to define functions for the 
rotated regions. Let 
	\[
		U_{\alpha} (\epsilon) \doteq R_0(\alpha) U_0 (\epsilon)  \; , 
		\qquad
		\alpha \in [ 0, 2 \pi) \; , 
	\]
and set
	\[
		z_\alpha (\tau,\widehat{\tau})
		\doteq R_0(\alpha) z(\tau,\widehat \tau) \; , 
		\qquad  (\tau,\widehat{\tau}) \in 
		V_0 (\epsilon)  \;  .  
	\]
For $\alpha$ fixed, $z_\alpha$ 
is a biholomorphic map from $V_0 (\epsilon)$ 
onto $U_{\alpha} (\epsilon) $.
With the same steps as above, we get an analytic extension of the
restriction $f_{| R_0(\alpha) W_1}$ of $f$  
to~$U_{\alpha} (\epsilon) $, which we denote by $f_{\alpha}$. 

\bigskip
\noindent
\emph{(Part IV: Uniqueness of the analytic continuations).}
Let $\alpha, \alpha' $ be two rational angles. 
We have to show that on the intersection $U_{\alpha} (\epsilon)
\cap U_{\alpha'} (\epsilon)$
the functions $f_{\alpha}$ and $f_{\alpha'}$ coincide. More precisely, 
if $ z_\alpha (\tau,\widehat{\tau}) = z_{\alpha'} (\tau',\widehat{\tau}')$, 
then 
	\begin{equation}
	\label{fa=fa'}
		f_{\alpha} \bigl( z_\alpha (\tau,\widehat{\tau}) \bigr)
		= f_{\alpha'} \bigl( z_{\alpha'} (\tau',\widehat{\tau}') \bigr) \; . 
	\end
	{equation}
Indeed, \eqref{fa=fa'} follows from the uniqueness principle for analytic 
functions, which holds due to the following three facts 
\begin{itemize}
\item [---] the $U_{\alpha}(\epsilon)$'s are tuboids (see 
Lemma~\ref{Jens-tube} below);  
\item [---] any point $ z_\alpha (\tau,\widehat{\tau}) 
= z_{\alpha'} (\tau',\widehat{\tau}') $ 
(and any point in a small neighbourhood of this point)
is connected to the boundary by a path lying entirely 
in $U_{\alpha} (\epsilon) 
\cap U_{\alpha'}(\epsilon)$
(this is Lemma~\ref{path-in-tube} below)  
\item [---]  for 
$z_\alpha (t,\hat{t}) = z_{\alpha'} (t',\hat{t}') \equiv x \in dS$, 
the boundary values coincide:
	\[
		f_{\alpha} \bigl( z_\alpha (t,\hat{t}) \bigr) = f(x) 
		= f_{\alpha'} \bigl( z_{\alpha'} (\tau',\widehat{\tau}') \bigr)  \; . 
	\]
\end{itemize}

\bigskip
\noindent
\emph{(Part V: An analytic continuation from $\mathcal{T}_+ 
		\cap \{  z \in dS_\mathbb{C} \mid \Re  z = 0 \}$ 
to $\mathcal{T}_+$).}
We first note that \eqref{f-extension} defines 
$f_{W_1,\widehat W}\big(z(\tau,\widehat \tau)\big)$ for the half-hyperbola
	\[
		\bigl\{ z \in dS_\mathbb{C} \mid 
		z = \Lambda_1 (-t) \left( \begin{smallmatrix} 
		- i r \\
		0 \\
		0
		\end{smallmatrix} \right) , t > 0 \bigr\}  \; . 
	\]
Applying the rotations $R_0(\alpha)$, $\alpha \in [0, 2 \pi)$, to this set, 
and taking the union, we find the region
	\begin{align}
		& \bigl\{ z \in dS_\mathbb{C} \mid 
		z = \Lambda  \left( \begin{smallmatrix} 
		- i r \\
		0 \\
		0
		\end{smallmatrix} \right) , \Lambda \in SO_0(1,2)  \bigr\}  
		\cap \bigl\{ z \in dS_\mathbb{C} \mid
		\Re z = 0 \bigr\} 
		\nonumber \\
		& \qquad \qquad \qquad \qquad
		\qquad \qquad
		= \mathcal{T}_+ \cap \bigl\{ z \in dS_\mathbb{C} \mid
		\Re z = 0 \bigr\} \; . 
	\label{img-mass-shell}
	\end{align}
Hence, the functions 
$f_{\alpha} \bigl( z_\alpha (\tau,\widehat{\tau}) \bigr)$
define an analytic extension to the subset 
\eqref{img-mass-shell} of $\mathcal{T}_+$.
The continuation across this subset is unique, 
as the set  \eqref{img-mass-shell}
is a two-dimensional (counting real dimensions)
subset of the four-dimensional 
tuboid $\mathcal{T}_+$ and the function is already 
extended to a complex neighbourhood.
\end{proof}	
 
\begin{lemma} 
\label{Jens-tube}
The set $U_{0}(\epsilon)$  is a tuboid over
$W_1$ in the sense of Definition~\ref{def-1.6.4}, 
with profile 
	\[
		\mathcal{P}
		= \bigcup_{x (t, \hat{t}) \in  W_1} 
		\bigl( x (t, \hat{t}), \mathcal{P}_{x (t, \hat{t})} \bigr)
	\] 
with 
	\[
		\mathcal{P}_{x (t, \hat{t})}
		\doteq 
		\big\{ \theta \Lgeo 
		x  (t, \hat{t})
		+ \widehat{\theta} \Lambda_{1}(t)
		\hat{\Lgeo} \Lambda_{\widehat{W}} (\hat{t}) o \mid
		\theta, \widehat{\theta} \in   (- \pi,0 )  \big\}  
	\]
where $x (t, \hat{t}) =  \Lambda_{1}(s) 
\Lambda_{ \widehat{W}} (t) o$, $\Lgeo = \Lgeo_{W_1}$ and 
$\hat{\Lgeo} = \Lgeo_{ \widehat{W}}$. 
\end{lemma}

\begin{proof}
Let us consider an open neighbourhood 
$\mathcal{N}_{TdS}\bigl(x (t, \hat{t}), 0\bigr)$ in $TdS$ 
of a point $x (t, \hat{t}) \in W_1$. For example, we may choose
$\theta,\widehat{\theta}\in(-\epsilon,\epsilon)$ in $\mathcal{P}_{x (t, \hat{t})}$, 
and define the local diffeomorphism  $\Xi \colon 
\mathcal{N}_{TdS}(x (t, \hat{t}), 0) \to dS_\mathbb{C}$ by  
	\begin{align*}
		& \Xi \big( x (t, \hat{t}) \, , \, \theta \Lgeo 
		x  (t, \hat{t})
		+ \widehat{\theta} \Lambda_{1}(t)
		\hat{\Lgeo} \Lambda_{\widehat{W}} (\hat{t}) o \big)  
		\doteq
		\Lambda_{1}(t+i \theta )
		\Lambda_{\widehat{W}} (\hat{t}+ i \widehat{\theta} ) o  \; . 
	\end{align*}
It is admissible in the sense of Definition~\ref{def-1.6.4}, in particular
	\[ 
		\frac{1}{i} \frac{d}{d\lambda} \Xi(x,\lambda y) \Bigl. 
		\Bigr|_{\lambda =0} = i y \; .
	\]
The proof of this fact uses
	\[
		\Lambda_{W}(t+i\theta) = \Lambda_{W}(t)\big( \Jgeo_W(\theta) + i \sin
		(\theta) \Lgeo_W \big) \quad \text{and} \quad \frac{1}{i}
		\frac{\rm d}{{\rm d}\theta} \Jgeo_W(\theta) 
		\Bigl. 
		\Bigr|_{\theta =0}=0 \; .  
	\]
\end{proof}

\begin{lemma}
\label{path-in-tube}
For every $\alpha, \alpha'$ such that $U_\alpha(\epsilon) \cap U_{\alpha'}(\epsilon)$ 
is non empty, it is connected and touches the real boundary $dS$.  
\end{lemma} 

\begin{proof}
It suffices to consider the case $\alpha'=0$ and $\alpha>0$.
In a first step, consider $\theta=0$: Let $U_0(0)$ 
be the set of $z$ in
the closure\footnote{The closure is needed as the points 
with $\theta=0$ are not in the
interior of $U_0(\epsilon)$.} 
of $U_0(\epsilon)$ such 
that $z=z(t,\hat{t} + i \widehat{\theta})$
(with $\theta=0$). Suppose, the point 
	\[ 
		z(t,\hat t + i \widehat \theta) \equiv 
		\Lambda_1(t)\big( \cos \widehat \theta + i \sin\widehat \theta \, \hat
		l \big) \widehat{\Lambda}(\hat t) o \doteq x+iy
	\]
is in $U_0(0) \cap U_\alpha(0)$. 
Then there are $t',\hat{t}'$ and $\widehat
{\theta}'$ such that   
	\begin{equation} 
		\label{eqz=Rz}
		z \bigl(t,\hat{t} + i \widehat{\theta} \bigr) 
		= R_0(\alpha) z \big(t',\hat{t}' + i \widehat{\theta}'\big) \; . 
	\end{equation}
Since $x\cdot x = - (r\cos\widehat \theta)^2$ and $y\cdot y =
(r\sin\widehat \theta)^2$ and the Lorentz square is conserved by the
rotation, $\widehat {\theta}'$ must be equal to $\widehat \theta$, and  
	\[
		\Lambda_1(t) \widehat{\Lambda}(\hat t) o 
		= R_0(\alpha)\,\Lambda_1(t')
		\widehat{\Lambda}(\hat{t}') o  
		\quad \text{and} \quad 
		\Lambda_1(t) \widehat{\Lambda}(\hat{t})\hat{l} o 
		= R_0(\alpha)\,\Lambda_1(t') \widehat{\Lambda}(\hat{t}') \hat l o \; .
	\]
It follows that, for every $s\in[0,1]$, 
	\[
		z(t,\hat t + is\widehat \theta) 
		= R_0(\alpha) z\big(t',\hat t' + is \widehat{\theta} \, \big)
		\doteq z(s) \; . 
	\]
Thus, the map $s\mapsto z(1-s)$, $s\in[0,1)$,  furnishes a path from
$z(t, \hat {t} + i\widehat \theta)$ to $z(t, \hat{t})$ within $U_0(0) \cap 
U_\alpha(0)$, which terminates at the real boundary of both 
$U_0(0) $ and~$U_\alpha(0)$.   

We have shown that every point $z$ in $U_0(0) \cap 
U_\alpha(0)$ 
(\emph{i.e.}, with $\theta=0$) is connected  to the real boundary of this region
by a path within that region. 
For $\epsilon>0$ sufficiently small, 
the same holds for $U_0(\epsilon) \cap 
U_\alpha(\epsilon)$,  
by continuity.  
\end{proof} 

The following result clarifies the relation between the  tuboid 
${\mathcal T}_+$, as described in Lemma~\ref{t+2}, and the tangent bundle $TdS$.  
 
\begin{lemma}[Bros \& Moschella \cite{BM}, p.~339]
\label{tds} 
Let $ {\mathcal P}^+_{ x}$ be defined by \eqref{px}. 
The map 
	\begin{align} 
		\label{tm}
				 ( x,  y ) & \mapsto \sqrt{1- \frac{y \cdot y}{r^2} } \;  x + i  y    
	\end{align}
defines a diffeomorphism from 
	\[
		\bigcup_{ x \in dS} \left(  x,  {\mathcal P}^+_{ x} 
			\cap \{  y \in V^+\mid  y \cdot y < r^2 \}  \right) 
	\]
onto  $ {\mathcal T}_+ \setminus \bigl\{  z \in dS_\mathbb{C} \mid \Re  z = 0 \bigr\} $.  
\end{lemma}

\section{The Euclidean sphere}
\label{sec:1.7}

As we have seen in Section \ref{SS:1.6.1}, the open upper and lower 
hemisphere $S_+$ and $S_-$ are 
contained in the tuboids~$\mathcal{T_-} $ and~$\mathcal{T_+} $, respectively. 
In the sequel, the \emph{Euclidean sphere}\index{Euclidean sphere}\footnote{We have
changed the notation; see  \eqref{euclidsphere} for comparison.}
	\[ 
		S^2 \doteq \bigl\{ \vec{{\tt x}} \in \mathbb{R}^{3} 
		\mid  {\tt x}_0^2 +  {\tt x}_1^2 + {\tt x}_2^2 = r^2  \bigr\} \; , 
	\]
embedded in ${\mathbb R}^3$, will play an important role in the 
construction of interacting theories. 
Hence it is worth while to recall a few  
basic facts. Let $\vec 0 =(0, 0, 0)$ denote 
the origin in ${\mathbb R}^3$. The  closed upper (resp.~lower) 
hemisphere\index{hemisphere} is 
\label{spherepluspage}
	\[
		\overline{S_\pm} \doteq \bigl\{ \vec{{\tt x}} \in S^2 
		\mid \pm {\tt x}_0 \ge 0 \bigr\}  \; . 
	\]
The equator\index{equator} $S^1 \doteq \bigl\{  \vec{{\tt x}} \in S^2 
\mid {\tt x}_0  =0 \bigr\} = \partial S_\pm$ 
forms the boundary of both $S_+$ and~$S_-$. Thus 
\label{equatoreuclidpage}
\label{eulidspherepage}
	\[ 
		S^1 = \overline{S_+}  \cap \overline{S_-} \; . 
	\]
The \emph{Euclidean time reflection}\index{time reflection (Euclidean)}  
	\begin{equation}
		\label{deftimerefl}
		T \colon ( {\tt x}_0, {\tt x}_1, {\tt x}_2)\mapsto (- {\tt x}_0, {\tt x}_1, {\tt x}_2)
	\end{equation}
maps~$S_\pm$ onto $ S_\mp$ and leaves $S^1$
invariant. $S^1$ itself is the \emph{disjoint} union 
	\[  
		I_+  \mathbin{\dot{\cup}} \bigl\{ ( 0,- r, 0) , ( 0,r, 0)  \bigr\} 
		\mathbin{\dot{\cup}} I_- \; ,  
	\]
with $I_\pm \doteq \bigl\{  \vec{{\tt x}} \in S^1 \mid   
\pm {\tt x}_2> 0 \bigr\}$ open half-circles. 
\index{half-circle}
Moreover, the Euclidean time reflection $T$ can be used to turn the 
Euclidean scalar product into 
the Minkwoski scalar product: 
	\begin{equation}
	\label{sp-e-m}
		\vec {\tt x} \, \cdot T \vec{\tt y} = {\tt x}_0 (- {\tt y}_0) 
		+ {\tt x}_1   {\tt y}_1 + {\tt x}_2   {\tt y}_2  \; . 
	\end{equation}
We will see in the sequel that this has important consequences. 

\bigskip

\goodbreak
We will now define two charts, which together provide an atlas for the sphere.

\subsection{Geographical coordinates}
\label{s-geo-chart}
\index{geographical coordinates}
The chart\footnote{If necessary, we restrict this map to $- \pi < \varrho <  \pi$, 
so that it provides a proper 
chart in the sense of differential geometry.}
	\[
			\begin{pmatrix}
					        {\tt x}_0 \\
						{\tt x}_1 \\
						{\tt x}_2 
			\end{pmatrix} 
				= \begin{pmatrix}
						r \sin \vartheta  \\
 						r \cos \vartheta \sin \varrho  \\
 						r \cos \vartheta \cos \varrho  \\
				\end{pmatrix} \; , \qquad - \frac{\pi}{2} <\vartheta <  \frac{\pi}{2} \;, 
				\quad  - \pi \le \varrho <  \pi \; , 
	\]
covers the sphere, except for the geographical poles\index{geographical pole}
$(\pm r, 0, 0) \in {\mathbb R}^3$. Refer to $(\vartheta, \varrho )$ as {\em geographical 
coordinates}\index{geographical coordinates}. The equator\index{equator} $S^1  
= \{ (\vartheta, \varrho ) \mid \vartheta =0 \}$ and the point $ (\vartheta, \varrho) 
\equiv (0, 0)$ is mapped to the origin $ \vec o =(0, 0,r)$. The restriction of the Euclidean metric
\index{Euclidean metric} to this chart is
	\[
		g = r^2 {\rm d} \vartheta \otimes {\rm d} \vartheta 
		+ r^2 \cos^2 \vartheta \; ({\rm d} \varrho 
		\otimes {\rm d} \varrho )
	\]
and
	\begin{align}
		\label{L1}
		\Delta
		&= |g|^{-1/2} \partial_\mu \Bigl( | g|^{1/2} g^{\mu \nu} \partial_\nu \Bigr)
		=  \frac{1}{r^2 \mathbb{cos}_\vartheta^2 } 
			\left( \Bigl( \mathbb{cos}_\vartheta \frac{\partial}{\partial \vartheta} \Bigr)^2
			+ \frac{\partial^2}{\partial \varrho^2}  \right) \; .
	\end{align}
The surface element \index{surface element} on $S^2$ is 
${\rm d}\Omega(\vartheta, \varrho ) = r^2 \cos \vartheta \,
{\rm d} \vartheta {\rm d} \varrho $. 

\subsection{Path-space coordinates}
\index{path-space coordinates}
The chart  
	\begin{equation}
		\label{path-space-coo}
		\begin{pmatrix}
			{\tt x}_0 \\
			{\tt x}_1 \\
			{\tt x}_2 
		\end{pmatrix} 
			= 	\begin{pmatrix}
 						r \sin \theta \cos \psi  \\
						r \sin  \psi  \\
						r \cos \theta \cos \psi  \\
				\end{pmatrix} \; , \qquad 0 \le \theta < 2 \pi  \; , 
				\quad - \frac{ \pi}{2} < \psi <  \frac{ \pi}{2} \; , 
	\end{equation}
covers the sphere with the exception of 
the two points $(0, \pm r, 0) \in {\mathbb R}^3$. We refer to this chart as {\em path-space 
coordinates}\index{path-space coordinates}. The point $(\theta, \psi)\equiv (0, 0)$ is mapped 
to the origin $\vec o =(0, 0,r)$. The restriction of the Euclidean metric to this chart  is
\label{Laplacepage}
	\[
	g =  \cos^2 \psi  \, ({\rm d} \theta \otimes {\rm d} \theta) 
	+ {\rm d} \psi \otimes {\rm d} \psi  
	\]
and
	\begin{equation}
	\label{L2}	
	\Delta
		= \frac{1}{r^2 \operatorname{\mathbb{cos}}_\psi^2} 
		\left( \frac{\partial^2}{\partial \theta^2}  
		+ \Bigl( \operatorname{\mathbb{cos}}_\psi 
		\frac{\partial}{\partial \psi} \Bigr)^2\right) .
	\end{equation}
The surface element \index{surface element} on $S^2$ is 
	$
		{\rm d}\Omega(\theta, \psi ) 
		= r^2 \cos \psi \,{\rm d} \theta {\rm d} \psi $.

\subsection{The Laplace operator}
The expressions in \eqref{L1} and \eqref{L2} both extend to the 
self-adjoint \emph{Laplace operator}\index{Laplace operator}  
$\Delta_{S^2}$ on $L^2(S^2, {\rm d} \Omega)$.  $- \Delta_{S^2}$~has non-negative
discrete spectrum and an isolated simple eigenvalue at zero with eigenspace the constants. 
As a consequence, the only smooth solution of the equation
	\[
		( - \Delta_{S^2} + \mu^2) f = 0 \; , \qquad \mu^2 >0 \; , 
	\]
is $f=0$, \emph{i.e.}, $f$ vanishes identically. 

\chapter{Space-time Symmetries}
\label{isometrygroup}

One of the objectives of this work is to emphasise the role 
space-time symmetries and their representations play in the 
construction of interacting quantum field theories. In this chapter, 
we will therefore discuss the symmetries of de Sitter space 
in some detail. 

\section{The isometry group of de Sitter space}

The isometry group of $( dS ,  g )$ is $O(1,2)$. Its linear action on the ambient
space $\mathbb{R}^{1+2}$ is given by $3\times 3$-matrices acting on vectors
$\left(\begin{smallmatrix}x_0\\ x_1 \\ x_2\end{smallmatrix}\right)\in\mathbb{R}^{1+2}$.  
The group 
	\[ 
		O(1,2) = O^\uparrow_+ (1,2)  \cup O^\downarrow_+ (1,2)  
			\cup O^\uparrow_- (1,2) 
				\cup O^\downarrow_- (1,2) 
	\]
has  four connected components~\cite{SW}, namely those (distinguished by $\pm$),
which preserve or change the 
orientation and those (distinguished by $\uparrow \downarrow$), which preserve or change the 
time orientation. Group elements, which preserve the orientation, are called {\em proper}. 
Lorentz transformations, which preserve the time orientation, are called 
{\em orthochronous}. The connected component containing the identity is the 
\index{orthochronous Lorentz group}\index{proper Lorentz group}\emph{proper, orthochronous 
Lorentz group}\index{Lorentz group}, denoted as $SO_0 (1,2) \equiv  O^\uparrow_+ (1,2)$. 
The group $SO_0(1,2)$ acts transitively on the de Sitter space $dS$. 

The \emph{isometry group}\index{isometry group} of the ambient 
space $\mathbb{R}^{1+2}$ is the Poincar\'e group~$E(1,2)$.
The stabiliser of the zero vector~$ \boldsymbol{0} \equiv (0,0,0)  
\in \mathbb{R}^{1+2}$ is the subgroup $O(1,2)$ of $E(1,2)$. 

\begin{lemma}
\label{lm:2.1.1}
The \index{group action} 
action of the group $O(1,2)$ splits $\mathbb{R}^{1+2}$ into \index{orbit} 
orbits\footnote{In other 
words, the sets $\partial V^+ \cup \partial V^-$, 
$dS$ and $H^+_m \cup H^-_m$ are \emph{$G$-sets}\index{$G$-set} 
for the group $G=O(1,2)$.}:
\begin{itemize}
\item[$i.)$] $\{ g \, \boldsymbol{0} \mid g \in O(1,2) \} = \{ \boldsymbol{0} \} $, \emph{i.e.}, 
the group $O(1,2)$  leaves the zero vector \index{zero vector} $\mathbf{0} =(0, 0, 0)$ invariant;
\item[$ii.)$]  $\bigl\{ g \left( \begin{smallmatrix} m \\ 0 \\ 0 \end{smallmatrix} \right) \mid 
g \in O(1,2) \bigr\} = H^+_m \cup H^-_m$, where
\label{masshyperboloid}
	\[
		H^\pm_m \doteq  \bigl\{ x \in \mathbb{R}^{1+2} 
		\mid x_0^2 - x_1^2 - x_2^2 = m^2 , \pm x_0 >0 \bigr\} \; .   
	\] 
More generally, the orbit of any point in the interior of the forward light-cone is a 
two-sheeted \emph{mass hyperboloid}\index{mass hyperboloid} 
$H^+_m \cup H^-_m$ for some mass $m>0$; 
\item[$iii.)$]  $\{ g o \mid g \in O(1,2) \} = dS$. More generally, the orbit 
of any point, which is space like to 
the zero vector $\boldsymbol{0}$, 
is a de Sitter space $dS_r$ of some radius $r>0$; 
\item[$iv.)$] $\bigl\{ g \left( \begin{smallmatrix} 1 \\ 0 \\ -1 \end{smallmatrix} \right) 
\mid g \in O(1,2) \bigr\} = (\partial V^+ \cup \partial V^-) 
\setminus \{ \boldsymbol{0} \}$. More generally, the orbit of any point, which 
is light-like to the zero vector  $\boldsymbol{0}$, 
is $(\partial V^+ \cup \partial V^-) 
\setminus \{ \boldsymbol{0} \}$. 
\end{itemize}
The Minkowski space $\mathbb{R}^{1+2}$ is the disjoint union of all of these sets; 
in $iii.)$ and $iv.)$ the union is over all $m>0$ and $r>0$, respectively.
\end{lemma}
 
\begin{proof}
If $X$ is $\{ \mathbf{0} \}$, $\partial V^+ \cup \partial V^-$, $dS$, or $H^+_m 
\cup H^-_m$,  then
	\[
	\Lambda (\Lambda' x ) = (\Lambda \circ \Lambda') x \in X \; , 
	\quad \Lambda, \Lambda ' \in O(1,2) \; , \quad x \in X \; . 
	\]
In particular, $\Lambda (\Lambda^{-1} x )= (\Lambda \circ \Lambda^{-1} ) x 
= x$ for all $x \in X$.
Moreover, the group $O(1,2)$ acts transitively
on $\partial V^+ \cup \partial V^-$, $dS$  and $H^+_m 
\cup H^-_m$:
	\begin{align*}
		\partial V^+ \cup \partial V^- & = \bigl\{ T^k R_0 (\alpha) \Lambda_1(t) \left( 
		\begin{smallmatrix}
		1\\
		0\\
		-1\\
		\end{smallmatrix} \right)
		\mid k= 0,1, \;  t \in \mathbb{R}, \alpha \in [0, 2 \pi) \bigr\} \; , \\
		dS & = \bigl\{ R_0 (\alpha) \Lambda_1(t) o \mid t \in \mathbb{R}, 
		\alpha \in [0, 2 \pi) \bigr\} \; , \\
		H^+_m \cup H^-_m & = \bigl\{ T^k R_0 (\alpha) \Lambda_1(t) \left( 
		\begin{smallmatrix}
		m\\
		0\\
		0\\
		\end{smallmatrix} \right)
		\mid k= 0,1, \;  t \in \mathbb{R}, \alpha \in [0, \pi) \bigr\} \; .
	\end{align*}
Here $T$ denotes the time reflection; see Section \ref{sec:reflections} below. 
\end{proof}

\subsection{The action of $SO_0(1,2)$ on the light-cone}
In the sequel, the action of $SO_0(1,2)$ on the forward light 
cone\footnote{In the second line, we have set $p_0 (t) = {\rm e}^{-t}$, 
$t \in \mathbb{R}$.} 
	\begin{align*}
		\partial V^+ & = \left\{ R_0 (\alpha) \Lambda_1 (t) \left( \begin{smallmatrix}
		1 \\  0 \\ -1 \end{smallmatrix} \right) \mid t \in \mathbb{R}, 
		\alpha \in [0, 2 \pi) \right\} \\
		& = \left\{ \left( \begin{smallmatrix} p_0	 \\ p_0 \sin \alpha \\ - p_0 \cos \alpha
		\end{smallmatrix} \right) \mid p_0 >0, \alpha \in [0, 2 \pi) \right\} 
	\end{align*}
will play an important role.  We therefore 
provide explicit formulas for the action of the boosts $\Lambda_{1}(t)$, $\Lambda_{2}(s)$ 
and the rotations $R_0(\beta)$ on $\partial V^+$: 
	\begin{equation} 
	\label{lambda1-s}
				\Lambda_{1}^{-1}(t)  
									\begin{pmatrix}
 								p_0 \\
								p_0 \sin \alpha   \\
								- p_0 \cos \alpha  
									\end{pmatrix}   
									= p_0
								\begin{pmatrix}
 								\cosh t + \sinh t \cos \alpha \\
								\sin \alpha   \\
								- \sinh t - \cosh t \cos \alpha  
									\end{pmatrix}    
				\end{equation}
and
	\begin{equation} 
	\label{lambda2-s}
					\Lambda_{2}^{-1}(s) 
					\begin{pmatrix}
 								p_0 \\
								p_0 \sin \alpha   \\
								- p_0 \cos \alpha  
									\end{pmatrix}  
									=
									p_0
									 \begin{pmatrix}
 								\cosh s -  \sinh s \sin \alpha \\
							       - \sinh s + \cosh s \sin \alpha  \\
								- \cos \alpha  
					\end{pmatrix} 
													 \;  . 
		\end{equation} 
Finally, 
	\begin{equation} 
	\label{R0-s}
				R_{0}^{-1}(\beta)  
									\begin{pmatrix}
 								p_0 \\
								p_0 \sin \alpha   \\
								- p_0 \cos \alpha  
									\end{pmatrix}   
									= p_0
								\begin{pmatrix}
 								1 \\
								\cos \beta \sin \alpha - \sin \beta  \cos \alpha \\
								- \sin \beta \sin \alpha - \cos \beta  \cos \alpha  
									\end{pmatrix}  \; .   
				\end{equation}
The equations \eqref{lambda1-s}, \eqref{lambda2-s} and \eqref{R0-s} 
will be used in Chapter \ref{ch:3}, 
where we will provide explicit formulas for the induced representations of the proper Lorentz group~$SO_0(1,2)$. 

\subsection{Reflections}
\label{sec:reflections}
The \emph{time reflection}\index{time reflection} $T$, 
the \emph{parity transformation}\index{parity transformation}
$P_1$ and the reflection $\Theta_{W_1}$ at the edge of the wedge $W_1$, 
\label{timerefl-page}
		\begin{equation}
		\label{theta-W}
			T\doteq  \begin{pmatrix}
					-1 &  0 &0 \\
					0 &  1 & 0  \\
					0  & 0 & 1   
 				\end{pmatrix}  \; , \quad 	P_1\doteq  \begin{pmatrix}
					1 &  0 &0 \\
					0 &  1 & 0  \\
					0  & 0 & -1   
 				\end{pmatrix} \; , 
				\quad \Theta_{W_1}:= P_1 T  \in \; O (1,2) \; ,
		\end{equation}
leave the Cauchy surface $S^1$ invariant. $\Theta_{W_1}$ is the  reflection at 
the edge of the wedge~$W_1$. Together $P_1$ and $T$ generate the 
\index{Klein four group} Klein four group $ \mathbb{Z}_2 \times \mathbb{Z}_2$ 
as $T^2=\mathbb{1}$ and~$P_1^2 = \mathbb{1}$.  

\bigskip
The reflection at the edge of an arbitrary  wedge $W = \Lambda W_1$,  is 
\label{PTwedgepage}
	\begin{equation*} 
	\label{PTwedge}
		\Theta_{ W } \doteq \Lambda\;  P_1 T \; \Lambda^{-1} \;, 
		\qquad \Lambda \in SO_0(1,2) \; .  
	\end{equation*} 
$\Theta_{ W }$ is an isometry of both ${\mathbb W}$ and $dS$. It 
preserves the orientation but inverts the
time orientation, in other words, $\Theta_{ W }$, just like $P_1 T$, 
is an element of $SO^\downarrow(1,2)$.  

\section{Horospheres}
\label{sec:2.2}

A \emph{horosphere}\footnote{Horospheres previously appeared in 
hyperbolic geometry. They are spheres of 
infinite radius with centres at infinity and different from hyperbolic hyperplanes.}
in a \index{symmetric space} symmetric space\footnote{A \emph{symmetric space} is a 
homogeneous space $G/H$ for a Lie group $G$ (see Section \ref{circle-mass-shell}) 
such that the stabilizer $H$ of a point 
is an open subgroup of the fixed point set of an involution of $G$. In the case of 
$G= SO_0(1,2)$, we can chose $H =\{ \Lambda_2 (s) \mid s \in \mathbb{R} \} $. 
The latter equals 
the subgroup of the fixed point set $\{ (0, r , 0 ), (0, -r , 0) \}$ of the involution $P_2T$.}
$G/H$ (of non-compact type) is an orbit of a maximally \index{unipotent 
group}\emph{unipotent}\footnote{A unipotent matrix is one such that 
$g - \mathbb{1}$ is a nilpotent matrix; \emph{i.e.}, $(g - \mathbb{1})^n$ is equal 
to the zero-matrix for some $n \in \mathbb{N}$.} subgroup of $G$. The importance 
of horospheres was 
emphasised by Gelfand and Gindikin, who have shown that the Fourier--Helgason 
transform on homogeneous spaces and the 
\index{horospherical Radon transform}
\emph{horospherical Radon transform}
\footnote{The horospherical Radon transform takes any function $f$ on 
a semi-simple symmetric space
(a homogeneous space for a semi-simple Lie algebra $\mathfrak{g}$) 
of non-compact type $X=G/H$ to a new function defined on the set 
of horospheres $\operatorname{Hor} X$.
This function is obtained by integrating $f$ over horospheres.} 
(introduced by Gelfand and Graev \cite{GG59}\cite{GGV})
are connected by the (commutative) \index{Mellin transformation}
\emph{Mellin transform} (see, e.g., \cite{Macfadyen-1, Macfadyen-2, Macfadyen-3}). 
We will discuss this topic further in Chapter \ref{ch:3}. 

\begin{lemma}
\label{stabil}
The {\em stabilizer} within the group $SO_0(1,2)$ of the point  
$\left( \begin{smallmatrix} 1 \\ 0 \\ -1 \end{smallmatrix} \right)  \in \partial V^+$ 
is the one-parameter group\footnote{One verifies that $D(q) D(q') =D(q+q')$ for 
all $q, q' \in \mathbb{R}$.}
		\begin{equation} 
			\label{1.5.1}
				D(q)  \doteq  \begin{pmatrix}
					 1+ \frac{q^2}{2} &  q &  \frac{q^2}{2}\\
					 q &  1& q  \\
					- \frac{q^2}{2} & - q & 1- \frac{q^2}{2}   
 					\end{pmatrix} , 
							\qquad 
				q \in \mathbb{R}\; .   
		\end{equation} 
It has the following properties:
\begin{itemize}
\item [$i.)$] it leaves the light ray $\lambda (1, 0, -1) $, $\lambda \in \mathbb{R}$, 
pointwise invariant; 
\item [$ii.)$] it is \emph{nilpotent}. In fact, 
	\[
		D(q) = 
		{\rm e}^{q( \Lgeo_2 - \Kgeo_0 )} 
		= \mathbb{1} + q ( \Lgeo_2 - \Kgeo_0 ) 
		+ \tfrac{q^2}{2} (  \Lgeo_2 - \Kgeo_0 )^2  \; ,  
		\qquad  q \in \mathbb{R} \; . 
	\]
The matrices $\mathfrak{l}_2$ and $\mathfrak{k}_0$ are given in Section \ref{sec:2.7};  
\item [$iii.)$] it leaves the half-spaces $\Gamma^+ (W_1)$ 
and $\Gamma^- (W_1')$ invariant. 
In particular, it leaves the two light rays 
	\[ 
		D(q) \begin{pmatrix}
				0 \\
				 \pm r \\
				  0 \end{pmatrix} 
				  = \begin{pmatrix} \pm r q \\
				   \pm r \\
				    \mp r q \end{pmatrix}
				  \; ,  \qquad  q \in \mathbb{R} \; ,
	\]
which form the intersection of $\Gamma^+ (W_1)$ with $\Gamma^- (W_1')$, invariant; 
\item [$iv.)$] it satisfies the \index{Anosov relations}
Anosov relations\footnote{See, \emph{e.g.}, \cite[Chapter 9.1.1, Equ.~(11)]{Vil}. 
See also \cite{Narn} for a discussion of some interesting consequences.} 
	\begin{align}
		\label{scalingH}
		\Lambda_1(-t) D(q) \Lambda_1(t) & = D \bigl( {\rm e}^t q \bigr)\; , 
		\qquad t, q \in \mathbb{R} \; , 
		\\
		P_1T \; D(q) \, (P_1T)^{-1} &= D ( -q )\; , \qquad q \in \mathbb{R} \; . 
		\nonumber
	\end{align}
\end{itemize}
\end{lemma}

\begin{proof}
These results are established by elementary computation. 
\end{proof}

\subsection{Coordinates for the half-space\index{half-space} $\Gamma^+ (W_1)$}
The boosts $\Lambda_{1}(t)$, $t \in \mathbb{R}$, together with the translations
$D(q)$, $ q \in \mathbb{R}$, give rise to the chart\footnote{These coordinates are  
called \emph{Lema\^itre-Robinson} coordinates 
\index{Lema\^itre-Robinson coordinates} in the physics literature, see \cite{Le}. 
In the mathematics literature
they are called \index{orispherical coordinates}
\emph{orispherical} coordinates, see \cite[Chapter 9.1.5, Equ.~(16)]{Vil}.}
	\begin{align}
	\label{XR}
	x ( \tau, \xi) & \doteq 	
		 D \left( \tfrac{\xi}{r} \right) \Lambda_{1} \left( \tfrac{\tau}{r} \right)  \begin{pmatrix}
										0 \\
										0  \\
										r   
									\end{pmatrix}
									=
									 \begin{pmatrix}
									  r \sinh \tfrac{\tau}{r} 
									  + \tfrac{\xi^2}{2r} {\rm e}^{\frac{\tau}{r}} \\
									   \xi {\rm e}^{\frac{\tau}{r}} \\
									   r \cosh \tfrac{\tau}{r} 
									   - \tfrac{\xi^2}{2r} {\rm e}^{\frac{\tau}{r}}
									\end{pmatrix}   
	\end{align}
for the interior of the half-space $\Gamma^+ (W_1)$. In particular, 
	\[ 
		D \left( \tfrac{\xi}{r} \right) \begin{pmatrix}
				0 \\
				 0 \\
				  r \end{pmatrix} 
				  = \begin{pmatrix} \tfrac{\xi^2}{2r} \\
				   \xi \\
				    r - \tfrac{\xi^2}{2r} \end{pmatrix}
				  \qquad \text{for}  \; \; \xi \in \mathbb{R} \; .
	\]
The metric takes the form 
	$ 
		g_{\upharpoonright \Gamma^+ (W_1)} 
		= {\rm d} \tau \otimes {\rm d} \tau 
		-  {\rm e}^{\frac{2\tau}{r}} {\rm d} \xi \otimes {\rm d} \xi \, $. 
\goodbreak

\subsection{Parabolas in $\Gamma^+ (W_1)$} For $\tau$ fixed, the map 
\eqref{XR} parametrizes the \emph{horosphere}\index{horosphere} (which actually is a parabola\index{parabola} in $\mathbb{R}^{1+2}$)
	\[ 
	 P_\tau \doteq \bigl\{ x(\tau, \xi) \mid \xi \in \mathbb{R} \bigr\} \subset dS \; . 
	\]
General horospheres result from taking the intersection of $dS$ with a plane whose 
normal\footnote{Note 
that the Lorentzian scalar product  $x \cdot p$, 
$x \in \mathbb{R}^{1+2}$, $p \in \partial V^+$, equals the Euclidean 
scalar product of $x$ with $Pp \in \partial V^+$, with 
$P = \operatorname{diag} (1, -1, -1)$ the space-reflection. 
The plane defined by $x \cdot p = 0$, $x \in \mathbb{R}^{1+2}$, 
$p \in \partial V^+$ fixed, contains
the point $x=p$.} vector~$p$ is light like, \emph{i.e.}, $p  \cdot p = 0$. 
In particular, the horospheres $P_\tau$, $\tau \in \mathbb{R}$,   
are given by   
	\begin{equation} 
	\label{horosphere}
	P_\tau =  \left\{ x \in dS
	\mid x \cdot \left( \begin{smallmatrix}
					1\\
					0  \\
					- 1   
 			\end{smallmatrix} \right) = r \, {\rm e}^{\frac{\tau}{r}} \right\} \; . 
	\end{equation}
General horospheres are of the form $R_0(\alpha)P_\tau$,  $ \tau \in \mathbb{R}$,  
for some $\alpha \in [0, 2\pi)$.

\subsection{Horospheric distances} \index{horospheric distance}
The \index{proper time} proper time-difference---given by \eqref{dlength2}---of 
the points $\Lambda_1 \bigl( \tfrac{\tau_1}{r} \bigr) o$ 
and $\Lambda_1 \bigl(\tfrac{\tau_2}{r} \bigr) o$ 
on the geodesic passing through the 
origin $o = (0, 0, r )$  is\footnote{``${\rm arcosh}$'' denotes the 
\emph{area cosinus hyperbolicus}, \emph{i.e.}, the inverse of the 
hyperbolic cosine. For the second equality, recall 
that $\cosh (x - y) = \cosh x \cosh y - \sinh x \sinh y$.}
	\[
	d \big(\Lambda_1 \bigl( \tfrac{\tau_1}{r} \bigr) o, \Lambda_1 
	\bigl( \tfrac{\tau_2}{r} \bigr) o \bigr)
	= r \; {\rm arcosh} \, \left( - \begin{pmatrix}
						\sinh \tfrac{\tau_1}{r}  \\
						0  \\
						\cosh \tfrac{\tau_1}{r}   
					\end{pmatrix} \cdot 
					\begin{pmatrix}
						\sinh \tfrac{\tau_2}{r} \\
						0  \\
						\cosh \tfrac{\tau_2}{r}   
					\end{pmatrix} \right) =    | \tau_1 - \tau_2 | \; . 
	\]
As it turns out, $| \tau_1 - \tau_2 |$ is the minimal distance of \emph{any} 
two time-like points on the horospheres 
$P_{\tau_1}$ and $P_{\tau_2}$, respectively: if 
	\begin{equation} 
		\label{a}
		x  = \bigl( r \sinh \tfrac{\tau_2}{r} + \tfrac{\xi^2}{2r} 
		{\rm e}^{\tfrac{\tau_2}{r}} , \xi {\rm e}^{\tfrac{ \tau_2}{r}} , 
		r \cosh \tfrac{\tau_2}{r} - \tfrac{\xi^2}{2r} {\rm e}^{\tfrac{\tau_2}{r}} \bigr)
	\end{equation}
is a point in $P_{\tau_2}$, then the minimal distance to time-like points in the horosphere~$P_{\tau_1}$, 
called the \emph{horospheric distance},  is given by\footnote{To show 
the second identity, one may use the invariance 
of the Minkowski scalar 
product, \emph{i.e.}, $D(q)x \cdot D(q') y= D(q-q')x \cdot y $ for 
all $q, q' \in \mathbb{R}$ and $x, y \in \mathbb{R}^{1+2}$.} 
	\begin{align}
	\label{distance-to-horosphere}
	d  \bigl(x, P_{\tau_1} \bigr) & \doteq r \; {\rm arcosh} 
	\bigl( \, \min_{y \in P_{\tau_1} } - \tfrac{x \cdot y}{r^2} \bigr)  \nonumber \\
	& 
	= |Ê\tau_1 - \tau_2| 
	= r \ln | \, \tfrac{x}{r} \cdot  p \left( \tfrac{\tau_1}{r} \right) |  
	\;  ,  
	\end{align}
with 
	\[
		p(t) \doteq 
				 \left( \begin{smallmatrix}
					{\rm e}^{-t}\\
					0  \\
					- {\rm e}^{-t}   
 			\end{smallmatrix} \right) = 
			\Lambda_1  (t) 
			 \left( \begin{smallmatrix}
					1\\
					0  \\
					- 1   
 			\end{smallmatrix} \right) \; , \qquad t \in \mathbb{R} \; . 
		\] 
Note that $P_\tau =  \left\{ x \in dS \mid \tfrac{x}{r} \cdot  
p \left( \tfrac{\tau}{r} \right) = 1 \right\} $. 

\section{The Cartan decomposition\index{Cartan decomposition} of $SO_0(1,2)$}
\label{sec:2.3}

If $g \in SO_0(1,2)$, then $g \left( \begin{smallmatrix} m \\ 0 \\ 0 \end{smallmatrix} \right) 
\in H_m^+ \cong SO_0(1,2) / SO(2)$ is of the form 
	\[ R_0(\alpha) \Lambda_{1}(t) \left( \begin{smallmatrix}
						 m   \\
						0 \\
						0  
					\end{smallmatrix} \right) \, , 
		\qquad \alpha \in  [0, 2 \pi) \, , \; t \in \mathbb{R} \, . 
	\]
It follows that this point can be carried back 
to the point $(m , 0, 0  )$ by the action of $\Lambda_1(-t) R_0 (-\alpha)$, \emph{i.e.}, 
	\begin{equation}
		\Lambda_1(-t) R_0 (-\alpha) \, g 
		\left( \begin{smallmatrix} m \\ 0 \\ 0 \end{smallmatrix} \right)
		= \left(\begin{smallmatrix} m \\ 0 \\ 0 \end{smallmatrix} \right) \; . 
	\end{equation}
But 
the stabiliser of the point $\left( \begin{smallmatrix} m \\ 0 \\ 0 \end{smallmatrix} \right)$ 
is the group $K 
= \bigl\{ R_0 (\alpha) \mid \alpha \in [0, 2 \pi) \bigr\} \cong SO(2)$. 
Thus there exists some $\alpha' \in [0, 2 \pi)$ such that 
	\begin{equation}
	\label{cartan}
		R_0 (- \alpha') \Lambda_1(-t)  R_0 (- \alpha ) \, g = \mathbb{1} \, . 
	\end{equation} 
The corresponding decomposition 
	\begin{equation}
	\label{Cartan-deco}
		SO_0(1,2) = KAK \; , 
	\end{equation}
with $K  \cong SO(2)$ and $A \cong SO(1,1)$, is called the \emph{Cartan decomposition}. 
Note that the decomposition \eqref{cartan} is not unique; see, 
\emph{e.g.}, \cite[Chapter 9.1.5]{Vil}.

\section{An alternative decomposition of $SO_0(1,2)$}

If $g \in SO_0(1,2)$, then $g \left( \begin{smallmatrix} m \\ 0 \\ 0 \end{smallmatrix} \right) 
\in H_m^+ \cong SO_0(1,2) / SO(2)$ can also be written as 
	\[ 
		 \Lambda_2 (s) \Lambda_{1}(\hat{t}) \Lambda_2 (-s) \Lambda_{1}( t)
		\left( \begin{smallmatrix}
						 m   \\
						0 \\
						0  
					\end{smallmatrix} \right) \, , 
		\qquad  t , \hat{t} \in \mathbb{R} \, ,  
	\]
with $s \ne 0$ a fixed real number. 
It follows that this point can be carried back 
to the point $(m , 0, 0  )$ by the action of 
$ \Lambda_1(-t) \Lambda_2 (s) \Lambda_{1}(-\hat{t}) \Lambda_2 (-s)$, \emph{i.e.}, 
	\begin{equation}
		\Lambda_1(-t) \Lambda_2 (s) \Lambda_{1}(-\hat{t}) \Lambda_2 (-s) \, g 
		\left( \begin{smallmatrix} m \\ 0 \\ 0 \end{smallmatrix} \right)
		= \left(\begin{smallmatrix} m \\ 0 \\ 0 \end{smallmatrix} \right) \; . 
	\end{equation}
But the stabiliser of the point 
$\left( \begin{smallmatrix} m \\ 0 \\ 0 \end{smallmatrix} \right)$ 
is the group $K 
= \bigl\{ R_0 (\alpha) \mid \alpha \in [0, 2 \pi) \bigr\} \cong SO(2)$. 
Thus there exists some $\alpha \in [0, 2 \pi)$ such that 
	\begin{equation}
	\label{cartan-2}
		R_0 ( - \alpha) \Lambda_1(-t) \Lambda_2 (s) 
		\Lambda_{1}(-\hat{t}) \Lambda_2 (-s) \,  \, g = 
		\mathbb{1} \, . 
	\end{equation} 
Identifying $SO_0(1,2) $ with $ \bigl\{ g^{-1} \mid g \in SO_0(1,2) \bigr\}$, 
the corresponding decomposition 
	\[
		SO_0(1,2) = KA A' \; , \qquad \Lambda_2(s) A \Lambda_2 (-s) 
	\]
with $K  \cong SO(2)$ and $A \cong SO(1,1)$, has been used in the 
proof of Theorem~\ref{flattube}.

\section{The Iwasawa decomposition\index{Iwasawa decomposition} of $SO_0(1,2)$}
\label{sec:2.4}

A brief inspections shows that every point $x \in H_m^+ $ can also be written in the form
	\begin{align}
	\label{XR2}
	x ( t, q) & = 	
		 D (q) \Lambda_{1} (t)  \begin{pmatrix}
										m \\
										0  \\
										0   
									\end{pmatrix}
									=
									 m \begin{pmatrix}
									  \cosh t + \tfrac{q^2}{2} {\rm e}^{t} \\
									   q {\rm e}^{t} \\
									  \sinh t - \tfrac{q^2}{2} {\rm e}^{t}
									\end{pmatrix}  \; . 
	\end{align}	
Now, if $g \in SO_0(1,2)$, then $g \left( \begin{smallmatrix} m \\ 0 \\ 0 \end{smallmatrix} \right)$ 
is of the form \eqref{XR2} for some \emph{unique} $t$ and $q$. It follows that 
this point can be carried back 
to the point $(m , 0,  0)$ by applying the transformation  $\Lambda_1(-t) D(-q)$, \emph{i.e.}, 
	\[
		\Lambda_1(-t) D(-q) \, g \left( \begin{smallmatrix} m \\ 0 \\ 0 \end{smallmatrix} \right) 
		= \left( \begin{smallmatrix} m \\ 0 \\ 0 \end{smallmatrix} \right)\; . 
	\]
As mentioned before, 
the stabiliser of the point $\left( \begin{smallmatrix} m \\ 0 \\ 0 \end{smallmatrix} \right)$ 
is the group $K \cong SO(2)$. 
Thus there exists some $\alpha \in [0, 2 \pi)$ such 
that $R_0 (- \alpha) \Lambda_1(-t) D(-q) \, g = \mathbb{1}$. 

Renaming $g \mapsto g^{-1}$ as well as the parameters $\alpha, t$ and $q$, we arrive 
at the so-called \emph{Iwasawa  decomposition}\footnote{This is, of course, just a 
particular case of  the Iwasawa decomposition:  given \emph{any} non-compact 
semi-simple Lie group $G$, 
one can choose a maximal compact sub\-group~$K$ and a suitable abelian 
subgroup $A$ such that 
any group element $g \in G$ can be written \emph{uniquely} as 
	\[ 
		g = kan
	\]
with $k \in K$, $a \in A$ and $n \in N$, where $N$ is a nilpotent sub\-group, 
normalised by $A$.  Recall that $A$ normalises $N$, if $a n a^{-1} \in N$ for 
all $a \in A$ and all $n\in N$; see 
also \eqref{scalingH}.} \cite{Iwa}:

\begin{lemma}
\label{iwa}
Any element $g \in SO_0(1,2)$ can be written as 
	\begin{align*} 
		\label{Iwaw}
		g 
			& = R_0(\alpha)  \Lambda_{1}(t) D(q)
			\\
			& =
				\begin{pmatrix}
					1 &  0 &0 \\
					0 &  \cos  \alpha & - \sin  \alpha  \\
					0  & \sin  \alpha & \cos  \alpha   
				\end{pmatrix}  
				\begin{pmatrix}
					\cosh t &  0 &\sinh t \\
					0 &  1&0  \\
					\sinh t & 0 & \cosh t   
 					\end{pmatrix} 
			\begin{pmatrix}
					1+ \frac{q^2}{2} &  q &  \frac{q^2}{2}\\
					q &  1& q  \\
					- \frac{q^2}{2} & -q  & 1- \frac{q^2}{2}   
 			\end{pmatrix}, 	
	\end{align*} 
with $\alpha \in [0,  2 \pi) $ and $ t,q \in \mathbb{R}$. 
\end{lemma}

The resulting decomposition, 
$G= KAN $, provides  
\begin{itemize}
\item[$i.)$] a maximal compact subgroup 
$K \cong SO(2)$, namely the group of rotations
$\{ R_0(\alpha) \mid  \alpha \in [0, 2 \pi) \}$ which keep the Cauchy surface invariant; 
\item[$ii.)$] a Cartan maximal abelian subgroup $A\cong (\mathbb{R}, +)$, 
which is given by the boosts 
$\{ \Lambda_1(t) \mid  t \in \mathbb{R} \}$ keeping the wedge $W_1$ invariant; 
\item[$iii.)$] a nilpotent
group $N \cong (\mathbb{R}, +)$,  which can be identified with the group of 
horospheric translations $\{ D(q) \mid  q \in \mathbb{R} \}$.
\end{itemize}

\section{The Hannabuss decomposition of $SO_0(1,2)$}
\label{sec:2.5}

A decomposition, which is closely related to the Iwasawa 
decomposition of the Lorentz group, was 
discovered by Takahashi \cite{T}. Its relevance in the present 
context was emphasised by Hannabuss \cite{Hanna}.

\begin{lemma}[Hannabus \cite{Hanna}]
Almost every element $g \in SO_0(1,2)$ can be written 
uniquely in the form of a product
	\begin{align*} 
		g 
			& = \Lambda_{2}(s) P^k \Lambda_{1}(t) D(q)
			\\
			& \quad \\
			& =
				\begin{pmatrix}
								\cosh s &  \sinh s &0 \\
								\sinh s &   \cosh s &0  \\
								0 & 0 & 1  
				\end{pmatrix} 
				 \begin{pmatrix}
						1 &  0 &0 \\
						0 &  (-1)^k & 0  \\
						0  & 0 & (-1)^k   
					\end{pmatrix} 
					\\
			& \qquad \qquad  \qquad \qquad \qquad  \times  
			\begin{pmatrix}
					\cosh t &  0 &\sinh t \\
					0 &  1&0  \\
					\sinh t & 0 & \cosh t   
 					\end{pmatrix}
			\begin{pmatrix}
					1+ \frac{q^2}{2} &  q & - \frac{q^2}{2}\\
					q &  1& -q  \\
					\frac{q^2}{2} & q  & 1- \frac{q^2}{2}   
 			\end{pmatrix}, 	
	\end{align*} 
with  $s, t, q \in \mathbb{R}$ and $k=\{0,1\}$, \emph{i.e.}, almost every 
element $g \in SO_0(1, 2)$
can be decomposed into a product, which consists of a 
Lorentz transformation $s \mapsto \Lambda_2(s)$, 
possibly a reflection $P^k$,
a time translation $t \mapsto \Lambda_1(t)$, and a 
spatial translation $q \mapsto D(q)$. 
\end{lemma}

\goodbreak
\begin{proof} Let $g \in G$ be given in its Iwasawa decomposition, \emph{i.e.}, 
	\[
		g=  R_0( \alpha)  \Lambda_1(t'') D(q'') \; , \qquad \alpha \in [0, 2 \pi), 
		\quad  t'',q''  \in \mathbb{R}\; . 
	\]
We will show that, unless $\alpha =  \tfrac{\pi}{2}$  or  
$ \tfrac{3\pi}{2}$, $R_0( \alpha)$ can be written in  the form 
	\begin{equation}
	\label{r0H}
		R_0( \alpha) = \Lambda_2(s) P^k D(-q') \Lambda_1(-t')  \; , 
		\qquad s, t', q' \in \mathbb{R}\; , \quad k = 0,1 \; . 
	\end{equation}
Taking \eqref{scalingH} into account,   this will imply that  
	\[
		g = \Lambda_2(s) P^k  \Lambda_1(\underbrace{t''-t'}_{t}) 
		D \bigl( \underbrace{q'' - {\rm e}^{(t''-t')} q' }_{q} 
		\bigr) \; . 
	\]
Thus it remains to establish \eqref{r0H}. Multiplying 	\eqref{r0H} 
with $\Lambda_1(t') D(q')$ from the right yields 
	\[
		\Lambda_2(s) P^k = R_0( \alpha)  \Lambda_1(t') D(q') \; , 
		\qquad s \in \mathbb{R}\; , \quad k = 0,1 \; . 
	\]
This is the Iwasawa decomposition of $\Lambda_2(s)P^k$, $k=0, 1$, 
which  is given by choosing 
	\[
		\begin{matrix}	& \cosh s = (-1)^k \cos^{-1}\alpha \; , 
		\qquad  & {\rm e}^{t'} = \tfrac{1}{| \cos \alpha | }  \; , \\ 
				& \sinh s  = (- 1)^{k+1}\tan \alpha \:,  \qquad &
				q' = (-1)^{k+1} \sin \alpha \; . 
		\end{matrix}
	\]
In fact, 
unless $\cos \alpha = 0$, 
	\begin{align*} 
		&\begin{pmatrix}
 				| \cos \alpha |^{-1}& - \frac{\sin \alpha}{| \cos \alpha|}  &0 \\
				-\frac{\sin \alpha}{| \cos \alpha|}  &  | \cos \alpha |^{-1}  &0  \\
				0 & 0 &   1    
		\end{pmatrix}  \begin{pmatrix}
 								1 &  0  &0 \\
								 0  &  (-1)^k  &0  \\
								0 & 0 &   (-1)^k     
		\end{pmatrix} \\
\\
		& \qquad =						\begin{pmatrix}
					1 &  0 &0 \\
					0 &  \cos \alpha & - \sin \alpha  \\
					0  & \sin \alpha & \cos \alpha   
				\end{pmatrix}  
\begin{pmatrix}
					 \tfrac{ | \cos \alpha |^{-1}   
					 + | \cos \alpha |}{2}  &  0 &  \tfrac{ | \cos \alpha |^{-1}   
					 - | \cos \alpha | }{2} \\
					0 &  1&0  \\
					 \tfrac{| \cos \alpha |^{-1}   - | \cos \alpha |}{2} & 0 &  
					 \tfrac{| \cos \alpha |^{-1}   + | \cos \alpha |}{2}   
 					\end{pmatrix} 
		\\
		&
		\qquad \qquad \times 	\begin{pmatrix}
					1+ \frac{\sin^2 \alpha }{2} &  (-1)^{k+1} \sin \alpha  
					& - \frac{\sin^2 \alpha }{2}\\
					(-1)^{k+1} \sin \alpha  &  1&  (-1)^{k}\sin \alpha   \\
					\frac{\sin^2 \alpha}{2} & (-1)^{k+1} \sin \alpha 
					& 1- \frac{\sin^2 \alpha }{2}   
 			\end{pmatrix}. 	
	\end{align*} 
with 
	\[
		k = \begin{cases} 0 & \text{if} \quad \cos \alpha >0 \; , \\
		1 & \text{if} \quad \cos \alpha <0 \; .  
		\end{cases} 
	\]
The exceptional group elements, which can not be represented in this form,  
contain the rotations 
$R_0 (\pm \frac{\pi}{2})$ in their Iwasawa decomposition.  
\end{proof}

The resulting 
decomposition of $G$ is of the form
	\begin{equation}
	 	\label{habu}
		G= A'AN \cup A' P AN \; , \qquad A' = 
		\bigl\{ \Lambda_2 (s)  \mid s \in \mathbb{R} \bigr\} \; . 
	\end{equation}
The spatial reflection $P$ is needed to account for elements whose Iwasawa decomposition 
contains a rotation $R_0(\alpha)$ with $\frac{\pi}{2} < \alpha< \frac{3 \pi}{2}$. 

\section{Homogeneous spaces, cosets and orbits}
\label{circle-mass-shell}

Consider a closed subgroup $H$ of a topological group $G$. 
The \index{homogeneous space}\emph{homogeneous space}~$G/H$ is the space of all 
\index{left cosets}\emph{left cosets} $gH,~g\in G$. Let $\mathbb{\Pi} \colon G \to G/H$ 
denote the canonical mapping defined by 
	\begin{equation}
	\label{quotient-map}
		\mathbb{\Pi} (g) = gH \; . 
	\end{equation}
By construction, each point $x = gH$ of $G/H$ remains fixed 
under the left action of the subgroup
$gHg^{-1} \cong H$. Hence, $H$ is the \emph{stability group} 
of the space $X=G/H$. 

We equip $G/H$ with the \emph{quotient 
topology}\index{quotient topology}, \emph{i.e.}, a set 
$O \subset G/H$ is open if 
	\[
		\mathbb{\Pi}^{-1}(O) \subset G
	\]
is open. By construction, $G$ acts \emph{transitively}\index{transitive 
group action} on $G/H$: 
	\[
		g \mathbb{\Pi} (g') = \mathbb{\Pi} (g g') \;  . 
	\]
We note that locally\footnote{It is well known that, even if $G$ is a connected Lie group, 
smooth (continuous) cross sections need not exist; however Mackey \cite{MA1} showed, 
using the theory of standard Borel spaces, that a Borel measurable 
cross section exists if $G$ is a separable (second countable) locally compact group; see 
also~\cite{Feldman-Greenleaf}.} there is a continuous section $\Xi \colon G/H \to G$, 
which satisfies 
	\[
		\mathbb{\Pi} \circ \Xi = \mathbb{1}_{G/H} \, .  
	\]
This will become relevant when we discuss induced representations in Chapter \ref{ch:3}.

\goodbreak
\begin{lemma} 
\label{surfaceG/H}
Let $G$ be the Lorentz group $SO_0(1,2)$.
Furthermore, let $H \subset   G  $ be the stabiliser of an arbitrary 
point $x \in \mathbb{R}^{1+2}$, 
whose orbit is $\mathbb{O}(x) \doteq \{ gx \in \mathbb{R}^{1+2} 
\mid g \in G \}$.  

Then there exists a bijective map $\mathbb{\Gamma} \colon G/H \to \mathbb{O}(x)$,
	\begin{equation}
	\label{mgamma}
		g H \mapsto g x \; , \qquad g \in G \; , 
	\end{equation}
such that 
	\begin{equation}
	\label{groupaction}
		 \mathbb{\Gamma} (gg'H) = g \mathbb{\Gamma}(g'H) 
		 \qquad \forall g, g' \in G \; ; 
	\end{equation}
\emph{i.e.}, $g (g' x) = (g \circ g') x$ for all $g, g' \in G$.
\end{lemma}

\begin{proof}
One easily verifies that  $\mathbb{\Gamma}$ is well-defined: 
if $g_1 H = g_2 H$,  then $g_1 = g_2 h$ for some $h \in H$, and since 
the $H$ leaves the point $x$ invariant, the map is well-defined. On the other hand, if 
	\[
		g_1	x = g_2 	x \; , 
	\]
then  $g^{-1}_2 g_1 $ fixes $x$ and thus must be in $H$. 
This implies $g_1 H = g_2 H$. Thus the map $\mathbb{\Gamma} \colon G/H 
\to \mathbb{O}(x) $ is bijective.
By construction, it satisfies \eqref{groupaction}. 
\end{proof}

In the following, we concentrate on homogeneous spaces $SO_0(1,2)/H$,
where the subgroup $H$ is the stabiliser of a specific point $x \in \mathbb{R}^{1+2}$. 

\subsection{The forward light-cone}

\label{surfaces} Let us first consider the case $x = \left( \begin{smallmatrix}
1 \\ 0 \\ -1 \end{smallmatrix} \right)$. According to Lemma \ref{stabil}, 
	\[
		 \left(\begin{smallmatrix}
		1 \\ 0 \\ -1 \end{smallmatrix} \right) = D(q)
		 \left( \begin{smallmatrix}
		1 \\ 0 \\ -1 \end{smallmatrix} \right)  \qquad \forall q \in \mathbb{R} \; . 
	\]
Since the group $\{ D(q) \mid q \in \mathbb{R} \}$ is nilpotent, 
it is usually denoted by the letter $N$. 
Thus the stabiliser of $x$ in $SO_0(1,2)$ is $H= N$. 

\begin{lemma} 
\label{repV+}
The homogeneous space $ SO_0(1,2)/N \cong \{ gN \mid g \in SO_0(1,2) \}$ 
can be naturally identified with $\partial V^+ \setminus \{(0, 0,0)\}$ by 
the map
	\[
		(  g N ) \mapsto
		g 	\left( \begin{smallmatrix}
							1 \\
							0 \\
							-1
						\end{smallmatrix} \right) \; , 
		\qquad g \in SO_0(1,2) \; .  
	\]
Moreover, $g (g' x) = (g \circ g') x$ for all $g, g' \in SO_0(1,2) $ 
and $x \in \partial V^+ \setminus \{(0, 0,0)\}$. 
\end{lemma}

\begin{proof}
By Lemma \ref{iwa}, the canonical mapping $\mathbb{\Pi} 
\colon  SO_0(1,2) \to SO_0(1,2)/N $ is given by
	\[
		g \mapsto R_0(\alpha) \Lambda_{1}(t) N  \; , \qquad \text{with} \quad g =  
		R_0(\alpha) \Lambda_{1}(t) D(q)  \; .  
	\]
Now recall that the map 
	\[
		(\alpha, t) \mapsto R_0 (\alpha) \Lambda_{1}(t) \left( \begin{smallmatrix}
						1 \\
						0 \\
						- 1
					\end{smallmatrix} \right) = R_0 (\alpha) \left( \begin{smallmatrix}
						{\rm e}^{-t} \\
						0 \\
						- {\rm e}^{-t}
					\end{smallmatrix} \right) \in  \partial V^+ \; , 
		\quad t \in \mathbb{R} \; , \; \; \alpha \in [0, 2 \pi) \; , 
	\]
provides coordinates for $\partial V^+ \setminus \{ (0,0,0) \}$. 
Note that $\Lambda_{1}(t) $ leaves the light ray connecting the 
origin $(0,0,0)$ and the point $(1, 0,-1)$ invariant.   
\end{proof}

\subsection{The mass hyperboloid} 
Next consider the point $x = \left( \begin{smallmatrix} m \\ 0 \\ 0 
\end{smallmatrix} \right)$. Clearly, 
	\[
		R_0(\alpha) \left( \begin{smallmatrix}
		m \\ 0 \\ 0 \end{smallmatrix} \right) = \left( \begin{smallmatrix}
		m \\ 0 \\ 0 \end{smallmatrix} \right)		
		\qquad \forall \alpha \in [0, 2 \pi) \; . 
	\]
In other words, the stabiliser of $x$ is $K \cong \{ R_0 (\alpha) \mid \alpha \in [0, 2 \pi) \}$. 

\begin{lemma} The coset space $SO_0(1,2)/K$ can be naturally identified with a 
\emph{two-fold covering}  of $H_m^+ $ by the map 
	\[
		(g K) \mapsto
		g 	\left( \begin{smallmatrix}
							m \\
							0 \\
							0
						\end{smallmatrix} \right) \; , 
		\qquad g \in SO_0(1,2) \; .  
	\]
\end{lemma}

\begin{proof} The rotations $K \cong SO(2)$ are the stabiliser of 
the point  $\left( \begin{smallmatrix}
							m \\
							0 \\
							0
						\end{smallmatrix} \right)$
in $SO_0(1,2)$. Note that 
	\[
			\Lambda_{1}(t)  
			\left( \begin{smallmatrix}
							m \\
							0 \\
							0
						\end{smallmatrix} \right) 
			 		= \left( \begin{smallmatrix}
						 m \cosh t \\
						0 \\
						m \sinh t
					\end{smallmatrix} \right) \in H_m^+  \; ,
		\qquad t \in \mathbb{R} \; . 
	\]
Applying the rotations $R_0(\alpha)$, 
$\alpha \in [0, 2 \pi)$, to $\Lambda_{1}(t) x$ results in a two-fold 
covering of the mass hyper\-boloid~$H^+_m$. 
Thus the result follows from Lemma \ref{surfaceG/H}.
\end{proof}

\subsection{De Sitter space} 
\label{SS:2.6.3}
Finally,  consider the case $x= o$.  
The boosts $\Lambda_2(t)$, $t \in \mathbb{R}$, form the 
stabiliser $A'$ of  the origin $o$. Moreover, the map
	\begin{equation}
	\label{ds-me}
	x(\alpha, t) \doteq R_0 (\alpha) \Lambda_{1} (t) o = R_0 (\alpha) \begin{pmatrix}
						r \sinh t  \\
						0 \\
						r \cosh t 
					\end{pmatrix}  ,  
	\end{equation}
with $\alpha \in [0, 2 \pi  )$ and $t \in \mathbb{R}$, 
provides coordinates\footnote{This should be compared with the chart introduced in 
\eqref{w1psi}, which only covers $\mathbb{W}_1$.} for $dS$. 

\begin{lemma} The coset space $G/ A' $ can be naturally identified with $dS$, 
by the map 
	\[
		(g A') \mapsto    
		g 	\left( \begin{smallmatrix}
							0 \\
							0 \\
							r 
						\end{smallmatrix} \right) \; , \qquad g \in G \; .  
	\]
Moreover, $g (g' x) = (g \circ g') x$ for all $g, g' \in G$ and $x \in dS$. 
\end{lemma}

\subsection{Circles}
\label{circ-sub} We next consider the case $H= AN $.
Clearly, $H$ leaves the light ray 
	\[
	 	\bigl\{ \lambda  \left( \begin{smallmatrix}
							1 \\
							0 \\
							-1
		\end{smallmatrix} \right) \mid \lambda >0 \bigr\}
	\] 
passing through the origin and the point $\left( \begin{smallmatrix}
							1 \\
							0 \\
							-1
						\end{smallmatrix} \right)$ invariant. 
It is therefore natural to identify the factor group $SO(2) \cong 
G/AN  $ with the 
projective space 
	\[
		\partial \dot V^+ = \bigl\{ \dot y \mid y \in \partial V^+ \bigr\} \; ,  
	\]
formed by the light rays  
	\[
		\dot y =  \{ \lambda y \mid \lambda >0 \} \; , \qquad y \in \partial V^+  \; .
	\]
Each light ray in $\partial \dot V^+$ intersects the 
circle\footnote{Obviously, one could as well consider the 
circles $\bigl\{ R_0 (\alpha) \left( \begin{smallmatrix}
							\lambda \\
							0 \\
							-\lambda
						\end{smallmatrix} \right) 
						\mid \alpha \in [0 , 2 \pi) \bigr\}$, $\lambda >0$.}  
	\begin{equation}
		\label{gamma-0}
		\gamma_0 \doteq \left\{ R_0 (\alpha) \left( \begin{smallmatrix}
							1 \\
							0 \\
							-1
						\end{smallmatrix} \right) 
		\mid \alpha \in [0 , 2 \pi) \right\} 
	\end{equation}
just once. Hence there is a one-to-one relation between 
points in $\gamma_0$ and elements of $\partial \dot V^+$. 
However, it should be emphasised that the boosts 
in $A= \{ \Lambda_1 (t) \mid t \in \mathbb{R} \}$ do not leave the 
point $\left( \begin{smallmatrix}
							1 \\
							0 \\
							-1
						\end{smallmatrix} \right)$ invariant.   

\subsection{Mass shells} 
\label{massshell-sub}
\index{mass shell}
Using the Hannabuss decomposition of 
$SO_0(1,2)$, almost every element $g$ in $SO_0(1,2)$ can be 
written in the form 
	\[
		g =  \Lambda_{2}(s) P^k \Lambda_{1}(t) D(q) \; , 
		\qquad   k \in \{0, 1\} \; . 
	\]
Thus, for each $m_0 >0$, almost all of the cosets 
$gAN$, $g \in SO_0(1,2)$,  
(recall that these cosets can be identified with the light rays in $\partial \dot V^+$) 
are in one-to-one correspondence 
to points on the two hyperbolas
	\begin{equation}
	\label{hypm0}
		{\gamma}_+ \; \dot \cup \; {\gamma}_- 
		\doteq \left\{ \Lambda_2 (s) \left( \begin{smallmatrix}
	m_0  \\
	0 \\
	m_0  
	\end{smallmatrix} \right) \mid s \in \mathbb{R} \right\} \cup
	\left\{ \Lambda_2 (s) P \left( \begin{smallmatrix}
	m_0  \\
	0 \\
	m_0  
	\end{smallmatrix} \right) \mid s \in \mathbb{R} \right\} \; . 
	\end{equation}
Note that $P\Lambda_2 (s)P= \Lambda_2 (-s)$ for all $s \in \mathbb{R}$. 

\begin{remark} It is convenient to choose a parametrization such that
	\[ 
		m_0 \cosh  s  
		= \sqrt{p_1^2 + m_0^2} \; , \qquad  m_0 \sinh s = p_1 \; . 
	\]
Then the measure is 
		${\rm d} s= \tfrac{ {\rm d} p_1 } { \sqrt{p_1^2 + m_0^2} }$
and $\Lambda_2 (s)$ is of the form
	\[ 
	\Lambda_2 (s) =   \begin{pmatrix}
	\frac{\sqrt{p_1^2 + m_0^2} }{ m_0 } & \frac{p_1}{ m_0} & 0 \\
	\frac{p_1}{ m_0 } &  \frac{\sqrt{p_1^2 + m_0^2}}{ m_0 } & 0 \\
	0 & 0 & 1
	\end{pmatrix}  \; , 
	\qquad 
	s = {\rm arcsinh}  \tfrac{p_1}{ m_0 }  \;  .
	\]  
The mass shells $\Gamma_+$ and $\Gamma_-$ will play an important role in 
Section~\ref{UIRm}. 
\end{remark}

\section{The complex Lorentz group}
\label{sec:2.7}

The \emph{generators}\index{generator} of the \emph{boosts}\index{boost} 
$\mathbb{R}\ni t \mapsto \Lambda_{1}(t)$, $\mathbb{R}\ni 
s \mapsto \Lambda_{2}(s)$
and the rotations $ [0, 2 \pi )\ni \alpha \mapsto R_{0}(\alpha)$ 
introduced in Section \ref{subsec:1.3} are 
	\begin{equation}
	\label{elle-1}
			\Lgeo_1  =  
			\begin{pmatrix}
							0 &  0 & 1 \\
							0 &  0 & 0  \\
							1 & 0 & 0   
				\end{pmatrix}  , 
		\; 	 
			\Lgeo_2  = 
			\begin{pmatrix}
							0 &  1 & 0 \\
							1 &  0 & 0  \\
							0 & 0 & 0   
			\end{pmatrix} 
		\;
				\text{and} \; \; 
			\Kgeo_0  =
			\begin{pmatrix}
							0 &  0 & 0 \\
							0 &  0 & -1  \\
							0  & 1 & 0   
			\end{pmatrix} ,  
		\end{equation}
respectively. The matrices $\Lgeo_1= \Lgeo_1^* $ and 
$ \Lgeo_2 = \Lgeo_2^* $ are \emph{symmetric}, the matrix 
$ \Kgeo_0 = -\Kgeo_0^*  $ is \emph{anti-symmetric}. They satisfy 
the $\mathfrak{so}(1,2)$ Lie algebra relations:
	\[
		[ \Lgeo_1 , \Lgeo_2 ] = \Kgeo_0 \; , 
		\qquad 
		[ \Kgeo_0 , \Lgeo_1 ] = - \Lgeo_2 \; , 
		\qquad 
		[ \Lgeo_2 , \Kgeo_0 ] = - \Lgeo_1 \; .  		
	\]
Instead of $\Lgeo_2$, we can also use $\mathfrak{d} \doteq \Lgeo_2 - \Kgeo_0$
to generate the $\mathfrak{so}(1,2)$ Lie algebra. The element 
$\mathfrak{d} \in \mathfrak{so}(1,2)$ satisfies the following relations: 
	\[
		[ \Lgeo_1 , \mathfrak{d} ] 		= \mathfrak{d}  \; , 
		\qquad 
		[ \Kgeo_0 , \Lgeo_1 ] =  \Kgeo_0 - \mathfrak{d} \; , 
		\qquad
		[ \mathfrak{d} , \Kgeo_0 ] =  - \Lgeo_1 \; .  		
	\]

\begin{remark}
Occasionally, we will also refer to the generators
	\label{Lambdaalphapage}
		\begin{equation} 
		\label{Lalpha}
			\Lgeo^{(\alpha)}  
			\doteq \cos \alpha \; 
			\Lgeo_1  
			+ \sin \alpha  \; 
			\Lgeo_2   \; , 
			\qquad \alpha \in [ 0, 2 \pi ) \; , 
		\end{equation} 
of the boosts  $ t \mapsto \Lambda^{(\alpha)} (t)$ defined in \eqref{Lambdaalpha}. 
Note that $ \Lgeo^{(0)}= \Lgeo_1  $ and 
$ \Lgeo^{(\pi/2)} = \Lgeo_2  $. 
\end{remark}

The \emph{Casimir operator} 
	\[
		\Cgeo^{\, 2} = - \Kgeo_0^{\, 2} + \Lgeo_1^{\, 2} + 
		\Lgeo_2^{\, 2} 
		= 2 \cdot \mathbb{1}_3 \; , 
	\]
with~$\mathbb{1}_3$ the unit $3 \times 3$-matrix, is an element in the center of the 
\emph{universal enveloping algebra}\footnote{The universal enveloping 
algebra of a Lie algebra $\Lgeo$ arises from the free tensor algebra 
over $\Lgeo$ (considered as a vector space) by factoring out the 
ideal generated by elements of the 
form $\{ X \otimes Y - Y \otimes X - [X, Y] \mid X, Y \in \Lgeo \}$.} 
of the Lie algebra $\mathfrak{so}(1,2)$. 
Note that for $\mathfrak{so}(3)$,  the Casimir operator 
equals $2 \cdot \mathbb{1}_3$ as well.

\subsection{The dual Lie algebra}
\label{ssec:2.7.1}
The time reflection $T$ on $dS$ not only turns the 
Minkowski scalar product into the 
Euclidean scalar product (see \eqref{m-e-sp} below), 
it also induces an 
involution $\mathbb{\theta} \in \operatorname{Aut}(\mathfrak{so}(1,2))$: 
	\begin{equation}
	\label{mathbb-theta}
		\mathbb{\theta}  \colon \mathfrak{g} \mapsto  
		\mathfrak{g}^\mathbb{\theta} 
		\doteq T \mathfrak{g} T \; , 
		\qquad \mathfrak{g} \in \mathfrak{so}(1,2) \; , 
	\end{equation}
As $\mathbb{\theta}$ is an automorphism 
of $\mathfrak{so}(1,2)$, the elements 
	\[
			\Lgeo_1^\mathbb{\theta} = - \Lgeo_1  =  
			\begin{pmatrix}
							0 &  0 & -1 \\
							0 &  0 & 0  \\
							-1 & 0 & 0   
				\end{pmatrix}  , 
		\;
			\Lgeo_2^\mathbb{\theta} = - \Lgeo_2  =
			\begin{pmatrix}
							0 &  -1 & 0 \\
							-1 &  0 & 0  \\
							0 & 0 & 0   
			\end{pmatrix}  , \; 
			\Kgeo_0^\mathbb{\theta} =  \Kgeo_0  =  
			\begin{pmatrix}
							0 &  0 & 0 \\
							0 &  0 & -1  \\
							0  & 1 & 0   
			\end{pmatrix} .  
		\]
still satisfy the $\mathfrak{so}(1,2)$ relations\footnote{Note that 
the matrices $\Lgeo^\mathbb{\theta}_1$ and $\Lgeo^\mathbb{\theta}_2$ 
are \emph{symmetric}, while the matrix $\Kgeo^\mathbb{\theta}_0$ 
is \emph{anti-symmetric}.} 
	\[
		[  \Lgeo_1^\mathbb{\theta} \, , \,  \Lgeo_2^\mathbb{\theta} ] =   \Kgeo_0 \; , 
		\qquad 
		[ \Kgeo_0 \, , \, \Lgeo_1^\mathbb{\theta} ] =    - \Lgeo_2^\mathbb{\theta} \; , 
		\qquad
		[ \Lgeo_2^\mathbb{\theta} \, , \, \Kgeo_0] =    - \Lgeo_1^\mathbb{\theta} \;  . 
	\]
Moreover, the matrices $\Lgeo_1^\mathbb{\theta}= (\Lgeo_1^\mathbb{\theta})^* $ and 
$\Lgeo_2^\mathbb{\theta}= (\Lgeo_2^\mathbb{\theta})^* $ are \emph{hermitian}, as they 
differ from $\Lgeo_1$ and $\Lgeo_2$, respectively, only by a minus sign. 

On the other hand, $i \Lgeo_1$, $i \Lgeo_2$ and $\Kgeo_0$ are all skew-hermitian 
on $\mathbb{R}^3$ with the Euclidean scalar product. 
They satisfy the $\mathfrak{so}(3)$ relations, \emph{i.e.}, the relations of  
the dual symmetric Lie algebra of $\mathfrak{so}(1,2)$:
	\[
		[  i \Lgeo_1  \, , \,  i \Lgeo_2 ] =  - \Kgeo_0 \; , 
		\qquad 
		[ \Kgeo_0 \, , \, i\Lgeo_1 ] =    - i \Lgeo_2 \; , 
		\qquad
		[ i \Lgeo_2 \, , \, \Kgeo_0] =    - i \Lgeo_1 \;  . 
	\]
Recall that if~${\mathfrak g}$ is a Lie algebra with an involution $\mathbb{\theta}$, 
then the decomposition (where~$\oplus$ indicates a direct sum of vector spaces)
	\begin{equation}
		\label{VR1}
		{\mathfrak g} 
		= \underbrace{ 
		\operatorname{ker} ( \mathbb{\theta} - \mathbb{1} )
		}_{ \doteq {\mathfrak k} } 
		\oplus \underbrace{ \operatorname{ker} (\mathbb{\theta} 
		+\mathbb{1})}_{\doteq {\mathfrak m}}
	\end{equation}
into eigenspaces of $\mathbb{\theta} $ shows that the subspace
${\mathfrak g}^* = {\mathfrak k} \oplus i{\mathfrak m} $
of the complexification of ${\mathfrak g}$, called the {\em dual 
symmetric Lie algebra} (see, \emph{e.g.}, \cite{KN}), 
is the {\em real} Lie algebra of a 
connected Lie group~$G^*$. This fact is known as \emph{Cartan duality}.

\subsection{A virtual representation of $\mathfrak{so}(1,2)$}
Let us consider the space $\mathbb{R}^{1+2}$ and define a new scalar product: 
	\begin {equation}
	\label{m-e-sp}
		\mathbb{R}^{1+2} \ni x, y \mapsto \lfloor x , y \rfloor 
		\doteq \underbrace{ - x \cdot T y }_{  = x_0 y_0 + x_1 y_1 + x_2 y_2} \; . 
	\end{equation}
As before $\cdot$ denotes the Minkowski product. Now compute 
	\begin{align*}
		\lfloor x , g y \rfloor &=  - x \cdot T g y \\
			& = - x \cdot \mathbb{\theta} (g ) T y \\
			& = - \mathbb{\theta} (g^{-1}) x \cdot  T y \\
			& = \lfloor \mathbb{\theta} (g^{-1}) x ,  y \rfloor \; , 
			\qquad g \in SO_0 (1,2) \; . 
	\end{align*}
Hence on the space	$\bigl( \mathbb{R}^{1+2} , 
\lfloor \, .\, , \, .\, \rfloor \bigr)$, we have 
	\[
		g^* = \mathbb{\theta} (g^{-1}) \, , \qquad g \in SO_0 (1,2) \;. 
	\]
In other words, on $\bigl( \mathbb{R}^{1+2} , \lfloor \, .\, , \, .\, \rfloor \bigr)$, 
we have
	\[
		\Lambda_1 (t)^* = \Lambda_1 (t) \; , \quad  \Lambda_2 (s)^* 
		= \Lambda_2 (s) \; , 
		\quad R_0 (\alpha)^* = R_0 (-\alpha) \; . 
	\]
Thus the map $\alpha \mapsto R_0 (\alpha)$  is a \emph{orthogonal 
representation}, and the 
operators  $\Lambda_1 (t) $ and $ \Lambda_2 (s)$ are \emph{symmetric} 
for all $t \in \mathbb{R}$ and 
$s \in \mathbb{R}$, respectively. 

\subsection{The exponential map for the dual symmetric Lie algebra}
We choose a neighbourhood $U^*$ of the identity $\mathbb{1}$ 
in $G^*= SO(3)$ in such a way that 
each $g \in U^*$ has a representation as
	\begin{equation}
	\label{group-law-1}
		g = {\rm e}^{ i  \mu } {\rm e}^{\alpha \Kgeo_0} \; , 
		\qquad \mu \in \mathfrak{m} \, . 
	\end{equation}
We next verify the group multiplication law: 
	\begin{align}
		{\rm e}^{ i \mu_1 } {\rm e}^{ \alpha_1 \Kgeo_0 } 
		{\rm e}^{ i \mu_2 } {\rm e}^{ \alpha_2 \Kgeo_0 }
		& = {\rm e}^{ i \mu_1 } \bigl( {\rm e}^{ \alpha_1 \Kgeo_0 } {\rm e}^{ i \mu_2 }
				{\rm e}^{ - \alpha_1 \Kgeo_0 }\bigr) 
				{\rm e}^{ \alpha_1 \Kgeo_0 } {\rm e}^{ \alpha_2 \Kgeo_0 }  
		\nonumber \\
		& = {\rm e}^{ i \mu_1} {\rm e}^{ i{\tt ad} \, 
		{\rm e}^{ \alpha_1 \Kgeo_0} ( \mu_2) }
				{\rm e}^{ (\alpha_1 + \alpha_2) \Kgeo_0}  \; , 
				\qquad \mu_1 , \mu_2 \in \mathfrak{m} \, .  				
	\label{group-law-2}
	\end{align}
The second inequality follows from the fact that for 
real $t \in \mathbb{R}$, we have 
	\[
		{\rm e}^{ \alpha_1 \Kgeo_0 } {\rm e}^{ t \mu_2 } 
		{\rm e}^{ - \alpha_1 \Kgeo_0 } 
		=  {\rm e}^{ t \, {\tt ad} \, {\rm e}^{ \alpha_1 \Kgeo_0} ( \mu_2) } \; , 
		\qquad t \in \mathbb{R} \; .  
	\]
Both sides can be continued analytically to $t=i$. In order to prove that 
\eqref{group-law-2} has the form \eqref{group-law-1}, we still have to show that 
	\[
		{\rm e}^{ i \mu_1 } {\rm e}^{ i \mu_2 }  
		=  {\rm e}^{  i \mu_3 } {\rm e}^{  \alpha \Kgeo_0 }  \; , \qquad 
		\mu_1, \mu_2, \mu_3 \in \mathfrak{m} \; , \quad \alpha \in [0, 2 \pi) \; .  
	\]
This can be seen as follows: there are maps $\mathbf{f}$ and $\mathbf{g}$, 
holomorphic on some polydisc $P \subset \mathbb{C}^2$
centred at the origin with values in  $\mathbb{C}^2$ and $\mathbb{C}$, 
respectively; \emph{i.e.}, 
	\[
		\mathbf{f} \colon P \to \mathbb{C}^2 \; , 
		\qquad 
		\mathbf{g} \colon P \to \mathbb{C} \; , 
	\]
such that for $( z_1, z_2) \in P$, 
	\begin{equation}
	\label{2.7.7}
		{\rm e}^{z_1 \mu_1}  {\rm e}^{z_2 \mu_2} 
		= \exp \left( \sum_{i=1}^2 \mathbf{f}_i (z_1, z_2) \Lgeo_i \right) 
		\exp \left( \mathbf{g} (z_1, z_2) \Kgeo_0 \right) 
	\end{equation}
see, \emph{e.g.}, \cite{Chev}. Since $ i \mathfrak{m}$ is a 
real subspace of $\mathfrak{so(3)}$, the $\mathbf{f}_i$ take 
purely imaginary values and the $\mathbf{g}$ real values if the 
complex numbers $z_1$ and $z_2$ are purely
imaginary. In this case, \eqref{2.7.7} just expresses the 
multiplication law of $SO(3)$ in a neighbourhood of $\mathbb{1}$. 

\subsection{Complexification}
\label{sec:2.8.4}
The \emph{complex  de Sitter group}\index{complex  de Sitter group} is defined as
the group 
	\[ 
		O_{\mathbb{C}} (1,2)  
		\doteq 	\bigl\{ \Lambda \in GL (3,\mathbb{C})  
				\mid \Lambda \,  \mathbb{g} \, \Lambda^T =  \mathbb{g} 
			\bigl\}   \; . 
	\]
The elements in $GL (3,\mathbb{C})$ are 
invertible $3 \times 3$ matrices with 
complex  entries and $\mathbb{g}$ is the metric 
on Minkowski space $\mathbb{R}^{1+2}$ given in \eqref{metrik}.
The group $O_{\mathbb{C}} (1,2) $ has two connected components 
(distinguished by the sign of $\det \Lambda$, which 
takes the values  $\det \Lambda= \pm 1$). Following standard terminology, we set  
	\[
		SO_{\mathbb{C}} (1,2) 
		\doteq	\bigl\{ \Lambda \in M_3 (\mathbb{C}) 
				\mid \Lambda \,  \mathbb{g} \, \Lambda^T =  \mathbb{g} \; , \; \det \Lambda = 1  
			\bigl\} \; .  
	\] 
Note that $SO_{\mathbb{C}} (3)$ is isomorphic to $SO_{\mathbb{C}} (1,2)$; the isomorphism from 
$SO_{\mathbb{C}} (1,2)$ to $SO_{\mathbb{C}} (3)$ is given by the map 
	\begin{equation}
	\label{so3c}
		\Lambda \mapsto \begin{pmatrix} - i & 0 & 0 \\
		0 & 1 & 0 \\
		0 & 0 & 1
		\end{pmatrix} \Lambda \begin{pmatrix} i & 0 & 0 \\
		0 & 1 & 0 \\
		0 & 0 & 1
		\end{pmatrix} .
	\end{equation}
In particular, 
	\begin{align*}
		& \begin{pmatrix} i x_0 \\ x_1 \\ x_2 
		\end{pmatrix}^T
		\begin{pmatrix}  -1 & 0 & 0 \\
		0 & 1 & 0 \\
		0 &0 & 1
		\end{pmatrix} 
		\begin{pmatrix} 
		 \cos \theta & 0 & i \sin \theta \\
		0 & 1 & 0 \\
		 i \sin \theta &0 & \cos \theta
		\end{pmatrix} 
		\begin{pmatrix} i y_0 \\ y_1 \\ y_2 
		\end{pmatrix}
		\\
		& \qquad \qquad = 
		\begin{pmatrix} x_0 \\ x_1 \\ x_2 
		\end{pmatrix}^T
		\begin{pmatrix} 
		- i & 0 & 0 \\
		0 & 1 & 0 \\
		0 & 0 & 1
		\end{pmatrix} 
		\begin{pmatrix}  \cos \theta & 0 & i \sin \theta \\
		0 & 1 & 0 \\
		 i \sin \theta &0 & \cos \theta
		\end{pmatrix}
		\begin{pmatrix} i & 0 & 0 \\
		0 & 1 & 0 \\
		0 & 0 & 1
		\end{pmatrix} \begin{pmatrix} y_0 \\ y_1 \\ y_2 
		\end{pmatrix}
		\\
		& \qquad \qquad = \begin{pmatrix} x_0 \\ x_1 \\ x_2 
		\end{pmatrix}^T
		\underbrace{ \begin{pmatrix} \cos \theta & 0 & - \sin \theta  \\
		0 & 1 & 0\\
		\sin \theta & 0 & \cos \theta
		\end{pmatrix} }_{= R_1(\theta)}  \begin{pmatrix} y_0 \\ y_1 \\ y_2 
		\end{pmatrix} .		
	\end{align*}
This result explains the appearance of this matrix in \eqref{eqBooW} below.
As expected, the map \eqref{so3c} maps $R_1(\theta)$ to $\Lambda_{1}(i\theta)$. 

\subsection{Rotations}
\label{SO3}
We now change to Euclidean coordinates 
	\[
		\vec x \equiv (x_0, x_1, x_2) \in S^2 \; , 
	\]
as suggested by the last computation.
The rotations, which leave the 
Euclidean sphere \eqref{euclidsphere} invariant,
form the subgroup $SO(3)$ of $SO_\mathbb{C} (1,2)$; the imaginary part in the square bracket on 
the right hand side of~\eqref{eqBooW} is in agreement with \eqref{euclidsphere}. 
The group of rotations\index{group of rotations} $SO(3)$  leaves the 
sphere $S^2$ invariant. We denote the generators of the rotations  
		\[ 
				R_{1}(\beta)  \doteq  \begin{pmatrix}
										 \cos \beta &  0 & - \sin \beta \\
										0 &  1&0  \\
										 \sin \beta & 0 & \cos \beta   
 								\end{pmatrix}  , \quad
				R_{2}(\gamma) \doteq \begin{pmatrix}
 								\cos \gamma &  - \sin \gamma &0 \\
								 \sin \gamma &  \cos \gamma &0  \\
								0 & 0 & 1  
									\end{pmatrix} , 
				\quad  \beta,\gamma\in [0, 2 \pi ) \; , 
		\] 
around the coordinate axis by 
	\[
			\Kgeo_1  
			=  
			\begin{pmatrix}
							0 &  0 & -1 \\
							0 &  0 & 0  \\
							1 & 0 & 0   
				\end{pmatrix}  , 
		\quad 	 
			\Kgeo_2  = 
			\begin{pmatrix}
							0 &  -1 & 0 \\
							1 &  0 & 0  \\
							0 & 0 & 0   
			\end{pmatrix} , 
		\]
respectively. The $R_{0}(\alpha)$, $R_{1}(\beta)$ and $R_{2}(\gamma)$ are all hermitian, hence the matrices $\Kgeo_0$, $\Kgeo_1$ and 
$\Kgeo_2$ are all skew-hermitian. The latter satisfy the following relations characteristic for $\mathfrak{so}(3)$:
	\[
		[ \Kgeo_1 \, , \, \Kgeo_2 ] =  - \Kgeo_0 \; , 
		\qquad 
		[ \Kgeo_0 \, , \, \Kgeo_1 ] =   - \Kgeo_2 \; , 
		\qquad
		[ \Kgeo_2 \, , \, \Kgeo_0 ] = - \Kgeo_1 \;  . 
	\]

\goodbreak

\begin{remark}
We denote by $R^{(\alpha)}$ the rotations generated by\footnote{This is in agreement with 
the definition of $\Lgeo^{(\alpha)}$ in \eqref{Lalpha}.}
 	\[
		\Kgeo^{(\alpha)} \doteq \cos \alpha \; \Kgeo_1 + \sin \alpha  \; \Kgeo_2\;  , \qquad \alpha \in [ 0, 2 \pi) \; ,
	\]
namely the rotations 
	\[
		R^{(\alpha)} (\theta) 
		= R_{0} (\alpha) R_{1} (\theta) R_{0} (-\alpha) \; , \qquad \alpha, \theta \in [0, 2\pi) \; , 
	\]
which leave the boundary 
points ${\tt x}_\alpha = (0, r \sin \alpha, r \cos \alpha)$ 
and $- {\tt x}_\alpha$ of the time-zero half-circles 
	$\label{halfcirclealphapage}
		I_\alpha = R_{0} (\alpha) I_+
	$
invariant. 
\end{remark}

\subsection{The Euclidean sphere}
If $g \in SO(3)$, then $g \vec{o} \in S^2 \cong SO(3) / SO(2)$ is of 
the form 
	\[ 
		R_0(\alpha) R_{1}(\beta) \vec{o} \, , 
		\qquad \alpha \in  [0, 2 \pi) \, , \quad \beta \in [- \tfrac{\pi}{2} , \tfrac{\pi}{2} ] \, , 
		\quad 
		\vec{o} = \begin{pmatrix} 0 \\ 0 \\ r \end{pmatrix} \; .  
	\]
This point can be carried back 
to the point $\vec{o}$ by the action of $R_1(-\beta) R_0 (-\alpha)$, \emph{i.e.}, 
	\begin{equation}
		R_1(-\beta) R_0 (-\alpha) \, g \vec{o} = \vec{o} \; . 
	\end{equation}
The stabiliser of the point $\vec{o}$ is the group $\bigl\{ R_2 (\gamma) 
\mid \gamma \in [0, 2 \pi) \bigr\} \cong SO(2)$. 
Thus there exists some $\gamma \in [0, 2 \pi)$ such that 
	\[
		g = R_0 (\alpha )R_1(\beta) R_2 ( \gamma ) \; , 
	\]
\emph{i.e.}, any element $ g \in SO(3)$ can be written as an ordered product of 
three rotations which keep the coordinate axes invariant. 

\chapter{Induced Representations for the Lorentz Group}
\label{ch:3}

The single most important method for generating representations of a locally compact group 
is inducing representations from subgroups. If $H$ is a closed subgroup of a locally compact 
group $G$ and $\pi$ is a unitary representation of $H$, then $\operatorname{ind}^G_H \pi$ 
is a unitary representation of $G$ that is constructed by combining the action 
of $\pi$ with the algebraic and measure-theoretic interrelation of $G$, $H$, and~$G/H$.
The definition of $\operatorname{ind}^G_H \pi$ in full generality is technical and somewhat challenging. 
Much of the complexity of the definition of $\operatorname{ind}^G_H \pi$ in the general case 
is due to measure-theoretic delicacy in the action of $G$ on the quotient space $G/H$.
Hence, before we specialise our discussion to the case of $SO_0(1,2)$, we will  
briefly review some measure theoretic aspects, as well as some other key elements, of 
the general theory of induced representations, following mostly\footnote{The reader may also 
consult~\cite{Bruhat, GG, Knapp, Kostant, Lipsman, Mack, Vara, Vil, VG, VGG}.} \cite{BaR} 
and \cite{Folland}.

\section{Integration on homogeneous spaces} 
\label{sec:3.1}

As a preliminary step, we recall the measure on 
locally compact groups introduced by Alfr\'ed Haar in 1933. 

\subsection{Haar measure} 
Let $G$ be a locally compact topological group. Then there exists a left invariant \emph{Haar 
measure}\index{Haar measure} $\mu_G$ on the $\sigma$-algebra of Borel sets of $G$. 
For a detailed account of the construction of the Haar measure we refer the reader 
to \cite[Chapter III.7]{FD}. This measure is unique up to a normalisation 
factor; see, \emph{e.g.}, \cite[Theorems~(2.10) and (2.20)]{Folland}. 

Our first objective is to provide explicit formulas for the Haar measure, tailored towards a 
semisimple Lie group $G$, whose Iwasawa decomposition 
	$G=KAN$
is known. As a first step, we reproduce \cite[Proposition 2.1]{Lang}:

\begin{proposition} 
\label{prop:3.1.1}
Let $H$ be a locally compact group with two closed subgroups $A$ and $N$, such 
that $A$ normalizes\footnote{$A$ is said to normalize $N$ if $ANA= N$.
For the case of $SO_0(1,2)$, see \eqref{scalingH}.} $N$, 
and such that the product
	\[
		A \times N \to AN = H
	\]
is a topological isomorphism. 
Denote the Haar measures on $A$ and $N$ by ${\rm d} a$ 
and ${\rm d} n$, respectively. Then the functional
	\[
		f \mapsto \int_N \int_A f (an) \, {\rm d} a {\rm d} n
		= \int_A \int_N f (an) \, {\rm d} n {\rm d} a \; , 
		\qquad 
		f \in C_0 (H) \; , 
	\]
specifies a (left invariant) Haar measure on $H$. 
\end{proposition}

Before proceeding any further, we recall how the left and the right invariant Haar measure are 
related to each other.

\subsection{Modular functions} 

In general, the left-invariant Haar measure $\mu_G$ 
(denoted in Proposition~\ref{prop:3.1.1} as 
well as in the sequel simply by ${\rm d} g$) 
on~$G$ is \emph{not} equal to the 
right-invariant Haar measure. However, there always 
exists a multiplicative $\mathbb{R}^+$-valued function\footnote{Actually, 
it is a continuous homomorphism into $\mathbb{R}^+$.} 
$\Delta_G$ on~$G$, called 
the \emph{modular function}\index{modular function} of $G$, such that
	\[
		\int_G {\rm d} \mu_G (g) \; f(g g') = \frac{1}{\Delta_G (g')} \int  {\rm d} \mu_G (g) \; f(g) 
		\qquad \forall g' \in G
	\]
and for every $\mu_G$-integrable function $f$ on $G$. 
The modular function relates the left- and the right-invariant Haar measure:
	\[
		\int_G {\rm d} \mu_G (g) \; f(g^{-1}) = \int_G {\rm d} \mu_G (g) \; f(g) \Delta_G (g^{-1})  \;  . 
	\]
In case $ \Delta_G (g)=1$ for all $g \in G$,
the left and the right Haar measure coincide, and~$G$ is called \emph{unimodular}.

\goodbreak
\begin{example}
\label{lm:3.1.2} 
Consider the action $\Delta_* \colon \mathfrak{sl}(2,\mathbb{R}) \to \mathbb{R}$ of the character 
$\Delta_{SL(2, \mathbb{R})}$
(given by the modular function)
on the Lie algebra of $SL(2, \mathbb{R})$. Since $\mathbb{R}$ is
abelian, 
	\[
		\Delta_* (  [ X, Y] ) = 0 \qquad \forall  X, Y \in \mathfrak{sl}(2,\mathbb{R}) \; . 
	\]
But every element of $\mathfrak{sl}(2,\mathbb{R})$ 
is of this form, hence $\Delta_*=0$. As $SL(2, \mathbb{R})$ is connected, this implies $ \Delta_{SL(2, \mathbb{R})} (g)=1$ 
for all $g \in SL(2, \mathbb{R})$. In other words, $SL(2, \mathbb{R})$ is unimodular.
\end{example}

We can now further explore the case discussed above \cite[Proposition 2.2]{Lang}:
 
\begin{proposition}
\label{prop:3.1.3}
Let $H$ be a locally compact group with two closed subgroups $A$ and $N$, such that $A$ normalizes $N$.
Then there exists a unique continuous homomorphism 
$\delta \colon A \to \mathbb{R}^+$ such that 
	\[
		\int_N f(a^{-1}na) \, {\rm d} n = \delta (a) \int_N f(n) \,  {\rm d} n \; , 
		\qquad f \in C_0(N) \; , 
	\]
or, in other words,  
	\[
		\int_N f(na) \, {\rm d} n = \delta (a) \int_N f(an) \,  {\rm d} n \; , 
		\qquad f \in C_0(H) \; , 
	\]
If $N$ is unimodular, then $\delta$ is the modular function on $H$, that is
	\[
		\Delta_H (h) = \Delta_H (an) = \delta (a) \; .
	\]
\end{proposition}

\begin{remark}
The first statement is immediate because the map 
	\[
		n \mapsto a^{-1} n a
	\] 
is a topological group automorphism of $N$, which preserves Haar measure up to a 
constant factor, by uniqueness of the Haar measure.
\end{remark}

\begin{example}
\label{lm:3.1.5} 
Consider the groups $A = \{ \Lambda_1(t) \mid t \in \mathbb{R} \}$, 
$N = \{ D(q) \mid q \in \mathbb{R} \}$ and $AN 
= \{ \Lambda_1(t) D(q) \mid t, q \in \mathbb{R} \}$. 
Compute, using \eqref{scalingH} and 
setting $q' = {\rm e}^{t} q $,
	\begin{align*}
		\qquad \quad
		\int \;  f \bigl(  \Lambda_1(-t)  D(q)  \Lambda_1(t) \bigr) \; {\rm d}q
		& = \int  \;  f \bigl(  D( {\rm e}^{t} q) \bigr) \;  {\rm d}q  
		= {\rm e}^{-t}  \int   \;  f \bigl(  D( q') \bigr) \;  {\rm d} q'  \; ;
	\end{align*}
hence
	\[
		\Delta_{AN} \bigl( \Lambda_1(t) D(q) \bigr) 
		= \delta \bigl( \Lambda_1(t)  \bigr) = {\rm e}^{-t} \, ,
		\qquad t, q  \in \mathbb{R} \, . 
	\]
Thus $AN$ is \emph{not} unimodular. Note that 
\begin{align*}
	[ \Lgeo_1, \Kgeo_0 - \Lgeo_2 ] 
	&  
				= \begin{pmatrix}
							0 &  1 & 0 \\
							-1 &  0 & -1  \\
							0 & -1 & 0   
				\end{pmatrix} = \Lgeo_2 - \Kgeo_0\; , 
\end{align*}
but $\Lgeo_1$ does \emph{not} arise from a Lie bracket. Thus the argument in 
Example \ref{lm:3.1.2} does \emph{not} apply in the present case. 
\end{example}

\subsection{The Haar measure for $SO_0(1,2)$}
We are now able to provide an explicit formula 
for the Haar measure \index{Haar measure} of the groups we are interested in, 
taking advantage of the existence of the Iwasawa decomposition. 
The following results is \cite[Proposition 2.3]{Lang}:

\begin{proposition}
\label{prop:2.3}
Let $G$ be a locally compact group with two closed subgroups, $H$, $K$ such that
	\[
		K \times H \to KH = G
	\]
is a topological isomorphism (but not necessarily a group isomorphism). Assume that $G$, $K$ 
are unimodular. Let ${\rm d}g$, ${\rm d}h$, ${\rm d}k$ be given Haar measures on 
$G$, $H$, $K$ respectively. Then there is a constant~$c$ such that for all $f \in C_0 (G)$, 
	\[
		\int_G f(g) \; {\rm d} g = c \int_H \int_K  f (kh) \; {\rm d} k {\rm d} h \; . 
	\]
If in addition $H = AN$ as in Proposition \ref{prop:3.1.1}, with $N$ unimodular, so
we have the product decomposition
	\[
		K \times A \times N \to  G \; ,
	\]
then
	\[
		\int_G f ( g ) \;  {\rm d} g 
		= c  \int_N \int_A \int_K   f ( k a n ) \delta^{-1} ( a ) \;  {\rm d} k {\rm d} a {\rm d} n  \; .	
	\]
The modular function $ \delta \colon A \to \mathbb{R}^+$ appearing on 
the r.h.s.~has been introduced in Prop.~\ref{prop:3.1.3}.  
\end{proposition}
	
\begin{example}
Using the \index{Cartan decomposition} Cartan decomposition,
the \emph{Haar measure}\index{Haar measure} 
${\rm d} g$ on the \index{Lorentz group} Lorentz group $SO_0(1,2)$ can be written as 
(see Lemma~\ref{iwa}) 
	\begin{equation}
	\label{hmeasure}
		{\rm d} g =  \frac{{\rm d} \alpha}{2 \pi}
		 \; \sinh t \, {\rm d} t \; \frac{{\rm d}  \alpha' }{2 \pi} \; , \qquad  
		g = R_0 (\alpha) \Lambda_1 (t) R_0 (\alpha')\; ,
	\end{equation}
with $\alpha , \alpha' \in [0, 2\pi)$ and $t \in \mathbb{R}$ \cite[Chapter 9]{Vil}. 
On the other hand, using the Iwasawa decomposition, ${\rm d} g$ 
takes the form (see Proposition~\ref{prop:2.3}) 
	\begin{equation}
	\label{hmeasure2}
		{\rm d} g = \frac{{\rm d} \alpha}{2 \pi} \,  {\rm e}^{-t} {\rm d} t \, {\rm d} q \; , 
		\qquad g =  R_0(\alpha) \Lambda_{1}(t) D(q)  \; ,  
	\end{equation}
with  $\alpha \in [ 0, 2\pi)$, $t, q \in \mathbb{R}$ and $k \in \{ 0, 1\}$. 
Note that the two expressions \eqref{hmeasure} and \eqref{hmeasure2} for ${\rm d} g$ differ by 
a constant. The expression coincides with formula (12) provided on p.~24 in \cite{Vilenkin}.
\end{example}

\subsection{Integration over cosets} \index{integration over cosets}
Averaging over a subgroup $H$ provides a linear map from $C_0(G)$ to $C_0(G/H)$: define, for $f \in C_0(G)$, 
a function $f^H$ on $G$ by
	\[
		f^H(g)=\int_H {\rm d} h \; f(gh)  \, , \qquad g \in G \, .
	\]
Since, by assumption, $f$ is uniformly continuous, $f^H$ is a continuous function on $G$. Moreover, left 
invariance of the Haar measure ${\rm d}h$ on $H$ yields 
	\[
		f^H(gh)= f^H(g)  \qquad \forall g \in G\, , \; \forall h \in H \, . 
	\]
Hence there exists a unique function on $C_0(G/H)$, 
denoted by~$f^\sharp$, such that
	\[
		f^\sharp (gH) \doteq f^H(g)= \int_H {\rm d} \mu_H (h) \; f(gh)  \qquad \forall g \in G.
	\]
We note that for every $ \phi \in C_0(G/H)$, there exists an $f \in C_0(G)$ 
such that $ f^\sharp = \phi$. If~$\phi \in  C_0^+(G/H)$, then f can be chosen in~$C_0^+(G)$ \cite[Proposition 1.9]{Kaniuth}.

\subsection{$G$-invariant measures} \index{$G$-invariant measure}
Given a Borel measure $\mu$ on $G/H$, the action of the group $G$
on $G/H$ gives rise to a family of measures $\{ \mu_g \mid g \in G \}$ on $G/H$:
let~$E$ be a Borel set in $G/H$ and let  
	\[
		g \cdot E \doteq \{ g \cdot x  \mid x \in E \}  
	\]
be the \emph{left translate} of $E$ by an element $g \in G$. It follows that there exists  
a new measure $\mu_g \equiv \mu (g \, . \, )$ on $G/H$ given 
by the formula 
	\[
		\mu_g(f) = \int_{G/H} {\rm d} \mu (g \cdot x)\;  f(x) = \int_{G/H} {\rm d} \mu (x)\;  f(g^{-1}\cdot x) \qquad \forall f \in C_0 (G/H) \; . 
	\]
In other words, $\mu_g(E) = \mu (g\cdot E)$ for every Borel set $E$ in $G/H$.
A \index{regular Borel measure} regular Borel measure $\mu$ on $G/H$ is called a $G$-\emph{invariant} measure, if 
	\[
		\mu_g = \mu \qquad \forall g \in G \, .
	\]
A criterium for the existence of a $G$-invariant measure on $G/H$ is presented next.
 
\begin{theorem} 
\label{in-me}
The homogeneous space \index{homogeneous space} 
$G/H$ admits\footnote{This is Theorem 2.49 in \cite{Folland} and Theorem 1.16 in \cite{Kaniuth}.} 
a nonzero positive $G$~invari\-ant regular Borel measure $\mu_{G / H}$, 
if and only if 
	\begin{equation}
	\label{modular function}
		\Delta_{G} \upharpoonright H = \Delta_H \; . 
	\end{equation}
If \eqref{modular function} holds, then the positive invariant measure $\mu_{G / H}$
is \emph{unique} up to multiplication by a positive constant. Moreover, one can normalize the 
invariant measure $\mu_{G/H}$ on $G/H$ such that for every $f$ in $C_0 (G)$,
	\begin{equation}
		\label{quotient measure}
		\int_{G / H} {\rm d} \mu_{G / H} (gH) \underbrace{ \int_H {\rm d} \mu_H (h)  \; f(gh) }_{= f^\sharp (gH)}
		= \int_G {\rm d} \mu_G (g) \; f(g)  \; , 
	\end{equation}
where $\mu_G$  and $\mu_H$ denote the Haar measures of $G$ and $H$, respectively.
\end{theorem} 

\begin{lemma}
Let $H \subseteq G$ be compact, then $\Delta_{G} \upharpoonright H = 1$. In particular, if $G$ is
compact, then $G$ is unimodular.
\end{lemma}

\begin{proof} As $\Delta_G$ 
is continuous, it follows that $\Delta_G (H)$ is a compact subgroup of~$(\mathbb{R}^+, \cdot )$
and hence equal to $\{1\}$.
\end{proof}

\goodbreak

\begin{examples}
\quad
\begin{itemize}
\item [$i.)$] 
Since $G=SO_0(1,2)$ and $H= SO(2)$ are unimodular, we have 
	\[
		\Delta_{G} (h) = \Delta_H (h) =1 \qquad \forall h \in H \; . 
	\]
Thus the homogeneous space $H_m^+ = G/H$ possesses 
a Lorentz invariant measure by virtue of Theorem \ref{in-me}. In fact, 
the restriction of the measure \eqref{hmeasure} to the mass hyperboloid $H_m^+ \, $,  
	\[
		{\rm d} \alpha \,  \sinh t \,  {\rm d} t \; , 
	\]
equals $2/m$-times the Lorentz invariant measure on the mass hyperboloid 
	\begin{align}
	\label{dSmeasure}
		\qquad \quad	\int  {\rm d}^3 p \; \theta (p_0) \delta (p_0^2 - p_1^2 - p_2^2  - m^2) 
		& = \int \rho {\rm d} \rho \, {\rm d} \alpha \,  {\rm d} p_0 \; \theta (p_0) \tfrac{ \delta \left(\rho - \sqrt{p_0^2 - m^2} \right)}
		{2 \sqrt{ p_0^2 - m^2} } \nonumber \\
		& = \frac{1}{2} \int   {\rm d} \alpha \,  {\rm d} p_0
	\end{align}
used by Bros and Moschella in \cite{BM}. This can be seen by setting $p_0= m \cosh t$, which implies 
	\[ 
		{\rm d} p_0 = m \sinh t \, {\rm d} t \; .
	\]
Note that we have changed coordinates in \eqref{dSmeasure}, 
setting $p_1\doteq \rho \sin \alpha$ and $p_2 \doteq \rho \cos \alpha$. 
\item [$ii.)$] 
The group $SO(1,1) \cong (\mathbb{R}, +)$ is also unimodular. Therefore the 
de Sitter space $dS = SO_0(1,2) / SO(1,1)$ allows a Lorentz invariant measure too. 
The measure  used by Bros and Moschella in \cite{BM}, 
	\begin{align}
	\label{hhmeasure}
		\qquad \qquad \int  {\rm d}^3 x \;   \delta (x_0^2 - x_1^2 - x_2^2 + r^2 ) 
		& = \int \rho {\rm d} \rho \, {\rm d} \psi \,  {\rm d} x_0 \;  \tfrac{ \delta \left(\rho - \sqrt{x_0^2 + r^2} \right)}
		{2 \sqrt{ x_0^2 + r^2} } \nonumber  
		= \frac{1}{2} \int_{dS}  r\, {\rm d} \psi \,  {\rm d} x_0 
	\end{align}
differs from the measure we will use, namely 
	\[ 
		{\rm d} \mu_{dS} \doteq {\rm d} x_0 \, r \, {\rm d} \psi 
	\]
by a factor two. Taking \eqref{ds-me} into account we find $x_0 = r \sinh t$ and 
	\[
		{\rm d} \mu_{dS} = r^2 \cosh t \, {\rm d} t  \, {\rm d} \psi \; . 
	\]

\item [$iii.)$] 
As $N \cong (\mathbb{R}, +)$ is unimodular,  the forward light cone 
	\begin{align*}
		\qquad \partial V^+ \setminus \{(0, 0, 0) \} & = \left\{  \left( \begin{smallmatrix} p_0 \\
		p_0 \sin \alpha \\
		- p_0 \cos \alpha
		\end{smallmatrix} \right) \mid p_0 >0 , \alpha \in [0, 2 \pi) \right\} 
		 \cong \bigl \{g N \mid g \in SO_0(1,2) \bigr\}
	\end{align*}
posses a Lorentz invariant measure by Theorem \ref{in-me}.
Setting $p_0 = {\rm e}^{-t }$, we find $ {\rm d} p_0 = {\rm e}^{-t} {\rm d} t $ and, consequently,
the invariant measure on the forward light cone  is given (up to normalisation) by the formula
\cite[Chapter 9.1.9, Equ.~(13)]{Vil}
	\[
	 	|p_0|^{-1} {\rm d} p_1 {\rm d} p_2 = {\rm d} p_0 {\rm d} \alpha \; , 
	\]
in agreement with taking the limit $m \to 0$ in \eqref{dSmeasure}.  
There one finds 
	\begin{equation}
	\label{v+measure}
		{\rm d} \mu_{\partial V^+} = \tfrac{1}{2} {\rm d} \alpha \, {\rm d} p_0 \; ; 
	\end{equation}
thus the difference is again a factor of two. 
\end{itemize}
\end{examples}

\goodbreak

If the condition \eqref{modular function} is not satisfied, there is no 
\index{$G$-invariant measure}
$G$-invariant measure on $G/H$.
However, there may well exist \index{quasi-invariant measure} quasi-invariant measures: 

\begin{definition}[Strongly quasi-invariant measures]
\index{strongly quasi-invariant measure}
A regular Borel measure $\mu$ on $G/H$ is called 
\begin{itemize}
\item[$i.)$] \emph{quasi-invariant}\index{quasi-invariant measure},
if, for any fixed $g \in G$,  the measure  $\mu$ and the measure
$\mu_g $
are mutually absolutely continuous;
\item[$ii.)$] \emph{strongly quasi-invariant}, if the Radon-Nikodym derivative
	\begin{equation}
	\label{lambda-function}
		\lambda_g (g'H) \doteq\frac{ {\rm d} \mu_g} { {\rm d} \mu}  (g'H)  \; ,  
	\end{equation}
interpreted as a 
function $\lambda_{\, . \,} (\, . \,) \colon G \times G/H \to \mathbb{R}^+  $,
is jointly continuous in~$g$ and $g'H$ for $g, g' \in G$.
\end{itemize}
\end{definition}

Quasi-invariant measures send null sets  into null sets under the action 
of $G$. Strongly quasi-invariant measures  are needed to ensure that 
the induced representations, which we will construct, are \emph{strongly continuous}. 
For the applications we are interested in, the existence of measures 
with continuous Radon-Nikodym derivatives is assured by the following result 
(Theorem 1, Chapter 4 in \cite{BaR}):  

\begin{theorem}[Existence of strongly quasi-invariant measures]
\label{th:3.1.12}
Let $G$ be a locally compact separable group, 
$H$  a closed subgroup of $G$. Then 
\begin{itemize}
\item[$i.)$] there exists a strongly quasi-invariant measure on $G/H \, $; 
\item[$ii.)$] the Radon-Nikodym derivative \eqref{lambda-function} satisfies the \emph{cocycle relation}\index{cocycle}
	\begin{equation}
	\label{RNC}
		\lambda_{g_1 g_2} (g H ) = \lambda_{g_1} (g_2g H) \lambda_{g_2} (gH)  		
		\qquad 
		\forall g_1, g_2, g  \in G \; ;  
	\end{equation}	
\item[$iii.)$] any two strongly quasi-invariant measures on $G/H$ are equivalent,
\emph{i.e.}, they are mutually absolutely continuous. 
\end{itemize}
\end{theorem}

Explicit formulas for the strongly quasi-invariant measures on $G/H$ can be found
by exploring the fact that these measures are closely related to
rho-functions on~$G$. The latter are used to transfer integrations between $G$ and $G/H$: 

\begin{definition}
A real-valued function $\rho$ on~$G$ is a \emph{rho-function}\index{rho-function} for $(G,H)$, if 
it is non-negative, continuous and satisfies
	\begin{equation}
	\label{rhofunct}
		\rho(gh) = \frac{\Delta_H(h)}{ \Delta_G (h)} \rho(g)
		\qquad \forall g \in G \, ,  \quad \forall h \in H \, . 
	\end{equation}
\end{definition}

The advantage of this definition is that rho-functions are easily encountered:
if $0\le f \in C_0^+ (G)$, then the function $\rho_f$, defined by 
	\[
		\rho_f (g) \doteq \int_H {\rm d}h \;  \frac{\Delta_G(h)}{ \Delta_H (h)} f (gh) \, , \qquad g \in G \, , 
	\]
is a rho-function for $(G,H)$; this is Proposition 1.12 in \cite{Kaniuth}. 
	
\begin{example} 
\label{rhocomputed}
Let $G= SO_0(1,2)$ and $AN = \{ \Lambda_1(t) D(q) \mid t, q \in \mathbb{R} \}$. 
As $G$ is unimodular, 
	\[
		\rho(gh) = \Delta_{AN} (h)  \rho(g) \qquad \forall h \in AN \; , \quad \forall g \in G \; . 
	\]
We have already seen (in Example \ref{lm:3.1.2} $ii.)$) that 
	\[
		\Delta_{AN} \bigl( \Lambda_1(t) D(q) \bigr) = {\rm e}^{-t} \; , \qquad t, q \in \mathbb{R} \; . 
	\]
According to \eqref{rhofunct}, the allowed rho functions satisfy
		\begin{equation}
		\label{a-rhof2}
			\rho\bigl( R_0 (\alpha) \underbrace{\Lambda_1(t) D(q) 
			\Lambda_1(t') D(q') }_{\Lambda_1(t+t') D({\rm e}^{t'} q +q')} \bigr) 
			= {\rm e}^{-t'} 		
			\rho\bigl( R_0 (\alpha) \Lambda_1(t) D(q) \bigr) 
		\end{equation}
for $t, t', q, q,  \in \mathbb{R}$ and $\alpha \in [0, 2 \pi)$. In particular, choosing $t=q= 0$ we find 
		\begin{equation}
		\label{rhof2}
			\rho\bigl( R_0 (\alpha) \Lambda_1(t') D(q') \bigr) 
			= {\rm e}^{-t'} \rho \bigl( R_0 (\alpha) \bigr)  \, ,
			\qquad t', q' \in \mathbb{R} \, . 
		\end{equation}
A possible choice is $\rho \bigl( R_0 (\alpha) \bigr)=1$; 
this yields $\rho\bigl( R_0 (\alpha) \Lambda_1(t') D(q') \bigr) = {\rm e}^{-t'} $.
		
\end{example}

The importance of rho-functions stems 
from the fact that they induce Borel measures on $G/H$ (Proposition~1.14 in \cite{Kaniuth}):

\begin{theorem}
[Radon-Nikodym derivatives] \index{Radon-Nikodym derivative}
Let $\rho$ be a rho-function for~$(G,H)$. It follows that
\begin{itemize}
\item[$i.)$] there exists a unique (up to a 
multiplicative constant) strongly quasi-invariant measure $\mu_\rho$ on $G/H$ such that 
	\begin{equation}
	\label{GHmeasurea}
		\int_{G/H} {\rm d} \mu_\rho  (gH) \; f^\sharp (gH)   
		= \int_G {\rm d} \mu_G (g) \;  \rho(g) f(g)  \qquad \forall f \in C_0(G) \; ;
	\end{equation}
\item[$ii.)$] the measures $\mu_{\rho, g} \equiv \mu_\rho (g \, . \, )$, $g \in G$, are all 
absolutely continuous to each other;
\item[$iii.)$] the Radon-Nikodym derivative\index{Radon-Nikodym derivative} is given by 
	\begin{equation}
	\label{lambda-function-2}
	\lambda_g (g'H)  = \frac{\rho (gg')}{\rho (g') } \; , \qquad g, g' \in G \; . 
	\end{equation}
\end{itemize}
\end{theorem}
	
\begin{remark}
Equation \eqref{GHmeasurea} is called \index{Weil's integration formula}
\emph{Weil's integration formula}.
Clearly, \eqref{GHmeasurea} should be compared with \eqref{quotient measure}.
\end{remark}

In fact, given a $rho$-function $\rho$, we can now specify a \emph{unique} 
quasi-invariant measure $\mu_\rho$ by providing an explicit 
expression for its Radon-Nikodym derivative \cite[Theorem~2.56]{Folland}:

\begin{remarks}
\label{rhot} \quad
\begin{itemize}
\item[$i.)$] If $ \mu_{\rho'}$ is another strongly quasi-invariant measure with 
$rho$-function  $\rho'$, then 
	\[
		{\rm d} \mu_{\rho'} = \tfrac{\rho'(g)}{\rho(g)} {\rm d} \mu_{\rho} \qquad  
		\forall g \in G \; . 
	\]
\item[$ii.)$] 
For the choice discussed in Example \ref{rhocomputed}, we find 
(using \eqref{lambda-function} and \eqref{rhof2})
	\begin{align}
		\lambda_{ g^{-1} } \bigl( R_0(\alpha') AN \bigr) 
		=\rho (R_0(\alpha) \Lambda_1(t) D(q)  ) 
		\nonumber 
		= {\rm e}^{-t} \; , 
	\label{rhog}
	\end{align}
where $t \in \mathbb{R}$ is one of the parameters in the Iwasawa decomposition 
$R_0(\alpha) \Lambda_1(t) D(q)$ of $g^{-1} R_0(\alpha')$, 
in agreement with \eqref{lambda-function-2}.  
\item[$iii.)$] The restriction of the Lorentz invariant measure on $\mathbb{R}^{1+2}$ to the forward light-cone 	
\[
\qquad \bigl\{ gN \mid g \in SO_0(1,2) \bigr\}
\cong
\left\{ (\alpha, {\rm e}^{-t}) \in S^1 \times \mathbb{R}^+  \mid \alpha \in [0, 2\pi), t \in \mathbb{R} \right\} \; ,
\]
given by 
	\begin{equation}
	\label{measure-lightcone}
		{\rm d} \mu_{\partial V^+} 
		=  \tfrac{1}{ 2}  {\rm d} \alpha {\rm d} p_0 \; , 
		\qquad p_0 = {\rm e}^{-t} \; , 
	\end{equation}
defines an invariant measure on $G/N$; see \eqref{v+measure}. 
\item[$iv.)$]  
The cosets 
	$
		\{gAN \mid g \in SO_0(1,2) \}
	$
can be identified either with a 
circle on the forward light cone (using the Iwasawa decomposition, see  Section \eqref{circ-sub}) or 
with a pair of mass-hyperbolas \index{mass-hyperbola}
on the forward light cone (using the Hannabus decomposition, see Section \eqref{massshell-sub}).
More generally, we may consider any contour $\gamma$ on the light cone $SO_0(1,2)/N$, 
which intersects almost every light ray of the light cone at one point. 
Following \cite[Chapter 9.1.9]{Vil}, we denote by ${\rm d} \mu_{\gamma}$ the unique 
measure on the contour ${\gamma}$ that satisfies 
\label{Gammacpluspage}
	\begin{equation}
	\label{d-mu-gamma}
		|p_0|^{-1} {\rm d} p_1 {\rm d} p_2 
		= {\rm d} \lambda {\rm d} \mu_{\gamma} (\eta) \; , 
		\qquad \eta \in {\gamma} \; , 
	\end{equation}
where $p= \lambda \eta$, $\lambda >0$, $\eta \in {\gamma}$. The measure ${\rm d} \mu_{\gamma}$ 
is a strongly quasi-invariant measure with Radon--Nikodym derivative  
	\begin{align}
	\label{RN}
	 	 \frac{{\rm d} \mu_{{\gamma},   g^{-1}  }}  
		{ {\rm d} \mu_{\gamma} } (g' AN) =   p_0		\; , 
		\qquad  \text{with} \quad g^{-1}g' =  R_0( \alpha) \Lambda_{1}(t) D(q) \; ,      
	\end{align}
and $p_0= {\rm e}^{-t}$,  in agreement with \cite[p.169, 170] {Knapp}.
It follows that if a function $f(p)$ on $\partial V^+$ is homogeneous\footnote{Given a homogeneous function on $\partial V^+$, 
one can define a function on the two hyperbolas introduces in \eqref{hypm0} by restriction. On the contrary, given a pair of functions 
$p_1 \mapsto (h_+ (p_1), h_-(p_1))$ on ${\gamma}_+ \; \dot \cup \; {\gamma}_-$, 
the map
	\[
	 	f(p) \doteq \begin{cases}
				\left(\tfrac{p_2}{ m_0 }\right)^s 
				h_+ \left(\tfrac{ m_0  p_1}{p_2}\right)  & \text{if $p_2 >0\; $, }\\ 
				\left( - \tfrac{p_2}{ m_0 }\right)^s  
				h_- \left(\tfrac{ m_0  p_1}{p_2}\right)   & \text{if $p_2 <0\; $, }
			\end{cases}
	\]
defines a homogeneous function of degree $s$ on the light cone $\partial V^+$, 
except for the set of measure zero consisting of the two 
light rays $\{ \lambda (1, \pm 1, 0) \mid \lambda \in \mathbb{R}^+ \}$; 
see \cite[Equ.~4.44]{BM}.} of degree $-1$, \emph{i.e.}, 
	\[
		f(\lambda p) = \lambda^{-1} f(p)\; , \qquad \lambda >0 \; , 
	\]
then the integral 
	\[
		\int_{\gamma} {\rm d} \mu_{\gamma} (\eta) f (\eta) 
	\]
does not depend on the choice of the contour ${\gamma}$
\cite[Proposition 10]{BM2}.
\end{itemize}
\end{remarks}

\section{Induced representations\index{representation}}
\label{sec:3.2}

Let $G$ be a locally compact, separable group, $H$ a closed subgroup, and let 
	\[
		\pi \colon H \to {\mathscr B}(\mathcal{H})
	\] 
be a representation of $H$ on some separable Hilbert space $\mathcal{H}$.  We denote the norm and 
the inner product on $\mathcal{H}$ by $\| u \|_\mathcal{H}$ and  $ \langle u , v \rangle_\mathcal{H}$, and denote by 
$C(G,\mathcal{H})$ the space of norm continuous vector valued functions from $G$ to $\mathcal{H}$. 

\goodbreak 

\begin{definition}
\label{fhg}
Let $\mathcal{F}^H(G, \pi)$ denote the set of 
functions $\eta \colon G \to \mathcal{H}$ that share the following 
properties: 
\begin{itemize}
\item [$i.)$] $\eta$ is continuous; 
\item [$ii.)$] the image $\mathbb{\Pi} ({\rm supp} \, \eta)$ 
of the support of $\eta$
under the map $\mathbb{H}$ introduced 
in~\eqref{quotient-map} is compact in $G/H$; 
\item [$iii.)$] for $g \in G$ and $h \in H$, 
	$\eta(gh) = \pi(h^{-1}) \eta(g)  $. 
\end{itemize}
\end{definition}

Note that if $\pi$ is unitary and $f \in \mathcal{F}^H(G, \pi)$, then $\| f (g) \|_{\scriptscriptstyle \mathcal{H} }$ depends only on 
the equivalence classes $gH$, $g \in G$. 
It is not difficult to find functions that satisfy the conditions $i.)$, $ii.)$ and $iii.)$; see, for example, Proposition 6.1 in \cite{Folland}:

\begin{proposition} 
If $f \colon G \to \mathcal{H}$ is continuous with compact support, then the function 
	\[
		\eta_f (g) \doteq \int_H {\rm d} h \;  \pi (h^{-1}) f(gh) 
	\]
belongs to $\mathcal{F}^H(G, \pi)$ and is uniformly continuous on $G$. Moreover, every element of $\mathcal{F}^H(G, \pi)$ is of the 
form $\eta_f$ for some $f \in C_0(G , \mathcal{H})$.
\end{proposition}

\goodbreak
Now let $\mu$ be a strongly quasi-invariant measure on $G/H$.
\begin{itemize}
\item[$i.)$] In case $\pi$ is unitary, 
	\begin{equation}
	\label{SP}
		\langle f (gh), f' (gh) \rangle_\mathcal{H} 
		= \langle f (g),  \underbrace{\pi^*(h^{-1}) \pi (h^{-1})}_{=\mathbb{1}} f' (g) \rangle_\mathcal{H} \; , 
	\end{equation} 
and therefore
	\begin{equation}
	\label{Spr}
		\langle f, f' \rangle_{\mu} \doteq \int_{G/H} 
		{\rm d} \mu(gH) \; \langle f (g), f' (g) \rangle_\mathcal{H}  \; , 
		\qquad  f, f' \in \mathcal{F}^H(G, \pi) \; , 
	\end{equation}
defines an \emph{inner product}\index{inner product} on $\mathcal{F}^H(G, \pi)$. The corresponding induced representations 
are called \emph{principal series} representations.

\goodbreak
\item[$ii.)$]  
In case $\pi$ is a non-unitary character of $H$, 
the scalar product \eqref{Spr} can \emph{not} be $G$-invariant. 
However, 
it turns out that it is possible to replace \eqref{Spr} by   
a \emph{new} inner product\index{inner product} on $\mathcal{F}^H(G, \pi)\, $, 
	\[
		\qquad \qquad \langle f, f' \rangle_{\mu} \doteq 
		\int_{G/H \times G/H} {\rm d} \mu(gH) {\rm d} \mu(g' H)\; 
		K (gH , g'H )\langle f (g), f' (g') \rangle_\mathcal{H}  \; . 
	\]
The kernel $K (\, . \, , \, . \, )$ has to be selected in 
such a way that it compensates the additional factor resulting from the non-unitarity 
of the representation $\pi$ of $H$. The ensuing unitary representations on the completion 
of $\mathcal{F}^H(G, \pi)$ are called the \emph{complementary series} representations 
(see, \emph{e.g.}, \cite[p.~32]{Lipsman}).

\end{itemize}

\begin{definition}
\label{Fmu}
Let $\mathcal{F}_\mu$ denote the completing of $\mathcal{F}^H(G, \pi)$ w.r.t.~the norm 
	\[
		\| f \|_\mu \doteq \sqrt{ \langle f, f\rangle_{\mu}} \; . 
	\] 
Let $\rho$ be a rho-function for~$(G,H)$ and let $\mu$ be the associated
strongly quasi-invariant measure on $G/H$ specified in 
\eqref{GHmeasurea}.
The \emph{induced representation}\index{induced representation} $ \Pi_\mu (g)$ on the Hilbert 
space~$\mathcal{F}_\mu$ is defined by setting
	\begin{align}
		\bigl( \Pi_\mu (g) f \bigr) (g') & \doteq \sqrt{  \lambda_{g^{-1}}(g'H)} \; f (g^{-1} g') \; , \qquad g, g' \in G \; , 
	\label{indrep}
	\end{align}
where $\lambda_g$  is the Radon-Nikodym derivative specified in \eqref{lambda-function-2}. 
\end{definition}

\goodbreak

\goodbreak
\begin{remarks}
\label{rm:3.2.2}
\quad
\begin{itemize}
\item[$i.)$]  The cocycle relation \eqref{RNC} ensures that \eqref{indrep} defines a representation of~$G$. 
\item[$ii.)$]
Given two quasi-invariant measures $\mu$ and $\mu'$, there exists a unitary operator 
from ${\mathcal F}_\mu$ to ${\mathcal F}_{\mu'}$ , which intertwines the representations 
$\Pi_\mu$ and $\Pi_{\mu'}$ (see Chapter 16, Proposition 4 in \cite{BaR}).
In other words, while $\Pi_\mu$ depends on the choice of the quasi-invariant measure $\mu$, 
its unitary equivalence class depends only on $\pi$.  
\item[$iii.)$] 
For the principal series, \eqref{indrep} implies that $\Pi_\mu (g)$ is unitary:
	\begin{align*}
		\qquad \quad \int_{G/H} {\rm d} \mu (g'H) \; 
		\| \bigl( \Pi_\mu (g) f \bigr) (g') \|^2_\mathcal{H} & = 
		\int_{G/H} {\rm d} \mu (g'H) \; 
		 \lambda_{g^{-1}} (g'H) \| f (g^{-1} g') \|^2_\mathcal{H} \\
		& = 
		\int_{G/H} {\rm d} \mu (g'H) \; \| f (g') \|^2_\mathcal{H} \; . 		
	\end{align*}
For the complementary series of $SO_0(1,2)$, we will show that $\Pi_\mu (g)$ is unitary by explicit computations.
\end{itemize}
\end{remarks}

\subsection{A first reformulation} The advantage of choosing $\mathcal{F}^H(G, \pi)$ as a starting point 
is that the induced representation \eqref{indrep} takes a relatively simple form. 
However, taking into account the conditions $ii.)$  and $iii.)$ in Definition \ref{fhg},
one may be tempted to reformulate the induced representation directly on functions 
in $C_0(G/H,\mathcal{H})$. 

In fact, if $G$ is second countable and separable, then there exist (see Lemma 1.1 in \cite{MA1})
a smooth global \emph{Borel section} \index{Borel section} (which is neither unique nor canonical)
	\[
		\Xi \colon G/H \to G \, , 
	\]
\emph{i.e.}, $\Xi (G/H )$ is a Borel set $M \subset G$, namely the image of $G/H$ under the map $\Xi$, 
that meets each coset in $G/H$ in exactly one point. It follows that each $g \in G$ can be written uniquely as 
	\begin{equation}
	\label{Mackey-decomp}
		g= g_M g_H \; , \qquad g_M  \in M \; , \; g_H \in H \; . 
	\end{equation}
Note that, by construction, 
	\begin{equation}
	\label{mackey-element}
		g_M= \Xi(\mathbb{\Pi} (g)) 
		\quad \text{and} \quad x 
		= \mathbb{\Pi} (\Xi (x)) \, ,  \quad x \in G / H \, .
	\end{equation}
Clearly, each $f\in\mathcal{F}^H(G, \pi) \, $ is completely determined by its restriction $f_{\upharpoonright M}$; 
and a quasi-invariant measure $\mu$ on $G/H$ yields a measure $\widetilde  \mu$ on $M$ by 
	\[
		\widetilde \mu (E) = \mu(\mathbb{\Pi}(E)) \; , 
		\qquad E \subset G \, .  
	\]
As before, $\mathbb{\Pi}$ denotes the canonical quotient map introduced in 
\eqref{quotient-map}.

\begin{lemma}
The map 
	\[
		f \mapsto f_{\upharpoonright M} \equiv \widetilde f 
	\]
gives a unitary identification of ${\mathcal F}_\mu$ and 
$L^2( M , \widetilde \mu, \mathcal{H})$, and under this identification the representation 
$\Pi_\mu$ given by \eqref{indrep} turns into 
	\begin{align}
		\bigl( \widetilde
		\Pi_\mu (g) \widetilde f \, \bigr) (m) & \doteq \sqrt{ \lambda_{g^{-1}} (mH)} \; 
		\pi\big((g^{-1} m)_H^{-1} \big)\, 
		\widetilde f \bigl((g^{-1} m)_M \bigr) \; ,   
	\label{indrepM}
	\end{align}
where $g \in G$, $m \in M$, and $\widetilde f \in L^2( M , \widetilde \mu, \mathcal{H})$.
\end{lemma}

\begin{proof} 
Inspecting \eqref{indrep}, it is sufficient to note that, for $f \in \mathcal{F}^H(G, \pi)$,
	\begin{align*}
		f (g^{-1} m) & = \pi\big((g^{-1} m)_H^{-1}\big)\, 
		f \bigl((g^{-1} m)_M\bigr) \\
		& = \pi\big((g^{-1} m)_H^{-1}\big)\, 
		f_{\upharpoonright M} \bigl((g^{-1} m)_M \bigr) \; . 
	\end{align*}
This result extends to $f \in {\mathcal F}_\mu$.
\end{proof}

Using \eqref{mackey-element}, one can replace $L^2( M , \widetilde \mu, \mathcal{H})$ by the Hilbert space 
$L^2 (G/H, \mu, \mathcal{H})$ of square integrable vector valued functions with domain in $G/H$:

\begin{definition}
Given a $f\in\mathcal{F}^H(G, \pi) \, $ and a smooth Borel section $\Xi$, 
we define a new function $f_{\Xi} \in C_0(G/H,\mathcal{H})$ associated to the section $\Xi$ by
	\begin{equation} 
		\label{eqFeqFGH}
			f_{\Xi} (x) \doteq f\big(\Xi(x)\big) \; ,  \quad x\in G/H \; ;
	\end{equation} 
this circumstances are described by the following commutative diagram:

\vskip 0.8cm

\begin{picture}(140,50)
    	\put(105,45){$G$} \put(140,10){\vector(-40,60){20}}
    	\put(180,45){$\mathcal{H}$} \put(115,40){\vector(20,-30){20}}
    	\put(140,0){$G/H$} \put(180,20){$f_{\Xi}$}
    	\put(160,10){\vector(20,30){20}}
    	\put(130,45){$\vector(30,0){40}$} \put(110,20){$ \mathbb{\Pi} $}
    	\put(135,25){$ \Xi$} \put(145,55){$f$}
\end{picture}
\end{definition}

\begin{remark}
Given a Borel section $\Xi$ and a function 
$\psi \in C_0(G/H,\mathcal{H}) $, we can recover a 
function $\psi^\Xi \in \mathcal{F}^H(G, \pi)$ by setting
	\begin{equation} 
		\label{eqFeqFGH2}
		\psi^\Xi  (g) \doteq \pi\big(g^{-1}\Xi(
		\mathbb{\Pi} (g))\big)\, \psi( 
		\mathbb{\Pi} (g)) \; .
	\end{equation}
Given a function $f\in\mathcal{F}^H(G, \pi) \, $, we find
	\begin{align*}
		(f_{\Xi})^\Xi (g) & =   \pi\big(g^{-1}\Xi( 
		\mathbb{\Pi} (g))\big)\, 
		f_{\Xi} (\mathbb{\Pi} (g)) \\
		& = \pi\big(g^{-1} \underbrace{\Xi( 
		\mathbb{\Pi}  (g))}_{=g_M} \big)\,  
		f\big(\underbrace{\Xi( 
		\mathbb{\Pi} (g))}_{=g_M}\big) = f(g) \qquad \forall g \in G \; . 
	\end{align*}
In the last equality, we have used $g^{-1}g_M = g_H^{-1}$ and 
property $iii.)$ of Definition~\ref{fhg}.
\end{remark}

The two maps \eqref{eqFeqFGH} and \eqref{eqFeqFGH2}
establish an isomorphism of linear space between~$\mathcal{F}^H(G, \pi)$ and 
$C_0(G/H;\mathcal{H})$. 
This isomorphism 
extends to the appropriate closures \cite[Ch.~16, Lemma 1]{BaR}:

\begin{lemma}
The space $ {\mathcal F}_\mu$ is isomorphic to the Hilbert space 
$L^2 (G/H, \mu, \mathcal{H})$. The isomorphism is given by the formula
	\begin{equation}
	\label{eqFeqFGH2b}
		f (g) = \pi ( g_H^{-1}) f_\Xi (\mathbb{\Pi}(g)) \; , 
	\end{equation}
where $g_H$ is the factor of $g$ in the Mackey decomposition \eqref{Mackey-decomp}.
\end{lemma}

\begin{proof} 
It follows from Remark 
\ref{rm:3.2.2} $i.)$ that the 
scalar product \eqref{SP} associates a positive valued function on $G/H$ to any $ h \in C_0(G/H,\mathcal{H})$: 
	\[
		\| h (x) \|_\mathcal{H} 
		\doteq \| h^\Xi \big(\Xi(x)\big) \|_\mathcal{H} \;  , \qquad x \in G/H \; .  
	\] 
Thus the norm $\| f\|_\mu$ associated to the scalar product~\eqref{Spr} 
in $\mathcal{F}^H(G, \pi)$ equals 
the $L^2$-product in $C_0(G/H, \mathcal{H})$: 
	\begin{align} 
		\label{eqScalProdGH} 
		\| f\|^2_\mu & = \int_{G/H} {\rm d} \mu(gH)\, \|f(g)\|^2_{\mathcal{H}}  \nonumber \\
		& = \int_{G/H} {\rm d} \mu(gH)\, 
		\| f_{\Xi} (\mathbb{\Pi} (g)) \|^2_{\mathcal{H}} 
		\nonumber
		\\
		& = \int_{G/H} {\rm d} \mu(x)\, \| f_{\Xi} (x) \|^2_{\mathcal{H}} 
		= \| f_{\Xi} \|^2_{L^2 (G/H, \mu, \mathcal{H})}
		\; . 
	\end{align}
Inspecting \eqref{Mackey-decomp}, we note that 
	\[
		g = \Xi(x)  g_H  \quad \text{with} \quad  x\doteq 
		\mathbb{\Pi} (g) \; . 
	\]
Thus $g^{-1}\Xi(x) = g_H^{-1}$, and therefore \eqref{eqFeqFGH2b} is 
just the extension of \eqref{eqFeqFGH2}.
\end{proof}

\begin{proposition}[Wigner representation]
\label{prop:3.2.9}
The map $ g \mapsto \Pi_{\Xi,\mu}(g)$ specified by setting\footnote{As before, 
we denote the action of $G$ in $G/H$
by a dot, $g\cdot (gH) \doteq (gg')H$.}
	\begin{equation} 
		\label{eqRepGH}
		\bigl( \Pi_{\Xi,\mu} (g) \psi \bigr) (x) \doteq 
		\sqrt{ \lambda_{g^{-1}} \big(\Xi(x)H\big)} \; 
		\pi\big(\Omega(g,x)\big)\; \psi (g^{-1}\cdot x) \; ,
	\end{equation}
where $x \in G/H$ and 
	\begin{equation}
	\label{Wigner-Rot}
		\Omega(g,x) \doteq \Xi(x)\,g \, \Xi(g^{-1}\cdot x) \, \in H 
	\end{equation}
is the so-called \emph{Wigner rotation}, provides a unitary 
representation of $G$ on $L^2(G/H, \mu , \mathcal{H})$.
\end{proposition}

\begin{proof} According to \eqref{indrepM}, we have, for $x \in G/H$, 
	\begin{align*}
		\bigl( \Pi_{\Xi,\mu} (g) f_\Xi \, \bigr) (x)  
		& 
		\doteq \sqrt{ \lambda_{g^{-1}} (\Xi(x)H)} \; 
		\underbrace{ \pi\big((g^{-1} \Xi(x))_H^{-1} \big)\, 
		f_{\Xi} \bigl(  \mathbb{\Pi} 
		\bigl(g^{-1} \Xi(x)\bigr) \bigr)}_{= f (g^{-1} \Xi(x))}\; .   
	\label{indrepM}
	\end{align*}
We would like to replace $ f (g^{-1} \Xi(x))$ by 
$ f \bigl(\Xi(g^{-1}\cdot x)\bigr)$. Note that for $x = g'H$
	\begin{align*}
	\bigl(g^{-1} \Xi(x) \bigr)_M & 
	= \Xi \bigl(\mathbb{\Pi} ( g^{-1} \Xi(x)) \bigr)  
	 = \Xi \bigl(\mathbb{\Pi} ( g^{-1} g'_M) \bigr) \\
	& =  \Xi( (g^{-1}g'_M)H)  
	 = \Xi(g^{-1}\cdot g'_M g'_H H)  = \Xi(g^{-1}\cdot x)\; . 
	\end{align*}
Thus, using the Mackey decomposition \eqref{Mackey-decomp}, 
	\begin{equation} 
		\label{eqWignerRot}
		g^{-1} \Xi(x) = \underbrace{\Xi(g^{-1}\cdot x)}_{(g^{-1} \Xi(x))_M}  
		\, \Omega(g,x)^{-1} \; ,
	\end{equation}
with $\Omega(g,x) \doteq \Xi(x)\,g \, \Xi(g^{-1}\cdot x) \, \in H \; $.
\end{proof}

Note that contrary to the induced representation \eqref{indrep}, 
the representation \eqref{eqRepGH} involves a certain 
degree of arbitrariness as it involves the choice of a section $\Xi$.

\begin{remark}
The isomorphism~\eqref{eqFeqFGH} intertwines the respective 
scalar products, and \hbox{(anti-)} unitary operators in $\mathcal{F}^H(G, \pi)$ go 
over into (anti-) unitary operators in $L^2(G/H; \mathcal{H})$. 
\end{remark}

We will now concentrate on the construction of unitary irreducible representations of $SO_0(1,2)$; 
representations of its two-fold covering group $SL(2, \mathbb{R})$ will be discussed elsewhere. 
Unitary irreducible representations of the Lorentz group $SO_0(1,2)$ (and its two-fold covering 
group $SL(2, \mathbb{R})$) were first\footnote{The derivation and classification of the 
representations of $SL(2, \mathbb{R})$ are due to Bargmann~\cite{Ba}; see also \cite{GGV}. 
Gelfand and Naimark \cite{GN}\cite{Nai} and Harish-Chandra \cite{HC0}
investigated the group~$SL(2, \mathbb{C})$.} derived in the form of  
\emph{multiplier representations}. However, we prefer to use the method of  
\emph{induced representations}, which was pioneered by Wigner \cite{W} 
and Gelfand \& Naimark \cite{GN}, and in the sequel, by Mackey~\cite{Mack}. 
While the multiplier representations emerged from 
Schur's theory of projective representations, the induced representations  
were inspired by earlier work of Frobenius (see, \emph{e.g.},~\cite{Vara}). 

\section{Reducible representations on the light-cone}
\label{RRLC}

A representation $\pi_{\nu}  \colon AN \to \mathbb{C}$ 
of the closed subgroup $AN$ of $SO_0(1,2)$ 
on~$\mathbb{C}$ is defined by lifting a character~$\chi_\nu$ of $A$, namely
	\[ 
		\chi_\nu	 \begin{pmatrix}
				 \cosh t  &  0 &\sinh t \\
     					  0  &  1 & 0  \\
 				  \sinh t &  0 & \cosh t   
				\end{pmatrix} 
 			= {\rm e}^{i \nu t} \; ,  
		\]
to $AN$: 
	\begin{equation}
	\label{indrepc} 
		\pi_\nu \big(\Lambda_1(t)D(q) \big) \doteq  {\rm e}^{i\nu t} \;  , 
		\qquad t, q \in \mathbb{R} \; .  
	\end{equation} 

\goodbreak

\begin{remarks}
\quad
\begin{itemize}
\item [$i.)$] in case $\nu \in \mathbb{R}$, the representation $\pi_{\nu} $ is unitary; 
\item [$ii.)$] in case $\nu$ is purely imaginary, 
the representation \eqref{indrepc} is {\em no longer} a unitary representation 
of $AN$ in~$\mathbb{C}$. 
\end{itemize}
\end{remarks}

\begin{definition} 
\label{def:3.3.2}
Let ${\mathfrak h}_{\nu,0}$ denote the functions in $C (SO_0(1,2))$ 
which satisfy 
	\[
		f(gh)= \pi_{\nu} (h^{-1}) f(g) \; \text{for} \; h \in AN
	\]
and $ \mathbb{\Pi} ({\rm supp} \, f) $ compact. 
\end{definition}

\label{h-nu-0-page}
\goodbreak
This definition implies that a function $f \in {\mathfrak h}_{\nu,0}$ depends only on
the cosets $gN$, $g \in SO_0(1,2)$, as	
  \begin{equation}
	\label{fcoset}
		f (gn) = f(g) \qquad \forall g \in SO_0(1,2)  \; , 
		 \; \forall n \in N \; ;  
	\end{equation}
it also satisfies  $f \bigl( g \Lambda_1(t)D(q) \bigr)  = 
{\rm e}^{- i \nu t}   f (g)$  for all $t, q \in \mathbb{R}$.
We will explore these facts further in the next subsection. 
And finally, if $\nu$ is purely imaginary, then 
	\begin{equation}
	\label{Compseries}
	\int_G {\rm d} g \, | f(gh) |^2 = \int_G {\rm d} g \,  \bigl(\pi_{\nu} (h) 
	 \overline{f} \bigr)(g) 
	\bigl(\pi_{-\nu} (h) f\bigr)(g)  = \int_G {\rm d} g \, | f(g) |^2
	\end{equation}
for all $h \in AN$ (despite the fact that \eqref{indrepc} is not a
unitary representation  
of $AN$ in~$\mathbb{C}$). Note that the dependence of the middle term
in~\eqref{Compseries} 
on $\nu$ drops out as long as $\nu$ is purely imaginary.

\begin{definition}
The representation $\Pi_\nu$ of $SO_0(1,2)$, induced from the  
representation~$\pi_{\nu}$ of the closed subgroup $AN$, is given by
	\begin{equation} 
		\label{eqIndRepSO12}
		\bigl( \Pi_\nu (g)f \bigr)(g')= \sqrt{  \lambda_{g^{-1}} (g'AN)}\,  f(g^{-1}g') \; , 
		\quad f\in {\mathfrak h}_{\nu,0} \; .
	\end{equation}
We will extend $\Pi_\nu$ to the closure of ${\mathfrak h}_{\nu,0}$ in Proposition \ref{Prop:2.1}.
\end{definition}

To compute explicit expressions for the representation~\eqref{eqIndRepSO12}, and  
for specific choices of $g \in SO_0 (1,2)$, one can take advantage of \eqref{fcoset}. According to 
Lemma~\ref{repV+} the map 
	\[
		(g N)   
		\mapsto  g 	\left( \begin{smallmatrix}
							1 \\
							0 \\
							-1
						\end{smallmatrix} \right) \; , \qquad g \in SO_0 (1,2) \; ,  
	\]
defines a bijection, which identifies the homogeneous space 
	\[
		\bigl\{ gN \mid g \in SO_0(1,2) \bigr\}
		= \bigl\{  R_0(\alpha)  \Lambda_{1}(t) N  \mid \alpha \in [0, 2\pi), t \in \mathbb{R}  \bigr\}
	\]
with the forward light cone
	\begin{equation}
	\label{X-lightcone}
		\partial V^+  \cong \left\{ (\alpha, {\rm e}^{-t}) \in S^1 \times \mathbb{R}^+  \mid \alpha \in [0, 2\pi), t \in \mathbb{R} \right\} \; . 
	\end{equation}
Setting $p_0 = {\rm e}^{-t}$, 
the action of $SO_0(1,2)$ on the forward light cone
$\partial V^+$ is given by~\eqref{lambda2-s}, \emph{i.e.},
\label{umLambdapage}
	\begin{align}
		\label{udrei} 
 			\Lambda_2 (s)^{-1}  (\alpha', p_0') 
			& =  \bigl( \alpha_2 \,  , \,  p_0' (\cosh s - \sinh s \sin \alpha') \bigr)
	\nonumber \\  
			\Lambda_1 (t)^{-1} (\alpha', p_0')  
		&= \bigl( \alpha_1  \,  , \,  p_0' (\cosh t - \sinh t \cos \alpha') \bigr) 
	 \nonumber \\
			R_{0}^{-1} (\alpha)   (\alpha', p_0')  
		&=  (\alpha' -  \alpha \,  , \, p_0')     \; , 
	 \nonumber \\
			P  (\alpha', p_0')  
		&
		= (\alpha'  + \pi \,  , \, p_0')     \; , 
	\end{align}
with
	\begin{align*}
	( \sin \alpha_2 , \cos \alpha_2)  & = \left( \tfrac{-\sinh s + \cosh s \sin \alpha'}
	{\cosh s -  \sinh s \sin \alpha'} \; , 
	  \tfrac{\cos \alpha'}{\cosh s -  \sinh s \sin \alpha'} \right) \; , \\
	(\sin \alpha_1 , \cos \alpha_1)  &= 
	\left( \tfrac{\sin \alpha'}{\cosh t - \sinh t \cos \alpha'} ,  \tfrac{-\sinh t 
	+ \cosh t \sin \alpha'}{\cosh t - \sinh t \cos \alpha'} \right) \; .
	\end{align*}
Using these formulas, we can investigate the pullback action on 
$C^\infty_0$ functions $f$ on the forward light cone. The maps 
	\begin{align*}
 			f (\alpha', p_0') & \mapsto f \bigl( \Lambda_2 (s)^{-1}  (\alpha', p_0') \bigr) \; , 
			\\ 
			f (\alpha', p_0') & \mapsto f \bigl(\Lambda_1 (t)^{-1} (\alpha', p_0') \bigr) \; , 
			\\ 
 			f (\alpha', p_0') & \mapsto f \bigl(\alpha' -  \alpha \,  , \, p_0')     \; , 
	\end{align*}
are differentiable, and resulting 
generators $\LV_2$, $\LV_1$ and $\KV_0$ take the form (see~\cite[\S 6a, Equ.~(6.4)]{Ba})
\label{bargmangeneratorpage}
	\begin{align}
		i \LV_2  &= \cos (\alpha) \;  \tfrac{\partial}{ \partial \alpha}
			+ \sin (\alpha) \; p_0  \tfrac{\partial}{ \partial p_0} \; , \nonumber \\  
		i \LV_1  &= \sin (\alpha) \; \tfrac{\partial}{ \partial \alpha} 
			- \cos (\alpha) \;    p_0  \tfrac{\partial}{ \partial p_0} \; , \nonumber \\ 
		i \KV_0 &= -  \tfrac{\partial}{ \partial \alpha} \; .
	\label{qqdq}
	\end{align}
Note that $\KV_0^2=-  \frac{\partial^2}{ \partial \alpha^2}$ is a positive operator. 
The eigenfunctions of $\KV_0$ on the light cone for the eigen\-value~$k$ are 
of the form $h (p_0) e_k$ with 
	\begin{equation}
	\label{ev-k0}
		e_k  (\alpha) 
		= \frac{{\rm e}^{ik\alpha}}{ \sqrt{\pi}} \; ,  \qquad k \in \mathbb{Z}  \; .
	\end{equation}
The generator of the horospheric translations is $i (\LV_2 - \KV_0)$.  
The  Casimir operator is \label{lightconecoordinateKGpage}
	\begin{equation}
		\label{casimir}
		\CV^2 = - \KV_0^2 + \LV_1^2 + \LV_2^2   \; . 
	\end{equation}

As there exists a Lorentz invariant measure on  
$\partial V^+$, namely the 
measure~${\rm d} \mu_{\partial V^+} $ introduced in \eqref{measure-lightcone}, 
one may extend the space of test functions 
from $C^\infty_0(\partial V^+) $
to $L^2 \bigl(\partial V^+,  {\rm d} \mu_{\partial V^+} \bigr)$
and consider the \emph{left regular representation} 
	\[
		f (p) \mapsto f (g^{-1} \cdot p) \; , 
		\qquad 
		f \in L^2 \bigl(\partial V^+, {\rm d} \mu_{\partial V^+} \bigr) \; . 
	\]
We can decompose\footnote{We will show in the next subsection
that the decomposition into irreducible representations inside the principal and 
the complementary series is indeed given by the Mellin transform for the 
corresponding Cartan 
subgroup.} this representation with the help of the 
Mellin transform \cite{Oberh2, Oberh} $\mathscr{M}
\colon L^2([0,\infty), {\rm d} x)\to L^2( (-\infty,\infty), {\rm d} \nu)$, 
	\[
		({\mathscr M}f)  (\nu)  \doteq \frac{1}{\sqrt{2\pi}}
		\int_0^{\infty} {\rm d} x \; x^{-\frac{1}{2}+i \nu} f(x) \; . 
	\]
Note that $\mathscr{M}$ is a unitary operator from $L^2([0,\infty), {\rm d} x)$ 
to $L^2( (-\infty,\infty), {\rm d} \nu)$.

\begin{proposition}
\label{prop:3.3.4}
Let $ g \in  L^2( [0,\infty), {\rm d} p_0 )$. Then 
	\begin{equation}
		\label{mellin}
		g(p_0) = \frac{1}{2\pi} \int_\mathbb{R} {\rm d}\nu \;
		 p_0^{-\frac{1}{2}-i\nu} \int_0^\infty dp_0' \; {p_0'}^{-\frac{1}{2}+i\nu} g(p_0')  \; .		
	\end{equation}
The integral w.r.t.~${\rm d} \nu$ is over the whole real axis.   
\end{proposition}

\goodbreak 
This result has a number of interesting consequences:
\begin{itemize}
\item[$ i.)$] 
The space $L^2(\partial V^+, {\rm d} \mu_{\partial V^+} )$ is the 
direct integral over $\nu \in\mathbb{R}$ of
the Hilbert spaces~$\widetilde {\mathfrak h}_\nu  $ consisting of homogeneous functions
	\begin{equation}
	\label{h-nu-tilde}
	  (p_0,\alpha)  \mapsto     p_0^{-\frac{1}{2} - i\nu}  h(\alpha)  
	 \end{equation}
of degree $-\frac{1}{2} - i\nu$. 
The scalar product in $\widetilde {\mathfrak h}_{\nu}$ is just the scalar 
product in $L^2(S^1, \, \frac{{\rm d} \alpha}{2})$;
\item[$ ii.)$] for the eigenvalue $\zeta^2= \frac{1}{4} + \nu^2$ of $\CV^2$, 
the eigenspace is 	
	\[
		{\mathfrak H}_{\zeta^2} := 
		\widetilde {\mathfrak h}_\nu   \oplus \widetilde {\mathfrak h}_{-\nu}   \; ; 
	\]
\emph{i.e.},  homogeneous functions of degree $s^+ $ 
and $s^- $ (see \eqref{dd1} below) both appear.
\item[$ iii.)$] the Casimir operator $\CV^2 
= - \KV_0^2 + \LV_1^2 + \LV_2^2$  for $SO_0(1,2)$ can be written as
	\begin{equation} 
	\label{casimir2}
		\CV^2 = - S (S+1) = - \partial_{p_0} p_0^2  \partial_{p_0}
        \; , \qquad  \quad S= p_0  \partial_{p_0} \; , 
	\end{equation} 
Its spectrum of the operator $\CV^2$ 
in $L^2(\partial V^+ , {\rm d} \mu_{\partial V^+} )$, with
is $[\frac{1}{4},\infty)$; 
\item[$ iv.)$] The latter equals \cite[Eq.~(6.5)]{Ba} the operator 
introduced in \eqref{casimir2}.
It is positive, since 
	\begin{align}
	 	\langle g, \CV^2 g \rangle 
		&=-\int_0^\infty {\rm d} p_0 \int_0^{2 \pi} 
		\frac{{\rm d} \alpha}{ 2 } \; \overline{g (p_0, \alpha)}\, 
	 	\partial_{p_0} p_0^2 \partial_{p_0} g (p_0, \alpha)   \\
	 	&= \int_0^\infty {\rm d} p_0 \int_0^{2 \pi} \frac{{\rm d} \alpha}{ 2 } \; 
	 	p_0^2 \, | \partial_{p_0} g (p_0, \alpha)|^2 \ge 0 \;  .  \quad \nonumber
	\end{align}
\end{itemize}
These facts are summarised in the following statement.
 
\begin{theorem}[Spectral theorem] 
\label{spectheo}
As an operator on $L^2 \bigl(\partial V^+, {\rm d} \mu_{\partial V^+} \bigr)$ with
domain~${\mathcal D}_\mathbb{R} (\partial V^+)$, the Casimir operator $\CV^2$ given in 
\eqref{casimir2} is essentially self-adjoint and positive. The positive square root of its 
self-adjoint extension, denoted by~$\CV$, has spectrum ${\rm Sp} (\CV) = [1/2,\infty)$. 
The corresponding spectral decomposition is
	\[
		L^2 \bigl(\partial V^+, {\rm d} \mu_{\partial V^+}  \bigr) 
		=  \int_{\frac{1}{2}}^\infty {\rm d}  \zeta^2 \; 
 			{\mathfrak H}_{\zeta^2} \; , \qquad 
         		{\mathfrak H}_{\zeta^2} \cong L^2 \bigl(S^1, 
         		\tfrac{{\rm d} \alpha}{2} \bigr)\otimes
  			\mathbb{C}^2 \; . 
	\]
\end{theorem}

\begin{remark} 
The eigenvalue equation $\zeta^2= -s(s+1)$ has the solutions
	\begin{equation} 
		\label{dd1} 
			s^\pm= -\frac{1}{2}  \mp i \nu \; , \quad \text{with} \quad \nu =  
			\begin{cases}
				i \sqrt{\frac{1}{4} -\zeta^2} & 
				\text{if $  0< \zeta^2 < 1 /4$ }  \, ,\\
				 \sqrt{\zeta^2 - \frac{1}{4} } & 
				 \text{if  $ \zeta^2 \ge 1 /4$ } \, .
			\end{cases} 
	\end{equation} 
Eq.~\eqref{casimir2} implies \cite[Eq.~(6.6b)]{Ba} that for $ \zeta^2 \ge 1 /4$ 
the generalised eigenfunctions for the eigenvalue $\zeta^2$ of~$C^2$
are homogenous functions of the form 
	\begin{equation}
	\label{not-fast-enough}
		(\alpha, p_0) \mapsto {p_0}^{-\frac{1}{2} - i \nu} f ( \alpha, 1) \; . 
	\end{equation}
In the next subsection, we will show that for $0< \zeta^2 <1/4$, 
the eigenfunctions of the operator \eqref{casimir2} are \emph{not} 
in $L^2 (\partial V^+, {\rm d} \mu_{\partial V^+} )$, as their decay 
in the variable $p_0$ is not fast enough  (see \eqref{not-fast-enough} 
below) to ensure the existence of the integral. Thus the unitary irreducible 
representations in the \emph{complementary series} (corresponding to 
$0< \zeta^2 < 1/4$) do not appear, if one decomposes the reducible representation 
on $L^2 \bigl(\partial V^+, {\rm d} \mu_{\partial V^+} \bigr)$ given by 
the push-forward.
\end{remark}

\section{Unitary irreducible representations on a circle lying on the lightcone}
\label{UIRc}

We have seen that $SO_0 (1,2)/ N$ can be identified 
with $\partial V^+$, while $SO_0 (1,2)/AN$ can be identified with the projective 
space formed by the light rays on the 
forward light cone, see Subsection \ref{circ-sub}.
Thus, considered as a topological space, we have 
	\[
		SO_0 (1,2)/AN \cong SO(2) \; .
	\]      
This can also be seen by considering the unique Iwasawa decomposition
$SO_0 (1,2)=KAN = SO(2)\, AN$. The projection $SO_0 (1,2)\to SO_0 (1,2)/AN$ is then given by 
	\[
		R_0(\alpha) \Lambda_1(t)D(q)\mapsto 
		R_0(\alpha) AN \; ,  
		\qquad \alpha \in [0, 2 \pi) \; , 
	\]
and the embedding of $SO(2)$ into $G$ can be considered as a global
smooth Borel section  \index{Borel section}
	\begin{align*}
		\Xi \colon SO_0 (1,2)/ AN & \to SO_0 (1,2)  \\
		R_0(\alpha) AN & \mapsto R_0(\alpha) \; .
	\end{align*}
We can now reformulate the induced represention~\eqref{eqIndRepSO12} such that it acts on the completion of 
$C (SO(2), \mathbb{C})$, following Proposition \ref{prop:3.2.9}.
For given $g\in SO_0 (1,2)$ and $R_0(\alpha')\in SO(2)$ there are unique
$\alpha, t$ and  $q$ such that
	\begin{equation}
		\label{aktq}
		g^{-1} R_0(\alpha') = R_0 (\alpha) \;  \Lambda_1(t) D(q)\; .  
	\end{equation}
Taking the class w.r.t.\ $AN$, this implies that  
	\[
		g^{-1} R_0(\alpha') AN =R_0(\alpha) AN
	\]  
in the sense of the action of $SO_0 (1,2)$ on $SO_0 (1,2)/AN$ 
and that 
	\[
		\Xi \bigl( g^{-1} R_0(\alpha')AN \bigr) = R_0(\alpha)\in SO_0 (1,2) \; .  
	\]
Eq.~\eqref{aktq} then implies that 
	\[
		\Lambda_1(t) D(q)=\Omega\big(g,R_0(\alpha')\big)^{-1} \; , 
	\]
see \eqref{Wigner-Rot}.

Let us denote by $\widetilde{\Pi}_\nu$ the representation 
living on $C(SO(2))$ equivalent to the induced
representation ${\Pi}_\nu$~\eqref{eqIndRepSO12}. 
According to \eqref{eqRepGH}, it acts as   
	\begin{align*} 
		\big(\widetilde{\Pi}_\nu (g) f \big)_{\upharpoonright K} (R_0(\alpha')) & = 
			\sqrt{ \lambda_{g^{-1}} \bigl( R_0 (\alpha') AN \bigr)} \;
			\pi_\nu \big(\Omega(g,R_0(\alpha')\big) \; f_{\upharpoonright K} \bigl( 
			\bigl( g^{-1} \cdot R_0(\alpha') \bigr)_{\upharpoonright K} \bigr) \\
		&= {\rm e}^{-\frac{1}{2}t} \;\pi_\nu \big( \Lambda_1(t)
			D(q)\big)^{-1} \; f_{\upharpoonright K} (R_0(\alpha)) \\
		&= {\rm e}^{(-\frac{1}{2}-i\nu)  t}	 
		f_{\upharpoonright K} \left( R_0 ( \alpha )\right)  \; . 
\end{align*}
Of course, $K= SO(2)$. 
We have used \eqref{aktq}, as well as 
	\[
		\lambda_{g^{-1}}  ( R_0 (\alpha') AN ) = {\rm e}^{- t}  	
		\quad
			\text{and} 
		\quad
		\pi_\nu \big( \Lambda_1(t) D(q)\big)^{-1}= {\rm e}^{-i\nu  t} \; . 
	\]

Identifying $SO(2)$ with the circle $\gamma_0$ 
introduced in \eqref{gamma-0} by setting 
	\[
	h (\alpha) \doteq f_{\upharpoonright K} (R_0 (\alpha)) \; , \qquad \alpha \in [0, 2 \pi) \; , 
	\]
the representation $\widetilde{\Pi}_\nu$ extends to a unitary representation on 
$L^2(\gamma_0,{\rm d}\mu_{\gamma_0})$, with
${\rm d}\mu_{\gamma_0}= \tfrac{{\rm d} \alpha}{2}$ 
the strongly quasi-invariant 
measure on $SO_0 (1,2)/AN \cong \gamma_0$;  
see the remark after Eq.~\eqref{eqRepGH}. 
Note that in the sequel we will often prefer to use the 
normalised measure $\tfrac{{\rm d} \alpha}{2 \pi}$ 
instead of ${\rm d}\mu_{\gamma_0}$.

\goodbreak
\begin{proposition}
\label{Prop:2.1}
Let $\widetilde {\mathfrak h}_\nu$ denote\footnote{The Hilbert space $\widetilde {\mathfrak h}_\nu$
for the case $\frac{1}{4} \le \zeta^2 $ first appeared in \eqref{h-nu-tilde}. }
the completion of $C (\gamma_0)$ with respect to one of the following norms:
\begin{itemize}
\item[$i.)$]
in case $0 < \zeta^2 < \frac{1}{4}$, 
define for $ \nu =  i \sqrt{\frac{1}{4} - \zeta^2}$ a 
norm on ${\mathfrak h}_{\nu,0}$ by setting
	\[ 
		\| h \|_\nu^2 \doteq 
			\int_{\gamma_0} \frac{{\rm d} \alpha}
			{2 \pi}  \; 
			\overline{  h (\alpha) }  \int_{\gamma_0} \frac{{\rm d} \alpha'}
			{2 \pi} \; \varrho_\nu (\alpha - \alpha') \,   h ( \alpha') \; ,
	\]
with 
	\begin{equation} 
		\label{dd2} 
		\varrho_\nu (\alpha) 
		\doteq  
		\frac{\Gamma (\frac{1}{2} + i \nu )}{\Gamma ( \frac{1}{2})
		\Gamma (  i \nu  )} \;   
		\left(\sin^2 \tfrac{\alpha}{2} \right)^{-\frac{1}{2} + i \nu }  \pi  \; , 
	\end{equation} 
\item[$ii.)$]
in case $\frac{1}{4} \le \zeta^2 $, 
define for $  \nu  =   \sqrt{\zeta^2 -
  \frac{1}{4}}$ a norm on ${\mathfrak h}_{\nu,0}$ by setting
	\[ 
		\| h \|^2 \doteq 
			\frac{1}{2 \pi} \int_{\gamma_0} {\rm d}   \alpha \;  | h ( \alpha)|^2 \; . 
	\]
\end{itemize}
It follows that for all $\zeta^2>0$ the operators $\widetilde \Pi_{\nu} (g)$, 
$g \in SO_0(1,2)$, extend from $C (\gamma_0)$  
to a unitary representation\footnote{This is in agreement 
with \eqref{not-fast-enough}.}
	\begin{align}
		\left( \widetilde u_{\nu} (g) h \right) (\alpha')  
		& = {\rm e}^{( - \frac{1}{2} - i\nu) t }
		h \left( \alpha \right)  \; 
	\label{FG1}
	\end{align}
of the Lorentz group $SO_0(1,2)$. 
The parameters $\alpha,  t, q$ on the r.h.s.~ 
are given by \eqref{aktq}.
\end{proposition}

\begin{proof}
The case $\nu \in \mathbb{R}$ follows from the discussion preceding the proposition. 
In the case $-\tfrac{1}{2} < i \nu < \tfrac{1}{2}$, note that the 
norm reads 
	\[
		\| h\|_\nu^2 = \langle h,  A_\nu   h\rangle_{L^2(\gamma_0)} \; , 
	\]
where $A_\nu$ is the operator acting on $C(\gamma_0)$ as  
	\[ 
		(A_\nu h) (\alpha) \doteq  \int_{\gamma_0} 
		\frac{ {\rm d} \alpha' }{2 \pi}  \; 
        			\varrho_\nu (\alpha-\alpha') h(\alpha') \; , 
			\qquad \alpha \in [0, 2 \pi) \; . 
	\]
We will show in Section \ref{sec:3.5} that this map intertwines $\widetilde{\Pi}_{\nu}$
and $\widetilde{\Pi}_{-\nu}$; see \eqref{intertwiner}. 
Using this fact and the fact that $\overline{\pi_\nu (an)}\,
\pi_{-\nu}(an) =1$ for all $an \in AN$, one verifies that 
$\widetilde\Pi_{\nu} (g) $, $g \in SO_0(1,2)$, is a unitary operator in
${\mathfrak h}_{\nu,0}$.  
\end{proof}

\goodbreak
\begin{remarks} 
\quad
\begin{itemize}
\item[$i.)$]
In Bargmann's classification \cite{Ba} of the unitary irreducible representations of $SO_0(1, 2)$, the 
\emph{principle series}\index{principle series} and the \emph{complementary series}\index{complementary series} 
are both denoted by $C_{\zeta^2}^0$. They are distinguished by the 
positive eigenvalue $\zeta^2$ of the Casimir 
operator $\CV^2$, with $\zeta^2$ being larger or equal {\em or} smaller than $1/4$.  
\item [$ii.)$] 
The integral kernels appearing in \eqref{intertwiner-1} were first derived by Barg\-mann \cite{Ba}. 
In the literature they are frequently written in the following alternative form:
	\begin{equation} 
	\label{Barg-factor}
		\varrho_\nu (\alpha) 
		=	\frac{\Gamma (\frac{1}{2} +  i \nu  )}{\Gamma ( \frac{1}{2})
		\Gamma (  i \nu )} \;  \left(\tfrac{1- \cos \alpha}{2} \right)^{-\frac{1}{2} + i \nu} \pi \; . 
	\end{equation} 
\item[$iii.)$]
Note that in case $\frac{1}{4} \le \zeta^2$, the norm does not depend on $\nu$. 
In \cite{Ba}, $i \nu$ is denoted as $\sigma$; see, for instance, Equ.~(6.6b) in \cite{Ba}.
\item [$iv.)$] The function $\varrho_\nu (\alpha-\alpha') $ is finite if
$\Re (i \nu)  \ge 1/2$. In case $0 <  \Re (i \nu) < 1/2$, this function is infinite for $\alpha = \alpha'$; 
however, its integral over $\alpha$ and $\alpha'$ remains finite. Hence, for $\zeta^2 > 1/4$ 
and $\nu$ as specified in Proposition~\ref{Prop:2.1}, the expression 
	\[
		f, g \mapsto	 \langle f ,  A_\nu g \rangle_{L^2( \gamma_0)} \; , 
	\]
is certainly well-defined for functions $f, g \in C (\gamma_0)$ (cf.~\cite[p.~618]{Ba}). 
\item [$v.)$] 
The bilinear form-valued function $\nu \to  \langle \, . \,  , 
A_\nu \, . \,  \rangle_{L^2({\tt S}^1, {\rm d} \alpha)}$ 
is meromorphic in $\mathbb{C}$. 
The poles of this function are the points \cite[Theorem 3]{KSt-1}
	\[
		i \nu = 0,   - \tfrac{1}{2}, - 1, - \tfrac{3}{2}, \ldots \; . 
	\]
\end{itemize}
\end{remarks}

Choosing $p_0 = 1$ in \eqref{udrei} and using the notation introduced in \eqref{aktq}, one finds 
(see Equ.~(4.41) and Equ.~(4.42) in \cite{BM})
	\begin{align}
		\label{udrei23} 
 			\bigl( \widetilde  u_{\nu} (\Lambda_2 (s))h \bigr)  (\alpha') 
		&= {\rm e}^{( - \frac{1}{2} - i\nu) t_2  } h (  \alpha_2 ) 
	\nonumber \\  
			\bigl( \widetilde u_{\nu} (\Lambda_1 (t))h \bigr)   (\alpha')  
		&= {\rm e}^{( - \frac{1}{2} - i\nu)  t_1 }  
		h ( \alpha_1 ) 
	 \nonumber \\
			\bigl( \widetilde u_{\nu} (R_{0}(\alpha))h \bigr)   (\alpha')  
		&= h (\alpha - \alpha' )  \; , 
	\end{align}
with
	\begin{align*}
		t_2   & = \ln (\cosh s - \sinh s \sin \alpha')  \; , 
		\qquad \quad
		{\rm e}^{i \alpha_2}   
		=  \tfrac{\cos \alpha' -i \sinh s + i \cosh s \sin \alpha'}{\cosh s -  \sinh s \sin \alpha'}  \; , 
		\\
		 t_1   & = \ln (\cosh t - \sinh t \cos \alpha') \; , 
		 \qquad \quad 
		{\rm e}^{i \alpha_1}   
		=  \tfrac{-\sinh t + \cosh t \sin \alpha' +i \sin \alpha' }{\cosh t  +  \sinh t \cos \alpha'}  \; .		  
	\end{align*}

\begin{theorem}[Bargmann, \cite{Ba}]
\label{TH-irr}
The representations $\widetilde u_{\nu}$ given by \eqref{FG1} are irreducible. 
\end{theorem}

\begin{proof} 
We have already seen in \eqref{not-fast-enough}
that in the representation $\widetilde u_{\nu}$ the Casimir 
operator is a multiple of the identity with eigenvalue $s^+ = - \tfrac{1}{2} - i \nu$. 
We will now show, by direct computation,  that the 
representation $\widetilde u_{\nu}$ is indeed irreducible.
Let $A$ be  a bounded linear operator on $\widetilde {\mathfrak h}_\nu$, which 
commutes with all $\widetilde u_{\nu} (g)$, $g \in SO_0(1,2)$. It follows \cite[p.~608]{Ba} that  
	\begin{align}
		\KV_0  A \,  f_k & = A \KV_0 \,  f_k  \; , \qquad f_k \doteq {p_0}^{-\frac{1}{2} - i \nu} e_k  \; , 
		\nonumber
		 \\
		\LV_i  A \, f_k & = A \LV_i  \, f_k  \; , 
		\qquad \; \; i = 1, 2 \; , \quad  k \in \mathbb{Z}  \; . \label{key-irred}
	\end{align}
Here the $e_k$'s are the eigenfunctions of $\KV_0$ introduced in~\eqref{ev-k0}.
The first equation implies that $A \, f_k  =  \alpha_{k}  \cdot  f_k   $ 
for some $ \alpha_{k} \in \mathbb{C}$. 
To explore the content of the second and third equation in \eqref{key-irred}, we introduce
the ladder operators $\LV_\pm = \LV_1 \pm \LV_2$. They satisfy
	\begin{align}
		\LV_+ \, f_k & = c_{k+1} \sqrt{ \zeta^2+ k (k+1)} \, f_{k+1} \; , \nonumber \\
		\LV_- \, f_k & = c_{k}^{-1} \sqrt{ \zeta^2+ k (k-1)} \, f_{k-1} \; ,    
	\label{ladder-op}
	\end{align}
with $| c_k | = 1$ some constants of absolute value $1$. Since $\LV_1 = \tfrac{1}{2} (\LV_+ + \LV_-)$ 
and $\LV_2 = \tfrac{i}{2} (\LV_- - \LV_+)$, we obtain from \eqref{ladder-op} a set of equations, which may 
be written in the form \cite[Equ.~(5.34)]{Ba}
	\[
		\LV_i f_k = \sum_{k'} \beta_{k,k'} {f_{k'}} \; , \qquad \beta_{k,k'} = \overline{\beta_{k',k}} \; , 
	\]
and where we can read off from \eqref{ladder-op} that 
$\beta_{k,k'}= 0$ if $| k - k' | >1 $ and  $\beta_{k,k'}\ne 0$ if $| k - k' | =1 $. 
We therefore obtain from the equations involving $\LV_1$ and $\LV_2$ in 
\eqref{key-irred}  equations of the form
	\[
		( \alpha_{k}  - \alpha_{k'} ) \beta_{k, k'} 
		= 0 \qquad \forall k, k'  \in \mathbb{Z} \; . 
	\]
A brief inspection shows 
that all $\alpha_{k} $ have to be equal to each other (for $\nu$ fixed), \emph{i.e.}, 
that $A = \alpha_{\nu, 0} \cdot \mathbb{1}$. 
\end{proof}

\section{Intertwiners}
\label{sec:3.5}

The intertwiners for $SL(2, \mathbb{R})$ were analysed by Kunze and Stein \cite{KS}, 
Knapp and Stein \cite{KSt-1, KSt-2}, 
as well as Sally~\cite{Sally}, using 
fractional transformations. However, it is not difficult to 
construct them directly using the induced representations 
constructed in Section \ref{UIRc}. 

\bigskip
Consider the map 
	\begin{equation}
	\label{intertwiner-1}
		h \mapsto  (A_{\nu} h)(\alpha) \doteq \int_{\gamma_0}   
		\frac{{\rm d} \alpha' }{2 \pi} \; \varrho_\nu (\alpha-\alpha') h(\alpha') \; , 
		\qquad \alpha \in [0, 2 \pi) \; , 
	\end{equation}
with $h \in C^\infty (S^1)$ and $\varrho_\nu (\alpha) $ given by \eqref{dd2}.
Choosing $h(\alpha') = {\rm e}^{i k \alpha'}$, $k \in \mathbb{Z}$, we notice that 
	\begin{align*}
		\int_{\gamma_0}   
		\frac{{\rm d} \alpha' }{2 \pi} \; \varrho_\nu (\alpha-\alpha') {\rm e}^{i k \alpha'}
		& =	\underbrace{ \left( \int_{\gamma_0}   
		\frac{{\rm d} (\alpha' -\alpha)}{2 \pi} \; 
		\varrho_\nu (\alpha-\alpha') {\rm e}^{-i k (\alpha-\alpha')}
		\right) }_{ = c_k (\nu)} {\rm e}^{i k \alpha} \; . 
	\end{align*}
Hence, as expected from the fact that $[ A_{\nu} , \KV_0 ] =0$, the operator 
$A_\nu$ shares a basis of eigenvectors with the angular momentum operator $\KV_0$:
using the vectors $e_k$ introduced in \eqref{ev-k0}, 
	\[
		A_\nu e_k =c_k (\nu) e_k   \qquad \forall k \in \mathbb{Z} \; . 
	\]
It remains to compute the coefficients: according to \cite[p.~397, 2007]{Grad}
	\[
		\int_0^\pi {\rm d} \alpha \; \sin^{ q -1 } \alpha \cos p \alpha
		= \frac{ 1}{2^{q -1} q } 
		\frac{ \pi \cos \frac{p \pi}{2} }{ B (\frac{q + p + 1}{2}  , \frac{q - p + 1}{2} )} \; , 
		\qquad \Re q >0 \; . 
	\]
Here $B(x,y) =\Gamma (x)\Gamma (y) / \Gamma (x+y)$ denotes Euler's beta function.
Setting $q -1  = -1 + 2 i \nu$, we find  
	\[
		\int_0^{2\pi} \frac{{\rm d} \alpha}{2 \pi}  \; \sin^{ -1 + 2i \nu } 
		\tfrac{\alpha}{2} \cos (2k \tfrac{\alpha}{2})
		= \frac{ 2^{-2- 2 i \nu}}{  i \nu } 
		\frac{  \cos k \pi }{ B (\frac{1}{2} + i \nu + k   , \frac{1}{2} + i \nu - k   )} \; , 
		\quad \Re (i \nu ) >0 \; . 
	\]
Hence, using \eqref{eq:gamma-4} in the fourth equality
and \eqref{eq:gamma-3} twice in the fifth equality, we find
	\begin{align*}
		c_k (\nu)
		& =	  \pi \;  \frac{\Gamma (\frac{1}{2} + i \nu )}{\Gamma ( \frac{1}{2})
		\Gamma (  i \nu )} \; 
		 \frac{ 2^{-2- 2 i \nu}}{  i \nu } 
		\frac{  \cos k \pi }{ B \bigl( \frac{1}{2} + i \nu + k   , \frac{1}{2} + i \nu - k \bigr)}   \; , 
		\\
		& =	 \frac{ 2^{- 2- 2i \nu} \pi }{ i \nu}  \; \frac{\Gamma (\frac{1}{2} + i \nu )}
		{\Gamma ( \frac{1}{2})
		\Gamma (  i \nu )} \;  
		\frac{\Gamma(1 + i 2 \nu) (-1)^k  }{ \Gamma(\frac{1}{2} + i \nu + k) 
		\Gamma(\frac{1}{2} + i \nu - k)}
		 \; , 
		\\
		& =	(-1)^k \; \frac{ 2^{- 2- 2i \nu} \pi }{ i \nu}  \; 
		\frac{\Gamma (\frac{1}{2} + i \nu )}{\Gamma ( \frac{1}{2})
		\Gamma (  i \nu )} \;  
		 \frac{ 2^{2 + i 2 \nu } \Gamma(\frac{1}{2} + i  \nu) \, i \nu \, 
		\Gamma( i  \nu)}{ \sqrt{\pi} \, \Gamma(\frac{1}{2} + i \nu + k) 
		\Gamma(\frac{1}{2} + i \nu - k)}
		 \; , 
		\\
		& =	\frac{(-1)^k \Gamma (\frac{1}{2} + i \nu ) 
		\Gamma(\frac{1}{2} + i  \nu)}{   \Gamma(\frac{1}{2} + i \nu + k) 
		\Gamma(\frac{1}{2} + i \nu - k)}
		 \; , 
		\\
		& = \frac{ \Gamma (\frac{1}{2} + i \nu )} 
		{( - \frac{1}{2} - i  \nu) \Gamma(-\frac{1}{2} - i  \nu)}
		\frac{( k - \frac{1}{2} - i \nu ) \Gamma(k - \frac{1}{2} - i \nu)}
		{   \Gamma(k +\frac{1}{2} + i \nu ) }
		\\
		& =	\frac{ \Gamma (\frac{1}{2} + i \nu )} 
		{\Gamma(\frac{1}{2} - i  \nu)}
		\frac{  \Gamma(k + \frac{1}{2} - i \nu)}{   \Gamma(k +\frac{1}{2} + i \nu ) }
		 \; .
		\end{align*}
This expression coincides with the
expression used by Sally \cite[Chapter II]{Sally}.

\goodbreak 
\begin{remarks}
\label{rm:3.5.1}
\quad
\begin{itemize}
\item [$i.)$] In case $\nu \in \mathbb{R}$, 
$\Gamma (\bar{z}) = \overline{\Gamma(z)}$ implies 
	\[
		| c_k (\nu) |=1 \qquad \forall  k\in \mathbb{Z} \; , \quad \nu  \in \mathbb{R} \; .
	\]
This is \cite[Lemma 2.3.1 (1)]{Sally}. In fact,  for $\nu \in \mathbb{R}$, the 
intertwiners are \emph{unitary} operators on $L^2 \bigl( \gamma_0, \frac{{\rm d} \alpha' }{2 \pi} \bigr)$; 
\emph{i.e.}, 
	\begin{equation}
	\label{A-iso}
		A_\nu^* A_\nu = \mathbb{1} \; , \qquad  \nu \in \mathbb{R} \; ; 
	\end{equation}
\item [$ii.)$] 
For $\Re (i \nu) \ge 0$, the integral for the coefficient $c_k(\nu)$ is well-defined\footnote{Using 
$\Gamma(z+1) = z \Gamma (z)$ and
the Stirling formula for the $\Gamma$-functions one concludes \cite[Equ.~(8.12)]{Ba}, \cite[p.~605]{Ner} 
that the coefficients $c_k (\nu)$ 
diverge as  
	\begin{equation}
	\label{divergence-ck}
		|k|^{2 i \nu} \left( 1 + O \left( \frac{1}{k}\right)\right) \; , \qquad k \to \infty \; . 
	\end{equation}
 } and 
	\[
		| c_k (\nu) | \le 1 \qquad \forall k \in \mathbb{Z} \; , \qquad 	 \Re (i \nu) \ge 0 \; .
	\]
Furthermore, the function $\nu \mapsto c_k (\nu)$ is analytic on the domain $\Re (i\nu) >0$ and 
continuous on $\Re \nu \ge 0$ for all $k \in \mathbb{Z}$ \cite[Lemma 2.3.1 (7)]{Sally}.   
Hence, for $\Re (i \nu) >0$, the intertwiner $A_\nu$ is a \emph{bounded} operator on 
$L^2 \bigl( \gamma_0, \frac{{\rm d} \alpha' }{2 \pi} \bigr)$. 
\item [$iii.)$] 
For $\Re (i \nu) <0$, the coefficients $c_k ( i \nu)$ are \emph{not} bounded; 
see \eqref{divergence-ck}. Hence, for $\Re (i \nu) <0$,
$A_\nu$ is an unbounded operator on $L^2 \bigl( \gamma_0, \frac{{\rm d} \alpha' }{2 \pi} \bigr)$ .

\item [$iv.)$] In case $\nu =  \pm i \sqrt{\frac{1}{4} - \zeta^2}$ with
$0 < \zeta^2 < \frac{1}{4}$, the sesquilinear form  
	\[
		\widetilde {\mathfrak h}_{\nu} \ni   h, h' \mapsto \int_{\gamma_0} 
		 \frac{{\rm d} \alpha}{2 \pi}  \; \overline{h(\alpha)} (A_\nu h')(\alpha)
	\] 
is positive definite~\cite{Ner}, and 
	\[
		c_k(\nu) >0 \; \qquad \forall k \in \mathbb{Z} \; , 
		\quad 0 < \pm  i \nu < \tfrac{1}{2} \; ,
	\]
and, consequently, \eqref{intertwiner-1}
defines a positive operator\footnote{The operator $A_\nu$  
{\em intertwines} the pullback action of 
$SO_0(1,2)$ on homogeneous functions of degree 
$-\frac{1}{2} + \sqrt{\frac{1}{4}-\zeta^2}$ and 
$-\frac{1}{2} - \sqrt{\frac{1}{4}-\zeta^2}$, respectively.}.
\end{itemize}
\end{remarks}
 
\begin{proposition}
\label{Prop:2.1.0}
Consider the representations described in  \eqref{FG1}. It follows that the 
map $A_\nu \colon \widetilde {\mathfrak h}_{\nu} \to \widetilde  {\mathfrak h}_{-\nu}$ 
defines a unitary operator~$A_\nu$, which {\em intertwines}\index{intertwiner} 
$\widetilde u_{\nu}$ and~$\widetilde u_{-\nu}$, \emph{i.e.},  
	\begin{equation}
	\label{intertwiner}
		A_\nu \widetilde u_{\nu} (g)  = \widetilde u_{-\nu} (g)  A_\nu  
		\qquad \forall g \in G \; . 
	\end{equation}
It follows that the representations for $\nu$ and $- \nu$
are unitarily equivalent both for the principal and the complementary 
series (see, \emph{e.g.}, \cite[p.~104]{Pukanszky}.). 
\end{proposition}

\goodbreak 

\begin{proof}
Inspecting \eqref{udrei23} we find that 
	\begin{equation} 
		\widetilde u_{-\nu} (R_{0}(\alpha)) = \widetilde u_{\nu} (R_{0}(\alpha)) \; , 
		\qquad \alpha \in [0, 2 \pi) \; . 
	\label{remark-identity}
	\end{equation}
Moreover, 
	\begin{align*}
	\int_{\gamma_0} \frac{ {\rm d} \beta' }{2 \pi} \; \varrho_\nu (\beta-\beta') \bigl( 
	\widetilde u_{\nu} (R_{0}(\alpha))  h\bigr)(\beta') 
	& = 
	\int_{\gamma_0} \frac{ {\rm d} \beta' }{2 \pi} \;  \varrho_\nu (\beta-\beta') h (\beta' - \alpha) 
	\nonumber \\
	& =  \widetilde u_{- \nu} (R_{0}(\alpha)) 
	\Bigl( \int_{\gamma_0} \frac{ {\rm d} \beta' }{2 \pi} \;  \varrho_\nu (\beta - \beta') h (\beta') \Bigr) \; . 
	\end{align*}
Thus it remains to show that
	\[
		A_\nu \widetilde  u_{\nu} (\Lambda_2 (s))  = 
		\widetilde  u_{-\nu} (\Lambda_2 (s))  A_\nu   
		\qquad \forall s \in \mathbb{R} \; . 
	\]
Compute, using \eqref{udrei23},
	\begin{align}
	\label{I-1}
	& \int_{\gamma_0} \frac{ {\rm d} \beta' }{2 \pi} \;  \varrho_\nu (\beta-\beta') 
	\bigl( \widetilde u_{\nu} (\Lambda_{2}(s)) h\bigr)(\beta') 
	\nonumber	\\
	& \qquad \qquad
	= \int_{\gamma_0} \frac{ {\rm d} \beta' }{2 \pi} \;  \varrho_\nu (\beta-\beta') \, 
	{\rm e}^{(-\frac{1}{2} - i\nu) t } 
	h \left(\alpha \right)  \; ,  
	\end{align}
with
	\begin{align*}
		{\rm e}^t   & = \cosh s - \sinh s \sin \beta'   \; , 
		\qquad \quad
		{\rm e}^{i \alpha}   
		=  \tfrac{\cos \beta' -i \sinh s + i \cosh s \sin \beta'}{\cosh s -  \sinh s \sin \beta'}  \;  .		  
	\end{align*}
Hence\footnote{Compute the derivative $\frac { {\rm d} } {{\rm d} \beta'} \ln 
\tfrac{\cos \beta' -i \sinh s + i \cosh s \sin \beta'}{\cosh s -  \sinh s \sin \beta'}$ by 
hand or use an application, \emph{e.g.}, www.derivative-calculator.net.},  
	\[
		\frac{ {\rm d}  \alpha(s, \beta') }{ {\rm d} \beta'} 
		= (\cosh s - \sinh s \sin \beta')^{- 1 } \; . 
	\]
Thus the r.h.s.~in \eqref{I-1} equals 
	\begin{align*}
	& \int_{\gamma_0} \frac{ {\rm d} \beta' }{2 \pi} \;  \varrho_\nu (\beta-\beta') 
	\bigl( \widetilde u_{\nu} (\Lambda_{2}(s)) h\bigr)(\beta') 
	\\
	& \qquad \qquad
	= \int_{\gamma_0} \frac{ {\rm d} \alpha}{2 \pi} \;  \varrho_\nu (\beta-\beta') \, 
	(\cosh s - \sinh s \sin \beta')^{\frac{1}{2} - i \nu}  h \left(\alpha \right)  \; ,  
	\end{align*}
Using
	\[
	\cosh s - \sinh s \sin \beta' = \bigl(\cosh s  + \sinh s \sin \alpha \bigr)^{-1}  \; , 
	\]
this allows us to reformulate equation \eqref{I-1}: 
	\begin{align}
	& \int_{\gamma_0} \frac{ {\rm d} \beta' }{2 \pi} \;  \varrho_\nu (\beta-\beta') 
	\bigl( \widetilde u_{\nu} (\Lambda_{2}(s)) h\bigr)(\beta')  
	\nonumber \\
	& \quad = \int_{\gamma_0} \frac{ {\rm d} \alpha }{2 \pi} \;  
	\varrho_\nu \bigl(\beta-\beta' (s, \alpha) \bigr) 
	\bigl(\cosh s  + \sinh s \sin \alpha \bigr)^{-\frac{1}{2} + i \nu }  
	h ( \alpha ) \; . 
	\label{I-1-1}
	\end{align}
The kernel $\varrho_\nu  \bigl(\beta-\beta' (s, \alpha) \bigr) $ can be rewritten using the formula
	\begin{align}
	\label{I-7}
	& \bigl( 1 - \cos \bigl( \beta - \beta' (s, \alpha) \bigr)\bigr)^{-\frac{1}{2} + i \nu}
	 \\
	& \qquad \qquad = \Bigl( 1 - \cos  \beta  \cos \bigl(\beta' (s, \alpha)\bigr) 
	- \sin  \beta  \sin \bigl(\beta' (s, \alpha)\bigr) \Bigr)^{-\frac{1}{2} + i \nu } \; . 	
	 \nonumber \\
	& \qquad \qquad = \Bigl(
	\tfrac{ \cosh s  + \sinh s \sin \alpha - \cos \beta \cos \alpha - \sin \beta \sinh s 
	- \sin \beta \cosh s \sin \alpha}
	{\cosh s + \sinh s \sin \alpha}  \Bigr)^{-\frac{1}{2} + i \nu }. 	
	\nonumber
	\end{align}
Inserting the expression for $\varrho_\nu  \bigl(\beta-\beta' (s, \alpha) \bigr) $
in \eqref{I-1-1} yields 
	\begin{align}
	& \int_{\gamma_0} \frac{ {\rm d} \beta' }{2 \pi} \;   
	\varrho_\nu (\beta-\beta') \bigl( \widetilde u_{\nu} (\Lambda_{2}(s)) h\bigr)(\beta')   
	\nonumber \\
	& \qquad = \frac{\Gamma (\frac{1}{2}  + i \nu ) \; \pi }{\Gamma ( \frac{1}{2})
		\Gamma ( i \nu )}  \; \;  
		\int_{\gamma_0} \frac{ {\rm d} \alpha }{2 \pi} \;  
	\label{I-9}
	\nonumber 
	\\
	& \qquad \qquad \times \Bigl( \tfrac{\cosh s  + \sinh s \sin \alpha - \cos \beta \cos \alpha 
	- \sin \beta \sinh s 
	- \sin \beta \cosh s \sin \alpha }{2} \Bigr)^{-\frac{1}{2} + i \nu }  
		h \bigl( \alpha \bigr) \; . \quad
	\end{align}
On the other hand, using \eqref{udrei}, we find   
	\begin{align}
	\label{I-9a}
	& \widetilde u_{-\nu} (\Lambda_{2}(s)) 
	\Bigl( \int_{\gamma_0} \frac{ {\rm d} \beta' }{2 \pi} \; 
	 \varrho_\nu (\beta -\beta')  h (\beta') \Bigr) \nonumber \\
	& \qquad = \bigl(\cosh s - \sinh s \sin \beta \bigr)^{-\frac{1}{2} + i \nu }  
		\nonumber \\
		& \qquad \qquad \qquad \qquad  \times	 
		\int_{\gamma_0} \frac{ {\rm d} \alpha }{2 \pi} \;  
		\varrho_\nu 
		\bigl( \arccos \bigl( \tfrac{\cos \beta}{\cosh s -  \sinh s \sin \beta} \bigr) -\alpha \bigr)  
		h  \bigl( \alpha \bigr)
 		\nonumber \\
	& \qquad = \bigl(\cosh s - \sinh s \sin \beta \bigr)^{-\frac{1}{2} + i \nu } 
	 \int_{\gamma_0} \frac{ {\rm d} \alpha }{2 \pi} \;  \varrho_\nu 
	\bigl( \beta' (s, \alpha) -\alpha \bigr)  h  \bigl( \alpha \bigr) \; . 
	\end{align}
Next we compute $\varrho_{\nu} \bigl(\beta' (s, \alpha)-\alpha) \bigr)$ using    
	\begin{align*}
	\bigl( 1 - & \cos \bigl( \beta' (s, \alpha)  - \alpha \bigr) \bigr)^{-\frac{1}{2} + i \nu }
	 \\
	& = \Bigl( 1 - \cos  \bigl(\beta' (s, \alpha)\bigr)   \cos \alpha 
					- \sin  \bigl(\beta' (s, \alpha)\bigr)  \sin \alpha \Bigr)^{-\frac{1}{2} + i \nu }   	
	 \nonumber \\
	& = \Bigl(
	\tfrac{ \cosh s  - \sinh s \sin \beta - \cos \beta \cos \alpha + \sin \alpha \sinh s 
	- \sin \beta \cosh s \sin \alpha}
	{\cosh s - \sinh s \sin \beta}  \Bigr)^{-\frac{1}{2} + i \nu } \, . 	
	\nonumber
	\end{align*}
Inserting this result into \eqref{I-9a} shows that 
\begin{align}
	& \int_{\gamma_0} \frac{ {\rm d} \beta' }{2 \pi} \;  \varrho_\nu (\beta-\beta') 
	\bigl( \widetilde u_{\nu} (\Lambda_{2}(s)) h\bigr)(\beta') 
	\nonumber \\
	& \qquad =  \widetilde u_{-\nu} (\Lambda_{2}(s)) 
	\Bigl( \int_{\gamma_0} \frac{ {\rm d} \beta' }{2 \pi} \;  
	\varrho_\nu (\beta - \beta') h (\beta') \Bigr)  \; . 
\end{align} 
Since $ R_0(\alpha)$ and $\Lambda_2(s)$ generate $SO_0(1,2)$, this verifies \eqref{intertwiner}. 
\end{proof}

\begin{remark}
\label{A-unitary}
Clearly  \eqref{remark-identity} implies
	\[ 
		\bigl[ A_\nu ,  \widetilde u_{\nu} (R_{0}(\alpha)) \bigr] 
		= 0 \; , \qquad \alpha \in [0, 2 \pi) \; .  
	\]
Therefore $A_\nu$ has diagonal form in the spectral representation of the generator 
of the rotations $\alpha \mapsto R_{0}(\alpha)$. 	  
In fact,
	\[ 
	 	\varrho_\nu (\alpha) = 
		\sum_{k\in \mathbb{Z}  }  c_k (\nu) {\rm e}^{ i k \alpha  } \; ,  
	\]
as one can easily verify:
	\[
		\int_{\gamma_0} \frac{ {\rm d} \alpha' }{2 \pi} \, 
		\overline{ \varrho_\nu (\alpha-\alpha') } {\rm e}^{i k' \alpha'}
		= \sum_{k\in \mathbb{Z}  } c_k (\nu) {\rm e}^{ i k \alpha}
		\int_{\gamma_0} \frac{ {\rm d} \alpha' }{2 \pi}
		 {\rm e}^{i(k'-k) \alpha'  } = c_{k'} (\nu) {\rm e}^{i k' \alpha} \; .
	\]
Hence, for $\nu \in \mathbb{R}$, 
	\begin{align*}
		\int_{\gamma_0} \frac{ {\rm d} \alpha' }{2 \pi} \, 
		\overline{ \varrho_\nu  (\alpha-\alpha') } \,  \varrho_\nu
		(\alpha'-\alpha'') 
		&= \int_{\gamma_0} \frac{ {\rm d} \alpha' }{2 \pi} \, \overline{
		\sum_k c_k (\nu) {\rm e}^{- i k (\alpha -\alpha')} 
		} \, \sum_j c_j (\nu)   {\rm e}^{- i j (\alpha' -\alpha'')}  
		\\
		& = \sum_{k,j} \overline{c_k (\nu)}   c_j (\nu)   
		\int_{\gamma_0} 
		\frac{ {\rm d} \alpha' }{2 \pi} {\rm e}^{ i k (\alpha -\alpha')} {\rm e}^{ - i j (\alpha' -\alpha'')} 
		\\
		& =  	\sum_k {\rm e}^{i k (\alpha - \alpha'')} 
		=  	2\pi \;  \delta(\alpha-\alpha'')
		\; .   
	\end{align*}
We have used that $| c_k (\nu) | = 1 $ 
for all $k \in \mathbb{Z}$  and $\nu \in \mathbb{R}$ 
(which follows from the properties of the $\Gamma$ function; 
see Remark~\ref{rm:3.5.1} i.). Note that 
$2\pi \;  \delta(\alpha-\alpha'')$ is the kernel of the identity operator with respect to the 
measure $\tfrac{{\rm d} \alpha''}{2\pi}$. 
In other words, the intertwiner $A_\nu$ is an isometric operator; see \eqref{A-iso}. 
\end{remark}

\section{The time reflection\index{time reflection}}
\label{sec:time reflection} 

Our next aim is to extend the unitary irreducible representations of
$SO_0(1,2)$ to (anti-)unitary representations of $O(1,2)$. 
We start with the induced representation~$ \Pi_\nu$ defined in  
\eqref{eqIndRepSO12}, and consider first the case  $\nu\in
\mathbb{R}$, \emph{i.e.}, $\zeta^2 \geq 1/4$. 
Let $\Pi_{\nu,0}(T)$ be the anti-linear map from
$\mathfrak{h}_{-\nu,0}$ to $\mathfrak{h}_{\nu,0}$ 
defined by 
	\begin{equation} 
		\label{eqUT0}
		(\Pi_{\nu,0}(T) f)(g) \doteq \overline{f(Pg)} \; ,  \quad
 		 f\in  \mathfrak{h}_{\nu,0}   \; ,  
	\end{equation}
where $P$ is the space-reflection, $P=R_0(\pi)\in SO_0(1,2)$. 
Since $\lambda_P( g AN )=1$, this is an isometric
map. Now use the intertwiner 
	$
		A_{\nu} \colon \mathfrak{h}_{\nu,0}  \to \mathfrak{h}_{-\nu,0}  
	$
between the representations $\Pi_\nu$ and $\Pi_{-\nu}$ and
define $\Pi_{\nu}(T) \doteq \Pi_{\nu,0}(T)\circ A_{\nu}$: 
	\begin{equation} 
		\label{eqInRepT}
		\big(\Pi_{\nu}(T) f\big)(g) \doteq \overline{( A_{\nu} f)(Pg)} \; , 
		\qquad  f \in \mathfrak{h}_{\nu,0} \; . 
	\end{equation}
Then one has 
	\[
		\Pi_{\nu}(T)\, \Pi_{\nu}(g)\, \Pi_{\nu}(T)^{-1}=
		\Pi_{\nu,0}(T) \, \Pi_{-\nu}(g)\, \Pi_{\nu,0}(T)^{-1}  
	\]
and 
	\[
		\big(\Pi_{\nu}(T)\, \Pi_{\nu}(g)\, \Pi_{\nu}(T)^{-1} \; f \big)(g')
		= \sqrt{ \lambda_{g^{-1}} (Pg' AN ) }\, f(Pg^{-1}P\,g') \; ,
	\]
while  on the other hand
	\[
		\big(\Pi_{\nu}(TgT^{-1}) \; f \big)(g')= \sqrt{ \lambda_{ Pg^{-1}P }(g' \, g AN ) }\, f \bigl( (PgP)^{-1}\,g' \bigr) \; , 
	\]
where it has been used that the adjoint action of $T$ on $SO_0(1,2)$ 
coincides with that of the space-reflection $P$, $TgT=PgP$.  Since
$\lambda_{Pg^{-1} P}(g' \,H)= \lambda_{g^{-1}} (Pg' \,H)$, this proves that 
	\[
		\Pi_{\nu}(T)\, \Pi_{\nu}(g)\, \Pi_{\nu}(T)^{-1}=\Pi_{\nu}(TgT^{-1}) \;  .
	\]
Thus, $\Pi_{\nu}(T)$ is a representer of $T$ which, in addition, 
is easily seen to be anti-unitary. 
 
Next, we wish to find the equivalent representer in the representation space 
$C_0(G/AN)$. The intertwiner $A_{\nu}$ corresponds uniquely to an
operator $\widetilde  {A_{\nu}}$ acting on $C_0(G/AN)$ 
by the equivalence~\eqref{eqFeqFGH}, 
	\[
		\widetilde{A}_{\nu} \tilde{f} \doteq \widetilde{A_{\nu}  f} \; , 
	\]
which intertwines the representations $\widetilde u_\nu$ and  $\widetilde u_{-\nu}$. 
Now this equivalence translates $\Pi_{\nu}(T)$ into the
anti-unitary operator $\widetilde u_{\nu}(T)$ in $C (SO(2))$ given by  
	\begin{align*} 
		\big(\widetilde u_{\nu}(T)\,\tilde f\big)(R_0(\alpha)) 
		& \doteq   \big(\widetilde{\Pi_{\nu}(T)\,f}\big)(R_0(\alpha))
		=  \big({\Pi_{\nu}(T)\,f}\big)(R_0(\alpha)) = 
		\overline{\big( A_{\nu}\,f\big) (P R_0(\alpha))}\\
		&= \overline{\big(\widetilde{ A_{\nu}\,f}\big) (P R_0(\alpha))}
		\; = \; \overline{\big(\widetilde A_{\nu} \,\tilde f\big) (P R_0(\alpha))} \; . 
\end{align*}
In the second and fourth equation we have used the fact that
	\[
		\Xi(R_0(\alpha))=R_0(\alpha) \in SO_0(1,2) \; , \qquad \forall R_0(\alpha) \in SO(2) \; , 
	\]
and that $PR_0(\alpha)\equiv R_0(\alpha+\pi)$ is a rotation. 
In short, $\widetilde u_{\nu}(T)$ acts on $C_0(SO(2))$ as 
	\begin{equation} 
		\big(\widetilde u_{\nu}(T)\, h \big)(R_0(\alpha)) = 
			\overline{\big(\widetilde A_{\nu}\,h \big) (P R_0(\alpha))} \; ,  
			\qquad \nu \in \mathbb{R} \; . 
	\end{equation}
Note that $\widetilde u_{\nu}(T)^2= A_{\nu}^* A_{\nu} = \mathbb{1}$.

\goodbreak
In the case  $\nu\in i\mathbb{R}$,  \emph{i.e.}, $0<\zeta^2< 1/4$, 
the anti-linear map $\Pi_{\nu,0}(T)$ defined above leaves
$\mathfrak{h}_{\nu,0}$ invariant, and we take this operator to be
the representer of $T$ in $\mathfrak{h}_{\nu,0}$. The proof of the 
representation property goes as above.  
We then define $\widetilde u_{\nu}(T)$ as the equivalent representer
in the  representation space $C (\gamma_0)$, namely, 
	\[
		\big(\widetilde u_{\nu}(T)\,  h  \big)(R_0(\alpha)) \doteq 
		 \overline{h(P R_0(\alpha))} \; . 
	\]
Anti-unitarity can be seen as follows: 
	\begin{align*}
		\|\widetilde u_{\nu}(T)h\|_\nu &= \big\langle \widetilde u_{\nu}(T)h \, , 
		\, A_{\nu} \,
		\widetilde u_{\nu}(T)h \big\rangle_{ L^2(\gamma_0, 
		 \frac{{\rm d} \alpha}{2 \pi}) } \\
		&=  \int_{SO(2)} \frac{d\alpha}{2\pi} \; h(PR_0(\alpha)) \, 
			\int_{\gamma_0} \frac{d\alpha'}{2\pi} \; 
			\varrho_\nu  (\alpha-\alpha')\overline{h(PR_0(\alpha'))} \\
		&=  \int_{\gamma_0} \frac{d\alpha'}{2\pi} \; \overline{h(R_0(\alpha'))}) \, 
			\int_{\gamma_0} \frac{d\alpha}{2\pi} \; 
			\varrho_\nu (\alpha-\alpha') h(R_0(\alpha)) \\
		&=  \big\langle h \, , \, A_{\nu} \, 
		h \big\rangle_{ L^2(\gamma_0, \frac{{\rm d} \alpha}{2 \pi}) } 
		= \|h\|_\nu \; . 
\end{align*}
In the fourth equation we have used the symmetry
$\varrho_\nu  (\alpha)=\varrho_\nu  (-\alpha)$. 

Note that the preceding discussion also shows that in both cases,
\emph{i.e.}, both for $\zeta^2<1/4$ and $\zeta^2 \geq 1/4$,  
the \emph{unitary} representer of the space-reflection $P$ is given by
	\[ 
		\big(\widetilde u_{\nu}(P)h\big)(R_0(\alpha)) =
		h(PR_0(\alpha))\equiv h(R_0(\alpha-\pi)) \; .
	\] 
In summary, we have shown: 

\begin{proposition}
\label{prop:3.6.1}
The anti-unitary operator 
$\widetilde  u_{\nu} (T) \colon  \widetilde {\mathfrak h}_\nu 
\to  \widetilde {\mathfrak h}_\nu$,  
	\begin{equation}
		\label{tilde-time-reflection}
			\big(\widetilde u_{\nu} (T)h\big)(\alpha) \doteq
				\begin{cases}  
                       			\overline{( A_{\nu} h) (\alpha-\pi) }
					& \text{if \ $1/4 \le \zeta^2 $\, ,  }\\
					\overline{h (\alpha-\pi) }
					& \text{if \  $0<  \zeta^2 < 1/4$\, .  }
				\end{cases}
	\end{equation}
is an anti-unitary representer of the time-reflection $T$
on $\widetilde{\mathfrak h}_\nu$. Together with $\widetilde u_{\nu} (P_2)$ it 
extends the representation $\widetilde u_{\nu}$ from $ SO_0(1,2)$ to $O(1,2)$. 
\end{proposition}

\section{Unitary irreducible representations on two mass shells}
\label{UIRm}

The representation of $SO_0(1,2)$ constructed in Section \ref{UIRc} is by far the one 
most commonly used. However, if one wants to see what happens in the limit of 
curvature to zero, one can take adavantage of the fact the circle used in 
Section~\ref{UIRc} can be replace by the two mass shells $\Gamma_+$ and 
$\Gamma_-$, which lie on the forward light cone. 

The details are as follows. Recall Definition \ref{def:3.3.2}. 
Clearly, $f \in {\mathfrak h}_{\nu,0}$ is determined almost everywhere  
(using the Hannabus decomposition) by $f (k' AN) $ with 
	\[
		k' \in K'= \left\{ \Lambda_2 (s) \mid s \in \mathbb{R} \right\} 
		\dot \cup \left\{ \Lambda_2 (s)P \mid s \in \mathbb{R} \right\} \; . 
	\]
If we identify the cosets $ k' N$, $ k' \in K' $ with 
the points in the forward light cone~$\partial V^+$, then (see Subsection \ref{massshell-sub})
the cosets $k' AN $, $k' \in K'$, will be identified with points in the two hyperbolas
	\begin{equation} 
		\label{eqGamma1}
		\Gamma_1 \doteq  \Gamma_+ \; \dot \cup \; \Gamma_-  
				= \bigl\{ \underbrace{ \Lambda_{2} (s) 
		\left( \begin{smallmatrix} m_0 \\ 0 \\  m_0  
		\end{smallmatrix}\right)}_{= p_+(s)} \mid s \in \mathbb{R} \bigr\} \; 
		\dot \cup \;  \bigl\{ \underbrace{\Lambda_{2} (s) P \left( \begin{smallmatrix} m_0 
		 \\ 0 \\  m_0  \end{smallmatrix}\right)}_{= p_- (s)}
		  \mid s \in \mathbb{R} \bigr\} \; . 
	\end{equation} 
By construction, the boosts $s \mapsto \Lambda_{2}(s)$, $s \in \mathbb{R}$, leave the 
curves  $\Gamma_\pm$ invariant. The restriction of the invariant 
volume form $\tfrac{{\rm d} \alpha}{2 \pi} \wedge {\rm d} p_0$ on $\partial V^+$ to
$\Gamma_\pm$ is~$\pm \tfrac{{\rm d} s}{2}$. It follows that 
$L^2(\Gamma_1,{\rm d} \mu_{\Gamma_1})$ consists of 
two copies of $L^2(\mathbb{R},\tfrac{{\rm d} s}{2})$:
	\begin{equation} 
	\label{eqUBM}
		L^2(\Gamma_1,{\rm d} \mu_{\Gamma_1})
		\cong L^2(\mathbb{R},\tfrac{{\rm d} s}{2})
		\oplus L^2(\mathbb{R}, \tfrac{{\rm d} s}{2} ) \; . 
	\end{equation} 
The generator of the boost $s \mapsto \widetilde u_{\nu} ( \Lambda_2(s)) $ 
on the Hilbert space \eqref{eqUBM},
\begin{equation}
\label{boostgenerator}
 - i 
\bigl( \tfrac{\partial}{\partial s} \oplus \tfrac{\partial}{\partial s} \bigr) \; ,
\end{equation}
has absolutely continuous spectrum on the whole real line. 
The reflections $P_1, P_2 $ and $P$ act on $\Gamma_1$:
	\[
	P_1 \Lambda_{2} (s)  p_\pm(0) 
	=  \left( \begin{smallmatrix} m_0 \cosh s \\
		m_0 \sinh s\\
		\mp m_0 
		\end{smallmatrix} \right) \; , \quad 
	P_2 \Lambda_{2} (s)  p_\pm(0) 
		= \left( \begin{smallmatrix} m_0 \cosh s \\
		- m_0  \sinh s \\
		\pm m_0 
		\end{smallmatrix} \right) = \left( \begin{smallmatrix} m_0 \cosh (-s) \\
		m_0 \sinh (-s)\\
		\pm m_0 
		\end{smallmatrix} \right) \; , 
	\]
and, consequently, 
	$
	P \Lambda_{2} (s)  p_\pm(0) 
		= \left( \begin{smallmatrix} m_0  \cosh (-s) \\
		 m_0 \sinh (-s)\\
		\mp m_0 
		\end{smallmatrix} \right)$. 

\begin{theorem} 
\label{th:3.7.1}
The induced representation \eqref{eqIndRepSO12} 
on $L^2(\Gamma_1,{\rm d} \mu_{\Gamma_1})$ is given by
	\begin{align}
		\left( \widetilde u_{\nu} (g) h_{+} \right) (s')  
		& = 
		\chi_{\frac{i}{2}-\nu} \bigl( \Lambda_1(t)\bigr) h_{(-)^j} (s) 
		= {\rm e}^{(-\frac{1}{2} - i \nu) t} h_{(-)^j} (s) \; ,  
	\label{GHK}
	\end{align}
where $h_\pm \in L^2(\mathbb{R},  \tfrac{{\rm d} s}{2})$ and
	\begin{equation}
	\label{sktq}
		\Lambda_2(s)P^j \Lambda_1(t) D(q) \doteq g^{-1} \Lambda_2(s')\; , 
		\quad j \in \{ 0, 1, \} \; , \quad s, t, q, s' \in \mathbb{R} \; . 
	\end{equation}
In particular, in case $g= \Lambda_2(s'')$, Equ.~\eqref{sktq}
yields $\Lambda_2(s) = \Lambda_2(s'')^{-1}  \Lambda_2(s') = \Lambda_2(s'-s'')$.
\end{theorem}

\begin{proof} 
If  $p \in \Gamma_1$ and $p_\pm = \Lambda_2 (s)  p_\pm (0)$, then the cosets $g AN$ can be 
identified with $\Gamma_1$, 
and we may thus consider 
	\[
	\bigl(  \Pi_{\nu}  (g) f \bigr)  (\Lambda_2 (s'))
	= \sqrt{ \lambda_{g^{-1}}  \bigl( \Lambda_2 (s') N \bigr)} \; f \bigl(  g^{-1}  \Lambda_2 (s') \bigr) 
	\]
Thus the induced representation takes the form \eqref{GHK}. 
\end{proof}

Changing the parametrisation of the curve $\Gamma_1$, we can write
	\[
		\Gamma_+ \; \dot \cup \; \Gamma_-  \cong \left\{ \left( \begin{smallmatrix}
		\sqrt{k^2 + m_0^2 } \\
		k \\
		m_0
		\end{smallmatrix} \right) \mid k \in \mathbb{R} \right\}
		 \dot \cup
		 \left\{ \left( \begin{smallmatrix}
		\sqrt{k^2 + m_0^2 } \\
		k \\
		- m_0
		\end{smallmatrix} \right) \mid k \in \mathbb{R} \right\}  \; .  
	\]
Thus all unitary irreducible representation of $SO_0(1,2)$ within the principal and the complementary series 
can be realised on the \emph{common} Hilbert space  
	\begin{align}
	\label{directsum}
		{\mathcal H} \doteq 
		{\mathcal H}_+ \oplus {\mathcal H}_-  
		& \cong L^2 \bigl(\mathbb{R}, \tfrac{\mathrm{d} k}{2 \sqrt{k^2 +m_0^2}} \bigr) 
		\oplus L^2 \bigl(\mathbb{R},  \tfrac{\mathrm{d} k}{2 \sqrt{k^2 +m_0^2}} 
		 \bigr)   \; .
	\end{align}
Once again, the restriction of the invariant volume 
form $\tfrac{{\rm d} \alpha}{2 \pi} \wedge {\rm d} p_0$ on $\partial V_+ $ 
to $\Gamma_\pm$ is $\pm \tfrac{\mathrm{d} k}{2 \sqrt{k^2 +m_0^2}}$.
The following formulas were first given (in a slightly different form) in \cite[p.~369]{BM}. 
Note that there is an $``m''$ missing in Equ.~(4.45) in \cite[p.~369]{BM}.

\begin{theorem}  
\label{th:3.7.2}
Let $m>0$ and let $\widetilde{u}_{m r}$ 
be the unitary irreducible 
representation of $SO_0(1,2)$ for the eigenvalue $\zeta^2 
= \frac{1}{4} + {m}^2 r^2$ of the Casimir operator $\CH^{ 2}
= - \KH_0^2 + \LH_1^2 + \LH_2^2$. 
Then the action of $\widetilde{u}_{m r}$ on an element $f = (f_+, f_-) \in 
{\mathcal H}_+ \oplus {\mathcal H}_- $ is given by  
	\begin{align*}
		\left( \widetilde{u}_{m r} (\Lambda_2  (s)) f \right) (k)  
		& =    
		f \bigl( k \cosh s  \mp 
		 \sqrt{k^2+ m_0^2} \sinh s \bigr)	\; ,  \qquad s \in \mathbb{R} \; ; 
	\\
			\left( \widetilde{u}_{m r} (\Lambda_1( \tfrac{t}{r})) f \right) (k)  
		& =  \left| \cosh \tfrac{t}{r} - \sqrt{ \tfrac{k^2}{m_0^2}+ 1} 
		\sinh \tfrac{t}{r}  \right|^{-\frac{1}{2} - i m r} \\
		& 
				\qquad \qquad \times (\mathscr{Q} f)  \biggl(  \tfrac{k} {\cosh \tfrac{t}{r} - \sqrt{ \frac{k^2}{m_0^2}+ 1} \; \sinh \tfrac{t}{r}} \biggr)
				\; ,  \qquad t \in \mathbb{R} \; ; 	
		\\
		\left( \widetilde{u}_{m r} (R_0(\tfrac{\alpha}{r})) f \right) (k)  
		& =  	\left| \tfrac{k}{m_0} \sin \tfrac{\alpha}{r} 
		+  \cos \tfrac{\alpha}{r} \right|^{-\frac{1}{2} - i m r} \\
		& 
				\qquad \qquad \times (\mathscr{P} f) 
				\left(  \tfrac{k \cos \tfrac{\alpha}{r} - m_0 
				\sin \tfrac{\alpha}{r}}{ \frac{k}{m_0} \sin \tfrac{\alpha}{r} 
				+ \cos \tfrac{\alpha}{r} } \right) 
				\; ,  \qquad \alpha \in [0, 2\pi \, r) \; , 
	\end{align*}
with 
	\[
		\mathscr{Q} f  = \begin{cases}  (f_+, f_-) & 
		\text{if $\;  m_0 \cosh \tfrac{t}{r} - \sqrt{k^2+ m_0^2} \sinh \tfrac{t}{r}  >   0 $} \; , \\
					       (f_-, f_+) & 
		\text{if $\;  m_0 \cosh \tfrac{t}{r} - \sqrt{k^2+ m_0^2} \sinh \tfrac{t}{r} <   0 $} \; , 
					\end{cases}
	\]
and	
	\[
		\mathscr{P} f  = \begin{cases}  (f_+, f_-) & 
		\text{if $\;  k \sin \tfrac{\alpha}{r} + m_0 \cos \tfrac{\alpha}{r} > 0 $} \; , \\
					       (f_-, f_+) & 
					       \text{if $\; k \sin \tfrac{\alpha}{r} + m_0 \cos \tfrac{\alpha}{r}  < 0$} \; . 
					\end{cases}
	\]
Note that $\widetilde{u}_{m r}(\Lambda_2 (s))$ depends on $m_0$, 
but \underline{not} on $m$ or $r$.
\end{theorem}

\begin{remark}
\label{rm:3.7.3}
We note that (see \eqref{1.5.1}) 
	\[
		D(q) \begin{pmatrix}
		\sqrt{k^2 + m_0^2 } \\
		\pm  k \\
		\pm m_0
		\end{pmatrix}= \begin{pmatrix}
		\sqrt{k^2 + m_0^2 } \; ( 1+ \frac{q^2}{2}) \pm  k q \pm  m_0 \frac{q^2}{2} \\
		\sqrt{k^2 + m_0^2 } \; q \pm  k   \pm  m_0 q  \\
		- \sqrt{k^2 + m_0^2 } \;  \frac{q^2}{2}  \mp k q \pm  m_0 (1- \frac{q^2}{2} )
		\end{pmatrix} . 
	\]
Thus
	\begin{align*}
		\left( \widetilde{u}_{m r} (D(\tfrac{q}{r})) f \right) (k)  
		& =  			\left| - \sqrt{k^2 + m_0^2 }  \; 
		\tfrac{q^2}{2m_0 r^2 } \mp  \tfrac{k}{m_0} \tfrac{q}{r} \pm    
					\bigl(1- \tfrac{q^2}{2r^2}  \bigr) \right|^{-\frac{1}{2} - i  m r} \\
		& 
				\qquad \qquad \times (\mathscr{W} f) 
				\left(   
						\tfrac {\sqrt{k^2 + m_0^2 } \; 
						\frac{q}{r} \pm   k   \pm  m_0  \frac{q}{r} }
							{- \sqrt{k^2 + m_0^2 } \;  
							\frac{q^2}{2m r^2 }  \mp  \frac{k}{m_0} \frac{q}{r}  
							 \pm    (1- \frac{q^2}{2r^2} ) } \right) \; ,   
	\end{align*}
with 
	\[
		\mathscr{W} f  = \begin{cases}  (f_+, f_-) 
		& \text{if $\;  - \sqrt{k^2 + m^2 } \;  \frac{q^2}{2m r^2 }   - \frac{k}{m} \frac{q}{r} 
		 +  (1- \frac{q^2}{2r^2} )  >   0 $} \; , \\
					       (f_-, f_+) 
		& \text{if $\;  - \sqrt{k^2 + m^2 } \; \frac{q^2}{2m r^2 }  + \frac{k}{m} \frac{q}{r} 
					        -  (1- \frac{q^2}{2r^2} ) <   0 $} \; . 
					\end{cases}
	\]
We note that for $|t|$ and $| \alpha|$ fixed, $f_+, f_- \in C^{\infty}_0 (\mathbb{R})$ 
and $r$ sufficiently large, 
	\[
		\mathscr{Q} f  = \mathscr{P} f = \mathscr{W} f = f \; ; 
	\]
\emph{i.e.}, the two components of $f$ remain separated. 
\end{remark}

\chapter{Harmonic Analysis on the Hyperboloid}
\setcounter{equation}{0}
\label{Harmanaly}

Harmonic analysis on symmetric spaces $X = G/H$ originated with the monumental work of 
Harish-Chandra \cite{HC1} -- \cite{HC9}. The subject  has been developed further in 
particular by Helgason~\cite{He} (for the case that the subgroup $H$ is compact\footnote{The 
necessary alterations in case $H$ fails to be compact, can be found in the work of 
Molchanov~\cite{Mo1, Mo2} and Faraut~\cite{Fa}.}) and the Russian school, see, 
\emph{e.g.},  \cite{GN, GG, GG59, Vil, VG, VGG}. In this work we will use a reformulation of the 
Fourier(-Helgason) transformation on de Sitter space, which emphasises the analyticity properties 
of the (generalized) Fourier transform. The latter is due to Bros and Moschella~\cite{BM2}.

\section{Plane waves}
\label{planwaves}

On the two-dimensional Minkowski space, the plane waves 
	\[
		(t, q) \mapsto {\rm e}^{i (t, q) \cdot (\sqrt{p_1^2+ m_0^2}, p_1) } \; , 
		\qquad p_1 \in \mathbb{R} \; \text{fixed} \;, 
	\] 
can be interpreted as 
improper common eigenvectors of  the space-time translation operator 
$\mathbb{R}^{1+1} \ni (t', q') \mapsto T(t', q')$. The generators of the translations, 
namely, the energy operator ${\it P}_0$ and the momentum operator~${\it P}_1$, 
act as multiplication operators on the plane waves: 
	\begin{align*}
		{\it P}_0 \; {\rm e}^{i \,   (t, q) \cdot (\sqrt{p_1^2+ m_0^2}, p_1) } & 
		= \sqrt{p_1^2+ m_0^2} \;  {\rm e}^{i \,  (t, q) \cdot (\sqrt{p_1^2+ m_0^2}, p_1)} \; , \\
		\qquad {\it P}_1 \;  {\rm e}^{i \,   (t, q) \cdot (\sqrt{p_1^2+ m_0^2}, p_1) } & 
		= p_1 \; {\rm e}^{i \,  (t, q) \cdot (\sqrt{p_1^2+ m_0^2}, p_1)} \; . 
	\end{align*}
In fact, the plane waves form an improper basis in the eigenspace of the Casimir operator 
	\[
		{\it M}_0^2 =  {\it P}_0^2- {\it P}_1^2   
	\] 
for the eigenvalue $m_0^2>0$. Note that the inner product 
$	 (t, q) \cdot (\sqrt{p_1^2+ m_0^2}, p_1) $ 
equals $m_0$ times the Euclidean distance of the point $(t, q)$ 
from the line passing through the origin whose normal vector is $(\sqrt{p_1^2+ m_0^2}, - p_1) $.

Now let us compare this with the situation on the two-dimensional
de Sitter space. 
The eigenfunctions of the Casimir operator on the light-cone $\partial V^+$
are homogeneous functions of degree $s=s^\pm$; see \eqref{dd1}.
Thus, in order to construct a {\em plane wave} on $dS \ni x$, 
one considers homogeneous functions of the scalar product 
	\begin{equation}
	\label{arbitrary-plane}
	x \cdot p = ( x + \lambda p + \mu q) \cdot p   \; , \qquad \lambda , \mu \in \mathbb{R} \; . 
	\end{equation}
In \eqref{arbitrary-plane} we have used \eqref{zero-plane}, with $q \in S^1$ such that $q \cdot p= 0$. 
Since $p \in \partial V^+$, $p \cdot p = 0$.

\begin{lemma}
Given a point $x \in  \Gamma^+ (W_1)$, the intersection of the 
plane\footnote{Of course, the planes \eqref{x-plane} for different $x \in \Gamma^+ (W_1)$ 
are all parallel to each other.}
	\begin{equation}
	\label{x-plane}
	\left\{ x  +  \left[ \lambda 
	 \left( \begin{smallmatrix} 1 \\ 0 \\ -1 
	 \end{smallmatrix} \right)
	  + \mu q \right]  \mid  \lambda , \mu \in \mathbb{R}  \right\}  \; , 
	 \qquad 
	q = \begin{pmatrix} 0 \\ r \\ 0 \end{pmatrix}  .
	\end{equation}
with the de Sitter space $dS$ is the horosphere
	\[
	 	P_{ \tau } = \left\{ y  \in dS  \mid  r {\rm e}^{\frac{\tau}{r}}  
		= y \cdot p_\circ  \right\} \; ,  
		\qquad 
		p_\circ = \left( \begin{smallmatrix} 1 \\ 0 \\ -1 
	 	\end{smallmatrix} \right)  \;, 
	\]
where $\tau \in \mathbb{R}$ is fixed by requesting  
$r {\rm e}^{\frac{\tau}{r}}  = x \cdot p_\circ  $. 
For each $x \in \Gamma^+ (W_1)$, the angle between 
the plane \eqref{x-plane} and the $x_0$-axis is $\pi/4$. 
\end{lemma}

\subsection{Holomorphy}
For $x$ in one of the two\footnote{We will soon integrate 
over $p\in \gamma_0$. As $p = R_0(\alpha) p_\circ$ 
rotates on the light cone $\partial V^+$, 
{\em all} light rays in $dS$ are affected.}
light rays forming the horosphere~$P_{- \infty}$, \emph{i.e.}, the 
intersection of~$dS$ with the plane \eqref{zero-plane}, the scalar $x \cdot p_\circ$ 
vanishes and powers with negative real part have to be defined in 
distributional sense. One possibility, which we will pursue, 
is to define them as the boundary values of analytic functions, using 
the \emph{principal value}\index{principle values} 
of the complex powers. The characterisation of the tuboid given 
in \eqref{tube-0} guarantees that the functions
	\[   
		z \mapsto  ( z \cdot  p)^{s} \, , 
		\qquad p \in \partial V^+ \; , 
	\]
are holomorphic both in ${\mathcal T}_+$ and ${\mathcal T}_-$. Their boundary values  
as~$ z \in dS_\mathbb{C}$ tends to $x \in dS$ 
from within the respective tuboids ${\mathcal T}_+$ and ${\mathcal T}_-$ of $dS$
are denoted as 
\label{planewavepage}
	\begin{equation}
		\label{eqPW}
		 x \mapsto   (  x_\pm \cdot  p )^{s} \; , \qquad x \in dS \; .
	\end{equation}
As expected, we encounter a discontinuity as $  \Im 
x_+ \cdot p \nearrow  0$ or $  \Im   x_- \cdot p  \searrow 0$, respectively.
Another way of denoting the function \eqref{eqPW} is \cite[Eq.~(45)]{BM2}
	\begin{equation}
		\label{eqPW-b}
		(x_\pm \cdot p)^{s} 
		= \mathbb{1}_{[0, \infty)} ( - x \cdot p ) \;  | x \cdot p |^{s} 
			+ {\rm e}^{ \pm i \pi s} \mathbb{1}_{(0, \infty)} 
			( x \cdot p )\;  | x \cdot p |^{s} \; , 
	\end{equation}
where $\mathbb{1}_{[0, \infty)}$ is the \emph{Heaviside step 
function}\index{Heaviside step function}, \emph{i.e.}, 
$\mathbb{1}_{[0, \infty)} (t) = 0$ if $t<0$ and $\mathbb{1}_{[0, \infty)} (t) = 1$ 
if $t \ge 0$. In case $\Re s > -1$, the singularity is 
integrable and the equality \eqref{eqPW-b}
holds in the sense of $L^1$-functions.  

\subsection{The wave equation}
An explicit computation\footnote{The \index{Laplace-Beltrami operator}
\emph{Laplace-Beltrami operator} 
$\square_{dS} =|g|^{-1/2}\partial_\mu g^{\mu\nu}|g|^{1/2}\partial_\nu$ on $dS$
is related to the \emph{D'Alembert operator} \index{D'Alembert operator}
$\square_{\mathbb{R}^{1+2}}$ for the ambient Minkowski space $\mathbb{R}^{1+2}$ 
by the following identity
	\[
		r^2 \square_{\mathbb{R}^{1+2}} 
		=  r^2 \square_{dS} + \mathcal{D}(\mathcal{D} + 1) \; , 
	\qquad 
	\text{where} 
	\quad
		\mathcal{D} = \sum_{\nu=0}^2 x_\nu\frac{\partial}{\partial x_\nu}  
	\]
is the \index{Cauchy-Euler operator} \emph{Cauchy-Euler operator}, 
\emph{i.e.}, the generator of the \emph{dilatation} subgroup 
of the \emph{conformal group}.} \cite[Eq.~(4.3)]{BM} shows that the
plane waves given in \eqref{eqPW} satisfy the eigenvalue equation
	\begin{align}
	\label{eqPW-new}
		\square_{dS} \left(  x_\pm  \cdot   p \right)^s  & =
		r^{-2} \underbrace{ \left(   \Kgeo_0^2 
		- \Lgeo_1^2 - \Lgeo_2^2  \right) }_{= \Cgeo^2}
		\left(  x_\pm  \cdot   p \right)^{s}
		\nonumber \\
		& = r^{-2} \bigl( \underbrace{ -s(s+1) }_{\zeta^2}  \bigr)
		\left(  x_\pm  \cdot   p \right)^s \; . 
	\end{align}
This is just the Klein--Gordon equation
	\begin{equation}
	\label{4.1.6}
		(\square_{dS} +\mu^2)\left(  x_\pm  \cdot   p \right)^{-\frac{1}{2} \mp  i mr}   = 0 \; , 
		\quad \hbox{with} \quad \mu^2 r^2 = \frac{1}{4} + m^2 r^2\; , 
	\end{equation}
on the de Sitter space $dS$.  
As one can seen from Equ.~\eqref{casimir2}, 
they also satisfy the Klein--Gordon equation on the forward 
light cone~$\partial V^+$ (see also \eqref{casimir2}):
	\[
			(\square_{\partial V^+} + \mu^2 ) \left(  x_\pm  \cdot  p \right)^s = 0 \; , \qquad -s(s+1) = 
		\mu^2 r^2  \; . 
	\]
Note that in contrast to the Minkowsi space case, the operators $\Kgeo_0, \Lgeo_1$ and $\Lgeo_2$ 
do not commute, so they can not be represented 
as commuting multiplication operators in Fourier space. 

\section{The Fourier--Helgason transformation}
\label{Fourier--HelgasonTransformation}

In Chapter 3 we have seen that unitary irreducible representations of $SO_0(1,2)$ are 
most conveniently constructed on the forward light cone $\partial V^+ =SO_0(1,2)/N$.
Formally\footnote{The precise statement is slightly more 
involved, as the integral in \eqref{graev} has to be defined carefully.}, the $L^2$-functions 
on the de Sitter space $dS$ and the $L^2$-functions on the light cone $\partial V^+$ are 
related by the \emph{horospheric Radon transform} (intoduced by Gelfand and Graev 
\cite{GG55, GG68}) 
	\begin{equation}
	\label{graev}
		f \mapsto  \left( \partial V^+ \ni p \mapsto \int_{dS} {\rm d}\mu_{dS} ( x) \; 
		f(x) \delta (x \cdot p -1) \right) \; , 
	\end{equation}
which maps functions on $dS= SO_0(1,2)/ SO(1,1)$ to functions on $\partial V^+$. 
Given functions on the light cone $\partial V^+$, we can proceed as in in Section \ref{RRLC}: 
the Mellin transform (which decomposes the delta function in \eqref{graev} into plane waves)  
decomposes them into homogeneous functions of $p \in \partial V^+$. As we have seen in the 
proof of Theorem~\ref{TH-irr}, the latter transform irreducibly under the action of~$SO_0(1,2)$.  
Thus, roughly speaking, by starting with $f(x)$ on $dS$, moving to the light 
cone $\partial V^+$ by using the horospherical transform and taking the Mellin transform, 
one can decompose $f(x)$ into components transforming irreducibly under the action of the 
symmetry group. 

\subsection{The Fourier--Helgason transforms} We are now able to present the generalisation 
of the Fourier transform suitable for the de Sitter space. 

\begin{definition}
\label{def:4.2.1}
Let $p \in \partial V^+ $ and  $s \in \{ z \in \mathbb{C}  \mid -z (z+1)> 0 \}$.
The {\em Fourier--Helgason transforms}~$ {\mathcal F}_\pm $  are 
defined~\cite[Eq.~(44), see also Definition 2]{BM2} by
\label{Fouriertransformpage}
	\begin{equation}
	\label{ftps}
 	 {\mathcal D} (dS)  \ni f \mapsto	\widetilde {f}_\pm (  p, s) 
		= \int_{dS} {\rm d} \mu_{dS} ( x ) \; f( x ) \; (  x_\pm \cdot  p )^s \; . 
	\end{equation}
\end{definition}

For $p$ fixed, the functions  $\widetilde {f}_\pm ( p , \, . \,)$
are holomorphic with respect\footnote{Note that a function  analytic in the strip $- 1 < \Re s < 0$ 
is uniquely determined by its values on one of the two symmetry axis given 
in \eqref{s1} and \eqref{s2}.} to $s$ in the strip $- 1 < \Re s < 0$ 
[Bros und Moschella~\cite{BM2}, Prop.~8.a]. 

\begin{lemma} 
\label{nu-analyicity}
The function
	\[
		\nu \mapsto \widetilde {f}_\pm \bigl(  p, -\tfrac{1}{2} - i \nu \bigr) 
	\]
is analytic in the open strip $\bigl\{ \nu \in \mathbb{C} \mid | \Im \nu | < \tfrac{1}{2} \bigr\}$.
\end{lemma}

For $s$ fixed, the two functions $\widetilde {f}_\pm ( \, . \, , s)$ are
continuous, homogeneous functions of degree $s$ on~$\partial V^+$. 
Together with \eqref{casimir2} this implies that the functions $\widetilde {f}_\pm (\, . \, , s)$ 
are eigenfunction of the Casimir operator $M^2$ on $\partial V^+$, iff $s$ lies on    
\begin{itemize}
\item[$i.)$]  the symmetry axis 
	\begin{equation}
	\label{s1} 
		s = -1/2 \mp i \nu \; , \qquad \nu = \sqrt{\mu^2 r^2 - \tfrac{1}{4}} = m r\in \mathbb{R}^+_0 \; , 
	\end{equation}
of the strip $- 1 < \Re s < 0$. This choice corresponds to  $\mu^2  = \tfrac{1}{4r^2}+ m^2 \ge \tfrac{1}{4r^2}$, 
\emph{i.e.}, to $m^2  \ge  0$; 
\item[$ii.)$] the symmetry axis 
	\begin{equation}
		\label{s2} 
		s = -1/2 \mp i  \nu \; , \qquad  \nu = i \sqrt{\tfrac{1}{4} - \mu^2 r^2} = i mr  \; , 
		\quad -\tfrac{1}{2r} < m \le 0 \; .
	\end{equation}
of the strip $- 1 < \Re s < 0$. This choice corresponds to  $0 < \mu^2  \le \tfrac{1}{4r^2} $.
\end{itemize}
Thus  the critical value~$\mu_c$, which separates the two cases, is 
	\[
		\mu_c=\sqrt{-s (s+1)}= \frac{1}{2r} \; . 
	\]
Note that the factor $(2r)^{-1}$ may be interpreted as a contribution to the particle
mass coming from the curvature of space-time (see, \emph{e.g.}, \cite{GaT, Urs}).

\begin{remark}
\label{splitting}
Taking advantage of \eqref{eqPW-b},  the Fourier--Helgason transforms~$ {\mathcal F}_\pm$ can  be written in the following form
(see \cite[Eq.~(50)]{BM2})
	\begin{align*}
 	\widetilde {f}_\pm (  p, s )
		& = \int_{ \{ x \in dS \mid x \cdot p >0 \}} {\rm d} \mu_{dS} ( x ) \; f( x ) \; |  x \cdot  p |^s \\
		& \qquad    + {\rm e}^{ \mp i \pi s}  
		\int_{ \{ x \in dS \mid x \cdot (-p) >0 \}} {\rm d} \mu_{dS} ( x ) \; f( x ) 
		\; |  x \cdot  p |^s \; . 
	\end{align*}
This identity is valid in the open strip $\{ \nu \in \mathbb{C} \mid | \Im \nu | < 1/2 \}$. The second term can be viewed as 
a continuous, homogeneous function of degree $s$ on~$\partial V^-$. 
\end{remark}

\section{The Plancherel theorem on the hyperboloid}
\label{sec:4.3}

\label{hardyspacepage}
Recall the tuboids defined  
in \eqref{chiral-tuboids} and denote by $H^2 ({\mathcal T}_+)$, 
$H^2 ({\mathcal T}_-)$, $H^2 ({\mathcal T}_\leftarrow)$ and 
$H^2 ({\mathcal T}_\rightarrow)$ the Hardy spaces of functions~$F(z)$ 
characterised by the following properties~\cite[Sect.~3.2]{BM2}\cite[Sect.~3.3]{Ne}: 
\begin{itemize}
\item [$i.)$] $F$ is holomorphic in the tuboid considered; 
\item[$ii.)$] the function $F$ admits  boundary values $f \in L^2 (dS, {\rm d} \mu_{dS})$ on $dS$;  
\item[$iii.)$] $F$ is `sufficiently regular at infinity in its domain' (in the sense made precise 
in \cite[p.~10]{BM2}).
\end{itemize}

\begin{theorem}[Bros \& Moschella~\cite{BM2}, Theorem 1]
\label{hardy}
Any given function $f \in L^2 (dS, {\rm d} \mu_{dS})$ admits a decomposition of the form
\label{hardyspacedecomposition}
	\begin{equation}
		\label{decompose}
 		f = f_+ + f_- + f_\leftarrow + f_\rightarrow \equiv \sum_{tub} f_{(tub)} \; , 
		\qquad (tub) = +, -, \leftarrow, \rightarrow  \; ,  
	\end{equation}
where $f_{(tub)} ( x ) \in L^2 (dS, {\rm d} \mu_{dS})$ is the boundary value of the function 
	\begin{equation}
		\label{ck}
 		F_{(tub)} ( z) 
		= \epsilon_{(tub)} \frac{1}{\pi^2} \int_{dS} {\rm d} 
		\mu_{dS} ( x) \; \frac{f( x)}{{( x -  z ) \cdot ( x -  z )} } \in H^2 ({\mathcal T}_{(tub)}) \; .
	\end{equation}
The sign function $\epsilon_{(tub)}$ takes the value $-1$ for  ${\mathcal T}_\pm$,  and
$+1$ for ${\mathcal T}_\leftarrow$ and~${\mathcal T}_\rightarrow$.
\end{theorem}

\goodbreak
\begin{remark}In Minkowski space $\mathbb{R}^{1+1}$, a similar decomposition can 
be gained by simply dividing the support of the Fourier transform $\widetilde {f}$ 
into the four cones $\{ (E, p ) \in \mathbb{R}^{1+1} \mid \pm E > |p|\}$ and 
$\{ (E, p ) \in \mathbb{R}^{1+1} \mid \pm p > |E|\}$. Note that for $m>0$ the boundary sets 
$\{ (E, p ) \in \mathbb{R}^{1+1} \mid \pm p = E \}$ are of measure zero. The inverse Fourier 
transform of each of these functions is then the boundary  of a function analytic 
in a tube. For the first two, the tube is $T^\pm = \mathbb{R}^2 \pm i \{ (x_0, x_1 ) 
\in \mathbb{R}^{1+1} \mid \pm x_0 > |x_1|\}$. 
The situation is similar for the two other cases. 
\end{remark}

The Cauchy kernel\footnote{This formula should be compared with the one given for the 
Wightman two-point function in Theorem \ref{prop:4.1} below.}
 on $dS_\mathbb{C}$ introduced in  \eqref{ck} arises as a limit of the function~\cite[Proposition~11]{BM2}
	\begin{equation}
		\label{ac}
		\frac{1}{( z' -  z ) \cdot ( z' -  z )} 
		= - \frac{\pi^2}{2} \int_0^\infty {\rm d} \mu_\pm (\nu)
			\int_{\gamma} {\rm d} \mu_{\gamma} (p )\;  
			( z \cdot p)^{- \frac{1}{2} + i \nu}   ( p \cdot  z')^{- \frac{1}{2} - i \nu} \; ,   
	\end{equation}
where $\gamma$ is a closed curve on the forward light cone $\partial V^+$, which encloses the origin. 
The integral in \eqref{ac}  is absolutely convergent for 
$( z,  z') \in {\mathcal T}_+ \times {\mathcal T}_-$ for~${\rm d} \mu_+$ and for 
$( z,  z') \in {\mathcal T}_- \times {\mathcal T}_+ $ for~${\rm d} \mu_-$, respectively. 
The measure ${\rm d} \mu_\pm (\nu)$ on $\mathbb{R}^+$ is 
 {\rm (see~\cite[Sect.~4.1]{BM2})}
	\[  
		\label{dmu}
		{\rm d} \mu_\pm (\nu) 
		= \frac{1}{2\pi^2} \; \frac{\nu \tanh \pi \nu}{ {\rm e}^{\pm \pi \nu} \cosh \pi \nu }  {\rm d} \nu  \; . 
	\]
Combine \eqref{ck}, \eqref{ac} and \eqref{ftps} to find the {\em inversion formula} \cite[Eq.~(80)]{BM2}
	\begin{equation}
	\label{inverse}
		F_\pm (  z) = - \int_0^\infty {\rm d} \mu_\pm (\nu) 
		\int_{\gamma} {\rm d} \mu_{\gamma} ( p )\;  
			( z \cdot  p)^{- \frac{1}{2} + i \nu} \;  
			\widetilde {f}_\pm \left(  p, - {\textstyle \frac{1}{2} } - i \nu \right) \; .
	\end{equation}
The functions $f_\pm (  x) $ introduced in \eqref{decompose} now appear as  boundary values of the 
holomorphic functions $F_\pm (  z)$, $z \in {\mathcal T}_\pm$. 

\begin{remark}
For every function $F_\pm \in H^2 ({\mathcal T}_\pm)$ the  transform 
$\widetilde {f}_\pm \left(  p, - {\textstyle \frac{1}{2} } - i \nu \right)$ 
vanishes \cite[Proposition~8]{BM2}. This follows from analyticity properties, which we will establish in 
Theorem~\ref{prop:4.1} below. A similar result holds true in the Minkowski space-time:
The functions $f$ on $\mathbb{R}^{1+d}$, which are boundary values of holomorphic functions 
in the tube ${\mathfrak T}_\pm = \mathbb{R}^{1+d} \mp i V^+ $ are the functions whose Fourier transforms 
	\[
	\widetilde f (k) = \frac{1}{(2\pi)^{\frac{1+d}{2}}}\int_{\mathbb{R}^{1+d}} {\rm d} y \; f(y) \; {\rm e}^{i   k \cdot y} \; , 
	\qquad f \in {\mathcal D} (\mathbb{R}^{1+d}) \; , 
	\]
have their support contained in the closure of either $V^+$ or $V^-$; see, \emph{e.g.},~\cite[Ch.~8]{Schwartz}. 
\end{remark}

\begin{theorem}[Molchanov \cite{Mo1}]
\label{plancherel}
For any pair of functions $f, g$ in $L^2(dS, {\rm d} \mu_{dS})$ and their corresponding 
decomposition given in \eqref{decompose}, one has the {\em Plancherel theorem}\footnote{These 
are the Eq.~(118) and Eq.~(119) in \cite{BM2}.}:
	\begin{align*}
 		& \int_{dS} {\rm d} \mu_{dS} ( x)\;  \overline{f_\pm(
            x )} g_\pm (  x )   
		= 
		\int_0^\infty {\rm d} \mu_\pm (\nu)
		\int_{\gamma} {\rm d} \mu_{\gamma} ( p ) \; \overline{\widetilde {f}_\pm  
		\left(  p, - {\textstyle \frac{1}{2} } - i \nu \right) } 
		\; \widetilde {g}_\pm \left(  p, - {\textstyle \frac{1}{2} } - i \nu \right)  . 
	\end{align*}
The measures ${\rm d} \mu_{dS}$ and ${\rm d} \mu_{\gamma}$ denote the 
Lorentz invariant measure on $dS$ and  
the restriction of the Lorentz invariant measure on $\partial V^+$ to
$ \gamma $.   
\end{theorem}

\section{Unitary irreducible representations on de Sitter space} 
\label{new-3.4}

Our basic strategy is to use, just as in Minkowski space (see, \emph{e.g.},~\cite{RS}), 
the restriction of the Fourier--Helgason transform 
${\mathcal F}_{+ \upharpoonright \nu} \colon   {\mathcal D}_{\mathbb{R}} (dS)
 \to  \widetilde {\mathfrak h}_\nu  \, $,  
\label{mass-shell-ft-page}
	\begin{align}
	\label{mass-shell-ft}
				 f  & \mapsto  \sqrt{ \frac{c_\nu {\rm e}^{-  \pi \nu}  r   }{\pi}} 	\; 		
				 	 \widetilde {f}_+  (  \, . \, , s^+) \doteq  {\widetilde f}_\nu \; , 
	\end{align}
to the ``upper mass shell'', with $s^+$ given by \eqref{dd1},
to define a (complex valued) semi-definite quadratic form 
	\begin{equation}
	\label{nuclear}
	{\mathcal D}_{\mathbb{R}} (dS) \ni f, g \mapsto 
	\langle \widetilde f_\nu , \widetilde g_\nu \rangle_{ \widetilde {\mathfrak h}_\nu }
	\end{equation}
on the test-functions.  (We will suppress the index $\nu$ when possible. For example, we 
will frequently write $\widetilde {{\mathfrak h}} (\partial V^+) $
instead of $\widetilde {{\mathfrak h}}_\nu (\partial V^+)$.) 
The value of the positive normalisation constant (see Harish-Chandra~\cite{HC1, HC2}) 
	\begin{equation}
	\label{cnu}
		c_\nu = -\frac{1}{2\sin  (\pi s^+ )} = 
		\frac{1}{2 \cos (i \nu \pi ) }  
	\end{equation}  
is chosen such that twice the imaginary part of the scalar product \eqref{nuclear} equals the 
value of the symplectic form $\sigma$ of the covariant classical dynamical system we will 
introduce in Section \ref{CCDS}; for further details, see \eqref{eqSplForm} and 
the discussion preceding \eqref{comfkt} below.
Using \eqref{eq:gamma-2}, one can show that
	\[
		c_\nu = \frac{\Gamma\left(1 + s^+ \right)\Gamma\left(1 +s^- \right)}{2\pi} = c_{-\nu} \; . 
	\]
We next state how ${\widetilde f}_\nu$ is related 
to ${\widetilde f}_{-\nu}$, for $0 < -i \nu < \frac{1}{2}$. 

\begin{proposition}[Faraut \cite{Fa}, Prop.~II.4]
\label{propo2.5}
Let $f \in {\mathcal D}_{\mathbb{R}} (dS)$ and $0 <  -i \nu < \frac{1}{2}$. Then
	\begin{align*} 
		\underbrace{ \int_{dS} {\rm d} \mu_{dS} (x) \; f (x) \; (  x_\pm \cdot  p )^{- { \frac{1}{2} } 
		- i \nu}}_{ = \widetilde{f}_\pm ( p , - \frac{1}{2} - i \nu)}
		&= \frac{\Gamma(\frac{1- i \nu}{2})}{\Gamma (\frac{3}{4}
			+\frac{i \nu}{2})}  \; \frac{\sqrt{\pi} \; 
			\Gamma(\frac{1}{2}+ i \nu)}{ 2^{-\frac{1}{2}+ i \nu} \, \Gamma( i \nu)}
			 \int_{\gamma} 
			 {\rm d} \mu_{\gamma} ( p') \; (  p \cdot p')^{-\frac{1}{2}+ i \nu} 
			\nonumber \\
		& \qquad \qquad \quad \times
			\underbrace{\int_{dS} {\rm d} \mu_{dS} (x) \; f (x) \; 
			(  x_\pm \cdot  p' )^{- { \frac{1}{2} } + i \nu} }_{= \widetilde{f}_\pm ( p' , - \frac{1}{2} + i \nu)}\;  ,  
	\end{align*}
where $\gamma$ is a closed curve on the 
forward light cone~$\partial V^+$,  which encloses the origin.
\end{proposition}

\begin{remark} The result of Faraut was pointed out to us by J.~Bros. Choosing 
$p  = (1, \cos \alpha,  \sin \alpha)$  and $ p'  = (1, \cos \alpha', \sin \alpha')$ 
we find 
	\[
		p  \cdot p' = 1- \cos (\upsilon-\upsilon') \; .   
	\]
Thus we have recovered the factor \eqref{Barg-factor} first introduced by Bargmann; 
see the definition of the intertwiner $A_{\nu}$.   
\end{remark}

\subsection{Real Hilbert Spaces}
The kernel of the quadratic form \eqref{nuclear} equals 
$\ker {\mathcal F}_{+ \upharpoonright \nu}$.  This  turns the 
$\mathbb{R}$-linear spaces $\mathcal{D}_{\mathbb{R}}(X) / 
\ker {\mathcal F}_{+ \upharpoonright \nu} \,$, $X = dS,  {\mathcal O }, W$, 
into {\em real} pre-Hilbert spaces
\label{FHtransformationmasspage}
\label{ophs1page}
	\begin{equation}
		\label{ophs2}
		{\mathfrak h}^\circ (X) \doteq \bigl( \mathcal{D}_{\mathbb{R}}(X) / 
		\ker {\mathcal F}_{+ \upharpoonright \nu} , 
		\Re \langle \, \widetilde \cdot , \widetilde \cdot \,  \rangle_{  \widetilde {\mathfrak h}_\nu } \bigr) \; , 
		\qquad X = dS,  {\mathcal O }, W \; .  
	\end{equation}
The completion of ${\mathfrak h}^\circ (X)$ 
defines the real Hilbert spaces 
${\mathfrak h} (X)$, $X = dS,  {\mathcal O }, W$. 
Their real valued scalar products are all given by the real part 
	\[
	\Re \langle f , g \rangle_{{\mathfrak h} (dS)}
	 \doteq \tfrac{1}{4} \Bigl( \| f + g \|_{{\mathfrak h} (dS)}
	- \| f - g \|_{{\mathfrak h} (dS)} \Bigr) \; 
	\]
of the {\em complex valued} scalar products  
	\begin{equation}
		\label{ophs2a}
		\langle [f] , [g] \rangle_{{\mathfrak h} (dS)} \doteq 
		\langle \, \widetilde f_\nu , \widetilde g_\nu  \,  
		\rangle_{ \widetilde {\mathfrak h}_\nu } \;   , 
		\qquad [f], [g] \in {\mathfrak k} (X)\; ,   
	\end{equation}
with $\nu = \nu(\mu)$ given by \eqref{dd1} with $\zeta^2 = \mu^2 r^2$. 

\bigskip

\subsection{Complex Hilbert Spaces}
Given the existence of the complex-valued scalar products \eqref{ophs2a}, the 
question now arises, whether these real Hilbert spaces can be interpreted as complex 
Hilbert spaces, \emph{i.e.}, whether or not they carry an intrinsic complex structure. The 
answer to this question depends on the choice of~$X \subset dS$: 

\begin{proposition}
In case $X = dS$,  the imaginary part of the
complex-valued scalar products \eqref{ophs2a} defines 
a weakly non-degenerate symplectic form, which gives rise to 
a  complex structure~${\mathscr I}
\colon {\mathfrak h} (dS) \to {\mathfrak h} (dS)$ on the real Hilbert 
space ${\mathfrak h} (dS)$ such that 
	\[
		\Re \langle  f , g \rangle_{{\mathfrak h} (dS)} 
		\doteq - \Im \langle  {\mathscr I} f , g \rangle_{{\mathfrak h} (dS)}    
		\qquad \forall   g \in {\mathfrak h} (dS) \; . 
	\]
Note that this is done without enlarging ${\mathfrak h} (dS)$.
\end{proposition}

\vskip -.5cm 

\begin{proof} 
We note that $f \in {\mathfrak h} (dS)$ is 
equal to the zero vector in ${\mathfrak h} (dS)$, if
	\[
		\Im \langle  f , g \rangle_{{\mathfrak h} (dS)} = 0 \qquad 
		\forall g \in{\mathfrak h} (dS)  \;  .
	\]
Thus the symplectic form $\Im \langle \, . \, ,  \, . \, \rangle_{{\mathfrak h} (dS)}  $ is 
weakly\footnote{The map $\flat \colon {\mathfrak h} (dS) \to {\mathfrak h} (dS)$ 
defined by $ \langle \flat (h) , g \rangle = \Im \langle h , g \rangle$ is injective;  
thus $\flat$ is called \emph{weakly} non-degenerate according to \cite{AM}.}  
non-degener\-ate\footnote{This statement is independent, 
but in agreement with \eqref{eqEf}.} on the dense sub\-space ${\mathfrak h}^\circ (dS)$. 
The second part of the statement is a particular case of  \cite[Theorem 3.1.19]{AM}; 
we present the argument given there: for $[g] \in {\mathfrak k} (dS) $ fixed, the map 
	\[
		[f] \mapsto  \Im
		\langle \, \widetilde f_\nu , \widetilde g_\nu  \,  
		\rangle_{ \widetilde {\mathfrak h}_\nu } \;   , 
		\qquad [f]  \in {\mathfrak k} (dS)\; ,
	\]   
is continuous. Applying the Riesz lemma to the real Hilbert space ${\mathfrak h} (dS)$, 
we conclude that there exists a \emph{bounded} 
operator $A \colon {\mathfrak h} (dS) \to {\mathfrak h} (dS)$ such that 
	\[
		\langle  A f , g \rangle_{{\mathfrak h} (dS)}    
		\doteq   \Im \langle  f , g \rangle_{{\mathfrak h} (dS)}
		\qquad \forall   g \in {\mathfrak h} (dS) \; . 
	\]
Since $\Im \langle  f , g \rangle_{{\mathfrak h} (dS)}$ is anti-symmetric, 
	\begin{equation}
	\label{I-star}
		A = - A^* \; .
	\end{equation}
Since $\Im \langle  \, . \, , \, . \,  \rangle_{{\mathfrak h} (dS)}$ is weakly 
non-degenerate, $A$ is injective. By construction, 
	\[
		-A^2 = A A^* \ge 0 \; . 
	\]
From \eqref{I-star} it follows that $A^2$ is injective. Let $P \ge 0$ be a symmetric 
non-negative square root of $-A^2$. By definition, $P$ is injective and has dense range, 
as $P = P^*$ is self-adjoint. 
Thus $P^{-1}$ is a well-defined (possibly) \emph{unbounded} operator. 
Set 
		\[ 
			{\mathscr I} = A P^{-1} 
		\]
so that $A= {\mathscr I}P$. From \eqref{I-star} and $P^2 = -A^2$, we find that 
	\[ 
		{\mathscr I}^* = - {\mathscr I} = {\mathscr I}^{-1} \; , 
	\]
${\mathscr I}$ is orthogonal, and ${\mathscr I}^2 = - \mathbb{1}$. Thus ${\mathscr I}$ 
is uniformly 
bounded on the (dense) range 
of $P$, so extends to an orthogonal operator defined on all of ${\mathfrak h} (dS)$. 
Moreover, ${\mathscr I}$ is symplectic since
	\[
		\Im \langle  {\mathscr I} f , {\mathscr I} g \rangle_{{\mathfrak h} (dS)}    
		\doteq   \Im \langle  f , g \rangle_{{\mathfrak h} (dS)}
		\qquad \forall   f, g \in {\mathfrak h} (dS) \; . 
	\]
It follows that ${\mathscr I}$ defines a 
complex structure: for $f \in {\mathfrak h} (dS)$ we have
	\begin{equation}
		\label{CS}
		(\lambda_1 + i \lambda_2) f  = \lambda_1 f  + \lambda_2 ({\mathscr I} f)  \; , 
		\qquad \lambda_1, \lambda_2 \in \mathbb{R} \; . 
	\end{equation}
This turns the real Hilbert space $\bigl( {\mathfrak h} (dS), \Re \langle \, . \, ,  \, . \, 
\rangle_{{\mathfrak h} (dS)}  \bigr)$  into a 
complex Hilbert space $\bigl({\mathfrak h} (dS), \langle \, . \, ,  
\, . \, \rangle_{{\mathfrak h} (dS)} \bigr)$. 
The scalar product $f, g \mapsto \langle f ,  g \rangle_{{\mathfrak h} (dS)}$  
is anti-linear in $f$ and linear in $g$ with respect to the complex 
structure~defined in \eqref{CS}. 
\label{umLambdapage2a}
\end{proof}

\begin{remark}
In case $X= {\mathcal O }$ (with ${\mathcal O }$ bounded) or $X=W$,  
the $\mathbb{R}$-linear subspaces ${\mathfrak h} ({\mathcal O })$ 
and ${\mathfrak h} (W)$ of~${\mathfrak h} (dS)$ are \emph{not} dense in ${\mathfrak h} (dS)$. 
We will later show that their \emph{$\mathbb{C}$-linear span} is dense in ${\mathfrak h} (dS)$. 
\end{remark}

\subsection{A representation of $O(1,2)$}
\label{sec:4.5.3}

We will now show that ${\mathfrak h} (dS)$ carries a representation of $O(1,2)$. 

\begin{proposition}
\label{UIR-FH}
There is a unitary representation $ u$ of $SO_0(1,2)$ on~${\mathfrak h}  (dS)$ such that, 
for $f \in {\mathcal D}_{\mathbb{R}} (dS)$,
	\begin{equation} 
		\label{u-oph}
		u (\Lambda) [f] =  [ \Lambda_*f ]  \; , \qquad \Lambda \in SO_0(1,2) \; ; 
	\end{equation}
and consequently, $u(\Lambda) {\mathfrak h}  ({\mathcal O}) 
=   {\mathfrak h}  (\Lambda {\mathcal O}) $, $\Lambda \in SO_0(1,2)$.  
In other words, $u$ acts geometrically on ${\mathfrak h}  (dS)$. 
\end{proposition}

\label{umLambdapage2b}

\begin{proof} In order to extend the push-forward 
from ${\mathfrak h}^ \circ (dS)$ to a unitary representation  
on ${\mathfrak h} (dS)$, we have to show that
$ \| [ \Lambda_* f ] \|_{{\mathfrak h} (dS)} = \| [f] \|_{{\mathfrak h} (dS)}  $. 
By construction, 
	\begin{align*}
 		\| [ \Lambda_* f ] \|_{{\mathfrak h} (dS)} & =
		\Bigl\| \int_{dS} {\rm d} \mu_{dS} ( x ) \; f( \Lambda^{-1} x ) \; 
		(  x_+ \cdot  p )^{s^+} \Bigr\|_{\widetilde {\mathfrak h} (\partial V^+)} \\
		&= \Bigl\| \int_{dS} {\rm d} \mu_{dS} ( x ) \; f( x ) \; 
		(  \Lambda x_+ \cdot  p )^{s^+} \Bigr\|_{\widetilde {\mathfrak h} (\partial V^+)}  \\
		&= \Bigl\| \int_{dS} {\rm d} \mu_{dS} ( x ) \; f( x ) \; 
		(  x_+ \cdot \Lambda^{-1} p )^{s^+}  \Bigr\|_{\widetilde {\mathfrak h} (\partial V^+)} \\
		&= \|  \widetilde u_\nu^+   (\Lambda)   \widetilde f_\nu  \|_{\widetilde {\mathfrak h} (\partial V^+)}
		=  \| \widetilde f_\nu  \|_{\widetilde {\mathfrak h} (\partial V^+)} = \| [f] \|_{{\mathfrak h} (dS)}\;   , 
		\end{align*}
where ${s^+}$ is given by \eqref{dd1} with $\zeta^2 = \mu^2 r^2$. 
\end{proof}

\begin{proposition} 
\label{prop:4.4.5}
Let $u(T)$ and $u(P_2)$ be defined by 
	\[ 
		u (T) [ f ]
		\doteq[ T_* f]  
		\; , 
		\qquad u (P_2)[f] \doteq  [ P_{2*}f ] \; , 
		\qquad f \in {\mathcal D}_{\mathbb{R}} (dS) \; .
	\]
The operators $u(T)$ and $u(P_2)$ extend to well-defined 
(anti-)unitary operators on $\mathfrak{h}(dS)$. 
\end{proposition}

\begin{proof}
Since $P_2$ leaves the forward light cone invariant, the argument given in the proof of 
Prop.~\ref{UIR-FH} applies to $u(P_2)$ without further alterations. In order to establish 
a similar result for the time reflection, we have to show that 
		\begin{equation}
		\label{u-u-T}
		\widetilde{u}_\nu (T) {\mathcal F}_{+ \upharpoonright \nu} f 
		= 
		 {\mathcal F}_{+ \upharpoonright \nu} T_* \overline{f} 
		\; , 
		\qquad f \in {\mathcal D}_{\mathbb{R}} (dS) \; .
		\end{equation}
Let us evaluate the r.h.s.~in \eqref{u-u-T}: 
	\[
		\big({\mathcal F}_{+ \upharpoonright \nu} T_*\bar{f}\big)(p) = 
		\int_{dS} {\rm d} \mu_{dS} ( x ) \; \overline{f( x )} \; (  (Tx)_+ \cdot  p )^{s^+} \; .
	\]
Using the fact that, for $t\in\mathbb{R}$, 
	\begin{equation} 
		\big(-(t\pm i\epsilon)\big)^{s}={\rm e}^{\mp i\pi s}(t\pm i\epsilon)^{s},
	\end{equation}
we write 
	\begin{align*}
		\big(Tx\cdot p + i\epsilon\big)^{s^+}  & \equiv \big(-x\cdot (-Tp)+ i\epsilon\big)^{s^+}   = 
		\Big(-\big(x\cdot (-Tp)- i\epsilon\big)\Big)^{s^+}  \\
		& = {\rm e}^{i\pi s^+}\, \big(x\cdot (-Tp)- i\epsilon\big)^{s^+}  
		\equiv
		{\rm e}^{i\pi s^+}\, \overline{\big(x\cdot (-Tp)+ i\epsilon\big)^{\overline{s^+}}} \; .
	\end{align*}
Now for $0<m< 1/2$ the number $s^+$ is real, hence 
(note that $P= -T$ leaves the light cone invariant) 
	\[
		\big({\mathcal F}_{+ \upharpoonright \nu} T_*\bar{f}\big)(p) =
		{\rm e}^{i\pi s^+}\, \overline{\big({\mathcal F}_{+ \upharpoonright \nu} {f}\big)(-Tp) } \; . 
	\]
For $\mu \geq 1/2r$, the complex conjugate of $s^+$ is $s^-$, hence
	\[
		\big({\mathcal F}_{+ \upharpoonright \nu} T_*\bar{f}\big)(p) = 
		{\rm e}^{i\pi s^+}\, \overline{\big({\mathcal F}_{- \upharpoonright \nu}{f}\big)(-Tp) } 
		\equiv 
		{\rm e}^{i\pi s^+}\, \overline{\big(A_\nu\,{\mathcal F}_{+ \upharpoonright \nu} {f}\big)(-Tp) } \; , 
	\]
where we have used that (see Proposition \ref{propo2.5})
$A_{\nu} \colon \widetilde{\mathfrak{h}}_\nu \to \widetilde{\mathfrak{h}}_{- \nu}  \,  $,  
	\begin{align*}
			 {\mathcal F}_{+ \upharpoonright \nu}  f & \mapsto {\mathcal F}_{- \upharpoonright \nu}    f 
			 \; . 
	\end{align*} 
Comparing this result with the corresponding result 
for~$\widetilde u_\nu^+ (T)$, and 
inspecting the definition~\eqref{tilde-time-reflection}, proves the claim. 
\end{proof}

\subsection{The Wightman two-point function}
Finally, we apply the nuclear theorem to the quadratic form \eqref{nuclear}. It follows that 
there exists a tempered distribution  ${\mathcal W}^{(2)} (  x_1 ,  x_2 )$ on $dS\times dS$ 
such that
\label{twopointpage}
	\begin{equation}
	\label{2-point}
	\int_{dS \times dS} {\rm d} \mu_{dS} 
	( x_1) {\rm d} \mu_{dS} (  x_2)   { f ( x_1 ) } 
		{\mathcal W}^{(2)} (  x_1 ,  x_2 ) g ( x_2) 
		\doteq \langle [f]  ,   [g] \, \rangle_{{\mathfrak h} (dS)}  \; .
	 \end{equation}
The distribution ${\mathcal W}^{(2)} (  x_1 ,  x_2 )$ is
called the {\em two-point function}.  

\goodbreak
\begin{theorem}[Bros and Moschella \cite{BM}, Theorem 4.1 \& 4.2] 
\label{prop:4.1}
The \index{Wightman two-point function}
Wightman two-point function ${\mathcal W}^{(2)} ( x_1,  x_2)$ is
a tempered distribution, which is the boundary value of the function
	\begin{equation}
		\label{tpf-1}
	\quad {\mathcal W}^{(2)}  ( z_1 ,  z_2) =  c_\nu \frac{{\rm e}^{-   \pi \nu} r } {  \pi} 
	\int_{\gamma} {\rm d} \mu_{\gamma} ( p )
		\;  (  z_1 \cdot  p )^{s^- } (  p \cdot  z_2)^{ s^+} 		 
	\end{equation}
defined and holomorphic for $( z_1,   z_2) \in {\mathcal T}_ + \times {\mathcal T}_-$. 
The boundary values of \eqref{tpf-1} are taken as $\Im z_1 \nearrow 0$ and $\Im z_2 \searrow 0$, 
$( z_1,   z_2) \in {\mathcal T}_ + \times {\mathcal T}_-$.
As before, the exponents $s^\pm $ are given by~\eqref{dd1} 
and for the measure ${\rm d} \mu_{\gamma_\circ} ( p )$ one has
	\begin{equation}
	\label{measure-on-circle} 
		{\rm d} \mu_{\gamma_\circ} ( p ) = \tfrac{{\rm d} \alpha} {2} ; 
	\end{equation}
in agreement with the normalisation used in \cite[Section 4.2]{BM2}. 
The circle $\gamma_\circ$ was introduced in \eqref{gamma-0}.
\end{theorem}

\begin{remark}
In Minkowski space, after Fourier transformation, the two-point function 
	\[ 
	{\mathcal W}^{(2)}_m (x, y) = \int_{\mathbb{R}^{1+d}} {\rm d} k \; \theta (k^0) \delta (k \cdot k-m^2) \; 
	{\rm e}^{-i    k \cdot x} {\rm e}^{i   k \cdot y}
	\]
is the boundary value of a holomorphic function as 
$x \in  \bigl( \mathbb{R}^{1+1} - i \bigl\{   x' \in \mathbb{R}^{1+1} \mid  | x'_0 | > |x'_1| \bigr\}
\bigr)$ and $y \in  \bigl( \mathbb{R}^{1+1} 
+ i \bigl\{   x' \in \mathbb{R}^{1+1} \mid  | x'_0 | > |x'_1| \bigr\} \bigr) $ 
approach the reals. For $x= (x_0, \vec{x}\, )$, $y = (y_0, \vec{y}\, )$ 
and $p = \bigl( \sqrt{{\vec k \, }^2+m^2}, \vec k\, \bigr)$ this yields  
	\begin{equation}
	\label{flat2point}
	{\mathcal W}^{(2)}_m (x_0, \vec{x}, y_0, \vec{y} \, )
	= \frac{1}{2\pi}\int_{\mathbb{R}^d}  \tfrac{ {\rm d} \vec k}
	{2 \sqrt{ {\vec k \, }^2+m^2} } \; {\rm e}^{i  \vec k ( \vec{x}- \vec{y} \, ) - i  (x_0- y_0) \sqrt{{\vec k \, }^2+m^2}}  \; . 
	\end{equation}
\end{remark}

\bigskip
A direct consequence of this result is the 
\index{one-particle Reeh-Schlieder theorem}
\emph{one-particle Reeh-Schlieder theorem}:

\begin{theorem}[Bros and Moschella \cite{BM},  Proposition 5.4]
\label{oprs}
Let ${\mathcal O}$ be an open region in $dS$. It follows that 
${\mathfrak h}({\mathcal O}) + i {\mathfrak h}({\mathcal O})$
is dense in ${\mathfrak h}(dS)$. 
\end{theorem}

\begin{proof} It is sufficient to show that if $[f] \in {\mathfrak h}^\circ (dS)$ is orthogonal 
to ${\mathfrak h}({\mathcal O}) + i {\mathfrak h}({\mathcal O})$, then~$[f] $ 
is the zero-vector. Consider the complex valued function  
\[
z \mapsto F(z) = c_\nu \frac{{\rm e}^{-   \pi \nu}  r } {\pi} 
\int_{\gamma} {\rm d} \mu_{\gamma} ( p )
		\;  (  z \cdot  p )^{s^- } \; {\widetilde f}_\nu (p)\; , \]
which is holomorphic within ${\mathcal T}_ +$. Assume that\footnote{Note that $[g] \in 
{\mathfrak h}^\circ({\mathcal O}) + i {\mathfrak h}^\circ({\mathcal O})$ for  $g \in {\mathcal D}_{\mathbb{C}} ({\mathcal O}) $, 
by linearity of \eqref{ftps}.}
\[
\langle [g], [f] \rangle_{{\mathfrak h}(dS)}
=\int  {\rm d} \mu_{dS} \; g (x) F(x_+) = 0 \qquad \forall g \in {\mathcal D}_{\mathbb{C}} ({\mathcal O}) \; . 
\]
This implies that $F(z)$ vanishes on its boundary (as $\Im z \nearrow 0$) in 
the open region ${\mathcal O}$. It follows that its boundary values vanish on $dS$. This means that 
$[ f ] $ is orthogonal to any vector in ${\mathfrak h}(dS)$; thus it is the zero-vector. 
\end{proof}

\begin{proposition}[Bros and Moschella \cite{BM}, Proposition 2.2] 
\label{prop:4.1b}
The Wightman two-point function ${\mathcal W}^{(2)} ( z_1,  z_2)$ can be analytically continued into the cut-domain 
	\[
		\Delta = dS_{\mathbb{C}} \times dS_{\mathbb{C}} \setminus \Sigma
	\]
where the cut $\Sigma$ is the set 
	\[
		\Sigma = \bigl\{ (z_1, z_2) \in dS_{\mathbb{C}} \times dS_{\mathbb{C}} \mid (z_1 -z_2) \cdot (z_1 -z_2)   \ge 0 \bigr\} \; . 
	\]
Within $\Delta$ the two point function is invariant under the transformations 
	\[
		{\mathcal W}^{(2)} ( z_1 ,  z_2) = {\mathcal W}^{(2)} ( \Lambda_*z_1 ,  \Lambda_* z_2)  	\; , 
		\qquad \Lambda \in SO_{\mathbb{C}} (1,2) \; . 	 
	\]
Moreover, the permuted Wightman function ${\mathcal W}^{(2)} ( x_2,  x_1)$ is the boundary value of 
the analytic function ${\mathcal W}^{(2)} ( z_2 ,  z_1)$ from its domain $\{ (z_1, z_2) \mid z_1 \in {\mathcal T}^+ , 
z_2 \in {\mathcal T}^- \}$. 
\end{proposition}

\begin{proof} Proposition \ref{UIR-FH} guarantees that the distribution ${\mathcal W}^{(2)} ( z_1,  z_2)$ 
is invariant under Lorentz transformations, \emph{i.e.}, if 
$\Lambda \in SO_0 (1,2) $. 
Invariance under  the complexified group then follows by analytic continuation in the group parameter. 
For further details see [Bros and Moschella \cite{BM}, Proposition 2.2]. 
\end{proof}

\begin{remark} The cut $\Sigma$ contains all pairs of points $(x, y) \in dS \times dS$, which are causal to each other. In other words, 
	\[ 
		\Sigma \cap (dS \times dS) = \bigl\{ (x, y) \in dS \times dS \mid y \in \Gamma^+ (x) \cup \Gamma^- (x) \bigr\} \; . 
	\]
\end{remark}

\begin{proposition}
[Proposition 12 \cite{BM2}]
\label{legendre}
The two-point function \eqref{tpf-1} can be expressed in terms of 
\index{Legendre functions}
Legendre functions: for $( z_1,   z_2) \in {\mathcal T}_ + \times {\mathcal T}_-$,
	\begin{equation}
		\label{tpf-2}
	{\mathcal W}^{(2)} ( z_1 ,  z_2) =  c_\nu \, P_{s^+} \left(  \tfrac{z_1 \cdot  z_2}{r^2} \right) 	\; , \qquad m> 0 \; . 	 
	\end{equation}
The boundary values of \eqref{tpf-2} can be taken as 
$\Im z_1 \nearrow 0$ and $\Im z_2 \searrow 0$. 
\end{proposition}

\begin{remark}
\label{Wdomain}
The image of the domain ${\mathcal T}_ + \times {\mathcal T}_-$ by the mapping 
\[ ( z_1,   z_2) \mapsto \tfrac{z_1 \cdot  z_2}{r^2} \]
is\footnote{See \cite[Proposition 3]{BM2}.} the cut-plane $\mathbb{C} \setminus (- \infty, -1]$: consider the following points
	\begin{align*}
		z_1 & = (i r \sin (u_1+i {\rm v}_1), 0, r \cos (u_1 +i{\rm v}_1))\; ,  \\
		z_2 & = (-i r \sin u_2, 0, r \cos u_2) \; ,  \qquad \quad \qquad \qquad  0< u_1, u_2 < \pi \; , \; \; {\rm v}_1\in \mathbb{R} \; . 
	\end{align*}
It follows that 
	\[
		\tfrac{z_1 \cdot  z_2}{r^2} = - \cos (u_1 + u_2 + i {\rm v}_1) \;, \qquad 0<u+u_2 < 2\pi \; , \quad {\rm v}_1 \in \mathbb{R} \; . 
	\]
Thus $\mathbb{C} \setminus (- \infty, -1]$ is contained in the image. 
The fact that  $\mathbb{C} \setminus (- \infty, -1]$ equals the image follows from an argument in the ambient space, see \cite[Proposition 3]{BM2}.
The region $\mathbb{C} \setminus (- \infty, -1]$
is exactly the domain of analyticity of the \emph{Legendre function}\index{Legendre function}~$P_{s^+} $. 
\end{remark} 

\begin{proof} Let $p \in \gamma_\circ 
= \bigl\{ (1, r \cos \alpha, r \sin \alpha) \in \partial V^+ \mid  - \pi \le  \alpha \le \pi \bigr\}$; 
see \eqref{gamma-0}. Because of the invariance 
properties of ${\mathcal W}^{(2)} (  z_1,  z_2)$, it is sufficient to consider the choice
$z_1 = (-i r \cosh \beta , 0 , i r \sinh \beta)$, $z_2 = (ir,0,0)$ such that $\tfrac{z_1 \cdot  z_2}{r^2} = \cosh \beta \in \mathbb{R}^+$. 
Hence\footnote{The second line in the following formula is exactly the one given in \cite[Eq.~(4.18)]{BM}.}
	\begin{align*}
		{\mathcal W}^{(2)} ( z_1 ,  z_2) & = c_\nu  \frac{{\rm e}^{-   \pi \nu} r}{\pi} 
		\int_{\gamma_\circ} {\rm d} \mu_{\gamma_\circ } ( p )
		\;  (  z_1 \cdot  p  )^{ s^-} (  p \cdot  z_2)^{ s^+} \\
		&=  c_\nu \int_{-\pi}^\pi \frac{{\rm d} \alpha}{2 \pi}
		\;   (  \cosh \beta + \sinh \beta \sin \alpha  )^{ s^-} \; . 	 
	\end{align*}
In the second equality we have used \eqref{measure-on-circle}. Finally, recall that according to \cite[Eq.~7.4.2]{Lebedev}  
	\[
	P_{s^+} (\cosh \beta ) = \frac{1}{\pi} \int_0^\pi \frac{{\rm d} \alpha }
	{(\cosh \beta - \sinh \beta \cos \alpha)^{s^+ +1} }\; , 
	\]
and $-s^+ -1 = s^-$. 
\end{proof}

To end this subsection, we add a result on vector valued analytic continuations
into the tube domain ${\mathcal T}_-$
(this is Proposition 5.3 in \cite{BM}):

\begin{proposition}[Bros \& Moschella]
\label{prop:2.2}
The vector-valued distribution\footnote{According to  
\eqref{mass-shell-ft} the integral with a test function $f \in \mathcal{D}(\mathbb{dS})$
yields a vector ${\widetilde f}_\nu \in \widetilde{\mathsf{h}}$.}
	\[ 
		\mathbb{dS} \ni 
		x \mapsto  \left( p \mapsto (  x_+ \cdot  p )^{- \frac{1}{2} - i \nu } \right) \; , 
		\qquad p \in \gamma_\circ \; ,
	 \]
is the boundary value of the vector-valued function 
	\begin{equation}
	\label{vvf} 
		{\mathcal T}_- \ni 
		z \mapsto  \left(p \mapsto (  z \cdot  p )^{- \frac{1}{2} - i \nu } \right) 
		 \; , 
	\end{equation}
which is strongly analytic in the tuboid ${\mathcal T}_- $.
\end{proposition}

This result illustrates Theorem~\ref{flattube}.

\section{The  Euclidean one-particle Hilbert space over the sphere}
\label{sec:4.5}

Proposition \ref{legendre} allows us to analytically  continue the Wightman two-point function 
introduced in (\ref{2-point}) from the circle (where they equal $c_\nu \, P_{- \frac{1}{2} - i \nu}  
		 \bigl( - \tfrac{  \vec{{\tt x}}\cdot \vec{{\tt y}} }{r^2} \bigr)$ for 
		 $\vec{{\tt x}} , \vec{{\tt y}} \in S^1 \subset dS$) to the 
Euclidean sphere: for $f, g \in C^\infty_{\mathbb{R}}(S^2)$, 
we define the \emph{covariance} $C$ by
\label{dualitypage} 	
	\begin{equation}
	\label{h-1}
		  C(f,g) \doteq \frac{r^2}{ 2}  \int_{S^2}  {\rm d} \Omega (\vec{{\tt x}})  
		  \int_{S^2}  {\rm d} \Omega (\vec{{\tt y}}) \; 
		{ f(\vec{{\tt x}}) } \, c_\nu \, P_{- \frac{1}{2} - i \nu}  
		 \bigl( - \tfrac{  \vec{{\tt x}} \cdot \vec{{\tt y}} }{r^2} \bigr) \,   g(\vec{{\tt y}})  \; ,  
	\end{equation}
where $\vec{{\tt x}} \cdot \vec{{\tt y}} $ now denotes the scalar product of the 
vectors $\vec{{\tt x}}, \vec{{\tt y}} \in S^2 \subset \mathbb{R}^3$. As in~\eqref{cnu}, the 
constant~$c_{\nu}$  appearing in \eqref{h-1} is given by
	\[
		c_{\nu} = -\frac{1}{2\sin  ( \pi (- \tfrac{1}{2} + i \nu))} = \frac{1} {2 \cos (i \nu  \pi )}  
	\]   
and, just as in \eqref{s1} and \eqref{s2},
	\[ 		
		\nu =  
			\begin{cases}
				i \sqrt{\frac{1}{4} -\mu^2 r^2} & \text{if \ $ 0< \mu < \tfrac{1}{2r} $}   \; , \\
				 \sqrt{\mu^2 r^2 - \frac{1}{4} } & \text{if \ $ \mu \ge \tfrac{1}{2r} $}  \; .
			\end{cases}  
	\] 
Allowing complex valued functions, it turns out that the map
	\[
		f, g \mapsto C(\overline{f},g)
	\] 
extends from $C^\infty(S^2) \times
C^\infty(S^2) $ to the scalar product of the Sobolev space
$\mathbb{H}^{-1} (S^2)$:

\begin{proposition}
\label{3.9a}
Let $\mathbb{H}^{-1} (S^2)$ denote the completion of $ C^\infty(S^2)$ with
respect to the norm\footnote{For $\mu =1$, this definition coincides with the
one provided in Appendix \ref{apC}.}  
	\[
		\| f \|_{\mathbb{H}^{-1} (S^2)} \doteq \left\langle f , (- \Delta_{S^2}
			+\mu^2)^{-1} f \right\rangle_{L^2( S^2, {\rm d} \Omega)}. 
	\] 
Then the kernel of the operator $(- \Delta_{S^2} + \mu^2)^{-1}$  is given by
	\begin{align*}
		 (- \Delta_{S^2} + \mu^2)^{-1}  ( \vec{{\tt x}}, \vec{{\tt y}}) & 
		 =   \frac{r^2}{ 2}  \; c_\nu \, P_{-\frac{1}{2} - i \nu}  
		 \bigl( - \tfrac{  \vec{{\tt x}} \cdot \vec{{\tt y}} }{r^2} \bigr)    \; ;   
	\end{align*}
 \emph{i.e.}, $\langle  f , g \rangle_{ \mathbb{H}^{-1} (S^2) } 
		= C(f,g) $ for $f, g \in C^{\infty}(S^2)$.  
\end{proposition}

\goodbreak
\begin{proof} We recall that the spherical harmonics (written here in geographical coordinates)
\index{spherical harmonics} 
	\begin{equation}
	\label{spherical harmonics}	
		Y_{l,k} (\vartheta, \rho)= \sqrt{ \frac{2l +1}{4 \pi} \frac{(l-k)!}{(l+k)!} } \; 
		P_l^k (\sin \vartheta) \,  {\rm e}^{i k \rho} \; , \qquad  - \frac{\pi}{2} < \vartheta <  \frac{\pi}{2}  \; , 
	\end{equation}
are orthonormal,  \emph{i.e.},   
	\[
	\int_{S^2} {\rm d} \Omega (\vartheta, \rho) \; 
	\overline{Y_{l',k'} (\vartheta, \rho)} Y_{l,k} (\vartheta, \rho) = r^2 \delta_{l, l'} \delta_{k, k'} \; , 
	\]
and satisfy 
	\[
	 \Delta_{S^2} Y_{l,k}  = - \frac{l (l+1)}{r^2} Y_{l,k}   \;  .  
	\]
Now consider two vectors $\vec{{\tt x}} \equiv \vec{{\tt x}} (\vartheta,\rho)$ 
and $\vec{{\tt y}} \equiv \vec{{\tt y}} (\vartheta',\rho')$ 
of length $ | \vec{{\tt x}} | = | \vec{{\tt y}} | = r$. It follows that 
	\begin{align*}
		 (- \Delta_{S^2} + \mu^2)^{-1}  (\vec{{\tt x}}, \vec{{\tt y}}) & =  r^2 \sum_{l=0}^\infty \sum_{k=-l}^l 
		\frac{\overline{Y_{l,k} (\vartheta', \rho')} Y_{l,k} (\vartheta, \rho)}{l(l+1) + \mu^2r^2} 
		 \\
		& =   \frac{r^2} {4 \pi}\sum_{l=0}^\infty 
				\frac{2l+1}{l(l+1) + \mu^2r^2} P_l \bigl( \tfrac{\vec{{\tt x}} \cdot \vec{{\tt y}}}{r^2} \bigr) 
		 \\
		& =   \frac{r^2} {4 \pi}\sum_{l=0}^\infty  
				\frac{2l+1}{l(l+1) + \mu^2r^2} 
				(-1)^l   P_l \bigl( - \tfrac{  \vec{{\tt x}} \cdot \vec{{\tt y}}}{r^2} \bigr) \; . 
	\end{align*}
In the second equality we have used the addition theorem \cite[p. 395]{WaWi}: 
	\[
	P_l \bigl( \tfrac{\vec{{\tt x}} \cdot \vec{{\tt y}}}{r^2} \bigr) = \frac{4 \pi}{2l+1}	\sum_{k=-l}^l 
		\overline{Y_{l,k} (\vartheta', \rho')} Y_{l,k} (\vartheta, \rho) \; . 
	\]
In the third equality we have used $	(-1)^l  P_l (z)  =  P_l (-z)$. 

The identity\footnote{This identity extends by analyticity from $\nu \in \mathbb{R}$ to the case
$\nu =i \sqrt{\frac{1}{4} -\mu^2r^2}$,  $ 0< \mu < 1 /2r$.} 
 \cite[Eq.~(23), page 205]{Neumann}
	\begin{align*}
	\int_{-1}^1 {\rm  d} z \; P_l (z) P_{-\frac{1}{2} - i \nu} (z)  & 
	= \frac{2 \cos (i \nu  \pi ) }{\pi}  \frac{(-1)^l}{(l+\frac{1}{2})^2 +  \nu^2} 	\\
	& = \frac{2 \cos (i \nu  \pi ) }{\pi}  \frac{(-1)^l}{l(l + l) +\mu^2r^2} \, , 
	\end{align*}
together with the fact that $\sqrt{\frac{2l +1}{2}} P_l$ is an orthonormal basis in $L^2([ -1, 1])$, implies that 
	\begin{align*}
	P_{-\frac{1}{2} - i \nu} (z)  
	& = \sum_{l=0}^\infty \left( \frac{2l +1}{2}\int_{-1}^1 {\rm  d} z' \; 
	P_l (z') P_{-\frac{1}{2} - i \nu} (z') ) \right) P_l (z) \\
	& = \frac{ \cos (i \nu  \pi ) }{\pi} \sum_{l=0}^\infty (-1)^l \frac{ 2l +1 }{l(l + l) + \mu^2r^2}  P_l (z) \; .  
	\end{align*}
Thus 
	\begin{align*}
		 (- \Delta_{S^2} + \mu^2)^{-1}  ( \vec{{\tt x}}, \vec{{\tt y}}) & 
		 =   r^2  \; \frac{P_{-\frac{1}{2} - i \nu}  
		 \bigl( - \frac{  \vec{{\tt x}} \cdot \vec{{\tt y}} }{r^2} \bigr)   } {4 \cos (i \nu \pi )}  \; .  
	\end{align*}
Comparing this result with \eqref{cnu} verifies the claim.
\end{proof}

For  $\mu >0$ fixed, one defines the Sobolev 
space $\mathbb{H}^{1} (S^2)$ as the completion of $ C^\infty  (S^2)$ in the norm
\label{sobolevpage}
	\[
		\| h \|^2_{ \mathbb{H}^{1} (S^2) } 
			= \left\langle h, (- \Delta_{S^2} +\mu^2)h \right\rangle_{L^2( S^2, {\rm d} \Omega)} \;  . 
	\]
The spaces $\mathbb{H}^{\pm 1} (S^2)$ are  $\mathbb{C}$-linear Hilbert spaces, 
	\[
		 | \langle f,g \rangle_{L^2( S^2, {\rm d} \Omega)} | \le \| f \|_{ \mathbb{H}^{1} (S^2) } 
		\| g \|_{ \mathbb{H}^{-1} (S^2) } \; , 
	\]
and
	\[
		C ^\infty (S^2) 
			\subset \mathbb{H}^1(S^2) \subset  L^2   (S^2, {\rm d} \Omega) 
				\subset \mathbb{H}^{-1} (S^2)  \; . 
	\]
The inner product extends to a bilinear pairing of $\mathbb{H}^1(S^2)$ and $\mathbb{H}^{-1}(S^2)$. 
In fact, $\mathbb{H}^1(S^2)$ and $\mathbb{H}^{-1}(S^2)$ 
are dual to each other with respect to this pairing,  and the map
$f \mapsto (- \Delta_{S^2} + \mu^2)f$ is unitary from $\mathbb{H}^1 (S^2)$ to $\mathbb{H}^{-1}(S^2)$.
For a compact subset $K \subset S^2$, 
we define a closed subspace $\mathbb{H}^{-1}_{\upharpoonright K} (S^2)$ of $\mathbb{H}^{-1}(S^2)$:
	\begin{equation}
		\label{square}
		\mathbb{H}^{-1}_{\upharpoonright K} (S^2) = \{ f \in \mathbb{H}^{-1}(S^2) \mid {\rm supp\,} f \subset K \}   \; . 
	\end{equation}
Here the support of a distribution $f \in \mathbb{H}^{-1}(S^2)$ is the 
complement of the largest open set on which $f$ vanishes. 
For the open half-spheres $S_\pm$, let $\mathbb{H}^1_0(S_\pm)$ be the closure of 
$C^\infty_{0} (S_\pm)$ in~$\mathbb{H}^1 (S^2)$. Dimock \cite[Lemma 1, p.~245]{D} has 
shown that
\label{detheo}  
	\begin{align*}
		\mathbb{H}^{-1} (S^2)  &
		=  \mathbb{H}^{-1}_{\upharpoonright \overline{S_\mp}} (S^2) \oplus (- \Delta_{S^2} + \mu^2) 
		\mathbb{H}^1_0 (S_\pm ) \; , 
		\nonumber \\ 
		\mathbb{H}^{-1} (S^2) &
		=  (- \Delta_{S^2} + \mu^2) \mathbb{H}_0^{1}(S_-) \oplus 
		\mathbb{H}^{-1}_{\upharpoonright S^1} (S^2) \oplus 
			(- \Delta_{S^2} + \mu^2) \mathbb{H}^1_0 (S_+) \; . 
	\end{align*}
The following result is Lemma 2 in \cite{D}. 
\goodbreak
\begin{lemma}[Dimock's Pre-Markov property] 
\label{dlemma} 
Let $e_0$ and $e_\pm$ denote the orthogonal projections from 
$\mathbb{H}^{-1}(S^2)$ onto $\mathbb{H}^{-1}_{\upharpoonright S^1} (S^2)$
and $\mathbb{H}^{-1}_{\upharpoonright \overline{S_\mp}} (S^2)$, respectively. 
Then 
	\[
			e_\mp   e_\pm  = e_0
			\qquad \hbox{on \; $\mathbb{H}^{-1} ( S^2 )$.}
	\]
Thus $\mathbb{H}^{-1}_{\upharpoonright S^1} (S^2) =   \mathbb{H}^{-1}_{\upharpoonright \overline{S_+}} (S^2) 
\cap   \mathbb{H}^{-1}_{\upharpoonright \overline{S_-}} (S^2) $.
\end{lemma}

We note that the origins of Dimock's work can be traced back to \cite{GRS} and even further to \cite{N1, N2, N3, N4}. 

\bigskip
The Sobolev space~$\mathbb{H}^{-1} (S^2)$ contains the distribution 
	\begin{equation} 
	\label{eqDeltaTensorh}
	 (\delta \otimes h) ( \vec{x}) \doteq   r^{-1} \delta (\vartheta)  h (0, \varrho) \; , 
	 \qquad h (0, \, . \,)\in C^\infty(S^1)\; , \qquad
	  \vec{x} \equiv  \vec{x} (\vartheta, \varrho) \; , 
	\end{equation}
using geographical coordinates, 
which is supported on $S^1$. If ${\rm supp\,} h$ does not contain $(0, \pm r, 0)$,
then (\ref{eqDeltaTensorh}) equals, as an element  in $  \mathbb{H}^{-1}(S^2)$,
	\begin{equation} 
	\label{eqDeltaTensorh'}
	 (\delta \otimes h) ( \vec{x})=	\delta (\theta)  \;  \frac{ h  (0,\rho )}{r \cos \rho }\;  +
	\delta (\theta - \pi)  \;  \frac{ h(\pi,\rho)}{r \cos \rho }\; , \qquad 
			  \vec{x} \equiv  \vec{x} (\theta, \rho) \; , 
	\end{equation} 
in path-space coordinates. 

\begin{lemma}
\label{3.9}
Consider distributions of the form \eqref{eqDeltaTensorh}.  It follows that the 
time-zero covariance\index{time-zero covariance}
	\begin{align}
		\label{a-tratra} 
			\langle \delta \otimes  h_{1} \,  , \, &   \delta 
					\otimes h_{2} \rangle_{\mathbb{H}^{-1} (S^2)}  
				\nonumber \\  			
				&=  \tfrac{1}{2}
				\int_{S^1} r \, {\rm d} \varrho  \int_{S^1} r \, {\rm d} \varrho' \;  \overline{h_{1}(\varrho)}    \; c_{\nu} \, 
				P_{-\frac{1}{2} + i \nu} (- \cos (\varrho - \varrho')) \, h_{2} (\varrho') 
	\end{align}
exists as a positive quadratic form on $C^\infty (S^1)$, which is invariant under  rotations around the  axis connecting the geographical poles. 
\end{lemma}

\begin{proof}
Recall Proposition \ref{3.9a}.
This allows us to compute
	\begin{align*}
		 & \left\langle \delta \otimes h_1, (- \Delta_{S^2} +\mu^2)^{-1}  \delta \otimes h_{2} 
		\right\rangle_{L^2( S^2, {\rm d} \Omega)}  
		 \\
		& \qquad = \tfrac{1} {2 } \, c_\nu \int_{S^1} r {\rm d} \rho'  \int_{S^1}r  {\rm d} \rho'  \; \overline{h_{1}(\rho')} 
		P_{-\frac{1}{2} + i \nu}  ( - \cos (\rho-\rho') )  		h_{2} (\rho)  \; .  
	\end{align*}
\goodbreak
We have used that for $\vec{x}= (0, \sin \rho, \cos \rho)$ and $\vec{y}= (0, \sin \rho', \cos \rho')$  
	\[
		\tfrac{\vec{x} \cdot \vec{y}}{r^2} = \cos ( \rho'-\rho) \; , 
	\]
as $\cos  \rho  \cos \rho' + \sin  \rho \sin \rho' = \cos ( \rho'-\rho)$. 
\end{proof}

\goodbreak

We complement this result with a similar one, which uses path-space coordinates. 

\begin{lemma}
\label{lm:4.5.4}
Consider distributions of the form \eqref{eqDeltaTensorh'}, 
with $\operatorname{supp} h_i \subset I_+$, 
$i =1,2$.  It follows that the 
time-zero covariance satisfies
	\begin{align}
		\label{tratra} 
			\langle \delta_{\theta_1} \otimes  h_{1} \,  , \, &   \delta_{\theta_2} 
					\otimes h_{2} \rangle_{\mathbb{H}^{-1} (S^2)}  
				\nonumber
				 \\  			
				&\qquad =  r \, \Bigl\langle   \cos_\psi  h_{1} ,  \frac{{\rm e}^{-   |\theta_{2}
						-\theta_{1}|  |\varepsilon  | }
					+ {\rm e}^{- (2 \pi-|\theta_{2}-\theta_{1}|) | \varepsilon |   }}{
						2 | \varepsilon | (1-{\rm e}^{- 2 \pi  | \varepsilon | })} 
					\cos_\psi  h_{2} \Bigr\rangle_{L^{2}(I_+ , 
					\frac{r {\rm d} \psi} {\cos \psi})} \, , 
	\end{align}
with $\varepsilon^2  =-  (\cos \psi \, \partial_\psi)^2 + \mu^2r^2 \cos^2 \psi $.  
\end{lemma}

\begin{proof}  
An approximation of the Dirac $\delta_{\theta'}$-function is given 
by $\delta_{k}$, $k\in {\mathbb N}$, with
	\begin{equation}
		\label{deltaka-new}
		\delta_{k}( \, . \, - \theta'  ) 
		=(2 \pi )^{-1}\sum_{|\ell|\leq k}{\rm e}^{i  \ell (  \, . \, - \theta') } 
		\chi_{[0,2\pi)} (\, . \, ) \; , 
		\qquad 		\theta \in [0, 2 \pi )  \; ,
	\end{equation}
and $\chi_{[0,2\pi)}$  the characteristic function of the interval $ [0, 2 \pi ) \subset {\mathbb R}$. 
Use 
	\[
		 (- \Delta
		 	+\mu^2)^{-1}
		 	=   \left( - \partial^2_\theta + \varepsilon^2 \right)^{-1} r^2 \cos^{2} \psi   
	\] 
and
	\[
		 \sum_{\ell' \in {\mathbb Z}} \int_0^{2 \pi} \frac{{\rm d} \theta} {2 \pi }  
		 \; {\rm e}^{i  \ell (\theta - \theta_1) } 
		 \left( - \partial^2_\theta + \varepsilon^2 \right)^{-1} 
		 {\rm e}^{i  \ell' (\theta - \theta_2)} 
		 =  \frac{{\rm e}^{i  \ell (\theta_{2}- \theta_{1})}}{
                   {\ell^2} + \varepsilon^2}  
	\]
to show that for $0\leq \theta_{1} , \theta_{2}< 2 \pi $ 
	\begin{align*}
			&   
			\lim_{k \to \infty} \lim_{k' \to \infty} \Bigl\langle  \delta_{k'}(.-\theta_{1})\otimes  h_{1}, 
				\bigl(-\Delta_{S^2}+m^2\bigr)^{-1} \delta_{k}(.-\theta_{2})\otimes
				 h_{2} \Bigr\rangle_{L^2 (S^2,  {\rm d} \Omega) } \nonumber \\ 
			& \quad  = \lim_{k \to \infty} (2 \pi)^{-1}\sum_{|\ell|\leq k} 
				\Bigl\langle   \cos_\psi^{-1}\, h_{1} , 
				\frac{r {\rm e}^{i  \ell (\theta_{2}- \theta_{1})}}{  {\ell^2} 
			+ \varepsilon^2} \, 
					\cos_\psi  \, h_{2}   \Bigr\rangle_{L^{2}(I_+,  r \cos \psi \, {\rm d} \psi)}
				\nonumber \\ 
			& \quad  = \lim_{k \to \infty} (2 \pi)^{-1}\sum_{|\ell|\leq k} 
					\Bigl\langle  \cos_\psi  h_{1}  ,  
					\frac{ r {\rm e}^{i  \ell (\theta_{2}- \theta_{1})}}{  {\ell^2} 
						+ \varepsilon^2} \,  \cos_\psi h_{2} 
						\Bigr\rangle_{L^{2}(I_+,  \cos \psi^{-1} \,r  {\rm d} \psi)} \; .
		\end{align*}
Next, let us recall 
the definition of $\varepsilon = \varepsilon_{\upharpoonright I_+}
+\varepsilon_{\upharpoonright I_-}$
from \eqref{vaepsdef}. 
On the half\-circle~$I_+ $ the operator $\varepsilon$ has,  
just like $\varepsilon^2$, purely a.c.~spectrum on all of~${\mathbb R}^+$. 
Especially, $\varepsilon$ does not have a discrete eigenvalue at zero.
Thus, despite the fact that $\varepsilon$ has no mass gap,  one can apply the Poisson 
sum formula (see, \emph{e.g.},~\cite{KL2}) 
	\[
		\frac{1}{2 \pi  } \sum_{\ell\in {\mathbb Z}}\frac{{\rm e}^{i \ell  \theta}}{ \ell^{2}+
			\varepsilon^{2}}= \frac{{\rm e}^{-| \theta | |\varepsilon |}+
			{\rm e}^{-(2 \pi -| \theta |)|\varepsilon|}}{ 2  |\varepsilon|  (
			1-{\rm e}^{- 2 \pi  |\varepsilon|  })}  \: \: \hbox{ for } \: \: 0\leq |  \theta |< 2 \pi \;  
	\]
to conclude that \eqref{tratra} holds.
\end{proof}

\goodbreak

Note that for $\theta=\pi$, we have 
	\begin{equation} 
		\label{eqDeltaPi}
		\delta_{\pi}\otimes  h = 	\delta \otimes {P_1}_* h \; , 
	\end{equation} 
where ${P_1}_*$ is the pull-back of the reflection at the
$x_0$-$x_1$ plane, \emph{i.e.},
	\[ 
		({P_1}_* h)(\psi) \doteq h(\pi-\psi) \; .
	\] 

\goodbreak
\begin{corollary}  
\label{cor:4.5.5}
Consider distributions of the form \eqref{eqDeltaTensorh'}. Then 
	\begin{align}
		\label{corwichtig}
		\|  \delta  \otimes  h \|_{\mathbb{H}^{-1} (S^2)}  
		= r  \bigl\langle |\cos_\psi| \,  h , \big( \tfrac{\coth  \pi |\varepsilon|}
			{2  |\varepsilon|   }
   		+
		\tfrac{	{P_1}_*}{2 \varepsilon \sinh \pi \varepsilon  }\big)\,  
		|\cos_\psi|\,  h   \bigr\rangle_{L^{2}(S^1 , 
                          \tfrac{r {\rm d} \psi}{|\cos \psi|}) }.  
	\end{align}
Note that there is no restriction on the support of the function $h$.
\end{corollary}

\begin{proof}
Write $h=h_++h_-$, where the support of $h_\pm$ is contained in
$I_\pm$, respectively. By Lemma~\ref{lm:4.5.4}, 
	\begin{align*}
		\bigl\langle \delta\otimes   h_+ ,\delta  \otimes h_+ \bigr\rangle_{\mathbb{H}^{-1} (S^2)}
		&= r \bigl\langle \cos_\psi  h_+   ,  \tfrac{\coth   \pi  |\varepsilon| }
			{2  |\varepsilon|   }  
			\cos_\psi h_+   \bigr\rangle_{L^{2}(S^1 ,
                          \frac{r {\rm d} \psi}{|\cos \psi|}) }.  
	\end{align*}
Now note that 
	\[
		\bigl\langle \delta\otimes   h_- ,\delta  \otimes h_- \bigr\rangle_{\mathbb{H}^{-1} (S^2)} =  
		\langle \delta \otimes {P_1}_* h_- ,\delta \otimes {P_1}_* h_-\rangle_{\mathbb{H}^{-1} (S^2)}
	\] 
and again apply Lemma~\ref{lm:4.5.4} to find a similar expression. 
For the mixed term, use Eq.~\eqref{eqDeltaPi} to write 
	\begin{align*}
		 \bigl\langle \delta\otimes   h_+ ,\delta  \otimes h_- \bigr\rangle_{\mathbb{H}^{-1} (S^2)}  
		&= \bigl\langle \delta_0\otimes h_+ ,\delta_\pi  
		\otimes {P_1}_*h_- \bigr\rangle_{\mathbb{H}^{-1} (S^2)}  
		\nonumber \\
		&= 
				r \bigl\langle \cos_\psi h_+ ,  \tfrac{ {\rm e}^{- \pi  |\varepsilon|}}
					{ |\varepsilon| (1- {\rm e}^{-2\pi  |\varepsilon|  })}  
					\cos_\psi {P_1}_*h_-   \bigr\rangle_{L^{2}(I_+ ,
                         			 \frac{r {\rm d} \psi}{|\cos \psi|}) }  \nonumber \\
		&= 
				r \bigl\langle \cos_\psi  h_+ ,  \tfrac{1}{2  |\varepsilon| \sinh \pi 
 					|\varepsilon|}\cos_\psi {P_1}_*h_-   \bigr\rangle_{L^{2}(I_+ ,
                          			\frac{r {\rm d} \psi}{|\cos \psi|}) } \; .  
	\end{align*}
The term $\bigl\langle \delta\otimes   h_- ,\delta  \otimes h_+ 
\bigr\rangle_{\mathbb{H}^{-1} (S^2)}$
yields a similar expression, with $h_+$ and $h_-$ interchanged. 
We can now put the four terms together. The proof of Eq.~\eqref{corwichtig}
is completed by  noting that $\varepsilon$ leaves
the two subspaces $L^{2}(I_\pm,\frac{r{\rm d} \psi}{|\cos \psi|})$ invariant. 
\end{proof}

The next lemma applies the projection $e_0$ to the characteristic function 
$\chi_{\overline{S^+}}$ of the upper hemisphere. In \eqref{bedingter-ew} 
the interacting vacuum vector  
is expressed by a conditional expectation of a 
Euclidean vacuum vector. As the conditional expectation respects the $n$-particle 
subspaces in Fock space and the only rotation invariant one-particle function 
on the sphere is the constant wave-function, the one-particle contribution to the 
interacting vacuum vector can be computed as follows. 

\begin{lemma} Let $h \in C^\infty (S^1)$ and let $\chi_{\overline{S_+}}$ denote the characteristic function of the closed upper hemisphere. Then 
\label{kappa}
	\begin{align*}
		\langle e_+ \chi_{\overline{S_+}} \; , &  e_- (\delta \otimes h) 
		\rangle_{\mathbb{H}^{-1} (S_2)} 
		=  \int_{\overline{S^+}} {\rm d} \Omega(\vec{x}) 
		\int_{\overline{S_-}} {\rm d} \Omega(\vec{y}) \, (\delta \otimes h) (\vec{x}) \, 
		P_{- \frac{1}{2} - i \nu} \bigl( - \tfrac{\vec{x} \cdot \vec{y}}{r^2} \bigr) \\
				& \qquad 
		\\
		& =  \underbrace{ \frac{ \sqrt{\pi} \, r}{ | \Gamma ( \frac{3}{4} + i \frac{\nu}{2} ) |^4 } 
		\frac{ (2\pi)^2}{  | \Gamma ( \frac{1}{4} + i \frac{\nu}{2} ) |^2 } }_{:= \kappa_0} \; 
		\\
		& 
		\qquad \qquad \qquad \qquad \times c_\nu
			\int_{S^1} r {\rm d} \psi \int_{S^1} r {\rm d} \psi'  \; h (\psi) \,  
			P_{-\frac{1}{2} - i \nu} (- \cos (\psi - \psi') )  \; . 
	\end{align*}
\end{lemma}

\begin{proof} 
The first identity follows from Proposition \ref{3.9a}.
For the second identity, we use geographical coordinates. As
	\[
		\tfrac{\vec{x} \cdot \vec{y}}{r^2} = \sin \theta \sin \theta' + \cos \theta \cos' \theta \cos (\psi - \psi')  \; , 
	\]
we find 
	\begin{align*}
		& \int_{\overline{S_+}} {\rm d} \Omega(\vec{x}) 
		\int_{\overline{S_-}} {\rm d} \Omega(\vec{y}) \,  
		(\delta \otimes h) (\vec{x}) \, 
		P_{- \frac{1}{2} - i \nu} ( - \tfrac{\vec{x} \cdot \vec{y}}{r^2} ) \\
		& \qquad = r^3
			\int^{\frac{\pi}{2}}_0  \cos \theta' {\rm d} \theta'
			\int_{S^1}  {\rm d} \psi \int_{S^1}  {\rm d} \psi' \; 
		h (\psi) \, P_{- \frac{1}{2} - i \nu} \bigl( -  \cos \theta' \cos (\psi - \psi') \bigr) \; . 
\end{align*}
A special case of \eqref{eq:addition-formula} is the following formula.
	\begin{align}
		P_s  \bigl( - & \cos(\psi-\psi') \cos \theta' \bigr)
		\label{crucial}
		 \\
		&	= \frac{ \sqrt{\pi} } { \Gamma {(\frac{1}{2} - \frac{s}{2})}  
		\Gamma {(\frac{s}{2} +1)} } 
		P_s(\sin(\psi-\psi')) \nonumber
		\\
		& \qquad + 2 \sum_{k=1}^\infty (-1)^k \frac{\Gamma(s -k + 1)}{\Gamma(s+k+1)} 
 				\cos(k \theta') \; P_s^k (0) P_s^k (\sin(\psi-\psi')) \;.
		\nonumber
	\end{align}
We have used 
$P_{s}(0) = \frac{ \sqrt{\pi} } { \Gamma {(\frac{1}{2} - \frac{s}{2})}  \Gamma {(\frac{s}{2} +1)} } 
= \frac{ \sqrt{\pi} } { \Gamma {(\frac{3}{4} + i \frac{\nu}{2})}  \Gamma {(\frac{3}{4} - i \frac{\nu}{2})} }$.
Next recall that 
	\[
		\Gamma(2z)= \frac{2^{2z-1}}{\sqrt{\pi}}\; \Gamma(z)\Gamma\left(z+\tfrac{1}{2}\right) \; . 
	\]
Thus
	\begin{align*}
		c_\nu  & = \frac{\Gamma\left(\frac{1}{2} - i \nu  \right) \Gamma\left(\frac{1}{2} + i \nu \right)}{2\pi}  \\
		& = \frac{\ \Gamma(\frac{1}{4} - i \frac{\nu}{2})
		\Gamma\left(\frac{3}{4} - i \frac{\nu}{2}\right)
		 \Gamma(\frac{1}{4} + i \frac{\nu}{2})
		\Gamma\left(\frac{3}{4} + i \frac{\nu}{2}\right) }{(2\pi)^2} 
		\; . 
	\end{align*}
When integrating out the $\theta'$ variable, only the first term on the r.h.s.~contributes. 
By shifting the integration in the $\psi'$ variable, the identity follows. 
\end{proof}

\begin{remark}
The integral over $\psi'$  can be computed using the formula 
	\[
		\lambda P_\lambda (x) =   x P_\lambda' (x) - P_{\lambda -1}' (x) \; , 
	\] 
which implies that 
\begin{align*}
	\lambda \int_0^1 {\rm d} x \; P_\lambda (x) & = \int_0^1 {\rm d} x \; x P_\lambda' (x) - P_{\lambda -1} (1) + P_{\lambda -1} (0)
	\\
	& = P_\lambda (1) - \int_0^1 {\rm d} x \; P_\lambda (x) - P_{\lambda -1} (1) + P_{\lambda -1} (0) \; . 
\end{align*}
Note that $P_\lambda (1) = P_{\lambda -1}(1) = 1$ and 
$
P_{\lambda}(0) = \frac{ \sqrt{\pi} } { \Gamma {(\frac{1}{2} - \frac{\lambda}{2})}  \Gamma {(\frac{\lambda}{2} +1)} } 
$. Thus 
	\[
		\int_{0}^1 {\rm d} u  \;  P_s(u) = \frac{ \sqrt{\pi} }{ (1+s) \Gamma ( 1- \frac{s}{2} ) \Gamma ( \frac{s}{2} + \frac{1}{2} ) } \; . 
	\]
\end{remark}

\goodbreak
\section{Unitary irreducible representations on the time-zero circle}
\label{Sect: canon-HS}

We now define a Hilbert space for functions supported on the time-zero circle~$S^1$.
As we have seen, the two-point function ${\mathcal W}^{(2)} (  x,  y)$ is analytic 
for~$x$ space-like to $y$. For $x= (0, r \sin \psi, r \cos \psi)$ and $y= (0, r \sin \psi' , 
r \cos \psi' )$, the (minimal) spatial distance $d$ define in \eqref{dlength} is given by
	\[
		d( x ,  y ) = 
		r \arccos \bigl( - \tfrac{x \cdot y }{r^2} )= | \psi'-\psi | \,  r \; . 
	\]
Thus ${\mathcal W}^{(2)} (  x,  y) =  c_\nu \, P_{s^+}  ( - \cos ( \psi'-\psi)) $; 
see \eqref{tpf-2}. 	 
Note that the singularity at $\psi = \psi' $ is integrable. 
This suggest the following definition. 

\begin{definition}
\label{zeit-null-Hilbert}
The completion of $C^\infty (S^1)$ with respect to the scalar product
	\begin{equation}
		\langle h , h' \rangle_{\widehat{{\mathfrak h}} (S^1)}  = c_\nu
				\int_{S^1} r \, {\rm d}\psi \int_{S^1} r \, {\rm d}\psi' \; \overline{h(\psi)} \, 
				P_{s^+}\big( - \cos(\psi' - \psi) \big) \, h'(\psi') \;.
				\label{eq:def-scalar-product-2}
	\end{equation}
is a Hilbert space, which we denote by $\widehat {{\mathfrak h}} (S^1)$.
As before, $s^+$ is given by \eqref{dd1}.
\end{definition}

\label{chfdpage}
\begin{remark}
\label{rm:4.7.2}
By construction,  
	\[
		\langle h , h' \rangle_{\widehat{{\mathfrak h}} (S^1)}  = 2 \langle \delta \otimes h , \delta \otimes h' 
		\rangle_{\mathbb{H}^{-1} (S^2) }  \;.
	\]
Thus $\widehat{{\mathfrak h}} (S^1) \cong \mathbb{H}^{-1}_{\upharpoonright S^1} (S^2)$; the two Hilbert spaces 
will be identified in the sequel. 
\end{remark}

\label{hs1-page}

\goodbreak 

\begin{proposition} 
\label{thhm}
The scalar product \eqref{eq:def-scalar-product-2} can be 
expressed as
	\[
		\langle h , h' \rangle_{\widehat{{\mathfrak h}} (S^1)}  = 
		\bigl\langle h  , \tfrac{1}{2\omega} 
		h' \bigr\rangle_{L^2( S^1, r {\rm d} \psi)} \; ,  
	\]
with $\omega $ a strictly positive self-adjoint operator on $L^2( S^1, r {\rm d} \psi)$ with Fourier coefficients 
	\begin{equation}
	\label{eq:omega-1}
		\widetilde {\omega}(k) =  \widetilde {\omega}(- k) =  \frac{k+s^+}{r} \, 
					 \frac{\Gamma \left( \frac{k+s^+}{2} \right)}{ \Gamma \left( \frac{k-s^+}{2} \right)}
			\frac{ \Gamma \left( \frac{k+1-s^+}{2} \right) }{ \Gamma \left( \frac{k+1+s^+}{2} \right)} >0  \; ,
	\end{equation}
for all $k  \in \mathbb{Z}$. 
\end{proposition}

\begin{remark}
\label{rm:4.6.4}
As we shall see in
(\ref{eq:tilde-omega-and-the-traditional-dispersion-relation}), 
for both the principal and complementary series one has
$\frac{1}{2} \big( \widetilde\omega(k) \widetilde\omega(k+1) +\widetilde\omega(k) \widetilde\omega(k-1)  \big)
= \frac{k^2}{{r}^2}  + \mu^2$
and, hence, we conclude that $\widetilde\omega(k)$ behaves 
for large $|k|$ as 
	\[ 
		\omega (k) \sim \sqrt{\frac{k^2}{{r}^2}  + \mu^2} \; , \qquad 1 \ll k \; , 
	\]
approaching
thus the well-known \emph{dispersion relation} of the Minkowski space-time.  
\end{remark} 

\goodbreak
\begin{proof} 
Set
	\begin{equation}
		P_{s^+}\big( - \cos(\psi' - \psi) \big) =  \sum_{k\in\mathbb{Z}} p_k \frac{{\rm e}^{ik (\psi' - \psi)}}{ \sqrt{2\pi r} } \; . 
			\label{eq:Fourrier-Pmu}
	\end{equation}
This yields
	\begin{align}
	 \langle h,  h'   \rangle_{\widehat{{\mathfrak h}} (S^1)} 
		& =  \sqrt{2\pi r} \, c_\nu  \sum_{k\in\mathbb{Z}} p_k 
				\overline{ \left(   \int_{S^1} r\,  {\rm d}\psi \, h(\psi)  \frac{{\rm e}^{-ik \psi}}{\sqrt{2\pi r} } \right) }
						\left( \int_{S^1}  r\, {\rm d}\psi' \,  h'(\psi')  \frac{{\rm e}^{-ik \psi'}}{\sqrt{2\pi r} } \right)
			\nonumber \\
		& =  \sqrt{2\pi r} \, c_\nu  \sum_{k\in\mathbb{Z}} p_k \, \overline{h_k} \, h'_k \;,
		\label{eq:Jhoitrytfsd}
	\end{align}
where $h_k$ and $h'_k$ are the Fourier coefficients of $h$ and $h'$, respectively. 
Comparing \eqref{eq:def-scalar-product-2}
with \eqref{eq:Jhoitrytfsd} we see that 
\begin{itemize}
\item [$a.)$]
$\omega$ is a diagonal operator w.r.t.~the orthonormal basis 
$\bigl\{e_k \in L^{2}(S^1, \, r {\rm d} \psi) 
	\mid e_k (\psi)=\frac{{\rm e}^{-ik \psi}}{\sqrt{2\pi r}} \; , k\in \mathbb{Z} \bigr\} $.
\item [$b.)$]
the Fourier coefficients 
$\widetilde \omega(k) \doteq \left\langle e_{k}, \; 
\omega e_k \right\rangle_{L^{2}(S^1, \, r {\rm d}\psi)} $ of $\omega$ are given by
	\begin{equation}
		\widetilde \omega(k) =  -\frac{2\sin(\pi s^+)}{ \sqrt{2\pi r} } \; \frac{1 }{r \, p_k} \;,
		\qquad k\in\mathbb{Z} \; . 
		\label{eq:definition-omegak-0}
	\end{equation}
\end{itemize}
Using Proposition \ref{w-coefficients}, we arrive at  \eqref{eq:omega-1}. 
In (\ref{eq:symmetry-of-tildeomega}) we will establish that
$\widetilde \omega(k)=\widetilde \omega(-k)$ for all $k\in\mathbb{Z}$.  
For the case of the principal series, one has
$s^\pm=-\frac{1}{2}\mp i\nu$, with~$\nu \in\mathbb{R}^+_0$, 
and\footnote{Here 
\eqref{eq:gamma-1} and $\eqref{eq:gamma-5}$ refer the to 
the identities stated at the beginning of Appendix C.}
	\[ 
		\Gamma\left(\tfrac{k+\frac{3}{2}-i\nu }{2}\right)
		\stackrel{(\ref{eq:gamma-1})}{=} 
		\tfrac{k-\frac{1}{2}-i\nu }{2}\Gamma\left(\tfrac{k-\frac{1}{2}-i\nu }{2}\right) 
	\]
implies, from (\ref{eq:omega-1}),
	\begin{align}
          \label{eq:def-tilde-omega}
		\widetilde \omega(k) & 
				 \stackrel{(\ref{eq:gamma-5})}{=} \; {r}^{-1}
						\left( \tfrac{ (k-1/2)^2 + \nu^2 }{4}\right)
				\frac{\bigl|\Gamma \bigl(\frac{k-\frac{1}{2}+i\nu}{2}\bigr) \bigr|^2}{
				\bigl|\Gamma \bigl( \frac{k+\frac{1}{2}+i\nu}{2} \bigr)\bigr|^2}
				\;,
	\end{align}
showing that $\widetilde \omega(k)>0$ for all $k \in \mathbb{Z}$. This positivity property 
also holds in the case of the complementary
series. There one has $\nu=i\sqrt{\frac{1}{4}-\zeta^2}$, with $0< \zeta^2 \leq 1/4$. 
Thus,  $-1/2<s^+\leq 0$.
Since $\widetilde \omega(k)=\widetilde \omega(-k)$ for all $k\in\mathbb{Z}$, it is enough to 
consider $k\geq 0$. We will show, see (\ref{eq:useful-1}), that $\widetilde\omega(k) 
\widetilde\omega(k+1) = r^{-2} k (k+1) + \mu^2 >0$. Hence, $\widetilde\omega(k+1)$ and
$\widetilde\omega(k)$ have the same sign and, therefore, in order to prove that 
$\widetilde\omega(k)>0$ for all~$k\geq 0$, it is 
enough to establish that $\omega(0)>0$. But, from (\ref{eq:omega-1}), one has
	\[
			\widetilde {\omega}(0) = {r}^{-1} \, s^+
					 \frac{\Gamma \left( \frac{s^+}{2} \right)}{ \Gamma \left( \frac{-s^+}{2} \right)}
			\frac{ \Gamma \left( \frac{1-s^+}{2} \right) }{ \Gamma \left( \frac{1+s^+}{2} \right)} 
			\; > \; 0 \;,
	\]
as $s^+<0$, and since $\Gamma(x)>0$ for all $x>0$ and $\Gamma(x)<0$ for all $x\in (-1, \; 0)$ 
(one has $\frac{s^+}{2} \in (-1/4, \; 0]$,  but $\frac{1-s^+}{2}$, $\frac{-s^+}{2} $ and $ \frac{1+s^+}{2}$ are all positive).
\end{proof}

\label{KShatpage}
\label{LShatpage}

We now turn to the action of $SO(1,2)$ on $\widehat {{\mathfrak h}} (S^1)$. 

\begin{theorem} 
\label{UIR-S1}
The rotations 
	\[
		\bigr(\widehat{u} (R_0(\alpha)) h \bigl) (\psi) 
		= h (\psi - \alpha) \; , \qquad \alpha \in [0, 2\pi) \; , \quad h \in \widehat{{\mathfrak h}}  (S^1) \; , 
	\]
and the boosts\footnote{Recall that $\mathbb{cos}_\psi$ denotes the multiplication 
operator \index{multiplication operator}
	$
		(\mathbb{cos}_\psi h)(\psi) \doteq \cos \psi  \, h(\psi)
	$
acting on functions of $\psi$.}
\label{Umhatpage}
\begin{equation} \label{eqUmhat}
			\widehat{u} (\Lambda_1(t)) 
			= {\rm e}^{i t \omega  r \, \mathbb{cos}_\psi  } \; , \qquad t \in \mathbb{R} \; , 
\end{equation}
generate a unitary representation of $SO_0(1,2)$ on $\widehat{{\mathfrak h}}  (S^1)$. 
\end{theorem}

\begin{proof}
The generator of the rotations  $ \KS_0 =- i \partial_\psi$ has purely discrete spectrum.  
Its normalized eigenfunctions are 
	\[
		\mathfrak{f}_k  (\psi)  
		= \sqrt{\frac{ \widetilde {\omega} (k)}{r \pi }}  \;  {\rm e}^{ ik\psi}
		\; ,  \qquad k \in \mathbb{Z}  \; .
	\]
The generators of the boosts,
	\begin{equation}
	\label{L1-L2}
		\LS_1 = \omega r \,  \mathbb{cos}_\psi  \; , \qquad 
		\LS_2 = \omega r \, \mathbb{sin}_\psi \; , 
	\end{equation}
satisfy the commutator relations 
	\[
		[ \KS_0 \, , \, \LS_1 ] =   i \LS_2 \; , 
		\qquad
		[ \LS_2\, , \, \KS_0] =  i \LS_1 \;  . 
	\]
The latter follow from
	\[
		[ - i \partial_\psi \, , \,  \omega r \,  \mathbb{cos}_\psi  ] = i \omega r \,\mathbb{sin}_\psi \; , 
		\qquad 
		[ \omega r \, \mathbb{sin}_\psi \, , \, - i \partial_\psi ] = i \omega r \,  \mathbb{cos}_\psi  \; . 
	\]
It remains to verify the remaining 
commutation relation $[ \LS_1 \, , \, \LS_2] = - i \KS_0 $. 
In order to do so, we consider the ladder operators
	\begin{equation}
	\label{ladderoperators}
		\LS_\pm = \LS_1  \pm i \LS_2  = \omega r \, {\rm e}^{ \pm i \psi }\; . 
	\end{equation}
We will show that 
	\[
		\langle  \mathfrak{f}_{k'} ,  
		\big[ \LS_+ ,  \LS_- \big] \mathfrak{f}_k  \rangle_{\widehat{{\mathfrak h}}  (S^1)}
		 = \ -2 \langle \mathfrak{f}_{k'} ,  \KS_0  \mathfrak{f}_k 
		 \rangle_{\widehat{{\mathfrak h}}  (S^1)} \; .
	\]
The latter is equivalent to  
	\begin{equation}
		\label{eq:venoirgtuuwp}
		\widetilde \omega(k)\big(\widetilde \omega(k-1) - \widetilde \omega(k+1) \big) 
		=  - \frac{2k} { {r}^2} \;, \qquad \forall k \in \mathbb{Z} \;.
	\end{equation}
In order to verify this identity, let us first consider only the $\Gamma$-factors occurring 
in~\eqref{eq:omega-1}. Define, for $k\in\mathbb{Z}$,
	\[ 
		w (k) \doteq  \frac{ \Gamma\left(\frac{k+s^+}{2}\right) }{
      							\Gamma\left( \frac{k-s^+}{2} \right)}
					\frac{ \Gamma\left(\frac{k-s^+ +1}{2}\right)
						}{\Gamma\left( \frac{k+s^+ +1}{2} \right) }\;.
	\]
One has
	\begin{align*}
		w (k+1) & =  
					\frac{ \Gamma\left(\frac{k+s^+ +1}{2}\right) }{
      							\Gamma\left( \frac{k-s^+ +1}{2} \right) }
					\frac{ \Gamma\left(\frac{k-s^+ +2}{2}\right)
						}{\Gamma\left( \frac{k+s^+ +2}{2} \right)}  
				 =  
				 	\frac{ \Gamma\left(\frac{k+s^+ +1}{2}\right) }{
     						 \Gamma\left( \frac{k-s^+ +1}{2} \right) }
					\frac{ \Gamma\left(\frac{k-s^+}{2}+1\right)
						}{\Gamma\left( \frac{k+s^+}{2} +1\right)}   \\
				& 
				=   
					\frac{k-s^+}{k+s^+} \;  \frac{ \Gamma\left( \frac{k+s^+ +1}{2} \right) }{
      										\Gamma\left( \frac{k-s^+ +1}{2} \right) }
									\frac{ \Gamma\left(\frac{k-s^+}{2}\right)
										}{\Gamma\left( \frac{k+s^+}{2} \right) }  
				= \; \frac{k-s^+}{k+s^+} \; \frac{1}{w (k)} \:,
	\end{align*}
as one easily verifies. Hence, 
	\begin{equation}
		w(k)w(k+1)=\frac{k-s^+}{k+s^+} \;  .
		\label{eq:JUytiuytuiu}
	\end{equation}
Since $\widetilde \omega(k)  =  \frac{{ (k +s^+) } }{r} \, w (k)$, we have
	\[ \widetilde \omega(k) \widetilde \omega(k+1) 
	= {r}^{-2} (k +s^+) (k +s^+ +1)  \frac{k-s^+}{k+s^+}
	= {r}^{-2} (k-s^+)(k +s^+ +1)  \;.
	\]
Thus, we get the two following useful relations:
	\begin{align}
		\label{eq:useful-1}
		\widetilde \omega(k) \widetilde \omega(k+1) & = {r}^{-2} (k-s^+)(k +s^+ +1) = r^{-2} k (k+1) + \mu^2  \; , \\
		\label{eq:useful-2}
		\widetilde \omega(k) \widetilde \omega(k-1) & = {r}^{-2} (k +s^+)(k-s^+-1) 
		= r^{-2} k (k-1) + \mu^2\;.
	\end{align}
The last one is obtained from the previous one by taking $k\to k-1$.
We note that 
	\begin{equation}
		           \label{eq:tilde-omega-and-the-traditional-dispersion-relation}
	\frac{1}{2} \Bigl( 
				\widetilde \omega(k) \widetilde \omega(k+1) +	\widetilde \omega(k) \widetilde \omega(k-1)  \Bigr) = \frac{k^2}{{r}^2}  + \mu^2 \; , 
	\end{equation}
which allows us to establish the usual dispersion relation in the limit $r \to \infty$. 

Relation~(\ref{eq:venoirgtuuwp}) can now be verified using \eqref{eq:useful-1}--\eqref{eq:useful-2}:
	\begin{align*}
		\widetilde \omega(k)\widetilde \omega(k-1) -\widetilde \omega(k)\widetilde \omega(k+1) & = {r}^{-2}
		(k +s^+)(k-s^+-1) \\
		& \qquad - {r}^{-2} (k-s^+)(k +s^+ +1)  =  - \frac{2k}{ {r}^{2}} \;,
	\end{align*}
as desired. We conclude that 
	\[
		[ \LS_+ , \LS_-] = - 2 \KS_0 \; , 
		\qquad [ {\mathfrak k}_0 , \LS_\pm ] = \pm \LS_\pm \; ,
	\]
in agreement with $[ \LS_1 \, , \, \LS_2] = - i \KS_0 $. 
We have thus arrived at a representation of the 
Lie algebra $\mathfrak{so}(1,2)$. Now, as  
$\widetilde {\omega}(k) \sim k $ for $k \to \infty$ and
	\[
		\LS_1 = \tfrac{1}{2} \bigl( \LS_+ + \LS_- \bigr) \; , 
		\qquad 
		\LS_2 = \tfrac{1}{2i} \bigl( \LS_+ - \LS_- \bigr) \; ,
	\]	
the formulas 
\begin{align*}
	 	\KS_0 \mathfrak{f}_k & = k 
		\mathfrak{f}_k \; , 
		\\
		\LS_+ \mathfrak{f}_k & 
		= \sqrt{ \widetilde {\omega}(k)\widetilde {\omega}(k+1)} \, 
		\mathfrak{f}_{k+1} \; , 
		\\
		\LS_- \mathfrak{f}_k & 
		=	\sqrt{\widetilde {\omega}(k)\widetilde {\omega}(k-1)  } \, 
		\mathfrak{f}_{k-1}  \; , 
	\end{align*}
imply that the eigenvectors $\mathfrak{f}_k$ of the 
angular momentum operator $\KS_0$ form a dense set of analytic vectors in separate 
real coordinate directions of the real Lie group $SO_0(1,2)$. According to 
\cite[Theorem 1]{FSSS}, this implies that the Lie algebra is integrable to a unique unitary 
group representation of $SO_0(1,2)$.  	
\end{proof}

\begin{lemma}
\label{lm:L2-invariance}
Let $\delta_n$, given by 
	\[
		\delta_n (\psi)= 
		\frac{1}{2 \pi r } \sum_{k = - n}^n  {\rm e}^{i k \psi}    
		= 
		\frac{1}{\sqrt{4 \pi r }} \sum_{k = - n}^n  		
		\frac{ \mathfrak{f}_k (\psi) }{\sqrt{ \widetilde {\omega}(k)} }  \; , 
	\]
be an approximation of the Dirac delta-function on the circle $S^1$ 
supported at the point $\psi =0$. Then 
	\[
		\lim_{n \to \infty} \langle g , \LS_2 \delta_n \rangle = 0  
	\]
for all $ g \in \widehat{{\mathfrak h}}  (S^1)$ of the form 
	\begin{equation}
	\label{ggg}
		g = \sum_{n = -N}^{n = N} \widetilde{g}(n) \,  \mathfrak{f}_n \; ,
		\qquad N \in \mathbb{N} \; ; 
	\end{equation}
despite the fact that $\lim_{n \to \infty} 
\| \delta_n \|_{\widehat{{\mathfrak h}}  (S^1)} = \infty$.
\end{lemma}

\begin{proof}
We compute 
	\begin{align*}
		\LS_2 \delta_n 
		 & = \frac{1}{2i \sqrt{4 r \pi}} \sum_{k = - n}^n  
		 \left( \sqrt{ \widetilde {\omega}  
		 (k+1)}  \, \mathfrak{f}_{k+1} 
			 -  \sqrt{ \widetilde {\omega}(k-1)} \, 
			 \mathfrak{f}_{k-1} \right)\; . 
		 \\
		 & = \frac{1}{2i \sqrt{4 r \pi}} \Bigl( \sqrt{ 
		  \widetilde {\omega}  (n+1)} \, \mathfrak{f}_{n+1} 
		 + \sqrt{  \widetilde {\omega}  (n)} \, \mathfrak{f}_{n} 
		 \\
		 & \qquad  \qquad \qquad
		 -   \sqrt{  \widetilde {\omega}  (-n)} \, 
		 \mathfrak{f}_{-n} 
		 - \sqrt{  \widetilde {\omega}  (-n-1)} \, 
		 \mathfrak{f}_{-n-1} \Bigr) \; . 
	\end{align*}
The explicit formula for $\LS_2 \delta_n$ given above
implies  that
	\begin{align}
	\lim_{n \to \infty} \langle g , \LS_2 
 	\delta_n \rangle_{\widehat{{\mathfrak h}}  (S^1)} 
	& = \frac{1}{2i \sqrt{4 r \pi}} \lim_{n \to \infty}  \Bigl( \sqrt{ 
		  \widetilde {\omega}  (n+1)} \, \widetilde{g} (n+1) 
		 + \sqrt{  \widetilde {\omega}  (n)} \, \widetilde{g} (n) 
		 \nonumber \\
		 & \qquad \qquad  \qquad \qquad
		 -   \sqrt{  \widetilde {\omega}  (-n)} \, \widetilde{g} (-n) 
		 - \sqrt{  \widetilde {\omega}  (-n-1)} \, 
		 \widetilde{g} (-n-1) \Bigr)  
	\nonumber \\
	& = 
	0
	\nonumber
	\end{align} 
for every fixed function $g \in \widehat{{\mathfrak h}}  (S^1)$ 
of the form \eqref{ggg}. 
\end{proof}

\goodbreak
\begin{corollary}
\label{cor:4.6.7}
The Casimir operator ${\CS\, }^2$ is represented by a multiple of the 
identity operator.  
\end{corollary}

\begin{remark}
We will prove in Proposition~\ref{Prop5.7} that the representation 
$\widehat{u}$ is unitarily equivalent to the unitary irreducible representation 
$\widetilde{u}$ on~$\widetilde{{\mathfrak h}}(\partial V^+)$. 
\end{remark}

\begin{proof} 
Casimir operator takes the form 
\label{Chatpage}
	\[		
		{\CS\, }^2  = - \KS_0^2 + \frac{1}{2} ( \LS_+ \LS_- + \LS_- \LS_+) \; . 
	\]
Its off-diagonal matrix elements vanish and the diagonal matrix elements equal~$\zeta^2$: 
	\begin{align*}
		\frac{ \langle \mathfrak{f}_k , \; 
		\CS^2  \mathfrak{f}_k \rangle_{\widehat{{\mathfrak h}}  (S^1)}}{
					\big\| \mathfrak{f}_k \big\|^2_{\widehat{{\mathfrak h}}  (S^1)} } 
		& =  -k^2 +\frac{{r}^2}{2}\big( \widetilde \omega(k)\widetilde \omega(k-1)
		+\widetilde \omega(k)\widetilde \omega(k+1) \big) \\ \; 
		& =
			-k^2 +\frac{1}{2} \big( (k +s)(k-s-1) + (k-s)(k +s+1) \big) \\
		&=  -s(s+1) = \tfrac{1}{4} + \nu^2 = \zeta^2 \; , 
	\end{align*}
as expected. In the second but last equality we have 
used \eqref{eq:useful-1}--\eqref{eq:useful-2}.
\end{proof}

For the statement of the next result, recall the definition of 
distributions in path-space coordinates provided in 
\eqref{eqDeltaTensorh'}.

\begin{lemma}
\label{coid}
For $h_{1}, h_{2}\in  {\mathcal D}  (I_+) $  and  $| \theta_{1} - \theta_{2}| \le   \pi $, 
\label{stcpage}
	\begin{equation}
		\label{a-coideq2}
		\bigl\langle \delta_{\theta_{1}}\otimes h_{1} , 
		\delta_{\theta_{2}}\otimes h_{2} \bigr\rangle_{\mathbb{H}^{-1}(S^2)}  
		=  \bigl\langle \delta \otimes h_{1} , 
		\delta \otimes {\rm e}^{- | \theta_2 - \theta_1 | \omega \mathbb{cos}_\psi } 
		h_{2} \bigr\rangle_{\mathbb{H}^{-1}(S^2)} \, . 
	\end{equation}
Here	$\delta_{\theta'} (\theta)= \lim_{k \to \infty}(2 \pi )^{-1}
\sum_{|\ell| \leq k}{\rm e}^{i  \ell (\theta - \theta') } 
   \; \chi_{[0,2\pi)}(\theta- \theta') $, $\theta, \theta' \in [0, 2 \pi )$,
and $\chi_{[0,2\pi)}$  the characteristic function of the interval $ [0, 2 \pi ) \subset {\mathbb R}$, 
while $\delta (\theta) \doteq \delta_{0} (\theta)$.
\end{lemma}

\begin{proof} 
According to Proposition \ref{3.9a},
	\begin{align*}
		& \langle  \delta_{\theta_{1}}\otimes h_{1} , \delta_{\theta_{2}}
		\otimes h_{2} \rangle_{ \mathbb{H}^{-1} (S^2) } \\
		& \qquad = \frac{r^2}{ 2}  \int_{S^2}  {\rm d} \Omega (\vec{\tt x})  
		\int_{S^2}  {\rm d} \Omega (\vec{\tt y}) \; 
		( \delta_{\theta_{1}}\otimes \overline{h_{1} } ) (\vec{\tt x}) \, 
		c_\nu \, P_{- \frac{1}{2} - i \nu} 
		 \bigl( - \tfrac{  \vec{\tt x} \cdot \vec{\tt y} }{r^2} \bigr) \,  
		 ( \delta_{\theta_{2}}\otimes h_{2} ) (\vec{\tt y})  \\
		& \qquad = \frac{r^2}{ 2}  \int_{S^2}  {\rm d} \Omega (\vec{\tt x})  
		\int_{S^2}  {\rm d} \Omega (\vec{\tt y}) \; 
		(\delta \otimes \overline{h_{1} } ) (\vec{\tt x}) \, c_\nu \, P_{- \frac{1}{2} - i \nu} 
		 \bigl( - \tfrac{  \vec{\tt x} \cdot \vec{\tt y} }{r^2} \bigr) \,  
		 ( \delta_{| \theta_1 - \theta_{2}| }\otimes h_{2} ) (\vec{\tt y}) \\
		& \qquad  = \bigl\langle  \delta 
			\otimes h_{1} , u (\Lambda_1(t)) (\delta \otimes h_{2})
			\bigr\rangle_{ \mathbb{H}^{-1} (S^2) } \Bigl|_{t = i | \theta_1 - \theta_{2}| }   \; ,  
	\end{align*}
where we have used 
	\begin{align*}
		& \frac{1}{r^2} 	\left( \begin{smallmatrix}
						r \sin \theta_1 \cos \psi_1  \\
						r \sin  \psi_1  \\
						r \cos \theta_1 \cos \psi_1  
					\end{smallmatrix} \right)
		 			\left( \begin{smallmatrix}
						r \sin \theta_2 \cos \psi_2  \\
						r \sin  \psi_2  \\
						r \cos \theta_2 \cos \psi_2  
					\end{smallmatrix} \right) \\
		& \qquad =  - \cos \psi_1 \cos \psi_2 ( \underbrace{ \sin  (-\theta_1) \sin \theta_2 
		+ \cos  (- \theta_1) \cos \theta_2}_{= \cos | \theta_1 - \theta_2 |} )
		- \sin \psi_1 \sin \psi_2 \\
		& \qquad = 	\left( \begin{smallmatrix}
						0  \\
						\sin  \psi_1  \\
						\cos \psi_1 
					\end{smallmatrix} \right)
					\left( \begin{smallmatrix}
						\sin | \theta_1 - \theta_2 |  \cos \psi_2  \\
						\sin  \psi_2  \\
						\cos | \theta_1 - \theta_2 | \cos \psi_2  
					\end{smallmatrix} \right) \, . 
	\end{align*}
and
	\[
		u ( \underbrace{\Lambda_1 (i \theta)}_{=R_1(\theta)}) (\delta \otimes h_2) 
		= (\delta_\theta \otimes h_2) \qquad 
		\forall h_{2}\in  {\mathcal D}  (I_+) \; ,  
		\quad \forall \theta \in [0, \pi] \; . 
	\]
The implementer of the boost $u(\Lambda_1(t))$ is defined in \eqref{u-oph}.
Now, 
	\begin{align*}
		& \bigl\langle  \delta \otimes h_{1} , u (\Lambda_1(t)) (\delta \otimes h_{2})
			\bigr\rangle_{ \mathbb{H}^{-1} (S^2) } \Bigl|_{t = i | \theta_1 - \theta_{2}| } \\
		& \qquad =  \bigl\langle  h_{1} , \widehat{u} (\Lambda_1(t)) h_{2} 
		\bigr\rangle_{ \widehat{{\mathfrak h}}  (S^1) } \Bigl|_{t = i | \theta_1 - \theta_{2}| } 
		=  \bigl\langle h_{1} , {\rm e}^{- | \theta_2 - \theta_1 | \omega \mathbb{cos}_\psi } 
		h_{2} \bigr\rangle_{ \widehat{{\mathfrak h}}  (S^1) } 
		\\ 
		& \qquad 
		=  \bigl\langle \delta \otimes h_{1} , 
		\delta \otimes {\rm e}^{- | \theta_2 - \theta_1 | \omega \mathbb{cos}_\psi } 
		h_{2} \bigr\rangle_{\mathbb{H}^{-1}(S^2)} \, .  
	\end{align*}
Note that, by linearity, this result can be extended to $h_2 \in \widehat{{\mathfrak h}} (S^1)$. 
\end{proof}

\begin{proposition} 
\label{th4.20}
For $h_1, h_2 \in {\mathscr D}(\omega)$, we have
	\[
	 \bigl\langle \omega  \, h_1  , \omega   \, 
	 h_2 \bigr\rangle_{\widehat{{\mathfrak h}} (S^1)} 
	 =   c_\nu  \int_{S^1} {\rm d}\psi \int_{S^1}  {\rm d}\psi' \; 
	 	\overline{h_1(\psi')} \, 
				P'_{s}\big( - \cos(\psi' - \psi) \big) \, h_2(\psi) \;.
	\]
Note that $C^\infty(S^1) \subset {\mathscr D}(\omega)$. 
In other words, the integral kernel $\omega(\psi, \psi')$ of $\omega$ 
on $L^2 (S^1, r {\rm d} \psi)$ equals
	\[
		\omega (\psi, \psi') = \frac{c_\nu}{r^2} \;  P'_{s}\big( - \cos(\psi' - \psi) \big) \; . 
	\]
\end{proposition}

For the proof of this statement we refer the reader to Proposition \ref{prop:E5}.

\section{Reflection positivity: from $SO(3)$ to $SO(1,2)$}

\label{sec:4.7}

According to Theorem \ref{UIR-S1}, the rotations 
$\alpha \mapsto {\rm e}^{- i \alpha \KS_0} $ and the boosts 
	\[
		\theta_1 \mapsto   {\rm e}^{i \theta_1 \LS_1 } \; , 
		\qquad \theta_2 \mapsto   {\rm e}^{i \theta_2 \LS_2 } \; , 
	\]
generate a unitary representation of $SO_0(1,2)$ on 
$\widehat{{\mathfrak h}} (S^1) \cong \mathbb{H}^{-1}_{\upharpoonright S^1} (S^2)$; see 
Remark~\ref{rm:4.7.2}. 
Hence the generators (explicit formulas were given in \eqref{L1-L2}) satisfy  
	\begin{equation}
	\label{drehimplusalgebra}
		[ \LS_1 \, , \, \LS_2] = - i \KS_0 \; , 
		\qquad 
		[ \KS_0 \, , \, \LS_1] =   i \LS_2 \; , 
		\qquad
		[ \LS_2\, , \, \KS_0] =  i \LS_1 \;  . 
	\end{equation}
It is important 
to understand how the representation of $SO_0(1,2)$ given above can be 
reconstructed\footnote{Due to Theorem \ref{martheo}, Nelson's reconstruction theorem 
(see \cite{N1}\cite{N2}\cite{N3}\cite{N4}) applies, and the more sophisticated 
reconstruction theorem of Osterwalder and Schrader \cite{OS1}\cite{OS2} is not necessary
for the present work.} from the geometric action $\mathbb{u}(g) \colon 
\mathbb{H}^{-1} (S^2) \to \mathbb{H}^{-1} (S^2)$,
\label{umathbfpage}
	\begin{equation}
	\label{u-kugel}
		 \bigl(\mathbb{u}(g) h\bigr) (\vec x) = h (g^{-1} \vec x) \; , 
		\qquad \vec x \in S^2 \; , \; \;  g \in SO(3)\; , 
	\end{equation}
of $SO(3)$ on $\mathbb{H}^{-1} (S^2)$. 
As we have seen in Section \ref{sec:2.7}, the key is to choose a neighbourhood $N$ of 
the identity $\mathbb{1}$ in $SO(3)$, which is invariant under the rotations $R_0(\alpha)$, 
$\alpha \in [0, 2\pi)$,  and a domain of definition $\mathscr{D}$ for the definition 
of a virtual representation $\pi$ of $SO(3)$ on $\mathbb{H}^{-1}_{\upharpoonright S^1} (S^2)$. 

\bigskip
We now provide the necessary definitions:

\subsection{The polar cap}
Given a neighbourhood $N$ of the identity $\mathbb{1} \in SO(3) $
which is invariant under the rotations $R_0$, we define
	\begin{equation}
		\label{D-N}
		\mathscr{D}_N \doteq \bigl\{ h \in 
		\mathbb{H}^{-1}_{\upharpoonright \overline{S_+} } (S^2)\mid h \; 
		\text{real valued and}   \;  
		 \mathbb{u}(g) h \in \mathbb{H}^{-1}_{\upharpoonright \overline{S_+} } (S^2) 
		 \quad \forall g \in N \bigr\} \; . 
	\end{equation}
We note that, due to the covariance property, 
	\[
		\mathscr{D}_N = 
			\bigcup_{ {\tt O} \subset  {\tt O}_N }  \mathbb{H}^{-1}_{\upharpoonright {\tt O} } (S^2) \; , 
	\]
where the \emph{polar cap} (using the geographical coordinates introduced in Section \ref{s-geo-chart})	
	\begin{equation}
		\label{O-N}
		{\tt O}_N = \left\{ \left( \begin{smallmatrix} r \sin \vartheta \\ 
		r \cos \vartheta \sin \varrho \\
		r \cos \vartheta \cos \varrho 
		\end{smallmatrix} \right) \in S^ 2 \mid 
		\tfrac{\pi}{2} - \delta_N < \vartheta < \tfrac{\pi}{2} \right\}
		\qquad \text{for some $\delta_N >0$} \; , 
	\end{equation}
As required, $N$ is  invariant under the rotations $R_0$. 
Furthermore, ${\tt O}_N$ is the largest subset of $\overline{S_+}$ 
whose image under an arbitrary $g \in N$ is still contained in 
$\overline{S_+}$. It is clear that if $N$ is sufficiently small, ${\tt O}_N$ 
contains an open neighbourhood. We define
	\begin{equation}
	\label{DDD}
		\mathscr{D} \doteq e_0 \mathscr{D}_N \;, 
	\end{equation}
where $e_0$ is the projection defined in Lemma \ref{dlemma}. 
We will later on show that $\mathscr{D}$ is \emph{total} 
in $\mathbb{H}^{-1}_{\upharpoonright S^1} (S^2)$. 

\subsection{The virtual representation of $SO(3)$} 
We define a homomorphism~$\wp$ from the neighbourhood $N$ 
of the identity $\mathbb{1} \in SO(3) $ to linear operators 
defined on the dense subspace $\mathscr{D}$ 
in $\mathbb{H}^{-1}_{\upharpoonright S^1} (S^2)$:  for $g \in N$, 
 	\begin{equation}
	\label{def-vr-so3}
		\wp ( g) \; e_0 h \doteq e_0 \; \mathbb{u} (g) h 
		\qquad \forall \, h \in \mathscr{D}_N \; . 
	\end{equation}

\begin{lemma}
\label{lm:4.7.1}
The map \eqref{def-vr-so3} is well-defined: if $e_0 h=0$ for $ h \in \mathscr{D}_N$, then 
$e_0 \; \mathbb{u} (g) h=0$ for all~$g \in N$. 
\end{lemma}

\begin{proof}
For the rotations which keep 
$S^1$ invariant, this is obvious as $e_0 $ is invariant under such rotations.
We may thus use the Cartan decomposition for $SO(3)$, and consider,
without restriction of arbitrariness, only the rotations $\theta' 
\mapsto \Lambda (i \theta') $, $\theta' \in [0, \theta)$.  Using 
	\[ 
		\mathbb{u} (\Lambda (i \theta')) \mathbb{u}(T) = 
		\mathbb{u} (T) \mathbb{u} (\Lambda (-i \theta')) \; ,
		\qquad \theta' \in [0, \theta) \; ,
	\]
we compute 
	\begin{align*}
		\| e_0 \; \mathbb{u} (\Lambda (i \theta')) h \|^2_{\mathbb{H}^{-1} (S^2) }
		& = \langle \mathbb{u} (\Lambda (i \theta')) h , 
		e_0 \; \mathbb{u} (\Lambda (i \theta')) h \rangle_{\mathbb{H}^{-1} (S^2) }
		\\
		& = \langle e_0 \mathbb{u} (\Lambda (i (\theta' - \theta''))) h , 
		\mathbb{u} (\Lambda (i (\theta'+\theta''))) h \rangle
		\\
		& \le \| e_0 \; \mathbb{u} (\Lambda (i (\theta' - \theta''))) 
		h \|_{\mathbb{H}^{-1} (S^2) } 
		\| h \|_{\mathbb{H}^{-1} (S^2) } 
		\; . 
	\end{align*}
Now choose $\theta''$ such that $n\theta''=\theta'$ for some positive
integer $n$, and such that $\theta''+\theta'\equiv (n+1)\theta''$ is
still smaller or equal to $\theta$. Iterating the above inequality $n$
times yields 
	\begin{align*} 
		\| e_0 \; \mathbb{u} (\Lambda (i \theta')) h \|^2_{\mathbb{H}^{-1} (S^2) }
		&\leq \| 
				e_0 \; \mathbb{u} (\Lambda (i (\theta'-n\theta'')) 
				h \|_{\mathbb{H}^{-1} (S^2) }^{\tfrac{1}{2^n}}  
				\; 
				\| h \|_{\mathbb{H}^{-1} (S^2) }^{\tfrac{1}{2}+\cdots+\tfrac{1}{2^n}} 
			\\
			& =   
			\| h \|_{\mathbb{H}^{-1} (S^2) }^{\tfrac{1}{2^n}} \; 
			\| h \|_{\mathbb{H}^{-1} (S^2) }^{\tfrac{1}{2}+\cdots+\tfrac{1}{2^n}}
			\, .
	\end{align*}
Thus, $e_0 h= 0$ implies $e_0 \; \mathbb{u} (\Lambda (i \theta')) h=0$ for all $\theta'<\theta$. 
By continuity, this fact extends to $\theta'=\theta$. 
\end{proof}

Next, we will establish the property characterising 
a \emph{virtual representation} \cite{FOS}:
 	\begin{equation}
	\label{vr-so3}
		\wp ( \mathbb{\sigma} (g) )^* \psi 
		= \wp (g^{-1}) \psi \qquad \forall \psi \in \mathscr{D} \; ,
	\end{equation}  
where the involution $\mathbb{\sigma}$ is given by 
	\begin{equation}
	\label{m-sigma}	
		\mathbb{\sigma} (g) = T g T \; , \qquad g \in SO(3) \; . 
	\end{equation}
This property follows from the following calculation:
	\begin{align}
		\langle h_1 , e_0  \mathbb{u}(g) 
		h_2 \rangle_{\mathbb{H}^{-1}(S^2)} 
		& = \langle \mathbb{u}(T) h_1 , \mathbb{u}(g) 
		h_2 \rangle_{\mathbb{H}^{-1}(S^2)} \nonumber \\
		& = \bigl\langle \mathbb{u}(g^{-1}) \mathbb{u}(T) h_1 ,  
		h_2 \bigr\rangle_{\mathbb{H}^{-1}(S^2)} \nonumber \\
		& = \bigl\langle \mathbb{u}(T) 
		\mathbb{u}(\mathbb{\sigma}(g^{-1}))  h_1 ,  
		h_2 \bigr\rangle_{\mathbb{H}^{-1}(S^2)} \nonumber \\
		& = \bigl\langle \mathbb{u}(\mathbb{\sigma}(g^{-1}))  h_1 ,  
		e_0 h_2 \bigr\rangle_{\mathbb{H}^{-1}(S^2)} \; . 
	\label{one-p-vr}
	\end{align}
As the group $SO(3)$ is generated by the rotations $R_0(\alpha)$, 
$\alpha \in [0, 2 \pi)$, and $R_1(\theta)$, $\theta \in [0, 2 \pi)$, 
the next step is to compute the local symmetric 
semigroup\footnote{See Lemma \ref{SemiGroupL} for further details.}
	\[
		R_1 (\theta) \mapsto \wp \bigl( R_1 (\theta) \bigr) \; . 
	\]
We claim that (due to \eqref{L1-L2}) 
	\begin{equation}
	\label{reconstructed-free-boost}
		\wp \bigl( R_1 (\theta) \bigr)^{**} = {\rm e}^{ -\theta r \,  \omega \operatorname{\mathbb{cos}}_\psi  } 
		= {\rm e}^{ -\theta \LS_1 }\; . 
	\end{equation}
This can be seen as follows: for $f \in \mathscr{D}_N$, we have
	\begin{align}
		 \langle T_* f,  f \rangle_{\mathbb{H}^{-1}(S^2)}  
		&  =  \int r \, {\rm d}  \theta  
		\int r \, {\rm d} \theta' \; \; 
		\langle \delta_{- \theta} \otimes f_\theta ,    \delta_{  \theta'}  
		\otimes f_{\theta'} \rangle_{\mathbb{H}^{-1}(S^2)} \nonumber \\
		&  =  
		\Bigl\| \delta \otimes \int_0^\pi  r \,  {\rm d}  \theta  \,  {\rm e}^{ -\theta r \,  \omega   
		\operatorname{\mathbb{cos}}  } f_{\theta}  
					\Bigr\|_{\mathbb{H}^{-1} (S^2)}^2 \, . 
		\label{neu}
	\end{align}
In the second identity, we have used Lemma \ref{coid}. 
Similarly, 
	\begin{align*}
		 \langle T_*  (\delta \otimes g_1 + \delta_\pi \otimes g_2),  
		 f \rangle_{\mathbb{H}^{-1}(S^2)}  
		&  =  
		\Bigl\langle (\delta \otimes g_1 + \delta_\pi \otimes g_2) ,  
		\int r \, {\rm d} \theta  
		\, ( \delta_{  \theta}  
		\otimes f_{\theta} ) \rangle_{\mathbb{H}^{-1}(S^2)} \nonumber \\
		&  =  
		\Bigl\langle g_1 \, , 
		 \int_0^\pi  r \,  {\rm d}  \theta  \,  {\rm e}^{ -\theta r \,  \omega   
		\operatorname{\mathbb{cos}}  } f_{\theta}  
					\Bigr\rangle_{\widehat{\mathfrak{h}}(S^1)} \, . 
		\\
		& \qquad  
		+ \Bigl\langle (P_1)_* g_2 \, , 
		 \int_0^\pi  r \,  {\rm d}  \theta  \,  {\rm e}^{ -\theta r \,  \omega   
		\operatorname{\mathbb{cos}}  } f_{\theta}  
					\Bigr\rangle_{\widehat{\mathfrak{h}}(S^1)}  
	\end{align*}
for all $g_1 , g_2 \in C^\infty_0 (I_+)$. 
Hence, 
	\begin{align*}
		e_0 f  &= e_0  \left( \delta \otimes \int_0^\pi  r \,  {\rm d}  \theta'  \,  
		{\rm e}^{ -\theta' r \,  \omega   
		\operatorname{\mathbb{cos}}_\psi  } f_{\theta'} \right) \\
		& =  \int_0^\pi  r \,  {\rm d}  \theta'  \,  {\rm e}^{ -\theta' r \,  \omega   
		\operatorname{\mathbb{cos}}_\psi  } f_{\theta'}  \; . 
	\end{align*}
Similarly, 
	\begin{align*}
		e_0 \mathbb{u}(R_1 (\theta))  f 
		&= e_0 \left( \delta \otimes \int_0^\pi  r \,  {\rm d}  \theta'  \,  {\rm e}^{ -(\theta+ \theta' )r \,  \omega   
		\operatorname{ \mathbb{cos} }_\psi  } f_{\theta'} \right)  \\
		& = {\rm e}^{ - \theta r \,  \omega   
		\operatorname{\mathbb{cos}}_\psi  }  \int_0^\pi  r \,  {\rm d}  \theta'  \,  {\rm e}^{ - \theta' r \,  \omega   
		\operatorname{\mathbb{cos}}_\psi  } f_{\theta'}   \\
		& = {\rm e}^{ - \theta r \,  \omega  \operatorname{\mathbb{cos}}_\psi  } \, e_0  f     
		\; . 
	\end{align*}
In other words, the map \eqref{def-vr-so3} equals 
	\[
		e_0  f
		\mapsto {\rm e}^{ - \theta r \,  \omega  \operatorname{\mathbb{cos}}_\psi  } e_0  f  \, . 
	\]
One can show that for any open set ${\tt O} \subset S_+$ the set 
$\bigl\{    e_0  f  \mid \operatorname{supp}  \Re f \subset {\tt O}  , \Im f = 0 \bigr\}$
is indeed an operator core for $ {\rm e}^{ - \theta r \,  \omega  \operatorname{\mathbb{cos}}_\psi  }$, 
see also Lemma \ref{eqEU(K)Om} below. Hence the identity \eqref{neu} verifies that the generator of 
the boost on $\mathbb{H}^{-1}_{\upharpoonright S^1} (S^2) $ is $\omega \operatorname{\mathbb{cos}}_\psi$.

The generator of the rotations $R_0(\alpha)$, $\alpha \in [0, 2 \pi)$, on $\mathbb{H}^{-1}(S^2) $ 
is (in geographical coordinates) 
	\[
		\KStwo_0 = 	-i   \partial_\rho \, ;  
	\]
and the action on $\mathbb{H}^{-1}_{\upharpoonright S^1} (S^2)$ is 
simply the restriction to this 
subspace. Thus 
	\[
		\wp \bigl( R_0 (\alpha) \bigr)^{**} = {\rm e}^{ i \alpha \KStwo_0} \; . 
	\]
As expected form \eqref{vr-so3}, we find:
	\begin{align*}
		\wp \bigl( \mathbb{\sigma}( R_0(\alpha) ) \bigr)^* \psi & = 
		\wp ( R_0(\alpha) )^* \psi = {\rm e}^{-  i \alpha \KS_0}
		= \wp \bigl( R_0(\alpha)^{-1} \bigr) \psi \; , \\
		\wp \bigl( \mathbb{\sigma}(R_1(\theta) ) \bigr)^* \psi & 
		= \wp ( R_1(-\theta) )^* \psi = \wp \bigl( R_1(\theta)^{-1} \bigr) \psi  \; \; ,  
	\end{align*}
for all $\psi \in \mathscr{D}$.

\begin{lemma}
\label{lm:4.8.1} 
The map $\wp$ defined above extends to a virtual representation 
	\[
		R \mapsto \wp (R) 
	\]
of $SO(3)$ in the sense of Fr\"ohlich, Osterwalder and Seiler \cite{FOS}, \emph{i.e.},   
$\wp$ is a local group homomorphism  
from $SO(3)$ into linear operators densely 
defined on $\mathbb{H}^{-1}_{\upharpoonright S^1} (S^2) $, with the following properties: 
\begin{itemize}
\item [$ i.)$] the map
	\[
		\alpha \mapsto \wp(R_0 (\alpha) ) 
	\]
is a continuous unitary representation of $SO(2)$ on $\mathbb{H}^{-1}_{\upharpoonright S^1} (S^2) $; 
\item [$ ii.)$]
there exists a neighbourhood $N$ of $\mathbb{1} \in SO(3)$, invariant under the 
rotations $R_0 (\alpha) $, $\alpha \in [0, 2 \pi)$, and a linear subspace $\mathscr{D}$, dense 
in $\mathbb{H}^{-1}_{\upharpoonright S^1} (S^2) $, such that 
\begin{itemize}
\item[---] $\mathscr{D} \subset \mathscr{D}(\wp(g))$ for all $g \in N$;  and 
\item[---] if $g_1, g_2$ and $g_1 \circ g_2$ are all in $N$, then
\label{virtreppage2}
	\begin{equation} 
		\label{49th}
		\wp (g_2) \Psi \in {\mathscr D}(\wp(g_1))\; , \qquad \Psi \in \mathscr{D}\; ,  
	\end{equation}
	and
	\[ 
		\wp(g_1) \wp(g_2) \Psi = \wp(g_1 \circ g_2) \Psi \; , \qquad \Psi \in \mathscr{D}\; ; 
	\]
\end{itemize}
\item[$ iii.)$] 
if $\ell \in {\mathfrak m}$\footnote{For the definition of ${\mathfrak m}$ see~\eqref{VR1}.}, 
$0 \le t \le 1$, and 
	\[
		{\rm e}^{- t \ell} \in N  \; ,  \qquad 0 \le t \le 1 \; , 
	\]
then $ \wp ({\rm e}^{-t \ell })$ is a hermitian operator defined on $\mathscr{D}$ and 
	\begin{equation}
	\label{52th}
		\operatornamewithlimits{s-lim}_{t \to 0} \wp ({\rm e}^{-t \ell}) \Psi = \Psi \; , \qquad \Psi \in \mathscr{D} \; . 
	\end{equation}
\end{itemize}
\end{lemma}

\begin{proof} The group relation follows directly from the definition of $\wp$: 
for all $\psi \in \mathscr{D} $, with $\mathscr{D}$ given by \eqref{DDD} and \eqref{D-N}, 
we have 
	\[ 
		\wp(g_1) \wp(g_2) \Psi = \wp(g_1 \circ g_2) \Psi \; , 
	\]
as any element of $SO(3)$ can be written as a product of the rotations $R_0$ and $R_1$, 
and both of them are given by the push-forward action. The choice of the domain ensures 
that the support of the transformed function remains in the upper hemisphere. 
\end{proof}

\noindent 
The main result in the theory of virtual representations is the following:

\begin{theorem}[Fr\"ohlich, Osterwalder, and Seiler \cite{FOS}] 
\label{FOS}
Let $\wp$ be a virtual representation of 
a symmetric space $(G, K, \mathbb{\sigma})$, with $K$ compact. 
Then $\wp$ can be analytically continued to a unitary representation $\wp^*$ of 
the dual symmetric Lie group $G^*$ (defined in Subsection~\ref{ssec:2.7.1}).
\end{theorem} 

\begin{remark}
\label{rm:4.8.3}
Inspecting the explicit formulas provided, it is clear that the virtual representation 
of $SO(3)$ discussed in Lemma \ref{lm:4.8.1} can be analytically continued
to the representation $\widehat{u}$ of $SO(1,2)$ constructed in Proposition \ref{UIR-S1}. 
\end{remark}

\section{Time-symmetric and time-antisymmetric test-functions}
\label{sec:4.8}

The restriction of the Fourier transform to the mass shell 
allows an extension from~${\mathcal D}_\mathbb{R}(dS)$ to distributions  supported 
on the time-zero circle.   
We shall identify~$dS$ with $\mathbb{R}\times S^1$ via the coordinate system 
	\begin{equation}
	\label{w1psitau}
	x (x_0,\psi) = \begin{pmatrix}
				x_0 \\
				\sqrt{r^2 +x_0^2} \; \sin  \psi  \\
				\sqrt{r^2 + x_0^2} \; \cos  \psi 
				\end{pmatrix} \in dS
	\end{equation}
and write $(f\otimes h)(x):=f(x_0)h(\psi)$ for $f\in{\mathcal D}(\mathbb{R})$ 
and $h\in{\mathcal D}(S^1)$ if $x=x(x_0,\psi)$. The metric on $dS$ is
	\begin{equation}
	\label{w1psitau-metrik}
	g = \frac{ r^2 }{r^2+x_0^2} \; {\rm d} x_0 \otimes {\rm d} x_0 
	- ( r^2+ x_0^2) \; {\rm d}  \psi \otimes {\rm d} \psi  
	\end{equation}
and $|g|= r^2 $. Thus ${\rm d} \mu_{dS} ( x)= {\rm d} x_0 r {\rm d}  \psi $. 

\begin{theorem} 
\label{st-kappa}
Let $h \in C^\infty_\mathbb{R} (S^1)$ and let $\delta_k$ be a sequence of absolutely 
integrable smooth functions, supported in a neighbourhood of the origin in $\mathbb{R}$, 
approximating the Dirac $\delta$-function. It follows that the limits 
	\begin{equation}
		\label{tz}
			\lim_{k\to \infty} \| [ \delta_k \otimes h ] \|_{{\mathfrak h} (dS)}
\quad \text{and} \quad 
			\lim_{k \to \infty}  \| [ n \, (\delta_k \otimes g ) ] \|_{{\mathfrak h} (dS)}
	\end{equation}
exist and equal $\| h \|_{ \widehat{{\mathfrak h}}  (S^1)} $ 
and $\|  \omega  \, g \|_{ \widehat{{\mathfrak h}}  (S^1)}$, respectively. 
Here $n \, ( \delta_k \otimes h)$ denotes the Lie derivative\footnote{Recall that 
 $\int_{\mathbb{R}^{1+d}} {\rm d} t {\rm d}  \vec{x}  \, \delta' (t) h( \vec{x} \, )  
{\rm e}^{i  (\omega t -  \vec p \cdot \vec{x} )} 
	= 	- i \omega
	\int_{\mathbb{R}^d} {\rm d} \vec{x}  \,  h( \vec{x} \, ) {\rm e}^{-i  \vec p \cdot \vec{x}}$. }
of $( \delta_k \otimes h)$ along the unit normal future pointing vector field~$n$. 
\end{theorem}

\begin{proof} 
According to Proposition \ref{legendre},
	\begin{align*}
	& \lim_{k, k' \to \infty}  \langle [ \delta_k \otimes h ] ,  [ \delta_{k'} \otimes h' ] \rangle_{{\mathfrak h}(dS)} 
	 \\ 	
	 &  \quad = \lim_{k, k' \to \infty}\int_{dS \times dS} {\rm d} \mu_{dS} ( x ) \; {\rm d} \mu_{dS} (  x')  \; 
	 \overline { (\delta_k  \otimes h) ( x) }  \,    
	{\mathcal W}^{(2)} (  x  ,  x' )    (\delta_{k'}   \otimes h') ( x' )   \\
					& \quad = c_\nu 
				\int_{S^1} r\, {\rm d}\psi \int_{S^1} r\, {\rm d}\psi' \; \overline{h(\psi)} \, 
				P_{s^+}\big( - \cos(\psi' - \psi) \big) \, h'(\psi') \\
				&  \quad = \langle h , h'  \rangle_{ \widehat{{\mathfrak h}}  (S^1)}  \;.
	\end{align*}
For the derivative of the delta-function, partial integration yields
	\begin{align}
	\label{deriv}
	& \lim_{k, k' \to \infty}  \langle [ \delta'_k \otimes h ] ,  [\delta'_{k'} \otimes h' ] \rangle_{{\mathfrak h}(dS)} 
	 \nonumber \\ 	
	 &  \quad = \lim_{k, k' \to \infty}\int_{dS \times dS} \kern -.3cm {\rm d} \mu_{dS} ( x ) \; {\rm d} \mu_{dS} (  x')  \; 
	 \overline { (\delta'_k  \otimes h) ( x) }  \,    
	{\mathcal W}^{(2)} (  x  ,  x' )    (\delta'_{k'}   \otimes h') ( x' )   \nonumber \\
					& \quad = c_\nu 
				\int_{S^1 \times S^1}  r^2 {\rm d}\psi {\rm d}\psi' 
				\int_{{\mathbb R} \times {\mathbb R}} {\rm d}x_0 {\rm d}x'_0 \; \delta(x_0)\delta(x'_0)
				\nonumber \\
				& 
				\qquad \qquad \qquad \qquad \qquad \qquad \qquad \times 
				\; \overline{h(\psi)} \, h'(\psi')
				\left( \tfrac{\partial}{\partial x_0 }\tfrac{\partial}{\partial x'_0 } P_{s}\left( \tfrac{x_+ \cdot x'_- }{r^2} \right)  
				\right)
				 \nonumber \\
					& \quad = c_\nu 
				\int_{S^1 \times S^1} r^2 {\rm d}\psi {\rm d}\psi' 
				\int_{{\mathbb R} \times {\mathbb R}} {\rm d}x_0 {\rm d}x'_0 \; \delta(x_0)\delta(x'_0)
				\nonumber \\
				& 
				\qquad \qquad \qquad \qquad  \qquad \qquad \times 
				\; \overline{h(\psi)} \, h'(\psi')
				 \tfrac{\partial}{\partial x_0 } \left( P'_{s}\left( \tfrac{x_+ \cdot x'_- }{r^2} \right) \tfrac{\partial}{\partial x'_0 } 
				\left( \tfrac{x_+ \cdot x'_- }{r^2} \right) \right)
				 \nonumber \\
					&  \quad = c_\nu 
				\int_{S^1} \, {\rm d}\psi \int_{S^1} \,  {\rm d}\psi' \; \overline{h(\psi)} \, 
								P'_{s}\big( - \cos(\psi' - \psi) \big) \, h'(\psi') 
				\nonumber  \\
				&   \quad = \langle \omega  \, h , \omega  \, 
				h'  \rangle_{ \widehat{{\mathfrak h}}  (S^1)}  \;.
	\end{align}
The second but last equality follows from 
	\begin{align*}
	 \tfrac{\partial}{\partial x'_0 }  \left( \tfrac{x_+ \cdot x'_- }{r^2} \right) 
	 & =  \tfrac{\partial}{\partial x'_0 } \left( \tfrac{x_0 x'_0}{\sqrt{r^2 + x_0^2} \sqrt{r^2 + {x'_0}^2} } 
	 - \tfrac{ \sqrt{r^2 + x_0^2} \sqrt{r^2 + {x'_0}^2} }{r^2} \cos (\psi- \psi') \right)
	 \\ 	
	 &  = \tfrac{r^2}{(r^2-{x'_0}^2)^{\frac{3}{2}}}
	 \tfrac{x_0 }{\sqrt{r^2 + x_0^2} }
	 - \tfrac{{\tt x_0}' \sqrt{r^2 + {\tt x_0}^2} }{ r^2 \sqrt{r^2 + {\tt x'_0}^2} }  \cos (\psi- \psi')   
	\end{align*}
and 
		\[
		\tfrac{\partial}{\partial x_0 }\tfrac{\partial}{\partial x'_0 } \left( \tfrac{x_+ \cdot x'_- }{r^2} \right)
	 	= \tfrac{r^2}{(r^2-{x'_0}^2)^{\frac{3}{2}}} \tfrac{r^2}{(r^2-{x_0}^2)^{\frac{3}{2}}}
	  	- \tfrac{ {\tt x_0}}{r^2 \sqrt{r^2 + {\tt x_0}^2} } \tfrac{  {\tt x'_0}}{ \sqrt{r^2 + {\tt x'_0}^2} } \cos (\psi- \psi') \; . 
	 \]
Thus 
	\[
	 \tfrac{\partial}{\partial x_0 } \left( \tfrac{x_+ \cdot x'_- }{r^2} \right)_{ \upharpoonright x_0=x'_0 = 0}
	 =  \tfrac{\partial}{\partial x'_0 } \left( \tfrac{x_+ \cdot x'_- }{r^2} \right)_{\upharpoonright x_0=x'_0 = 0}
	 = 0 \; , \quad 
	 \tfrac{\partial}{\partial x_0 }\tfrac{\partial}{\partial x'_0 } 
	 \left( \tfrac{x_+ \cdot x'_- }{r^2} \right)_{ \upharpoonright x_0=x'_0 = 0}
	= \frac{1}{r^2} \; . 
	\]
The last equality in \eqref{deriv} follows from Proposition \ref{th4.20}.
\end{proof}	 

The functions $  \delta \otimes h $ and $\delta' \otimes h$ provide examples of test-functions, which are symmetric  
and anti-symmetric, respectively, under time-reflection. 
In fact, the time-reflection $T$  induces a conjugation\footnote{An anti-linear   
isometry~$C$ satisfying  $C^2= 1$ is called a {\em conjugation}. } $\kappa
\doteq u(T) $ on~${\mathfrak h} (dS)$, as the map 
$f \mapsto T_* f $ leaves the kernel of ${\mathcal F}_{+ \upharpoonright \nu}$  invariant.
The subspace consisting of functions  invariant under time-reflection is 
	\begin{equation}
	\label{kappat}
 		{\mathfrak h}^\kappa (dS) = \bigl\{ f \in {\mathfrak h} (dS) \mid   
		u(T)  f = f \bigr\}  \; . 
	\end{equation}
One can decompose
any testfunction into a symmetric and an anti-symmetric part with respect to time-reflections: 
	\[
		f= \tfrac{1}{2} ( f + \kappa f) + \tfrac{1}{2} ( f - \kappa f)  \; , \qquad f \in {\mathfrak h} (dS) \;  .
	\]
For $[f] , [g] \in {{\mathfrak h}^\circ} (dS) \cap {{\mathfrak h}^\kappa} (dS)$, 
polarisation  yields
	\[
		\langle  [f] , [g] \rangle_{{\mathfrak h} (dS)}  
			= \langle [ T_* f ] , [ T_* g ] \rangle_{{\mathfrak h} (dS)} 
			= \langle [g] , [f] \rangle_{{\mathfrak h} (dS)} \; . 
	\]
Since ${{\mathfrak h}^\circ} (dS)$ is dense in ${\mathfrak h} (dS)$,  this implies 
\[ 
\Im \langle f, g \rangle_{{\mathfrak h} (dS)}  = 0 \quad \text{for all $f, g \in {\mathfrak h}^\kappa (dS)$}.
\]
Thus ${\mathfrak h}^\kappa (dS)$ is a $\mathbb{R}$-linear subspace of ${\mathfrak h} (dS)$.

\section{Fock space}
\label{Fockspace}

Consider\footnote{We follow \cite[Vol.~II]{BR}.} a  Hilbert space ${\mathfrak h} $ with 
scalar product $\langle\, . \, , \, . \, \rangle$.
Let $ \Gamma^{(n)} ( {\mathfrak h} ) $, $n \in \mathbb{N}$, 
be the n-fold totally symmetric tensor product $\otimes_s$ of~${\mathfrak h} $ with itself. The elements of $ \Gamma^{(n)} ( {\mathfrak h} ) $
are of the form
	\[
		P_+ ( f_1 \otimes \ldots \otimes f_n) 
		\doteq\frac{1}{n!} \sum_\pi  f_{\pi_1} \otimes f_{\pi_2} \otimes \ldots \otimes  f_{\pi_n}\; , 
		\qquad f_1, \ldots, f_n \in {\mathfrak h} \; .
	\]
The sum is over all 
permutations $\pi \colon (1, 2, \ldots, n) \mapsto (\pi_1 , \pi_2 , \ldots , \pi_n)$. 
In other words, the \emph{symmetrisation operator} $P_+$  takes care of the necessary symmetrisation required. 

\subsection{Bosonic Fock space}
\label{sec:4.9}

The symmetric Fock-space 
$\Gamma ( {\mathfrak h} ) $ over ${\mathfrak h} $ is  the 
direct sum of the $n$-particle spaces:
	\[
		\Gamma ( {\mathfrak h} )  \doteq\oplus_{n= 0}^{\infty}  \Gamma^{(n)} ( {\mathfrak h} ) \;  , 
	\]
with $ \Gamma^{(0)} ( {\mathfrak h} ) \doteq \mathbb{C}$.  
The vectors with finitely many components unequal to zero form a dense subspace  
	\begin{equation}
	\label{gamma-circ}
		\Gamma^\circ ({\mathfrak h} ) \doteq {\rm Span} \, 
		\Bigl\{ \oplus_{n= 0}^{N} \Gamma^{(n)} 
		( {\mathfrak h} ) \mid  N \in \mathbb{N} \Bigr\}  
	\end{equation}
in~$\Gamma ( {\mathfrak h} )$. The vector $\Omega_\circ \doteq (1,0, 0, \ldots)$ is called the Fock vacuum vector. 

\subsection{Creation and annihilation operator}
For $f\in  {\mathfrak h} $, define the {\it creation operator} ${ a}^* ({f}) \colon \Gamma^\circ ( {\mathfrak h} )
\to \Gamma^\circ ( {\mathfrak h} )$ by  
	\[
		{ a}^* ({f}) \Psi^{(n)} \doteq   \sqrt {n+1} \, \, 
		P_+ ( f \otimes  \Psi^{(n)})  
		\;  , \qquad \Psi^{(n)} \in \Gamma^{(n)} ( {\mathfrak h} ) \; .
	\]
The operator ${ a} ({f}) $ denotes the adjoint of  ${ a}^* ({f})$, and is called the \emph{annihilation operator}. 
Both ${ a}({f})$ and ${ a}^* ({f})$ are defined on  $\Gamma^\circ ( {\mathfrak h} )$ and can be extended 
to densely defined closed, unbounded operators on 
$\Gamma ( {\mathfrak h} )$. 
 
The map ${f} \mapsto { a}^* ({f})$ is linear, while
the map ${f}  \mapsto { a} ({f})$ is anti-linear. They satisfy the \emph{canonical commutation relations}:
	\[
		\bigl[ { a} ({f}) , { a} ({g})  \bigr] = \bigl[ { a}^* ({f}) , { a}^* ({g}) \bigr] = 0 
	\]
and
	\[
		\bigl[ { a} ({f}) , { a}^* ({g})  \bigr] 
 		= \langle f    ,  g \rangle  \qquad \forall  {f}, {g}  \in   {\mathfrak h}  \; . 
 	\]
By applying the creation operators ${ a}^* ({f})$ to $\Omega_\circ$ we 
get $\Gamma^\circ ({\mathfrak h} )$ and by closure
all of~$\Gamma ({\mathfrak h} )$:
	\[
		{ a}^* ({f}_1) \ldots { a}^* ({f}_n) \Omega_\circ 
		= \sqrt {n!} \, \Bigl(  {f}_1    \otimes_s \ldots \otimes_s {f}_n \Bigr) 
		\in \Gamma^{(n)} ( {\mathfrak h} )  \; , \quad {f}_1 , \ldots, {f}_n \in  {\mathfrak h} \; .
	\]

\subsection{Bosonic field operators}
The symmetric operator ${ a}^*({f}) + { a}({f}) $ is essentially self-adjoint 
on~$\Gamma^\circ ( {\mathfrak h} )$, its closure is denoted by 
	\[ 
		\Phi_F (f) \doteq \frac{1}{\sqrt{2}} \bigl ( { a}^*({f}) + { a}({f}) \bigr)^- \;  . 
	\]
The field operators $\Phi_F (f)$ satisfy the commutation relations 
	\[
		[ \Phi_F (f) ,  \Phi_F (g)]= i \,  \Im \langle f , g \rangle \; ,
		\qquad f , \:g \in {\mathfrak h} \; .
	\]
        
\subsection{Weyl operators in Fock space}
\label{WOFS}
The operators ${ W}_F ({f}) \doteq {\rm e}^{ i  \Phi_F (f) }$ satisfy 
	\[ 
		{  W}_F ({f}) {  W}_F ({g} )  
		=  { \rm e}^{   i \Im \langle g ,   f \rangle } {  W}_F ( {f}  + {g} )  \;  ,\qquad f , \:g \in {\mathfrak h} \;  .
	\]
The \emph{Weyl operators} are related to the exponentials ${\rm e}^{i { a}^*  ( f ) } $ and 
${\rm e}^{i { a} ( f ) } $ by
	\[
		 {  W}_F (f) = {\rm e}^{i  {   a}^*(f)}\; {\rm e}^{i  {  a}(f)} {\rm e}^{-\frac12 \|f \|^2}     
	\]
on $\Gamma^\circ ( {\mathfrak h} )$. In particular,  
$\langle \Omega_\circ , {W}_F (f) 
\Omega_\circ \rangle = {\rm e}^{-\frac12 \|f \|^2}$.

\subsection{Subalgebras} The operators $\{ {W}_F (f) \mid f \in {\mathfrak h} \}$ 
generate all of $ {\mathcal B} \bigl(\Gamma ({\mathfrak h} )\bigr)$. Subalgebras emerge, when we consider 
$\mathbb{R}$-linear subspaces of ${\mathfrak h}$: let ${\mathfrak d}_{\alpha}$ be a family of $\mathbb{R}$-linear 
subspaces\footnote{$\mathbb{R}$-linear \emph{closed} subsets.} of ${\mathfrak h}$
and let ${\mathfrak d}_\alpha^{\perp}$ denote the symplectic complement with respect to the 
symplectic form $\sigma(h_{1}, h_{2})= 2 \Im \langle h_{1}, h_{2} \rangle_{\mathfrak h}$.

\begin{theorem}[Araki \cite{A1}, Theorem 1] 
\label{araki}
Let ${\mathfrak M}({\mathfrak d})\subset {\mathfrak M}({\mathfrak h})$ denote the   
von Neumann algebra generated by $\{W_F(h) \mid h\in {\mathfrak d}\}$.
Then
	\begin{align}
	\bigcap_{\alpha} {\mathfrak M}({\mathfrak d}_{\alpha} )
	 = {\mathfrak M}\left(\cap_{\alpha} {\mathfrak d}_{\alpha} \right)  \; ,
	\qquad
		\label{et.2n}
		\bigvee_{\alpha} {\mathfrak M}({\mathfrak d}_{\alpha}) 
		=  {\mathfrak M}\left(\vee_{\alpha} {\mathfrak d}_{\alpha} \right)   \; , 
	\end{align}
and ${\mathfrak M}({\mathfrak d}^{\perp})$ is equal to the commutant of the set $\{W_F(h) \mid h\in {\mathfrak d}\}\, $.
\end{theorem}

\goodbreak

\begin{theorem}[Leyland, Roberts and Testard \cite{LRT}, Theorem I.3.2]
\label{th:4.9.2}
Let ${\mathfrak M}({\mathfrak d})\subset {\mathfrak M}({\mathfrak h})$ denote the   
von Neumann algebra generated by $\{W_F(h) \mid h\in {\mathfrak d}\}$.
Then
\begin{itemize}
\item[$ i.)$] $\Omega_\circ$ is cyclic for ${\mathfrak M}({\mathfrak d}) $ if and only if ${\mathfrak d} + i {\mathfrak d}$ is dense in 
${\mathfrak h}$;
\item[$ ii.)$] $\Omega_\circ$ is separating for ${\mathfrak M}({\mathfrak d})$ 
if and only if $ {\mathfrak d}  \cap i  {\mathfrak d}  = \{0 \}$;
\item[$ iii.)$] ${\mathfrak M}({\mathfrak d})$ is a factor if and only if~${\mathfrak d}  \cap  {\mathfrak d}^\perp = \{ 0 \}$.
\end{itemize}
\end{theorem}

Next, assume that $\Omega_\circ$ is cyclic and 
separating for ${\mathfrak M}({\mathfrak d})$.
Let $S$ denote the closure of the 
operator $S_\circ \colon {\mathfrak M}({\mathfrak d}) \Omega_\circ \to 
{\mathfrak M}({\mathfrak d}) \Omega_\circ \, $,  
	\[
		S_\circ A \Omega_\circ := A^* \Omega_\circ \; . 
	\]
We call $S$ the Tomita operator for the pair 
$( {\mathfrak M}({\mathfrak d}), \Omega_\circ )$. 

\begin{theorem}[Eckmann and Osterwalder~\cite{EO}]
\label{EckOsw}
Let ${\mathfrak M}({\mathfrak d})\subset {\mathfrak M}({\mathfrak h})$ denote the   
von Neumann algebra generated by $\{W_F(h) \mid h\in {\mathfrak d}\}$.
Then the Tomita operator $S$ is the second qunatization of the closed, densely defined, 
conjugate linear operator $s$ over ${\mathfrak h}$ defined as
	\begin{equation} 
		\begin{matrix}
			s \colon & {\mathfrak d} & + & i {\mathfrak d}  
				& \to& {\mathfrak d} & + & i {\mathfrak d}  \\
				& f & + & i g & \mapsto & f & - & i g
						\end{matrix} \; \; . 
	\end{equation}
Moreover, if $s = j \delta$ is the polar decomposition of $s$, then
	\[
		J = \Gamma (j) \; , \qquad \Delta = \Gamma(\delta) \; , 
	\]
and $S = J \Delta$ the polar decomposition of $S$.
\end{theorem}

\goodbreak 
\subsection{Second quantisation}
Given a \emph{selfadjoint operator} $H$ acting on the one-particle space ${\mathfrak h} $, 
one can define operators $H_n$ acting
on the $n$-particle space $\Gamma^{(n)} ({\mathfrak h})$ by setting $H_0 \doteq 0$ and
	\[
		H_n \bigl( P_+ ( f_1 \otimes \ldots \otimes f_n) \bigr) \doteq
		P_+ \Bigl( \sum_{i=1}^n f_1 \otimes f_2 \otimes \ldots \otimes H f_i 
		\otimes \ldots \otimes f_n \Bigr) 
	\]
for all $f_i \in {\mathscr D} (H) \subset {\mathfrak h} $.  The operator $H_n $ extends to a 
selfadjoint operator on~$\Gamma^{(n)} ( {\mathfrak h})$. 
The direct sum of all $H_n$ is symmetric and therefore closable, and essentially selfadjoint, because there exists 
a dense set of analytic vectors, which is formed by the finite sums of symmetrised products of analytic 
vectors of $H$. The selfadjoint closure of the direct sum $\oplus_{n \in \mathbb{N}_\circ} H_n$ of  $H_n$ 
is called the second quantisation of~$H$.  It is denoted  by 
	\[
		{\rm d}\Gamma (H) \doteq \overline { \oplus_{n \in \mathbb{N}_\circ} H_n } \; . 
	\]

If $U$ is a {\em unitary operator} acting on ${\mathfrak h}$, then  $U_n$ 
acting on  $\Gamma^{(n)} ({\mathfrak h})$ is defined by 
	\[
		U_n \bigl( P_+ ( f_1 \otimes \ldots \otimes f_n) \bigr) 
		\doteq
		P_+ \bigl( U f_1 \otimes U f_2 \otimes \ldots  \otimes U f_n \bigr) \; , 
		\quad U_0 \doteq \mathbb{1}  \; , 
	\]
and by continuous extension. The second quantisation of $U$ is  
	\begin{equation}
	\label{gamma-u}
		\Gamma (U) \doteq \oplus_{n \in \mathbb{N}_\circ} U_n \;  . 
	\end{equation}
$\Gamma (U)$ is a unitary operator acting on $\Gamma ({\mathfrak h})$. 
If $t \mapsto U_t  = {\rm e}^{i t  H } $ is a strongly continuous group of unitary operators on ${\mathfrak h}$, 
then $\Gamma (U_t) = {\rm e}^{i t {\rm d}\Gamma (H)} $ holds on $\Gamma({\mathfrak h})$. 

\part{Free Quantum Fields}

\chapter{Classical Field Theory}
\label{ch:5}
\setcounter{equation}{0}

In this chapter, we will describe the \emph{classical dynamical systems} associated to the Klein-Gordon
equation on the de Sitter space. Each of them consists of a symplectic space together with an action of 
the Lorentz group in terms of symplectic maps. There are two equivalent classical dynamical systems, 
namely the \emph{covariant} (described in Section \ref{CCDS}) and the \emph{canonical} 
(described in Section \ref{CaCDS}), and they are connected by a symplectic map 
(given in Proposition \ref{cauchy-symplectic}), which encodes the solution 
of the \emph{Cauchy problem} (Theorem \ref{cauchyproblem}). 

Given a $C^\infty$-function $f$ with compact support, the fundamental solution $\mathbb{E}$ of the 
Klein-Gordon equation provides a $C^\infty$-solution $\mathbb{E}f$ (Theorem \ref{fundamental}). 
In fact, since $dS$ has a compact Cauchy surface, \emph{any} smooth solution of the Klein-Gordon 
equation is of this form (Theorem \ref{solutions}) and there is a \emph{one-to-one} correspondence 
between smooth solutions of the Klein-Gordon equation and equivalence classes of $C^\infty$-functions 
with compact support. These equivalence classes form a symplectic space $({\mathfrak k}(dS), \sigma)$, 
whose symplectic form $\sigma$ is given by the distributional bi-solution of the Klein-Gordon 
equation $\mathscr{E} ( x ,  y ) $, \emph{i.e.}, the kernel of $\mathbb{E}$. The dynamics on this symplectic 
space is given by the geometric action of the Lorentz group. More precisely, the pullback provides an 
action of $O(1,2)\ni \Lambda$ in terms of symplectic maps ${\mathfrak z}(\Lambda)$ 
on~$({\mathfrak k}(dS), \sigma)$ (see Proposition~\ref{Prop4-10}), thereby giving rise to the covariant 
dynamical system $ ({\mathfrak k}(dS), \sigma , {\mathfrak z}(\Lambda))$.

In an effort to provide explicit formulas, we may restrict the 
covariant dynamical 
system $ ({\mathfrak k}(dS), \sigma , {\mathfrak z}(\Lambda))$
to the subsystem $ ({\mathfrak k}(\mathbb{W}_1), \sigma , {\mathfrak z}(\Lambda_1(t))$. 
This is done in Section \ref{sec:5.4}. Note that the boost $t \mapsto \Lambda_1(t)$ 
leaves the double wedge $\mathbb{W}_1$ invariant. We express 
(see Lemma \ref{PropWedge})
the bi-solution $\mathscr{E} ( x ,  y ) $ in terms of a self-adjoint 
operator $\varepsilon$ (the generator 
of the boost) acting on the Hilbert 
space $L^2(S^1, \, | \cos \psi |^{-1}  \, r {\rm d} \psi)$
associated to the time-zero circle. The operator $\varepsilon$ takes 
only positive spectral values 
on $I_+$ and negative spectral values on~$I_-$. Although these 
explicit expressions 
for $\mathscr{E} ( x ,  y ) $ are valid only within a part of the support 
of the solutions of the Klein-Gordon 
equation, they are useful, as their regularity properties allow to 
extend $\mathbb{E}$ to sharp-time 
test-functions. In fact, it will be shown (in Section \ref{CCDS}  
and Section~\ref{CaCDS}) that every 
equivalence class in the symplectic space ${\mathfrak k}(dS)$ 
arises as the image of a sharp-time test-function. 
 
The canonical dynamical system associated to the Klein-Gordon 
equation is introduced in 
Section~\ref{CaCDS}; see, in particular, Proposition \ref{nocheinlabel}. 
It is described in some more detail in Proposition~\ref{porp4.13}.
As mentioned before, the symplectic map relating the canonical and the 
covariant dynamical system 
is given by the solution of the Cauchy problem. 

In Section \ref{SET}, we will briefly discuss the conserved currents associated to the   
(non-linear) Klein-Gordon equation on de Sitter space. This is most easily done by considering 
the classical {\em Lagrangian density}
	\begin{equation}
	\label{LagDen}
		{\mathcal L}(\mathbb{\Phi}) =  \frac{1}{2}  d \mathbb{\Phi} \wedge * d \mathbb{\Phi}  
		- \frac{\mu^2}{2} \mathbb{\Phi} * \mathbb{\Phi}  - P( \mathbb{\Phi})  * 1  \; , 
	\end{equation}
which gives rise to the non-linear Klein-Gordon equation; see Section \ref{sec:5.1}.
Here~$\mathbb{\Phi}$ is a real valued scalar field and the  polynomial $P$ is bounded 
from below 
and $\mu >0$ is the mass parameter\footnote{How the 
constant $\mu$ appearing in \eqref{3.25} is 
related to the physically observable mass of a particle on de Sitter space will be 
investigated in \cite{Urs}.} appearing in \eqref{4.1.6}. 
 
\section{The classical equations of motion}
\label{sec:5.1}

Let ${\rm K}$ be a compact submanifold of $dS$. The action associated to 
the Lagrangian density \eqref{LagDen} and $K$ is 
	\begin{align}
	\label{action}
	S({\rm K}, \mathbb{\Phi}) & 
	=  \frac{1}{2} \int_{\rm K}  \Bigl(  d \mathbb{\Phi} \wedge * d \mathbb{\Phi}  
		-   \mu^2 \mathbb{\Phi} * \mathbb{\Phi}  \Bigr) 
		-   \int_{\rm K}  P( \mathbb{\Phi})  * 1  \; . 
	\end{align}
The (non-linear) Klein--Gordon equation can be recovered by demanding 
that for every such ${\rm K}$,
the action $S({\rm K}, \mathbb{\Phi}) $ is stationary with respect to smooth variations 
$\mathbb{\Phi} \mapsto \mathbb{\Phi} + \delta \mathbb{\Phi} $ of the field 
$\mathbb{\Phi} $ that vanish on the boundary $\partial K$ of $K$.  In other words, 
we require that
	\[
			0 = \frac{\delta S({\rm K}, \mathbb{\Phi})}{\delta \mathbb{\Phi}(y) } 
	\]
for every such ${\rm K}$. 
The resulting \emph{Euler-Lagrange equation}
	\begin{equation}
		\label{equation-motion}
		d \frac{\partial {\mathcal L}}{\partial (d \mathbb{\Phi})} 
		-  \frac{\partial {\mathcal L}}{\partial \mathbb{\Phi}}  = 0
	\end{equation}
is the \emph{equation of motion}\index{equation of motion}
	\[
		{\rm d} * {\rm d} \mathbb{\Phi} + \left( \mu^2 \mathbb{\Phi} +
		P' ( \mathbb{\Phi})  \right) * 1 = 0 
	\]
on $dS$. We recall that the co-differential\footnote{Here $m$ denotes 
the dimension of the manifold  
and $E_p$ denotes the set of covariant, totally 
anti-symmetric tensors of $p$-th degree.}
$\delta \colon E_p \to E_{p-1}$
	\[
		\delta \doteq * d * (-1)^{m(p+1)+s} \; , 
		\qquad \text{with} \quad (-1)^s = \frac{ \det g }{ | \det g | }  \; , 
	\]
can be coupled with the exterior derivative ${\rm d}$ to construct the Laplace-Beltrami 
operator\footnote{In local coordinates, the Laplace-Beltrami operator 
$\square_{dS}$ equals $ |g|^{-1/2}\partial_\mu g^{\mu\nu}|g|^{1/2}\partial_\nu $,  
with $ |g|\equiv | {\rm det} g| $. }
	\[
		\square_{dS} = d \delta + \delta d \; . 
	\]
As $\Phi \in E_0$, we have $\delta \mathbb{\Phi}=0$ and therefore 
the equation of motion may be rewritten as 
\label{squarepage}
	\begin{equation}
		\label{3.25}
		(\square_{dS}+\mu^2) \mathbb{\Phi}  = - P'  (\mathbb{\Phi} ) \; , 
		\qquad \mathbb{\Phi} \in C^\infty (dS) \;  ,   \quad \mu>0 \; .
	\end{equation}
See \cite{GaT, Urs} for a discussion of several interpretations of~$\mu$ found in the literature. 

In the sequel, we keep $\mu>0$ fixed, and although almost all quantities we encounter 
depend on $\mu$, we will suppress this dependence on $\mu$ in the notation. 

\section{Conservation laws}
\label{SET}

The advantage of the Lagrangian formulation is that any one-parameter subgroup, which leaves 
the Lagrangian density invariant, gives rise to a conservation law\footnote{In \cite[p.~269]{FHN}, 
the authors have chosen $K= W_1$, and thus the action $S({\rm K}, \mathbb{\Phi}) $ yields
	\[
		S(W_1, {\mathbb \Phi}) = \frac{1}{2} 
		\int_{W_1} r^2 \cos \psi \, {\rm d} \psi  \, {\rm d} t  \; 
		\Bigl(  \underbrace{r^{-2} \cos^{-2} \psi \bigl( \tfrac{ \partial {\mathbb \Phi} }{\partial t}\bigr)^2}_{
		= \bigl( \tfrac{ \partial {\mathbb \Phi} }{\partial x_0}\bigr)^2}  - 
		r^{-2} \bigl( \tfrac{ \partial {\mathbb \Phi} }{\partial \psi}\bigr)^2 - \mu^2 {\mathbb \Phi}^2 
		- 2  P (\mathbb{\Phi}) \Bigr) \; .
	\]
The invariance with respect to translations of $t$ yields the conserved quantity 
\begin{align*}
	L_{1 \upharpoonright I_+ } & =  
	\int_{I_+} r^2  \cos \psi \, {\rm d} \psi \;  {T^0}_{0} \; . 
\end{align*}
In the last equation we have used  $n= r^{-1} \cos^{-1}\psi \partial_t$ and $n \mathbb{\phi} = \mathbb{\pi}$. 
Here $n$ denotes unit normal, future pointing vector field, 
restricted to the Cauchy surface~$S^1$.}. 
The variation of the 2-form ${\mathcal L}$ is 
	\[
		\delta {\mathcal L} =  \delta \mathbb{\Phi} \wedge 
		\left[ \tfrac{\partial {\mathcal L} }{ \partial \mathbb{\Phi}}  
		-  d \tfrac{\partial {\mathcal L} }{ \partial (d \mathbb{\Phi}) } \right] 
		+ d \Bigl(  \delta\mathbb{\Phi} \wedge  
		\tfrac{\partial {\mathcal L} }{ \partial (d \mathbb{\Phi}) } \Bigr) \; , 
	\]
and the equations of motion \eqref{equation-motion} imply that 
	\[
		\delta {\mathcal L} = 
		d \Bigl(  \delta\mathbb{\Phi} \wedge  \tfrac{\partial {\mathcal L} }{ \partial (d \mathbb{\Phi}) } \Bigr) \; . 
	\]
If the variation\footnote{The \emph{interior product} $(\omega, X) \to i_X \omega$ is a mapping 
from $E_p \times T_0^1$ to $E_{p-1}$ (here the elements of~$T_0^1$ are vector fields). It is linear in both factors and 
determined by 
\begin{itemize}
\item [$i.)$] $i_X \omega= ( \omega \mid X )$ for $\omega \in E_1$; and
\item [$ii.)$] $i_X (\omega \wedge \nu)= (i_X \omega) \wedge \nu + (-1)^p \omega \wedge i_X  \nu$ for $\omega \in E_p$. 
\end{itemize}} 
results from a Lie derivative $L_X = d \circ i_X + i_X \circ d $, with $X$ some vector field, then 
	\[
		\delta {\mathcal L} = L_X {\mathcal L} = d ( i_X {\mathcal L}) 
	\]
as the exterior derivative of the 2-form ${\mathcal L}$  vanishes in two-dimensional space. It follows that 
	\[
		 \delta\mathbb{\Phi} \wedge  \frac{\partial {\mathcal L} }{ \partial (d \mathbb{\Phi}) }
		- i_X {\mathcal L}
	\]
is a closed 1-form, or, using the Hodge $*$,  a \emph{conserved current}. 

\begin{theorem}[Noether]
\label{noether} 
Let $X$ be a vector field with $\delta \mathbb{\Phi} = L_X \mathbb{\Phi}$ and 
$\delta {\mathcal L} = L_X {\mathcal L}$. It follows that
	\[
		d \left[  L_X \mathbb{\Phi} \wedge \tfrac{\partial {\mathcal L}}{\partial (d \mathbb{\Phi})} 
		- i_X {\mathcal L} \right] =:  d *  T_X = 0   \; . 
	\]
\end{theorem} 

The Killing vector fields on $dS$ are given by $\partial_t $ (within the double cone $\mathbb{W}_1$, 
using the coordinates introduced in \eqref{w1psi}) and $\partial_\psi $. 
If one integrates $* T_{\partial_t} $ (respectively, 
$* T_{\partial_\psi} $) over the space-like surface $S^1$, than one finds
the conserved quantities\footnote{The integral of an $n$-from $\omega$ with compact support over an 
$n$-dimensional sub-manifold $N$ of a manifold of dimension $m$ is defined as the integral of the restriction of 
$\omega$ to $N$:
	\[
		\int_N \omega \doteq \int_N \omega_{| N} \; . 
	\]
Here the restriction $\omega_{| N}$ of the form $\omega$ is defined by introducing local coordinates so that 
$x_{n+1} = x_{n+2} = \cdots = x_m = 0$, then defining  ${\rm d}{x_{n+1}}_{| N} = {\rm d}{x_{n+2}}_{| N} 
= \cdots = {\rm d}{x_m}_{| N}  = 0$ and
letting the restriction commute with $+$ and $\wedge$.} 
	\[
		L^{\scriptscriptstyle Class.}_1 
		= \int_{S^1} * T_{\partial_t}  \quad \text{and} \quad
		K^{\scriptscriptstyle Class.}_0 
		= \int_{S^1}  * T_{\partial_\psi} \; , 
	\]
which generate the $\Lambda_1$-boosts 
and the rotations around the $x_0$-axis, respectively.  

We find it worth while to derive explicit expressions.
As $L_X \mathbb{\Phi} = i_X d \mathbb{\Phi} \in E_0$, we find \cite[Ch. 8.1.19, Exercise 4]{Thirring}
	\[
		* T_X =    \bigl( i_X d \mathbb{\Phi} \bigr)\wedge * d \mathbb{\Phi} - \tfrac{1}{2}
		i_X \bigl(  d \mathbb{\Phi}  \wedge * d \mathbb{\Phi} - 
		  \mu^2 \mathbb{\Phi} * \mathbb{\Phi} \bigr) + P( \mathbb{\Phi}) i_X * 1  \; . 
	\]
By definition, $ i_{\partial_t} \circ d \mathbb{\Phi}  
= \partial_t \mathbb{\Phi} $. 
Rewriting $\partial_t\mathbb{\Phi}$ as 
	\[
		\partial_t\mathbb{\Phi}= r\cos\psi \, n \mathbb{\Phi} \equiv r\cos\psi \, \mathbb{\pi} \; , 
	\]
where $n$ is the future directed normal vector field to the time-circle $x_0 =0$. Hence
	\begin{align}
		L^{\scriptscriptstyle Class.}_1  
		& = \int_{S^1} \Bigl( L_{\partial_t} \mathbb{\Phi} \wedge \tfrac{\partial {\mathcal L}}{\partial (d \mathbb{\Phi})} 
		- i_{\partial_t} {\mathcal L} \Bigr)
		\nonumber \\
		& =  \int_{S^1}   r\cos\psi
		\;  \mathbb{\pi} * {\rm d} \mathbb{\Phi}  
		- \frac{1}{2} \int_{S^1}   
		i_{\partial_t} \bigl( d \mathbb{\Phi} \wedge * d \mathbb{\Phi} \bigr)
				+ \int_{S^1} \Bigl(\frac{\mu^2}{2} \mathbb{\Phi}^2  + P( \mathbb{\Phi}) \Bigr)i_{\partial_t} * 1 
		\nonumber \\
		& =   \int_{S^1}   r\cos\psi
		\;  \mathbb{\pi}  \partial_{x_0} \mathbb{\Phi}  * {\rm d}x_0 
		\nonumber \\
		& \qquad  \qquad - \frac{1}{2} \int_{S^1}   
		i_{\partial_t} \bigl( d \mathbb{\Phi} \wedge * d \mathbb{\Phi} \bigr)
				+ \int_{S^1} \Bigl( \frac{\mu^2}{2} \mathbb{\Phi}^2  + P( \mathbb{\Phi}) \Bigr) i_{\partial_t} * 1 
		\nonumber \\
		& 
		= \int_{S^1} r^2 \, \cos\psi \; {\rm d} \psi \;  
							\Bigl( \frac{\mathbb{\pi}^2}{2}  + \frac{1}{2r^2}(\partial_\psi\mathbb{\Phi})^2   +
							\frac{\mu^2}{2} \mathbb{\Phi}^2 +  P(\mathbb{\Phi})\Bigr) \; . 
	\end{align}
In the last equation, the Hodge $*$ contributes a factor $|g|^{1/2} = r$. 
We have used that 
	\[
		\partial_{x_0} \mathbb{\Phi}  * {\rm d}x_0 = \mathbb{\pi} \, r {\rm d} \psi 
		\quad 
		\text{and}
		\quad \int_{S^1} * {\rm d} \psi = 0 \; . 
	\]
Similarly, we compute
the formula for the angular momentum   
	\begin{align}
		K^{\scriptscriptstyle Class.}_0 
		 & = \int_{S^1}  \Bigl( L_{\partial_\psi} \mathbb{\Phi} \wedge \tfrac{\partial {\mathcal L}}{\partial (d \mathbb{\Phi})} 
		- i_{\partial_\psi} {\mathcal L} \Bigr)
		\nonumber \\
		& =  \int_{S^1}   \partial_\psi \mathbb{\Phi} * d \mathbb{\Phi}  
		\nonumber \\
		& =    \int_{S^1} \,  r \, {\rm d} \psi \;   \mathbb{\pi} \, (\partial_\psi \mathbb{\Phi})  \; .
		\label{angular-momentum}
	\end{align}
Both in the first equality and in the second equality in \eqref{angular-momentum} we have used 
that ${{\rm d}x_0}_{| S^1} =0$. We will encounter similar expressions for the quantum fields 
in Section~\ref{stress-energy-tensor}; see, in particular, Lemma \ref{Q-conserved-quantities}.

\bigskip
A convenient coordinate system to derive explicit expressions 
for the stress-energy tensor on de Sitter space
was introduced in \eqref{w1psitau}. The corresponding metric 
was given in \eqref{w1psitau-metrik}. The conserved currents 
	\[
		T_{0} \equiv T_{\partial_{x_0}} \; , \qquad T_{1}  \equiv T_{\partial_\psi}  \; , 
	\]
can now be expressed in terms of the \emph{classical stress-energy tensor} 
$T_{\upsilon \nu}$,   
	\[
		T_{\upsilon} = T_{\upsilon \nu} {\rm d}x^\nu \; ,  
		\qquad x^0 \equiv x_0 \, , \quad x^1 \equiv \psi \; ,
	\]
given by
\begin{align*}
	{T^{\upsilon}}_{\nu} & = \partial^{\upsilon} \mathbb{\Phi} 
	\partial_\nu \mathbb{\Phi}  -  g^{{\upsilon}\kappa} 
	g_{\kappa {\nu}}{\mathcal L}(\mathbb{\Phi}) \\
        &= \partial^{\upsilon} \mathbb{\Phi} \partial_\nu \mathbb{\Phi} 
        - \tfrac{1}{2} {\delta^{{\upsilon}}}_{\nu} 
	\bigl(  \partial^\kappa \mathbb{\Phi}\partial_\kappa \mathbb{\Phi} \bigr)
	+  {\delta^{\upsilon}}_{\nu}  \Bigl( \frac{ \mu^2}{2} \mathbb{\Phi}^2  
	+ P (\mathbb{\Phi}) \Bigr) 	\; .  
\end{align*}
The tensor ${T^{\upsilon}}_{\nu}$ 
describes the flux of the ${\upsilon}$-th component of the 
conserved energy-momentum vector across a surface 
with constant $x_\nu$ coordinate (see, \emph{e.g.}, \cite[p.~35]{Thirring}). 
In particular, 
	\begin{align*}
		{T^0}_{0}  &= \frac{1}{2} \left( \mathbb{\pi}^2 
		+  r^{-2} \bigl( \partial_\psi  \mathbb{\Phi} \bigr)^2 + 
					\mu^2 \mathbb{\Phi}^2 \right)  + P  (\mathbb{\Phi} ) \; ,  
	\end{align*}
with $\mathbb{\pi} = \tfrac{\partial}{\partial x_0} \mathbb{\Phi}$, and, 
using $g^{11}= r^{-2}$, 
	\begin{equation}
	\label{T01}
		{T^{1}}_{0} = r^{-2} \partial_\psi \mathbb{\Phi} \, 
		\partial_{x_0} \mathbb{\Phi} \,
		= r^{-2} \partial_\psi \mathbb{\Phi} \, \mathbb{\pi}  \; .   
	\end{equation}
Hence
	\[
		L^{\scriptscriptstyle Class.}_1  
		= \int_{S^1} \, r^2 \, \cos\psi \; {\rm d} \psi \;    {T}_{00}  
		\qquad \text{and} \qquad 
		K^{\scriptscriptstyle Class.}_0 
		= \int_{S^1} \,   r \; {\rm d} \psi \; {T}_{10} \; .
	\]

\begin{remark} 
\label{classical-energy}
Integrating ${T}_{00}$ over the time-zero circle $S^1$ yields a positive quantity, 
	\[
		\int_{S^1} r\, {\rm d} \psi \; {T}_{00} (\psi)   > 0 \; ,  
	\]
which may be interpreted as the \emph{energy} for the classical $P(\varphi)_2$ model on 
the \emph{Einstein universe} (see, \emph{e.g.}, \cite{Fe1}\cite{Fe2}), \emph{i.e.}, 
the space-time of the form $S^1 \times \mathbb{R}$ with the metric induced from the 
ambient Minkowski space $\mathbb{R}^{1+2}$. 
\end{remark}

Although there are interesting results concerning the  \emph{non-linear} Klein-Gordon equation 
in two space-time dimensions (see, \emph{e.g.}, \cite{Delort-1, Delort-2, Delort-3, Delort-4, 
Delort-5, GV, Lebeau, Yagdjian, Yagdjian-2}), we will concentrate on free fields for the rest of this chapter. 
The reader interested in the case of interacting quantum fields may want to consult Lemma~\ref{lm:10.2.1}.

\section{The covariant classical dynamical system}
\label{CCDS}

As mentioned in Section \ref{geodesics}, the de Sitter space-time $dS$ is  globally hyperbolic. Thus the
inhomogeneous Klein--Gordon equation
	\begin{equation}
		 \label{4.3n}
			 (\square_{dS}+\mu^2)  \mathbb{\Phi}   
		 =  f \; , \qquad f \in{\mathcal D}_\mathbb{R} (dS) \; , 
	\end{equation}
has smooth solutions, which are uniquely specified by fixing their support properties  
(see \cite{C-B,D77, L, Lich}):

\begin{theorem}
\label{fundamental}
There exist unique operators 
	\[
		\mathbb{E}^{\pm} \colon{\mathcal D}_{\mathbb{R}} (dS) \to C^\infty (dS) 
	\]
such that  $\mathbb{E}^\pm f $ is a solution of the \index{inhomogenous Klein-Gordon equation}
inhomogenous equation \eqref{4.3n} with
	\[
		{\rm supp\,}  (\mathbb{E}^\pm f)   \subset \Gamma^\pm ( {\rm supp\,} f) \quad  \text{and} \quad 
		{\rm supp\,} ( \mathbb{E}^\pm f ) \cap \Gamma^\mp ( {\rm supp\,} f) \;  \text{compact}. 
	\] 
\end{theorem}

The $C^\infty$-functions $\mathbb{E}^{\pm}f$ are called the \emph{retarded}\index{retarded solution} and the 
\emph{advanced solution}\index{advanced solution} 
of the equation \eqref{4.3n}, respectively. 
The difference between the retarded and the advanced solution of the inhomogeneous equation \eqref{4.3n}, 
namely 
	\begin{equation}
	\label{Ef}
	\mathbb{\Phi}=\mathbb{E} f \; , \qquad \text{with} \quad  \mathbb{E} = \mathbb{E}^{+} - \mathbb{E}^{-} \; , 
	\end{equation}
is a solution of the homogenous Klein--Gordon equation \eqref{3.25}.

\begin{remark}
For comparison, we briefly recall the situation on Minkowski space $\mathbb{R}^{1+1}$.
After Fourier transformation, the inhomogeneous equation
	\[
		(\square_{\mathbb{R}^{1+1}} + m^2)G = - \delta \; , 
	\]
takes the simple form $(- {\tt P}_0^2 + {\tt P}_1^2 + m^2) \widetilde {G} = -1$; the latter has the 
\emph{retarded} and \emph{advanced propagators} \index{retarded propagator}\index{advanced propagator} 
as its solution. In other words, 
	\[
		{\mathscr G}_{adv}( {\tt x}, {\tt y} ) = \lim_{\epsilon \downarrow 0} \frac{1}{(2 \pi)^2} 
			\int {\rm d}^2 p \frac{ {\rm e}^{- i {\tt p} \cdot ({\tt x}- {\tt y})}}
			{({\tt p}_0 - i \epsilon)^2 - {\tt p}_1^2 - m^2} \; , 
	\]
and
	\[
		{\mathscr G}_{ret}( {\tt x}, {\tt y} ) = \lim_{\epsilon \downarrow 0} \frac{1}{(2 \pi)^2} 
			\int {\rm d}^2 p \frac{ {\rm e}^{- i {\tt p} \cdot ({\tt x}- {\tt y})}}
			{({\tt p}_0 + i \epsilon)^2 - {\tt p}_1^2 - m^2} \; . 
	\]
The difference 
	\[
		{\mathscr G}( {\tt x}, {\tt y} ) \doteq {\mathscr G}_{ret}( {\tt x}, {\tt y} ) - {\mathscr G}_{adv}( {\tt x}, {\tt y} )
	\] 
is a \emph{bi-solution}\index{bi-solution} of the Klein-Gordon equation. 
\end{remark}

\bigskip
As we will see next, {\em any} smooth solution of \eqref{3.25} with $P(\lambda) =0$
is of the type~\eqref{Ef}.

\goodbreak
\begin{theorem}[B\"ar, Ginoux and Pf\"affle \cite {BGP},Theorem 3.4.7]
\label{solutions}
\quad
\begin{itemize}
\item[$ i.)$]Any smooth solution $\mathbb{\Phi}$ of the free Klein--Gordon equation \eqref{3.25} 
may be written in the form 
	\[
	\mathbb{\Phi} = \mathbb{E} f \; , \qquad \text {for some} \; f \in {\mathcal D}_{\mathbb{R}}(dS) \; ; 
	\]
and, given any neighbourhood \footnote{In Section~\ref{sec:10.2-new}  
we demonstrate that for the ${\mathscr P}(\varphi)_2$ 
model on the de Sitter space, the expectation values of all observables can 
be predicted from the expectation values of observables, 
which can be measured within an {\em arbitrarily small} time interval. Thus 
the interacting quantum theory on the de Sitter space satisfies the Time-Slice
Axiom~\cite{CF}.}  ${\mathscr N}$  
of a Cauchy surface ${\mathcal C}$, one may choose such an $f \in {\mathcal D}({\mathscr N})$. 
\item[$ ii.)$] We have 
	\[
		\ker \mathbb{E} = (\square_{dS}+\mu^2) {\mathcal D}_{\mathbb{R}} (dS) \; . 
	\]
In consequence, the {\em space of smooth real-valued solutions} 
$\mathbb{E}  {\mathcal D}_{\mathbb{R}} (dS)$ is in one-to-one correspondence  
with the space of equivalence classes
\label{kldypage-a}
	\begin{equation*}
	{\mathfrak k}(dS) \doteq  
	{\mathcal D}_{\mathbb{R}} (dS) / 
		(\square_{dS}+\mu^2) {\mathcal D}_{\mathbb{R}} (dS)  \; .  
	\end{equation*}
\end{itemize}
\end{theorem}

\goodbreak
Taking advantage of the properties $ i.)$ and $ ii.)$, we 
can define a projection
	\begin{align*}
			\mathbb{P} \colon {\mathcal D}_\mathbb{R}(dS) 
					& \to   {\mathfrak k} (dS) \nonumber \\
			 f & \mapsto   [f] \; . 
	\end{align*}
The one-to-one correspondence mentioned above now takes the form 
	\[ 
	{\mathfrak k} (dS) \ni [f] \longleftrightarrow \mathbb{f} \in \mathbb{E}  {\mathcal D}_{\mathbb{R}} (dS)  \; , 
	\]
with $[f]  \doteq \{ f + (\square_{dS}+\mu^2)h \mid h \in {\mathcal D}_{\mathbb{R}}(dS) \}$
and $\mathbb{f} \doteq \mathbb{E} f$. 

\begin{definition}
\label{symp-sub}
Subspaces of $ {\mathfrak k} (dS) $ associated to open space-time regions ${\mathcal O} \subset dS$ 
are defined by restricting $\mathbb{P}$ to ${\mathcal D}_\mathbb{R}({\mathcal O})$, \emph{i.e.}, 
	\[
			{\mathfrak k} ({\mathcal O}) 
			\doteq{\mathcal D}_\mathbb{R}({\mathcal O})/ \ker \mathbb{E}  \;. 
	\]
\end{definition}

${\mathfrak k} ({\mathcal O})$ will be used in Chapter \ref{2Q} to define {\em local} von Neumann algebras.

\bigskip
Embedding $ C^\infty (dS)$ into ${\mathcal D}'(dS)$ 
(see \cite[Sect.~2]{Fe1}\cite{Fe2}), the map~$\mathbb{E}  \colon{\mathcal D}_{\mathbb{R}} (dS) \to C^\infty (dS) $ 
gives rise to a bi\-distribution ${\mathcal E}$ on $dS\times dS$, 
	\begin{align} 
		\label{integralkernel}
			{\mathcal E}  ( f,g) \; &
			\doteq \int {\rm d} \mu_{dS} ( x) \;  f (x)( \mathbb{E} g)(x) 
			\nonumber \\
				&\doteq 
			\int {\rm d} \mu_{dS} ( x) {\rm d} \mu_{dS} ( y) \; 
			f ( x ) \mathscr{E} ( x ,  y )  
			g ( y  )  \; ,  
	\end{align}
antisymmetric in $f, g \in{\mathcal D}_\mathbb{R} 
(dS)$, whose kernel $\mathscr{E} ( x ,  y ) $, called the \emph{fundamental 
solution}\index{fundamental solution}, is a weak bisolution for the Klein--Gordon equation, 
\label{ccCfpage}
\label{vnpage}
	\begin{equation}
		 \label{cmzero} 
		  {\mathcal E}  \left( (\square_{dS}+\mu^2)
		  f,g \right)  
		  = {\mathcal E}  \left( f, 
		  (\square_{dS}+\mu^2)g \right) = 0 \; , 
	\end{equation}
with \index{initial data}
initial data\footnote{Micro-local analysis shows  that $\mathscr{E}$ and
its normal derivatives can be restricted to 
${{\mathcal C}} \times {{\mathcal C}}$, see~\cite{Hoerm}.}
	\begin{align}
		{\mathscr{E}}_{ \upharpoonright {\mathcal C} \times {\mathcal C} } 
				& =  0 \; , \label{eqESigma} \\  
				(n_\ell \mathscr{E})_{ \upharpoonright {{\mathcal C}} 
				\times {{\mathcal C}}}
				&=  -\delta_{{\mathcal C}} \; .  
				\label{eqESigma2}				
	\end{align}
Here $n_\ell $ denotes the vector field $n$ acting on  
the left variable $ x$ in $\mathscr{E} ( x,  y)$
and $\delta_{{\mathcal C}}$ is the integral kernel of the unit operator with 
respect to the 
induced measure on the Cauchy surface ${{\mathcal C}}$. 

\begin{remark}
Note that
	$
	\overline { {\mathcal W}^{(2)} (  x_1,  x_2) } = {\mathcal W}^{(2)} (  x_2,  x_1)
	$,
and
	\begin{equation}
	\label{comfun}
		{\mathcal W}^{(2)} (  x_1,  x_2) - {\mathcal W}^{(2)} (  x_2,  x_1) 
		=  2 i \Im {\mathcal W}^{(2)} (  x_1,  x_2) \; .
	\end{equation}
The commutator function $2  \Im {\mathcal W}^{(2)} (  x_1,  x_2)$ 
is an anti-symmetric distribution on $dS \times dS$, which satisfies the Klein--Gordon equation in both entries, 
with initial conditions described in (\ref{eqESigma}) and (\ref{eqESigma2}).  
In fact, for $x_1,x_2$ space-like, this 
is obvious and for $x_1=x_2$ this follows from \cite[page 199]{Lebedev}:
	\[
	P_{s^+} (-1 +i0) - P_{s^+} (-1 -i0) = 2 i \sin s^+ \pi \; . 
	\]
In other words, 
	\[
	{\mathcal W}^{(2)} (  x_+,  x_-) - {\mathcal W}^{(2)} (  x_-,  x_+) =  c_\nu \cdot 2 i \sin s^+ \pi = - i \; . 
	\]
The constants introduced in
\eqref{mass-shell-ft} were chosen to ensure that 
	\[
		\frac{\partial}{\partial x_0} 2\Im {\mathcal W}^{(2)} (  x,  y) = -  \delta_{S^1} \; .  
	\]
As before, $\delta_{S^1}$ is the integral kernel of the unit operator with 
respect to the induced measure on $S^1$.  It follows that 
	\begin{equation}
		\label{comfkt}
			\mathscr{E} (  x_1,  x_2) = 2 \Im {\mathcal W}(  x_1,  x_2) \;  
	\end{equation}
is the kernel of the {\em commutator function} defined in  \eqref{integralkernel}. 
To show that $ \mathscr{E} (  x_1,  x_2) $ as given in (\ref{comfkt})
is invariant under the rotations~$R_{0}(\alpha)$, $\alpha \in [0, 2 \pi)$, choose a 
circle~$\gamma_0$  
on $\partial V^+$ with $p_0 = 1$  in~\eqref{tpf-1}. Rotation invariance of the propagator  
now follows from $ z_1 \cdot  R_0 p = R_0^{-1}z_1 \cdot   p$ and rotation 
invariance of the measure ${\rm d} \mu_{\gamma_0} 
= \frac{{\rm d} \alpha}{2}$; see~\eqref{lambda2-s}. 
\end{remark}

The map 
	\[
		{\mathcal D}(dS) \ni f \mapsto \mathbb{f} = \mathbb{E} f 
	\]
can now be viewed as  a convolution\footnote{On Minkowski space,  Fourier transformation
converts a convolution 
in position space to a multiplication in momentum space.
For the situation on $dS$, see Section \ref{Harmanaly}.} of a test function $f$ 
with the kernel $\mathscr{E}$, \emph{i.e.}, 
	\begin{equation}
		\label{feb-1}
			\mathbb{f} ( x ) 
				\doteq \int {\rm d} \mu_{dS} ( y ) \; \mathscr{E}( x ,  y ) f ( y ) \; , 
			\qquad  f \in{\mathcal D}_\mathbb{R} (dS) \; .  
	\end{equation}
The domain of $\mathbb{E}$ extends to distributions of the form 
	\begin{align}
	 \label{f-dist}
	 f (x) \equiv (\delta \otimes h) (x) &=	\delta (x_0)   h  (\psi ) \;  , 
	  \\
	\label{g-dist}
	g (x) \equiv (\delta' \otimes h) (x) &= \delta'  (x_0)  h  ( \psi )\;  , 
	\end{align} 
with $ h\in{\mathcal D}_{\mathbb{R}}(S^1)$ and $x \equiv x ( x_0, \psi)$, using the coordinates 
introduced in \eqref{w1psitau}. The Lorentz invariant measure is ${\rm d} \mu_{dS} (x_0,\psi) 
= {\rm d} x_0\,   r   {\rm d} \psi $. 

Eq.~(\ref{feb-1}) implies  that $\mathbb{f} (x) =0$ for all $ x \in dS$,  iff  
	\begin{equation}
		 \label{eqEf}
		f \in 
		\ker {\mathcal E}  
		\doteq \bigl\{ f \in{\mathcal D}_\mathbb{R} (dS) 
		\mid  {\mathcal E}  (g, f)= 0 \; \; 
		\forall g \in{\mathcal D}_\mathbb{R} (dS) \bigr\}\;  .
	\end{equation}
In other words, $\ker \mathbb{E} = \ker {\mathcal E} $. Consequently, the 
bidistribution ${\mathcal E}$ provides a \emph{non-degenerated} symplectic form \index{symplectic form}
$\sigma$ on the space of solutions ${\mathfrak k} (dS)$:
\label{sigmapage}
	\begin{equation}
	 	\label{eqSplForm}
		\qquad \qquad  \qquad \qquad 
		\sigma \bigl( [f] ,  [g] \bigr)
		\doteq {\mathcal E} (f,g) \; , 
		\qquad  f, g 
		\in{\mathcal D}_\mathbb{R}(dS)\; . 
	\end{equation}
As a consequence of \eqref{cmzero}, the right hand side does not dependent on the choice 
of the representatives in the equivalence classes $[f]$ and $[g]$.  
Thus $({\mathfrak k} (dS), \sigma)$ is a \index{symplectic vector space} symplectic vector space.

\goodbreak
\begin{lemma} 
\label{Lm3.8}
Let $f \in{\mathcal D}_\mathbb{R}({\mathcal O})$, 
${\mathcal O}\subset dS$ an open region. Then 
$\mathbb{f} = \mathbb{E} f$ is a solution of the Klein--Gordon equation with
	\[
		{\rm supp\,} (\mathbb{f} ) 
		\subset\Gamma^+ ({\mathcal O}) 
		\cup \Gamma^- ({\mathcal O}) \; .  
	\]
In particular, if ${\mathcal O} \subset W$, then 
${\rm supp\,} (\mathbb{f} ) 
\subset dS \setminus \overline{\, W'}$.
\end{lemma}

\begin{proof}
The support properties of~$\mathbb{E}^{\pm}$ force 
${\mathcal E} ( f,g)$  to vanish, 
whenever the support of $f$ is space-like separated from that 
of $g$. Thus, for $ y \in dS$ fixed,  the distribution 
$ x \mapsto \mathscr{E} ( x ,  y )$
has support in  $\Gamma^+ ( y) \cup \Gamma^- ( y ) $. 
The final statement follows from  this fact as well.
\end{proof}

\goodbreak 
Exploring the one-to-one correspondence between 
${\mathfrak k} (dS) \ni [f]$ and $ \mathbb{f} \in \mathbb{E}  {\mathcal D}_{\mathbb{R}} (dS) $, 
this result can be rephrased in the following way.  

\begin{lemma} 
\label{Lm4-9}
Let  $f \in{\mathcal D}_\mathbb{R}({\mathcal O})$, ${\mathcal O}\subset dS$ a bounded open region, and 
$g \in{\mathcal D}_\mathbb{R}({\mathcal O}')$, where ${\mathcal O}'$ denotes the space-like complement
of ${\mathcal O}$. Then 
	\begin{equation}
	\label{def:sigma}
	 		\sigma \bigl(   [f]  , [g] \bigr)= 0 \; .  
	\end{equation}
\end{lemma}
 
\begin{lemma}
Let $\mu^2  = \frac{1}{4r^2} + m^2$, \emph{i.e.}, $\nu^2 = m^2 r^2$. It follows that 
	\[
		\ker {\mathcal F}_{+ \upharpoonright \nu} = {\ker {\mathbb P}} 
		= (\square_{dS}+\mu^2) {\mathcal D}_{\mathbb{R}} (dS) \; . 
	\]
\end{lemma}

\begin{proof}
If $f \in \ker {\mathbb P}$, 
then \eqref{cmzero} implies that
there exists $g \in {\mathcal D}_{\mathbb{R}} (dS)$ 
such that $f = (\square_{dS}+\mu^2) g$. 
Evaluate $ {\mathcal F}_{+ \upharpoonright \nu} \bigl( (\square_{dS}+\mu^2) g \bigr)$ 
using the definition of the Fourier--Helgason transform 
(see  \eqref{eqPW-new}) and  
	\[ 
		(\square_{dS}+\mu^2) (  x_+  \cdot  p )^{s^\pm} = 0
	\]
for $s^\pm$ given by \eqref{dd1} with $\zeta^2= \mu^2 r^2$. This shows that  
$\ker {\mathcal F}_{+ \upharpoonright \nu} \supset {\ker {\mathbb P}}$. The inclusion 
$\ker {\mathcal F}_{+ \upharpoonright \nu} \subset {\ker {\mathbb P}}$ follows
from the fact that 
${\mathcal E} (f, g) = 2 \Im \langle \widetilde f , \widetilde g \rangle_{ \widetilde{\mathfrak h} 
(\partial V^+)} \; $; see \eqref{comfkt} above. 
\end{proof}

\begin{proposition}
\label{Prop4-10}
The symplectic space $({\mathfrak k}(dS), \sigma)$ carries a representation 
	\[
		\Lambda \mapsto {\mathfrak z} (\Lambda)\; ,  \qquad \Lambda \in O(1,2) \; , 
	\]
of the Lorentz group, defined by  ${\mathfrak z} (\Lambda) [f] = [ \Lambda_* f]$. 
\label{ccTpage}
\end{proposition}

\begin{proof}
The group of isometries $\Lambda \in O(1,2)$ of $dS$ gives rise to a
group of symplectic transformations $\Lambda \mapsto T_\Lambda$ on $({\mathfrak k} (dS), \sigma)$  
induced by the push-forward~$\Lambda_*$, which maps
\label{lambdasternpage}
	\begin{equation}
		\label{eqTt}
		f+\ker \mathbb{E} \mapsto \Lambda_* f+\ker \mathbb{E}\, \;  . 
	\end{equation}
The map \eqref{eqTt} is well-defined, because $g \in \ker \mathbb{E}$ 
implies $\Lambda_* g \in \ker \mathbb{E}$.  
\end{proof}

\begin{definition}
\label{Def4-11} The triple  $({\mathfrak k}(dS) , \sigma, {\mathfrak z} )$  is  the 
\emph{covariant classical dynamical system}\index{covariant classical dynamical system}
associated to the homogeneous Klein--Gordon equation \eqref{3.25}. 
\label{kldypage}
\end{definition}

\section{The restriction of the KG equation to a (double) wedge}
\label{sec:5.4}

Our next objective is to provide an explicit formula for $\mathscr{E} ( x ,  y )$. In some sense, 
it is sufficient to solve this problem in the causal dependence region of a half-circle: given an 
arbitrary point $ x \in dS$~and the Cauchy surface $ S^1$, there exists a wedge $W^{(\alpha)}
= R_0(\alpha)W_1$, which contains both $x$ and $\Gamma^- ( x) \cap S^1$ (or, if $x$ lies in the 
past of~$S^1$, $\Gamma^+ ( x) \cap S^1$). On the other hand,  all the formulas we will derive in 
this section naturally extend to the double-wedge $\mathbb{W}^{(\alpha)}=W^{(\alpha)} \cup 
W^{(\alpha + \pi)}$, so it is natural to state them in their extended form.  

In order to keep the notation simple, we work out explicit expressions for the double 
wedge $\mathbb{W}_1$ in the chart \eqref{w1psi} for $ x   \equiv  x (t,\psi)$ and 
$ y  \equiv  y (t',\psi')$.  (The points $ \psi = \pm \frac{\pi}{2}$ in this chart 
correspond  to the points  $ (0, \pm r, 0) \in dS$.) However, we would like to emphasize that 
all computations in this subsection can be carried out for arbitrary 
double wedges $\Lambda \mathbb{W}_1$, $\Lambda \in SO_0(1,2)$. 

Let us recall from \eqref{varepsilon} that, restricted to the \index{double wedge} 
double wedge $\mathbb{W}_1 \, $, the Klein--Gordon operator takes the form 
	\[
			\square_{\mathbb{W}_1}+\mu^2
				=  \frac{1}{r^2 \cos^2 \psi}\,(\partial_t^2+  \varepsilon^2) \; , 
	\]
with 
	\[
		\varepsilon^2  \doteq  - (\cos \psi  \, \partial_\psi)^2 + (\cos \psi )^2 \, \mu^2 r^2 \; . 
	\]

\paragraph{\it Notation.} If $P$ is a pseudo-differential  operator on $L^2(S^1, \, | \cos \psi |^{-1}  \, r {\rm d}\psi)$,    
define its kernel~$P (\psi, \psi') $ for all $h \in L^2(S^1, \, | \cos \psi |^{-1}  \, r {\rm d} \psi) \cap{\mathscr D} (P)$, 
for which the following expressions exist, by
	\[ 
	(P h) (\psi)  = \int_{S^1} \frac{ r {\rm d}\psi' }{ | \cos \psi' |} \; P (\psi, \psi') h (\psi')   
		 		 = \int_{S^1} \frac{ {\rm d} l (\psi') }{| \cos\psi' |^2}\;  P (\psi, \psi') h (\psi')  \; . 
	\] 
${\rm d} l (\psi')$ was defined in \eqref{new-surface}.

If $P$ is hermitian  with domain  
${\mathscr D} \subset L^2(S^1, \, | \cos \psi |^{-1}  \, r {\rm d}\psi)$, then 
	\[
		P(\psi, \psi') = \overline{P(\psi', \psi)}\; , \qquad \psi, \psi' \in S^1 \; . 
	\]
In the next lemma, $ | \varepsilon |^{-1} \sin(\varepsilon(t-t')) $ is considered as such a
pseudo-differential  operator on $L^2(S^1, \, | \cos \psi |^{-1}  \,  r {\rm d}\psi)$. 

\bigskip

\begin{lemma} \label{PropWedge} Use the coordinates $ x (t, \psi) 
= \Lambda_{1} ( t ) \, R_0 ( - \psi ) o$,  
$ t \in \mathbb{R}$, $\psi \in [0, 2\pi)$, introduced in \eqref{w1psi}. Then 
	\begin{equation}
		 \label{eqPropWedge'}
		 	\mathscr{E} (   x,  y )  
		 		= - r \left( \frac{\sin(\varepsilon(t-t')) } 
				{ | \varepsilon | } \right)  (\psi,\psi') \;   . 
	\end{equation}
\end{lemma}

\begin{remarks} 
\label{rm:5.4.2}\quad
\begin{itemize}
\item [$i.)$] Note that equation~(\ref{comfkt}) extends the formula 
for the propagator given 
in \eqref{eqPropWedge'} from $\mathbb{W}_1$  to  $dS$. 
\item [$ii.)$] For $ h\in{\mathcal D}_{\mathbb{R}} \left(I_+\right)$
one can extend the domain of $\mathbb{E}$ to distributions of the form 
	\begin{align}
	 \label{sharp-timetestfunction}
	 f (x) \equiv (\delta \otimes h) (x) &=	\delta (t)  \;  \frac{ h (\psi )}{ r \cos \psi }\;  , 
	 \nonumber \\
	g (x) \equiv (\delta' \otimes h) (x) &= 
	 \left( \frac{\partial_t}{ r \cos \psi } \delta \right) (t)  \;  
	\frac{ h (\psi ) }{ r \cos \psi }\;  , 
	\end{align} 
with $x \equiv x (t, \psi)$, using the coordinates introduced in \eqref{w1psi}, and 
	\[ 
		{\rm d} \mu_{\mathbb{W}_1 } (t,\psi) = r^2  {\rm d} t\,  {\rm d} \psi   \cos \psi  \; . 
	\]  
The properties of the convolution ensure that $\mathbb{f} ,  \mathbb{g}$ are 
$C^\infty$-solutions of the Klein--Gordon equation~\eqref{3.25}, whose 
support is contained in~$dS \setminus \overline{\, W'}$. Within the 
region~$\mathbb{W}_1$ these solutions are given by
	\begin{align}
		\label{5.4.11a}
				\mathbb{f} ( x  )   
					&=  -   \frac{ \sin(\varepsilon t)}{|\varepsilon|} \, 
					\cos \psi  \cdot h (\psi )   \; , 
		\\
		\label{5.4.11b}
				\mathbb{g} ( x  )   
					&=   \frac{ \cos  (\varepsilon  t)}{  r } \  
					 h (\psi )  \; . 
	\end{align} 
\end{itemize}
\end{remarks}

\begin{proof}
For $f, g \in{\mathcal D}_\mathbb{R}(\mathbb{W}_1)$, set $f_t(\psi) 
\doteq f(t,\psi)$ and $g_{t'}(\psi') \doteq g(t',\psi')$. 
Clearly, $f_t, g_t \in L^2(S^1, \, | \cos \psi |^{-1}  \, r {\rm d}\psi)$.
Consider 
	\begin{equation}
		\label{3.41}
		{\mathcal E}_{{\mathbb W}_1} ( f,g)  
   			 \doteq   - \int r^4 \, {\rm d} t \, {\rm d} t' 
			 \left\langle \mathbb{cos}_\psi^2  {f}_t \,, \,
				\tfrac{\sin(\varepsilon(t-t')) } { | \varepsilon | }\, 
				\mathbb{cos}_\psi^2  
				g_{t'}\right\rangle_{L^2(S^1, \, | \cos \psi |^{-1}  \, r {\rm d}\psi)} \; ,
	\end{equation}
with  $\,  \mathbb{cos}_\psi$  the multiplication operator by  $\cos\psi$. 
Clearly,  ${\mathcal E}_{{\mathbb W}_1}$  is anti-symmetric with respect to permutation of $f$ and~$g$. 
Moreover, according to \eqref{varepsilon}  
	\begin{align*} 
	 & {\mathcal E}_{{\mathbb W}_1}\big(f,(\square_{\mathbb{W}_1}+\mu^2) h\big)   \nonumber \\
	 	 &=  {\mathcal E}_{{\mathbb W}_1} \left( f, r^{-2} \mathbb{cos}^{-2}_\psi
		(\partial_t^2+\varepsilon^2) h\right) 
		\label{eqEKGE}\\ 
	&= - \int r^4 
		 \, {\rm d} t {\rm d} t'\left\langle  \mathbb{cos}_\psi^2  {f}_t \,,\,
	\tfrac{\sin(\varepsilon(t-t'))}{ | \varepsilon |}
		r^{-2} (\partial_{t'}^2+\varepsilon^2)h_{t'}\right\rangle_{L^2(S^1, \, 
		| \cos \psi |^{-1}  \, r {\rm d}\psi)}  \; , 	\nonumber	
	\end{align*}
where $h_{t'} (\psi) \doteq h(t', \psi)\in{\mathcal D}_{\mathbb{R}} \bigl(S^1 
\setminus \{ -\frac{\pi}{2}, \frac{\pi}{2} \} \bigr)$. Now  
	\begin{equation*}
		\label{seceq}
		\int {\rm d} t' \sin(\varepsilon t') \partial_{t'}^2h_{t'} 
		= \int {\rm d} t'  \big(\partial_{t'}^2 \sin(\varepsilon t')\big) h_{t'}
		= \int {\rm d} t'  \sin(\varepsilon t')(-\varepsilon^2)  h_{t'} 
	\end{equation*}
by partial integration and using $(\partial_{t'} h)_{t'}= \partial_{t'}(h_{t'})$.  
Thus  
	\[ 
		{\mathcal E}_{{\mathbb W}_1} \big(f,(\square_{\mathbb{W}_1}+\mu^2) 
		h\big) = 0 \; . 
	\]
A similar argument can be used to show 
${\mathcal E}_{{\mathbb W}_1} \big((\square_{\mathbb{W}_1}+\mu^2) h, f\big) = 0 $. It follows that
the kernel $\mathscr{E}_{{\mathbb W}_1} ( x ,  y )$, defined by    
	\begin{equation*}
		\label{integralkernel2}
				  \int {\rm d} \mu_{dS} ( x) 
				  {\rm d} \mu_{dS} ( y) \; 
			f ( x ) \mathscr{E}_{{\mathbb W}_1} ( x ,  y )  
			g ( y  ) 
			\doteq {\mathcal E}_{{\mathbb W}_1}( f,g)  \; ,  
	\end{equation*}
is anti-symmetric and satisfies the Klein--Gordon equation in both entries. 
Furthermore, the kernel of $\mathscr{E}_{{\mathbb W}_1}$ is indeed 
given by \eqref{eqPropWedge'}:
	\begin{align*} 
		\label{integralkernel3}
			{\mathcal E}_{{\mathbb W}_1}( f,g) 
   			 & =    - \int r^4  {\rm d} t \, {\rm d} t' \left\langle  
			 \mathbb{cos}_\psi^2 \,  {f}_t \,, \,
				\tfrac{\sin(\varepsilon(t-t')) } { | \varepsilon | }\, 
				\mathbb{cos}_\psi^2 \, g_{t'}
				\right\rangle_{L^2(S^1, \, | \cos \psi |^{-1}  \, r {\rm d}\psi)} 
					\nonumber 			\\ 			
				& =    - \int r^3  {\rm d} t \, {\rm d} t' \int  
				\tfrac{r\, {\rm d} \psi }{| \cos \psi |} \, 
				\cos^{2} \psi \, \, {f}_t (\psi) 
				\\
			& \qquad \times
			\int \frac{r^2 \, {\rm d} \psi' }{|\cos \psi'|}\left( 
			\tfrac{\sin(\varepsilon(t-t')) } { | \varepsilon | }\right) (\psi, \psi')  
				\cos^{2} \psi' \, g_{t'}(\psi')   \nonumber \\ 
			&=  -  \int r {\rm d} t \, {\rm d} l (\psi)   \int  r {\rm d} t' \, {\rm d} l (\psi') \; 
			{f}_t (\psi) \; r \left(\tfrac{\sin(\varepsilon(t-t'))}
			{| \varepsilon |} \right) (\psi,\psi') \;  g_{t'}(\psi') \; . 
			\nonumber
	\end{align*}
In the last equation we used \eqref{new-surface}, \emph{i.e.}, 
${\rm d} l (\psi) = r \, | \cos \psi | \, {\rm d} \psi $. Thus 
	\begin{equation*}
		\label{3.46}
		\mathscr{E}_{{\mathbb W}_1} \left( x ,  y \right) 
			= - r \left( \tfrac{\sin(\varepsilon(t-t'))}{ | \varepsilon | }\right) 
			(\psi,\psi') \; ,  
	\end{equation*}
using $ x   \equiv  x (t,\psi)$ and $ y  \equiv  y (t',\psi')$, 
is a bi-solution of the homogeneous Klein-Gordon equation in the 
double wedge ${\mathbb W}_1$.

It remains to verify the initial conditions \eqref{eqESigma} and \eqref{eqESigma2}. 
Clearly, $\mathscr{E}_{{\mathbb W}_1}$ satisfies \eqref{eqESigma} with ${\mathcal C}= S^1$. 
Thus we can concentrate on  \eqref{eqESigma2}. 
The unit normal future pointing vector field on $S^1 \setminus \{ - \frac{\pi}{2}, \frac{\pi}{2}\}$ is  
	\begin{equation}
		\label{eqnSigma}
		n(t,\psi) = \, r^{-1} \mathbb{cos}_\psi^{-1} \; \partial_t \; . 
	\end{equation}
In $I_-$ the vector field $\partial_t$ is past 
directed and $\cos\psi<0$, thus equation (\ref{eqnSigma}) holds for 
both half-circles $I_+$ and $I_-$. From \eqref{3.41} read off 
	\begin{equation}
		\label{3.49}
		r^{-1} \partial_t \mathscr{E}_{{\mathbb W}_1} \bigl(  x(t,\psi);  y (0,\psi')
		\bigr)_{\upharpoonright t=0}= -  
		\left( \frac{\varepsilon}{|\varepsilon|}  \,  \mathbb{1} \right) (\psi,\psi') \; , 
	\end{equation}
where 
	\[
		\mathbb{1}(\psi,\psi') = \tfrac{1}{r} |\cos\psi|\, \delta(\psi-\psi')
	\] 
is the kernel of the unit 
in $L^2 \left(S_1 \setminus \{ - \frac{\pi}{2}, \frac{\pi}{2}\} ,
|\cos\psi|^{-1} r {\rm d} \psi \right)$. 
Now $\,  \mathbb{cos}_\psi^{-1}\varepsilon|\varepsilon|^{-1}
=|\,  \mathbb{cos}_\psi|^{-1}$.
Hence the r.h.s.~in \eqref{3.49} is $- r^{-1} \cos\psi \, \delta(\psi-\psi')$ 
and  (\ref{eqnSigma}) implies
	\begin{equation*}
		\label{eqdtESigma}
		n_{\rm \ell}\, {\mathcal E}  \bigl( x (t,\psi);  y (0,\psi')
		 \bigr)_{\upharpoonright t=0}
		 = - \tfrac{1}{r} \delta(\psi-\psi')  
		 = \delta_{S^1} (\psi,\psi')\; .   
	\end{equation*}
As before, $\delta_{S^1}$ 
is the kernel of the unit with respect to the induced line element 
$r {\rm d}\psi$ on~$S^1$, 
see Equation~(\ref{new-eqVolInd}). 
Thus \eqref{eqESigma2} holds, 
and hence, by the uniqueness result mentioned, $\mathscr{E}_{{\mathbb W}_1} 
=\mathscr{E}_{\upharpoonright \mathbb{W}_1}$  
and ${\mathcal E}_{{\mathbb W}_1} = {\mathcal E}_{\upharpoonright \mathbb{W}_1}$ 
within the double wedge~$\mathbb{W}_1$.
\end{proof}

\goodbreak
Thus, for $f\in{\mathcal D}_\mathbb{R}(\mathbb{W}_1 )$,   
$x \equiv x (t,\psi) \in \mathbb{W}_1 $ and $f_{t'}(\psi) \doteq f (x(t', \psi))$,
	\begin{equation} 
		\label{pmf-2}
				\mathbb{f} ( x  )   
					= -  \int r \, {\rm d} t' 
					\Bigl( \frac{ \sin(\varepsilon(\, t-t'))}{|\varepsilon|} \, 
					\mathbb{cos}_\psi^2  f_{t'}\Bigr)  (\psi) \;  . 
	\end{equation} 
Note that \eqref	{pmf-2} describes $\mathbb{f} $ only on a proper 
subset of its support, namely the intersection of its support with $\mathbb{W}_1$. 

\section{The canonical classical dynamical system}
\label{CaCDS}

Let $(n \, \mathbb{\Phi}) _{\upharpoonright {\mathcal C}}$ denote the Lie 
derivative of $\mathbb{\Phi}$ along the unit normal, future pointing vector field~$n$, restricted to 
the Cauchy surface~${\mathcal C}$.

\begin{theorem}[Dimock \cite{D77}, Theorem 1]
\label{cauchyproblem}
 Let ${\mathcal C} \subset dS$ be a Cauchy surface and let~$(\mathbb{\phi}, \mathbb{\pi}) \in  
 C^\infty ({\mathcal C}) \times C^\infty ({\mathcal C}) $. 
Then there exists a unique $\mathbb{\Phi} \in C^\infty (dS)$ satisfying the 
homogeneous Klein--Gordon 
equation~\eqref{3.25} with Cauchy data
	\begin{equation}
		\label{3.26} 
		\mathbb{\Phi}_{\upharpoonright {\mathcal C}} = \mathbb{\phi} \; , 
		\quad (n \mathbb{\Phi})_{\upharpoonright {\mathcal C}} = \mathbb{\pi} \; . 
	\end{equation}
Furthermore, ${\rm supp\,} \mathbb{\Phi} \subset \bigcup_\pm 
\Gamma^\pm ({\rm supp\,} \mathbb{\phi} \cup {\rm supp\,} \mathbb{\pi} ) $. 
\end{theorem}

\goodbreak 

\begin{remark}
For functions in the \emph{Sobolev space}\index{Sobolev space} $\mathbb{H}^2_{\rm \, loc}(dS)$, 
this is the classical existence and uniqueness theorem of Leray \cite{L}. 
\end{remark}

If we choose the time-zero circle $S^1$ for our Cauchy surface $ {\mathcal C}$, then the space of Cauchy data,
\label{hatckpage}
	\begin{equation*}
		\label{calSHat}
		\widehat {\mathfrak k} (S^1) 
		\doteq C^\infty_\mathbb{R}(S^1)\times C^\infty_\mathbb{R}(S^1) \; ,
	\end{equation*}
together with the {\em canonical symplectic form} 
	\begin{equation}
		\label{eqSplFormHat}
		\widehat \sigma\big((\mathbb{\phi}_1,\mathbb{\pi}_1),(\mathbb{\phi}_2,\mathbb{\pi}_2)\big)
		\doteq \langle \mathbb{\phi}_1,\mathbb{\pi}_2 \rangle_{L^2(S^1, \, r\,  {\rm d} \psi ) }-  
		\langle\mathbb{\pi}_1,\mathbb{\phi}_2 \rangle_{L^2(S^1, \, r \, {\rm d} \psi)} \; , 
	\end{equation}
forms a symplectic space $( \widehat {\mathfrak k} (S^1), \widehat \sigma)$. 
As before (see  \eqref{new-eqVolInd}), the line element on $S^1$ is $r \, {\rm d} \psi $. 
The right hand side in \eqref{eqSplFormHat} is zero, if $(\mathbb{\phi}_1,\mathbb{\pi}_1)$ 
and $(\mathbb{\phi}_2,\mathbb{\pi}_2)$ have disjoint support. 
 
\begin{proposition}
\label{nocheinlabel} 
The symplectic space $(\widehat {\mathfrak k} (S^1),  \widehat \sigma)$ carries a 
representation 
	\[
		\Lambda  \mapsto \widehat  {\mathfrak z} ({\Lambda}) \; ,
		\qquad \Lambda \in O(1,2) \; , 
	\]
defined  by  
\label{widehatccTpage}
	\begin{equation}
		\label{eqTtHat}
		\widehat {\mathfrak z} (\Lambda) (\mathbb{\phi},\mathbb{\pi})  
		\doteq\big(  ( \Lambda_* \mathbb{\Phi} )_{ \upharpoonright S^1} \; ,
		( n  \Lambda_* \mathbb{\Phi})_{\upharpoonright S^1}\big) \; ,
	\end{equation}
where $\mathbb{\Phi}$ is the unique $C^\infty$-solution of the Klein--Gordon 
equation~\eqref{3.25} with Cauchy data given by \eqref{3.26}.
The triple  $\bigl(\, \widehat {\mathfrak k}(S^1), \widehat \sigma, 
\widehat  {\mathfrak z}  \bigr)$  is the 
{\em canonical classical dynamical system} 
associated to the homogeneous Klein--Gordon equation \eqref{3.25}. 
\end{proposition}

\begin{proof}This result follows directly from Theorem
  \ref{cauchyproblem} and  the invariance of the Klein--Gordon operator
under the adjoint push-forward action of $O(1,2)$. 
\end{proof}

\begin{proposition}
\label{cauchy-symplectic}
The map
	\begin{align*}
	 	{\mathbb  T} \colon ( {\mathfrak k}(dS),  
		\sigma,  {\mathfrak z} (\Lambda) )
		&\to (\widehat {\mathfrak k}(S^1), \widehat \sigma, 
		\widehat  {\mathfrak z} (\Lambda) ) \\
	 	 [f]  & \mapsto   ( \mathbb{f}_{\upharpoonright S^1} \; ,  
		 (n \mathbb{f})_{\upharpoonright S^1} )
		\equiv (\mathbb{\phi} ,  \mathbb{\pi}) 
	\end{align*}
is symplectic and 
	\begin{equation}
		\label{T-s-s}
		{\mathbb  T} \circ {\mathfrak z} (\Lambda) = 
		\widehat {\mathfrak z} (\Lambda) \circ {\mathbb  T} \; . 
	\end{equation}
\end{proposition}

\begin{proof} Let $f, g \in {\mathcal D}_{\mathbb{R}}(dS)$. Then Stokes' theorem 
implies (see\footnote{Note that Dimock's operator $E$ differs from our conventions by a sign, as can be seen 
by comparing Corollary 1.2 in \cite{D77} with \eqref{eqESigma2}.} \cite[Lemma~A.1]{D77}) that 
	\begin{align*}
		\sigma ([f], [g]) & = {\mathcal E} (f,g) 
		\\
		&= \int_{dS} {\rm d} \mu_{dS}(x) \; f (x) (\mathbb{E} g) (x)   \\
		& = \int_{S^1}  r \, {\rm d} \psi \; \Bigl(  
		(\mathbb{E} f)_{\upharpoonright S^1} (\psi) (n \mathbb{E}  g)_{\upharpoonright S^1} (\psi) 
		-
		( n  \mathbb{E} f)_{\upharpoonright S^1} (\psi) (\mathbb{E} g)_{\upharpoonright S^1} (\psi)  \Bigr) \; 
		\\
		& =   
		\langle \mathbb{f}_{\upharpoonright S^1}, 
		(n \mathbb{g})_{\upharpoonright S^1} \rangle_{L^2(S^1, \, r \, {\rm d} \psi ) }
		-
		\langle (n \mathbb{f})_{\upharpoonright S^1}, 
		\mathbb{g}_{\upharpoonright S^1} \rangle_{L^2(S^1, \,  r\, {\rm d} \psi)} 
		\\
		& =  \widehat \sigma\Bigl( \bigl( \mathbb{f}_{\upharpoonright S^1},
		(n \mathbb{f})_{\upharpoonright S^1} \bigr), \bigl( \mathbb{g}_{\upharpoonright S^1},
		(n \mathbb{g})_{\upharpoonright S^1} \bigr)\Bigr) \; .
	\end{align*}
Thus $\mathbb{T}$ is symplectic. Equation \eqref{T-s-s} follows by inspecting the action of 
${\mathfrak z} (\Lambda) $ and~$\widehat {\mathfrak z} (\Lambda) $, respectively. 
\end{proof}

The canonical projection  
\label{widehatcxC}
	\begin{align}
		\label{eqCovCan}
		\widehat {\mathbb{P}} \colon  \; \;{\mathcal D}_{\mathbb{R}} (dS) 
		& \to  \widehat {\mathfrak k} (S^1)  \nonumber  \\ 
		f  & \mapsto   
		\widehat f := \bigl( \mathbb{f}_{ \upharpoonright S^1}, 
		\left( n \, \mathbb{f} \right)_{\upharpoonright S^1} \bigr) 
	\end{align}
maps a smooth, real valued function $f \in{\mathcal D}_\mathbb{R} 
(dS) $ with compact support
to the Cauchy data of a $C^\infty$-solution $\mathbb{f}$ of the Klein--Gordon equation~\eqref{3.25}. 

\begin{remark}
For the special case $f\in{\mathcal D}_\mathbb{R}(\mathbb{W}_1 )$, 
Eq.~\eqref{eqCovCan} reads\footnote{See also Remark \ref{rm:5.4.2} $ii.)$.}
	\begin{align}  
		\label{pmfhut}
			\mathbb{f}_{ \upharpoonright S^1}(\psi)  	
 &=    \int r \, {\rm d} t' \Bigl( \tfrac{ \sin(t' \varepsilon )}{|\varepsilon|} \, 
			 	\mathbb{cos}_\psi^2  f_{t'}\Bigr)  ( \psi) \; ,  \\ 
		\label{pmfhutableitung} \quad
			(n \, \mathbb{f} )_{\upharpoonright S^1}
			(\psi) 
			&= - \tfrac{1}{ r |\cos  \psi|} \int r \, {\rm d} t' \,  \left( \cos(t' \varepsilon) \, 
			\mathbb{cos}_\psi^2  f_{t'}\right)  ( \psi) 
			\; ,  
	\end{align}
where $f_{t}(\psi):=f(  x (t, \psi)) $,
using again $\,  \mathbb{cos}_\psi^{-1}\varepsilon|\varepsilon|^{-1}=|\,  \mathbb{cos}_\psi|^{-1}$.
\end{remark}

Using the coordinates introduced in \eqref{w1psitau}, we can now extend $\widehat {\mathbb{P}}$ 
to the class of distributions given in \eqref{f-dist} and \eqref{g-dist}. Using \eqref{comfkt},
the properties of the convolution \eqref{feb-1} ensure that there exist  
$C^\infty$-solutions $\mathbb{f} , \mathbb{g}$ of the Klein--Gordon equation~\eqref{3.25} with Cauchy data:
	\begin{equation}
		\label{sharp-timetestfunction-phipi3}
		\big( \mathbb{f}_{ \upharpoonright S^1}, 
							 (n \, \mathbb{f} )_{\upharpoonright S^1})
  		 = (0, - h) \equiv  (\mathbb{\phi}, \mathbb{\pi}) \; , 
	\end{equation}
and, by partial integration, 
	\begin{equation}
	\label{sharp-timetestfunction-phipi4}
	\big(  \mathbb{g}_{ \upharpoonright S^1}, 
							 (n \, \mathbb{g})_{\upharpoonright S^1})
 		  = (  h, 0) \equiv  (\mathbb{\phi}, \mathbb{\pi}) \; . 
	\end{equation}
All elements in~$\widehat {\mathfrak k}(S^1)$ are linear combinations of the Cauchy data 
arising from  sharp-time testfunctions $f, g$ of the form described above.

\bigskip
For open intervals~$I \subset S^1$ define 
	\begin{equation*}
		\label{calSHat2}
		\widehat {\mathfrak k} (I ) 
		\doteq{\mathcal D}_\mathbb{R}(I )\times{\mathcal D}_\mathbb{R}(I ) \; . 
	\end{equation*}
We next discuss, how the localisation properties of the solutions of the Klein-Gordon equation 
manifest themselves on the space $\widehat {\mathfrak k} (S^1)$.

\begin{proposition}
\label{fsol}
Let $\widehat f \in \widehat {\mathfrak k} (I)$, $ I \subset S^1$. Then 
	\[
		\widehat{\mathfrak z} (\Lambda) \, \widehat f \in 
		\widehat {\mathfrak k} \; \Bigl( \bigl( \Gamma^{+}(\Lambda I ) 
		\cup \Gamma^{-}(\Lambda I) \bigr) \cap S^1 \Bigr) \; .
	\]
\end{proposition}

\begin{proof} Let $[f] = {\mathbb  T}^{-1} \, \widehat f \; $ be the element in ${\mathfrak k}(dS)$ associated 
to the smooth solution~$f$ of the Klein--Gordon equation with Cauchy data given by $\widehat f$. 
According to~\eqref{T-s-s},
	\[
		{\mathbb  T}^{-1} \bigl( \, \widehat{\mathfrak z} (\Lambda) \widehat f  \, \bigr) 
		= {\mathfrak z} (\Lambda) [f] \; . 
	\]
The smooth solution of the Klein--Gordon equation associated to $ {\mathfrak z} (\Lambda) [f]$ has support in 
$\Gamma^{+}(\Lambda I ) \cup \Gamma^{-}(\Lambda I) $; thus
the Cauchy data of the solution associated to $ {\mathfrak z} (\Lambda) [f]$ have support in 
$\bigl( \Gamma^{+}(\Lambda I ) \cup \Gamma^{-}(\Lambda I) \bigr) \cap S^1$.
\end{proof}

\goodbreak

Next, we provide explicit formulas for the 
transformations $\widehat {\mathfrak z} (R_0(\alpha))$, 
$\alpha \in  [0, 2\pi )$, and $\widehat  {\mathfrak z} (\Lambda_1(t))$, 
$t \in\mathbb{R}$. Both maps were 
defined in Proposition~\ref{nocheinlabel}.

\begin{proposition} 
\label{porp4.13} 
The  rotations $\widehat {\mathfrak z} (R_0(\alpha))$, $\alpha \in  [0, 2\pi )$, which map
	\[
		\bigl( \mathbb{\phi} (\psi), \mathbb{\pi}(\psi) \bigr) 
		\mapsto  \bigl(\mathbb{\phi} (\psi - \alpha) , \mathbb{\pi} (\psi - \alpha) \bigr) , 
		\qquad \alpha \in  [0, 2\pi ) \; , 
	\]
and the boosts $\widehat  {\mathfrak z} (\Lambda_1(t))$, $t \in\mathbb{R}$, which map
	\begin{equation}
		\label{3.77}
		(\mathbb{\phi}, \mathbb{\pi}) 
		\mapsto  
		  (\mathbb{\phi}_t , \mathbb{\pi}_t )  
		  \; ,
	\end{equation}
with
	\begin{align}
		\label{varphi-pi-t} 
			\mathbb{\phi}_t  
			&=  \cos(\varepsilon t)\mathbb{\phi}
					- \sin(\varepsilon t) \,
                                        \varepsilon^{-1} \,  \mathbb{cos}_\psi \,
                                        \mathbb{\pi}   \nonumber	 \\ 
			\mathbb{\pi}_t &=   
		( r \,  \mathbb{cos}_\psi)^{-1}\big(\varepsilon\sin(\varepsilon t)\mathbb{\phi} 
              + \cos(\varepsilon t) \,  \mathbb{cos}_\psi \; \mathbb{\pi} \big) \; ,  
	\end{align}
generate the representation $ \Lambda  \mapsto \widehat  {\mathfrak z} (\Lambda) $ 
of $SO_0(1,2)$ introduced in (\ref{eqTtHat}).
The 	points $(\mathbb{\phi} (\pm\frac{\pi}{2}) , \mathbb{\pi} (\pm\frac{\pi}{2}) )$ are fixed
points of the map  $t\mapsto (\mathbb{\phi}_t (\psi) , \mathbb{\pi}_t (\psi))  $. 
The representers of the reflections $P_1$ and $ T$ are 
	\begin{align}  
		\label{eqJCan}
		\widehat  {\mathfrak z} (P_1 ) \colon  (\mathbb{\phi},\mathbb{\pi})& \mapsto  
		\bigl( (P_1)_*\mathbb{\phi}, (P_1)_*\mathbb{\pi} \bigr) \; ,  \\ 
		\label{eqJCan2}
		\widehat  {\mathfrak z} (  T) \colon  (\mathbb{\phi},\mathbb{\pi})& \mapsto 
		 ( \mathbb{\phi}, - \mathbb{\pi} ) \; .  
	\end{align}
\end{proposition}

\begin{proof} Recall \eqref{3.32} and consider the boosts
$t \mapsto \Lambda_1 (t)$, acting on the Cauchy data on~$S^1$.  
Now combine  (\ref{eqTtHat}) and  the definition of $\Lambda_*$ to conclude that the boosts
$ \widehat {\mathfrak z} (\Lambda_1 (t)) (\mathbb{\phi},\mathbb{\pi}) $, $ t \in\mathbb{R}, $ 
are determined by  $  \mathbb{\Phi}_{ \upharpoonright S^1 \cup \, \mathbb{W}_1} $, 
where  $\mathbb{\Phi}$ is the solution of the Klein--Gordon equation \eqref{3.25} specified 
in Theorem~\ref{cauchyproblem}. Evaluate~\eqref{3.77} with care---write it out explicitly and 
take advantage of the fact that $\varepsilon^2$ is a differential operator which satisfies 
$(\varepsilon^2 h)  (\psi\pm\frac{\pi}{2}) = O(\psi)$ as $\psi \to 0$ 
for $h \in C^\infty (S^1)$, just as $\cos (\psi\pm\frac{\pi}{2})$---to show that the map 
$t\mapsto (\mathbb{\phi}_t (\psi) , \mathbb{\pi}_t (\psi))  $ 
is well-defined  for $\psi  \to \pm \frac{\pi}{2}$ 
and 
	\[
		\textstyle{
		\bigl(\mathbb{\phi}_t (\pm \frac{\pi}{2}) , \mathbb{\pi}_t (\pm \frac{\pi}{2}) \bigr) = 
		\bigl(\mathbb{\phi} (\pm \frac{\pi}{2}), \mathbb{\pi}(\pm \frac{\pi}{2}) \bigr) }\qquad 
		\forall t \in\mathbb{R} \; .
	\]
(This ensures that $\mathbb{\phi}_t$ and $\mathbb{\pi}_t$ are both well-defined despite the fact that
the coordinate system is degenerated at $\psi = \pm \frac{\pi}{2}$.) It remains to 
construct $\widehat {\mathfrak z} (\Lambda_1 (t))$ in the 
space-time region~$\mathbb{W}_1$. On~$\mathbb{W}_1$, the Klein--Gordon equation \eqref{3.25} reads 
	\begin{equation}
		\label{3.79}
		\frac{1}{r^2 \cos^2 \psi}\,(\partial_t^2+\varepsilon^2) \mathbb{\Phi}=0 \; , 
	\end{equation}
using \eqref{varepsilon}. The real valued solution  of  \eqref{3.79} in the region 
$\mathbb{W}_1$ with  Cauchy data   (see  \eqref{eqnSigma})
	\begin{equation}
		\label{3.80}
			\mathbb{\phi}  = \mathbb{\Phi}_{ \upharpoonright S^1} \; , 
			\qquad 
			\mathbb{\pi} = \frac{1}{r \cos \psi} \, (\partial_t \mathbb{\Phi})_{ \upharpoonright S^1} \; ,
	\end{equation}
 is   
	$
		\mathbb{\Phi}  (t,\cdot )= \cos(\varepsilon t)\mathbb{\phi} + 
 		 \sin(\varepsilon t) \, \varepsilon^{-1} \,  \mathbb{cos}_\psi\, \mathbb{\pi} $. 
Hence 
	\[
		\mathbb{\phi}_t  \equiv (\Lambda_1(t)_*\mathbb{\Phi}) _{ \upharpoonright S^1} 
		\qquad \text{and} \qquad
 		\mathbb{\pi}_t \equiv n (\Lambda_1(t)_*  \mathbb{\Phi}) _{ \upharpoonright S^1} \;   
	\]
are determined by 
	\begin{align*}
		(\Lambda_1(t)_*\mathbb{\Phi})(t',\cdot ) 
		&= \cos(\varepsilon (t'-t))\mathbb{\phi} + 
 				 \sin(\varepsilon (t'-t)) \, \varepsilon^{-1} \,  \mathbb{cos}_\psi \mathbb{\pi} \; ,\\ 
		(\partial_{t'} \Lambda_1(t)_*\mathbb{\Phi})(t',\cdot ) 
		&= -\varepsilon\sin(\varepsilon (t'-t))\mathbb{\phi} + 
 			 \cos(\varepsilon (t'-t)) \,  \mathbb{cos}_\psi \mathbb{\pi} \; . 
	\end{align*}
$\varepsilon^2 $ maps $ C_\mathbb{R}^\infty (S^1)$ to the functions in $C_\mathbb{R}^\infty (S^1 )$, 
which vanish at $\psi =\pm \frac{\pi}{2}$. Thus $\widehat  {\mathfrak z} (\Lambda_1(t))$ maps
	\[	
		C^\infty_\mathbb{R}(S^1)\times C^\infty_\mathbb{R}(S^1) 
		\mapsto C^\infty_\mathbb{R}(S^1)\times C^\infty_\mathbb{R}(S^1) \; . 
	\] 
The boosts $\Lambda_1(t)$, $t \in\mathbb{R}$, together with the rotations $R_0(\alpha)$, 
$\alpha \in [0, 2 \pi)$, generate $SO_0(1, 2)$. 
(The group relations were established in Proposition~\ref{nocheinlabel}.)

$(P_1)_*$ and $ T_*$  commute with the restriction to $S^1$, and $(P_1)_*$ 
commutes with $n$ while $ T_*$  anti-commutes with $n$. On the time-zero 
circle $S^1$, the spatial reflection $P_1$ acts as 
	\[
		P_1 \colon \psi \mapsto 
		\pi - \psi \;, \qquad \psi \in [ -\tfrac{\pi}{2}, \tfrac{3 \pi}{2}) \; . 
	\]
Thus \eqref{eqJCan} and \eqref{eqJCan2} follow. 
\end{proof}

\goodbreak
The reflection at the edge of the wedge $W_1$
	\[
		(P_1T)_* \colon \; g (x_0, x_1,  x_2) 
		\mapsto g (-x_0, x_1,  -x_2) \; , \qquad g \in{\mathcal D} (dS) \; , 
	\]
gives rise to a double classical system in the sense of Kay~\cite{Kay1}: 

\goodbreak 
\begin{proposition}
\label{ssdd} The symplectic space $\widehat {\mathfrak k} 
\bigl(S^1 \setminus \{ \pm \frac{\pi}{2} \}\bigr) $ is the direct sum of  $\widehat {\mathfrak k} (I_+ )$ 
and~$ \widehat {\mathfrak k} (I_- )$. Moreover, 
\begin{itemize}
\item [$ i.)$] 	$
		\widehat \sigma (  \widehat f ,  \widehat g ) = 0  
	$
 for all $\widehat{f} \in \widehat {\mathfrak k} (I_+ )$ and $\widehat{g} \in \widehat {\mathfrak k} (I_- )$;  
\item [$ ii.)$] the maps $\widehat  {\mathfrak z} (\Lambda_1 (t)) $, $t \in\mathbb{R}$, leave the 
subspaces $\widehat {\mathfrak k} (I_+ )$ and $\widehat {\mathfrak k} (I_- )$ invariant;
\item [$ iii.)$] $\widehat  {\mathfrak z} (P_1T)$ is an anti-symplectic involution,   which satisfies
	\[
		\widehat  {\mathfrak z} (P_1T)  \, \widehat {\mathfrak k} (I_+ ) 
		= \widehat {\mathfrak k} (I_- ) 
		\quad
		\text{and}
		\quad
		\Bigl[ \, \widehat  {\mathfrak z} (\Lambda_1 (t))  \, , 
		\, \widehat  {\mathfrak z} (P_1T) \Bigr]=0 \quad \forall t \in\mathbb{R} \; . 
	\]
\end{itemize} 
Thus $\bigl( \, \widehat {\mathfrak k} \bigl(S^1 
\setminus \{ \pm \frac{\pi}{2} \}\bigr) ,  \widehat  {\mathfrak z} (\Lambda_1 ) , 
\widehat  {\mathfrak z} (P_1T) \bigr)$ is a double classical linear dynamical system 
in the sense of~\ref{dcldsy}.
\end{proposition}

\bigskip
\goodbreak 
In other words, the following diagram commutes:
\vskip -.5cm

\begin{picture}(200,140)

\put(130,100){$\longrightarrow$}
\put(60,100){$\bigl(\, \widehat {\mathfrak k} (I_+ ), \widehat \sigma \bigr)$}
\put(125,110){$\widehat  {\mathfrak z} (P_1T)$}
\put(35,70){$ {\widehat  {\mathfrak z} (\Lambda_1 (t))}$}
\put(210,70){$\widehat {\mathfrak z} (\Lambda_1 (t))$}
\put(125,50){$\widehat  {\mathfrak z} (P_1T)$}

\put(170,100){$ \bigl(\, \widehat {\mathfrak k} (I_- ), \widehat \sigma\bigr)$}

\put(60,40){$\bigl(\, \widehat {\mathfrak k} (I_+ ), \widehat \sigma \bigr)$}

\put(170,40){$\bigl(\, \widehat {\mathfrak k} (I_- ), \widehat \sigma \bigr)\, .$}

\put(130,40){$\longrightarrow$}

\put(85,85){\vector(0,-3){20}}
\put(200,85){\vector(0,-3){20}}

\end{picture}
\vskip - 1.8cm
\quad

\chapter{Quantum One-Particle Structures}
\setcounter{equation}{0}

Given a classical dynamical system for the Klein--Gordon equation on the de Sitter space (in 
either the covariant or the canonical formulation) there is a \emph{unique one-particle quantum 
system} associated to it, characterised by the \emph{geodesic KMS condition}\index{geodesic 
KMS condition}. The importance of one-particle structures has been emphasised by Kay; see, 
for example,~\cite {Kay3}. They allow us to identify different realisations of the same quantum 
field theory on the level of the one-particle Hilbert space and one-particle dynamics, \emph{i.e.}, 
before second quantisation is carried out. The aim of this chapter is to show how the 
\emph{covariant formulation} pioneered by Bros and Moschella~\cite{BM}
is related to the \emph{canonical approach} favoured\footnote{The coordinates \eqref{w1psi} 
used in \cite{FHN} are convenient for the description of the boost $\Lambda_1(t)$, $t \in \mathbb{R}$, 
associated to the wedge $W_1$, but are rather awkward for a description of the rotations $R_0 (\alpha)$, 
$\alpha \in [0, 2\pi)$, which leave the Cauchy surface~$S^1$ invariant, as they are singular at the 
fixed points for the boosts $t \mapsto \Lambda_1(t)$. To resolve these problems, we present a novel 
elementary realisation (associated to the Cauchy surface $S^1$) of the unitary irreducible representations 
of $SO_0(1,2)$ for both the principal and the complementary series; see Theorem \ref{UIR-S1}.} by 
Figari, Høegh-Krohn and Nappi~\cite{FHN}. In particular, we will show that, just as in Minkowski space, 
sharp-time test functions (for the classical Klein-Gordon equation) can be used to identify the covariant 
and the canonical formulations; see Proposition \ref{Prop5.7}. 

\section{The covariant one-particle structure} 
\label{new-3.4b}

As we have seen, the Hilbert space ${\mathfrak h} (dS)$ carries an
(anti-)unitary irreducible  
representation $u$ of $O(1,2)$. 
Although we do not indicate it in the notation, the 
Hilbert space ${\mathfrak h} (dS)$ depends on $\nu$, 
which itself depends on $\mu$ by \eqref{s1} 
and \eqref{s2}; see Section~\ref{new-3.4}.  

\label{umLambdapage2}

\begin{theorem} 
\label{1PStrucHe} 
Consider the identity map 
\label{covonepartpage}
	\begin{align*}
	\label{K1PStrucHe}
		K  \colon  \qquad {\mathfrak k}(dS) 
		& \rightarrow  {\mathfrak h} (dS) \nonumber \\
		  [f] 
		  & \mapsto   [f]\; , \qquad f  \in {\mathcal D}_{\mathbb{R}} (dS) \; . 
	\end{align*}
It follows that the triple $ \bigl(K , {\mathfrak h} (dS), u \bigr) $
is a \index{de Sitter one-particle structure}
\emph{de Sitter one-particle structure} for the classical dynamical system 
$\bigl( {\mathfrak k}  (dS) , \sigma, {\mathfrak z}   \bigr)$. In other words, 
\begin{itemize}
\item [$ i.)$] $K$ defines a symplectic map from $({\mathfrak k}  (dS ) , \sigma)$ 
to $\left({\mathfrak h} (dS), \Im \langle \, . \, , 
\, . \, \rangle_{{\mathfrak h} (dS)} \right)$ and the image  
of ${\mathfrak k}  (dS )$ is dense in ${\mathfrak h} (dS)$;
\item [$ ii.)$] the (anti-) unitary representation $u$ of $O(1,2)$ on ${\mathfrak h} (dS)$ 
defined in Proposition~\ref{UIR-FH} satisfies 
	\begin{equation*} 
		\label{eqUHe}
		 u (\Lambda)\circ K = K \circ  {\mathfrak z} (\Lambda)  \;, 
		 \qquad \Lambda \in O(1,2)\; ;  
	\end{equation*}
\item [$ iii.)$]  for any wedge $W$, the   
\emph{pre-Bisogano Wichmann property} 
\index{Bisogano Wichmann property} holds:
	\begin{equation} 
		\label{5.1}
		K {\mathfrak k}(W)  \subset 
		{\mathscr D} \bigl( u (\Lambda_W ( i \pi)) \bigr) \; ,
	\end{equation}
and
	\begin{equation}
		\label{5.4}
		u (\Lambda_W ( i \pi)) [g] 
		= u (\Theta_{W}) [g] \; , 
		\qquad [g] \in K {\mathfrak k}(W) \; , 
	\end{equation}
where $\Theta_{W}$ is the reflection on the edge of the 
wedge $W$.
\end{itemize}
\end{theorem}

\begin{proof} $K$ is well-defined, 
as $\ker {\mathcal F}_{+ \upharpoonright \nu} = {\ker {\mathbb P}}$.

\begin{itemize}
\item[$ i.)$] It follows from 	
	\begin{equation*}
	\label{comfkt-3}
	\mathscr{E} (  x_1, x_2) = 2 \Im {\mathcal W}^{(2)} (  x_1,  x_2) \;  
	\end{equation*} 
that $K$ is symplectic; see \eqref{comfkt}. Since ${\mathfrak h}^\circ (dS)$ 
is dense in ${\mathfrak h} (dS)$,  the density statement follows; 
\item[$ ii.)$]  Both ${\mathfrak z} (\Lambda)$ and $u (\Lambda)$ are induced 
by the pullback action 
on the test functions: for $f \in {\mathcal D}_{\mathbb{R}} (dS)$
	\begin{align*}  
		\qquad K \circ  {\mathfrak z} (\Lambda)\; [f] &
		= K \circ {\mathbb P} (\Lambda_* f)
		=  [\Lambda_* f] 
		\nonumber \\
		&= u (\Lambda) [f]  
		= u (\Lambda)\circ K \, [f]  \;, 
		 \quad \Lambda \in O(1,2)\; .  \quad
	\end{align*} 
The second but last identity follows from the definition of the Fourier--Helgason transform, 
see \eqref{mass-shell-ft}, and Proposition \ref{UIR-FH}.

\item[$ iii.)$]  One can read off from \eqref{eqBooW} that
	\[
		\Lambda_1( t+i \pi ) = \Lambda_1( t) TP_1 \; . 
	\]
Now, let $f, g \in  {\mathcal D}_{\mathbb{R}} (W_1)$.  Lemma  \ref{lemma1.2} and Theorem \ref{prop:4.1} 
together imply that the map 
	\begin{align*}
	 	t \mapsto & \langle [f] ,
				u (\Lambda_W ( t)) [g] \rangle_{\mathfrak{h}(dS)} \\
				& = \langle [f] , [\Lambda_1 (t)_* g] \rangle_{\mathfrak{h}(dS)} 
		\\
		& =
			\int_{dS \times dS} {\rm d} \mu_{dS} ( x_1) {\rm d} \mu_{dS} (  x_2)   { f ( x_1 ) } 
				{\mathcal W}^{(2)} (  x_1 ,  x_2 ) g ( \Lambda_1^{-1}(t) x_2) \\
				& =
			\int_{dS \times dS} {\rm d} \mu_{dS} ( x_1) {\rm d} \mu_{dS} (  x_2)   { f ( x_1 ) } 
				{\mathcal W}^{(2)} (  x_1 ,  \Lambda_1 (t) x_2 ) g (  x_2) 
	\end{align*}
allows an analytic continuation into the strip $\{ t \in \mathbb{C} \mid 0< \Im t <  \pi \, r \}$ with continuous boundary values. 
The boundary values are 
	\begin{align*}
	 	 \langle [f] , & \; [\Lambda_1 (i \pi r)_* g] \rangle_{\mathfrak{h}(dS)} \\
		& =
			\int_{dS \times dS} {\rm d} \mu_{dS} ( x_1) {\rm d} \mu_{dS} (  x_2)   { f ( x_1 ) } 
				{\mathcal W}^{(2)} (  x_1 ,  TP_1 x_2 ) g ( x_2) \\
				& =
			\int_{dS \times dS} {\rm d} \mu_{dS} ( x_1) {\rm d} \mu_{dS} (  x_2)   { f ( x_1 ) } 
				{\mathcal W}^{(2)} (  x_1 ,  x_2 ) g ( TP_1 x_2) \\
			& = \langle [f] ,u (TP_1) [g] \rangle_{\mathfrak{h}(dS)} \; . 
	\end{align*}
This identity holds for the \emph{total set} of vectors $\bigl\{ [f] \in \mathfrak{h}(dS) \mid f \in  {\mathcal D}_{\mathbb{R}} (W_1) \bigr\}$. It follows 
that the identity  \eqref{5.4}  holds. 
\end{itemize}
\end{proof}

The space ${\mathfrak h}(W)$ is a \index{standard subspace} \emph{standard subspace}, \emph{i.e.}, 
a $\mathbb{R}$-linear subspace in ${\mathfrak h}(dS)$ such that ${\mathfrak h}(W)
+ i {\mathfrak h}(W)$ is dense in ${\mathfrak h}(dS)$ and ${\mathfrak h}(W)  \cap i {\mathfrak h}(W)  = \{0\}$. 
Thus one can define, following Eckmann and Osterwalder \cite{EO} (see also \cite{LRT} as well as
Theorem~\ref{EckOsw}), a closeable 
anti-linear operator 
	\begin{equation} 
		\begin{matrix}
			s_{\scriptscriptstyle W} \colon & {\mathfrak h}(W) & 
				+ & i {\mathfrak h}(W)  
				& \to&  {\mathfrak h}(W) & 
				+ & i {\mathfrak h}(W)  \\
				& f & + & i g & \mapsto & f & - & i g
						\end{matrix} \; \; . 
	\end{equation}
The polar decomposition of its closure $\overline {s}_{\scriptscriptstyle W} 
= j_{\scriptscriptstyle W} \delta_{\scriptscriptstyle W}^{1/2}$
provides 
\begin{itemize}
\item[---] an anti-unitary involution (\emph{i.e.}, a conjugation) $j_{\scriptscriptstyle W}$; 
\item[---] a $\mathbb{C}$-linear, positive operator $\delta^{1/2}_{\scriptscriptstyle W} $.  
\end{itemize}

\bigskip
According to Theorem \ref{1PStrucHe} $iii.)$ we have 
	\[
		u (TP_1) ( [f] + i [g]) = u (\Lambda_1 ( i \pi)) ([f] +i [g])\; , 
		\qquad [f] , [g] \in {\mathfrak h}^\circ (W_1) \; , 
	\]
Since $u (TP_1)$ is idempotent and anti-linear, this implies   
	\begin{equation}
	\label{6.1.4} 
		( [f] -  i [g]) = u (TP_1) u (\Lambda_1( i\pi)) ( [f] + i [g]) \; , 
		\qquad [f] , [g] \in {\mathfrak h}^\circ (W_1) \; . 
	\end{equation}
The left hand side coincides with $s_{\scriptscriptstyle W_1} ( [f] + i [g])$. 

\begin{theorem}[One-particle Bisognano-Wichmann theorem]
\label{one-p-BW}
The one-particle Tomita operator $s_{\scriptscriptstyle W_1}$ has the polar decomposition 
	\begin{equation} 
		\label{eqSPolarDecomp}
			s_{\scriptscriptstyle W_1} = u 
			\bigl(\Theta_{W_1}\bigr)\, 
			u\bigl(\Lambda_1( i\pi) \bigr) \; .
	\end{equation}
Here $\Theta_{W_1}$ denotes the reflection at the edge of the wedge; 
see \eqref{theta-W}. 
\end{theorem}

\begin{proof} 
The space ${\mathfrak h}^\circ (W)$ is invariant under $u (\Lambda_1(t))$ and therefore is a core  for
$u(\Lambda_1(i\pi))$. Therefore equation \eqref{6.1.4} implies that
$s_{\scriptscriptstyle W_1}$ has the polar decomposition~\eqref{eqSPolarDecomp}. 
\end{proof}

\begin{corollary} The
the quadrupel $ \bigl(K , {\mathfrak h} (\mathbb{W}), u(\Lambda_W (\tfrac{.}{r})), 
u (\Theta_{W})\bigr) $, with $W$ an arbitrary wedge, forms a double $2\pi r$-KMS one-particle 
structure for the classical double dynamical system 
$\bigl( {\mathfrak k}  (\mathbb{W} ) , \sigma, {\mathfrak z} (\Lambda_W (\tfrac{.}{r})),  {\mathfrak z} (\Theta_{W})  \bigr)$ in the 
sense of~\ref{dbops}.
\end{corollary}

\begin{proof}
We verify the properties listed in~\ref{dbops}: 
\begin{itemize}
\item [$ a.)$] ${\mathfrak h} (\mathbb{W})$ is a complex Hilbert space;  in fact, it equals ${\mathfrak h} (dS)$, 
see Proposition~\ref{WdSprop} below.
\item [$ b.)$] The map $K \colon  {\mathfrak k} (\mathbb{W}) \to {\mathfrak h} (\mathbb{W})$ 
is  $\mathbb{R}$-linear and symplectic; see Theorem~\ref{1PStrucHe}~$i.)$. Moreover,   
	\[
		K {\mathfrak k} (W) 
		+ i \, K  {\mathfrak k} (W) 
		= {\mathfrak h}^\circ (W) +    i  \, {\mathfrak h}^\circ (W)  
	\]
is dense in ${\mathfrak h} (dS) = {\mathfrak h} (\mathbb{W})$. This follows from Theorem \ref{oprs}. 

\item [$ c.)$] 
$t \mapsto u (\Lambda_W(t)) $ is a strongly continuous  one-parameter group of unitaries, and 
	\begin{equation}
		u (\Lambda_W(t)) \circ K  
		= K  \circ   {\mathfrak z} (\Lambda_W(t) )   \; . 
		\label{eqUeps2}
	\end{equation}
This is a special case of Theorem \ref{1PStrucHe} $ii.)$. By construction, 
the generator of the boost $ t \mapsto \Lambda_W(t)$ is  unitarily equivalent to 
\eqref{boostgenerator}. 
It has no zero eigenvalue and 
$\mathbb{C}$-linearity together with \eqref{5.1} implies 
	\[
	 \bigl( K {\mathfrak k} (W) + i \,   K {\mathfrak k} (W) \bigr) 
		\subset {\mathscr D} \bigl( u (\Lambda_W(i \pi)) \bigr) \; ; 
	\]
\item [$d.)$] $u (\Theta_{W})$ is a conjugation, and 
	\[
		u (\Theta_{W}) \circ K  =
                K  \circ  {\mathfrak z} (\Theta_{W}) \; .  
	\] 
The pre-Bisognano-Wichmann condition (see \cite[p. 75]{Kay3}) holds:  
	\[
		u (\Lambda_W(i \pi)) \, K [f] =  u (\Theta_{W}) \, 
		 K [f] \; , \qquad [f] \in {\mathfrak k}(W_1) \; .
	\]
Both properties follow from Theorem \ref{1PStrucHe} $ii.)$ and $iii.)$.  
\end{itemize}
\end{proof}

\goodbreak

We will next describe two auxiliary one-particle structures associated to the Cauchy data. The canonical one-particle structures 
will be presented in Section~\ref{co-ps}.

\goodbreak
\section{One-particle structures with positive and negative energy}
\label{posneg}

Let $\widehat {\mathfrak  d} (S^1)$ be the completion of $ {\mathcal D} 
\left(S^1 \setminus \{ - \frac{\pi}{2}, \frac{\pi}{2}\} \right)$ w.r.t.~the
scalar product 
\label{hfdpage}
	\begin{equation}
		\label{eqSkalProd}
		\langle h_1,h_2 {\rangle}_{\widehat {\mathfrak  d} (S^1)}
		 \doteq \langle h_1,(2 | \varepsilon| )^{-1}h_2 
		\rangle_{L^2(S^1, \, |\cos \psi |^{-1} r {\rm d} \psi)} \; ,   
	\end{equation}
for $ h_1, h_2 \in {\mathcal D} 
\left(S^1 \setminus \{ - \frac{\pi}{2}, \frac{\pi}{2}\} \right)$.  
Let $\widehat {\mathfrak  d} (I_\pm)$ be the completion of 
${\mathcal D}_{\mathbb{C}} (I_\pm)$ with respect to the 
scalar product~\eqref{eqSkalProd}. Then
	\[
		\widehat {\mathfrak  d} (S^1) = \widehat {\mathfrak  d} (I_+) \oplus \widehat {\mathfrak  d} (I_-) \; . 
	\]
This follows from Eq.~\eqref{4.9} and Lemma \ref{Lm3.5}. 

\begin{proposition} 
\label{OnePartdS} 
Let $\widehat{K}_\infty \colon \widehat {\mathfrak k}  (S^1) \to 
\widehat {\mathfrak  d} (S^1) $ be the map given by 
	\begin{equation}
		\label{eqKdotf} 
			\bigl( \widehat{K}_\infty (\mathbb{\phi}, \mathbb{\pi}) \bigr) (\psi)  
				\doteq   \cos \psi \; \mathbb{\pi} (\psi)-i
                                \,(\varepsilon \mathbb{\phi}) (\psi)  \; . 
	\end{equation}
Then $\bigl( \widehat{K}_\infty,\widehat {\mathfrak  d} (S^1) , {\rm e}^{it\varepsilon } \bigr)$
forms a one-particle structure in the sense of Definition~\ref{def:A.1}
for the classical dynamical system $\bigl(\, \widehat{\mathfrak k} (S^1) ,
\widehat \sigma, \widehat   {\mathfrak z}(\Lambda_1(t))\bigr)$.
\end{proposition}

\begin{proof} The map \eqref{eqKdotf} is well-defined  for 
$(\mathbb{\phi}, {\mathbb \pi}) \in C^\infty_{\mathbb{R} } (S^1) 
\times C^\infty_{\mathbb{R} } (S^1)$. 
This follows from the fact that $\varepsilon^2$ is a differential operator, which satisfies 
	\[ 
		(\varepsilon^2 h)  (\psi \pm \tfrac{\pi}{2}) = O(\psi)\quad \text{for $\psi \to 0$} \; , 
	\]
and for $h\in C^\infty (S^1)$, just as $\cos (\psi\pm\frac{\pi}{2})$.  In other words, 
there exists a positive constant $M$ such that for all sufficiently small 
values of $|\psi|$, the absolute value of the l.h.s.~is at most $M \cdot |\psi|$. 
Now, use that  $\varepsilon|\varepsilon|^{-1}
=\,  \mathbb{cos}_\psi |\,  \mathbb{cos}_\psi |^{-1}$ equals  $1$ 
on $L^2( I_+, r {\rm d} \psi )$ and $-1$ on $L^2( I_- , r {\rm d} \psi )$ to show
	\begin{align*} 
		& \qquad 2 \Im \langle   \, \widehat{K}_\infty (\mathbb{\phi}_1, {\mathbb \pi}_1),
		\widehat{K}_\infty (\mathbb{\phi}_2, {\mathbb \pi}_2)  {\rangle}_{\widehat {\mathfrak  d} (S^1)}
		\nonumber \\ 
			&\qquad \qquad =  2\Im  \bigl\langle   \,  \mathbb{cos}_\psi  \,
                         {\mathbb \pi}_1-i\varepsilon \, \mathbb{\phi}_1    \, , \,  \tfrac{1}{2 |\varepsilon| } \;
			( \,  \mathbb{cos}_\psi  \, {\mathbb \pi}_2-i \varepsilon  \,
                         \mathbb{\phi}_2  ) \bigr\rangle_{L^2(S^1, \,   |\cos \psi|^{-1}  r {\rm d} \psi)}  \qquad \qquad \nonumber \\  
  			& \qquad \qquad =   \langle \mathbb{\phi}_1   , {\mathbb \pi}_2  \rangle_{L^2(S^1, \,    r {\rm d} \psi)} 
				 -    \langle    {\mathbb \pi}_1   ,   \mathbb{\phi}_2    \rangle_{L^2(S^1, \, r {\rm d} \psi)}     \; . \qquad \qquad 
	\end{align*} 
Thus $\widehat{K}_\infty$ is symplectic.
Moreover, $\widehat{K}_\infty$ intertwines $\widehat {\mathfrak z} (\Lambda_1(t))$ and
${\rm e}^{it\varepsilon}$:  according to~(\ref{varphi-pi-t})  
	\[
		\widehat  {\mathfrak z} (\Lambda_1(t)) (\mathbb{\phi},{\mathbb \pi})=(\mathbb{\phi}_t,{\mathbb \pi}_t)
	\]
with
	\begin{align*} 
		\mathbb{\phi}_t (\psi) 
			&= \big(\cos(\varepsilon t)\mathbb{\phi} 
				- \sin( \varepsilon t) \varepsilon^{-1} \,  \mathbb{cos}_\psi  \;  {\mathbb \pi}\big) (\psi)  \; , \nonumber \\  
		{\mathbb \pi}_t (\psi) 
			&=  \cos^{-1} (\psi)   \big(\varepsilon\sin(\varepsilon t)\mathbb{\phi}  
			+ \cos(\varepsilon t) \,  \mathbb{cos}_\psi  \;  {\mathbb \pi} \big)(\psi)  \; . 
	\end{align*} 
Consequently, 
	\begin{align*} 
		\widehat{K}_\infty \circ \widehat  {\mathfrak z} (\Lambda_1(t)) (\mathbb{\phi},{\mathbb \pi})
			& =  \,  \mathbb{cos}_\psi  \, {\mathbb \pi}_t - i\varepsilon  \,
                \mathbb{\phi}_t   \nonumber \\  
			& =  \big(\varepsilon\sin(\varepsilon t)\; \mathbb{\phi}   + \cos(\varepsilon t) \;
				\,  \mathbb{cos}_\psi  \, {\mathbb \pi}  \big)  \nonumber  \\ 
			&   \qquad \qquad -i \varepsilon  \, \bigl(  \cos(\varepsilon t) \; \mathbb{\phi}   
					- \varepsilon^{-1} \sin( \varepsilon t) \; \,  \mathbb{cos}_\psi  \,  {\mathbb \pi}    \bigr)   
					\nonumber \\  
			& =   \cos (\varepsilon t)  \bigl( \,  \mathbb{cos}_\psi  \, 
			{\mathbb \pi} -i\varepsilon \, \mathbb{\phi}  \bigr)
			+ i  \sin (\varepsilon t)  \bigl(\,  \mathbb{cos}_\psi  \, 
			{\mathbb \pi} -i\varepsilon \, \mathbb{\phi}  \bigr)  \nonumber \\
			&= {\rm e}^{it\varepsilon} (\, \widehat{K}_\infty (\mathbb{\phi}, {\mathbb \pi}))   \; . 
	\end{align*}
Since $|\cos \psi|^{-1}$ is finite  away from the boundary $\partial I_+$ of~$I_+$,  the 
set~$\widehat{K}_\infty \bigl( \, \widehat {\mathfrak k}  (S^1 )  \bigr) $ is dense 
in $\widehat {\mathfrak  d} (S^1)$. 
\end{proof}

\begin{proposition} \label{restrictedOnePartdS} 
Consider the one-particle structure $\bigl(\widehat{K}_\infty,\widehat {\mathfrak  d} (S^1) , 
{\rm e}^{it\varepsilon} \bigr)$. It follows that 
\begin{itemize} 
\item [$ i.)$] 
the restricted structure $ \bigl(\widehat{K}_\infty, \widehat {\mathfrak  d} (I_+) ,
{\rm e}^{it\varepsilon_{\upharpoonright I_+ }}\bigr)$
is a positive energy one-particle  structure for $\bigl(\, \widehat {\mathfrak k} (I_+) ,
\widehat \sigma, \widehat  {\mathfrak z} (\Lambda_1(t)) \bigr)$,
 \emph{i.e.}, 
\begin{itemize}
\item[---] the group 
$t \mapsto {\rm e}^{it\varepsilon_{\upharpoonright I_+ }}$ has a  
positive  generator   $\varepsilon_{\upharpoonright I_+ } \ge 0$;
\item[---] 
$  \widehat{K}_\infty\, \widehat {\mathfrak k}   (I_+)$ is dense 
in $\widehat {\mathfrak  d} (I_+)$. 
\end{itemize}
\item [$ ii.)$] 
the restricted structure $\bigl(\widehat{K}_\infty, \widehat {\mathfrak  d} (I_-) , 
{\rm e}^{it\varepsilon_{\upharpoonright I_- }}\bigr)$
is a negative energy one-particle  structure for $\bigl(\,\widehat {\mathfrak k} (I_-) ,
\widehat \sigma, \widehat  {\mathfrak z} (\Lambda_1(t))\bigr)$, 
\emph{i.e.}, 
\begin{itemize}
\item[---] the group 
$t \mapsto {\rm e}^{it\varepsilon_{\upharpoonright I_- }}$ has a 
 negative generator $\varepsilon_{\upharpoonright I_- } \le 0$;
\item[---] 
$  \widehat{K}_\infty\, \widehat {\mathfrak k}   (I_-)$ is dense in 
$\widehat {\mathfrak  d} (I_-)$. 
\end{itemize}
\item [$ iii.)$] the parity and time-reflections are represented (anti-) unitarily, namely
	\begin{align}
		\label{5.46}
		\widehat{K}_\infty \circ\widehat {\mathfrak z} (P_1) &= - (P_1)_*  \circ \widehat{K}_\infty  \; ;  
		\\
		\label{5.47}
		\widehat{K}_\infty \circ\widehat {\mathfrak z} (T) &= - C  \circ \widehat{K}_\infty  \; , 
	\end{align}
where 
	\[
		(Ch)(\psi)\doteq \overline{h(\psi)} \; , \qquad h \in C^\infty (S^1) \; ,
	\]
extends to $ \widehat {\mathfrak  d} (S^1)$;
\item [$ iv.)$] zero is not an eigenvalue of $\varepsilon$; thus the one-particle structures 
given in $i.)$ and~$ii.)$ are unique, up to unitary equivalence. 
\end{itemize}
\end{proposition}

\begin{proof} $i.)$ and $ ii.)$ follow from \eqref{vaepsdef} as well as the final statement in the proof of 
Proposition \ref{OnePartdS}. For $ iii.)$ use that $({P_1})_*$ anti-commutes with~$\varepsilon$ and
with the multiplication operator $\,  \mathbb{cos}_\psi $. Eq.~\eqref{5.46} follows from
	\[
		\widehat{K}_\infty \bigl( (P_1)_*\mathbb{\phi}, (P_1)_*{\mathbb \pi} \bigr) 
		= -(P_1)_*\,  \mathbb{cos}_\psi  {\mathbb \pi} + i (P_1)_* \varepsilon \mathbb{\phi}
	\] 
and Eq.~\eqref{5.47} follows from 
$\widehat{K}_\infty (\mathbb{\phi}, -{\mathbb \pi}) = - \overline{(\,  \mathbb{cos}_\psi  
{\mathbb \pi} - i \varepsilon \mathbb{\phi})} $. Finally, $iv.)$ follows from Lemma \ref{Lm3.5} 
and Proposition \ref{Kay Th}.
\end{proof}

\begin{proposition}   \label{Jeins}
The operator 
	$
		j \doteq C (P_1)_*
	$
acting on $\widehat {\mathfrak  d} (S^1)$ is an anti-unitary involution (\emph{i.e.}, a conjugation), which implements
the $P_1 T$ transformation and anti-com\-mutes with the generator
$\varepsilon$ of the boosts $t \mapsto \Lambda_1 (t)$: 
	\begin{align} 
		j \circ \widehat{K}_\infty &= \widehat{K}_\infty \circ \widehat {\mathfrak z} (P_1 T) \; ,\label{eqJKKP1T}  \\
		j\, \varepsilon &=-\varepsilon \, j \; .  \label{eqJeps} 
	\end{align}
\end{proposition}

Note that Eq.~\eqref{eqJeps} and anti-linearity imply 
	\begin{equation}
	\label{eqJUt} 
		{\rm e}^{it\varepsilon} \; j=j  \; {\rm e}^{it\varepsilon}\qquad\text{and}
	\qquad j \, |\varepsilon| =|\varepsilon| \, j \; .  
	\end{equation}
	
\begin{proof} 
Clearly, $({P_1})_*$ commutes with $\varepsilon^2$ and hence with
its positive square root $|\varepsilon|$. Now~$\varepsilon$ may be
written 
	\begin{equation}
	\label{varepsilon-chi}
		\varepsilon = |\varepsilon|(\chi_{I_+}- \chi_{I_-}) \; ,
	\end{equation}
where $\chi_{I_\pm}$ denotes multiplication by the characteristic
function of $I_\pm$. Since 
	\[
		({P_1})_* \circ \chi_{I_\pm}
		= \chi_{I_\mp}\circ ({P_1})_*
	\] 
and pointwise complex conjugation commutes with $\varepsilon$, this proves~\eqref{eqJeps}. 
Equation~\eqref{eqJKKP1T}
follows from Proposition \ref{restrictedOnePartdS} $iii.)$. 
\end{proof}

\section{One-particle KMS structures}
\label{KMSops}

Define the $\mathbb{R}$-linear map $\widehat{K}_\beta \colon\widehat {\mathfrak k}(S^1)
\to\widehat {\mathfrak  d} (S^1) $, $\beta >0$,  by 
	\begin{equation*}
		\widehat{K}_\beta (\mathbb{\phi},{\mathbb \pi}) \doteq
			\big((1+ \rho_\beta )^{{\frac{1}{2}}}+\rho_\beta^{\frac{1}{2}}\, j \big) \;
 			\widehat{K}_\infty (\mathbb{\phi},{\mathbb \pi}) 
	\end{equation*}
with 
	\[ 
		\rho_\beta \doteq \frac{{\rm e}^{-\beta|\varepsilon|}}{ 1
		 -  {\rm e}^{-\beta|\varepsilon|} } 
		\quad \text{and} \quad 
		(1+\rho_\beta) = \frac{1}{ 1 -  {\rm e}^{-\beta|\varepsilon|} } \; .
	\]
The domain of $\rho_\beta^{1/2}$ and $(1+\rho_\beta)^{1/2}$ contains 
${\mathscr D} (|\varepsilon|^{-1/2})$, 
as can be seen from the elementary bound \cite[\S A2]{Kay1}
	\[
		 0 < 
		\sqrt{\frac{ {\rm e}^{-\lambda} }{ 1- {\rm e}^{-\lambda} } }\; , \; 
		\sqrt{\frac{ 1 }{ 1- {\rm e}^{-\lambda} } } 
		\le const. \cdot \max (1, \lambda^{-1/2}) \; , 
		\qquad \lambda \in \mathbb{R}^+ \; . 
	\]

Note that $\widehat{K}_\beta$ is {\em not} the Araki-Woods 
map $K_{\scriptscriptstyle \rm AW}$
discussed in~\ref{Kbeta}, as $K_{\scriptscriptstyle \rm AW}$ 
would map 
$\widehat {\mathfrak k}(I_+)$ to $\widehat {\mathfrak  d} (I_+) 
\oplus \overline{\widehat {\mathfrak  d} (I_+)}$.

\begin{proposition} 
\label{OnePart}  
The quadruple $\bigl( \widehat{K}_\beta , \widehat {\mathfrak  d} (S^1), {\rm e}^{i \frac{.}{r} \varepsilon},  j \bigr) $
is a double $\beta r $-KMS one-particle structure for the classical 
double dynamical system 
\[ \bigl(\, \widehat {\mathfrak k}  (S^1 ) , \widehat \sigma, 
\widehat  {\mathfrak z} (\Lambda_1 (\tfrac{.}{r})), \widehat {\mathfrak z} (P_1 T)  \bigr)
\]
in the sense of Kay, see~\ref{dbops}.  
\end{proposition}

\begin{proof}
Let $\widehat{\mathbb{\Phi}}_i=(\mathbb{\phi}_i,{\mathbb \pi}_i) \in \widehat {\mathfrak k}(S^1)$, $i=1,2$ and denote the scalar 
product in $\widehat {\mathfrak  d} (S^1)$ just by~$\langle \, . \, ,\, . \,  \rangle$.  Then 
	\begin{align} 
	\Im	\langle \, \widehat{K}_\beta \widehat{\mathbb{\Phi}}_1,
	\widehat{K}_\beta \widehat{\mathbb{\Phi}}_2\rangle &= 
	\Im \big\{	\langle \, \widehat{K}_\infty \widehat{\mathbb{\Phi}}_1,
	(1+\rho_\beta) \widehat{K}_\infty \widehat{\mathbb{\Phi}}_2\rangle 
		+ \langle \, j  \widehat{K}_\infty \widehat{\mathbb{\Phi}}_1, 
		\rho_\beta j \widehat{K}_\infty \widehat{\mathbb{\Phi}}_2\rangle \big\} 
		\nonumber  \\
	&= \Im \big\{ \langle \, \widehat{K}_\infty \widehat{\mathbb{\Phi}}_1,
	(1+\rho_\beta) \widehat{K}_\infty \widehat{\mathbb{\Phi}}_2\rangle 
		+ \overline{ \langle \, \widehat{K}_\infty \widehat{\mathbb{\Phi}}_1,
		 \rho_\beta \widehat{K}_\infty \widehat{\mathbb{\Phi}}_2 \rangle}\big\} \nonumber  \\
		&= 
		\Im\big\{ \langle \, \widehat{K}_\infty \widehat{\mathbb{\Phi}}_1,(1+\rho_\beta) 
		\widehat{K}_\infty \widehat{\mathbb{\Phi}}_2\rangle 
		- \langle \, \widehat{K}_\infty \widehat{\mathbb{\Phi}}_1, \rho_\beta \widehat{K}_\infty 
		\widehat{\mathbb{\Phi}}_2\rangle\big\} \nonumber \\
		&= \Im\langle \,\widehat{K}_\infty \widehat{\mathbb{\Phi}}_1, 
		\widehat{K}_\infty \widehat{\mathbb{\Phi}}_2\rangle
			=\frac{1}{2}\,\widehat\sigma(\widehat{\mathbb{\Phi}}_1,\widehat{\mathbb{\Phi}}_2) \; . 
		\label{eqImbeta} 
	\end{align}
In the third equality we used that $\Im z = - \Im \overline{z}$ for any $z \in \mathbb{C}$.

\goodbreak
Now verify the properties listed in Definition~\ref{dbops}: 
\begin{itemize}
\item [$i.)$] $\widehat {\mathfrak  d} (S^1)$ is a complex Hilbert space;  
\item [$ii.)$] the map $\widehat{K}_\beta \colon \widehat {\mathfrak k}
 \left(S^1\right) \to \widehat {\mathfrak  d} (S^1)$ is  $\mathbb{R}$-linear and symplectic, as can be 
seen from Eq.~\eqref{eqImbeta}. Moreover,   
	\begin{align*}
		\qquad & 
		\widehat{K}_\beta \;  \widehat {{\mathfrak k} } (I_+)   
		+ i \; \widehat{K}_\beta \; \widehat {{\mathfrak k} } (I_+) 
		\nonumber \\ 
		&\quad =    \big((1+\rho_\beta)^{{\frac{1}{2}}}+\rho_\beta^{\frac{1}{2}}\,j \big)  
		\widehat{K}_\infty\, \widehat {{\mathfrak k} } (I_+) 
			+ i \big((1+\rho_\beta)^{{\frac{1}{2}}}+\rho_\beta^{\frac{1}{2}}\,j \big) 
			\widehat{K}_\infty\, \widehat {{\mathfrak k} } (I_+)
		\nonumber \\ 
		&\quad =    (1+\rho_\beta)^{{\frac{1}{2}}}  \bigl( \widehat{K}_\infty \, \widehat {{\mathfrak k} } (I_+) 
		+i \widehat{K}_\infty \, \widehat {{\mathfrak k} } (I_+)  \bigr)
		\nonumber \\ 
		&	\qquad \qquad	+ \rho_\beta^{\frac{1}{2}}  \bigl( 
		\widehat{K}_\infty \circ \widehat {\mathfrak z} (P_1 T) \widehat {{\mathfrak k} } (I_+) 
		+ i \widehat{K}_\infty \circ \widehat {\mathfrak z} (P_1 T) \widehat {{\mathfrak k} } (I_+)  \bigr) 
				\nonumber \\ 
		&\quad =    (1+\rho_\beta)^{{\frac{1}{2}}}   \bigl( \widehat{K}_\infty \, \widehat {{\mathfrak k} } (I_+) 
		+i \widehat{K}_\infty \, \widehat {{\mathfrak k} } (I_+)  \bigr)
 			+ \rho_\beta^{\frac{1}{2}}  \bigl( \widehat{K}_\infty  \,  \widehat {{\mathfrak k} } (I_-) 
			+ i \widehat{K}_\infty \, \widehat {{\mathfrak k} } (I_-) \bigr)\; .  
	\end{align*}
It follows from Proposition \ref{restrictedOnePartdS} $i.)$ and $ii.)$ that this set is dense 
in $\widehat {\mathfrak  d} (S^1)$. We have also used \eqref{eqJKKP1T} and the fact 
that $(1+\rho_\beta)$ and $\rho_\beta$ are strictly positive, and therefore invertible on 
$\widehat{K}_\infty \, {\mathcal D}_{\mathbb{C}} \left(S^1 \setminus \{ - \frac{\pi}{2}, 
\frac{\pi}{2}\} \right)$. 

\item [$iii.)$] 
$t \mapsto {\rm e}^{it \varepsilon } $ is a strongly continuous  group of unitaries, and \eqref{eqJUt} implies
	\begin{align}
		{\rm e}^{it \varepsilon } \circ \widehat{K}_\beta  
		&= {\rm e}^{it  \varepsilon}\circ 
		\big((1+\rho_\beta)^{{\frac{1}{2}}}+\rho_\beta^{\frac{1}{2}}\,j \big) \circ  
		\widehat{K}_\infty \nonumber \\
		&=
			\big((1+\rho_\beta)^{{\frac{1}{2}}}+\rho_\beta^{\frac{1}{2}}\,j \big) \circ 
			{\rm e}^{it \varepsilon}\circ \widehat{K}_\infty  \nonumber \\
		&= \widehat{K}_\beta  \circ \widehat  {\mathfrak z} (\Lambda_1(t) ) \; .  
		\label{eqUeps}
	\end{align}
Let $\widehat{\mathbb{\Phi}}=(\mathbb{\phi},{\mathbb \pi})$, where $\mathbb{\phi}$ and
${\mathbb \pi}$ have compact supports in the open half-circle $I_+$. Then 
	\[
		\varepsilon \, \widehat{K}_\infty \,  \widehat{\mathbb{\Phi}} 
		=|\varepsilon|\, 
		\widehat{K}_\infty \, \widehat{\mathbb{\Phi}} 
		\qquad \text{and} \qquad \varepsilon \,  j \, \widehat{K}_\infty \, \widehat{\mathbb{\Phi}} 
		= -|\varepsilon| \, j \, \widehat{K}_\infty
		\, \widehat{\mathbb{\Phi}} \; . 
	\]
This implies   
	\begin{align*} 
		{\rm e}^{-\beta\varepsilon/2}\, \widehat{K}_\beta \widehat{\mathbb{\Phi}} 
		&= 
			{\rm e}^{-\beta\varepsilon/2}
			\big((1+\rho_\beta)^{{\frac{1}{2}}} 
			+\rho_\beta^{\frac{1}{2}}\,j\big)\,  \widehat{K}_\infty \, \widehat{\mathbb{\Phi}} \nonumber \\ 
		& =  {\rm e}^{-\beta|\varepsilon|/2}(1+\rho_\beta)^{{\frac{1}{2}}} \,\widehat{K}_\infty\, \widehat{\mathbb{\Phi}}+ 
				{\rm e}^{\beta|\varepsilon|/2}\, \rho_\beta^{\frac{1}{2}}\,j \, \widehat{K}_\infty \, \widehat{\mathbb{\Phi}}
		\\
		&=
			\left(\tfrac{{\rm e}^{-\beta|\varepsilon|}}
			{1-{\rm e}^{-\beta|\varepsilon|}}\right)^{\frac{1}{2}} \,\widehat{K}_\infty \, 
			\widehat{\mathbb{\Phi}} 
			+ \left(
			\tfrac{1}
			{1-{\rm e}^{-\beta|\varepsilon|}} \right)^{\frac{1}{2}}\,
			j  \, \widehat{K}_\infty \, \widehat{\mathbb{\Phi}}  \nonumber \\
		&= \rho_\beta^{\frac{1}{2}} \, \widehat{K}_\infty \, \widehat{\mathbb{\Phi}} + 
		(1+\rho_\beta)^{\frac{1}{2}} j \, \widehat{K}_\infty \, \widehat{\mathbb{\Phi}} \; . 
	\end{align*}
Thus (by linearity) 
	\begin{equation}
	 \label{can-geo}
	 \widehat{K}_\beta \, \widehat {\mathfrak k}   (I_+ ) 
	 + i \,\widehat{K}_\beta \, \widehat {\mathfrak k}   (I_+ )
		\subset {\mathscr D} \bigl( {\rm e}^{-\beta\varepsilon / 2 } \bigr) \; ; 
	\end{equation}
Moreover, according to Lemma \ref{Lm3.5},
zero is not an eigenvalue of the generator $ \varepsilon$. 
\item [$iv.)$] $j $ is a conjugation, and  
	\[
		j \circ \widehat{K}_\beta  =
                \widehat{K}_\beta  \circ \widehat {\mathfrak z} (P_1T) \;  
	\] 
by Lemma~\ref{Jeins} and the fact that $j$ commutes with $\rho_\beta$. 
The KMS condition holds: we have already seen that 
	\begin{align*} 
		{\rm e}^{-\beta\varepsilon/2}\, \widehat{K}_\beta \widehat{\mathbb{\Phi}} 
		&= \rho_\beta^{\frac{1}{2}} \, \widehat{K}_\infty \, \widehat{\mathbb{\Phi}} + 
		(1+\rho_\beta)^{\frac{1}{2}} j \, \widehat{K}_\infty \, \widehat{\mathbb{\Phi}} =  j \, 
			\widehat{K}_\beta \,\widehat{\mathbb{\Phi}} \;  .
	\end{align*}
\end{itemize}
\end{proof}

\begin{lemma} 
\label{lm:6.3.2}
Let $h\in {\mathcal D}_{\mathbb{C}} \left(S^1 \setminus \{ - \frac{\pi}{2}, 
\frac{\pi}{2}\} \right)$. Then 
\begin{align}
\bigl\| \bigl((1+& \rho_\beta)^{{\frac{1}{2}}}+\rho_\beta^{\frac{1}{2}}\, (P_1)_* \bigr) h \bigr\|_{\widehat {\mathfrak  d} (S^1)}^2 
 \\
& =  \bigl\langle 
		h \, , 
		\tfrac{1}{2 |\varepsilon| }\, \bigl(\coth \tfrac{\beta\varepsilon}{2}+\tfrac{(P_1)_*}
					{\sinh \frac{ \beta |\varepsilon|}{2}} \bigr) 
		h  	\bigr\rangle_{L^{2}(S^1, \frac{r {\rm d} \psi}{|\cos \psi|})} \; . 
\nonumber
\end{align}
\end{lemma}

\begin{remark}
We note that this lemma may be of interest, if $\beta \ne 2\pi$. The special case
$\beta = 2\pi$ is treated in the next subsection.
\end{remark}

\begin{proof} Write $h=h_+ + h_-$, where the support of $h_\pm$ is contained in
$I_\pm$, respectively. Note that $(P_1)_*$ commutes with $|\varepsilon|$ and with the
multiplication operator~$|\,  \mathbb{cos}_\psi |$. It follows that  
	\begin{align*} 
		 \bigl\| \bigl((1+ \rho_\beta)^{{\frac{1}{2}}}+\rho_\beta^{\frac{1}{2}} (P_1)_* \bigr) 
		 h_\pm \bigr\|_{\widehat {\mathfrak  d} (S^1)}^2 
			 &= \bigl\langle  
					 h_\pm \,  , (1+2\rho_\beta)  
					h_\pm \bigr\rangle_{\widehat {\mathfrak  d} (S^1)} 
					\\
			&= \bigl\langle 
			h_\pm  \,  ,  \frac{\coth  \tfrac{ \beta}{2} |\varepsilon| }
			{2  |\varepsilon|   }  
			h_\pm   \bigr\rangle_{L^{2}(S^1 ,  
			\frac{r {\rm d} \psi}{|\cos \psi|}) } \; . 
	\end{align*}
For the mixed terms, we find 
	\begin{align}
		\bigl\langle 
		 \bigl( (1+ &\rho_\beta)^{ \frac{1}{2} }
		 +\rho_\beta^{ \frac{1}{2} }\, (P_1)_* \bigr)  h_+ \, , 
		\bigl((1+ \rho_\beta)^{ \frac{1}{2} }
		+\rho_\beta^{ \frac{1}{2} }\, (P_1)_* \bigr)
		h_-   \bigr\rangle_{ \widehat {\mathfrak  d} (S^1) }
 		\nonumber \\
		&=  \bigl\langle 
					h_+ \,  ,  \tfrac{{\rm e}^{- \frac{\beta}{2} |\varepsilon|} }
					{ |\varepsilon| (1- {\rm e}^{- \beta |\varepsilon|  })}  
					(P_1)_*h_-   \bigr\rangle_{L^{2} (I_+ ,
                        			 \frac{ r{\rm d} \psi}{|\cos \psi|}) }  \nonumber \\
		&= \bigl\langle 
				h_+  \, ,  \tfrac{1}{2  |\varepsilon| \sinh \frac{\beta}{2} 
 					|\varepsilon|}
					(P_1)_*h_-   \bigr\rangle_{L^{2}(I_+ ,
                          			\frac{r {\rm d} \psi}{|\cos \psi|}) } \; .  
	\end{align}
We have used the identities $1+2\rho_\beta=\coth \frac{\beta}{2} |\varepsilon|$ and 
	\[
		2(\rho_\beta (1+\rho_\beta ))^{\frac{1}{2}}
		= \bigl(\sinh \tfrac{\beta}{2} |\varepsilon| \bigr)^{-1}  \; . 
	\]
The term with $h_+$ and $h_-$ interchanged yields a similar
expression. Putting together the four terms, and noting that $\varepsilon$ leaves
the subspaces $L^{2}(I_\pm,\frac{r{\rm d} \psi}{|\cos \psi|})$ invariant,  
completes the proof.
\end{proof}

\section{The canonical one-particle structure}
\label{co-ps}

It was recognised by Borchers and Buchholz~\cite{BoB} that the 
proper, ortho\-chron\-ous Lorentz group $SO_0(1,2)$ 
can be unitarily implemented iff~$\beta$ is equal to the 
Hawking\footnote{In the present context, the temperature $T= 2 \pi r$
was first derived by Figari, H\o egh-Krohn  and Nappi~\cite{FHN}. The article by Hawking was submitted soon afterwards.}
temperature $2 \pi r$ \cite{Ha, Se1, Se2}. In fact, we will now show that if~$\beta = 2 \pi r$, then the unitary map 
	\begin{align*} 
		\mathbb{v} \colon \widehat {\mathfrak  d} (S^1) & \to \widehat{{\mathfrak h}}  (S^1) \\
		h & \mapsto 
		\tfrac{1}{\sqrt{r}}  \; 
		|\,  \mathbb{cos}_\psi |^{-1}\, \big( \rho_{2 \pi}^{\frac{1}{2}} (P_1)_* - (1+\rho_{2 \pi})^{\frac{1}{2}}  \big) h \; , 
	\end{align*}
allows us to implement the rotations $R_0 (\alpha)$, $\alpha \in [0, 2 \pi)$, 
in the double $(2 \pi r)$-KMS 
one-particle structure introduced in Proposition \ref{OnePart}. 
The Hilbert space $\widehat{{\mathfrak h}}  (S^1) $
was introduced in Definition~\ref{zeit-null-Hilbert}.

\begin{proposition}
\label{Prop5.6}
The operator $\mathbb{v}$ is unitary, \emph{i.e.}, 
	\[
		\| \mathbb{v} h \|_{\widehat{{\mathfrak h}}  (S^1)} = \| h \|_{\widehat {\mathfrak  d} (S^1)} \; . 
	\]
Its inverse $ \mathbb{v}^{-1} \colon \widehat{{\mathfrak h}}  (S^1)
\to  \widehat {\mathfrak  d} (S^1)$ is given by 
	\begin{equation}
  		\label{eq2piH1}
			 \mathbb{v}^{-1}  
			= - \sqrt{r}  \big( (1+\rho_{2 \pi})^{\frac{1}{2}} +
			\rho_{2 \pi}^{\frac{1}{2}}
			(P_1)_*\big) \,|\,  \mathbb{cos}_\psi |   \, . 
	\end{equation}
\end{proposition}

\begin{proof} 
Let $h\in \widehat{{\mathfrak h}}  (S^1)$. Using again  that $(P_1)_*$ commutes 
with $|\varepsilon|$ and with the multiplication operator~$|\,  \mathbb{cos}_\psi |$, we find 
	\begin{align} 
		\| \mathbb{v}^{-1} h\|_{\widehat {\mathfrak  d} (S^1)}^2 
			&=   r \,  
			\bigl\| \big((1+ \rho_{2 \pi})^{{\frac{1}{2}}}
			+\rho_{2 \pi}^{\frac{1}{2}}\,(P_1)_* \big) \,|\,  \mathbb{cos}_\psi | \, 
			h \bigr\|_{\widehat {\mathfrak  d} (S^1)}^2    
					\nonumber
					\\
				& =   r \, 
					\bigl\langle |\,  \mathbb{cos}_\psi | \, h \,,\, (1+2\rho_{2 \pi})
					  |\,  \mathbb{cos}_\psi | \, h \bigr\rangle_{\widehat {\mathfrak  d} (S^1)} 
					\nonumber \\
					&
					\qquad  + 2  r \, 
					  \bigl\langle |\,  \mathbb{cos}_\psi | \, h \,,\, 
					  (\rho_{2 \pi}(1+\rho_{2 \pi}))^{\frac{1}{2}} (P_1)_* |\,  \mathbb{cos}_\psi |  \, h 
					\bigr\rangle_{\widehat {\mathfrak  d} (S^1)} 
										\nonumber \\
			&=  r \,   \bigl\langle  h,\tfrac{1}{2  |\varepsilon| } 
					\bigl(\coth   \pi |\varepsilon| +  
					\tfrac{(P_1)_*}{\sinh \pi |\varepsilon | }\bigr) |\,  \mathbb{cos}_\psi | \, h  
					\bigr\rangle_{L^{2}(S^1, r {\rm d} \psi)} 
					\nonumber
					\\
			&=   \|  h \|^{2} _{\widehat{{\mathfrak h}} (S^1)} 
				\; . \label{eqMagicFormC} 
	\end{align}
Just as in the proof of Lemma~\ref{lm:6.3.2}, we
have again used the identities $1+2 \rho_{2 \pi}=\coth \pi |\varepsilon|$ and 
	\[
		2( \rho_{2 \pi} (1+\rho_{2 \pi}))^{\frac{1}{2}}= (\sinh\pi |\varepsilon|)^{-1}  \; .  
	\]
The last equality in \eqref{eqMagicFormC} follows from Corollary~\ref{cor:4.5.5}, 
which together Theorem~\ref{st-kappa} yields
\begin{equation}
		\langle h , h \rangle_{\widehat{{\mathfrak h}} (S^1)}  
		=  r \, \bigl\langle  h,\tfrac{1}{2  |\varepsilon| } 
					\bigl(\coth   \pi |\varepsilon| +  
					\tfrac{(P_1)_*}{\sinh \pi |\varepsilon | }\bigr) |\,  \mathbb{cos}_\psi | \, h  
					\bigr\rangle_{L^{2}(S^1, r {\rm d} \psi)} \; ,  
					\label{eqMagicFormB} 
	\end{equation}
with $\varepsilon^2  =-  (\cos \psi \, \partial_\psi)^2 +(\cos \psi)^{2}\mu^2r^2$.  
\end{proof}

\begin{remark} 
Inspecting Proposition~\ref{thhm} and
applying the polarisation identity to \eqref{eqMagicFormB}, we find  
	\begin{equation}
	\label{bounded-magic}
		\omega^{-1} = \tfrac{1}{ |\varepsilon| } \, 
		\bigl(\coth\pi |\varepsilon|+\tfrac{(P_1)_*}{\sinh\pi|\varepsilon|} \bigr) \,
				| r \,  \mathbb{cos}_\psi | \, .   
	\end{equation}
Note that \eqref{bounded-magic} is an identity of bounded operators, 
with the right hand side initially defined 
on smooth functions, using arguments similar to those used in the 
proof of Proposition \ref{porp4.13} to define
$(\sinh\pi|\varepsilon|)^{-1} |   \mathbb{cos}_\psi |$ on smooth functions. 

Hence, the unbounded operator $\omega $ introduced  
in Proposition~\ref{thhm} satisfies (for example, 
on the dense set of smooth functions in $L^2(S^1, \, r{\rm d}\psi)$) 
the operator identity 
	\begin{equation}
		\label{eqMagicForm}
 		\omega =   | r \,  \mathbb{cos}_\psi|^{-1}\,|\varepsilon|\,  
		\bigl(\coth\pi |\varepsilon|
		-  
		\tfrac{(P_1)_*}{\sinh\pi|\varepsilon|} \bigr)  \; ,  
	\end{equation}
which follows from \eqref{bounded-magic} and the 
identity  $\cosh^2 \psi - 1 = \sinh^2 \psi$.
\end{remark}

\begin{proposition}
\label{prop:6.4.3}
Consider the map 
\label{Kmhatpage}
	\begin{align*}
	\label{K1PStrucHe}
		 \widehat{K} \colon  \qquad \widehat{\mathfrak k}(S^1) 
		& \rightarrow  \widehat{{\mathfrak h}}  (S^1) \nonumber \\
		 (\mathbb{\phi},\mathbb{\pi})  		  & \mapsto   \tfrac{1}{\sqrt{r}}  ( - \mathbb{\pi} +  i \,\omega
		 r \, \mathbb{\phi}) \; .
	\end{align*}
It follows that the quadruple 
	\[
		\bigl(\widehat{K} , \widehat{{\mathfrak h}}  (S^1), {\rm e}^{i   
		t \omega \,  \mathbb{cos}_\psi } , 
		C (P_1)_* \bigr) 
	\]
forms a double $2\pi r $-KMS one-particle structure for the classical 
double dynamical system 
$\bigl( \widehat {\mathfrak k}  \left(S^1 \setminus \{ - \frac{\pi}{2} , 
\frac{\pi}{2} \} \right) , \widehat \sigma, \widehat {\mathfrak z} (\Lambda_1( \frac{t}{r}) ),  
\widehat {\mathfrak z} (P_1T )  \bigr)$ in the sense
of~\ref{dbops}, unitarily equivalent to 
$\bigl(\, \widehat{K}_{2 \pi} , \widehat {\mathfrak  d} (S^1), {\rm e}^{i \frac{t}{r} \varepsilon},  j \bigr) $, 
in agreement with Theorem~\ref{ThB1}.
\end{proposition}

\begin{proof} We first show that $\widehat{K} =  \mathbb{v} \circ \widehat{K}_{2 \pi} $.   
Using $j = C (P_1)_*$, where $(C h)(\psi)\doteq \overline{h(\psi)}$, one gets  
	\begin{align*}
		& \quad \; \;  \mathbb{v} \circ 
		\widehat{K}_{2 \pi} (\mathbb{\phi},\mathbb{\pi}) 
		\\
		& \qquad =   \mathbb{v} \circ 
			\big((1+\rho_{2 \pi})^{{\frac{1}{2}}}+\rho_{2 \pi}^{\frac{1}{2}}\,j \big) \;
 			\widehat{K}_\infty (\mathbb{\phi},\mathbb{\pi}) 
			\nonumber \\
		&\qquad = - \tfrac{1}{\sqrt{r}}  \, |\,  \mathbb{cos}_\psi |^{-1}\,
 			\big( (1+\rho_{2 \pi})^{\frac{1}{2}} - \rho_{2 \pi}^{\frac{1}{2}} (P_1)_*\big)
			\big((1+\rho_{2 \pi})^{{\frac{1}{2}}}+\rho_{2 \pi}^{\frac{1}{2}}\,j \big) \;
 			\widehat{K}_\infty (\mathbb{\phi},\mathbb{\pi}) 
			\nonumber \\
		&\qquad =	 - \tfrac{1}{\sqrt{r}}  \, |\,  \mathbb{cos}_\psi |^{-1}\,
		\Big( 1+\big(\rho_{2 \pi}-(\rho_{2 \pi }(1+\rho_{2 \pi}))^{\frac{1}{2}}
				(P_1)_*\big)(1-C ) \Big) \,
		\widehat{K}_\infty (\mathbb{\phi},\mathbb{\pi}) \; .
			\nonumber  				
	\end{align*}
Taking $1+2\rho_{2 \pi} =\coth \pi  |\varepsilon|$ and 
$2(\rho_{2 \pi} (1+\rho_{2 \pi} ))^{\frac{1}{2}}= (\sinh\pi  |\varepsilon|)^{-1}$ into account, we find 
	\[
		\mathbb{v} \circ \widehat{K}_{2 \pi} (\mathbb{\phi}, 
		\mathbb{\pi} ) 
		=  \begin{cases}  - \tfrac{1}{\sqrt{r}}  
		\, |\,  \mathbb{cos}_\psi |^{-1}\, 
		\widehat{K}_\infty (\mathbb{\phi},\mathbb{\pi})&
		\text { if } \widehat{K}_\infty
			(\mathbb{\phi},\mathbb{\pi})  \in 
			\widehat {\mathfrak  d} (S^1, {\mathbb{R}})\; ,\\
				-   \sqrt{r}   \, 
				\omega \, \varepsilon^{-1}\, 
				\widehat{K}_\infty (\mathbb{\phi},\mathbb{\pi}) 
				& \text{ if }  \widehat{K}_\infty
				(\mathbb{\phi},\mathbb{\pi}) 
				\in i \widehat {\mathfrak  d} (S^1, {\mathbb{R}})\, \nonumber .
				\end{cases}  
	\]
In the last equation we have used~\eqref{eqMagicForm} 
and $(P_1)_* \varepsilon = - \varepsilon (P_1)_*$. 
By $\widehat {\mathfrak  d} (S^1, {\mathbb{R}})$ we have denoted the real 
subspace of real valued functions 
in $\widehat {\mathfrak  d} (S^1)$. We will 
use $\widehat{K}_\infty (\mathbb{\phi}, \mathbb{\pi}) 
=  \,  \mathbb{cos}_\psi  \,\mathbb{\pi}  -i  \varepsilon \mathbb{\phi} $ to prove that
	\begin{equation}
	\label{can-geo2}
		\widehat{K} =  \mathbb{v} \circ \widehat{K}_{2 \pi} \; . 
	\end{equation}  
It remains to show that the unitary map $\mathbb{v}$ satisfies 
	\[ 
	  	\mathbb{v}  \circ \varepsilon \circ \mathbb{v}^{-1} 
		= \omega  \, r \, \mathbb{cos}_\psi  \quad 
	 	\text{and} \quad
	 	\mathbb{v} \circ j\circ \mathbb{v}^{-1}  
	  =  C (P_1)_*  \quad \text{on $\widehat{{\mathfrak h}}  (S^1)$} \, . 
	 \]
Using again $(P_1)_* \varepsilon = - \varepsilon (P_1)_*$,  we can verify the first of these two 
identities: 
	\begin{align*}
			 \mathbb{v}  \circ \varepsilon \circ  \mathbb{v}^{-1} 
			&=    |\,  \mathbb{cos}_\psi |^{-1}\,
 			\bigl( (1+\rho_{2 \pi})^{\frac{1}{2}} 
			- \rho_{2 \pi}^{\frac{1}{2}} (P_1)_*\bigr) 
						\varepsilon 
			\bigl( (1+\rho_{2 \pi})^{\frac{1}{2}}+\rho_{2 \pi}^{\frac{1}{2}}  (P_1)_*\bigr)\,
						|\,  \mathbb{cos}_\psi | 
						\nonumber \\
			& =  |\,  \mathbb{cos}_\psi |^{-1}\, 
				\bigl(  (1+2 \rho_{2 \pi})  
				-  2 \rho_{2 \pi}^{\frac{1}{2}} (1+\rho_{2 \pi})^{\frac{1}{2}} (P_1)_* 
				\bigr) \, \varepsilon  \, |\,  \mathbb{cos}_\psi |
			\nonumber \\
			& =  |\,  \mathbb{cos}_\psi |^{-1}  
			\big( (1+2\rho_{2 \pi})  
			- 2\rho_{2 \pi}^{\frac{1}{2}}(1+\rho_{2 \pi})^{\frac{1}{2}} (P_1)_*\big) 
			\;  | \varepsilon | \, \mathbb{cos}_\psi  
			\nonumber \\
			&= |\,  \mathbb{cos}_\psi |^{-1}  
				| \varepsilon | \,
				\bigl(\coth \pi  |\varepsilon|
		 	 		-  \tfrac{(P_1)_*}{\sinh\pi |\varepsilon|} \bigr) 
				\;  \mathbb{cos}_\psi   
			\nonumber \\
			&= \omega r  \,  \mathbb{cos}_\psi  \; . 
	\end{align*}
In the second but last equality we have used the identity \eqref{eqMagicForm} 
as well as 
	\[
	  	\varepsilon \,|\,  \mathbb{cos}_\psi |  = | \varepsilon | \, \mathbb{cos}_\psi  \; . 
	\]
The second identity follows from the fact that $j  =C (P_1)_* $ 
commutes with the multiplication operator $|\,  \mathbb{cos}_\psi |$ 
and with $|\varepsilon|$: 
	\[ 
	  	\mathbb{v} \circ j \circ \mathbb{v}^{-1}
		=  C (P_1)_*  \quad \text{on $\widehat{{\mathfrak h}}  (S^1)$} \, . 
	 \]
We have thus established unitary 
equivalence of the two double $2\pi r$-KMS 
one-particle structure under consideration, in agreement with Theorem~\ref{ThB1}.

It is now straight forward to verify that $\bigl(\widehat{K} , \widehat{{\mathfrak h}}  (S^1), 
{\rm e}^{i t \omega \,  \mathbb{cos}_\psi } , C (P_1)_* \bigr)$ forms a double $2\pi r$-KMS 
one-particle structure for the classical double dynamical system 
$\bigl( \widehat {\mathfrak k}  \left(S^1 \setminus \{ - \frac{\pi}{2} , 
\frac{\pi}{2} \} \right) , \widehat \sigma, \widehat {\mathfrak z} (\Lambda_1(\frac{.}{r}) ),  
\widehat {\mathfrak z} (P_1T )  \bigr)$ in the sense of~\ref{dbops}:   
	\begin{align}
		 \widehat{K}\circ \widehat {\mathfrak z} (\Lambda_1 (t))  
		& =  \mathbb{v} \circ \widehat{K}_{2 \pi } 
		\circ \widehat {\mathfrak z} (\Lambda_1 (t)) \nonumber \\
		& = \mathbb{v}  \circ {\rm e}^{it \varepsilon} \;  
		\widehat{K}_{2 \pi } \nonumber \\
		& =  {\rm e}^{it\, \omega r \,  \mathbb{cos}_\psi } \circ \mathbb{v}
		\circ  \widehat{\mathbb k}_{2 \pi r} \nonumber \\
		& =  \widehat{u}(\Lambda_1(t)) \circ \widehat{K} \; , 
		\label{canonical-boosts}
	\end{align}
see Eq.~\eqref{eqUeps}; and 
	\begin{equation}
		\label{reflections}
		\widehat{K} \circ \widehat {\mathfrak z} (P_1T) 
		= \mathbb{v} \circ \widehat{K}_{2 \pi r} \circ 
		\widehat {\mathfrak z} (P_1T)
		= \mathbb{v} \circ j \circ \widehat{K}_{2 \pi r}
		=C (P_1)_* \circ \widehat{K}  \; . 
	 \end{equation}
This also shows that $\widehat{u} (P_1T) = C (P_1)_* $ is anti-unitary. 
\end{proof}
 
\begin{theorem} 
\label{1PStrucHe2} 
The triple $  \bigl(\widehat{K} , \widehat{{\mathfrak h}}  (S^1), \widehat{u} \bigr) $
is a {\em de Sitter  one-particle 
structure} for the canonical classical dynamical system 
$\bigl( \, \widehat{\mathfrak k}  (S^1 ) , \widehat\sigma, \widehat{\mathfrak z}   \bigr)$ introduced 
in Proposition~\ref{nocheinlabel}. In other words, 
\begin{itemize}
\item [$ i.)$] $\widehat{K}$ defines a symplectic map from $(\widehat{\mathfrak k}  (S^1 ) , \widehat\sigma)$ 
to $\bigl( \, \widehat{{\mathfrak h}}  (S^1), \Im \langle \, . \, , 
\, . \, \rangle_{\widehat{{\mathfrak h}}  (S^1)} \bigr)$ and $\widehat{K} \, \widehat{\mathfrak k}  (S^1 )$ 
is dense in $\widehat{{\mathfrak h}}  (S^1)$;
\item [$ ii.)$] 
there exists a unique (anti-) unitary representation of $O(1,2)$ satisfying 
	\begin{equation} 
		\label{eqUHe2}
		 \widehat{u} (\Lambda)\circ \widehat{K} = \widehat{K} \circ \widehat {\mathfrak z} (\Lambda) \; . 
	\end{equation}
Moreover, $\widehat{u} (R_{0}(\alpha)) = R_{0}(\alpha)_*$ for $\alpha \in [0, 2 \pi)$;
\item [$ iii.)$]  for any half-circle\footnote{Given the fact that we consider $\widehat{{\mathfrak h}}  (S^1)$,  
it is more natural to specify a half-circle $I_\alpha= R_{0}(\alpha) I_+$. Recall that $W^{(\alpha)} = I_\alpha''$.}
$I_\alpha$, the pre-Bisognano-Wichmann property \cite[p.~75]{Kay3} holds: 
	\begin{equation} 
		\label{5.3b}
		 \widehat{K}  \, \widehat{{\mathfrak k}} (I_\alpha)  \subset {\mathscr D} \bigl( \widehat{u} 
		 (\Lambda_{W^{(\alpha)}} ( i \pi)) \bigr) \; ,
	\end{equation}
and
	\begin{equation}
		\label{5.4b}
		\widehat{u} (\Lambda_{W^{(\alpha)}} ( i \pi)) h	= \widehat{u} (\Theta_{W^{(\alpha)}}) h \; , 
		\qquad h \in \widehat{K} \, \widehat{\mathfrak k}(I_\alpha) \; . 
	\end{equation}
\end{itemize}
\end{theorem}

\goodbreak
\begin{proof} \quad 
\begin{itemize}
\item [$ i.)$] 	
Clearly, $C^\infty (S^1) + i \omega  C^\infty (S^1)$ is dense in $\widehat{{\mathfrak h}}  (S^1)$. 
To verify that $\widehat{K}$ is a symplectic map, compute
\begin{align*}
	\qquad \quad 2 \Im \langle \widehat{K}(\mathbb{\phi}_1,\mathbb{\pi}_1) , 
	\widehat{K}(\mathbb{\phi}_2,\mathbb{\pi}_2) \rangle_{\widehat{{\mathfrak h}}  (S^1)}
	& = 2 \Im \langle -\mathbb{\pi}_1 +  i \,\omega r \, \mathbb{\phi}_1  , -\mathbb{\pi}_2 
	+  i \,\omega r \, \mathbb{\phi}_2  \rangle_{\widehat{{\mathfrak h}}  (S^1)}
	\\
	& =   \langle \mathbb{\phi}_1,\mathbb{\pi}_2 \rangle_{L^2(S^1, \, r {\rm d} \psi ) } - \langle \mathbb{\pi}_1   , 
		\mathbb{\phi}_2  \rangle_{L^2(S^1, \, r {\rm d} \psi)} 
	\\
	&= \widehat \sigma\big((\mathbb{\phi}_1,\mathbb{\pi}_1),(\mathbb{\phi}_2,\mathbb{\pi}_2)\big) \; . 
\end{align*}
\item[$ ii.)$]
For $\Lambda=
R_{0}$ a rotation, we have 
	\begin{align*}
		(\widehat{u}(R_{0}) \circ \widehat{K})(\mathbb{\phi},\mathbb{\pi})&
		= (R_{0}) _* \left(-\mathbb{\pi}+ i\omega r \, \mathbb{\phi} \right) 
		=- (R_{0})_*\mathbb{\pi} +  i\omega r \,  (R_{0})_*\mathbb{\phi}
		\nonumber \\
		&= \widehat{K} \bigl( (R_{0})_*\mathbb{\phi}, (R_{0})_*\mathbb{\pi} \bigr) 
		= \bigl( \widehat{K} \circ \widehat{{\mathfrak z}} (R_{0})\bigr) (\mathbb{\phi},\mathbb{\pi})  \; , 
	\end{align*}
since $\omega$ commutes with the pullback $(R_{0})_*$ of a rotation. 
For the boosts, see \eqref{canonical-boosts}; and for the reflections, 
see \eqref{reflections}.

\item [$ iii.)$] for $(\mathbb{\phi},\mathbb{\pi}) \in  \widehat{\mathfrak k}(I_\alpha)$, 
the identity \eqref{can-geo2} assures that \eqref{5.3b} holds. Moreover,  
	\begin{align*}
		\widehat u (\Lambda_W ( i \pi)) \widehat{K} (\mathbb{\phi},\mathbb{\pi}) &=
		\widehat K \circ \widehat {\mathfrak z} (\Lambda_W ( i \pi)) \,  (\mathbb{\phi},\mathbb{\pi})
		\nonumber \\
		&= \widehat K \circ  \widehat {\mathfrak z} (\Theta_{W}) \, (\mathbb{\phi},\mathbb{\pi})
		= \widehat u (\Theta_{W}) \, \widehat{K} (\mathbb{\phi},\mathbb{\pi})  \; , 
	\end{align*}
which demonstrates \eqref{5.4b}. The first equality follows from \eqref{can-geo}.
\end{itemize}
\end{proof}

\goodbreak

\begin{proposition} 
\label{Prop5.7}
There exists a unitary map ${\mathfrak U}$  
from $\widehat{{\mathfrak h}}  (S^1)$ to ${\mathfrak h} (dS)$, 
which  intertwines  the representations $\widehat{u} (\Lambda)$ and 
$ u (\Lambda)$, $ \Lambda \in O(1,2)$, and the one-particle structures.
In other words, the following diagram commutes:

\begin{picture}(250,140)

\put(60,80){$\nearrow$}
\put(0,70){${\mathcal D}_{\mathbb{R}} (dS)$}
\put(220,70){${\mathfrak U} $}
\put(130,70){${\mathbb T} $}
\put(50,95){$\widehat  {\mathbb P}$}
\put(50,45){$ {{\mathbb P}}$}
\put(60,55){$\searrow$}

\put(90,110){$ (\widehat {\mathfrak k} (S^1 ), \widehat {\mathfrak z})$}
\put(160,110){$\longrightarrow$}
\put(190,110){$(\widehat{{\mathfrak h}}  (S^1), \widehat{u})$}

\put(90,30){$({\mathfrak k} (dS), {\mathfrak z})$}
\put(160,30){$\longrightarrow$}
\put(190,30){$({\mathfrak h} (dS), u)\; .$}

\put(165,40){$K$}
\put(165,120){$ \widehat{K}$}

\put(120,85){\vector(0,-3){20}}
\put(210,85){\vector(0,-3){20}}

\end{picture}

\vskip -.8cm
\noindent
\end{proposition}

\goodbreak
\begin{proof} The existence of ${\mathfrak U}$ follows from the uniqueness of 
the de Sitter one-particle structure. The latter is a direct consequence of the uniqueness of 
the $(2 \pi r)$-KMS structure for the double wedge, see Theorem~\ref{ThB1}. 
\end{proof}

The following result shows that the functions introduced in 
Theorem \ref{st-kappa} are the most general
elements in ${\mathfrak h}^\kappa(dS)$ and its symplectic 
complement~${\mathfrak h}^\kappa(dS)^{\perp}$, respectively. 

\begin{theorem}
\label{tztf}
Let $I\subset S^1$ be an open interval (or $I= S^1$). 
If $h, g \in {\mathcal D}_\mathbb{R} (I)$, then
\begin{itemize}
\item[$ i.)$] $\delta \otimes h \in {\mathfrak h}^\kappa(dS) \cap 
{\mathfrak h}(\mathcal{O}_I)$ and $h \in \widehat{{\mathfrak h}}  ( I)$ is real valued; 
\item[$ ii.)$] $\delta' \otimes g \in {\mathfrak h}^\kappa(dS)^{\perp}
\cap {\mathfrak h}(\mathcal{O}_I)$ and $i \omega g \in \widehat{{\mathfrak h}}  ( I)$
has purely imaginary values. 
\end{itemize}
Further,
\begin{itemize}
\item[$ iii.)$] for every time-symmetric function $f \in {\mathcal D}_\mathbb{R} ({\mathcal O}_I)$ 
there exists a function $h \in {\mathcal D}_\mathbb{R} (I)$ such that $[f] = [\delta \otimes h] $; 
\item[$ iv.)$] for every anti-time-symmetric function $e \in {\mathcal D}_\mathbb{R} ({\mathcal O}_I)$ 
there exists a function $g \in {\mathcal D}_\mathbb{R} (I)$ such that $[e] = [n(\delta \otimes g)] $.  
\end{itemize}
\end{theorem}

\goodbreak

\begin{remark}
The statements $iii.)$ and $iv.)$ imply that there is a one-to-one relation between the image of
time-symmetric (time-antisymmetric) testfunctions in ${\mathfrak h}(dS)$ and real (purely imaginary) 
valued functions in~$\widehat{{\mathfrak h}} (S^1)$. The Minkowski space case of this result
is proven in \cite[Vol.~II p.~217]{RS}. It also follows directly by differentiation from Eq.~\eqref{flat2point}.
\end{remark}

\begin{proof} Let $h, g \in {\mathcal D}_\mathbb{R} (I)$. 
\begin{itemize}
\item [$i.)$] By assumption, the function $h$ lies in $\widehat{{\mathfrak h}}  (I)$
and is real valued; in particular, it has support in $I$. As we have seen, this 
implies that $\delta \otimes h \in {\mathfrak h}({\mathcal O}_I)$.  
Thus it remains to prove 
that $\delta \otimes h \in {\mathfrak h}^\kappa(dS)$  
(see~\eqref{kappat}).  
This can be shown by approximation the delta function 
with a sequence of functions which are all 
symmetric around the origin. 

\item [$ii.)$] 
By assumption, the function $g$ is real valued and 
hence $i \omega g$ takes purely imaginary values.
Inspecting the definition of $\widehat{{\mathfrak h}} (I)$, we 
conclude that $i \omega g \in 
\widehat{{\mathfrak h}} (I)$. Thus it 
remains to prove that $\delta' \otimes g \in {\mathfrak h}^\kappa(dS)^{\perp}$. 
This can be shown by approximation the derivative of the delta 
function with a sequence of functions which are all anti-symmetric around the origin. 

\item [$iii.)$]  For every time-symmetric function $f \in {\mathcal D}_\mathbb{R} ({\mathcal O}_I)$, 
the $C^\infty$-solution $\mathbb{f}$ of the Klein--Gordon equation is time-symmetric. This implies 
that  $(n \mathbb{f})_{\upharpoonright S^1} $ vanishes. On the other hand, 
we can define $h \doteq \mathbb{f}_{\upharpoonright S^1} $. It then follows from 
Theorem \ref{cauchyproblem} that $[f] = [\delta \otimes h]$.

\item [$iv.)$] For every anti-time-symmetric function $e \in {\mathcal D}_\mathbb{R} ({\mathcal O}_I)$,  
the corresponding $C^\infty$-solution $\mathbb{e}$ of the Klein-Gordon equation is 
anti-time-sym\-metric. This implies that $ \mathbb{e}_{\upharpoonright S^1} $ vanishes. On 
the other hand, we can define $g \doteq - (n \mathbb{e})_{\upharpoonright S^1} $. It then 
follows from Theorem \ref{cauchyproblem} that $[ e]  = [n(\delta \otimes g)] $. 
\end{itemize}
\end{proof}

\section{Localisation}
\label{sec:6.5}

We have seen for $ h\in {\mathcal D} (S^1)$ and
	\begin{align*} 
		 f (x ) \equiv (\delta \otimes h) (x ) &=	
		 \delta (x_0)  \;  h  ( \psi ) \;  , 
	 \nonumber \\
	g (x ) \equiv (\delta' \otimes h) (x ) &= 
	\left( \frac{\partial}{\partial x_0 } \delta \right) (x_0)  \;   h  (\psi ) \;  , 
	\end{align*} 
with $x  \equiv x  (t, \psi)$ the coordinates introduced in \eqref{w1psitau}, 
the  Cauchy data for the corresponding solutions $\mathbb{f}, \mathbb{g}$ of the 
Klein--Gordon equation are:
	\begin{align*}
		\big( \mathbb{f}_{ \upharpoonright S^1}, 
							 (n \, \mathbb{f} )_{\upharpoonright S^1})
   &= (0, - h) \equiv  (\mathbb{\phi}, \mathbb{\pi}) \; , 
\nonumber \\
	\big( \mathbb{g}_{ \upharpoonright S^1}, 
							 (n \, \mathbb{g} )_{\upharpoonright S^1})
 		  &= (  h, 0) \equiv  (\mathbb{\phi}, \mathbb{\pi}) \; . 
	\end{align*}
Together with $\widehat{K}\,(\mathbb{\phi},\mathbb{\pi}) 
= - \mathbb{\pi} +  i \,\omega r \, \mathbb{\phi}$ this gives 
	\begin{align*}
		\widehat{K}\, \big( \mathbb{f}_{ \upharpoonright S^1}, 
							 (n \, \mathbb{f} )_{\upharpoonright S^1})
   		&= h  \; , 
\nonumber \\
	\widehat{K}\, \big( \mathbb{g}_{\upharpoonright S^1}, 
							 (n \, \mathbb{g} )_{\upharpoonright S^1})
 		  &= i \omega r \, h  \; , 
	\end{align*}
both elements\footnote{As mentioned before,  $C^\infty (S^1) \subset {\mathscr D}(\omega)$.} 
of $\widehat{{\mathfrak h}}  (S^1)$. This suggest the following definition. 

\begin{definition}
\label{local-h-hat}
For $I$ a simply connected open interval in $S^1$, we define a  {\em real}
subspace of $\widehat{{\mathfrak h}} (S^1)$ by  
	\begin{equation*}
		\widehat{{\mathfrak h}}  ( I) 
		\doteq
		\overline{ \bigl\{ h \in \widehat{{\mathfrak h}}  (S^1)   \mid  
		{\rm supp\,}  \Re h \subset I \, ,    
		\, {\rm supp\,}  \omega^{-1}\Im h  \subset I \bigr\} }\; .
	\end{equation*}
\end{definition}

Clearly, $\widehat{{\mathfrak h}}  (J)$ is in the symplectic complement 
of $\widehat{{\mathfrak h}}  ( I) $ if $J \subset S^1 \setminus I$. This follows
directly from the definition: for $h\in \widehat{{\mathfrak h}}  ( I) $ and 
$g \in \widehat{{\mathfrak h}}  ( J)$, we have  
	\[
		 \Im \langle h, g \rangle_{\widehat{{\mathfrak h}}  (S^1)} 
		=  \langle  \Re h, \omega^{-1} \Im g \rangle_{L^2(S^1, \, r {\rm d} \psi)}
		-
		 \langle  \omega^{-1}\Im h
		 ,  \Re g \rangle_{L^2(S^1, \, r {\rm d} \psi)} = 0  \; . 
	\]

\begin{remark}
\label{rm:6.5.4}
Let $\chi_{I}$ denote the multiplication operator for the characteristic 
function of an open interval $I \subset S^1$. 
The operator 
	\begin{equation}
	\label{propfreeboost}
		\omega r \,  \mathbb{cos}_\psi  
		= \omega r \,  \mathbb{cos}_\psi \circ \chi_{I_+} 
		+ \omega r \,  \mathbb{cos}_\psi \circ \chi_{I_-} 
	\end{equation}
is the sum of a positive operator $\omega r \,  \mathbb{cos}_\psi \circ \chi_{I_+} $ 
vanishing on functions in $\widehat{{\mathfrak h}}  (S^1)$ with support in 
$\overline{I_-}$, and a negative operator $\omega r\,  
\mathbb{cos}_\psi \circ \chi_{I_-}$ vanishing 
on functions in $\widehat{{\mathfrak h}}  (S^1)$ with support in $\overline{I_+}$. 
Both operators have absolutely continuous spectrum. It is, however, 
important to stress that (see Definition~\ref{local-h-hat})
	\[
	 	\widehat{{\mathfrak h}}  (I_+)  \ne 
		\bigl\{ h \in \widehat{{\mathfrak h}}  (S^1)  \mid 
		\operatorname{supp} h \subset I_+ \bigr\} \; . 
	\]
As a consequence, there exists $h \in \widehat{{\mathfrak h}}  (I_+)$ 
such that 
	\[ 
		(\omega r \,  \mathbb{cos}_\psi \circ \chi_{I_+} ) h 
		= (\omega r \,  \mathbb{cos}_\psi ) (\chi_{I_+} \circ h)
		\ne (\omega r \,  \mathbb{cos}_\psi ) h   \; . 
	\]
\end{remark}

We are now in a position to specify the unitary operator $\mathfrak{U}$ 
introduced in Proposition \ref{Prop5.7}.

\begin{proposition}
\label{prop:4.10.5}
The unitary map $\mathfrak{U} \colon \widehat{{\mathfrak h}}  (S^1) \to {\mathfrak h}(dS)$ 
is the linear extension of the map
	\begin{equation}
	\label{s1-to-dS}
		h_1 + i \omega r h_2 \mapsto [\delta \otimes h_1] + 
		[\delta' \otimes h_2] \; ,
	\end{equation}
which respects the local structure, \emph{i.e.}, the 
restrictions $\mathfrak{U}_{\upharpoonright I}$ 
maps $\widehat{{\mathfrak h}}  (I)$ to ${\mathfrak h}({\mathcal O}_I)$, 
with ${\mathcal O}_I= I''$ the causal completion of $I \subset S^1$. 
\end{proposition}

\begin{proof} We have seen  that the image of $[\delta \otimes h_1] + 
		[\delta' \otimes h_2]$ is dense in~${\mathfrak h}(dS)$. Moreover,
	\[
		\| h_1 + i \omega r h_2 \|_{\widehat{{\mathfrak h}}  (S^1)} 
		= \|  [\delta \otimes h_1] + 
		[ \delta' \otimes h_2 ] \|_{{\mathfrak h}(dS)} \; . 
	\]
The result now follows by linear extension. The local part follows from the discussion proceeding 
Definition \ref{local-h-hat}. 
\end{proof}

\begin{corollary} 
\label{cor:6.5.4}
Let $I \subset S^1$ be an open interval. Then
$\widehat{{\mathfrak h}} (I) +i \widehat{{\mathfrak h}} (I)$
is dense in~$\widehat{{\mathfrak h}} (S^1)$.

\end{corollary}

\begin{proof} 
Assume that $I$ is sufficiently small 
such that $\mathcal{O}_I$ is a bounded space-time region. 
(The general case follows from isotony once this special case has been established.) 
The result now follows directly from Proposition \ref{oprs}:
	\[
		\overline{\widehat{{\mathfrak h}} (I) +i \widehat{{\mathfrak h}} (I)} 
		= \mathfrak{U}^{-1} \overline{ {\mathfrak h} ({\mathcal O}_I) +i {\mathfrak h} ({\mathcal O}_I)}
		= \mathfrak{U}^{-1} {\mathfrak h} (dS) = \widehat{\mathfrak h} (S^1)\; . 
	\]
(A direct proof might be based on arguments similar to those given in \cite{V}. However, 
we have not fully investigated this question.)
\end{proof}

\begin{corollary}
\label{WdSprop}
For any double wedge ${\mathbb W}$, we have ${\mathfrak h} ({\mathbb W})={\mathfrak h} (dS)$.
\end{corollary}

\begin{proof} 
Inspecting the definition (given in \eqref{ophs2}) 
and \eqref{feb-1}, we conclude that~${\mathfrak h} (dS)$ 
arises by taking the closed span 
of the equivalence classes $\{ [f], [g] \}$,  
	\begin{align*}
	 f (x) \equiv (\delta \otimes h) (x) &=	\delta (x_0)   h  (\psi ) \;  , 
	  \\
	g (x) \equiv (\delta' \otimes h) (x) &= \delta'  (x_0)  h  ( \psi )\;  , 
	\end{align*} 
with $ h\in{\mathcal D}_{\mathbb{R}}(S^1)$, 
while ${\mathfrak h} ({\mathbb W}_1)$ is spanned by 
similar equivalence classes with $ h\in {\mathcal D}_{\mathbb{C}} 
\left(S^1 \setminus \{ - \frac{\pi}{2} , 
\frac{\pi}{2}\} \right)$. However, ${\mathfrak h} ({\mathbb W}_1)^\perp = \{0 \}$, 
and consequently, 
${\mathfrak h} ({\mathbb W}_1)={\mathfrak h} (dS)$ as both spaces are closed. 
The general result follows 
from ${\mathbb W}= \Lambda {\mathbb W}_1$ for some $ \Lambda \in SO_0(1,2)$.
\end{proof}

The following result can be interpreted as demonstrating finite speed 
of propagation for the free field in the canonical formulation.

\begin{proposition} 
\label{halpha}
Let $I$ be an open subset in $S^1$. The unitary operator 
${\rm e}^{i t \omega r \,  \mathbb{cos}_\psi }$ maps $\widehat{{\mathfrak h}}  (I)$ to
	\[
		\widehat{{\mathfrak h}}  \Bigl( \bigl( \Gamma^+(\Lambda_1(t)I) 
		\cup \Gamma^-(\Lambda_1(t)I)  \bigr) \cap S^1 \Bigr) \; . 
	\]
In particular, the unitary group $t \mapsto {\rm e}^{it (\omega r \,  
\mathbb{cos}_\psi )_{\upharpoonright I_\pm}}$ leaves $\widehat{{\mathfrak h}}  (I_\pm)$ invariant.
\end{proposition}

\begin{proof} This is a direct consequence of the fact that ${\rm e}^{it \omega \,  
\mathbb{cos}_\psi }$ implements the Lorentz boost $\Lambda_1(t)$. As can be seen from 
Proposition \ref{UIR-FH} and Proposition \ref{prop:4.10.5}, the latter act geometrically 
on ${\mathfrak h} (dS)$, \emph{i.e.}, a testfunction supported at $I \subset S^1 \subset dS$ 
is mapped to a testfunction supported at $\Lambda_1(t) I \subset dS$. 
This result extends by continuity to~$\widehat{{\mathfrak h}}  (I)$. 
\end{proof}

Next, let us associate $\mathbb{R}$-linear subspaces to arbitrary 
wedges $W = \Lambda W_1$:
	\[
		\widehat{\mathfrak{h}}_W \doteq \widehat u(\Lambda) \widehat{\mathfrak{h}} (I_+) \; ,
		\qquad		
		\Lambda \in SO_0(1,2) \; . 
	\]
Since $\widehat u \bigl(\Lambda_1(t) \bigr) \widehat{\mathfrak{h}}(I_+) = 
\widehat{\mathfrak{h}}(I_+)$ for all $t \in \mathbb{R}$, 
$\widehat{\mathfrak{h}}_W$ is well-defined.  
The following result shows that modular localization extends the localization map given 
by the Cauchy data.

\begin{corollary}
\label{Prop-ii.4}
For $I$ a bounded open interval of length $ | I |  \le \pi \, r$ in $S^1$ there holds 
	\begin{equation}
	\label{cauchy-localization}
		\qquad
		\bigcap_{ \mathcal{O}_I \subset W} \widehat{\mathfrak{h}}_W
		= \widehat{\mathfrak{h}}(I) \;  , 
	\end{equation}
where $\mathcal{O}_I = I''$ denotes the \emph{causal completion} of the interval $I$ in $dS$. 
\end{corollary}

\begin{proof} This result is Proposition 2.5 in \cite{MJ-2}. For the convenience of the reader, 
we provide a sketch of the proof. 
Let $I$ be an interval as stated in the corollary. As $\mathcal{O}_I$
is causally complete and bounded,  
	\[
		\bigcap_{ \mathcal{O}_I \subset W} W = \mathcal{O}_I 
		= R_0(\alpha) W_1 \cap R_0(\beta) W_1  \; , \qquad \alpha, \beta \in [0, 2\pi) \; .
	\]
Inspecting the definitions, 
we find that   
	   \[
		\widehat{\mathfrak{h}} (I)
		= \widehat{\mathfrak{h}} \bigl(R_0(\alpha) I_+\bigr) \cap
			\widehat{\mathfrak{h}} \bigl( R_0(\beta) I_+ \bigr)
		= \widehat{\mathfrak{h}} ( I_\alpha ) \cap
			\widehat{\mathfrak{h}} (  I_\beta )  
			\supseteq \bigcap_{ \mathcal{O}_I \subset W} \widehat{\mathfrak{h}}_W \; .  
	\]
The inclusion on the r.h.s.~is due to the fact that 
$W (\alpha)$ and $W (\beta)$ are wedges containing $\mathcal{O}_I$. 

Next, we assume that $W$ is an arbitrary wedge which contains~$\mathcal{O}_I$. 
The $\mathbb{R}$-linear subspace associated to the \emph{opposite} wedge $W'$ is 
	\[
		\widehat{\mathfrak{h}}_{W'} = u\bigl( \Lambda^{(\beta 
            )}(t)\bigr)
	  \widehat{\mathfrak{h}}  ( I_\alpha ) \;  
	\]
for suitable $\alpha, \beta$ and~$t$.
Hence,  by Proposition~\ref{halpha} (finite speed of propagation\footnote{This 
statement has been shown in Proposition~\ref{halpha} for $\Lambda_1(t)$, 
but also holds for $\Lambda^{(\beta)}(t)$ as
$u\big(R_0(\alpha)\big) \;  \widehat{\mathfrak{h}}(I) =  \left\{ h \in \mathcal{H} \mid
\operatorname{supp} \Re h \subset R_0(\alpha)I \, ,  \; \operatorname{supp}
\omega^{-1}\Im h \subset R_0(\alpha) I \right\}$ for 
all $\alpha \in [0, 2\pi)$.}), 
	\[ 
		\widehat{\mathfrak{h}}_{W'} \subset \widehat{\mathfrak{h}}( J ) 
		\quad \text{with} \quad J  = \Gamma (W') \cap S^1 \; ,
	\]
where $\Gamma(M)= \Gamma^+ (M) \cup \Gamma^- (M)$ is the
domain of dependence of a set~$M$. Note that $W'$ is space-like to $I$, since $W$ 
contains~$\mathcal{O}_I$. 
Hence, $J$ is in the interior $I^c \doteq S^1 \setminus \overline{I} $
of the complement of $I$ within~$S^1$. Thus, $\widehat{\mathfrak{h}}_{W'}
\subset \widehat{\mathfrak{h}} (I^c) $. Wedge duality now implies  	
$\widehat{\mathfrak{h}}_{W} 
= \bigl(\widehat{\mathfrak{h}}_{W'}\bigr)' \supseteq  \widehat{\mathfrak{h}}(I) $.  
This verifies~\eqref{cauchy-localization}. 
\end{proof}

\section{Standard subspaces of $\widehat{\mathfrak{h}}(S^1) $}

Let $C_{\mathbb{R}}^\infty(S^1)$ denote the \emph{real-valued} smooth functions 
on the circle. We equip $C_{\mathbb{R}}^\infty(S^1)$ with two new inner products  
	\begin{align}
		\label{4.5}
		\langle f, g \rangle_{-\frac{1}{2}} 
		& = \sum_{k \in \mathbb{Z}} \; \frac{ \overline{\widetilde{f}(k)} g(k) }
		{\widetilde{\omega}(k)} \;  , 
		\\
		\label{4.6}
		\langle f, g \rangle_{\frac{1}{2}} 
		& = \sum_{k \in \mathbb{Z}} \; 
		\widetilde{\omega}(k)
		\overline{\widetilde{f}(k)} g(k)   \;  .
	\end{align}
The asymptotic behaviour of $\widetilde{\omega}(k)$ as
$| k | \to \infty$ (see Remark~\ref{rm:4.6.4}) ensures that the 
completion of $C_{\mathbb{R}}^\infty(S^1)$ 
with respect to~\eqref{4.5} and \eqref{4.6} coincide as sets with the real Sobolev spaces 
$\mathbb{H}^{-\frac{1}{2}} (S^1)$ and $\mathbb{H}^{\frac{1}{2}} (S^1)$, respectively.  
However, the scalar products $\langle \, . \, , \, . \, \rangle_{\pm \frac{1}{2}}$ differ slightly 
from those
of $\mathbb{H}^{\pm \frac{1}{2}} (S^1)$; see Appendix B. This can be rectified by 
the unitary mapping 
	\begin{align*}
		\left( \mathbb{H}^{\pm \frac{1}{2}} (S^1), 
		\langle \, . \, , \, . \, \rangle_{\pm \frac{1}{2}} \right)
		& \to \left( \mathbb{H}^{\pm \frac{1}{2}} (S^1), 
		\langle \, . \, , \, . \, \rangle_{\mathbb{H}^{\pm
                    \frac{1}{2}} (S^1)} \right)  
		\\
		\sum_{k \in \mathbb{Z}} 
		\widetilde{h}(k) \, {\rm e}^{i k \psi}  & \mapsto 
		\sum_{k \in \mathbb{Z}} \frac{\sqrt{1+k^2}}{\sqrt{\widetilde{\omega}(k)}}  \; 
		\widetilde{h}(k) \, {\rm e}^{i k \psi} \; .  
	\end{align*}
According to Remark~\ref{rm:4.6.4}, the factor $\frac{\sqrt{1+k^2}}
{\sqrt{\widetilde{\omega}(k)}} >0$, $k \in \mathbb{Z}$, 
converges to a constant as $| k | \to \infty$. Nevertheless, the orthogonality relation 
\emph{is} effected by the choice of the scalar product. 
In the sequel, the symbols $\mathbb{H}^{\pm \frac{1}{2}} (S^1)$ are used to 
denote $\bigl( \mathbb{H}^{\pm \frac{1}{2}} (S^1), 
		\langle \, . \, , \, . \, \rangle_{\pm \frac{1}{2}} \bigr)$.

As 
	\[
		\langle f,f \rangle_{ -\frac{1}{2} }
		\le
		\frac{1}{\widetilde{\omega}(0)} \; \langle f,f \rangle_{ L^2 (S^1, {\rm d}\psi) }
		\le 
		\frac{1}{\widetilde{\omega}(0)^2} 
		\langle f,f \rangle_{ \frac{1}{2} }  \; , 
	\]
the \emph{identity mapping} $f \mapsto f $, $f \in  C^\infty(S^1)$, extends to 
\emph{bounded injections} 
	\[
		\mathbb{H}^{\frac{1}{2}} (S^1)
		  \xhookrightarrow{ \imath_1 }  L^2 (S^1, {\rm d}\psi)
		\xhookrightarrow{ \imath_2 } \mathbb{H}^{-\frac{1}{2}} (S^1) \; . 
	\]
The mapping
	\[	
		f \mapsto \omega^{1/2} f \; , 
		\qquad f \in C_\mathbb{R}^\infty(S^1) \; , 
	\]
induces \emph{unitary mappings} of $\mathbb{H}^{\frac{1}{2}} (S^1)$ 
onto $L^2_\mathbb{R} (S^1, {\rm d}\psi)$ and $L^2_\mathbb{R} (S^1, {\rm d}\psi)$ 
onto~$\mathbb{H}^{-\frac{1}{2}} (S^1)$, while the map 
	\[
		f \mapsto \omega f \; , 
		\qquad f \in C_\mathbb{R}^\infty(S^1) \; , 
	\] 
induces a \emph{unitary mapping} of  $\mathbb{H}^{\frac{1}{2}} (S^1)$ 
onto~$\mathbb{H}^{-\frac{1}{2}} (S^1)$. The extended mappings are denoted by 
$\omega^{1/2} \imath_1$, $\omega^{1/2}  \imath_2$, 
and $\omega  \imath_2  \imath_1$. One also finds
	\begin{equation}
	\label{add-111}
		\| f \|_{L^2} \le \frac{1}{\widetilde{\omega} (0)} \, \| \omega^{1/2} f \|_{L^2} 
		= \frac{1}{\widetilde{\omega} (0)} \, \|  f \|_{ \frac{1}{2}  } \; . 
	\end{equation}

\begin{remark} As the function $\mathbb{R} \ni k \mapsto 
\sqrt{\widetilde{\omega}(k)}$ is \emph{concave}\footnote{Actually, 
this property only needs to hold for $|k|$ large, and there it follows from 
Remark~\ref{rm:4.6.4}.}, 
and $\widetilde{\omega}(0) > 0$, 
	\begin{align*}
		\sqrt{\widetilde{\omega} ( \lambda k)} 
		& \geq \lambda \sqrt{\widetilde{\omega} (k)}
		+(1-\lambda) \sqrt{\widetilde{\omega}(0)}
		\geq \lambda \sqrt{\widetilde{\omega}(k)}  
		\qquad \forall  0 \le \lambda \le 1 \; . 
	\end{align*}
It follows that the function $\mathbb{R} \ni k \mapsto 
\sqrt{\widetilde{\omega}(k)}$ is \emph{subadditive} on $[0,\infty )$: 
	\begin{align*}
		 \sqrt{\widetilde{\omega}( k)}+\sqrt{\widetilde{\omega}( k')} & 
		 = \sqrt{\widetilde{\omega} \left(\lambda (k+k')\right) }
		 + \sqrt{\widetilde{\omega} \left(\lambda' (k+k')\right) }
		 \\
		 & \geq  
		 \sqrt{\widetilde{\omega}(k+k')} \; , 
		 \qquad k,k'\in [0,\infty ) \; , 
	\end{align*}
with $ \lambda = \frac {k}{k+k'} $ and  $ \lambda' = \frac {k'}{k+k'}$; thus 
$\lambda + \lambda'= 1$. Consequently, 
	\begin{align}
	\label{Araki-Ungleichung}
		\sqrt{\widetilde{\omega}( k_1)}  
		& \le \sqrt{\widetilde{\omega} ( k_1 - k_2)} + \sqrt{\widetilde{\omega}(k_2)}\; ,
		\qquad k_1, k_2 , ( k_1 - k_2) \in \mathbb{N}_0 \; . 
	\end{align} 
\end{remark}

\begin{lemma}[Araki \cite{A1a}]
\label{lm:6-6-6} 
The multiplication with a $C_\mathbb{R}^\infty(S_1)$-
function $\chi(\psi)$ defines a bounded operator 
in~$\mathbb{H}^{\frac{1}{2}} (S^1)$.
\end{lemma}

\begin{proof} Using \eqref{Araki-Ungleichung}, 
we find
	\begin{align*}
		& \bigl\| \chi f \bigr\|_{\mathbb{H}^{\frac{1}{2}} ( S^1)} 
		= \bigl\| \omega^{1/2} ( \chi f) \bigr\|_{L^2(S^1, {\rm d} \psi)} 
		\\
	 	& = \left\| \sqrt{\widetilde{\omega( \, . \,  )} } \sum_{k' \in \mathbb{Z}} 
		\;  \widetilde{\chi}(\, . \,  - k') \widetilde{f}(k') 
		\right\|_{\ell^2 (\mathbb{Z})}
		\le 
		 \left\|  \sum_{k' \in \mathbb{Z}} 
		\sqrt{\widetilde{\omega( \, . \,  )} } \; \bigl| \widetilde{\chi}(\, . \,  - k') 
		 \widetilde{f}(k') \bigr|
		\right\|_{\ell^2 (\mathbb{Z})}
		\\
		& \le 
		\left\| \;    \sum_{k' \in \mathbb{Z}}
		\sqrt{\widetilde{\omega} ( \, . \,  - k')} 
		\; \bigl|  \widetilde{\chi}(\, . \,  - k')\widetilde{f}(k') \bigr|  
		+  \sum_{k' \in \mathbb{Z}} \sqrt{\widetilde{\omega}(k')}
		\;  \bigl|  \widetilde{\chi}(\, . \,  - k')\widetilde{f}(k') \bigr|   
		\; \right\|_{\ell^2 (\mathbb{Z})} 
		\\
	\end{align*}
	\begin{align*}
		&  \le 
		\left\|    \sum_{k' \in \mathbb{Z}} \;  \bigl| \widetilde{\omega^{1/2}\chi}(\, . \,  - k') 
		\bigr| \; 
		\bigl| \widetilde{f}(k') \bigr| 
		\right\|_{\ell^2 (\mathbb{Z})}
		+ \left\| \sum_{k' \in \mathbb{Z}} \;  \bigl| \widetilde{\chi}(\, . \,  - k') \bigr|  \; 
		\bigl| \widetilde{\omega^{1/2}f}(k') \bigr|
		\right\|_{\ell^2 (\mathbb{Z})} 
		\\
		&  = 
		\left\|  \left( \sum_{m \in \mathbb{Z}} \;  {\rm e}^{im \, . \, }
		\;  | \widetilde{f} \, (m) | \right)
		\left( \sum_{k \in \mathbb{Z}} \;  {\rm e}^{ik \, . \, }
		\;  \widetilde{\omega^{1/2}|\chi|} \, (k)  \right)
		\right\|_{L^2(S^1, {\rm d} \psi)}
				\\
		& 
		\qquad  \qquad  \qquad  \qquad  \qquad  \qquad  \qquad   
		+ \left\| 
		\left( \sum_{m \in \mathbb{Z}} \;  
		{\rm e}^{im \, . \,} \bigl| \widetilde{\omega^{1/2}f}(m) \bigr| \right)
		\left( \sum_{k \in \mathbb{Z}} \;  
		{\rm e}^{ik \, . \,} \bigl| \widetilde{\chi}(k) \bigr| \right)
		\right\|_{L^2(S^1, {\rm d} \psi)} 
		\\
		&  \le 
		\left\|
		\sum_{k \in \mathbb{Z}} \;  {\rm e}^{ik \, . \, }
		\;  \sqrt{\widetilde{\omega}(k)}  \, 
		 \bigl| \widetilde{\chi}(k) \bigr|
		\right\|_{L^\infty(S^1, {\rm d} \psi)} 
		\left\|   f  \right\|_{L^2(S^1, {\rm d} \psi)}
		\\
		& 
		\qquad  \qquad  \qquad  \qquad  \qquad  \qquad  \qquad   
		+ \left\| \sum_{k \in \mathbb{Z}}\;  {\rm e}^{ik \, . \, }
		\; \bigl| \widetilde{\chi}(k) \bigr| \right\|_{L^\infty(S^1, {\rm d} \psi)}
		\bigl\|  \omega^{1/2}f  \bigr\|_{L^2(S^1, {\rm d} \psi)} 
		\\
		& \le
		A \| f \|_{L_2(S^1, {\rm d} \psi)} + B \| \omega^{1/2} f \|_{L_2(S^1, {\rm d} \psi)} \; , 
	\end{align*}
where 
	\[ 
		A \doteq \sup_{\psi \in S^1}
		 \left| \sum_{k \in \mathbb{Z}} \;  {\rm e}^{ik \psi }
		\;  \sqrt{\widetilde{\omega}(k)}  \bigl| \widetilde{\chi}(k) \bigr| \right| 
		= \sum_{k \in \mathbb{Z}} \;  
		\;  \sqrt{\widetilde{\omega}(k)} \,  \bigl| \widetilde{\chi}(k) \bigr| 
	\]
and 
	\[ 
		B \doteq \sup_{\psi \in S^1}
		 \left| \sum_{k \in \mathbb{Z}} \;  {\rm e}^{ik \psi }
		 \bigl| \widetilde{\chi}(k) \bigr| \right| 
		= \sum_{k \in \mathbb{Z}} \;    \bigl| \widetilde{\chi}(k) \bigr| \; .
	\]  
We note that $A$ and $B$ are finite: since $\chi \in C^\infty(S^1)$, the 
Fourier coefficients 
$\widetilde{\chi}(k)$ satisfy \cite[Prop.~6.3.1]{BSimon}
	\[
	  	(1 + |k|)^n \widetilde{\chi}(k) \to 0 \qquad \text{as}
		\quad k \to \infty 
		\qquad \forall n \in \mathbb{N} \; . 
	\]
Hence 
	\[
		\bigl\|  \widetilde{\omega^{1/2} \chi} \bigr\|_{\ell^1} 
		= \sum_k  \sqrt{\widetilde{\omega}(k)} \; 
		\bigl| \widetilde{\chi}(k) \bigr|
		< \infty \; , 
		\quad \text{and} \quad 
				\bigl\|  \widetilde{\chi} \bigr\|_{\ell^1} 
		= \sum_k  
		\bigl| \widetilde{\chi}(k) \bigr|
		< \infty \; . 
	\]
Recalling~\eqref{add-111}, we conclude that the multiplication by 
$\chi(\psi)$ defines a bounded operator 
in~$\mathbb{H}^{\frac{1}{2}} (S^1)$.
\end{proof}

\subsection{Subspaces of $\mathbb{H}^{\pm \frac{1}{2}}( S^1)$ associated to
measurable subsets of $S^1$}
Let $X$ be a \emph{measurable subset} of $S^1$ and let 
$L^2(X, {\rm d}\psi)$ be the subspace of $L^2(S^1, {\rm d}\psi)$ 
consisting of functions which vanish outside of $X$. We define
	\begin{align}
		\mathbb{H}^{\frac{1}{2}}( X) &
		\doteq \imath_1^{-1} \bigl( L^2(X, {\rm d}\psi) 
		\cap \imath_1 \mathbb{H}^{\frac{1}{2}}(S^1) \bigr)  \; , 
		\\
		\mathbb{H}^{-\frac{1}{2}}( X )
		& \doteq \overline{ \imath_2 \bigl( L^2(X, 
		{\rm d}\psi)\bigr) }^{\, \mathbb{H}^{-\frac{1}{2}}( S^1 )}  \; . 
		\label{5.2}
	\end{align}
\emph{Isotony} follows directly from the definition:  
	\[
		\mathbb{H}^{-\frac{1}{2}} (X_1) \subset \mathbb{H}^{-\frac{1}{2}} (X_2) \; , 
		\quad 
		\mathbb{H}^{\frac{1}{2}}(X_1) \subset \mathbb{H}^{\frac{1}{2}}(X_2) \; , 
		\qquad  X_1 \subset X_2 \; . 
	\]

\goodbreak
\begin{lemma}[Araki \cite{A1a}]
\label{lm:1.1}
Let $X \subset S^1$ be a measurable subsets of $S^1$. Then 
\begin{itemize}
\item[$i.)$] the orthogonal complement of $\mathbb{H}^{\frac{1}{2}}( X)$
in $\mathbb{H}^{\frac{1}{2}}(S^1)$ is
	\begin{equation}
		\mathbb{H}^{\frac{1}{2}}( X)^\perp 
		= \imath_1^{-1} \imath_2^{-1}\omega^{-1}\mathbb{H}^{-\frac{1}{2}} ( X^c)
		\; , 
		\label{oc-pi}
	\end{equation}
\item[$ii.)$]  the orthogonal complement of $\mathbb{H}^{-\frac{1}{2}}( X)$ 
in $\mathbb{H}^{-\frac{1}{2}}(S^1)$ is
	\begin{equation}
		\mathbb{H}^{-\frac{1}{2}} ( X)^\perp 
		= - \omega \imath_2 \imath_1 \mathbb{H}^{\frac{1}{2}}( X^c) \; . 
		\label{oc-varphi}
	\end{equation}
(The sign is a convention and 
does not effect the statement as this is a statement about real subspaces.)\end{itemize}
\end{lemma}

\begin{proof} Since $L ^2 (X, {\rm d} \psi)^\perp = L ^2 (X^c, {\rm d} \psi)$
		in $L^2 (S^1 , {\rm d} \psi) $,
	\[
		\mathbb{H}^{\frac{1}{2}} (X^c) 
		= \bigl\{ g \in \mathbb{H}^{\frac{1}{2}} (S^1) \mid 
		\langle f, \imath_1 g \rangle_{L^2} = 0 \; \forall f 
		\in L^2 (X, {\rm d}\psi) \bigr\} \; . 
	\]
Since, for $f  \in L^2 (X, {\rm d}\psi)$ and
$g \in \mathbb{H}^{\frac{1}{2}} (S^1)$, 
	\[
		\langle f , \imath_1 g \rangle_{L^2 (S^1)} = 0
		\quad \Leftrightarrow \quad
		\langle \imath_2 f, \omega \imath_2 
		\imath_1 g \rangle_{\mathbb{H}^{-\frac{1}{2}} (S^1)}  = 0 \; , 
	\]
we have 
	\[ 
		(\imath_2 L^2 (X))^\perp = - \omega \imath_2 
		\imath_1  \mathbb{H}^{\frac{1}{2}} (X^c) 
	\] 
in $\mathbb{H}^{-\frac{1}{2}} (S^1)$; 
the minus sign is irrelevant. 
Thus \eqref{oc-varphi} holds. 
Taking the orthogonal complement in $\mathbb{H}^{-\frac{1}{2}} (S^1)$, we also have
	\[
		\mathbb{H}^{-\frac{1}{2}} (X^c) 
		= - \omega \imath_2 \imath_1 \mathbb{H}^{\frac{1}{2}}(X)^\perp  \; , 
	\]
which implies \eqref{oc-pi}. 
\end{proof}

\goodbreak 

\begin{lemma}[Araki \cite{A1a}]
\label{Lm-2}
Let $X_\alpha \subset S^1$ be a family of measurable subsets of $S^1$. Then 
\quad 
\begin{itemize}
\item[$i.)$]  \emph{additivity} holds for the $\mathbb{H}^{-\frac{1}{2}}(X_\alpha)$'s: 
	\begin{equation}
	\label{5.4a}
		\bigvee_\alpha \mathbb{H}^{-\frac{1}{2}} (X_\alpha) 
		= \mathbb{H}^{-\frac{1}{2}} \Bigl( \bigcup_\alpha X_\alpha \Bigr) \; . 
	\end{equation}
The symbol $\bigvee_\alpha$ denotes the 
closure of the $\mathbb{R}$-linear span;
\item[$ii.)$]  the \emph{intersection} of the $\mathbb{H}^{\frac{1}{2}}( X_\alpha)$'s equals  
	\[ 
		\bigcap \mathbb{H}^{\frac{1}{2}} (X_\alpha) 
		=  \mathbb{H}^{\frac{1}{2}} \left( \bigcap X_\alpha \right)  \; . 
	\]
\end{itemize}
\end{lemma}

\begin{proof}
Property $i.)$ can be seen as follows: if the set $X_\alpha$'s are disjoint, then 
any $f \in L^2 (S^1, {\rm d}\psi)$ can be split uniquely as 
	\[
		f = \sum_\alpha f_\alpha \; , 
		\qquad 
		f_\alpha (\psi) = \chi_\alpha(\psi) f(\psi) \; , 
	\]
where $\chi_\alpha$ is the \emph{characteristic function} of $X_\alpha$ and
the sum over $\alpha$ converges in the strong topology of $L^2 (X,
{\rm d}\psi)$. Hence, 
	\[
		L^2 (X, {\rm d}\psi) = \bigvee_\alpha L^2(X_\alpha , {\rm d}\psi) \; , 
		\qquad 
		X = \bigcup_\alpha X_\alpha \; . 
	\]
Since $\imath_2$ is bounded,
	$ \imath_2 \bigl( L^2 (X, {\rm d}\psi) \bigr) 
		= \bigvee_\alpha \imath_2 \bigl( L^2 (X_\alpha , {\rm d}\psi) \bigr) $ . 
Hence, \eqref{5.4a} follows by extending this result to the closure in 
$\mathbb{H}^{-\frac{1}{2}} (S^1)$. 	
Property $ii.)$ follows from $i.)$ and Lemma~\ref{lm:1.1}.
\end{proof}		

\subsection{Regularity from the inside and the outside}

We now consider an \emph{open} subset $I$ of $S^1$. We define
approximations from the outside 
	\[
		\mathbb{H}^{\pm \frac{1}{2}}_\downarrow (I) 
		\doteq  \bigcap_{\overline{I} 
		\subset J } 
		\mathbb{H}^{\pm \frac{1}{2}} ( J ) \; , 
	\]
where the intersection is over open sets $J \supset \overline{I}$ of $S^1$, 
and approximations from the inside 
	\[
		\mathbb{H}^{\pm \frac{1}{2}}_\uparrow  (I) 
		\doteq  \bigvee_{K \subset I} \mathbb{H}^{\pm \frac{1}{2}} (K) \; , 
	\]
where the union is over closed subsets $K \subset I $ of $S^1$. In all cases,
\emph{isotony} is a direct consequence of the definitions: if $I_1 \subset I_2$, then 
	\begin{align}
		\mathbb{H}^{\pm \frac{1}{2}}_\downarrow (I_1) 
		& \subset \mathbb{H}^{\pm \frac{1}{2}}_\downarrow (I_2)  \; , 
		\qquad
		\mathbb{H}^{\pm \frac{1}{2}}_\uparrow (I_1) 
		\subset \mathbb{H}^{\pm \frac{1}{2}}_\uparrow (I_2)  \; . 
		\label{5.10}
	\end{align}
By construction, \emph{the approximation from the outside contains the approximation 
from the inside}:
	\begin{align} 
	\label{inside-outside}
			\mathbb{H}^{\pm \frac{1}{2}}_\uparrow (I) 
			\subset 
			\mathbb{H}^{\pm \frac{1}{2}}_\downarrow (I)
			\; . 
	\end{align}
If $I_1, I_3$ are open intervals such that $I_1\subset
\overline{I_3}$, then there exists  
another open interval~$I_2$ with 
	\[ 
		\overline{I_1} \subset I_2  \quad \text{and} \quad 
		\overline{I_2} \subset I_3 \; , 
	\] 
due to the separation theorem, and, consequently, 
	\[
		\mathbb{H}^{\pm \frac{1}{2}}_\downarrow (I_1)
		\subset 
		\mathbb{H}^{\pm \frac{1}{2}}_\uparrow (I_3)  \; .  
	\] 
Thus, if an interval is properly contained in a slightly larger one, then 
\emph{the approximation from the inside for larger interval contains the approximation 
from the outside of the smaller one.}

\begin{lemma}[Araki \cite{A1a}]
\label{lm:1.3}
Let $I$ be an open subset of $S^1$. Then
	\begin{equation}
	\label{5.12a}
		\mathbb{H}^{\frac{1}{2}}_\downarrow (I) = \mathbb{H}^{\frac{1}{2}}(I) \; .  
	\end{equation} 
\end{lemma}

\begin{proof}
Since $\bigcap_{J \supset \overline{I}} L^2(J, {\rm d}\psi ) 
= L^2(\overline{I}, {\rm d}\psi) = L^2(I, {\rm d}\psi) $, we find
	\[ 
		\imath_1  \bigl( \mathbb{H}^{\frac{1}{2}}_\downarrow (I) \bigr) 
		= \imath_1 \left( \mathbb{H}^{\frac{1}{2}} (I) \right) \; . 
	\]
Since $\imath_1$ is an injection, the identity \eqref{5.12a} holds. 
\end{proof}

\begin{lemma}[Araki \cite{A1a}]
\label{lm:1.4}
Let $I$ be an open subset of $S^1$. 
Then we have the following result concerning orthogonal complements:
	\begin{align}
		 \mathbb{H}^{\frac{1}{2}}_\uparrow (I)^\perp 
		& = \imath_1^{-1} \imath_2^{-1}\omega^{-1}
		\mathbb{H}^{-\frac{1}{2}}_\downarrow (\overline{I}^c) \; , 
		\nonumber \\
		\mathbb{H}^{-\frac{1}{2}}_\uparrow (I)^\perp 
		& = - \omega \imath_2 \imath_1 \mathbb{H}^{\frac{1}{2}}_\downarrow (\overline{I}^c)		
		\;  .  
		\label{5.14}
	\end{align} 
\end{lemma}

\begin{proof} Let $I$ be an on open interval and $K$ be a closed interval. Then
	\[
		{\overline{I}}^{\, c} 
		= \operatorname{int} \left( I^c \right) \; , 
		\quad 
		\left(\operatorname{int} K\right)^c =  \overline{K^c}\; . 
	\]
Applying 	\eqref{oc-pi} and \eqref {oc-varphi}, we obtain~\eqref{5.14}. 
\end{proof}

\begin{lemma}[Araki \cite{A1a}]
\label{lm-A-6.6.7}
Let $I$ be an open subset of $S^1$. Then
	\begin{equation}
	\label{5.12b}
		\mathbb{H}^{-\frac{1}{2}}_\uparrow (I) = \mathbb{H}^{-\frac{1}{2}} (I)  \; .  
	\end{equation} 
\end{lemma}

\begin{proof}
By using Lemma~\ref{lm:1.3} and Lemma~\ref{lm:1.4}, we have~\eqref{5.12b}. 
\end{proof}

\begin{lemma}[Araki \cite{A1a}]
Let $I_\alpha$ be a family of open subsets of $S^1$ and 
let $I= \bigcup_\alpha I_\alpha$, then 
	\begin{equation}
		 \bigvee_\alpha \mathbb{H}^{\frac{1}{2}}_\uparrow (I_\alpha) 
		= \mathbb{H}^{\frac{1}{2}}_\uparrow (I) \;  .  
		\label{5.14a}
	\end{equation} 
\end{lemma}

\begin{proof}
Since $I_\alpha \subset I$, isotony implies
	\[
		\bigvee_\alpha \mathbb{H}^{\frac{1}{2}}_\uparrow (I_\alpha) 
		\subset 
		\mathbb{H}^{\frac{1}{2}}_\uparrow  (I) \; . 
	\]
Now if $f \in \mathbb{H}^{\frac{1}{2}}_\uparrow (I) $, then there exists a closed 
interval $K \subset I$ such that 
	\[
		\operatorname{supp} f \subset K \subset I  \; . 
	\] 
As $\operatorname{supp} f$ is compact, the support of $f$ can be covered 
by a finite number of intervals 
	\[
		\operatorname{supp} f  \subset \bigcup_{I= 1, \ldots , n}  I_{\alpha_i}  \; . 
	\]
It follows that there exist functions $\chi_i \in C^\infty (S^1)$ such that 
	\[
		\operatorname{supp} \chi_i \subset I_{\alpha_i} \; , 
		\qquad 
		\sum_{i=1}^n \chi_i = 1 	
		\quad \text{on} \;  \operatorname{supp} f \; . 	
	\]
According to Lemma~\ref{lm:6-6-6}, multiplication 
with the functions $\chi_i \in  C^\infty (S^1)$ is a bounded 
operator on $\mathbb{H}^{\frac{1}{2}} (S^1)$. Hence,  we have 
	\[	
		f = \sum_{i=1}^n \chi_i f \; , 
		\quad \text{with} \; \; \chi_i f \in \mathbb{H}^{\frac{1}{2}} \bigl( I_{\alpha_i} \bigr) \; . 
	\]
Thus~\eqref{5.14a} follows, as $ f \in \bigvee_{i= 1, \ldots , n} 
\mathbb{H}^{\frac{1}{2}}_\uparrow (I_{\alpha_i})$.  
\end{proof}

\goodbreak
\begin{proposition}[Araki \cite{A1a}]
\label{prop:6.6.9}
Let $I$ and $I_\alpha$ be open subsets satisfying 
$\operatorname{int} \overline{I} = I$ and
$\operatorname{int} \overline{I}_\alpha = I_\alpha$. 
It follows that 
\begin{itemize}
\item [$i.)$] the outer and inner approximations coincide. 
\begin{align}
\mathbb{H}^{-\frac{1}{2}}_\downarrow (I)  
& = \mathbb{H}^{-\frac{1}{2}}_\uparrow (I) \; ,
\qquad
\mathbb{H}^{\frac{1}{2}}_\downarrow (I)
 = \mathbb{H}^{\frac{1}{2}}_\uparrow (I) \; ;
	\label{5.15}
\end{align}
\item [$ii.)$] the outer approximation for boundary points is the 
zero vector (and nothing else):
\begin{align}
	\mathbb{H}^{-\frac{1}{2}}_\downarrow ( \partial I) 
	= \{ 0 \} \; , 
	\qquad
	\mathbb{H}^{\frac{1}{2}}_\downarrow ( \partial I) 
	= \{ 0 \} \; ; 
	\label{5.16}
\end{align}

\item [$iii.)$] additivity holds for $\mathbb{H}^{-\frac{1}{2}}_\uparrow$ 
and $\mathbb{H}^{\frac{1}{2}}_\uparrow$: 
\begin{align}
	\mathbb{H}^{-\frac{1}{2}}_\uparrow ( I_1) \vee 
	\mathbb{H}^{-\frac{1}{2}}_\uparrow (I_2) & = 
	\mathbb{H}^{-\frac{1}{2}}_\uparrow \bigl( \operatorname{int} 
	\overline{I_1 \cup I_2} \bigr) \; , 
	\nonumber
	\\
	\mathbb{H}^{\frac{1}{2}}_\uparrow (I_1) \vee 
	\mathbb{H}^{\frac{1}{2}}_\uparrow ( I_2 ) 
	& = 
	\mathbb{H}^{\frac{1}{2}}_\uparrow (\operatorname{int} \overline{I_1 \cup I_2}) \; . 
	\label{5.17}
\end{align}
\end{itemize}
\end{proposition}

\begin{remark} 
\label{rm-6.6.10}
Combining \eqref{5.15}
with Lemma~\ref{lm:1.3} and Lemma~\ref{lm-A-6.6.7} yields
\begin{align}
\mathbb{H}^{-\frac{1}{2}}_\downarrow (I)  
& = \mathbb{H}^{-\frac{1}{2}} (I)
= \mathbb{H}^{-\frac{1}{2}}_\uparrow (I) \; ,
\qquad
\mathbb{H}^{\frac{1}{2}}_\downarrow (I)
= \mathbb{H}^{\frac{1}{2}} (I)
 = \mathbb{H}^{\frac{1}{2}}_\uparrow (I) \; .
\end{align}
\end{remark}

\begin{proof}
The first equation of \eqref{5.17} follows from \eqref{5.12b},   
\eqref{5.4a}, \eqref{5.2}  and
	\[
		L^2 ( I_1 \cup I_2 ) = L^2 \bigl( \operatorname{int} ( \overline{I_1 \cup I_2} ), 
		{\rm d} \psi \bigr)  \; . 
	\]
By \eqref{5.10} and \eqref{5.14}, 
	\[
		\mathbb{H}^{\frac{1}{2}}_\downarrow (\partial I) \subset 
		\mathbb{H}^{\frac{1}{2}}_\downarrow (I)
		\cap \mathbb{H}^{\frac{1}{2}}_\downarrow (\overline{I}^c)
		= \imath_1^{-1} \imath_2^{-1}\omega^{-1}
		\bigl( \mathbb{H}^{-\frac{1}{2}}_\uparrow (\overline{I}^c)
		\vee \mathbb{H}^{-\frac{1}{2}}_\uparrow (I) 
		\bigr)^\perp \; . 
	\]
Hence, the second equation of \eqref{5.16} follows from the first of
\eqref{5.17}. 
Similarly, the first equation of \eqref{5.16} follows from the second
of \eqref{5.17}. The latter is addressed below.

We now show the non-trivial inclusion\footnote{The inclusion 
$\mathbb{H}^{\frac{1}{2}}_\downarrow (I)
\supset \mathbb{H}^{\frac{1}{2}}_\uparrow (I)$ was established in 
\eqref{inside-outside}.}
of the second equation in \eqref{5.15}, which may be written in the form
	\begin{equation*}
		\bigcap_{J \supset \overline {I} }  \mathbb{H}^{\frac{1}{2}}  ( J)
		\subset \bigvee_{K \subset I } \mathbb{H}^{\frac{1}{2}} ( K) \; ,
	\end{equation*}
where the union is over closed subsets $K \subset I $ of $S^1$. 
The most general element on the left hand side is a function
	\[
		f \in \mathbb{H}^{\frac{1}{2}} (S^1)
		\qquad \text{with} \quad \operatorname{supp} f \subset \overline{I} \; . 
	\]
We have to show that $f$ can be approximated with elements in 	
$\mathbb{H}^{\frac{1}{2}} (S^1)$ whose (compact) support lies in the open 
interval $I$. 
Lemma~\ref{lm:6-6-6} allows us to decompose 
$f$ into  a sum of two functions $\chi_\ell f , \chi_r f \in \mathbb{H}^{\frac{1}{2}} (S^1) $, 
whose support does not contain the right or the left boundary point of $I$, respectively.  
In both cases, there exist two angles $\psi_\ell, \psi_r \in  S^1$ (in the neighbourhood of zero)
such that 
	\[
		\operatorname{supp} \Bigl( (\chi_\ell f) ( \, . \, - \lambda \psi_\ell ) \Bigr) \, , \; 
		\operatorname{supp} \Bigl( (\chi_r f) ( \, . \, - \lambda \psi_r ) \Bigr)
		\subset I \; 
	\]
for sufficiently small $\lambda >0 $. Now 
	\[ 
		\bigl\|  (\chi_\ell f) (\, . \, - \lambda \psi_\ell) 
		- \chi_\ell f  \, \bigr\|_{\mathbb{H}^{\frac{1}{2}} (S^1)}^2
		= \sum_{k \in \mathbb{Z}} \, | 1 - {\rm e}^{ i \lambda \psi_\ell k} |^2 \, 
		| \widetilde{\chi_\ell f} (k) |^2 \widetilde{\omega} (k) \; , 
	\]
where (due to Lemma~\ref{lm:6-6-6}) 
$ \sum_{k \in \mathbb{Z}} | \widetilde{\chi_\ell f} (k) |^2 \widetilde \omega (k) < \infty$ 
and $ | 1 - {\rm e}^{ i \lambda \psi_\ell k} | $ is uniformly bounded and tends to zero 
uniformly on finite sets of $k$'s  in $\mathbb{Z}$ as $\lambda \to 0 $. 
Hence, 
	\[
		\lim_{\lambda \to 0} (\chi_\ell f )( \, . \, - \lambda \beta) = \chi_\ell f  
		\qquad \text{in} \quad \mathbb{H}^{\frac{1}{2}} (S^1) \; . 
	\]
A similar argument holds for $\chi_r f$. Finally, we note that the two equations 
of~\eqref{5.15} are equivalent because of \eqref{5.14}. This proves \eqref{5.15}. 

For the second equation of \eqref{5.17}, we have to prove that
any
	\[
		f \in \mathbb{H}^{\frac{1}{2}}_\uparrow \bigl( \operatorname{int} 
		\overline{I_1 \cup I_2} \bigr)
	\]
is a limit of a sum of functions in $\mathbb{H}^{\frac{1}{2}}_\uparrow (I_1)$
and $\mathbb{H}^{\frac{1}{2}}_\uparrow (I_2)$.
Anticipating Lemma~\ref{lm-6.6.11} we may assume $f \in C^\infty(S^1)$ with
$\operatorname{supp} f \subset \operatorname{int} 
		\overline{I_1 \cup I_2}$. 
The nontrivial problem 
occurs at the boundary points $\partial I_1 \cap \partial I_2$, 
where $I_1$ and $I_2$ occupy the opposite sides of the boundary point. 
The idea is to multiply $f$ with a sequence of functions $\chi_\alpha$, 
	\[
		\chi_\alpha (\psi) = ( 1 - h_\alpha ) (\psi-\psi_\circ) \; , \qquad \alpha > 1\; , 
	\]
with	 
	\[
		h_\alpha (\psi) \doteq \gamma(\alpha)^{-1}    
		\sum_{k \in \mathbb{Z}} \; \frac{{\rm e}^{i \psi k}}{ (\omega( k))^{ \alpha}}   \; , 
	\qquad
		\gamma(\alpha) \doteq \sum_{k \in \mathbb{Z}} \;  
		(\omega( k))^{ - \alpha}  \;  , \qquad \alpha > 1\; ,  
	\]
which vanish at $ \{ \psi_\circ \}
= \partial I_1 \cap \partial I_2$, but nevertheless converge in 
$\mathbb{H}^{\frac{1}{2}} (S^1)$ to the constant function equal to $1$.
In the limit $\alpha \to 1$, $\gamma(\alpha) \to \infty$, and 
$h_\alpha$ tends to $0$ in~$\mathbb{H}^{\frac{1}{2}}(S^1)$. 
It follows that the smooth functions 
	\[
		f_\alpha (\psi) \doteq \chi_\alpha(\psi) f(\psi) 
	\]
vanish at the boundary $\partial I_1 \cap \partial I_2$ and can therefore be 
decomposed into two functions $ f _\alpha = f^{(1)}_\alpha + f^{(2)}_\alpha$
with support in $I_1$ and $I_2$, respectively, 
and such that 
	\[
		\lim_{\alpha \to 1} \| f^{(1)}_\alpha + f^{(2)}_\alpha 
		- f \|^2_{\mathbb{H}^{\frac{1}{2}}_\uparrow (\operatorname{int} 
		\overline{I_1 \cup I_2})} = 0\; . 
	\] 
\end{proof}

\begin{lemma} Let $I$ be an open subset in $S^1$. Then
\label{lm-6.6.11}
	\begin{equation}
	\label{LM-41}
		\overline {  \imath_1^{-1} C^\infty(S^1)
		\cap \mathbb{H}^{\frac{1}{2}}_\uparrow (I)
		} = \mathbb{H}^{\frac{1}{2}}_\uparrow (I) \; . 
	\end{equation}
\end{lemma}

\begin{proof}
The functions $f \in \mathbb{H}^{\frac{1}{2}}_\uparrow (I)$ with compact 
support in $I$ are dense in $\mathbb{H}^{\frac{1}{2}}_\uparrow (I)$. Now, let 
$\delta_\lambda \in C^\infty (S^1)$, $\delta_\lambda (\psi) = 0$ for $| \psi | > \lambda$,  
$\delta_\lambda \ge 0$, and
	\[
		\int_{S^1} {\rm d} \psi \; \delta_\lambda (\psi) = 1 \; ,  
	\]
be an approximation of Dirac delta function. 
Then the convolution 
	\[
		( \delta_\lambda   * f ) (\psi) = \int_{S^1} {\rm d} \psi' \; 
		\delta_\lambda (\psi -\psi') f ( \psi' ) 
	\]
is in $\imath_1^{-1} C^\infty(S^1)
\cap \mathbb{H}^{\frac{1}{2}}_\uparrow (I)$ for sufficiently small $\lambda$
and as the coefficients $\widetilde{\delta_\lambda} (k)$ tend to~$1$ uniformly on 
compacts as $\lambda \to 0$, we have 
	\[ 
		\lim_{\lambda \to 0} ( \delta_\lambda   * f ) = f \; . 
	\]
Hence we have \eqref{LM-41}. 
\end{proof}

\begin{proposition} The real subspace $\widehat{\mathfrak h} ( I)$ 
Is regular from the inside and from the outside, \emph{i.e.},  
\label{prop:6.5.2a}
	\begin{equation}
	\bigvee_{\overline {J} \subset I } \widehat{\mathfrak h}( J)
	= \widehat{\mathfrak h} ( I) = \bigcap_{J \supset \overline {I} }  
	\widehat{\mathfrak h} ( J)\; ,
	\label{ekg5}
	\end{equation}
The symbol $\bigvee$ denotes the 
closure of the $\mathbb{R}$-linear span.
\end{proposition}

\begin{proof}
  Inspecting Definition~\ref{local-h-hat}, we find
  \begin{equation}
	\label{h-i}
		\widehat{\mathfrak{h}}(I) = \mathbb{H}^{- 1/2} (I) 
		+ i \omega \,  \mathbb{H}^{1/2} (I) \; , \qquad I \subset S^1 \; . 
	\end{equation} 
Hence, \eqref{ekg5} follows from Proposition~\ref{prop:6.6.9} $i.)$ 
and Remark~\ref{rm-6.6.10}.
\end{proof}

\begin{proposition} Let $I \subset S^1$ be an open interval. Then 
the space $\widehat{\mathfrak{h}}(I)$ is a \index{standard subspace} 
\emph{standard subspace}, \emph{i.e.}, 
a $\mathbb{R}$-linear subspace in $\widehat{\mathfrak{h}}(S^1)$ such 
that $\widehat{\mathfrak{h}}(I)
+ i \widehat{\mathfrak{h}}(I)$ is dense 
in~$\widehat{\mathfrak{h}}(S^1)$ and $\widehat{\mathfrak{h}}(I) 
\cap i \widehat{\mathfrak{h}}(I)  = \{0\}$. 
\end{proposition}

\begin{proof}
The first part of the statement has been established in Corollary~\ref{cor:6.5.4}.
It remains to show that 
$\widehat{\mathfrak{h}}(I) \cap i \widehat{\mathfrak{h}}(I)  = \{0\} $. 
This follows from \emph{anti-locality}, see \cite{V, Masuda}: 
the function
	\[
		h_k (\eta, \psi) \doteq {\rm e}^{-i \eta \omega  r \mathbb{cos}_\psi} 
		c_k {\rm e}^{i k \psi } \; , 
		\qquad k \in \mathbb{Z}\setminus \{0 \} \; , 
	\]
satisfies the equation (see \eqref{L1-L2})
	\[
		\bigl( \partial_{\eta}^{2} - \LS_1^{\, 2}  \bigr) 
		h_k (\eta, \psi) =0 \; , 
		\qquad \LS_1 = \omega r \,  \mathbb{cos}_\psi  \; .	
	\]
Moreover, if $h_k (0, \psi)$
is in the orthogonal complement of the complex linear span 
of~$\widehat{\mathfrak{h}}(I)$, then 
	\[
		\langle h_k (0, \, . \,) , g \rangle  = 0
	\]
and 
	\[
		0= \langle h_k (0 , \, . \,)  , i \omega r \, \mathbb{cos}_\psi
		g \rangle =
		\frac{{\rm d} }{{\rm d}\eta} \langle h_k(\eta  \, . \,) , 
		g \rangle  \Bigl|_{\eta = 0} \Bigr.  
		\label{(2.3)}
	\]
for all (complex-valued) $g \in C^\infty_0 (I)$. 
It now follows from Proposition~\ref{halpha} that for every open set $I_\circ \subset I$ 
whose closure still lies in $I$ there exists some $\eta_\circ $ 
with $0< \eta_\circ < \tfrac{\pi}{2}$ such that
	\[
		\langle h_k(\eta , \, . \,) , g_\circ \rangle_{\mathfrak{h}} 
		= 0 \qquad  \forall \eta \in (-\eta_\circ , \eta_\circ )  
		\label{(2.7)}
	\]
and for all (complex-valued) $g_\circ \in C^\infty_0 (I_\circ)$. 
Since $h_k(0 , \, . \,)= c_k {\rm e}^{i k \psi}$ is an analytic vector for the 
map $\kappa \mapsto \exp(\kappa\, \omega r \, \mathbb{cos}_\psi)$, 
we conclude that
	\[
		\langle {\rm e}^{ \kappa r\omega \mathbb{cos}_\psi} 
		h_k(0 , \, . \,) , g_\circ \rangle  = 0 
		\qquad \forall \kappa \in \mathbb{R}   
	\]
and for all (complex-valued) $g_\circ \in C^\infty_0 (I_\circ)$. It follows that 
	\[
		{\rm e}^{ \kappa \omega r \, \mathbb{cos}_\psi} 
		h_k(0 , \, . \, )_{ \upharpoonright \psi \in I_\circ } = 0 
		\qquad \forall \kappa \in \mathbb{R} \;  .  
	\]
If $I_\circ$ is an open interval, this can only be satisfied if $c_k=0$. 
We thus infer that $\omega$ is \emph{anti-local}, \emph{i.e.}, 
for any open
interval $I \subset S^1$ and $g \in \mathscr{D}(\omega) \cap \widehat{\mathfrak{h}}(S^1) $, 
	\begin{equation}
	\label{anti-local}
		\omega g_{\upharpoonright I} = 0
		\quad \text{and} \quad g_{\upharpoonright I} = 0 
		\qquad \Rightarrow \qquad g = 0 \; . 
	\end{equation}
Finally we recall Definition~\ref{local-h-hat}:
for $I$ a simply connected open interval in $S^1$, 
	\begin{align*}
		\widehat{{\mathfrak h}}  ( I) \cap i \widehat{{\mathfrak h}}  ( I) 
		& =
		\overline{ \bigl\{ h \in \widehat{{\mathfrak h}}  (S^1)   \mid   
		{\rm supp\,}   h \subset I \, ,    
		\, {\rm supp\,}  \omega^{-1}  h  \subset I \bigr\} }\; .
	\end{align*}
Thus \eqref{anti-local} implies that $\widehat{{\mathfrak h}}  ( I) 
\cap i \widehat{{\mathfrak h}}  ( I) = 0$. 
\end{proof}
\goodbreak

\begin{proposition}[Figliolini \& Guido~\cite{FiGu1}]
\label{Figolini-Guido}
The number $1$ is in the spectrum $\sigma (\delta_I)$ of the one-particle 
modular operator $\delta_I$ associated to $\widehat{\mathfrak h} (I)$, but not in the 
point spectrum~$\sigma_p (\delta_I)$.
\end{proposition}

\begin{proof}
Let us define two subspaces 
	\[
		\mathcal{D}_{\pm} \doteq \mathbb{H}^{\pm 1/2} ( I ) 
		+ \mathbb{H}^{\pm 1/2} ( I^c ) \; , 
	\] 
which represent \emph{decoupled functions}, and which 
are dense in $\mathbb{H}^{\pm 1/2} ( S^1 ) $, 
respectively (Proposition~\ref{prop:6.6.9} $iii.)$
or Proposition 2.4 in \cite{FiGu1}). This allows us to define 
projections $P_\pm  \colon \mathcal{D}_\pm \mapsto \mathbb{H}^{\pm 1/2} ( S^1 )$, 
	\[
		P_{\pm \upharpoonright \mathbb{H}^{\pm 1/2} (I) } = \mathbb{1} \; , 
		\qquad 
		P_{\pm \upharpoonright H^{\pm 1/2} ( I^c )} = 0 \; . 
	\]
Because $\mathbb{H}^{\pm 1/2} ( I ) \cap \mathbb{H}^{\pm 1/2} ( I^c ) = \{ 0 \}$ 
--- this property was established in \eqref{5.16} ---  and 
$\mathbb{H}^{\pm 1/2} ( I ) + \mathbb{H}^{\pm 1/2} ( I^c )$ is dense, $P_\pm$ is 
a well defined, densely defined 
closed operator and it is idempotent \cite{FiGu1}. 

Next, define operators $A_{\pm}\colon \omega^{\mp 1} \mathcal{D}_{\mp}
\cap \mathbb{H}^{\pm 1/2} (I) \to \mathbb{H}^{\mp 1/2} (I)$ by setting
	\[
		A_{+}  = \bigl( P_{-} \omega
		\bigr) _{\upharpoonright \mathbb{H}^{1/2} (I)} \; , 
		\qquad
		A_{-}  = \bigl( P_{+} \omega^{-1} 
		\bigr)_{\upharpoonright \mathbb{H}^{-1/2} (I)} \; , 
	\]
respectively. According to \cite[Theorem 2.10]{FiGu1}, the operators $A_{\pm}$ are 
densely defined closed operators and 
	\[
		(A_\pm)^* = A_\mp \; . 
	\]
Now, let $s_I$ be the Tomita operator associated to the standard 
subspace $\widehat{\mathfrak{h}}(I)$ given in 
\eqref{h-i} by Theorem~\ref{EckOsw}. 
Combine the operators $A_+$ and $A_-$ to an operator valued matrix 
	\[
		B\colon \omega D_+ \cap \mathbb{H}^{-1/2}(I) 
		\oplus \omega^{- 1}D_{ -}  \cap 
		\mathbb{H}^{1/2}(I)  \to \mathbb{H}^{-1/2} (I) 
		\oplus \mathbb{H}^{1/2} (I)
	\]
defined by 
	\[
		B = \begin{pmatrix}
		0 & i A_{+}
		\\
		- i A_{-} & 0 
		\end{pmatrix} .   
	\]
Then $\delta_I = s_I^* s_I$ 
can be written as \cite[Theorem 3.6]{FiGu1}
	\begin{equation}
	\label{d-I}
		\delta_I = \frac{B+1}{B -1} \, . 
	\end{equation}
The anti-locality property of $\omega$ (see \cite{V, Masuda}) implies 
that $1 \notin \sigma_p (B)$ \cite[Theorem~3.5]{FiGu1}, 
hence this expression is well defined.  

Moreover, equation \eqref{d-I} shows that $1\not\in\sigma_{p}(\delta_I)$.  
On the other hand, using the explicit formula for $B$, one concludes 
that $B$ is unbounded, hence $1$ is in the spectrum of~$\delta_I$. 
\end{proof}

\chapter{Local Algebras for the Free Field}
\label{2Q}
\setcounter{equation}{0}

Given the symplectic space ${\mathfrak k}(dS)$, one can define a \emph{$C^*$-algebra} 
(the \emph{Weyl algebra}), which ensures \emph{bosonic particle statistics} and 
contains\footnote{To be precise, we eventually have to take the weak closure of the 
\emph{local} \emph{$C^*$-algebra} w.r.t.~a \emph{distinguished folium of states} in 
order to ensure that all the projection operators are contained.} the observables of the quantum 
field theory. As we were able to isolate subspaces~${\mathfrak k}({\mathcal O})$ associated to 
open and bounded regions ${\mathcal O} \subset dS$ (see Definition~\ref{symp-sub}), 
the Weyl algebra can be enriched with a \index{localisation map}\emph{localisation map}, 
giving rise to a \index{net of local $C^*$-algebras}\emph{net of local $C^*$-algebras}.
Moreover, the symplectic transformations ${\mathfrak z} (\Lambda)$, 
$\Lambda \in O(1,2)$, acting on ${\mathfrak k}(dS)$, lift to a \emph{covariant action} of $ O(1,2)$ 
in terms of \emph{automorphisms} on the net of local $C^*$-algebras.

Up to this point, our method resembles the construction of a Haag--Kastler net describing free 
bosons on Minkwoski space. But de Sitter space does not have a globally time-like Killing vector field.  
Hence there  is no \index{global time evolution} global time evolution (in terms of a one-parameter 
group of automorphisms) and no natural notion of energy, and, consequently, one can not require that 
the vacuum state is a state of minimal energy. One may still require that a \index{de Sitter vacuum state} 
\emph{de Sitter vacuum state} is invariant under the action of the Lorentz group. But in itself, this 
requirement does \emph{not} guarantee the necessary stability of the physical system. One may postulate 
that the short distance behaviour of the two-point function should be just as it is on Minkowski space. This 
is the so-called \emph{Hadamard condition}, which was reformulated (and renamed as \emph{microlocal 
spectrum condition}) by Radzikowski \cite{Raz1}\cite{Raz2} as a requirement for the \emph{wave front set} 
of the two-point function. While Hadamard states can be constructed on a variety of curved space-times 
(see \cite{KW}\cite{GM} and references therein) and are widely accepted as possible physical states of  
\emph{non-}interacting\footnote{The Hadamard condition plays an essential role in perturbative quantum 
field theory \cite{BrFr}, but this does not necessarily ensure its relevance in a non-perturbative setting.} 
quantum field theories, the relevance of the \index{Hadamard condition} Hadamard condition for interacting 
theories is less evident. One may therefore continue the search for other criteria to select de Sitter vacuum 
states. The aforementioned \index{stability properties} \emph{stability properties} (against small 
\index{adiabatic perturbations} adiabatic perturbations) are well-known in the context of quantum statistical 
mechanics, where they lead to analyticity properties of the $n$-point functions. Similar analyticity properties 
can be formulated on de Sitter space~\cite{BB}: for the free field, the so-called the \emph{geodesic KMS 
condition} \index{geodesic KMS condition} proposed by Borchers and Buchholz is equivalent to the Hadamard 
condition. In contrast to the latter, the geodesic KMS condition allows a physical interpretation, which may 
very well hold for interacting theories: the \index{Unruh effect} Unruh effect~\cite{Unruh} says that an observer 
following a time-like geodesic on the de Sitter space will observe a \emph{temperature}, if he carries along a 
small measurement device, see~\cite{DeBeM} for further details. 

\section{The covariant net of local algebras on $dS$}

Let $({\mathfrak k} , \sigma)$ be a symplectic space. The unique $C^*$-algebra ${\mathfrak W} 
({\mathfrak k}, \sigma)$ generated by nonzero elements ${\rm W}({\mathfrak f})$, ${\mathfrak f} 
\in {\mathfrak k}$, satisfying 
\label{weylalgebrapage}
	\begin{align}
		{\rm W} ({\mathfrak f}_1){\rm W} ({\mathfrak f}_2) 
		&= {\rm e}^{- i \sigma ( {\mathfrak f}_1,{\mathfrak f}_2 ) /2} \; {\rm W} ({\mathfrak f}_1+{\mathfrak f}_2) \; ,
		\nonumber \\ 
		\label{weylalgebra} 
		{\rm W} ^*({\mathfrak f}) &= {\rm W} (-{\mathfrak f}) \; ,  \qquad {\rm W} (0)=\mathbb{1} \; ,  
	\end{align}
is called the {\em Weyl algebra} associated to $({\mathfrak k} , \sigma)$;
see, \emph{e.g.}, \cite{BR}. In case ${\mathfrak k}$ is a Hilbert space, 
we suppress the dependence on the symplectic form given by twice the imaginary part of the scalar product. 

We now turn to the covariant dynamical system $({\mathfrak k}(X), \sigma, {\mathfrak z} (\Lambda))$ constructed 
in Section \ref{CCDS} and set 
	\[
		{\mathfrak W}(X) \equiv {\mathfrak W}({\mathfrak k}(X), \sigma) \; , \qquad X= {\mathcal O}, W, dS \; . 
	\]
The symplectic transformations ${\mathfrak z} (\Lambda)$, $\Lambda  \in O(1,2)$ 
acting on~${\mathfrak k} (dS)$ (see Proposition~\ref{Prop4-10}) give rise to a group of automorphisms 
$\alpha^\circ \colon \Lambda \mapsto \alpha_\Lambda^\circ$,   
\label{alphapage}
	\begin{equation}
	\label{auto-circ}
		\alpha_\Lambda^\circ (W ( [f] )) \doteq W ( {\mathfrak z} (\Lambda) [f] ) \; , 
		\qquad  [f] \in {\mathfrak k} (dS) \; , 
	\end{equation}
acting on ${\mathfrak W}(dS)$. The automorphisms~$\alpha^\circ $ respect the local structure:
	\[
		\alpha_\Lambda^\circ \bigl({\mathfrak W}({\mathcal O})\bigr)
		={\mathfrak W}(\Lambda {\mathcal O}) \;  , \qquad {\mathcal O} \subset dS \; . 
	\]
The map $ \alpha^\circ \colon \Lambda \mapsto \alpha_\Lambda^\circ$ fails to be 
\index{strongly continuous automorphisms}\emph{strongly continuous} in the $C^*$-norm; 
thus strictly speaking $\bigl({\mathfrak W}(dS), \alpha^\circ \bigr)$ is not a $C^*$-dynamical system. 

\begin{definition}
\label{def:7.1.1}
The pair $\bigl({\mathfrak W}(dS), \alpha^\circ \bigr)$ is called the {\em covariant quantum 
dynamical system} \index{covariant quantum dynamical system} associated to the Klein--Gordon 
equation on the de Sitter space.
\end{definition}
\label{cqds-page}

As suggested in the introduction to this chapter, we will now use the \index{geodesic KMS condition} 
\emph{geodesic KMS condition} to characterise \emph{de Sitter vacuum states}. 
Let  $ \alpha \colon \Lambda \mapsto \alpha_\Lambda$ be a representation (in terms of automorphisms) 
of $SO_0(1,2)$ on the $C^*$-algebra ${\mathfrak W}(dS)$. 

\begin{definition}
\label{vs}
A normalised positive linear functional $\omega$  is called a {\em de Sitter 
vacuum state} \index{de Sitter vacuum state} for the quantum dynamical system 
$\left({\mathfrak W} (dS) , \alpha  \right) $, if  
\begin{itemize}
\item[$ i.)$] $\omega$ is invariant under the action of the proper, orthochronous 
Lorentz group $SO_0(1,2)$, \emph{i.e.}, 
	\[
		\omega = \omega \circ \alpha_\Lambda  \qquad \forall \Lambda \in  SO_0(1,2)\; ;
	\]
\item [$ ii.)$] $\omega$ satisfies the {\em geodesic KMS condition\/}: for every wedge 
$W = \Lambda W_1$,  $\Lambda \in  SO_0(1,2)$, the restricted (partial) state 
$\omega_{\upharpoonright {\mathfrak W}(W)}$ satisfies the KMS-condition at inverse 
temperature $2 \pi r$ with respect to the one-parameter  group 
	\[
		t \mapsto \Lambda_{W}\bigl( \tfrac{t}{r} \bigr)
	\]
of boosts, which leaves the wedge $W$ invariant. In other words: for each pair  $[f],  [g] 
\in {\mathfrak k}(W)$, there exists a function~${\mathfrak F}_{f,g}(\tau)$ holomorphic in the strip 
	\[
	{\mathbb S}_{2 \pi r} =\{ \tau\in {\mathbb C} \mid 0  < \Im \tau <  2\pi r \}
	\] 
and continuous on~$\overline{{\mathbb S}_{2 \pi r}  } $ such that
	\begin{align*}
		{\mathfrak F}_{f,g}(t)&= \omega_{\upharpoonright {\mathfrak W}( W )} 
		\bigl({\rm W}([f])\alpha_{\Lambda_W (\frac{t}{r})} ({\rm W}([g])) \bigr) 
			\nonumber \\
		{\mathfrak F}_{f, g}(t + i 2\pi r)&=
		\omega_{\upharpoonright {\mathfrak W}( W)}
		\bigl(\alpha_{\Lambda_W (\frac{t}{r})} ({\rm W}([g])){\rm W}([f]) \bigr)   \qquad \forall  t\in {\mathbb R} \; .
	\end{align*}
\end{itemize}
\end{definition}

It is sufficient to verify the geodesic KMS condition 
for {\em one} wedge, as the invariance property then 
implies that it holds for any wedge. The {\em de Sitter 
vacuum state} for the free quantum field is presented next.

\begin{theorem} 
\label{th2.5} The state $\omega^\circ$ on ${\mathfrak W}( {dS})$ given by 
\label{freevacuumstatepage}
	\begin{equation}
		\label{statedef}
		\omega^\circ({\rm W}([f]))
		= {\rm e}^{-\frac{1}{2} \| K [f] \|_{ {\mathfrak h}
                    (dS) }}, \qquad  f \in {\mathcal D}_{\mathbb{R}}
                ({dS}) \; , 
	\end{equation}
is the {\em unique} de Sitter vacuum state for the 
pair~$\bigl({\mathfrak W}( dS ), \alpha^\circ \bigr)$,
which is $C^2$ and primary, and whose one-point function is zero.

Moreover, the GNS representation $\bigl(\pi^\circ, {\mathcal H} (dS)\bigr)$ 
associated to the pair $\bigl({\mathfrak W}( dS) , 
\omega^\circ \bigr) $ is (unitarily equivalent to) the  Fock
representation (see Section \ref{Fockspace})  
over the one-particle space ${\mathfrak h} (dS)$, \emph{i.e.}, 
\label{HdS-page}
	\[
 		\pi^\circ ({\rm W}( [f] )) 
		= {\rm W}_F ( [f]) \; , \qquad {\mathcal H} (dS) 
		= \Gamma ({\mathfrak h}(dS)) \; .
 	\]
\end{theorem}

\begin{remarks}
\quad
\begin{itemize}
\item[$i.)$]   See Theorem~\ref{1PStrucHe} for the definition of map $K$ appearing in \eqref{statedef}.
\item[$ii.)$]   We recall that a state is said to be of class $C^2$ if 
in the corresponding GNS representation $\pi_\omega$ of the Weyl
algebra the functions 
	\begin{equation} \label{eqC2}
		t \mapsto \pi_\omega \bigl({\rm W}(t [f] )\bigr) \; ,
                \qquad t \in \mathbb{R} \; ,  
	\end{equation}
are $C^2$ around $t=0$, for all $f \in {\mathcal D}_{\mathbb{R}} ({dS})$. 
It then follows that they are $C^2$ on the entire real line~\cite{Kay4}.
In this case the $1$- and $2$-point functions 
of $\omega$ are defined by 
	\begin{align*}  
		\omega_1([f]) & \doteq \langle \Omega_\omega, \phi_\omega(f)
				\Omega_\omega\rangle \; , 
		\\ 
		\omega_2\big([f],[g]\big) & \doteq 
		\langle \Omega_\omega, \phi_\omega(f) \phi_\omega(g) \Omega_\omega\rangle \; ,
	\end{align*}
where $\phi_\omega(f)$ denotes the generator of the unitary
group~\eqref{eqC2} and $\Omega_\omega$ is the GNS vector in the GNS Hilbert space 
${\mathcal H}_\omega$ implementing $\omega$. 
\item[$iii.)$]
If $\omega$ is a primary $C^2$ de Sitter vacuum state whose one-point
function $\omega_1\neq 0$, then it differs from $\omega^\circ$ by a Bogoliubov
automorphism\footnote{This corresponds to 
$\phi_\omega(f) =  \phi_{\omega^\circ}(f) + \omega_1([f])\mathbb{1}$.}
	\begin{equation} 
		\label{eqBogoliubov}
    		\omega = \omega^\circ \circ \varrho,\quad \text{ where }
    					\varrho\big(W([f])\big) \doteq e^{i\omega_1([f])} \,W([f]).
	\end{equation}
The corresponding GNS representation is unitarily equivalent to $\pi^\circ$. 
\end{itemize}
\end{remarks}

\begin{proof}
Clearly, \eqref{statedef} defines a positive linear functional on the
Weyl algebra.  
Invariance under the action of $SO_0(1,2)$
can be seen as follows:
	\begin{align*}
		\omega^\circ \bigl( \alpha_\Lambda^\circ ({\rm W}([f])) \bigr)
		& = \omega^\circ \bigl( ({\rm W}(  {\mathfrak z}  (\Lambda) [f])) \bigr)
		= {\rm e}^{-\frac{1}{2} \| u (\Lambda) [f] \|_{ {\mathfrak h} (dS) }} \\
		& = {\rm e}^{-\frac{1}{2} \|  [f] \|_{ {\mathfrak h} (dS) }} 
		= \omega^\circ \bigl( {\rm W}([f]) \bigr) \qquad \forall \Lambda \in SO_0(1,2) \; . 
	\end{align*}
Next, let us analyze the KMS condition:  for $[f], [g] \in {\mathfrak k} (W)$, we have 
	\begin{align*}
		{\mathfrak F}_{f,g}(t)&= \omega^\circ
		\Bigl({\rm W}([f]){\rm W}\bigl( u \bigl( \Lambda_W
                (\tfrac{t}{r}) \bigr) [g]) \bigr) \Bigr)  
			\nonumber \\
		& =  {\rm e}^{- i \Im \langle [f] , u (\Lambda_W
                          (\frac{t}{r}))  [g] \rangle } \;  
		\omega^\circ \Bigl({\rm W} \bigl( [f] + u ( \Lambda_W
                (\tfrac{t}{r}) )  [g] \bigr) \Bigr) 
			\nonumber \\
		& =  {\rm e}^{- i \Im \langle u (\Lambda_W
                          (\frac{t}{2r})) [f] , u (\Lambda_W
                          (\frac{t}{2r}))  [g] \rangle } \;  
		\omega^\circ \Bigl({\rm W} \bigl( u ( \Lambda_W
                (\tfrac{t}{r}) )  [g]  + [f]  \bigr) \Bigr) 
	\end{align*}
Inspecting \eqref{5.4b}, we conclude that 
	\[
		\Bigl\langle u \bigl(\Lambda_W \bigl(\tfrac{t+ i \pi}{2r}\bigr)\bigr) [f] , 
		u \bigl(\Lambda_W \bigl(\tfrac{t+ i \pi}{2r} \bigr)\bigr)  [g] \Bigr\rangle
		= \bigl\langle  u (\Lambda_W (\tfrac{t}{r}))  [g] ,  [f] \bigr\rangle \; . 
	\]
Hence,
	\begin{align*}
		{\mathfrak F}_{f, g}(t + i 2\pi r)&=
		\omega^\circ
		\bigl(\alpha_{\Lambda_W (\frac{t}{r})} 
		({\rm W}([g])){\rm W}([f]) \bigr)   \qquad \forall  t\in {\mathbb R} \; .
	\end{align*}
This is the KMS condition. 

Now, if the one-point function of $\omega$ vanishes, then its two-point function 
coincides with that of $\omega^\circ$; see Lemma~\ref{lm:7.1.5} below.
But then Eq.~\eqref{statedef} implies that $\omega^\circ$ is the
\emph{liberation} of $\omega$ in the sense of \cite[Definition~5.6]{Kay4}, \emph{i.e.}, 
	\[
		\omega^\circ\big(W([f])\big) = 
				{\rm e}^{-\frac{1}{2}  \omega_2([f],[f])} \; , 
				\qquad  f \in {\mathcal D}_{\mathbb{R}}  ({dS}) \; . 
	\] 
Thus, $\omega$ is a $C^2$ state whoses liberation, $\omega^\circ$, is a
pure state on 
	\[
		\pi^\circ \bigl({\mathfrak W}( dS ) \bigr) = \mathcal{B}\bigl( {\mathcal H} (dS)\bigr) \; , 
	\] 
as on $\mathcal{B}\bigl( {\mathcal H} (dS)\bigr)$
the notions of pure states and vector states coincide. The remarkable Theorem~6.1 
in~\cite{Kay4} then asserts that $\omega$ coincides with its
liberation, i.e., $\omega=\omega^\circ$. This proves uniqueness. 

Finally, an inspection of Subsection~\ref{WOFS} 
shows that the GNS representation~$\pi^\circ$ associated to the pair $\bigl({\mathfrak W}( dS) ,
\omega^\circ \bigr) $ is (unitarily equivalent to) the  Fock
representation.
\end{proof}

\begin{lemma}
\label{lm:7.1.5}
Let $\omega$ be a de Sitter vacuum state which is $C^2$ and primary. 
Then its two-point function of $\omega$ is related to that of
$\omega^\circ$ by 
\begin{equation}\label{eqOmega2Omega20}
\omega_2([f],[g]) =\omega_2^\circ([f],[g])  + \omega_1([f])\, \omega_1([g]).
\end{equation}
\end{lemma}
\begin{proof}
We consider a fixed wedge $W$ and denote by $f_t$ the push-forward  of $f$
under the corresponding boost.  
Both $\omega$ and $\omega^\circ$ are KMS states for this dynamics. 
By a standard argument as in the proof of Theorem~1b in~\cite{Kay1},
this implies that the difference of
the two-point functions is independent of the boost variable, \emph{i.e.}, 
for all $t \in \mathbb{R}$ there holds
	\begin{equation} 
		\label{eq2PtFct}
		\omega_2([f_t],[g]) =
		\omega_2^\circ([f_t],[g]) + s([f], [g]) \qquad
		\forall f,g\in \mathcal{D}_{R}(W) \; , 
	\end{equation}
where $s$ is a symmetric bilinear form on the classical space. 

We now show that $s$ is the product of one-point functions.
To this end, note that the symplectic form satisfies
	\[
		 \sigma([f_t],[g]) \to 0  \qquad \text{as} \qquad 
		 |t| \to 0 \; . 
	\] 
(This follows for example from the fact that $\sigma$ is
the imaginary part of the scalar product in the
one-particle space of $\omega^\circ$, where the generator of the boosts
has no zero eigenvalue.)
By the argument given in  \cite[Vol.~II, p.~41]{BR}),
this implies that the Weyl algebra is
asymptotically abelian for the boosts, namely
	\[
		\bigl\| \big[W([f_t]), W([g]) \big] \bigr\|  \leq |\sigma([f_t],[g])| \to 0 
		\qquad \text{for} \quad  t\to \infty \; . 
	\]
Since $\omega$ is primary, this fact implies the cluster property, \emph{i.e.}, 
for all $f,g \in \mathcal{D}_R(W)$	\[
		\omega\bigl(W([f_t] W([g]) \bigr) - \omega \bigl( W([f]) \bigr)
		\omega \bigl( W([g]) \bigr)
		\to 0
		\qquad \text{for} \quad  t\to \infty \; ; 
	\]
see \cite[Theorem 3.2.2]{H} or \cite[Bem.\ (3.3,8)]{Thirring-4}.
Taking derivatives, one finds  
	\[
		\omega\big(\phi_\omega[f_t] \phi_\omega[g]\big) \to 
		\omega(\phi_\omega[f]) \omega(\phi_\omega[g]) \equiv\omega_1([f])\,\omega_1([g])
		\qquad \text{for} \quad  t\to \infty \; . 
	\]
Since $\omega_2^\circ([f_t],[g])$ goes to zero for large $t$ (which
follows from the same argument, but also from the fact that the corresponding 
generator has no zero eigenvalues), Equ.~\eqref{eq2PtFct} implies that
the constant $s([f],[g])$ is just $\omega_1([f])\,\omega_1([g])$. This
completes the proof of the lemma.
\end{proof} 

The automorphisms \eqref{auto-circ} are (anti-)unitarily 
implemented in the Fock representation $\pi^\circ$ by the operators 
	\begin{equation}
	\label{U-Lambda} 
		U_\circ  (\Lambda) \doteq \Gamma (u(\Lambda))\; , \qquad \Lambda \in O(1,2) \; , 
	\end{equation}
defined in \eqref{gamma-u} and associated to the 
unitary irreducible representation \eqref{u-oph} of the Lorentz group: for $f \in {\mathcal D}_{\mathbb{R}}(dS)$,
	\begin{equation}
	\label{u-circ-F}
		\pi^\circ \bigl( \alpha_\Lambda^\circ ( {\rm W}( [f] ) ) \bigr) 
		= U_\circ  (\Lambda) {\rm W}_{F} ( [f] ) U_\circ  (\Lambda)^{-1} \; , 
		\qquad \Lambda \in O(1,2) \; . 
	\end{equation}	
We will denote the generators of the strongly continuous one-parameter groups
\label{LFockdSpage} 
	\[
		t \mapsto U_\circ (\Lambda_1(t)) \; , 
		\quad s \mapsto U_\circ (\Lambda_2(s)) \quad \text{and} \quad  
		\alpha \mapsto U_\circ (R_0(\alpha)) 
	\]
by $\LFockdS^\circ_1$, $\LFockdS^\circ_2$ and~$\KFockdS_0$. 
The right hand side in \eqref{u-circ-F} extends to arbitrary elements 
in the weak closure $\pi^\circ \bigl({\mathfrak W} (dS) \bigr)''$ of ${\mathfrak W} (dS)$. 
We denote this extension of the automorphism 
$\alpha_\Lambda^\circ$ by the same letter. In particular, for $f \in {\mathfrak h}(dS)$, 
we write 
 	\begin{equation}
	\label{aut-ext}
		\alpha_\Lambda^\circ ( {\rm W}_F( f) ) \bigr) 
		\doteq U_\circ  (\Lambda) {\rm W}_{F} ( f)  U_\circ  (\Lambda)^{-1} \; , 
		\qquad \Lambda \in O(1,2) \; . 
	\end{equation}
The GNS vector can be used to extend the free de Sitter vacuum state $\omega^\circ$ 
to the weak closure $\pi^\circ \bigl({\mathfrak W} (dS) \bigr)''$:
	\begin{equation}
	\label{state-ext}
	\omega^\circ (A) \doteq \langle \Omega_\circ , A \Omega_\circ \rangle \; , 
	\qquad A \in \pi^\circ \bigl({\mathfrak W} (dS) \bigr)'' \; . 
	\end{equation}
Here $\Omega_\circ$ denotes the vacuum vector in the Fock space ${\mathcal H}(dS)$. 

\begin{proposition} 
\label{prop:7.1.6}
The state \eqref{state-ext} satisfies the geodesic KMS condition 
with respect to the automorphisms \eqref{aut-ext}. 
\end{proposition}

\begin{proof} 
The geodesic KMS property for $\left({\mathfrak W} (W) , \alpha_{\Lambda_W}  \right) $
is part of Theorem \ref{th2.5}.
The fact that the KMS property 
extends to the weak closure is a standard result, see, \emph{e.g.}, \cite[Corollary 5.3.4]{BR}. 
\end{proof}

The \emph{local von Neumann algebras} \index{local von Neumann algebras}
for the free covariant field are defined by setting
	\begin{equation}
	\label{free-loc-cov-alg}
		{\mathscr A}_\circ ({\mathcal O}) \doteq  \pi^\circ \bigl({\mathfrak W}
		({\mathcal O}) \bigr)'' \; ,  \qquad {\mathcal O} \subset dS \; . 
	\end{equation} 
It follows from Theorem \ref{th2.5} that the algebra ${\mathscr A}_\circ ({\mathcal O}) $ 
is equal to the von Neumann algebra generated 
by ${\rm W}_{F} (f )$, $f \in {\mathfrak h}({\mathcal O})$. From now on we will suppress 
the subscript and simply write ${\rm W} (f )$ instead of ${\rm W}_{F} (f )$.
\label{AO-page}

\bigskip
By construction, the net of local algebras \index{net of local algebras} satisfies some key properties:

\begin{theorem}[ The Haag--Kastler axioms \index{Haag--Kastler axioms} for the free field]
\label{th:7.1.5}
The net of local algebras 
${\mathcal O} \mapsto {\mathscr A}_\circ ({\mathcal O})$
satisfies 
\begin{itemize}
\item [$i.)$] \emph{isotony}, \emph{i.e.}, \index{isotony}
	\[
		{\mathscr A}_\circ ({\mathcal O}_1) \subset {\mathscr A}_\circ ({\mathcal O}_2) 
		\qquad 
		\text{if ${\mathcal O}_1 \subset {\mathcal O}_2$} \; ; 
	\] 
\item [$ii.)$] \emph{locality}, \emph{i.e.}, \index{locality}
	\[
		\bigl[ {\mathscr A}_\circ ({\mathcal O}_1) , {\mathscr A}_\circ ({\mathcal O}_2) \bigr] = \{ 0 \} 
		\quad 
		\text{if ${\mathcal O}_1$ and ${\mathcal O}_2$ are space-like separated} \; ; 
	\] 
\item [$iii.)$] \emph{covariance}, \emph{i.e.}, \index{covariance}
	\[
		\alpha_\Lambda^\circ \bigl({\mathscr A}_\circ ({\mathcal O}) \bigr) 
			= {\mathscr A}_\circ (\Lambda {\mathcal O})\, ,
				\qquad \Lambda \in O(1,2) \; ; 
	\]
\item[$iv.)$]  {\em (additivity)}, \emph{i.e.},
for $X$ a double cone or a wedge, there holds
	\begin{equation} 
		\label{Additivity} 
		{\mathscr A}_\circ(X) = \bigvee_{{\mathcal O\subset X}} {\mathscr A}_\circ({\mathcal O}) \; .
	\end{equation}
The right hand side denotes the von Neumann algebra generated by the 
local algebras associated to double cones ${\mathcal O}$ contained in $X$.
(It thus makes sense to define~${\mathscr A}_\circ(X)$ for arbitrary regions
$X$ by Eq.~\eqref{Additivity}.) 
\item [$iv'.)$] \emph{weak additivity}, \emph{i.e.}, \index{weak additivity}
for each open region ${\mathcal O} \subset dS$ there holds
	\[
		\bigvee_{\Lambda \in SO_0 (1,2) } {\mathscr A}_\circ  (\Lambda {\mathcal O}) 
		\doteq {\mathscr A}_\circ  (dS) \quad ( \;  = \mathscr{B}( \mathcal{H} (dS))  \; )\; ; 
	\]
\item [$v.)$]  
the \emph{time-slice axiom}\footnote{See \cite{CF} for a discussion 
of this axiom.}, \emph{i.e.},  
\index{time-slice axiom} 
for $I$ an interval on a geodesic Cauchy surface and $I''$ its causal completion, 
let $\Xi \subset I'' $ be a neighbourhood of $I$. Then 
	\[
		\mathscr{A}( \Xi ) = \mathscr{A}(I'') \; ,  
	\] 
where both algebras are defined via Eq.~\eqref{Additivity}. 
In particular, the algebra of observables located within an 
arbitrary small time--slice coincides  with the algebra of all observables.
\item [$vi.)$] the \emph{de Sitter vacuum state} \index{de Sitter vacuum 
state} $\omega^\circ$ defined in 
\eqref{state-ext} is \emph{invariant} under the action of $O(1,2)$, \emph{i.e.}, 
	\[
		\omega^\circ = \omega^\circ \circ \alpha^\circ_\Lambda  
		\qquad \forall \Lambda \in  O (1,2)\; .
	\]
Moreover, $\omega^\circ$ satisfies the {\em geodesic KMS condition} 
\index{geodesic KMS condition} with respect to $\alpha^\circ$. 
\end{itemize}
\end{theorem}

In addition to these properties, the net of local algebras 
shares the following key property with the 
corresponding net on Minkowski space: 

\begin{theorem} 
\label{th:7.1.8}
The Reeh-Schlieder property \index{Reeh-Schlieder property} holds, \emph{i.e.},
	\[
		\overline{ {\mathscr A}_\circ ({\mathcal O}) \Omega } = \mathcal{H} (dS) \; , 
	\]
if ${\mathcal O}$ contains an open subset. 
\end{theorem}

\begin{proof}
This is a direct consequence of the one-particle Reeh-Schlieder theorem due to Bros and 
Moschella \cite{BM}; see Theorem \ref{oprs}.
\end{proof}

\begin{proposition}
Let $ f_j \in  {\mathfrak h}  (W_1)$, $1\leq j\leq n$ and set
	\[
		G \bigl(t_{1}, \dots, t_{n}; {\rm W}( f_{1}), \dots, {\rm W}( f_{n}) \bigr)
		\doteq  \omega^\circ \Bigl(
		\alpha_{\Lambda_1(t_1)} ({\rm W}( f_{1})) 
		\cdots
		\alpha_{\Lambda_1(t_n)} ({\rm W}( f_{n}))  \Bigr) \; . 
	\]
It follows that the function 
	\[ 
		(t_{1}, \dots, t_{n})\mapsto 
		G\bigl(t_{1}, \dots, t_{n}; {\rm W}( f_{1}), \dots, {\rm W}( f_{n})\bigr) 
	\] 
is holomorphic in the set
	\[
		I_{2 \pi r}^{n+}= \bigl\{(\tau_{1}, \dots, \tau_{n})\in {\mathbb C}^{n} 
			\mid \Im\tau_{i}< \Im\tau_{i+1},  \; \; \Im\tau_{n}- \Im\tau_{1}<2\pi r \bigr\} \; ,
	\]
and continuous on $\overline{I_{2 \pi r}^{n+}}$. The  holomorphic extension is
	\begin{equation}
	\label{r-tau}
		(\tau_{1}, \dots, \tau_{n})\mapsto \prod_{i=1}^{n}{\rm e}^{- \frac{ 1}{2}  \| f_{i} \|_{ \mathfrak{h}(dS)} }
		\prod_{1\leq i< j\leq n} { {\rm e}^{- R_{ \frac{t}{r} } 
		(  f_{i},  f_{j})  }}_{ |_{ t = (\tau_{j}- \tau_{i})   }  }\; .
	\end{equation}
where
	\begin{align*}
		R_{\frac{t}{r}} (   f_i ,  f_j )
			&= \bigl\langle \, f_{i} \, , \, u \bigl( \Lambda_1 (\tfrac{ t }{r}) \bigr) 
			f_{j} \bigr\rangle_{\mathfrak{h}(dS)} 
	\end{align*}
and the symbol $|_{t =  (\tau_{j}-\tau_{i})} $ in \eqref{r-tau} indicates the 
analytic extension of this function. 
\end{proposition}

\begin{proof}
We compute  
	\begin{align*}
		 &G\bigl(t_{1}, \dots, t_{n}; {\rm W}( f_{1}), \dots, {\rm W}(f_{n})\bigr)
			= \omega^\circ \Bigl(\prod_{j=1}^{n} {\rm W} \bigl( u (\Lambda_1( \tfrac{t_j}{r}) f_{j} ) \bigr)  \Bigr)
			\nonumber \\
			&\qquad =  \Bigl( \prod_{1\leq i< j\leq n}{\rm e}^{- i  \Im \langle u (\Lambda_1 (\frac{t_i}{r}) ) f_{i} \, , \,  
			u (\Lambda_1(\frac{t_j}{r})) f_{j}  \rangle_{\mathfrak{h}(dS)} } \Bigr)
			\omega^\circ \Bigl({\rm W} \bigl(\sum_{j=1}^{n} u \bigl(\Lambda_1\bigl(\tfrac{t_j}{r} \bigr)\bigr) f_{j} \bigr) \Bigr) 
			\nonumber \\
			&\qquad = \Bigl(\prod_{1\leq i< j\leq n}{\rm e}^{- i   
			\Im  \langle \,  f_{i} \, , \,  u(\Lambda_1 (\frac{t_{j}-t_{i}}{r} )  f_{j}  \rangle_{\mathfrak{h}(dS)} } \Bigr)
					\; {\rm e}^{-\frac{1}{2} \| \sum_{i=1}^{n} u(\Lambda_1 (\frac{t_{i}}{r} )  f_{i} \|_{ \mathfrak{h}(dS)}} 
					\nonumber \\
			&\qquad = \Bigl( \prod_{1\leq i< j\leq n}
					{\rm e}^{- R_{\frac{t_{j}- t_{i}}{r}} (   f_{i} ,  f_{j})} \Bigr)
					\Bigl(
					\prod_{i=1}^{n}{\rm e}^{-\frac{1}{2} \| f_{i} \|_{ \mathfrak{h}(dS)} }\Bigr)  \; .
	\end{align*}
In the last equality we have used that 
	\begin{align*}
		\Bigl\| \sum_{i=1}^{n} u \bigl(\Lambda_1 \bigl(\tfrac{t_{i}}{r} \bigr)\bigr)
		 f_{i} \Bigr\|_{ \mathfrak{h}(dS)} & =  \sum_{i=1}^{n} \| f_{i} \|_{ \mathfrak{h}(dS)} 
		 \\
		 & \qquad + 2 \sum_{i < j} \Re \Bigl\langle  f_{i} \, , \,  
			u (\Lambda_1(\tfrac{t_j - t_i}{r})) f_{j} \Bigr\rangle_{\mathfrak{h}(dS)} \; . 
	\end{align*}
For $ f_1,  f_2\in {\mathfrak h} (W_1)$ the 
function $t\mapsto R_{t/r} (   f_1,   f_2)$ allows a holomorphic
extension to the strip $\{ \tau \in {\mathbb C} \mid 0 < \Im \tau < 2\pi r\}$.  
\end{proof}

\begin{corollary} 
\label{identification-dS-EdS}
Let $f_i = \delta \otimes h_i$ with $h_i \in {\mathcal D}_{{\mathbb R}}(I_+ )$, $ i= 1, \ldots, n$. 
It follows that 
	\[
		G\bigl( i \theta_{1}, \dots,  i\theta_{n}; {\rm W} ( f_1), \dots, {\rm W} (  f_n ) \bigr)
		=\prod_{1\leq i, j\leq n}{\rm e}^{-\frac{1}{2} C_{|\theta_{i}-\theta_{j}|} (   h_{i},  h_{j}) } \; , 
	\]  
where, for  $0\leq \theta_{1}, \theta_{2}< 2 \pi $,  the covariance  $C_{|\theta_{i}-\theta_{j}|} (   h_{i},  h_{j})$ is defined by
	\begin{align}
		\label{coideq2}
			&C_{|\theta_1 -\theta_2|} (h_1, h_2) \doteq 
 			C \bigl(\delta_{\theta_{1}}\otimes 
			h_{1} , \delta_{\theta_{2}}\otimes h_{2} \bigr) \; . 
	\end{align}
$C ( \, . \, , \, . \, )$ has been defined in \eqref{h-1}, see also Lemma~\ref{coid}. 
\end{corollary}

\begin{proof}
This result follows from Lemma \ref{coid}. Note that the product in \eqref{r-tau} involved only terms with $i<j$. This 
condition was dropped in the expression in Corollary \ref{identification-dS-EdS}, and this is only possible for purely 
imaginary $\tau$'s and real-valued functions $h_i$. 
\end{proof}

\section{The canonical net of local $C^*$-algebras on $S^1$}
\label{sec:7.2}

It is convenient to also consider the Weyl algebra associated to the classical canonical dynamical system: set 
	\[
		\widehat {\mathfrak W} (I) \doteq {\mathfrak W} \bigl( \, \widehat {\mathfrak k} (I), \widehat {\sigma} \bigr) \; , 
		\qquad I \subseteq S^1 \; ,  
	\]
and let $\Lambda \mapsto \widehat {\mathfrak z} (\Lambda)$ be the  action of $O(1,2)$ 
on~$\widehat {\mathfrak k} (S^1 )$; see Proposition~\ref{nocheinlabel}. Define a group of automorphisms 
$\widehat \alpha^\circ \colon \Lambda \mapsto \widehat \alpha_\Lambda^\circ$ acting on 
$\widehat {\mathfrak W} (S^1 )$
by
	\[
		\widehat\alpha_\Lambda^\circ (\widehat {\rm W} (\widehat f ))
		\doteq \widehat {\rm W} \big(\widehat {\mathfrak z} (\Lambda) \widehat f \, \big) \; , 
		\qquad \widehat f \in \widehat {\mathfrak k} (S^1 ) \; , \qquad \Lambda \in O(1,2) \; .  
	\]
Just as in Minkowski space, the localisation properties are less evident in the canonical formulation: 
let $I \subseteq S^1$. Then 
	\[
		\widehat\alpha_\Lambda^\circ \bigl( \, \widehat {\mathfrak W} (I) \bigr) \subset 
		\widehat {\mathfrak W} \bigl( \bigl( \Gamma^{+}(\Lambda I ) \cup \Gamma^{-}(\Lambda I) \bigr) \cap S^1 \bigr) \; . 
	\]
This statement is a direct consequence of Proposition \ref{fsol}.

\goodbreak

\begin{definition} 
\label{def:7.2.1}
The pair $\bigl(\widehat {\mathfrak W} (S^1 ), \widehat \alpha^\circ \bigr)$ is the 
{\em canonical quantum dynamical system} associated to the Klein--Gordon equation on the de Sitter space.
\label{Wcqds-page}
\end{definition}

As $C^*$-algebras, the Weyl algebras $\widehat {\mathfrak W} (S^1)$ and  ${\mathfrak W} (dS)$ are isomorphic, and 
can be identified using the map (see Proposition \ref{nocheinlabel})
	\[	
		\widehat W ( \widehat f ) \mapsto  W([f]) \; , \qquad f \in {\mathcal D}_{\mathbb{R}}(dS) \;  . 
	\]
Moreover, for $f \in {\mathcal D}_{\mathbb{R}}(dS)$ we have  (see Proposition \ref{Prop5.7})
	\[
		{\rm e}^{-\frac{1}{2} \| \widehat K \widehat f \|_{\widehat {\mathfrak h} (S^1) }}
		= {\rm e}^{-\frac{1}{2} \| [f] \|_{ {\mathfrak h} (dS) } } .
	\]
Consequently, the state
	\begin{equation}
	\label{129a}
	\widehat \omega^\circ  \bigl( \widehat {{\rm W}} (\widehat f ) \bigr) 
	\doteq {\rm e}^{ - \frac{1}{2} \| \widehat K \widehat f \|_{\widehat{\mathfrak h} (S^1) } }\; , \qquad f \in {\mathcal D}_{\mathbb{R}}(dS) \; , 
	\end{equation}
describes the {\em same} (we will clarify exactly in which sense) state as the one given in Theorem \ref{th2.5}.

\begin{theorem}
\label{th:7.2.2}
\label{canonicalfreevacuumstatepage}
The state \eqref{129a} is the unique  normalised positive linear functional on 
$\widehat{\mathfrak W} \bigl( S^1 )$, which satisfies the 
following properties:
\begin{itemize}
\item[$ i.)$] $\widehat \omega^\circ$ is invariant under the action of $SO_0(1,2)$, \emph{i.e.}, 
	\[
		\widehat \omega^\circ = \widehat \omega^\circ \circ
		\widehat{\alpha}^\circ_\Lambda  \qquad \forall \Lambda \in  SO_0(1,2)\; ;
	\]
\item [$ ii.)$] $\widehat \omega^\circ$ satisfies the {\em geodesic KMS condition\/}: for every half-circle 
$I_\alpha$  the restricted (partial) state 	
	\[ 
	\widehat \omega_{\upharpoonright \widehat{\mathfrak W}( I_\alpha)}^\circ
	\]
satisfies the KMS-condition at inverse temperature $2 \pi r$ with respect to the one-parameter  group 
$t \mapsto \Lambda^{(\alpha)}(\frac{t}{r})$ of boosts. 
\end{itemize}
\end{theorem}

\begin{proof}
Property $i.)$ follows from the definition; property $ii.)$ follows from 
Corollary~\ref{identification-dS-EdS} and the properties of the time-zero 
covariance.  
\end{proof}

\label{widehat-Fock-space-page}

As in the covariant description, it is convenient to take the weak closure for the 
pair $\bigl( \, \widehat{\mathfrak W} \bigl( S^1 ), \widehat \omega^\circ \bigr)$ in the 
GNS representation $(\widehat{\pi}^\circ, \widehat{\mathcal H} (S^1), \widehat{\Omega}_\circ )$. 
The latter
is (unitarily equivalent to) the  Fock representation over the one-particle 
space~$\widehat {\mathfrak h} (S^1)$,~\emph{i.e.}, 
	\[
 		\widehat{\pi}^\circ \bigl( \, \widehat {\rm W}( \widehat f ) \bigr) 
		= \widehat{\rm W}_F ( \widehat {K}  \widehat f ) \; , \qquad \widehat{\mathcal H} (S^1) 
		= \Gamma \bigl( \, \widehat{\mathfrak h}(S^1) \bigr) \; .
 	\]
Once again, the GNS vacuum vector $\widehat{\Omega}_\circ$ can be used to extend $\widehat \omega$ 
to the weak closure: 
\[
	\widehat \omega (W_F (h)) \doteq \langle \widehat{\Omega}_\circ , \widehat{\rm W}_F 
	(h) \widehat{\Omega}_\circ \rangle 
	= {\rm e}^{-\frac{1}{2} \| h \|_{ \widehat{\mathfrak h} (S^1) } } \; , \qquad h  \in \widehat{\mathfrak h} (S^1) \; .
\]
The local von Neumann algebras for the free canonical field are defined by 
	\begin{equation}
	\label{vN-RI}
			{\mathcal R} (I) \doteq  \pi^\circ \bigl( \,
		\widehat{\mathfrak W} (\widehat {\mathfrak k}(I) )\bigr)'' \; ,  \qquad I \subset S^1 \; , \; \text{open}\; . 
	\end{equation}

The automorphism $\widehat{\alpha}^\circ_\Lambda$ are implemented in the GNS representation by 
unitaries $\widehat{U}_\circ ( \Lambda )$, which satisfy
	\[
		\widehat{U}_\circ ( \Lambda ) {\rm W}_F ( \widehat {K}  \widehat f ) \widehat{\Omega}_\circ 
		= {\rm W}_F ( \widehat {K}  \widehat {\Lambda_* f })
		\widehat{\Omega}_\circ
		\quad 
		\text{and} 
		\quad 
		\widehat{U}_\circ ( \Lambda ) \widehat{\Omega}_\circ = \widehat{\Omega}_\circ \; . 
	\]
By construction, $\widehat{U}_\circ ( \Lambda ) = \Gamma \bigl( \widehat{u} ( \Lambda )\bigr)$. Hence, 
the generators of the boosts are 
\label{KFockhatpage}
\label{LFockhatpage}
	\begin{equation}
	\label{widehat-L}
		\LFock_1^\circ = {\rm d} \Gamma ( \omega \, r\, \mathbb{cos})		
		\qquad 
		\text{and} 
		\qquad 
		\LFock_2^\circ = {\rm d} \Gamma ( \omega \, r \, \mathbb{sin}) \; ,
	\end{equation}
and the generator of the rotations is $\KFock_\circ = {\rm d} \Gamma ( - i \partial_\psi ) $. 
Hence
the \emph{modular objects} \index{modular objects} for the pair $\bigl( \mathcal{R}(I_+), \widehat{\Omega}_\circ \bigr)$ are 
	\begin{equation}
		\label{s1-reflections}
		\Delta^{(0)}_\circ = {\rm  e}^{- \pi  \LFock_1^\circ} 
		\quad \text{and} \quad
		J_\circ^{(0)} = \Gamma (\widehat{u} (P_1T)) \; , 
	\end{equation}
respectively. For the definition of $\widehat{u} (P_1T)$ see \eqref{reflections}.  

\bigskip
The local algebras share a number of interesting properties: 

\label{RI-page}

\goodbreak

\begin{proposition} \quad
\label{l6.1}
\begin{itemize}
\item[$i.)$] The local von Neumann algebras for the canonical free field are regular from the inside and regular from the outside:
	\[
		\bigcap_{ J \supset \overline {I} } 
		{\mathcal R} (J)={\mathcal R}  (I)
		= \bigvee_{ \overline{J}\subset I }   {\mathcal R}(J)\; ; 
	\]
\item[$ii.)$]  The net $I \mapsto {\mathcal R}(I)$ of local von Neumann 
algebras for the canonical free field is {\rm additive}: \index{additivity}
	\[
		{\mathcal R}(I)= \bigvee_{J_i} {\mathcal R}(J_i) \qquad 
		\hbox{ if }  I = \cup_i J_i \;  .  
	\]
Moreover, 
	\[
	{\mathcal R}(S^1) = {\mathcal B} \bigl(\Gamma (\widehat{\mathfrak h} (S^1))\bigr)\; , 
	\qquad {\mathcal R}(S^1)' = {\mathbb C} \cdot \mathbb{1} \; ; 
	\]
\end{itemize}
\end{proposition}

\begin{proof}
The algebra ${\mathcal R} (I) $ is equal to the von Neumann algebra generated 
by $\widehat {\rm W}_{F} (h)$, $h \in \widehat{\mathfrak h}(I)$.
Moreover, according to Proposition~\ref{prop:6.5.2a},
	\begin{equation}
	\bigcap_{J \supset \overline {I} }  \widehat{\mathfrak h} ( J)
	=\bigvee_{\overline {J} \subset I } \widehat{\mathfrak h}( J)
	= \widehat{\mathfrak h} ( I) \; ,
	\end{equation}
which together with Proposition \ref{araki}  implies $i.)$ and~$ii.)$. 
\end{proof}

\begin{remark}
\label{cor2}
A special case of $i.)$ is the following: let $I$ be an open interval contained in a half-circle. 
Then 
	\[
	{\mathcal R} (I) = \bigcap_{I \subset I_\alpha} 
	{\mathcal R} ( I_\alpha ) \; , 
	\]
where the $I_\alpha$'s are the half-circles containing $I$.
\end{remark}

\begin{theorem}
\label{th:7.2.4}
For each open  interval $I \subset S^1$, 
the local observable algebra ${\mathcal R}(I)$ is $*$-isomorphic to the
unique hyper-finite factor of type~{\rm III}$_1$. 
\end{theorem}

\begin{proof}
It has been proved in \cite[Prop.~4.5]{FiGu2} that if $1$ is in the spectrum of the one-particle 
modular operator $\delta_I$ associated to $\widehat{\mathfrak h} (I)$, but not in the 
point spectrum, then the second quantization algebra ${\mathcal R}(I)$
is a type III$_{1}$ factor.  
Hence, the result follows from Proposition~\ref{Figolini-Guido}.
\end{proof}

\begin{theorem}
[Finite speed of propagation]
\label{fst-theorem}
Let $I \subset S^1$ 
be an open interval. Then \index{finite speed of propagation}
	\begin{equation}
		\widehat 
		\alpha^{\circ}_{\Lambda^{(\alpha)} (t)}
 		\colon  
		{\mathcal R} (I)\hookrightarrow 
		{\mathcal R} \bigl( I (\alpha , t)  \bigr) \; .
		\label{e6.1f}
	\end{equation}
(Recall that $I (\alpha , t)$ has been defined in Proposition~\ref{ialpha}.)
\end{theorem}

\begin{proof} 
The statement follows from Proposition~\ref{halpha}. 
\end{proof}

Next, let ${\mathscr A}^{(\alpha)}_{r}(I_\alpha)$  denote the
von Neumann algebra generated by 
	\[
		\left\{ \widehat {\alpha}^{\circ}_{\Lambda^{(\alpha)} (t )} (A) 
		\mid A\in {\mathcal U} (I_\alpha), \; |t |<r \right\} \; , 
	\]
where ${\mathcal U} (I_\alpha)$ denotes the abelian $C^*$-algebra generated by Weyl 
operators $ \widehat{\rm W}_F (h)$, $h \in \widehat {\mathfrak h}(I_\alpha)$ real valued.
Then
\label{loc-nc-algpage}
	\begin{equation}
	\label{A-I-R}
			\bigcap_{r>0} {\mathscr A}^{(\alpha)}_{r}(I_\alpha)= {\mathcal R} (I_\alpha) \;  .
	\end{equation}
(This is a special case of Theorem \ref{p6.1} below). This suggests to identify 
the local non-commutative von Neumann algebra ${\mathcal R} (I)$
with the intersection of the von Neumann algebras ${\mathscr A}^{(\alpha)}_{r}(I)$, $r>0$, generated by 
	\[
		\left\{ \widehat {\alpha}^{\circ}_{\Lambda^{(\alpha)} (t )} (A) 
		\mid A\in {\mathcal U} (I), \; |t |<r \right\}.
	\]
There is however the question, whether this definition depends on $\alpha$. This is not the case, 
as will be shown next.

\begin{theorem} 
\label{p6.1}
{\rm (Time-slice property)}. 
Let ${\mathscr A}^{(\alpha)}_{r}(I)$, $r>0$, be the 
family of von Neumann algebras defined above. Then
	\begin{equation}
		\label{intersectBa2}
		{\mathcal R} (I) = \bigcap_{r>0} {\mathscr A}^{(\alpha)}_{r}(I)
		\qquad I  \subset S^1 \; .
	\end{equation}
In particular,  the r.h.s.~in \eqref{intersectBa2} does not dependent on $\alpha$.
\end{theorem}

\begin{proof} The following argument is similar to the one given in the proof 
of \cite[Theorem 6.5]{GeJII}. We first prove that $\bigcap_{r>0} {\mathscr A}^{(\alpha)}_{r}(I)
\subset {\mathcal R} (I)$. Using ${\mathcal U} (I)\subset {\mathcal R} (I)$ and 
finite speed of propagation  
(Theorem~\ref{fst-theorem}), we see that 
	\[
	 {\mathscr A}^{(\alpha)}_{r}(I)\subset {\mathcal R} \bigl( I (\alpha, r) \bigr)\; 
	 \qquad \forall r > 0 \; . 
	 \]
According to Proposition \ref{l6.1} the von Neumann 
algebras ${\mathcal R} (I)$, $I \subset S^1$, are 
regular from the outside. This implies~$\bigcap_{r>0} 
{\mathscr A}^{(\alpha)}_{r}(I)\subset {\mathcal R} (I)$.

Let us now prove that ${\mathcal R} (I) \subset \bigcap_{r>0} {\mathscr A}^{(\alpha)}_{r}(I)$. 
Using that the local time-zero algebras are regular from the inside (Proposition \ref{l6.1}), it suffices to 
show that for each~$\overline{J}\subset I$ there exists some positive real number $r \ll 1$ such that  
	\begin{equation}
		\label{e6.02}
		{\mathcal R} (J)\subset {\mathscr A}^{(\alpha)}_{r}(I) \; .
	\end{equation}
To this end we fix $I$ and $J$ with $\overline{J}\subset I$ and set
$\delta={\frac{1}{2}}{\rm dist}(J, I^{c})$. We first note that 
	\begin{equation}
		\label{e6.03}
		{\rm e}^{i t \LFock_\circ^{(\alpha)}  }A{\rm e}^{-i t \LFock_\circ^{(\alpha)}  } \in {\mathscr A}^{(\alpha)}_{r}(I) \; , 
		\quad A\in
		{\mathcal U} (J) \; , \quad |t|<  r <\delta \; . 
	\end{equation}
Clearly, the Weyl operators $\widehat{\rm W}_F (h)$, 
$h \in \widehat {\mathfrak h}(S^1)$ real valued, 
belong to~${\mathcal U} (J)$ if ${\rm supp\,} h  \subset J$ and 
hence to ${\mathscr A}^{(\alpha)}_{r}(I)$. 
Now~(\ref{e6.03}) implies 
	\begin{align}
		\label{e6.04}
		\widehat {\alpha}^{\, \circ}_{\Lambda^{(\alpha)} (t)} \bigl( \widehat{\rm W}_F (h) \bigr) 
		&= \widehat{\rm W}_F \bigl({\rm e}^{i t \omega r \, \cos_{\psi +\alpha}   }h \bigr) 
		\in {\mathscr A}^{(\alpha)}_{r}(I)\; , \qquad |t|<r \; . 
	\end{align}
Hence 
	\[
	 \widehat{\rm W}_F \bigl(t^{-1}({\rm e}^{i t \omega r \, \cos_{\psi +\alpha} }h -h) \bigr)
	\in {\mathscr A}^{(\alpha)}_{r}(I)\; , \qquad 
	|t  |<\epsilon\; . 
	\]
Letting $t \to 0$ and using the fact that the map $h\mapsto \widehat{\rm W}_F (h)$ is continuous for the strong
operator topology, we obtain that $\widehat{\rm W}_F (i \omega \, r  \cos_{\psi +\alpha}  h)
\in {\mathscr A}^{(\alpha)}_{r}(I)$. But any vector 
	\[
		h \in 	 \bigl\{ h \in \widehat{\mathfrak h} (S^1)   \mid  
	{\rm supp\,} \left( \Re h \, ,  \, \omega^{-1}\Im h \right) \subset J \times J \bigr\}  
	\]
can be approximated in norm by vectors of the form 
	\[
	h_{1}+ i \omega \, r \cos_{\psi +\alpha} h_{2} \; , 
	\]
with ${\rm supp\,} h_{i}\in J$, $i=1,2$, real and $\cos_{\psi +\alpha} h_{2}\in {\mathscr D}(\omega)$. 
Thus for all $h\in \widehat{\mathfrak h} (J)$ 
the operators $\widehat{\rm W}_F (h)$ belong to ${\mathscr A}^{(\alpha)}_{r}(I)$ and hence 
${\mathcal R} (J) \subset {\mathscr A}^{(\alpha)}_{r}(I)$.  
\end{proof} 

\section{Euclidean fields and the net of  local algebras on $S^2$}
\label{sec:7.3}

In close analogy to the Fock space over $dS$,  
we now introduce a \index{Euclidean Fock space} {\em Euclidean Fock space} $\bbH \doteq \Gamma(\mathbb{H}^{-1} (S^2))$ 
over $\mathbb{H}^{-1} (S^2)$:
\label{fockpage}
	\[
		\bbH  \doteq\oplus_{n= 0}^{\infty}  \;  \mathbb{H}^{-1} (S^2)^{\otimes_s^n} \;  , 
		\qquad  \mathbb{H}^{-1} (S^2)^{\otimes_s^0} \doteq \mathbb{C} \; ,  
	\]
and with $ \mathbb{H}^{-1} (S^2)^{\otimes_s^n}$ the n-fold totally symmetric tensor product 
of~$\mathbb{H}^{-1} (S^2) $ with itself.  
Again, the coherent vectors 
	\[
		\Gamma( h) = \oplus_{n=0}^\infty \frac{1}{\sqrt{n!}} \underbrace{h \otimes_s \cdots \otimes_s h}_{n-{\rm times}}  \; ,
		\qquad 
		h \in 	\mathbb{H}^{-1} (S^2) \; , 	  
	\]
form a total set in $\bbH$. 
The vector $\Omega_\circ \doteq \Gamma(0)$ is called the Fock vacuum. 

\subsection{Weyl operators}
For $h, g \in \mathbb{H}^{-1} (S^2)$, the relations
	\begin{equation}
	\label{weyl-s2}
		\mathbb{V}  (h) \mathbb{V} (g) 
		= {\rm e}^{- i \Im \langle h ,  g \rangle } \mathbb{V} (h+g) \; , \qquad 
		\mathbb{V}  (h) \Omega_\circ ={\rm e}^{-\frac{1}{2} \| h \|^2}  \Gamma( i  h) \; , 
	\end{equation}
define unitary operators, called the {\em Weyl operators} for the sphere.
They satisfy 
	\[ 
		\mathbb{V}^*(h) = \mathbb{V} (-h) 
		\quad \hbox{and} \quad \mathbb{V} (0)= \mathbb{1} \; . 
	\]
The scalar product and the norm in the exponents in \eqref{weyl-s2} refer to the Hilbert 
space~$\mathbb{H}^{-1} (S^2)$.

\subsection{Rotations} As the vectors of the form $\mathbb{V}  (h) \Omega_\circ$, 
$h \in \mathbb{H}^{-1} (S^2)$, are total in the Euclidean Fock space~$\bbH$, 
the push-forward  
\index{push-forward} defines an action of the rotations on $\bbH$: 
	\[
		\mathbb{U}_\circ (R) \mathbb{V} (h) \Omega_\circ = \mathbb{V} (R_* h) \Omega_\circ \; , \qquad R \in SO(3) \; . 
	\]
We will denote the adjoint action of $\mathbb{U}_\circ (R)$ on $\mathcal{B} (\bbH) $ by 
$\mathbb{\alpha}^\circ_R$. Note that, by construction, $\mathbb{U}_\circ (R) = \Gamma (\mathbb{u}_\circ (R))$, 
$R \in SO(3)$; see \eqref{u-kugel}.

\subsection{Von Neumann Algebras} Let ${\tt O}$ be a \emph{compact} subset in $S^2$. We will 
denote the $C^*$-algebra generated by the set\footnote{It would have made sense to define local 
subspaces of $\mathbb{H}^{-1} (S^2)$ appearing in \eqref{square} accordingly. See, for comparison, 
also Definition \ref{local-h-hat}. If one wants define von Neumann algebras associated to \emph{open} 
sets, additional care is necessary, see the definition following \eqref{square}.} 
	\begin{equation}
	\label{eoal}
		\bigl\{ \mathbb{V}  (h) \mid \operatorname{supp} ( \Re h , ( - \Delta_{S^2} + \mu^2)^{-1} \Im h ) 
		\subset {\tt O}  \times {\tt O} \bigr\}  
	\end{equation}
by $\mathcal{E}({\tt O})$. For a distribution $h$, 
$\Re h$ and $\Im h$ are defined by duality. 
Its weak closure will be denoted by  $\mathscr{E}({\tt O})$. This definition is motivated by the 
following fact: if ${\tt O}$ contains an interval $I \subset S^1$, then $\mathscr{E}({\tt O})$ contains 
the algebra~${\mathcal R} (I)$ introduced in \eqref{vN-RI}, \emph{i.e.}, 
	\begin{equation}
	\label{eoal-I}
		I \subset {\tt O} \cap S^1 \qquad \Rightarrow \qquad {\mathcal R} (I) \subset \mathscr{E}({\tt O}) \; . 
	\end{equation} 
As we will see next, the definition \eqref{eoal} gives rise to an interesting structure:

\begin{theorem} [Euclidean Haag--Kastler axioms for the free field] 
\label{th:7.3.1}
\index{Euclidean Haag--Kastler axioms}
The net of local algebras ${\tt O} \mapsto {\mathscr E} ({\tt O})$ satisfies \index{net of local Euclidean algebras}
\begin{itemize}
\item [$i.)$] \emph{isotony}, \emph{i.e.}, \index{isotony}
	\[
		{\tt O}_1 \subset {\tt O}_2  
		\qquad \Rightarrow \qquad  \mathscr{E}({\tt O}_1) \subset \mathscr{E}({\tt O}_2 )  \; ;  
	\]
\item [$ii.)$] \emph{locality}, \emph{i.e.}, \index{locality}
	\[
		{\tt O}_1  \cap {\tt O}_2 = \emptyset 
		\qquad \Rightarrow \qquad \bigl[ \mathscr{E}({\tt O}_1) , \mathscr{E}({\tt O}_2 ) \bigr] = \{ 0 \} \; ,  
	\]
\emph{i.e.}, the von Neumann algebras associated to \emph{disjoint regions} on the sphere \emph{commute}. 
\item [$iii.)$] \emph{covariance}, \emph{i.e.},  the rotations on the sphere, given by \index{covariance}
	\[
		\mathbb{\alpha}^\circ_R \bigl( \mathbb{V} (h) \bigr) \doteq \mathbb{V} (R_*h) \; , 
		\qquad h \in \mathbb{H}^{-1} (S^2) \; ,
		\qquad R \in SO(3) \; ,  
	\]
act \emph{covariantly}, \emph{i.e.}, 
	\[
		\mathbb{\alpha}^\circ_R \bigl( \mathscr{E}({\tt O}) \bigr) \subset \mathscr{E}(R_* {\tt O}) \; . 
	\]
\end{itemize}
\end{theorem}

\begin{proof}
Consider, for ${\tt O}$ compact, the $\mathbb{R}$-linear subspace
	\begin{equation}
	\label{eh-1}
		\bbH^{(1)} ( {\tt O} ) \doteq \bigl\{ h \in \mathbb{H}^{-1} (S^2) \mid \operatorname{supp} 
		\bigl( \Re h , ( - \Delta_{S^2} + \mu^2)^{-1}   \Im h \bigr) \subset {\tt O}  \times {\tt O} \bigr\} \; .  
	\end{equation}
By definition, the map 
	\begin{equation}
	\label{eho}
		{\tt O} \mapsto \bbH^{(1)} ( {\tt O}  ) \; ,  \qquad   {\tt O} \subset S^2 \; , 
	\end{equation}
satisfies
	\[
		\bbH^{(1)} ( {\tt O}_1 ) \subset \bbH^{(1)} ( {\tt O}_2 ) \qquad \text{if} \quad  {\tt O}_1 \subset {\tt O}_2 \; , 
	\]
\emph{i.e.}, the map \eqref{eho} preservers inclusions.  

Next we claim that $\bbH^{(1)} ( {\tt O}_2 )$ is in the symplectic complement of $\bbH^{(1)} ( {\tt O}_1 ) $ if 
${\tt O}_1 \cap {\tt O}_2 = \emptyset $.  This can be seen as follows: for $h \in \bbH^{(1)} ( {\tt O}_1 ) $ and $g \in \bbH^{(1)} ( {\tt O}_2 )$, we have  
	\begin{align*}
		\Im \langle h, g \rangle_{\mathbb{H}^{-1} (S^2)} & =  \langle  \Re h, ( - \Delta_{S^2} + \mu^2)^{-1} \Im g \rangle_{L^2(S^2, \,  {\rm d} \Omega)} 
		\\
		& \qquad \qquad -
		 \langle  ( - \Delta_{S^2} + \mu^2)^{-1} \Im h,  
		 \Re g \rangle_{L^2(S^2, \,  {\rm d} \Omega)} \; .
	\end{align*}
Inspecting the definition \eqref{eh-1}, we conclude that $\Im \langle h, g \rangle_{\mathbb{H}^{-1} (S^2)}=0$. 

Finally, let us inspect how these subspaces behave under the 
rotations, specified by setting, for $f \in  C^\infty (S^2)$, 
	\[
		\underbrace{ ({\rm e}^{-i \theta_1 K_1 } f ) }_{ \doteq \mathbb{u} 
		( R_1(\theta_1))} (x) = f \bigl( R_1^{-1}(\theta_1) x\bigr) \; , 
		\quad 
		\underbrace{ ({\rm e}^{-i \theta_1 K_2 } f ) }_{ \doteq \mathbb{u} ( R_2(\theta_2))} (x) 
		= f \bigl( R_2^{-1} (\theta_2) x \bigr) \; , 
	\]
By definition, this implies that the rotations act \emph{covariantly}, \emph{i.e.}, 
	\[
		\mathbb{u} ( R) \bigl( \bbH^{(1)} ( {\tt O}  ) \bigr)  = \bbH^{(1)} ( R {\tt O}  ) \; , \qquad R \in SO(3) \; . 
	\]
Just as in the proof of Theorem \ref{th:7.1.5}, the net of von Neumann algebras ${\tt O} \mapsto \mathscr{E} ( {\tt O})$ 
inherits isotony, locality and covariance from the net \eqref{eho}.
\end{proof}

\subsection{Euclidean free fields}
For each $f \in \mathbb{H}^{-1} (S^2)$, the map 
$\mathbb{R} \ni \lambda  \mapsto  \mathbb{V} (\lambda f)$, $ \lambda \in \mathbb{R} $,
defines a strongly continuous one-parameter group of unitary operators. Thus
	\[ 
		\mathbb{\Phi}  (f) := \Bigr. - i \frac{{\rm d}}{ {\rm d} \lambda} \mathbb{V} (\lambda f) \Bigl|_{ \lambda=0} \;  , 
	\]
defines a Euclidean field operator, which could have also been defined in terms of 
Euclidean creation and annihilation operators; 
see Section \ref{Fockspace}. For further details on Euclidean Fock spaces 
the reader may consult \cite{JJM}. 

\subsection{Schwinger functions}
The vacuum expectation value\footnote{The relevance of the Euclidean Green's functions was first 
emphasised by Schwinger \cite{Schwinger} (and soon afterwards by Nakano \cite{Nakano}) 
and for this reason, they are also called \emph{Schwinger functions}. \index{Schwinger functions}
The Schwinger functions on the sphere are invariant under the 
action of the  rotation group $SO(3)$.} 
of a Euclidean field operator $\mathbb{\Phi}  (f)$, $f \in \mathbb{H}^{-1}(S^2)$,  is zero, 
and the \index{Euclidean two-point function} Euclidean two-point function coincides with 
the scalar product of the test functions in $\mathbb{H}^{-1}(S^2)$:   
	\[
		 \langle \Omega_\circ   \, , \,  
		 \mathbb{\Phi} (f) \mathbb{\Phi} (g)  \Omega_\circ  \rangle =  
			\langle \overline{ f }, g \rangle_{\mathbb{H}^{-1}(S^2)} \; .
	\] 
More generally,  
	\begin{equation}
		\langle \Omega_\circ  \, , \,  \mathbb{\Phi}(f)^{p} \Omega_\circ  \rangle =
			\left\{
				\begin{array}{l}
					0 \, , \qquad \qquad \qquad \qquad \; \; \,  p\hbox{ odd}\\
					(p-1)!! \,  \| f \|_{\mathbb{H}^{-1} (S^2)}^p  \, ,\quad p\hbox{ even}
				\end{array}
			\right. 
			\; ,
		\label{e1.0}
	\end{equation}
with $n!!= n(n-2)(n-4)\cdots 1$. The existence of Euclidean {\em sharp-time fields}   
\label{stfpage}	
	\begin{equation}
 		\mathbb{\Phi} (\theta, h) = \mathbb{\Phi}(
		\delta  (\, .\, - \theta)\otimes h) \; , \qquad   h \in C^\infty_{\mathbb{R}}( I_+) \, , 
		\label{e1.4}
	\end{equation}
now follows from the fact that $\mathbb{H}^{-1}(S^2)$ contains the distributions (\ref{eqDeltaTensorh}).

\begin{proposition} 
Let $\mathbb{V} (\delta \otimes h_i) = {\rm e}^{i \mathbb{\Phi} (0,h_i)}$, 
with $h_i \in {\mathcal D}_{{\mathbb R}}(I_+)$, $ i= 1, \ldots, n$. 
It follows that 
	\begin{align*}
		& \bigl\langle \Omega_\circ , \mathbb{U}_\circ \bigl(R_1( \tfrac{t_{n}}{r} ) \bigr) 
		\mathbb{V} (\delta \otimes h_n) \mathbb{U}_\circ \bigl(R_1( \tfrac{t_n - t_{n-1}}{r}  ) \bigr) 
		\cdots \mathbb{U}_\circ \bigl(R_1( \tfrac{t_2 -t_{1}}{r}  ) \bigr) 
		\mathbb{V} (\delta \otimes h_1) \Omega_\circ \bigr\rangle
		\\
		& \qquad \qquad \qquad \qquad
		=\prod_{1\leq i, j\leq n}{\rm e}^{-\langle \delta_{\theta_i} \otimes h_i , \delta_{\theta_j} \otimes h_j \rangle_{\mathbb{H}^{-1}(S^2)} } \; , 
	\end{align*}
where $\langle \delta_{\theta_i} \otimes h_i , \delta_{\theta_j} \otimes h_j \rangle_{\mathbb{H}^{-1}(S^2)}$
depends only on $h_i, h_j$ and $|\theta_{i}-\theta_{j}|$; see~\eqref{coid}. 
\end{proposition}

\begin{proof} 
By definition, 
	\begin{align*}
		\langle \Omega_\circ  ,  {\rm e}^{i \mathbb{\Phi} (\theta _n, h_n)}     \cdots  
		{\rm e}^{i \mathbb{\Phi} (\theta _1, h_1)} \Omega_\circ \rangle  
		&=   \langle \Omega_\circ  ,   {\rm e}^{i \mathbb{\Phi}\bigl( \sum_{i=1}^n 
		\delta (\, . \, - \theta_i )\otimes h_i \bigr)} \Omega_\circ \rangle \nonumber \\
		&= 
		\prod_{1\leq i, j\leq n}{\rm e}^{-\langle \delta_{\theta_i} \otimes h_i , \delta_{\theta_j} \otimes h_j \rangle_{\mathbb{H}^{-1}(S^2)} } \; . 
	\end{align*}  
We note that we have assumed that the $h_i$, $i=1, \ldots, n$, are real valued; 
so as in Corollary \ref{identification-dS-EdS} we do \emph{not} insist that $i<j$.
\end{proof}

It was Symanzik \cite{Symanzik-a}\cite{Symanzik-b}\cite{Symanzik-c}, who first realised that 
\index{Schwinger functions}
Schwinger functions have a remarkable positivity property, 
which allows one to define a probability measure (using Minlos' theorem). In the sequel, significant progress was made by 
Nelson \cite{N3}\cite{N4}, who was able to isolate a crucial property of Euclidean fields (the \emph{Markov property}).

\subsection{The Markov property}
The one-particle projections on the Sobolev space $\mathbb{H}^{-1} (S^2)$ give rise to projections on the 
Fock space~$\bbH$: set 
	\begin{equation}
	\label{e-E}
		E_\pm \doteq  \Gamma (e_\pm)  \; , \qquad 
		E_0 \doteq  \Gamma (e_0) \; .		
	\end{equation}
We denote the corresponding closed subspaces $E_\pm \bbH$ and $E_0 \bbH$ of $\bbH$ by $\bbH_\pm$ 
and $\widehat{\mathcal H}(S^1)$, respectively.  Note that 
these subspaces are \emph{neither} orthogonal to each other \emph{nor} does their union  span $\bbH$. 

\begin{theorem}[Dimock] 
\label{pre-martheo}   
The  \emph{Markov property}, \emph{i.e.}, 
	\begin{equation}
	\label{Markov}
		E_\pm E_\mp = E_0  \, , 
	\end{equation}
holds on {\rm $\bbH$}.
\end{theorem}

\begin{proof}
\bigskip
This result is Theorem 1 in \cite{D}; it follows directly  from 
	\[
		\Gamma (e_\pm e_\mp) = \Gamma (e_\pm)\Gamma (e_\mp) \; .
	\]
\end{proof}

\begin{remark}
The Markov property for the sphere is satisfied, iff  for any function of the Euclidean field in 
$S_\pm$, \emph{conditioning} to the fields in $S_\mp$ (\emph{i.e.}, applying the 
projections $E_\mp$) is the same as conditioning to the fields in $\partial S_\pm=S^1$. 
Thus the {\em Markov property} \index{Markov property} implies that the time-zero quantum 
fields acting on the vacuum vector generate the \emph{physical Hilbert space} $E_0 \bbH$.
The latter will be identified with $\widehat{\mathcal{H}} (S^1) 
= \Gamma \bigl(\, \widehat{\mathfrak{h}} (S^1) \bigr)$ in the sequel.
\end{remark}

In order to recover quantum fields on the de Sitter space, one has to somehow undo the analytic continuation of the Green's functions. 
The key property needed to establish a \emph{reconstruction theorem}, \index{reconstruction theorem}
called \emph{reflection positivity} \index{reflection positivity} (sometimes also called Osterwalder--Schrader positivity), 
is a direct consequence\footnote{In many cases, reflection positivity is easier to verify than the Markov 
property which itself is not needed in more general formulations of the reconstruction theorem.}
 of the Mar\-kov property discussed above (see Theorem~\ref{martheo} below). 

\subsection{Reflection positivity}\label{ref-pos-subsec}
The \emph{Euclidean time reflection} \index{Euclidean time reflection} diffeomorphism $T$ (see (\ref{deftimerefl}))
induces a map $T_*$ on $C^\infty (S^2)$, which extends to a \emph{unitary}
operator on~$\mathbb{H}^{-1} (S^2) $, denoted by the same symbol. Set 
	\begin{equation}
	\label{eucl-tr} 
		\mathbb{U}_\circ (T) \doteq \Gamma (T_*) \, . 
	\end{equation}   
For $f \in \mathbb{H}^{-1}(S^2)$, the Euclidean time reflection yields
	\[
		\mathbb{U}_\circ (T)  \mathbb{\Phi} (f) \mathbb{U}_\circ (T)^{-1}  
		= \mathbb{\Phi} (T_* f) \, . 
	\]

\begin{theorem}[Dimock \cite{D}, Theorem 1, p.~247] 
\label{martheo}   
$\mathbb{U}_\circ (T)$ is a {\em unitary} operator on~{\rm $\bbH$}, which satisfies 
\begin{itemize} 
\item[$ i.)$] $\mathbb{U}_\circ (T) E_\pm = E_\mp \mathbb{U}_\circ (T)$; 
\item[$ ii.)$] $\mathbb{U}_\circ (T) E_0 = E_0 \mathbb{U}_\circ (T) = E_0$; 
\item[$ iii.)$]  $\langle \mathbb{U}_\circ (T) \Psi  ,   \Psi    \rangle_{\bbh} 
= \| E_0 \Psi \|^2_{\bbh} \ge 0 $ for {\rm $\Psi \in  \bbH_+$}. 
\end{itemize}
The property $\langle \mathbb{U}_\circ (T) \Psi  ,   \Psi    \rangle_{\bbh}  \ge 0 $, 
{\rm $\Psi \in  \bbH_+$}, is called \emph{reflection positivity}. 
\end{theorem}

\begin{proof} Parts $ i.)$ and $ii.)$ follow directly from the definitions.
The  {\em Markov property} (\ref{Markov}) implies\footnote{The 
present formulation is due to Dimock \cite[Theorem 2, p.~248]{D}.} that, 
for $\Psi \in \bbH_+$,  property $ iii.)$  holds.
\end{proof}

We note that an alternative proof of reflection positivity for Riemannian manifolds with a suitable symmetry 
(the sphere being in the class considered) was given in \cite{GRS}.

\bigskip
In an operator algebraic setting, it is important to realize that reflection positivity is a property of 
an Euclidean state. However, the Euclidean Reeh-Schlieder theorem (Theorem~\ref{E-RS}) implies 
that\footnote{We note that, in general, a test function $h$ of an element $ \mathbb{V}  (h) 
\in \mathscr{E} ( \overline{ S_+ } ) $ may have an imaginary part $\Im h$ which 
is \emph{not} supported in $\overline{ S_+ }$, see~\eqref{eoal}. In contrast,
the imaginary part of the test functions of the elements in 
${\mathscr U}( \overline{S_+}) $ vanishes.}	
	\[
		\overline{ \mathscr{E} ( \overline{ S_+ } )  \Omega } = \bbH \; .
	\]
Thus we will have to work with appropriate subalgebras. 
Let $\mathsf{U}({\tt O})$, ${\tt O} \subset S^2$  compact, denote the 
abelian\footnote{Since the imaginary part of the scalar product in $\mathbb{H}^{-1}(S^2)$ vanishes 
for real-valued functions, these are abelian von Neumann algebras.}  von Neumann algebra 
generated by the Weyl operators $\mathbb{V}(h)$ with $h$ real valued and 
$\operatorname{supp} h \subset {\tt O}$. Denote the weak closure of 
$\mathsf{U}({\tt O})$ by~$\mathscr{U}({\tt O})$. 
Then, by construction,  
	\[
		\bbH	= \overline{ {\mathscr U}(S^2) \Omega_\circ  }   \; , \quad
		\bbH_\pm  \doteq \overline{ {\mathscr U}( \overline{S_\pm}) \Omega_\circ  }   \quad
		\hbox{and} \quad 
		\widehat {\mathcal{H}} (S^1)  = \overline{ {\mathscr U}(S^1) \Omega_\circ  \; . }
	\]
This result allows us to rephrase reflection positivity (as established in 
Theorem~\ref{martheo} $iii.)$) 
as a condition on an Euclidean state: 

\begin{definition}
\label{def:7.3.7}
A normal state $\eta$ over the \emph{abelian} von Neumann algebra $\mathscr{U} ( \overline{ S_+ } )$ 
is called \emph{reflection positive}, if \index{reflection positive states}
	\[
		\eta \bigl( \,  \underbrace{ \mathbb{U}_\circ (T) A^*  
		\mathbb{U}_\circ (T) }_{ = \, \mathbb{\alpha}^\circ_T (A^*) }  A \, \bigr)
		\ge 0 \qquad \forall  A \in \mathscr{U} ( \overline{ S_+ } ) \; . 
	\]
\end{definition}

Using this definition, Theorem \ref{martheo} $iii.)$ can be rephrased as follows: 

\begin{corollary}
\label{cor:7.3.8}
The Fock vacuum vector $\Omega_\circ$ induces a reflection positive state on the 
algebra $\mathscr{U} ( \overline{ S_+ } )$. 
\end{corollary}

\section{The reconstruction of free quantum fields on de Sitter space}

Since we have already established the reconstruction theorem on the one-particle level
(see Remark \ref{rm:4.8.3}), it is sufficient to sketch how it can be formulated in the operator 
algebraic language using second quantisation.

\bigskip
{\bf Reconstruction of the rotations.}
If $A\in \mathscr{U}(S^1)$, then $[ A , E_0] = 0$. Moreover, the rotations $\gamma \mapsto \mathbb{\alpha}^\circ_{R_0(\gamma)}$ 
leave $\mathscr{U}(S^1)$ invariant. Since $\mathscr{U}(S^1)\Omega_\circ$ is dense in $\widehat{\mathcal{H}}(S^1)$,  it follows that
	\begin{equation}
		\label{knullos-2}
		{\rm e}^{i \alpha \KFock_0}   A  \Omega_\circ 
		\doteq  \mathbb{\alpha}^\circ_{R_0(\gamma)} (A) \Omega_\circ \; ,
		\qquad A \in \mathscr{U}(S^1) \;  , 
	\end{equation}
extends to a strongly continuous unitary representation of the rotation group $SO(2)$ 
on the Hilbert space $\widehat{\mathcal{H}}(S^1)$. 
In geographical coordinates, 
	\[
		\KFock_0 = {\rm d} \Gamma \left( -i   \partial_\rho   \right) \, . 
	\]
Note that in \eqref{knullos-2}, we could have\footnote{This statement follows from \eqref{eoal-I}.}
replaced $\mathscr{U}(S^1)$ by any of the local algebras~${\mathcal R} (I)$, where $I$ is an open interval in $S^1$. 
The result would have been the same. 

\bigskip
{\bf Reconstruction of the boosts.}
To reconstruct the boosts, we proceed in several steps:  given 
a (sufficiently small) $R_0$-invariant 
neighbourhood $N$ of the identity $\mathbb{1} \in SO(3) $, 
we define
	\[
		\mathbb{D}_N \doteq \bigl\{ \mathbb{V} (h) \Omega \in \bbH_+ \mid
		 \mathbb{u}(g) h \in \mathbb{H}^{-1}_{\upharpoonright \overline{S_+} } (S^2) 
		 \quad \forall g \in N \bigr\} \;  
	\]
and 
	\[
		\mathbb{D} \doteq E_0 \mathbb{D}_N \;, 
	\]
where $E_0$ is the projection defined in \eqref{e-E}. We note that $\mathbb{D}_N 
= \Gamma(\mathcal{D}_N)$, with $\mathcal{D}_N$ defined in \eqref{D-N}.
We will soon show that $\mathbb{D}$ is \emph{total} in~$\widehat{\mathcal{H}}(S^1)$. 
We next define a homomorphism~$\mathcal{P}_\circ$ from the neighbourhood $N$ of the identity 
$\mathbb{1} \in SO(3) $ to linear operators defined on the dense subspace $\mathscr{D}$:  for $g \in N$, 
	\begin{equation}
	\label{hat-wp}
		\mathcal{P}_\circ ( g) \; E_0 \mathbb{V} (h) \Omega_\circ \doteq E_0 \; 
		\Gamma \bigl( \mathbb{u} (g) \bigr) \mathbb{V} (h) \Omega_\circ 
		\qquad \forall \, \mathbb{V} (h) \Omega_\circ \in \mathbb{D}_N \; . 
	\end{equation}
It can be deduced from Lemma~\ref{lm:4.7.1} that $\mathcal{P}_\circ$
is well-defined. Alternatively, the proof of Lemma~\ref{lm:9.1.1} could be used as well. 
As expected, $\mathcal{P}_\circ$ satisfies the property characterising 
\index{virtual representations} \emph{virtual representations} \cite{FOS}:
 	\begin{equation}
	\label{vr-so3}
		\mathcal{P}_\circ ( \mathbb{\sigma} (g) )^* \psi 
		=\mathcal{P}_\circ (g^{-1}) \psi \qquad \forall \psi \in \mathscr{D} \; ,
	\end{equation}  
where the \index{involution}
involution $\mathbb{\sigma}$ is still given by $\mathbb{\sigma} (g) = T g T$ for $ g \in SO(3) $. 
Moreover, for $R_0(\gamma) \in N$, the rotation defined in $\widehat{\mathcal{H}} (S^1)$ by \eqref{knullos-2} coincides with one 
given by~\eqref{hat-wp}, as
	\[	
		E_0 \mathbb{V} (h) \Omega =  \mathbb{V} (h) \Omega \qquad \text{if} \quad \mathbb{V} (h) \in \mathscr{U}(S^1) \; . 
	\]
We now claim that, due to \eqref{reconstructed-free-boost},
	\[
		\mathcal{P}_\circ \bigl( R_1 (\theta) \bigr)^{**} = 
		{\rm e}^{ -\theta  \,  {\rm d} \Gamma (	 \LS_1 ) } 
		= 
		{\rm e}^{ -\theta  
		\LFock_\circ^{(0)} 
		} \; . 
	\]
This can be seen by inspecting the one-particle computation, and using that 
	\[
		\Gamma \bigl( \mathbb{u} (g) \bigr) \mathbb{V} (h) \Omega_\circ = 
		\mathbb{V} ( \mathbb{u} (g)h) \Omega_\circ \; . 
	\]
In summary, we find:
	\begin{align*}
		\mathcal{P}_\circ \bigl( \mathbb{\sigma}( R_0(\alpha) ) \bigr)^* \psi  &  
		= \mathcal{P}_\circ \bigl( R_0(\alpha)  \bigr)^* \psi  
		=\mathcal{P}_\circ \bigl( R_0(\alpha)^{-1} \bigr) \psi \; , \\
		\mathcal{P}_\circ \bigl( \mathbb{\sigma}(R_1(\theta) ) \bigr)^* \psi  &= 
		\mathcal{P}_\circ \bigl( R_1(-\theta)  \bigr) \psi 
		=\mathcal{P}_\circ \bigl( R_1(\theta)^{-1} \bigr) \psi  \; \; ,  
	\end{align*}
for all $\psi \in \mathscr{D}$.

\bigskip
{\bf Reconstruction of the reflection at the edge of the wedge.}
The modular conjugation $J_{W_1}$ can be reconstructed from the Euclidean as well: set
	\begin{equation}
		\mathcal{P} (P_1) E_0 A \Omega_\circ \doteq 
		E_0\, \mathbb{\alpha}^\circ_{P_1}(A^*) \Omega_\circ \; , 
		\qquad 
		A  \in {\mathscr U}\bigl( \overline{S_+} \bigr) \; . 
		\label{eqJOmE}
	\end{equation}
Here $\mathbb{\alpha}^\circ_{P_1}$ is given 
by the adjoint action of $\mathbb{U}_\circ (P_1)$. 

\begin{lemma}
The anti-linear map $\mathcal{P} (P_1)$ is equal to the modular
conjugation $\widehat{U}_\circ (P_1T)$ 
for the pair $\bigl( \mathcal{R} (I_+), \Omega_\circ\bigr)$. 
\end{lemma}

\begin{proof}
Taking \eqref{A-I-R} into account, it is sufficient to prove that 
for  $A_1,\ldots, A_n \in {\mathscr U}(I_+)$ and $t_1, \ldots , t_n \in \mathbb{R}$, there holds the relation
	\begin{equation} 
		\label{eqJLMod} 
		\widehat{U}_\circ(P_1T)\, \alpha^\circ_{t_1}(A_1) \cdots \alpha^\circ_{t_n}(A_n)\Omega_\circ 
			= {\rm e}^{-\pi \LFock_\circ^{(0)}}\, \alpha^\circ_{t_n}(A_n^*)\cdots 
			\alpha^\circ_{t_1}(A_1^*) \Omega_\circ \; .  
	\end{equation}  
This can be seen following argument first given in~\cite[Thm.~12.1]{KL1}:   
let $\theta_1,\ldots,\theta_n\in[0,\pi]$ with 
$\sum_{k=1}^n\theta_k\leq \pi$. Then    
	\begin{align*}  
		{\rm e}^{-\theta_1 \LFock_\circ^{(0)}}A_1 \cdots {\rm e}^{-\theta_n \LFock_\circ^{(0)}} A_n\,\Omega_\circ 
		&= E_0 \,\mathbb{\alpha}^\circ_{\theta_1}(A_1)\mathbb{\alpha}^\circ_{\theta_1+\theta_2}
			(A_2)\cdots \mathbb{\alpha}^\circ_{\theta_1+\cdots+\theta_n}(A_n)\, \Omega_\circ \, .  
	\end{align*}
We now apply $\widehat{U}_\circ (P_1T)$ using 
	\[
		\widehat{U}_\circ (P_1T) E_0 A  \Omega_\circ  \doteq E_0\, 
		\mathbb{\alpha}^\circ_{P_1}(A^*) \Omega_\circ \; ,
		\qquad A\in \mathscr{U}(\overline{S^+})  \;  ,   
	\]
and the relation   
	\[
		P_1\circ R_1(\theta)= R_1(\pi-\theta)\circ T,
	\] 
as well as the time-reflection invariance  $\mathbb{\alpha}^\circ_T(A_k)=A_k$, and conclude  
	\begin{align*} 
		\widehat{U}_\circ(P_1T)\, & {\rm e}^{-\theta_1 \LFock_\circ^{(0)} }\,  A_1 
		\cdots {\rm e}^{-\theta_n \LFock_\circ^{(0)} } A_n\,\Omega_\circ \\
			& = E_0 \,\mathbb{\alpha}^\circ_{\pi-\theta_1-\cdots-\theta_n}(A_n^*) \cdots
				\mathbb{\alpha}^\circ_{\pi-\theta_1}(A_1^*)\cdots \, \Omega_\circ \\ 
			&= {\rm e}^{-(\pi-\sum_1^n\theta_k) \LFock_\circ^{(0)}}\, A_n^*  
				E_0 \,\mathbb{\alpha}^\circ_{\theta_n}(A_{n-1}^*)\mathbb{\alpha}^\circ_{\theta_n+\theta_{n-1}}(A_{n-2}^*)
				\cdots \mathbb{\alpha}^\circ_{\theta_n+\cdots+\theta_{2}}(A_{1}^*)  \Omega_\circ
				\\  
			& = {\rm e}^{-(\pi-\sum_1^n\theta_k) \LFock_\circ^{(0)}}\, A_n^* {\rm e}^{-\theta_n \LFock_\circ^{(0)}}  
			\cdots A_2^*{\rm e}^{-\theta_2 \LFock_\circ^{(0)} }A_1^* \,\Omega_\circ \, . 
	\end{align*}
By analytic continuation (observe that $\widehat{U}_\circ(P_1T)$ is anti-linear) this implies 
	\begin{align*} 
		\widehat{U}_\circ(P_1T)\, & {\rm e}^{is_1 \LFock_\circ^{(0)} }\, A_1 \cdots 
		{\rm e}^{is_n \LFock_\circ^{(0)} } A_n\,\Omega_\circ \\
			& = 
		{\rm e}^{-\pi \LFock_\circ^{(0)} }\, {\rm e}^{i\sum_{k=1}^n s_k \LFock_\circ^{(0)}}\, 
		A_n^* {\rm e}^{-is_n \LFock_\circ^{(0)} }  
		\cdots A_2^*{\rm e}^{-is_2 \LFock_\circ^{(0)} }A_1^* \,\Omega_\circ \, .   
	\end{align*}
Defining $t_1\doteq s_1$ and $t_k\doteq s_k-s_{k-1}$ for $k=2,\ldots,
n$, we find $\sum_{k=1}^n s_k = t_n$
hence this is just the desired relation~\eqref{eqJLMod}.  
\end{proof}

\begin{lemma}
\label{lm:7.4.2} 
The map $\mathcal{P}_\circ$ defined in \eqref{hat-wp} extends to a virtual representation 
	\[
		R \mapsto\mathcal{P}_\circ (R) 
	\]
of $SO(3)$ in the sense of Fr\"ohlich, Osterwalder and Seiler \cite{FOS}, \emph{i.e.},   
there is a local group homomorphism $\mathcal{P}_\circ$ from $SO(3)$ into linear operators densely defined on $\widehat{\mathcal{H}}(S^1) $, 
with the following properties: 
\begin{itemize}
\item [$ i.)$] the map
	\[
		\alpha \mapsto\mathcal{P}_\circ(R_0 (\alpha) ) 
	\]
is a continuous unitary representation of $SO(2)$ on $\widehat{\mathcal{H}}(S^1)$; 
\item [$ ii.)$]
there exists a neighbourhood $N$ of $\mathbb{1} \in SO(3)$, invariant under the rotations $R_0 (\alpha) $, $\alpha \in [0, 2 \pi)$, 
and a linear subspace $\mathscr{D}$ 
(namely $E_0 \mathbb{D}_N$), 
dense in $\widehat{\mathcal{H}}(S^1)$, 
such that 
\begin{itemize}
\item[---] $\mathscr{D} \subset \mathscr{D}(\mathcal{P}_\circ(g))$ for all $g \in N$;  and 
\item[---] if $g_1, g_2$ and $g_1 \circ g_2$ are all in $N$, then
\label{virtreppage2}
	\begin{equation} 
		\label{49th}
		\mathcal{P}_\circ (g_2) \Psi \in {\mathscr D}(\mathcal{P}_\circ(g_1))\; , \qquad \Psi \in \mathscr{D}\; ,  
	\end{equation}
	and
	\[ 
		\mathcal{P}_\circ(g_1)\mathcal{P}_\circ(g_2) \Psi =\mathcal{P}_\circ(g_1 \circ g_2) \Psi \; , \qquad \Psi \in \mathscr{D}\; ; 
	\]
\end{itemize}
\item[$ iii.)$] 
if $\ell \in {\mathfrak m}$, $0 \le t \le 1$, and
	\[
		{\rm e}^{- t \ell} \in N  \; ,  \qquad 0 \le t \le 1 \; , 
	\]
then $\mathcal{P}_\circ ({\rm e}^{-t \ell })$ is a hermitian operator defined on ${\mathfrak D}$ and 
	\begin{equation}
	\label{52th}
		\operatornamewithlimits{s-lim}_{t \to 0}\mathcal{P}_\circ ({\rm e}^{-t \ell}) \Psi = \Psi \; , \qquad \Psi \in \mathscr{D} \; . 
	\end{equation}
\end{itemize}
\end{lemma}

One may now appeal to Theorem \ref{FOS}, due to Fr\"ohlich, Osterwalder, and Seiler.
But inspecting the explicit formulas provided, it is clear that the virtual representation 
$\mathcal{P}_\circ$ of $SO(3)$ can be analytically continued to the representation of 
$SO(1,2)$ defined in \eqref{U-Lambda}. 

\begin{remark}
Given the local algebras $\mathscr{U} (I)$, $I \subset S^1$, and the unitary representation of $SO(1,2)$, 
Theorem \ref{p6.1} tells us that we will recover the Haag--Kastler net of the free field in its canonical formulation. 
Using the unitary equivalence between the canonical formulation and the covariant  formulation, the Haag 
Kastler net 
	\[
		\mathcal{O} \mapsto \mathscr{A}_\circ (\mathcal{O}) \; , \qquad \mathcal{O} \subset dS \;  , 
	\]
discussed in Theorem \ref{th:7.1.5} is recovered as well. 
\end{remark}

\begin{lemma} 
\label{eqEU(K)Om}
Let $0< \theta < \pi/2$. 
Then the set $\mathscr{U}(K_\theta) \Omega_\circ$, 
	\begin{equation}
	\label{K-theta}
		K_\theta \doteq S_+\cap R_1(-\theta)S_+ \,  ,  \; ,  
	\end{equation}
is a quantization domain\index{quantization domain}~\cite{HJJ}; \emph{i.e.}, the set
	\begin{equation}
		\label{d-theta}
		{\mathscr D}_\theta  \doteq E_0 \mathscr{U} (K_\theta) \Omega_\circ 
	\end{equation}
is dense in $\widehat{\mathcal{H}}(S^1)$. 
\end{lemma}

\begin{remark}
\label{rm:7.4.4} 
In fact, if $K$ is any open set in $S_+$, then the image of 
$\overline{\mathscr{U}(K) \Omega_\circ}$
under $E_0$ is dense in  $\widehat{\mathcal H} (S^1)$.  
\end{remark}

\begin{proof} In order to show that ${\mathscr D}_{\theta}$ is dense in  $\widehat{\mathcal H} (S^1)$, 
it is sufficient to show that 
if  $\Psi  \perp {\mathscr D}_{\theta}$ is a vector in the orthogonal complement of ${\mathscr D}_{\theta}
\subset \widehat{\mathcal H} (S^1)$, then it equals the zero-vector. We have already seen that ${\mathcal U} (S^1) \Omega_\circ$ is 
dense in $\widehat{\mathcal H} (S^1)$. Thus it is sufficient to show that	
	\[
		\langle \Psi , {\rm e}^{i\mathbb{\Phi} (h)} \Omega_\circ \rangle = 0 
		\qquad \forall \, {\rm e}^{i\mathbb{\Phi} (h)}  \in {\mathcal U} (S^1) \;  ,  
	\]
as this would imply that $\Psi $ is the zero-vector. Moreover,  
	\[
	{\mathcal U} (S^1) = \bigvee_{ I \subset S^1} \,  {\mathcal U} ( I)
	\]
with $\cup I$ a covering of $S^1$ in terms of open intervals. 
For the covering we choose sufficiently many circle segments 
	\[
	I_{\alpha, \theta + \epsilon} 
	= \bigl\{ \vec{x} \in I_\alpha \mid {\rm dist} (\vec{x} , \partial I_\alpha ) > \theta +\epsilon \bigr\} \; ,
	\qquad 0 < \epsilon \ll \theta \; , 
	\]
of equal size, consisting of points in the interior of the half-circle $I_\alpha$, 
which are more than $\theta + \epsilon$ away from the end points of the half-circle. 	

Now consider, for $h \in {\mathcal D}_{\mathbb R} (I_{\alpha, \theta + \epsilon} )$ fixed,  the analytic function
	\begin{equation}
		\label{f24} 
		z \mapsto 
		\langle \Psi , {\rm e}^{-z \LFock_\circ^{(\alpha)}  } {\rm e}^{i\mathbb{\Phi} (h)} \Omega_\circ \rangle  \;  , 
		\qquad \{ z \in {\mathbb C} \mid 0 < \Re z  < \pi\} \;  . 	
		\end{equation} 
By construction there exists an open interval $J$ (whose size depends on $\epsilon$) such that (see \eqref{K-theta})
	\[
		R^{(\alpha)} (\theta') I_{\alpha, \theta + \epsilon} \subset \; 
		\label{K-theta}
		\bigcap_{ \alpha' 
		\in [ 0, 2 \pi) } R_0 (\alpha') K_\theta \; ,
	\qquad 
	\theta' \in J \; , 
	\]
and consequently
	\[
	 {\rm e}^{-\Re z  \LFock_\circ^{(\alpha)} } {\rm e}^{i\mathbb{\Phi} (h)} 
	 \Omega_\circ \in {\mathscr D}_{\theta} \; ,
	 \qquad 
	\Re z  \in J \; , 
	\] 
and, since $\Psi  \perp {\mathscr D}_{\theta}$,
the analytic function \eqref{f24} vanishes on an open line segment in 
the interior of its domain, and is therefore identical zero. 
Since it is continuous on its boundary, the lemma follows. 
\end{proof}

\part{Interacting Quantum Fields}

\chapter{The Interacting Vacuum}
\label{ch:8}

In case one would like to construct an interacting quantum field theory, one usually has 
to face two problems: infrared and ultraviolet problems. Physical\footnote{It is useful to 
distinguish between infrared problems which are related to observable effects, and infrared 
problems which may arise by choosing a particular approximation or coordinate system. 
While the latter may very well appear on de Sitter space, we claim that the former are absent.}
infrared problems stem from infinite volume effects and they should not arise on de Sitter 
space. The ultraviolet problems are related to the short distance behaviour of the theory, 
and these short distance properties are not effected by a constant 
background curvature. Hence they are the same as in Minkowski space. 

In this section, we will study polynomial interactions. These are unbounded, but due to 
the properties of the covariance $C$, they are elements of certain $L^p$-spaces.
The short distance behaviour of the covariance $C$ has been studied by 
De Angelis, de Falco and Di Genova in \cite{AFG}. In the following section, we briefly 
present their findings. 

\section{Short-distance properties of the covariance}
\label{sec:10.2}

The covariance $C$ introduced in \eqref{h-1} 
can be expressed \cite{wald} in terms of the well-known  
heat kernel $p(t, \vec{x}, \vec{y})$ for diffusion constant $1/ \mu$ on the sphere~$S^2$ : 
	\[
		C (\vec{x}, \vec{y}) = \int_0^\infty {\rm d}t \; {\rm e}^{-t \mu^2} p(t, \vec{x}, \vec{y}) \; . 
	\]
This integral representation allows one to introduce the 
\emph{multi-scale decomposition} (see, \emph{e.g.}, \cite{Benfatto}) of the covariance 
$C(\vec{x},\vec{y})$, which may be written as 
	\begin{equation}
	\label{cov-without-cutoff}
		C(\vec{x},\vec{y})= \sum_{l=0}^\infty C_l (\vec{x},\vec{y}) \; , 
	\end{equation}
where, for some fixed constant $\gamma>1$,
	\[
		C_l (\vec{x},\vec{y}) = \int_0^\infty {\rm d}t \; \left( {\rm e}^{-t \gamma^{2l} \mu^2} 
		- {\rm e}^{-t \gamma^{2l+2} \mu^2}  \right) p(t, \vec{x}, \vec{y})
	\]
is the kernel of the operator
	\[
		C_l  = \frac{1}{- \Delta_{S^2} +  \mu^{2}\gamma^{2l}  } 
		- \frac{1}{- \Delta_{S^2} +  \mu^{2}\gamma^{2l+2} } \; . 
	\]
An approximation of the covariance \eqref{cov-without-cutoff} with tame ultraviolet behaviour can be defined 
by adding up  only finitely many terms:
following \cite{AFG}, we introduce the regularized covariance  
	\begin{equation}
	\label{cov-with-cutoff}
		C^{(k)} (\vec{x},\vec{y})= \sum_{l=0}^{\log_\gamma k - 1} C_l (\vec{x},\vec{y}) \; , 
	\end{equation}
where, of course, $k$ is such that $\log_\gamma k = n$ ranges\footnote{Obviously, $\log_\gamma k = n$ 
is equivalent to $\gamma^n =k$, and both $\gamma$ and $k$ can be chosen to be integers. }
on the positive integers. The map $C^{(k)}$ represents the covariance of the field with length 
cutoff $\gamma (\mu k)^{-1}$, the analog, 
in the flat case, of a momentum cutoff of order $ \mu k$. In this sense, $C^{(k)}$~compares 
with the~$\delta_k C \delta_k$ in the equations 
(LR1), (LR2), and (LR3) of \cite[p.~160--161]{GJ}.

\goodbreak
The ultraviolet behaviour of  \eqref{cov-without-cutoff}  can now be studied by controlling the properties of 
the cut-off covariances \eqref{cov-with-cutoff} as $k \to \infty$. 

\begin{theorem}[De Angelis, de Falco and Di Genova \cite{AFG}]
\label{c-log}
Let $1 \le q < \infty$. With the notation introduced above,  we have   
	\begin{itemize}
	\item [$i.)$]  $\sup_{\vec{x} \in S^2} \| C( \vec{x}, \, . \, ) \|_{L^q(S^2, {\rm d} \Omega)}   < + \infty \; $; 
	\item [$ii.)$]  $\| C^{(k)} ( \, . \, , \, . \, ) - C ( \, . \, , \, . \, ) \|_{L^q(S^2 \times S^2, {\rm d} \Omega \otimes {\rm d} \Omega )} 
	=  O(k^{-2/q} ) \; $;
	\item [$iiii.)$] $ \sup_{\vec{x} \in S^2}  C^{(k)} ( \vec{x}, \vec{x} ) 
	=  O( \log_\gamma k) \; $ for $k \to \infty$.  
	\end{itemize}
\end{theorem}

\begin{remark}
The logarithmic nature of the singularity of the covariance
$C(\vec{x}, \vec{y})$  as $\vec{x}$ approaches $\vec{y}$, 
	\[
		C(\vec{x},\vec{y}) \sim \frac{1}{2\pi} \log  \mu d(\vec{x},\vec{y}) \; , 
	\]
follows from the asymptotic behaviour of the heat kernel \cite{Bellaiche}:
let $d(\vec{x},\vec{y}) $ be the geodesic distance between $\vec{x}$ and $\vec{y}$. Then
	\[
		p(t, \vec{x},\vec{y}) \sim \frac{{\rm e}^{- \frac{(d(\vec{x},\vec{y}))^2}{4t}}} {4 \pi t} 
		\qquad \text{as} \quad t \downarrow 0 \; ,  
	\]
uniformly on all compact sets in $S^2 \times S^2$,  which do not intersect the  cut locus of~$S^2$. 
\end{remark}

\section{(Non-)Commutative $L^p$-spaces}
\label{sec:8.2}

In this work, we emphasize that Euclidean quantum field theories may be formulated 
on a Hilbert space.  $L^p$-estimates are an important ingredient in this approach.
Although we will only discuss polynomial 
interactions affiliated to the abelian algebra $\mathscr{U}(S^1)$, we find it worth while to briefly review the general setting. 

Among the many approaches to non-commutative $L^p$-spaces, 
Araki and Masuda's  approach~\cite{AM} is best suited for our purposes, so we now
present their definitions. Consider a general $(\sigma$-finite) von Neumann algebra 
$\mathcal{M}$ in \emph{standard form} acting on some Hilbert space $\mathcal{H}$ 
with cyclic and separating vector $\Omega_\circ$. For $2 \le p \le \infty$, we define 
	\[
		L^p (\mathcal{M}, \Omega_\circ) 
		\doteq \bigl\{ \Psi \in \bigcap_{\Omega \in \mathcal{H}} \mathscr{D} 	
		\bigl( \Delta_{\Omega, \Omega_\circ}^{\frac{1}{2} - \frac{1}{p}} 
		\bigr)  \mid  \| \Psi \|_p < \infty \bigr\},
	\]
where
	\begin{equation}
	\label{AM-1.4}
		\| \Psi \|_p = \sup_{ \| \Omega \|= 1} \| 
		\Delta_{\Omega, \Omega_\circ}^{\frac{1}{2} - \frac{1}{p}} \Psi \| \; .  
	\end{equation}
In particular, $L^2 (\mathcal{M}, \Omega_\circ) = \mathcal{H}$ 
and $L^\infty (\mathcal{M}, \Omega_\circ) = \mathcal{M}  \Omega_\circ $. 

For $1\le p <2$,  $L^p (\mathcal{M}, \Omega_\circ)$ is 
defined as the \emph{completion} of $\mathcal{H}$ with respect to the norm
	\begin{equation}
	\label{AM-1.5}
		\|\Psi\|_p=
		\inf \bigl\{ \|\Delta_{\Omega,\Omega_\circ}^{\frac{1}{2}-\frac{1}{p}} \Psi \|  \mid
		\|\Omega\|=1, s^{\mathcal{M}}(\Omega)\ge s^{\mathcal{M}}(\Psi) \bigr\} .
	\end{equation}
Here $s^{\mathcal{M}}(\Omega)$ denotes the smallest projection 
in $\mathcal{M}$ which leaves $\Omega$ invariant and
 $\| \Delta_{\Omega,\Omega_\circ}^{\frac{1}{2}-\frac{1}{p}} \Psi \|$
is defined to be $+\infty$, if $\Psi$ is not in the domain of 
$\Delta_{\Omega,\Omega_\circ}^{\frac{1}{2}-\frac{1}{p}}$. 
We note that any $\Psi \in \mathcal{H}$ is in 
$\mathscr{D} \bigl( \Delta_{ \Psi, \Omega_\circ}^{\frac{1}{2}-\frac{1}{p}} \bigr)$
if $ 1 \le p \le 2$ \cite[Lemma 7.1]{AM}. 

\begin{lemma} (Araki \& Masuda \cite[p.~340]{AM}). 
For $a \in \mathcal{M}$ and 
$\Psi \in L^p (\mathcal{M}, \Omega_\circ) \cap \mathcal{H}$, 
	\[
		a \Psi \in L^p (\mathcal{M}, \Omega_\circ)
	\quad 
	\text{and} 
	\quad	
		\| a \Psi \|_p \le \| a \| \| \Psi \|_p  \, . 
	\]
Therefore the multiplication of $a \in \mathcal{M}$ can be defined for any 
$\Psi \in L^p (\mathcal{M}, \Omega_\circ)$ by continuous extension.
\end{lemma}

\subsection{$L^p$-spaces for abelian von Neumann algebras}
\label{sec:8.2.1}
Next, we specialise these definitions to the case that the von Neumann algebra is abelian.  
Let $K$ be the spectrum of the unital abelian $C^*$-algebra ${\mathcal U}$
with a faithful state $\omega_\circ $. 
Then $K$ is a (weak$^*$) compact Hausdorff space
and  
	\[
		C(K) \cong {\mathcal U} \, . 
	\]
The GNS vector $\Omega_\circ \cong 1$ gives rise to a 
probability measure $\mathrm{d} \nu$ over~$K$,
	\begin{equation}
	\label{U-K}
		 L^\infty (K, \mathrm{d} \nu) \cong {\mathscr U} \doteq \pi_{\omega_\circ} 
			( \mathsf{U})''\, , 
		  \qquad \text{and} \qquad L^2 (K, \mathrm{d} \nu) 
		 \cong \overline{ {\mathscr U} \Omega_\circ } \doteq \mathcal{H} \; . 
	\end{equation} 
The restriction of a normal state $\omega$ on $\mathscr{B}(\mathcal{H})$ to ${\mathscr U}$
gives rise to a positive functional~$\omega_{ \scriptscriptstyle \upharpoonright {\mathscr U} }$, 
represented by a \emph{unique} positive function $\Omega (k)$ in~$L^1 (K , \mathrm{d} \nu)$. 

The \emph{relative} modular operator $S_{\Omega, \Omega_\circ}$
is given in its spectral representation: 
	\[
		S_{\Omega, \Omega_\circ} (k) 
		a(k) \underbrace{ \Omega_\circ (k) }_{= 1} 
		= \overline{a(k)} \Omega(k) \, , \qquad k \in  K \, , 
		\quad a \in L^\infty (K , \mathrm{d} \nu) \, . 
	\]
It has the polar decomposition $S_{\Omega, \Omega_\circ} = J_{\Omega, \Omega_\circ}
\Delta^{1/2}_{\Omega, \Omega_\circ}$, where 
	\[
		\bigl( \Delta_{\Omega, \Omega_\circ}^{1/2} a \Omega_\circ \bigr) (k) 
		=a(k) \Omega(k) 
	\]
and
	\[	
		\bigl( J_{\Omega, \Omega_\circ}   a  \Omega_\circ \bigr) (k) 
		= \chi_{\operatorname{supp} \Omega}(k)
		\frac{\Omega(k)}{ | \Omega(k) | } \, \overline{a(k)}  \, .  
	\]
In particular, if $\Omega(k) \ge 0$, \emph{i.e.}, $\Omega \in 
\mathscr{P}^\natural ( \mathscr{U}, \Omega_\circ)$, then $J_{\Omega, \Omega_\circ}= J$
is complex conjugation and $\Delta_{\Omega, \Omega_\circ}^{1/2}$ is given as a 
multiplication operator by the function $\Omega (k)$, \emph{i.e.}, 
	\[
		(\Delta_{\Omega, \Omega_\circ}^{1/2} \Omega_\circ ) (k)
		= \Omega (k) \underbrace{ \Omega_\circ (k) }_{= 1} \, , 
		\qquad 
		k \in K \,, 
		\qquad 
		\Omega \in 
		\mathscr{P}^\natural_{\Omega_\circ} \, . 
	\]	
Since $\Omega (k) \in L^2 (K, \mathrm{d} \nu)$, we 
have $| \Omega(k)|^{\frac{p-2}{p}} \in
L^{\frac{2p}{p-2}} (K, \mathrm{d} \nu)$ and therefore
	\[
		\int_K {\rm d} \nu (k) \; | \Omega (k) |^{\, 1 - \frac{2}{p}}  \;  
		| \Psi (k) |^2 < \infty
	\]
for all $ \Psi (k) \in L^q (K, \mathrm{d} \nu)$ with 
	\[
		\tfrac{2}{q}  = 1 - \left(1 - \tfrac{2}{p} \right) \, , 
		\quad
		\text{i.e.} \, , 
		\quad 
		q = p \, ,
	\]
using the H\"older inequality (for commutative $L^p$spaces).

\goodbreak
Now recall that for $2 \le p \le \infty$, we have\footnote{Here $| \Omega|$ denotes the multiplication 
operator associated to the function $k \mapsto | \Omega (k) |$.} 
	\[
		L^p (\mathscr{U}, \Omega_\circ) 
		\doteq \bigl\{ \Psi \in \bigcap_{\Omega \in \mathcal{H}} \mathscr{D} 	
		\bigl(  | \Omega|^{1 - \frac{2}{p}} 
		\bigr)  \mid  \| \Psi \|_p < \infty \bigr\},
	\]
where
	\begin{equation}
	\label{AM-1.4}
		\| \Psi \|_p =  \sup_{ \| \Omega \|_{L^2  (K, \mathrm{d} \nu)} = 1}  
		\; \; \int_K {\rm d} \nu (k) \; \;  
		|\Omega(k)|^{1 - \frac{2}{p}} \; | \Psi (k) |^2 \; .  
	\end{equation}
For $1\le p <2$,  $L^p (\mathcal{M}, \Omega_\circ)$ is  the \emph{completion} 
of $\mathcal{H}$ with respect to the norm
	\begin{equation}
	\label{AM-1.5}
		\|\Psi\|_p 
		=  
		\inf_{ \| \Omega \|_{L^2  (K, \mathrm{d} \nu)} = 1
		\atop \operatorname{supp} 
		\Omega \supseteq \operatorname{supp} \Psi}  
		\; \; \int_K {\rm d} \nu (k) \; \;  
		|\Omega(k)|^{1 - \frac{2}{p}} \; | \Psi (k) |^2 \; . 
	\end{equation}
Note that the smallest projection in $\mathscr{U}$, which leaves $\Omega$
invariant is the characteristic function $\chi_{\operatorname{supp} \Omega}$
for the support $\operatorname{supp} \Omega$ of the function
$k \mapsto \Omega (k)$, \emph{i.e.}, 
	\[
		s^{\mathcal{M}}(\Omega) = \chi_{\operatorname{supp} \Omega}
		\; . 
	\]  
It follows that the essentially bounded multiplication operators acting on 
$L^2 (K, \mathrm{d} \nu)$ are identified with elements in 
	\[
		{\mathscr U}  \cong L^\infty ( {\mathscr U} , \Omega_\circ)\, , 
	\]
whereas the operators of multiplication with functions in $L^p (K, \mathrm{d} \nu)$, 
$p=2, 3, \ldots$,  represent operators which, when applied to 
$\Omega_\circ \cong 1$ yield an element in 
	\[
		L^p( {\mathscr U} , \Omega_\circ) \cong L^p (K, \mathrm{d} \nu) \, . 
	\]
In particular, we will denote the spaces introduced in \eqref{U-K} by $L^{\infty}( {\mathscr U} , 
\Omega_\circ)$ and $L^{2}( {\mathscr U} , \Omega_\circ)$, respectively. 
Hence, if $A$ is an unbounded operator affiliated to the abelian von Neumann 
algebra $\mathscr{U}$ then 
	\[
		A \in L^p( {\mathscr U} , \Omega_\circ) \Longleftrightarrow 
		\langle \Omega_\circ,  | A |^p \Omega_\circ \rangle < \infty \; , 
	\]
as can be seen by expecting the definitions above.

\subsection{$L^p$ estimates for Eucildean fields}
\label{sec:8.2.2}
We will now indicate, how the general theory can be applied to the Euclidean field on the sphere. 
The Euclidean time-reflection $\mathbb{U}_\circ (T)$ introduced in \eqref{eucl-tr} induces an 
automorphism of ${\mathscr U}(S^2)$, which extends to an isometry of 
$L^{p}({\mathscr U}(S^2), \Omega_\circ  )$, $1\leq p<\infty$.  

It follows from (\ref{e1.0}) that 
	\[
		\mathbb{\Phi}(f) \in \bigcap_{1\leq
				p<\infty}L^{p}( \mathscr{U}(S^2) , \Omega_\circ  )
	\] 
and ${\rm e}^{\mathbb{\Phi}(f)}
\in L^{1}( \mathscr{U}(S^2), \Omega_\circ  ) $ 
if $ f\in   C^\infty  (S^2) $. The map
	\begin{equation}
		\label{e1.6bb}
			\begin{matrix}
				& \mathbb{H}^{-1} (S^2)  & \to  & \bigcap_{1\leq
				p<\infty}L^{p}( \mathscr{U}(S^2) , \Omega_\circ  ) \\
				& f & \mapsto & \mathbb{\Phi}(f) \\
			\end{matrix}  \,  
	\end{equation}
is continuous for each $p$ 
and the cylindrical functions $F \bigl(\mathbb{\Phi}(f_{1}), 
\dots, \mathbb{\Phi}(f_{n}) \bigr)$, 
$f_{i}\in \mathbb{H}^{-1} (S^2)$, $F$ a Borel function 
on~${\mathbb R}^{n}$, $n\in
{\mathbb N}$, are dense\footnote{Recall that the collection of 
all cylindrical functions is a closed 
sub-algebra in the algebra of continuous functions which 
contains the identity and separates the 
points; thus by the Stone-Weierstrass Theorem the cylindrical 
functions are dense in the continuous functions; the latter are 
dense in $L^2$.}
in $L^{p}(\mathscr{U}, \Omega_\circ )$ for  $1\leq p<\infty$. 

In fact, it follows from \eqref{e1.0} that the  {\em sharp-time fields} 
	\begin{equation}
 		\mathbb{\Phi} (\theta, h) = \mathbb{\Phi}(
		\delta  (\, .\, - \theta)\otimes h) \; , \qquad   h \in C^\infty_{\mathbb{R}}( I_+) \, , 
		\label{e1.4}
	\end{equation}
exist as elements
of $L^{p}({\mathscr U}(S^2),  \Omega_\circ )$, $1 \le p < \infty$.
To simplify the notation, we will sometimes write $\varphi (h)$ for the \emph{time-zero fields} $\mathbb{\Phi} (0, h)$.

\section{The Euclidean interaction}
\label{sec2.2}
{\em Normal ordering} with respect to the covariance 
$C( f, g) = \langle \overline{f} , g \rangle_{\mathbb{H}^{-1}(S^2)}$ 
is defined by 
\label{wickpage}
	\begin{equation}
		\label{wick}
			{:} \, \mathbb{\Phi}(f)^{n}\, {:}  
				\doteq \sum_{m=0}^{[n/2]}\frac{n!}{m!(n-2m)!}
					\mathbb{\Phi}(f)^{n-2m} \Bigl(-\tfrac{1}{2}  \| f \|^2_{\mathbb{H}^{-1}(S^2)}  \Bigr)^{m} \; , 
					\quad f \in  \mathbb{H}^{-1} (S^2) \; , 
	\end{equation}
where $[ \, . \, ]$ denotes the integer part. Normal  ordering according to \eqref{wick} coincides 
with Wick ordering with respect to the Fock vacuum on $\bbH$, and therefore 
	\begin{equation}
	\label{nnord}
		{:} \, \mathbb{\Phi}(f)^{n} \, {:} = \frac{1}{2^{n/2}} 
							\sum_{j=0}^n \begin{pmatrix} 
								n \\ 
								j
							\end{pmatrix} 
						a^*(f)^j a(f)^{n-j} \; , \qquad f \in  \mathbb{H}^{-1} (S^2) \; . 
	\end{equation}

Normal-ordering of point-like fields in $\bbH$ is  ill-defined (\emph{i.e.}, one cannot replace $f\in
\mathbb{H}^{-1} (S^2)$ in (\ref{wick}) by a two-dimensional Dirac $\delta$-function), 
but integrals over normal-ordered point-like fields \emph{can} be defined. In order to prove this, 
we define an approximation of the two-dimensional $\delta$-function. Working in the path-space 
coordinates introduced in \eqref{path-space-coo}, let, for~$m \in {\mathbb N}$,
	\begin{equation}  
	 \label{delta-m}
	 \delta^{(2)}_{m}(\, .-\theta, \, .- \psi)
	 =	\frac{1}{r^2}
	 \sum_{\ell = 0}^m \sum_{k= - \ell}^{\ell} \overline{Y_{\ell, k} (\theta, \psi)} 
	 Y_{\ell, k} (\, . \, , \, . \, ) \;   , 
	 \qquad \vec{x} \equiv \vec{x} (\theta, \psi) \in S^2 \; , 		
	 \end{equation}
if $\vec{x} \ne (0, \pm r , 0)$. The sequence $ \bigl\{ \delta^{(2)}_m \bigr\}_{m \in {\mathbb N}}$ 
approximates the  two-dimensional  Dirac $\delta$-function as $m \to \infty$. Note that $\delta^{(2)}$ 
is  supported at the point $(0,0,r) \in S^2$  and $\int_{S^2} {\rm d} \Omega \, \delta^{(2)}  = 1$.

\begin{theorem}[Ultraviolet renormalization]
\label{uvtheo}
For $n \in {\mathbb N}$ and $f\in L^{2}(S^2, {\rm d} \Omega)$
the following  limit exists  in $\kern -.2cm \bigcap \limits_{1\leq p<\infty} \kern -.2cm 
L^{p}( \mathscr{U}(S^2) , \Omega_\circ )$:
	\[
		\lim_{m \to \infty}\int_0^{2 \pi} r \, {\rm d} \theta \int_{-\pi/2}^{\pi/2} r \cos \psi \, 
		{\rm d} \psi \;  f(\theta,\psi) \; 
		{:}\, \mathbb{\Phi} \bigl(  \delta^{(2)}_{m}(\, .-\theta, \, .- \psi)  \bigr)^n \, {:}  \; .
	\]
It is denoted by $\int_{S^2} {\rm d} \Omega \; f(\theta,\psi)  \; {:} \, \mathbb{\Phi}( \theta,\psi)^{n} \, {:} \; $. 
\end{theorem}

\begin{proof} 
Applying \eqref{nnord}, we find 
	\begin{align}
		{:} \, \mathbb{\Phi}(\delta^{(2)}_{m})^{n} \, {:} & = \frac{1}{ 2^{\frac{n}{2}} r^{2n} } 
							\sum_{j=0}^n \begin{pmatrix} 
								n \\ 
								j
							\end{pmatrix} 
	 \sum_{\ell_1 = 0}^m \sum_{k_1= - \ell_1}^{\ell_1} 
	 \cdots
	 \sum_{\ell_n = 0}^m \sum_{k_n= - \ell_n}^{\ell_n} \nonumber \\
	 & \qquad  \overline{Y_{\ell_1, k_1} (\theta, \psi) } \cdots \overline{Y_{\ell_j, k_j} (\theta, \psi)} 
	Y_{\ell_{j+1}, k_{j+1}} (\theta, \psi) \cdots Y_{\ell_{n}, k_{n}} (\theta, \psi) \nonumber \\
	& \qquad 	\quad \times	a^*(Y_{\ell_1, k_1}) \ldots a^*(Y_{\ell_j, k_j}) 
						a(Y_{\ell_{j+1}, k_{j+1}} ) \ldots a(Y_{\ell_{n}, k_{n}}) \; . 
	\label{sum-m-1}
	\end{align}
Using
	$
		\overline{ Y_{\ell, k} (\theta, \psi)} = (-1)^k Y_{\ell, - k} (\theta, \psi) 	$,
we find that (see, \emph{e.g.}, \cite{SHK} or \cite[Sect.~6]{DG} for a recent survey) 
	\[
		P^{(n)}_m (f) \doteq \int_0^{2 \pi} r \, {\rm d} \theta \int_{-\pi/2}^{\pi/2} r \cos \psi \, {\rm d} \psi \;  f(\theta,\psi) \; 
		{:} \, \mathbb{\Phi} \bigl(  \delta^{(2)}_{m}(\, .-\theta, \, .- \psi)   \bigr)^n \, {:}   
	\]
is a linear combination of Wick monomials of the form 
	\begin{align}
		\sum_{j=0}^{n}  \begin{pmatrix} 
								n \\ 
								j
							\end{pmatrix}  & \sum_{\ell_1 = 0}^m \sum_{k_1= - \ell_1}^{\ell_1} 
	 \cdots
	 \sum_{\ell_n = 0}^m \sum_{k_n= - \ell_n}^{\ell_n}\;  (-1)^{\sum_{i=1}^j k_i} 
		w^{(n)} (\ell_{1}, k_1, 
		\ldots, \ell_{n}, k_{n}) 
		\nonumber \\
		& \qquad  \qquad  \qquad \qquad  \qquad  \qquad
		\qquad   \times		a^{*}_{\ell_1, k_1} \cdots a^{*}_{\ell_j, k_j}
				a_{\ell_{j+1}, - k_{j+1}} \cdots a_{\ell_{n}, - k_{n}} \; , 
	\label{sum-m-2a-a}
	\end{align}
where $a^{(*)}_{\ell_i, k_i} \equiv a^{(*)} (Y_{\ell_i, k_i})$ and 
	\begin{align*}
		w^{(n)} (\ell_{1}, & k_1, \ldots,   \ell_{n}, k_{n}) \\
		& = \frac{1}{ 2^{\frac{n}{2}} r^{2n} } 
		\int_0^{2 \pi} r \, {\rm d} \theta \int_{-\pi/2}^{\pi/2} r \cos \psi \, {\rm d} \psi \;  f(\theta,\psi) \; \prod_{i=1}^n
			 \overline{Y_{\ell_i, k_i} (\theta, \psi)}   \; .
	\end{align*}
Next we apply the Wick monomials \eqref{sum-m-2a-a} 
to the Fock vacuum $\Omega_\circ  \in \bbH$; 
see also Definition \ref{sobolev-S2}. Only the term with $j=n$ contributes in the sum over $j$. Thus it is sufficient to 
estimate the norm of 
	\[
		P^{(n)}_m(f)   \,  \Omega_\circ
	\]
in the $n$-particle subspace $\bbH^{(n)} = \Gamma^{(n)} ({\mathbb H}^{-1} (S^2))$: 
	\begin{align*}
	 	& \| P^{(n)}_m(f)  \, 
		\Omega_\circ \|_{{\rm \bbh}^{(n)}} 
		\\
		&=
		\left\| \sum_{\ell_1 = 0}^m \sum_{k_1= - \ell_1}^{\ell_1} \ldots
			\sum_{\ell_n = 0}^m \sum_{k_n= - \ell_n}^{\ell_n}\;  
	 	 \; \frac{  w^{(n)} (\ell_{1},  k_1, \ldots, \ell_{n}, k_{n}) 
		 \left( Y_{\ell_1, k_1} \otimes_s \cdots \otimes_s Y_{\ell_n, k_n} \right) } 
		{  \prod_{i =1}^n \left( \ell_i + \tfrac{1}{2} \right) }  \; 
		\right\|_{ L^2} .    
	\end{align*}
We have chosen $r=1$ and $\mu=1$ (the general case follows by 
rescaling) and used Definition~\ref{sobolev-S2} to replace the 
norm in $\Gamma^{(n)} ({\mathbb H}^{-1} (S^2)) $ by an $L^2$-norm. We can now  
use the bound\footnote{It follows from $\left( \prod_{p=1}^{n } \lambda_p \right)^{1/n}  
\le \sum_{p=1}^n \lambda_p $, applied to $\lambda_p = \prod_{i\ne p} 
\left( \ell_i  + \tfrac{1}{2}  \right)^{-\frac{n}{n-1}}$.}, \cite[Lem\-ma~6.1]{DG}
	\begin{equation}
	\label{sum-n}
		\prod_{i=1}^{n }  \left( \ell_i   + \tfrac{1}{2}   \right)^{-1} 
		\le \sum_{p=1}^n \Bigl( \prod_{i\ne p}  \left( \ell_i 
		  + \tfrac{1}{2}  \right)^{-\frac{n}{n-1}} \Bigr) \, , 
	\end{equation} 
which allows us to perform $n-1$ double sums, as the sums 
$ \sum_{\ell_i}  \left( \ell_i   + \tfrac{1}{2} \right)^{-\frac{n}{n-1}}$ are finite.
Hence, using \eqref{delta-m} to remove the sums and \eqref{H-p}, we find 
	\begin{align*}
	 	& \lim_{m \to \infty} \| P^{(n)}_m(f)   \, 
		\Omega_\circ \|_{{\rm \bbh}^{(n)}} \\
		& \quad \le \sum_{p=1}^n 
		\underbrace{ \lim_{m \to \infty}
	 \Bigl\|  \int_{S^2} r^2 \cos \psi {\rm d} \psi {\rm d} \theta \;  f(\theta,\psi) \; 
	\underbrace{\sum^m_{ \ell_p = 0} 
	\sum_{k_p= - \ell_p}^{\ell_p} \overline{Y_{\ell_p, k_p} (\theta, \psi)} 
	 Y_{\ell_p, k_p} (\, . \, , \, . \, ) }_{ = \frac{1}{r^2}	 \delta^{(2)}_{m}(\, .-\theta, \, .- \psi)}
	 \Bigr\|_{ L^2(S^2, {\rm d} \Omega) }}_{ \| f \|_{ L^2(S^2, {\rm d} \Omega) } } 
	\\
	&
	\qquad 
	\qquad 
	\qquad 
	\qquad 
	\qquad 
	\qquad 
	\qquad 
	\times
	\prod_{i\ne p} \underbrace{ \lim_{m \to \infty} \underbrace{ \Bigl\|	
	\sum_{\ell_i = 0}^m \sum_{k_i= - \ell_i}^{\ell_i} \frac{ \overline{Y_{\ell_i, k_i} (\theta, \psi)} 
	 Y_{\ell_i, k_i} (\, . \, , \, . \, ) }{\left( \ell_i 
		 + \tfrac{1}{2}  \right)^{\frac{n}{n-1}}} 
	 \Bigr\|_{ L^2(S^2, {\rm d} \Omega) } }_{ =  
	 \|  \delta^{(2)}_{m}(\, .-\theta, \, .- \psi) \|_{ \mathbb{H}^{-\frac{n}{n-1}}	(S^2) }} }_{< \infty}
	 \\
	 & \quad \le {\rm const.} \; \;  \| f \|_{ L^2(S^2, {\rm d} \Omega) } \; . 
	\end{align*}
We have used that convolution with the $\delta$-distribution yields the identity map
	\[
		 \lim_{m \to \infty} f * \delta_m  
		 = f \; ,  \qquad f\in L^{2}(S^2, {\rm d} \Omega) \; , 
	\]
and that, inspecting Equ.~\eqref{H-p}, the Dirac $\delta$ distribution 
has finite norm:  
	\begin{equation}
		\lim_{m \to \infty} \| \delta^{(2)}_{m} \|_{ \mathbb{H}^{-\frac{n}{n-1} }}
		= \underbrace{ \Bigl( \, \sum_{\ell = 0}^\infty \sum_{k= - \ell}^{ \ell } 
	 	\left( \ell + \tfrac{1}{2} \right)^{-\frac{2n}{n-1}}  \Bigr)^{1/2} 
		}_{ = \| \delta \|_{{\mathbb H}^{-\frac{n}{n-1}} (S^2)} }
		 < \infty 
		\; . 
	\end{equation}
The approximations $\delta^{(2)}_{m} $ of the $\delta$-distribution 
was specified in \eqref{delta-m}.

We conclude that $P^{(n)}_m (f) \Omega_\circ$ 
converges to a vector $P^{(n)}_\infty (f) \Omega_\circ$ in $\bbH^{(n)}$,
or equivalently that $P^{(n)}_m (f)$ converges to $P^{(n)}_\infty (f)$ in 
$L^{2}( \mathscr{U}(S^2) , \Omega_\circ )$. Since $P^{(n)}_m (f) \Omega_\circ$ 
is a finite particle vector, it follows from a standard argument (see, \emph{e.g.},
\cite[Theorem~1.22]{S} or \cite[Lemma 5.12]{DG}) that 
	\[
		P^{(n)}_m (f) \to P^{(n)}_\infty (f)\in L^{p}( \mathscr{U}(S^2) , \Omega_\circ )
	\] 
for all $1\leq p<\infty$.
\end{proof}

If ${\mathscr P}={\mathscr P}(\lambda)$ is a real valued polynomial, then 
\label{interactionspherepage}
	\[
		\int_{S^2}   {\rm d} \Omega \; f(\theta,\psi) \; {:} {\mathscr P}(\mathbb{\Phi}( \vec{x} )) {:}   \; , 
		\qquad \vec{x}  \equiv  \vec{x} (\theta,\psi) \in S^2 \; , 
	\]
is well defined, by linearity, for $f \in L^2 (S^2, {\rm d} \Omega)$.
On subsets $K \subset S^2$ with non-empty interiors the interaction is defined by
	\begin{equation}
	\label{L-Additivity}
		Q (K) \doteq  \int_{K} {\rm d} \Omega \;  {:} {\mathscr P} (\mathbb{\Phi}( \vec{x} )) { :}  \; ,
		\qquad K \subset S^2 \; .
	\end{equation}
$Q (K)$ is a densely defined operator, but it is unbounded from below \cite{Jaffe}, even 
though ${\mathscr P}(\lambda)$, $\lambda \in \mathbb{R}$, is by assumption 
bounded from below.

Consider a foliation of the upper hemisphere in terms of half-circles
	\[
		R_1 (\theta) I_+ \; , \qquad 0< \theta < \pi \; .
	\]
Before treating the interaction itself, we show that the Euclidean field allows a foliation. 
Note that one can combine \eqref{coid} 
and \eqref{e1.6bb} to show that the map
	\begin{equation}
		\label{e1.6bbb}
		\begin{matrix} 
			& S^{1}\times C^\infty_{{\mathbb R}} \left(I_+ \right) &
					\to & \bigcap_{1\leq p<\infty}
					L^{p}( \mathscr{U}(S^2) ,  \Omega_\circ  )  \\
			& (\theta,  h) & \mapsto & \mathbb{\Phi} ( \theta, h) \\ 
		\end{matrix}
	\end{equation}
is continuous. 

\begin{lemma}
\label{1.0b}
The following identity holds on $\bigcap_{1\leq p<\infty}L^{p}( \mathscr{U}(S^2) , \Omega_\circ  )$:
	\begin{equation}
	\label{fieldfoliation}
		\int_{S^{1}}  r  \, {\rm d}  \theta  \,   \mathbb{\Phi}  
		(\theta,  \operatorname{\mathbb{cos}}  f_{\theta})=\mathbb{\Phi}(f)\; , 
		\quad f_{\theta} \equiv f (\theta, \, . \, ) \in C^\infty_{{\mathbb R}} 
		\left( I_+\right)  \; , \; \; f\in C^\infty_{{\mathbb R}} (S^2) \; . 
	\end{equation}
\end{lemma}

\begin{proof}
Set $\delta_\theta \equiv \delta (\, .-\theta )$. It follows that, for $f\in C^\infty_{{\mathbb R}} (S^2)$, the map
	\[
		\begin{matrix} 
			&  S^1 & \to &  \mathbb{H}^{-1}(S^2) \\ 
			& \theta & \mapsto & \delta_\theta \otimes f_{\theta} 
		\end{matrix}
	\]
is continuous. Hence, by (\ref{e1.6bbb}) the map
	\begin{align*}
		  S^1 & \to  \bigcap_{1\leq p<\infty}L^{p}( \mathscr{U}(S^2) , 
		  \Omega_\circ )  \nonumber \\  
		 \theta & \mapsto  \mathbb{\Phi} \bigl( \delta_\theta 
		 \otimes  f_{\theta} \bigr) 
	\end{align*}
is continuous and $\bigl\| \mathbb{\Phi} \bigl( \delta_\theta \otimes   
f_{\theta}\bigr) \bigr\|_{L^{p} ( \mathscr{U}(S^2) , \Omega_\circ ) }$ is bounded.  
Therefore 
	\[
		 \int_{0}^{2\pi} r \, {\rm d} \theta \; 
		 \mathbb{\Phi} \bigl( \delta_\theta \otimes  
		 \operatorname{\mathbb{cos}}  f_{\theta} \bigr)  
	\]
is well defined as an element of $\bigcap_{1\leq p<\infty}
L^{p}( \mathscr{U}(S^2) , \Omega_\circ )$. Moreover,  
	\begin{align*}
 		\int_{0}^{2\pi} r   \,  {\rm d} \theta  \;  \mathbb{\Phi} \bigl(  \delta_\theta  \otimes  
		\operatorname{\mathbb{cos}}   f_{\theta}\bigr)
		&= \mathbb{\Phi} \Bigl(\int_{0}^{2\pi} r  \,   {\rm d}  \theta \;  
		(\delta_\theta \otimes \operatorname{\mathbb{cos}} f_{\theta} ) \Bigr)
		= \mathbb{\Phi}( \delta * f) \; ,
	\end{align*}
where the convolution product $*$ acts only in the variable $\theta$. Use  \eqref{eqDeltaTensorh'}  and 
	\[
		 \delta  *   f 
		 =  \int_{0}^{2\pi}  r {\rm d} \theta \; r^{-1} \delta  (\, .\, - \theta)    f(\theta , \, .\, ) 
		 =  f 
	\]
in~$\mathbb{H}^{-1}(S^2)$ to obtain from (\ref{e1.6bb}), also using \eqref{eqDeltaTensorh'},
	\[
		\int_{0}^{2\pi} r   \,  {\rm d}  \theta  \; \mathbb{\Phi} \bigl( \underbrace{ \delta_\theta \otimes    
		\operatorname{ \mathbb{cos}}  f_{\theta}}_{= r^{-1} \delta  (\, .\, - \theta)    f(\theta , \, .\, ) }  \bigr)\, 
		=
		\mathbb{\Phi}(f) \quad \hbox{in} \:  \bigcap_{1\leq p<\infty}L^{p}( \mathscr{U}(S^2) , \Omega_\circ ) \; .
	\] 
\end{proof}

The results of this section will be used  to define Feynman-Kac-Nelson vectors  in Section \ref{arpermodaut}. 

\bigskip

An approximation of the Dirac $\delta$-function 
on $L^2(S^1, r {\rm d} \psi)$ 
is given by $\delta_{k}$, $k\in {\mathbb N}$, with
	\begin{equation}
		\label{deltaka}
		\delta_{k}( \, . \, - \psi ) 
		=(2 \pi r)^{-1}
		\sum_{|\ell|\leq k}{\rm e}^{i  \ell ( \, . \, - \psi) } 
		\; , 
		\quad \psi  \in [0, 2 \pi )  \; .
	\end{equation}
These functions have already appeared in \eqref{deltaka-new}.

\begin{theorem}
\label{wickooo} 
The following limit exists  
in $\bigcap_{1\leq p<\infty}L^{p}( \mathscr{U}(S^1) , \Omega_\circ )$:
	\[
		\lim_{k\to \infty}\int_{S^1} r {\rm d} \psi \;  h (\psi)\,  {:}\mathbb{\Phi}(0, 
		\delta_{k}(.-\psi))^{n}  {:}  \; , 
		\qquad h \in L^2 ( S^1, {\rm d} \psi) \; .  
 	\]
It is denoted by $\int_{S^1}  r {\rm d} \psi \;  h (\psi)\,  {:}\mathbb{\Phi}(0, \psi)^{n} {:} \; $.
\end{theorem}

\begin{proof} 
Applying \eqref{nnord}, we find 
	\begin{align}
		{:}\mathbb{\Phi}(0, 
		\delta_{k}(.-\psi))^{n}  {:}  & = \frac{1}{ 2^{\frac{n}{2}} }
							\sum_{j=0}^n \begin{pmatrix} 
								n \\ 
								j
							\end{pmatrix} 
							\sum_{|\ell_1 |\leq k} \cdots \sum_{|\ell_n |\leq k} 
							\frac{ {\rm e}^{i ( - \ell_1 - \, \ldots  - \ell_j+\ell_{j+1} 
							+ \, \ldots + \ell_n)  \psi }}
							{(2\pi r)^n}
							\nonumber \\
	& \qquad  \times	a^*( \delta \otimes {\rm e}^{i  \ell_1 \, . \, }) \ldots 
						a^*( \delta \otimes {\rm e}^{i  \ell_j \, . \, }) 
						a( \delta \otimes {\rm e}^{i  \ell_{j+1} \, . \, }) \ldots 
						a(\delta \otimes {\rm e}^{i  \ell_{n} \, . \, }) \; . 
	\label{sum-m-1-b}
	\end{align}
Thus
	\begin{align*}
		P^{(n)}_k (0,h) & \doteq  \int_{S^1} r {\rm d} \psi \;  h (\psi)\,  {:}\mathbb{\Phi}(0, 
		\delta_{k}(.-\psi))^{n}  {:}  
	\end{align*}
is a linear combination of Wick monomials of the form 
	\begin{align}
		\sum_{j=0}^{n}  \begin{pmatrix} 
								n \\ 
								j
							\end{pmatrix}  
		& \sum_{|\ell_1 |\leq k} \cdots \sum_{|\ell_n |\leq k}
							\;   
		w^{(n)} (\ell_{1},  \ldots, \ell_{n} ) \, a^{*}_{\ell_1} \cdots a^{*}_{\ell_j}
				a_{\ell_{j+1}} \cdots a_{\ell_{n}} \; , 
	\label{sum-m-2a}
	\end{align}
where $a^{(*)}_{\ell_i} \equiv a^{(*)} (\delta \otimes {\rm e}^{i  \ell_{i} \, . \, })$ and 
	\begin{align*}
		w^{(n)} (\ell_{1}, \ldots,   \ell_{n}) 
		& = \frac{1}{ 2^{\frac{n}{2}} } 
		\int_{S^1} r {\rm d} \psi \;  h (\psi)\,  \; 
		\frac{ {\rm e}^{i ( - \ell_1 - \, \ldots  - \ell_j+\ell_{j+1} 
		+ \, \ldots + \ell_n)  \psi } }{ (2\pi r)^{n} }   \; .
	\end{align*}
Next we apply the Wick monomials \eqref{sum-m-2a} 
to the Fock vacuum $\Omega_\circ  \in \bbH$; 
see also Definition \ref{sobolev-S2}. Only the term with $j=n$ 
contributes in the sum over $j$. Thus it is sufficient to 
estimate the norm of 
	\[
		P^{(n)}_k (0,h)  \,  \Omega_\circ
	\]
in the $n$-particle subspace $\bbH^{(n)} = \Gamma^{(n)} ({\mathbb H}^{-1} (S^2))$: 
	\begin{align}
	 	& \| P^{(n)}_k  (0,h)  \, 
		\Omega_\circ \|_{{\rm \bbh}^{(n)}} 
		\nonumber \\
		&=
		\Bigl\| \sum_{|\ell_1 |\leq k} \cdots \sum_{|\ell_n |\leq k}  
	 	w^{(n)} (\ell_{1},   \ldots, \ell_{n} ) \left( (\delta \otimes {\rm e}^{i  \ell_1 \, . \, }) 
		 \otimes_s \cdots \otimes_s (\delta \otimes {\rm e}^{i  \ell_n  \, . \,  }) \right)    \; 
		\Bigr\|_{ \Gamma^{(n)} ({\mathbb H}^{-1} (S^2))}      
		\nonumber \\
		&=
		\Bigl\| \sum_{|\ell_1 |\leq k} \cdots \sum_{|\ell_n |\leq k}  
	 	w^{(n)} (\ell_{1},   \ldots, \ell_{n} ) \left( {\rm e}^{i  \ell_1  \, . \,  } 
		 \otimes_s \cdots \otimes_s {\rm e}^{i  \ell_n  \, . \,  } \right)    \; 
		\Bigr\|_{ \Gamma^{(n)} (\widehat{\mathfrak{h}} (S^1))} . 
		\nonumber \\
		\label{8.3.14}   
	\end{align}
Inspecting Definition~\eqref{sob-circ} 
and Remark~\ref{rm:4.6.4}, and using the fact that
	\[
		\| \, . \,  \|_{\widehat{\mathfrak{h}} (S^1)} 
		\le {\rm const.} \cdot \| \, . \, \|_{\mathbb{H}^{-1/2} (S^1)} \; , 
	\]
we can now  
use the bound\footnote{It follows from $\left( \prod_{p=1}^{n } \lambda_p \right)^{1/n}  
\le \sum_{p=1}^n \lambda_p $, applied to $\lambda_p = \prod_{i\ne p} 
\left( \ell_i^{2}  + 1 \right)^{ -\frac{n}{2(n-1)}}$.}, \cite[Lem\-ma~6.1]{DG}
	\begin{equation}
	\label{sum-n}
		\prod_{i=1}^{n }  \left( \ell_i^{2}   
		+ 1\right)^{ -1/2} 
		\le \sum_{p=1}^n \Bigl( \prod_{i\ne p}  \left( \ell_i^{2} 
		  + 1\right)^{ -\frac{n}{2(n-1)}} \Bigr) \, , 
	\end{equation} 
which allows us to perform $n-1$ double sums, as the sums 
$ \left( \ell_i^{2}  + 1\right)^{-\frac{n}{2(n-1)}}$ are finite.
Hence, starting from using \eqref{deltaka} 
to remove the sums and \eqref{H-p}, we find 
	\begin{align*}
	 	& \lim_{k \to \infty} \| P^{(n)}_k(0,h)   \, 
		\Omega_\circ \|_{{\rm \bbh}^{(n)}} \\
		& \quad \le \sum_{p=1}^n 
		\underbrace{ \lim_{k \to \infty}
	 \Bigl\|	 \int_{S^1} r {\rm d} \psi \;  h (\psi) \; 
	\underbrace{  
	\sum_{|\ell_p |\leq k}  \frac{ {\rm e}^{i  \ell_p (\, . \, -\psi )}}
	{2\pi \, r} }_{ =  \delta_{k}(\, .- \psi)}
	 \Bigr\|_{ L^2(S^2, {\rm d} \Omega) }}_{ \| h \|_{ L^2(S^1, {\rm d} \psi) } } 
	\\
	&
	\qquad 
	\qquad 
	\qquad 
	\qquad 
	\qquad 
	\qquad 
	\qquad 
	\times
	\prod_{i\ne p} \underbrace{ \lim_{k \to \infty} \underbrace{ \Bigl\|	
	\sum_{|\ell_i |\leq k} \frac{ {\rm e}^{i  \ell_i (\, . \, -\psi )} }{ \left( \ell_i^{2} 
		  + 1 \right)^{  -\frac{n}{2(n-1)}}} 
	 \Bigr\|_{ L^2(S^1, {\rm d} \psi) } }_{ =  
	 \|  \delta_{k}( \, .- \psi) \|_{ \mathbb{H}^{-\frac{n}{2(n-1)}}	(S^1) }} }_{< \infty}
	 \\
	 & \quad \le {\rm const.} \; \;  \| h \|_{ L^2(S^1, {\rm d} \psi) } \; . 
	\end{align*}
Once again, we have chosen $r=1$ and $\mu=1$ (the general case follows by 
rescaling) and used Definition~\ref{sob-circ} to replace the 
norm in $\Gamma^{(n)} (\widehat{\mathfrak{h}} (S^1))$ by an $L^2$-norm.
We have also used that 
	\[
		\delta (\, . \,) \in \mathbb{H}^s (S^1) \qquad \forall s < - \tfrac{1}{2} \; . 
	\]
Just as in the proof of Theorem \eqref{uvtheo},
we conclude that $P^{(n)}_k (0,h) (f) \Omega_\circ$ 
converges to a vector $P^{(n)}_\infty (0,h) \Omega_\circ$ in $\bbH^{(n)}$,
or equivalently that $P^{(n)}_k (0,h)$ converges to $P^{(n)}_\infty (0,h)$ in 
$L^{2}( \mathscr{U}(S^1) , 
\Omega_\circ )$. Since $P^{(n)}_k (0,h) \Omega_\circ$ 
is a finite particle vector, it follows from a standard argument (see, \emph{e.g.},
\cite[Theorem~1.22]{S} or \cite[Lemma 5.12]{DG}) that 
	\[
		P^{(n)}_k (0,h) \to P^{(n)}_\infty (0,h) \in L^{p}( 
		\mathscr{U}(S^1) , \Omega_\circ )
	\] 
for all $1\leq p<\infty$.
\end{proof}

Thus, for ${\mathscr P}$ a real valued polynomial,  the expression
	\label{vcospage}
	\begin{equation}
		\label{bbv-interaction}
		V (h) = \int_{S^1}  r {\rm d} \psi \;  h(\psi) \, {:} {\mathscr P}(\mathbb{\Phi}(0,\psi)) {:} \; ,
		 \qquad h \in L^2 (S^1, r {\rm d} \psi)\; ,
	\end{equation}
is well defined, by linearity.  
The  interaction\footnote{The interaction $V^{(\alpha)} $ 
is defined by rotating $V^{(0)}$ around the $x_0$-axis by an angle $\alpha$.}
	\begin{equation}
		\label{v0newdefinition}
		V^{(0)}   \doteq V ( \mathbb{cos} \, \chi_{ I_+  } ) \; , 
	\end{equation}
with $\chi_{ I_+ }$  the characteristic function of the interval
$I_+ \subset S^1$, can be considered 
as a self-adjoint operator affiliated to the abelian von
Neumann algebra $\mathscr{U} (S^1) $ acting on the Hilbert space~${\mathcal H}$.  

The $(2 \pi)$-periodic one-parameter group 
$ [0, 2 \pi)   \ni \theta  \mapsto  R^{(\alpha)} (\theta)_* $ induces a representation 
	\begin{equation}
		\label{wawa}
		\theta \mapsto \mathbb{U}_\circ^{(\alpha)} (\theta) 
	\end{equation}
of $SO(2)$ in terms of  automorphisms of $\mathscr{U}(S^2)$, 
which extends to a strongly
continuous representation in terms of isometries of $L^{p}( \mathscr{U}(S^2) , \Omega_\circ )$. 

\begin{theorem} 
\label{2.2}
For $h\in L^{2} \bigl( I_+ \, , r {\rm d} \psi \bigr)$ and $g \in C^\infty (S^{1})$,
	\begin{align}  
		\int_{S^{1}} r \, {\rm d} \theta \; g(\theta) \; \mathbb{U}_\circ^{(0)} (\theta) & 
		V ( \operatorname{\mathbb{cos}} h)  
		\mathbb{U}_\circ^{(0)} (-\theta) \nonumber \\
				& = \int_{S^2} {\rm d} \Omega \;   g(\theta)  h(\psi) \; {:}{\mathscr P}(\mathbb{\Phi}(\theta,\psi)) {:} 
		\label{e1.11}
	\end{align}
as unbounded operators on the Euclidean Fock space $ \bbH$. 
In particular, 
	\[ 
		Q(S^2)= 
		\int_{S^2} {\rm d} \Omega \; {:}{\mathscr P}(\mathbb{\Phi}(\theta,\psi)) {:} 
				= \int_{S^{1}} r \, {\rm d} \theta \; V^{(0)} ( \theta) \, ,  
	\]
where $V^{(0)} ( \theta) \doteq \mathbb{U}_\circ^{(0)} ( \theta ) V^{(0)} \mathbb{U}_\circ^{(0)} ( - \theta )$. 
\end{theorem}

\begin{proof} We consider path-space coordinates and follow an argument given in~\cite{GeJII}: 
let $A \in L^{p}( \mathscr{U}(S^1) , \Omega_\circ )$ for some
$1\leq p<\infty$ and $g \in C^\infty_{{\mathbb R}} (S^{1})$. Then
	\[ 
		\int_{S^{1}} r \, {\rm d} \theta \; g(\theta)\mathbb{U}_\circ^{(0)} (\theta) 
		A \mathbb{U}_\circ^{(0)} (- \theta) 
	\]
belongs to $L^{p}( \mathscr{U}(S^2) , \Omega_\circ )$.
Together with Theorem~\ref{wickooo} this implies that the functions given in~(\ref{e1.11}) 
are in $L^{p}( \mathscr{U}(S^2) , \Omega_\circ )$. Next prove that they are identical: by linearity, 
one may assume that ${\mathscr P}(\lambda)= \lambda^{n}$.
Use Theorem~\ref{wickooo} and the identity~\eqref{wick} to derive
	\[
		\int_{S^2} {\rm d} \Omega \; g(\theta)\,h(\psi)  \, {:} {\mathscr P}(\mathbb{\Phi}(\theta,\psi)) {:} \; 
		=\lim_{(k, k')\to \infty} F(k, k')\hbox{ in }L^{p}( \mathscr{U}(S^2) , \Omega_\circ ) \; ,
	\]
where
	\begin{align*}
		F(k, k') &= 
		\sum_{m=0}^{[n/2]} \frac{ n!  }{m!(n-2m)!} 
		\Bigl( -\frac{1}{2}  \| \delta^{(2)}_{k, k'} \|_{\mathbb{H}^{-1}(S^2)}^2	
		\Bigr)^{m}
		\\
		& \qquad	\qquad \times	
		\int_{S^2} 	{\rm d} \Omega \,    g(\theta)  h(\psi) \,
		\mathbb{\Phi} \bigl(\delta_{k}( \, .\, -\theta)\otimes  \delta_{k'}( \, .\, -\psi) \bigr)^{n-2m} 
	\end{align*}
and $\delta^{(2)}_{k, k'} = \delta_{k} \otimes \delta_{k'} $ provides an
approximation of the Dirac $\delta$-function $\delta^{(2)}$ on~$S^2$. 
By replacing both $\delta \otimes h_1$ and $\delta \otimes h_2$ by $\delta^{(2)}_{k, k'}$ in the 
proof of Lemma~\ref{3.9}, one can show that
	\[
		\lim_{k\to \infty}  \| \delta^{(2)}_{k, k'} \|_{\mathbb{H}^{-1}(S^2)}^2 
		=   \| \delta \otimes \delta_{k'}  \|_{\mathbb{H}^{-1}(S^2)}^2   \; , 
		\qquad k' \in \mathbb{N} \, . 
	\]
The definition of sharp-time fields in (\ref{e1.4})  implies that 
	\[
 		\lim_{k\to \infty}F(k, k')=  \int_{S^{1}} r \, {\rm d}
		\theta \; g(\theta)V_{k'}(\theta, \operatorname{\mathbb{cos}} h) \: \hbox{ in } \: 
		L^{p}( \mathscr{U}(S^2) , \Omega_\circ ) \; , 
	\]
where
	\begin{align*}
		V_{k'} (\theta, \operatorname{\mathbb{cos}} h) &= \sum_{m=0}^{[n/2]} \frac{n!}{m!(n-2m)!} 
		\Bigl(-\frac{1}{2}  \| \delta \otimes \delta_{k'}  \|_{\mathbb{H}^{-1}(S^2)}^2  \Bigr)^{m} \nonumber \\
		& \qquad \qquad \times 
		\int_{- \pi/2}^{\pi /2} r \cos \psi \,  {\rm d} \psi \;  h(\psi)  
		\mathbb{\Phi} \bigl(\theta, \delta_{k'}( \, .\, -\psi) \bigr)^{n - 2m}\; .    
	\end{align*}
Note that $V_{k'}(\theta, \operatorname{\mathbb{cos}} h)
		=\mathbb{U}_\circ^{(0)}  (\theta)V_{k'}(0, \operatorname{\mathbb{cos}} h)
		\mathbb{U}_\circ^{(0)}(-\theta) $.
By Theorem~\ref{wickooo}    
	\[
		\lim_{k'\to \infty}V_{k'}(0, \operatorname{\mathbb{cos}} h)
		= \int_{- \pi/2}^{\pi /2}  r \cos \psi \, {\rm d} \psi  \;  
		h(\psi)\, {:}\, {\mathscr P}(\mathbb{\Phi}(0, \psi)) \, {:}     
	\]
in $L^{p}( \mathscr{U}(S^2) , \Omega_\circ )$ and hence 
	\[
	\lim_{k' \to \infty} \underbrace{ \int_{S^{1}} r {\rm d} \theta \, g(\theta)
	V_{k'}(\theta, \operatorname{\mathbb{cos}} h) }_{ =: G(k') }
	= \int_{S^{1}} r {\rm d} \theta \; g(\theta) \; \mathbb{U}_\circ^{(0)} (\theta) V ( \cos _\psi  h) 
	\mathbb{U}_\circ^{(0)} (- \theta) 
	\]
in $L^{p}( \mathscr{U}(S^2) , \Omega_\circ ) $. In summary, we have shown that 
\begin{align*}
	\lim_{k,k'\to \infty}& F(k, k')  \text{ exists} \, ,\\
	\lim_{k\to \infty}\; & F(k', k)= G(k') \text{ exists }\forall k\in {\mathbb N} \, ,\\
	\lim_{k'\to \infty}\; & G(k') \text{ exists}\, .
\end{align*}
It follows that $\lim_{k,k'\to \infty} F(k, k')  = \lim_{k'\to \infty}\;  G(k')$ in 
$L^{p}( \mathscr{U}(S^2) , \Omega_\circ )$. 
\end{proof}

\goodbreak 

\section{The interacting vacuum vector}
\label{sec:8.4}

Up to this point, the sign of the coupling constant did not matter. However, in order to 
guarantee existence of the expression \eqref{intL1}, the interaction has to be repulsive  
for large field strengths. 

\begin{theorem}[Lemma 3.15 in \cite{SHK}]
\label{th:8.4.1}
If, in addition, the real-valued polynomial ${\mathscr P}$ is bounded 
from below, then the function 
	\[
		{\rm e}^{- tQ (K)}  \in L^1( \mathscr{U}(S^2) , \Omega_\circ ) 
		\qquad \forall t>0 \; . 
	\]
In particular, one has 
	\begin{equation}
	 	\label{intL1}
		{\rm e}^{- Q (S^2)}  \in L^{1}( \mathscr{U}(S^2) , \Omega_\circ ) \; . 
	 \end{equation}
The operator $Q (K )$ is defined in \eqref{L-Additivity}. 
\end{theorem}

\begin{proof}
This is Lemma 3.15 in \cite{SHK}. See also Proposition 3.18 in \cite{SHK} as well as
\cite{G,Se}. 
\end{proof}

Hence, the following \emph{rotation invariant} vector is well-defined:
	\begin{equation}	 
		\label{pmeasure}
			\frac{ {\rm e}^{- \frac{1}{2} Q (S^2) }  \Omega_\circ }{
		\| {\rm e}^{- \frac{1}{2} Q (S^2) } \Omega_\circ  \|  } \in \bbH \; . 
	\end{equation} 
The latter should be compared with the \emph{Euclidean interacting vacuum vector} 
	\begin{equation}	 
		\label{E-int-vec}
		\mathbb{\Omega} \doteq \frac{  {\rm e}^{-  Q \left(S_+ \right)} \Omega_\circ } 
		{ \|  {\rm e}^{-  Q \left(S_+ \right)} \Omega_\circ \| } \in \bbH_+  \, .  
	\end{equation} 
\emph{Additivity} of the Lebesgue integral in \eqref{L-Additivity} 
as well as the fact that the Lebesgue measure of $S^1$ is zero
then ensures that 
	\[
		\frac{\langle \mathbb{\Omega} , 
		A \mathbb{U}_\circ (T) \mathbb{\Omega} \rangle}
		{\langle \mathbb{\Omega} ,  \mathbb{U}_\circ (T) \mathbb{\Omega} \rangle}
		= \frac{ \langle {\rm e}^{- \frac{1}{2} Q (S^2) }  \Omega_\circ , 
		A {\rm e}^{- \frac{1}{2} Q (S^2) }  \Omega_\circ \rangle}{
		\| {\rm e}^{- \frac{1}{2} Q (S^2) } \Omega_\circ  \|^2  }
		\qquad \forall A \in \mathscr{U}(S^2) \; .
	\]   
The state induced by the vector 	
\label{intvacuumpage}
	\begin{equation}
	\label{intvacuum} 
		\widehat{\Omega}  
		\doteq \frac{ E_0 \mathbb{\Omega} }
		{ \left\|  E_0  \mathbb{\Omega} \right\| }
		 \in \widehat{\mathcal H} (S^1)  
	\end{equation}
will be identified as the {\em interacting de Sitter vacuum} in the sequel. 
We note that there is no need to distinguish a wedge with edge on $S^1$ 
to define the vector (\ref{intvacuum}).

\bigskip
We next show that $\mathscr{U} (S^1) \widehat{\Omega}$
is dense in $\widehat{\mathcal H} (S^1)$;  see also \cite[15.4 Remark]{KL1}.

\begin{theorem} 
\label{th:8.4.2}
The vector $\widehat{\Omega}$ is 
cyclic and separating for ${\mathscr U}(S^1)$.  
\end{theorem}

\begin{proof} 
As ${\mathscr U}(S^1)$ is a maximal abelian von Neumann algebra
in $\mathcal{B}\bigl(\widehat{\mathcal H} (S^1)\bigr)$, 
${\mathscr U}(S^1)$ coincides with its commutant 
in $\mathcal{B}\bigl(\widehat{\mathcal H} (S^1)\bigr)$. Thus the 
vector $\widehat{\Omega}$ is 
cyclic for~${\mathscr U}(S^1)$ if and only if it is separating for~${\mathscr U}(S^1)$. Now, 
for $0 \ne A \in {\mathcal U}(S^1)$, 
we have 
	\begin{align*} 
		0  < \langle \Omega_\circ \, , \, A^*A
		{\rm e}^{- Q \left(S^2\right)} \Omega_\circ \rangle
		 & = \langle A {\rm e}^{- Q \left(S_-\right)} \Omega_\circ \, , \, 
		 A {\rm e}^{- Q \left(S_+\right)} \Omega_\circ \rangle
		\\
		 & = \langle E_- A {\rm e}^{- Q \left(S_-\right)} \Omega_\circ \, , \, 
		 E_+ A {\rm e}^{- Q \left(S_+\right)} \Omega_\circ \rangle
		 \\
		 & = \langle A E_0{\rm e}^{- Q \left(S_-\right)} \Omega_\circ \, , \, 
		 E_0 A {\rm e}^{- Q \left(S_+\right)} \Omega_\circ \rangle
		 \\
		 & \le \| A E_0{\rm e}^{- Q \left(S_-\right)} \Omega_\circ \|  
		 \; \| A E_0  {\rm e}^{- Q \left(S_+\right)} \Omega_\circ \| \; .
	\end{align*} 
Hence $\widehat{\Omega}$ is separating for~${\mathscr U}(S^1)$,
as $0 \ne A \in {\mathscr U}(S^1) $ implies
$ A E_0  {\rm e}^{- Q \left(S_+\right)} \Omega_\circ \ne 0 $.
\end{proof}

We note that $E_0$ can be used to define a conditional expectation 
in the sense of Ohya and Petz \cite[p.~71]{OP}.
Hence there exists an unbounded operator ${\rm e}^{-  Q  (S^1) }$ 
affiliated to $\mathscr{U}(S^1)$ and given by  
	\begin{equation}
	\label{bedingter-ew}
	  {\rm e}^{-  Q  (S^1) } \doteq E_0 \, {\rm e}^{-  Q (S_+ )} E_0
		\qquad \text{on} \quad  \widehat{\mathcal H} (S^1)  \; .  
	\end{equation}
The operator $Q  (S^1)$ has to be distinguished from $ V (\chi_{S^1})$, 
defined in \eqref{bbv-interaction}. 
Note that, by construction ${\rm e}^{-  Q  (S^1) }$ is invariant under the 
adjoint action of the rotations
$\widehat{U}_\circ (R_0 (\alpha))$, $\alpha \in [ 0 , 2\pi)$. 

\bigskip
We finally add two results concerning the time-zero formulation. For a proof 
of the first one, we once again refer the reader to Lemma 3.15 in \cite{SHK}.
The second theorem (Theorem 5.11 in \cite{DG}) 
is based on a theorem of Nelson (Theorem 1.17 in~\cite{SHK}).

\begin{theorem}
\label{thm:8.4.3}
Let $I$ be an open interval contained in $I_+$. 
Assume the real-valued polynomial ${\mathscr P}$ is bounded from below, 
then the function 
	\begin{equation}
	 	\label{intL1}
		{\rm e}^{- t V (\chi_I \cos)}  \in L^{1}( \mathscr{U}(S^1) , \Omega_\circ ) \; , 
		\qquad \forall t >0 \;.  
	 \end{equation}
The operator $V (\chi_I \cos)$ is defined by \eqref{bbv-interaction}. 
\end{theorem}

The relevance of this result can be seen from the following statement, which 
is well known in the area of hyper-contractive semi-groups.

\begin{theorem}
\label{thm:8.4.4}
Let $\mathfrak{h}$ be a Hilbert space with a conjugation~$c$. 
Let $a$ be a selfadjoint operator on $\mathfrak{h}$ with 
	\[
		a \ge m > 0 \; , \qquad [ a, c] = 0 \; . 
	\]
Let $L^2( \mathscr{U}(S^1) , \Omega_\circ )$ be a representation of the Fock space $\Gamma(\mathfrak{h})$, 
and let $V$ be a real function in $L^p( \mathscr{U}(S^1) , \Omega_\circ )$, for some $p >2$, and 
$ {\rm e}^{- t  V }  \in L^{1}( \mathscr{U}(S^1) , \Omega_\circ ) $ for all $t >0$.  

Then
the operator sum $d\Gamma (a) + V$ is essentially selfadjoint on 
$\mathcal{D}\bigl(d\Gamma (a)\bigr) \cap \mathcal{D}(V)$ and
	\[
		d\Gamma (a) + V \ge - C \; , 
	\]
where $C$ depends only on $m$. 
\end{theorem}

\goodbreak

In our applications, the conjugation will be induced by the time-reflection $T$.

\chapter{The Interacting Representation of $SO(1,2)$}
\label{interactingdesitter}

\setcounter{equation}{0}

So far we have seen that the model describing \emph{non-interacting} bosons on the de 
Sitter space admits an analytic continuation to the Euclidean sphere and that one can reconstruct 
the former from the latter. We will now discuss, how a reflexion positive, rotation invariant state on the 
sphere gives rise to an interacting quantum theory on de Sitter space. 

\section{The reconstruction of the interacting boosts}
\label{sec:9.1}

Our aim in this section is to define a family of operators 	
	\begin{equation} 
		\label{eqDefPa}
		\mathcal{P} \bigl(R_1(\theta)\bigr) E_0 A \mathbb{\Omega}  
		\doteq E_0\, \mathbb{\alpha}^\circ_{R_1 (\theta)}(A) \mathbb{\Omega} \; , 
		\qquad A\in \mathscr{U}(K_\theta)  \;  ,  
	\end{equation} 
where  
	\begin{equation}
		\label{k-theta}
		K_\theta \doteq S_+\cap R_1(-\theta)S_+ \,  , \qquad 0\leq \theta \leq \pi \; .  
	\end{equation}
Hence, as the domain of $\mathcal{P} \bigl(R_1(\theta)\bigr)$ we take 
	\begin{equation}
	\label{d-theta}
		\mathfrak{D}_\theta \doteq E_0 \mathscr{U} (K_\theta) \mathbb{\Omega} \; . 
	\end{equation}
Note that \eqref{eqDefPa} differs from \eqref{hat-wp}  by 
replacing~$\Omega_\circ$ by the 
Euclidean interacting vacuum vector $\mathbb{\Omega}$ defined in \eqref{E-int-vec}.
				
\begin{lemma} 
\label{lm:9.1.1}
The operators $\mathcal{P} \bigl(R_1(\theta)\bigr)$, 
$0\leq \theta \leq \pi$, are well-defined. 
\end{lemma}

\begin{proof}
We first re-write the scalar product in terms of the Fock vacuum
vector $\Omega_\circ$: for $A_1,A_2 \in \mathscr{U}(S_+)$ there holds
	\begin{align}  
		\label{eqScalarProd} 
		\langle E_0 A_1 {\rm e}^{-  Q \left(S_+ \right)} \Omega_\circ \, 
		,\,E_0 A_2  {\rm e}^{-  Q \left(S_+ \right)} \Omega_\circ  \rangle 
		& = \langle {\rm e}^{- Q ({S_+})} E_+ A_1 \Omega_\circ, \,   
		 \mathbb{U}_\circ (T) {\rm e}^{- Q ({S_+})} 
		 E_+ A_2 \Omega_\circ \rangle \nonumber \\
		 & = 
		\langle E_+  A_1 \Omega_\circ, \,  {\rm e}^{- Q ({S^2})}
		 \mathbb{U}_\circ (T) E_+ A_2 \Omega_\circ \rangle \, . 
	\end{align}
From here, the proof goes in complete analogy with Section~8, part $c.)$, in~\cite{KL1}:
we have to show that, for $A\in \mathscr{U}(K_\theta)$, 
	\[
		E_0 A \mathbb{\Omega}=0
		\qquad \Rightarrow \qquad 
		E_0 \mathbb{\alpha}^\circ_{R_1(\theta')}(A) \mathbb{\Omega}=0
		\quad \text{for} \; \;  0<\theta' \leq \theta \; , 
	\] 
as in the proof of~\cite[Lemma 8.2]{KL1}. Assume $\theta'<\theta$; 
the result for $\theta' \le \theta$ will follow by continuity. 
Let $\theta''\in[0, \theta']$ be such that $\theta'+\theta''\leq
\theta$. Then, by Eq.~\eqref{eqScalarProd}, 
	\begin{align*} 
		\| & E_0 \mathbb{\alpha}^\circ_{R_1(\theta')} (A)
		{\rm e}^{-  Q \left(S_+ \right)} \Omega_\circ \, \|^2 \\
			& = \langle E_+ \mathbb{U}_\circ^{(0)} (\theta') A\Omega_\circ, \,  
			{\rm e}^{- Q ({S^2})}  \mathbb{U}_\circ (T) 
			E_+ \mathbb{U}_\circ (R_1(\theta'))A\Omega_\circ \rangle \\ 
 			& = \langle \mathbb{U}_\circ^{(0)} (-\theta'') 
			E_+ \mathbb{U}_\circ^{(0)} (\theta') A\Omega_\circ \, , \, 
				\mathbb{U}_\circ^{(0)} (-\theta'') {\rm e}^{- Q ({S^2})}  
				\mathbb{U}_\circ (T) E_+ \mathbb{U}_\circ^{(0)} (\theta')
				A \Omega_\circ \rangle \; .
	\end{align*}
Now since $\mathbb{\alpha}^\circ_{R_1 (\theta')} (A)$ 
and $\mathbb{\alpha}^\circ_{R_1(\theta'-\theta'')}(A)$ are in ${\mathscr U}(S_+)$,
the unitary $\mathbb{U}_\circ^{(0)} (-\theta'')$ commutes  
with $E_+$ on the left hand side of the scalar product. On the right hand side, observe 
that $\mathbb{U}_\circ^{(0)} (-\theta'')$  commutes with ${\rm e}^{- Q ({S^2})}$ and satisfies 	
	\[
		\mathbb{U}_\circ^{(0)} (-\theta'') \mathbb{U}_\circ (T)
		= \mathbb{U}_\circ(T) \mathbb{U}_\circ^{(0)} (\theta'') \, . 
	\]
Since $\mathbb{\alpha}^\circ_{\theta'+\theta''}(A)$ is still in ${\mathscr U}(S_+)$, 
one arrives at 
 	\begin{align*}
		\| E_0 \mathbb{\alpha}^\circ_{R_1(\theta')} & (A)
		{\rm e}^{-  Q \left(S_+ \right)} \Omega_\circ \|^2
		\\
		& = \langle E_+ \mathbb{U}_\circ^{(0)} (\theta'-\theta'')A\Omega_\circ \, , \, 
		{\rm e}^{- Q ({S^2})} \mathbb{U}_\circ (T) E_+ \mathbb{U}_\circ^{(0)} (\theta'+\theta'') A\Omega_\circ \rangle \,  . 
	\end{align*}
Hence
	\begin{align*} 
		\| E_0 \mathbb{\alpha}^\circ_{R_1(\theta')} (A) \mathbb{\Omega}\|^2 & = 
			\langle E_0 \mathbb{\alpha}^\circ_{R_1(\theta'-\theta'')}(A)\mathbb{\Omega} \, ,\, 
				E_0 \mathbb{\alpha}^\circ_{R_1(\theta'+ \theta'')}(A)\mathbb{\Omega} \rangle \\
		& \leq \| E_0 \mathbb{\alpha}^\circ_{R_1(\theta'-\theta'')}(A) \mathbb{\Omega}\| \, \; 
		 \| A \| \| \mathbb{\Omega} \| \, .
	\end{align*}
Now choose $\theta''$ such that $n\theta''=\theta'$ for some positive
integer $n$, and such that $\theta''+\theta'\equiv (n+1)\theta''$ is
still smaller or equal to $\theta$. Iterating the above inequality $n$
times yields 
	\begin{align*} 
		\|E_0 \mathbb{\alpha}^\circ_{R_1(\theta' )}(A)\mathbb{\Omega}\| 
			&\leq \bigl(  \| A \| \| \mathbb{\Omega} \|  \bigl)^{\tfrac{1}{2}+\cdots+\tfrac{1}{2^n}} 
				\, \| E_0 \mathbb{\alpha}^\circ_{R_1(\theta'-n\theta'')}(A) \mathbb{\Omega} \|^{\tfrac{1}{2^n}}  \\
			& = \bigl(  \| A \| \| \mathbb{\Omega} \|  \bigl)^{\tfrac{1}{2}+\cdots+\tfrac{1}{2^n}} \, 
			\| E_0A \mathbb{\Omega} \|^{\tfrac{1}{2^n}} \, .
	\end{align*}
Thus, $E_0A \mathbb{\Omega} =0$ implies $E_0 \, \mathbb{\alpha}^\circ_{R_1(\theta' )}(A)
\mathbb{\Omega}=0$ for all $\theta'<\theta$. 
By continuity, this fact extends to $\theta'=\theta$. 
\end{proof}

As one may expect, the maps $\mathcal{P}\bigl(R_1(\theta)\bigr)$ define a \index{symmetric local 
semigroup} \emph{symmetric local semigroup} in the sense of Fr\"ohlich~\cite{FOS} and Klein \&
Landau \cite{KL1,KL2}.  

\begin{proposition}
\label{SemiGroupL}
The family $\bigl(\mathfrak{D}_\theta ,\mathcal{P}\bigl(R_1(\theta)\bigr) \bigr)_{\theta\in [0,\pi]}$ 
forms a symmetric local semigroup, \emph{i.e.}, 
\begin{itemize}
\item [$i.)$] for each $\theta\in[0,\pi]$, the set $\mathfrak{D}_\theta $ is a linear subset 
of $\widehat{\mathcal H}(S^1)$. The union   
	\[
		\bigcup_{0 < \theta \le \pi} \mathfrak{D}_\theta  
	\] 
is dense in $\widehat{\mathcal H}(S^1)$ and $\mathfrak{D}_\theta 
\supset \mathfrak{D}_{\theta'} $ if $\theta \le \theta' $;
\item [$ii.)$] 
for each $\theta\in[0,\pi]$, $\mathcal{P}\bigl(R_1(\theta)\bigr)$ is a linear operator 
on $\widehat{\mathcal H}(S^1)$ with domain $\mathfrak{D}_\theta $  and
	\[
		\mathcal{P}\bigl(R_1(\theta')\bigr) \mathfrak{D}_\theta  \subset \mathfrak{D}_{\theta-\theta'} 
		\quad  \hbox{for} \quad 0 \le \theta' \le \theta \le \pi \;  ;
	\]
\item [$iii.)$]  
$\mathcal{P}\bigl(R_1(0)\bigr)  = \mathbb{1}$, and the \emph{semi-group property} \index{semi-group 
property} 
	\[
	 	\mathcal{P}\bigl(R_1(\theta)\bigr)\mathcal{P}\bigl(R_1(\theta' )\bigr)  
		=\mathcal{P}\bigl(R_1(\theta+ \theta')\bigr) 
	 \]
holds on $\mathfrak{D}_{\theta+\theta'}$ for $\theta, \theta', \theta + \theta' \in [0, \pi] $; 
\item [$iv.)$] 
$\mathcal{P}\bigl(R_1(\theta)\bigr)$ is {\em symmetric}, i.e., 
	\[
	 	\langle \Psi, \mathcal{P}\bigl(R_1(\theta)\bigr) 
		\Psi' \rangle 
	 	= \langle \mathcal{P}\bigl(R_1(\theta)\bigr)
		\Psi , \Psi' \rangle \quad  
		\forall \Psi , \Psi' \in \mathfrak{D}_\theta   \; , 
		\quad 0 \le \theta \le \pi  \; ;
	\] 
\item [$v.)$]  
the map $\theta \mapsto\mathcal{P}\bigl(R_1(\theta)\bigr)$ is {\em weakly continuous}, 
\emph{i.e.}, if $ \Psi \in \mathfrak{D}_{\theta'}$, $0 \le \theta' \le \pi$, then 
	\[
	 	\theta \mapsto \bigl\langle \Psi,\mathcal{P}\bigl(R_1(\theta)\bigr) \Psi \bigr\rangle
	 \] 
is a continuous function  for $ 0 < \theta < \theta'$.
\end{itemize}
\end{proposition}

\begin{proof}
The symmetric local semigroup property is shown as in the proof of~\cite[Lem\-ma~8.3]{KL1}. 
First, note that for $\theta<\pi$ the intersection $K_\theta$ contains an open set in $S_+$. 
As ${\rm e}^{- Q (S_+)} >0$, the set $\mathfrak{D}_\theta $ is dense in
$\widehat{\mathcal H}(S^1)$ by Lemma~\ref{eqEU(K)Om}, which implies property $i.)$.  
The properties $ii.)$ and $iii.)$ are satisfied by construction. Symmetry, property $iv.)$, 
follows from Eq.~\eqref{eqScalarProd}  
using 
	\[ 
		\mathbb{U}_\circ^{(0)} (\theta)\mathbb{U}_\circ^{(0)} (T)
		= \mathbb{U}_\circ^{(0)} (T) \mathbb{U}_\circ^{(0)} (-\theta) \; , 
		\qquad - \pi < \theta < \pi \; , 
	\]
and the fact that 
	\[
		\bigl[ \mathbb{U}_\circ^{(0)} (\theta) \, , \, {\rm e}^{- Q ({S^2})} \bigr] = 0 \, . 
	\]
Finally, strong continuity of
the map $\theta \mapsto \mathbb{U}_\circ^{(0)} (\theta)$
implies 
that $\theta\mapsto\mathcal{P}\bigl(R_1(\theta)\bigr)$ is weakly continuous.  
\end{proof}

It is remarkable that  a local symmetric semi-group has a unique self-adjoint generator:

\begin{theorem}[Fr\"ohlich \cite{F80}; Klein \& Landau \cite{KL2}]  
\label{klein-l-f}
\label{K-L--F}
Let $\bigl(P (\theta), {\mathscr D}_{\theta}\bigr)$ be 
a local symmetric semigroup, acting on a Hilbert space ${\mathcal H}$. Then
there exists a unique self-adjoint operator $L$, the {\em generator} of the local
symmetric semigroup $\bigl(P (\theta), {\mathscr D}_{\theta}\bigr)$
on~${\mathcal H}$, such that 
	\[
	 P(\theta') \Psi = {\rm e}^{-\theta' L} \, \Psi  \; ,   
	\qquad \Psi \in {\mathscr D}_{\theta} \; , \quad 0\leq \theta'\leq \theta \; . 
	\]
\end{theorem}

\label{interacting-boost-page}

We denote the generator the local symmetric semigroup 
$\bigl(\mathfrak{D}_\theta ,\mathcal{P}\bigl(R_1(\theta)\bigr) \bigr)_{\theta\in [0,\pi]}$
introduced in Proposition \ref{SemiGroupL} by $\LFock^{(0)}$, \emph{i.e.},
	\begin{equation}
	\label{l-L-0}
		\mathcal{P}\bigl(R_1(\theta)\bigr) = {\rm e}^{- \theta \LFock^{(0)}  } \, . 
	\end{equation}
The generators of the local symmetric semigroups 
$\bigl( U_\circ (R_0 (\alpha)) \mathfrak{D}_\theta ,
\mathcal{P}\bigl(R^{(\alpha)}(\theta)\bigr) \bigr)_{\theta\in [0,\pi]}$
will be denoted by $\LFock^{(\alpha)}$,  $\alpha \in [0, 2\pi)$.

\section{A unitary representation of the Lorentz group}
\label{sec5.8}

Any rotation in $SO(3)$ may be written as a product of rotations
leaving the $x_0$-axis or the $x_1$-axis, respectively, invariant. Hence, it is sufficient to 
set 
	\[
		\mathcal{P} (R_0 (\alpha) )  \doteq \widehat{U}_\circ (R_0 (\alpha) ) \, ,  
		\quad  \alpha \in [0, 2 \pi) \, ,  
	\]
and to extend this to a local group homomorphism from a neighbourhood of the 
identity $\mathbb{1} \in SO(3)$ to linear operators acting on $\widehat{\mathcal{H}}(S^1)$. 
We will denote the generator of $\widehat{U}_\circ (R_0 (\alpha) )$ by $\KFock_\circ$, \emph{i.e.},
	\[
		\widehat{U}_\circ (R_0 (\alpha) ) = {\rm e}^{i \alpha \KFock_\circ} 
		\, ,  \qquad  \alpha \in [0, 2 \pi) \, . 
	\]
Since $R_0(\alpha)$ preserves $S_+$, we have
	\begin{equation} 
		\label{eqDefP}
		\widehat{U}_\circ (R_0 (\alpha) ) E_0 A \mathbb{\Omega}  
		= 
		E_0\, \mathbb{\alpha}^\circ_{R_0 (\alpha)}(A) \mathbb{\Omega} \; , 
		\qquad A\in \mathscr{U}(\overline{S^+})  \;  .  
	\end{equation} 
This follows from the fact that $\mathbb{U}_\circ (R_0 (\alpha)) \mathbb{\Omega} 
= \mathbb{\Omega}$ and $\widehat{U}_\circ (R_0 (\alpha)) 
 \Omega_\circ  = \Omega_\circ $, as well as 
	\[
		\widehat{U}_\circ (R_0 (\alpha)) E_0 
		=  E_0 \mathbb{U}_\circ (R_0 (\alpha)) \, ,  \quad  \alpha \in [0, 2 \pi) \, . 
	\]
Similarly, $\mathbb{U}_\circ (P_1) \mathbb{\Omega} 
= \mathbb{\Omega} $ and $\widehat{U}_\circ (T) E_0 \mathbb{\Psi}
= E_0 \mathbb{\Psi}$ for all $\mathbb{\Psi}  \in  \bbH$.
Moreover, $P_1$ preserves the upper hemisphere
$S_+$ and therefore
	\begin{equation}
	\label{eqJOmE-2}
		\widehat{U}_\circ (P_1T) E_0 A \mathbb{\Omega}  
		= E_0\, 
		\mathbb{\alpha}^\circ_{P_1}(A^*) \mathbb{\Omega} \; ,
		\qquad A\in \mathscr{U}(\overline{S^+})  \;  .   
	\end{equation}
Moreover, the group $R \mapsto \mathbb{U}_\circ(R)$,  
$R \in SO(3)$, acts continuously on  $\bbH$ and $E_0$ is bounded and therefore 
continuous. Thus, given a neighbourhood $N$ of the identity $\mathbb{1} \in SO(3) $, 
the vector valued function
	\[
		N \ni R \mapsto \mathcal{P} (R) \Psi
	\]
is continuous for each\footnote{According to Remark~\ref{rm:7.4.4} 
the set $ {\mathscr D}_{N} $ is dense in $\widehat{\mathcal H}(S^1)$.} 
 	\begin{equation}
	\label{DD-N}
		\Psi \in \mathscr{D}_N \doteq E_0 \, \mathscr{U} ( {\tt O}_N ) 
		\mathbb{\Omega} \; , 
	\end{equation}
where $\mathscr{U} ( {\tt O}_N ) $ is the abelian algebra generated by the Weyl 
operators with test functions in the set $\mathscr{D}_N$ introduced in \eqref{D-N}, 
\emph{i.e.}, $\mathbb{R}$-valued testfunctions in 
$\mathbb{H}^{-1}_{\upharpoonright \overline{S_+} } (S^2)$ whose support lies in the 
polar cap ${\tt O}_N$ specified in \eqref{O-N}. In particular, if $\ell \in {\mathfrak m}$ and
	\[
		{\rm e}^{- t \ell} \in N  \; ,  \qquad 0 \le t \le 1 \; , 
	\]
then $ \mathcal{P} ({\rm e}^{-t \ell })$, $0 \le t \le 1$,  is a hermitian operator defined 
on $\mathscr{D}_N$ and 
	\begin{equation}
	\label{52th}
		\operatornamewithlimits{s-lim}_{t \to 0} \mathcal{P} ({\rm e}^{-t \ell}) \Psi 
		= \Psi \; , \qquad \Psi \in \mathscr{D}_N \; . 
	\end{equation}
The following result thus follows from the theory of virtual representations developed by 
Fr\"ohlich, Osterwalder, and Seiler \cite{FOS}.  

\label{umospage}
\begin{theorem} 
\label{uo}
The self-adjoint operators $\KFock_0 $, $\LFock_1 \doteq \LFock^{(0)} $ and 
$\LFock_2 \doteq \LFock^{(\pi/2)}$ generate a unitary representation 
$\Lambda \mapsto \widehat{U} (\Lambda)$ 
of $SO_0(1,2)$ on~$\widehat{\mathcal H}(S^1)$.
\end{theorem}

\begin{proof} It is sufficient to show that $\mathcal{P}$ defines a virtual 
representation of  $SO(3)$ on~$\widehat{\mathcal H}(S^1)$ and 
then apply Theorem \ref{FOS}; see also Lemma \ref{lm:7.4.2}. 
Thus we have to show that if $R, R'$ and $R \circ R'$ are all in 
some neighbourhood $N$ of the identity $\mathbb{1} \in SO(3) $, which is invariant under 
the rotations $R_0 (\alpha) $, $\alpha \in [0, 2 \pi)$, then
\label{virtreppage2}
	\begin{equation} 
		\label{49th}
		\mathcal{P} (R') \Psi \in {\mathscr D}(\mathcal{P}(R))\; , \qquad \Psi \in \mathscr{D}_N \; ,  
	\end{equation}
	and
	\begin{equation} 
		\label{50th}
		\mathcal{P}(R) \mathcal{P}(R') \Psi = \mathcal{P}(R \circ R') \Psi \; , \qquad \Psi \in \mathscr{D}_N \; , 
	\end{equation}
where $\mathscr{D}_N$ was defined in \eqref{DD-N}. 

Let us first show \eqref{49th}.
Recall the definition of $\mathbb{\sigma}$ from \eqref{m-sigma}. 
Similar to~\eqref{one-p-vr}, we have, for $R \in N$ and $A, A' \in \mathscr{U} ( {\tt O}_N)$, 
	\begin{align*}
		\langle E_0 A \mathbb{\Omega} , E_0 \mathbb{U}_\circ(R)  
		A' \mathbb{\Omega}\rangle
		& = \langle \mathbb{U}_\circ (T) A \mathbb{\Omega} ,  \mathbb{U}_\circ(R)  
		A' \mathbb{\Omega}\rangle \nonumber \\ 
		& = \langle \mathbb{U}_\circ(R^{-1})\mathbb{U}_\circ (T) A \mathbb{\Omega} ,    
		A' \mathbb{\Omega}\rangle \nonumber \\ 
		& = \langle \mathbb{U}_\circ (T) \mathbb{U}_\circ(\mathbb{\sigma} (R^{-1})) 
		A \mathbb{\Omega} ,    A' \mathbb{\Omega}\rangle \nonumber \\ 
		& = \langle E_0 \mathbb{U}_\circ(\mathbb{\sigma} (R^{-1})) A \mathbb{\Omega} ,  
		E_0  A' \mathbb{\Omega}\rangle 
		\,  .	
	\end{align*} 
Now one can use the Schwarz inequality to show that 
	\[	
		E_0  A' \mathbb{\Omega} = 0 \qquad \Rightarrow 
		\qquad E_0 \mathbb{U}_\circ(R)  A' \mathbb{\Omega} = 0 \; . 
	\]
Thus, for each $R \in N  \subset SO(3)$, the map 
	\begin{align*}
	\mathcal{P} (R) \colon \qquad {\mathscr D}_N \quad & \to  \quad \widehat{\mathcal H} (S^1)
		\nonumber \\
	 E_0 \Psi  & \mapsto   E_0  \mathbb{U}_\circ(R) \Psi \;  
	\end{align*}
is well-defined, and hence \eqref{49th} follows from 
	\[
		\mathbb{\alpha}^\circ_{R} \bigr( \mathbb{\alpha}^\circ_{R'}(A) \bigl) 
		= \mathbb{\alpha}^\circ_{R\circ R'}(A)  \in \mathscr{U}( \overline{S^+}) \; . 
	\]
Finally, let us verify \eqref{50th}. If $\Psi \in \mathscr{D}_N$ and $R , R' \in N$ 
as well as $R \circ R' \in N $, then 
	\[
		E_0  \mathbb{U}_\circ(R)\mathbb{U}_\circ(R') \Psi = 
		E_0 \mathbb{U}_\circ (R \circ R') \Psi \; . 
	\]
Thus the group property 
	\[
		\mathcal{P}(R_1)\mathcal{P}(R_2)=\mathcal{P}(R_1 \circ R_2)
	\] 
holds on $\mathscr{D}_N$. In summary, $\mathcal{P}$ is a virtual representation 
of $SO(3)$ on~$\widehat{\mathcal H}(S^1)$.
\end{proof}

\begin{definition}
\label{def:6}
Given the unitary representation $\widehat{U}$ of $SO_0(1,2)$ on the 
Hilbert space $\widehat{\mathcal H}(S^1)$, we define the corresponding 
automorphisms: set 
\label{autintseinspage}
	\[
	 	\widehat{\alpha}_{\Lambda} ( A) 
		= \widehat{U} (\Lambda) \, A \, \widehat{U} (\Lambda)^{-1} \; , 
	\qquad
	A \in \mathscr{B} (\widehat{\mathcal H}(S^1)) \; , \quad \Lambda \in SO_0(1,2) \; .  	
	\]
We call this group of automorphisms the {\em interacting dynamics}.
\end{definition}

\begin{lemma}
For  $A_1,\ldots, A_n$ in ${\mathscr U}(I_+)$ and $t_1, \ldots , t_n \in \mathbb{R}$, 
there holds the relation
	\begin{align} 
		\label{eqJLModa} 
		\widehat{U}_\circ  (P_1T) \, &
		\widehat{\alpha}_{\Lambda_1(t_1)} (A_1)  \cdots
		\widehat{\alpha}_{\Lambda_1(t_n)} (A_n)\Omega 
		\nonumber
		\\
		& 	= {\rm e}^{-\pi \LFock^{(0)}}\, 
			\widehat{\alpha}_{\Lambda_1(t_n)}
			(A_n^*)\cdots \widehat{\alpha}_{\Lambda_1(t_1)}
			(A_1^*) \Omega \; .  
	\end{align}  
\end{lemma}

\begin{proof}
This proof resembles the one of~\cite[Thm.~12.1]{KL1}.   
Let $\theta_1,\ldots,\theta_n\in[0,\pi]$ with 
$\sum_{k=1}^n\theta_k\leq \pi$. Then    
	\begin{align*}  
		{\rm e}^{-\theta_1 \LFock^{(0)}}A_1 & \cdots {\rm e}^{-\theta_n \LFock^{(0)}} A_n\,
		\widehat{\Omega} 
		\\
		&= E_0 \,\mathbb{\alpha}^\circ_{R_1(\theta_1)}(A_1)
		\mathbb{\alpha}^\circ_{R_1(\theta_1+\theta_2)}
			(A_2)\cdots \mathbb{\alpha}^\circ_{R_1(\theta_1+\cdots+\theta_n)}
			(A_n)\,\mathbb{\Omega} \, .  
	\end{align*}
We now apply $ \widehat{U}_\circ  (P_1T)$ using Eq.~\eqref{eqJOmE-2}, 
and use the relation   
	\[
		P_1\circ R_1(\theta)= R_1(\pi-\theta)\circ T \; ,
	\] 
as well as the time-reflection invariance  $\mathbb{\alpha}^\circ_T(A_k)=A_k$, and conclude  
	\begin{align*} 
		\widehat{U}_\circ (P_1T)
		\, & {\rm e}^{-\theta_1 \LFock^{(0)} }\,  A_1 \cdots {\rm e}^{-\theta_n \LFock^{(0)} } A_n\,
		\widehat{\Omega}\\
			& = E_0 \,\mathbb{\alpha}^\circ_{R_1(\pi-\theta_1-\cdots-\theta_n)}(A_n^*) \cdots
				\mathbb{\alpha}^\circ_{R_1(\pi-\theta_1)}(A_1^*) \mathbb{\Omega}\\ 
			&= {\rm e}^{-(\pi-\sum_1^n\theta_k) \LFock^{(0)}}\, A_n^*  
				E_0 \,\mathbb{\alpha}^\circ_{R_1(\theta_n)}(A_{n-1}^*)
				\mathbb{\alpha}^\circ_{R_1(\theta_n+\theta_{n-1})}(A_{n-2}^*)
				\\
			& \qquad \qquad 
			\cdots \mathbb{\alpha}^\circ_{R_1(\theta_n+\cdots+\theta_{2})}(A_{1}^*) 	
			\mathbb{\Omega}
				\\  
			& = {\rm e}^{-(\pi-\sum_1^n\theta_k) \LFock^{(0)}}\, A_n^* 
			{\rm e}^{-\theta_n \LFock^{(0)}}  \cdots 
			A_2^*{\rm e}^{-\theta_2 \LFock^{(0)} }A_1^* \,
			\widehat{\Omega} \, . 
	\end{align*}
By analytic continuation (observe that $U(P_1T)$ is anti-linear) this implies 
	\begin{align*} 
		\widehat{U}_\circ (P_1T)
		\, & {\rm e}^{is_1 \LFock^{(0)} }\, A_1 \cdots {\rm e}^{is_n \LFock^{(0)} } A_n\,
		\widehat{\Omega} \\
			& = 
	{\rm e}^{-\pi \LFock^{(0)} }\, {\rm e}^{i\sum_{k=1}^n s_k \LFock^{(0)}}\, 
		A_n^* {\rm e}^{-is_n \LFock^{(0)} }  
		\cdots A_2^*{\rm e}^{-is_2 \LFock^{(0)} }A_1^* \,
		\widehat{\Omega} \, .   
	\end{align*}
Defining $t_1\doteq s_1$ and $t_k\doteq s_k-s_{k-1}$ for $k=2,\ldots,
n$, we find $\sum_{k=1}^n s_k = t_n$ hence this is just the desired 
relation~\eqref{eqJLModa}.  
\end{proof}

\begin{theorem}
\label{9.2.4}
The anti-linear operator $\widehat{U}_\circ (P_1T)\, {\rm e}^{-\pi \LFock^{(0)}}$ is the 
Tomita operator for the pair $\bigr( \mathcal{R}_{\rm int}(I_+),
\widehat{\Omega} \bigl)$, where  
	\begin{equation}
	\label{A-L-W}
		\mathcal{R}_{\rm int} (I_+)
		 \doteq \bigvee_{t \in {\mathbb R}} 
		\left( \widehat{\alpha}_{\Lambda^{(0)} (t)} \bigl( \mathscr{U} (I_+)   \bigr) \right) \; .
	\end{equation}
\end{theorem}

In the next section (see Corollary~\ref{coll:9.3.3}), we will verify 
that $\mathcal{R}_{\rm int} (I_+) = \mathcal{R}  (I_+) $. 

\goodbreak

\section{Perturbation formulas for the boosts}
\label{arpermodaut}

Let $V^{(0)}$ be the interaction defined in~\eqref{v0newdefinition}. It follows from 
Theorem~\ref{wickooo} and \cite[Lemma~3.15]{SHK}, respectively, that
	\[
	V^{(0)} \in L^{1}(\mathscr{U}(I_+), \Omega_\circ ) \quad \text{and} \quad 
	{\rm e}^{- 2 \pi V^{(0)}}\in L^{1}(\mathscr{U}(I_+), \Omega_\circ ) \; . 
	\]
Hence  the Feyman-Kac-Nelson vectors 
	\[
		{\rm e}^{-\int_{0}^{\theta'}{\rm d} \theta'' \; V^{(0)}(\theta'') } \Omega_\circ \; ,
		\quad 0 \leq \theta' \leq \pi \;  ,
	\]
belong to $\bbH_+$. Now, define a new map, for  $0\leq \theta'\leq \theta $, by setting
	\begin{align*}
				\mathcal{P}_{\rm int}(\theta')\colon  {\mathscr D}_{\theta} 
				& \to \widehat{\mathcal H}(S^1) 
			 \\
				E_0 A \Omega_\circ & \mapsto E_0  
				\underbrace{ {\rm e}^{-\int_{0}^{\theta'}{\rm d} \theta'' \; 
				V^{(0)}(\theta'') }  \mathbb{U}_\circ 
				\bigl(R_1(\theta')\bigr) }_{=: \mathbb{U}^{(0)} (\theta') } A \Omega_\circ   \; , 
				\qquad A \in \mathscr{U} (K_\theta)  \; .  
	\end{align*}
Viewed as an element of $L^p({\mathscr U} (I_+) , \Omega_\circ) 
\cong L^p (K, \mathrm{d} \nu)$, 
	\begin{equation}
		\label{V>0}
		{\rm e}^{- \int_0^\pi  {\rm d} \theta \;  V^{(\alpha)} (\theta)  } > 0  \; , 
		\qquad \nu-a.e. \; . 
	\end{equation}
Hence, the unbounded operators $\mathbb{U}^{(0)} (\theta')$ are \emph{invertible} and they satisfy 
	\[
		\mathbb{U}^{(0)} (\theta') A  
		\mathbb{U}^{(0)} (\theta')^{-1} 
		= \mathbb{\alpha}^\circ_{R_1 (\theta')} (A) \qquad \forall \theta \in [0, 2 \pi) \; , 
	\]
and for all $A \in \mathscr{U} (S^2)$, as $\bigl[ {\rm e}^{-\int_{0}^{\theta'}{\rm d} \theta'' \; 
				V^{(0)}(\theta'') } , A \bigr] = 0 $ for all $A \in \mathscr{U} (S^2)$. 

\begin{theorem}
$\bigl(\mathcal{P}_{\rm int} (\theta), {\mathscr D}_{\theta}  \bigr)$ is
a local symmetric semigroup on~$\widehat{\mathcal H}(S^1)$.  
\end{theorem}

\begin{proof}
For a proof of this statement see \cite[Lemma 15.3]{KL1}. 
\end{proof}

\label{HFockhatpage}

We denote the generator of local symmetric semigroup $\bigl(\mathcal{P}_{\rm int} (\theta), 
{\mathscr D}_{\theta}  \bigr)$ by $\HFock^{(0)}$. By construction, 
	\[
		{\rm e}^{- \theta' \HFock^{(0)}} E_0 A \Omega_\circ = E_0 \mathbb{U}^{(0)} (\theta') A \Omega_\circ
		\; , \qquad A \in \mathscr{U} (K_\theta)\; , \quad \theta' \le \theta \; . 
	\]
In the sequel, we will also need the operators 
	\[ 
		\HFock^{(\alpha)} \doteq \widehat{U}_\circ (R_0 (\alpha) ) 
		\HFock^{(0)} \widehat{U}_\circ^{-1} (R_0 (\alpha) )
		\; , \qquad
		J^{(\alpha)} \doteq \widehat{U}_\circ (R_0 (\alpha) ) J^{(0)} \widehat{U}_\circ^{-1} (R_0 (\alpha) ) \; . 
	\]

\subsection{Operator sums} As we will see next, the properties of the interaction ensure 
that key operator sums are well defined despite the fact that they involve two 
operators which are both unbounded from both below and above.

\begin{theorem}
\label{th5.2} Set\footnote{Theorem~\ref{wickooo} ensures 
existence of \eqref{halberkreis}.}
	\begin{equation}
		\label{halberkreis}
			V^{(\alpha)} 
			=  \int_{I_\alpha}  r {\rm d} \psi \; \cos (\psi + \alpha) \, 
			{:}{\mathscr P}(\mathbb{\Phi} (0, \psi)) {:}  \: .
	\end{equation}
It follows that
\begin{itemize}
\item[$ i.)$]
\label{keyresult1}
the operator sum $\LFock_\circ^{(\alpha)} +V^{(\alpha)}$ is essentially self-adjoint on 
${\mathscr D} \bigl(\LFock_\circ^{(\alpha)} \bigr) \cap {\mathscr D} \bigl(V^{(\alpha)} \bigr)$ and 
	\begin{equation}
	\label{Araki1}
			\overline{\LFock_\circ^{(\alpha)}+V^{(\alpha)}} 
		= \HFock^{(\alpha)} \; , \qquad \alpha \in [ \, 0, 2 \pi  )\; ;
	\end{equation}
\item[$ ii.)$]
the operator $\HFock^{(\alpha)}- J^{(\alpha)} V^{(\alpha)} J^{(\alpha)}$ is essentially self-adjoint 
on the domain 
	\[
		{\mathscr D} \bigl(\HFock^{(\alpha)} \bigr) \cap {\mathscr D} \bigl(J^{(\alpha)} V^{(\alpha)} J^{(\alpha)} \bigr)
	\]
and the closure equals $\LFock^{(\alpha)} $, 
	\[
		\overline{\HFock^{(\alpha)}- J^{(\alpha)} V^{(\alpha)} 
		J^{(\alpha)}} = \LFock^{(\alpha)} \; ,  \qquad \alpha \in  [ \, 0, 2\pi ) \; , 
	\]
where $\LFock^{(\alpha)} = \widehat{U}_\circ (R_0(\alpha)) \LFock^{(0)} \widehat{U}_\circ (R_0(-\alpha))$ and 
$\LFock^{(0)} $ is defined in Equ.~\eqref{l-L-0}.
Moreover, $\LFock^{(\alpha)}\widehat{\Omega} = 0$;  
\item[$ iii.)$]
The operator sum $\LFock_\circ^{(\alpha)}+V (\mathbb{cos}_\alpha)$ 
is essentially self-adjoint on the natural domain
${\mathscr D} \bigl(\LFock_\circ^{(\alpha)} \bigr) \cap {\mathscr D} \bigl(V (\mathbb{cos}_\alpha) \bigr)$ 
and its closure equals 
	\[
	 \overline{\LFock_\circ^{(\alpha)}+V (\mathbb{cos}_\alpha) } = \LFock^{(\alpha)}  \; , 
	\] 
where $V (\mathbb{cos}_\alpha)$ was defined in (\ref{bbv-interaction}).  
\end{itemize}
\end{theorem}

Note that the integration in (\ref{bbv-interaction})  is over the 
whole circle~$S^1$, while the integration in~(\ref{halberkreis}) is restricted 
to the halfcircle $I_\alpha$. 

\begin{proof} 
The proofs of these results rely on results from the literature:
\begin{itemize}
\item [$ i.)$]  Essential selfadjointness follows from the results on local symmetric semigroups by 
Fr\"ohlich \cite{F80} and Klein and Landau \cite{KL1}\cite{KL2}. Equation \eqref{Araki1} is shown as in \cite{KL2}:
First, verify it on the sets $D_\theta \doteq E_0 \mathscr{U} (K_\theta) \Omega_\circ$, with $ 0< \theta < \pi$. But 
$\bigcap_\theta D_{0< \theta < \pi}$ is a core for $\HFock^{(\alpha)}$.
\item [$ ii.)$] Since ${\rm e}^{- 2 \pi V^{(\alpha)}}\in L^{1}(\mathscr{U}(S^2), \Omega_\circ)$ and
	\[
		V^{(\alpha)} \in L^{p}( \mathscr{U}(I_\alpha), \Omega_\circ ) \; ,\quad
		{\rm e}^{-\pi V^{(\alpha)}}\in L^{q} ( \mathscr{U}(I_\alpha), \Omega_\circ ) \; , 		
	\]
with $p^{-1}+ q^{-1}=\frac{1}{2} $ and $ 2\leq p, q\leq \infty$, property 
$ii.)$ follows from \cite[Theorem~7.12]{GeJ}.
\item [$iii.)$] This result follows from the fact that $J^{(\alpha)}$ implements 
the space-reflec\-tion 
$P^{(\alpha)} = R_{0} (\alpha) P_1 R_{0} (\alpha)^{-1}$
on $\mathscr{U} (S^1)$ and $\cos (\frac{\pi}{2} + \psi)  = - \cos (\frac{\pi}{2} - \psi)$. 
Thus
	\[
	 V (\mathbb{cos}_\alpha) 
	 = \overline{ V^{(\alpha)} - J^{(\alpha)} V^{(\alpha)} J^{(\alpha)} }  \; . 
	\] 
The statement now follows from property $ii.)$. 
\end{itemize}
\end{proof}

\begin{remark}
\label{rm:9.3.3}
It can be seen from Theorem~\ref{th5.2} $ii.)$ and $iii.)$ that the 
interacting boosts do not respect the 
$n$-particle sectors of Fock space, the latter get mixed up rapidly. 
In this sense, the $\mathscr{P}(\varphi)_2$ 
model on de Sitter space shows particle creation and annihilation. 
However, this statement is with respect 
to \emph{naked} particles, while \emph{physical} particles should 
be \emph{dressed}, they should carry a 
particle-antiparticle cloud with them as they move through de Sitter space. 
Whether such particles are 
stable and even whether there are entities (particles) which show particle 
like behaviour in collisions needs 
further investigation.  
\end{remark}

\begin{corollary}
\label{coll:9.3.3}
The algebras $\mathcal{R} (I_\alpha)$ and $\mathcal{R}_{\rm int} (I_\alpha)$ 
defined in \eqref{A-L-W} coincide. 
\end{corollary}

\begin{proof}
As consequence of \eqref{Araki1} and 
	\begin{equation}
		\label{hosaut}
			\widehat{\alpha}_{\Lambda^{(\alpha)} (t)} (A)
			= {\rm e}^{ i t \HFock^{(\alpha)} } A {\rm e}^{- i t \HFock^{(\alpha)} } \qquad \forall
				A \in \mathcal{R} (I_\alpha) \; ,
	\end{equation}
the \emph{Trotter product formula} applies and yields
	\[
		\widehat{\alpha}_{\Lambda^{(\alpha)}}  ( A) 
			= \lim_{n \to \infty}  \left( {\rm e}^{i  \frac{t}{n} V^{(\alpha)} } 
			{\rm e}^{i  \frac{t}{n} \LFock_\circ^{(\alpha)} } \right)^n
				A \left( {\rm e}^{-i  \frac{t}{n} V^{(\alpha)} } 
				{\rm e}^{-i  \frac{t}{n} \LFock_\circ^{(\alpha)} } \right)^n  \; , 
			\quad A \in \mathscr{U} (I_\alpha)  \; .
	\]
$\mathcal{R} (I_\alpha)=\mathcal{R}_{\rm int} (I_\alpha)$, 
as ${\rm e}^{i t V^{(\alpha)} } \in \mathscr{U} (I_\alpha)$ for $t \in {\mathbb R}$. 
\end{proof}

\subsection{Properties of the interacting vacuum vector} We can now provide a list of 
key properties, which the vacuum vector $\widehat{\Omega}$ satisfies:

\begin{theorem} 
\label{th5.2} Given the same expression for the interaction $V^{(\alpha)}$ as in the previous 
theorem, we have 
\begin{itemize}
\item[$ i.)$]
the Fock vacuum vector $\Omega_\circ $ belongs to
${\mathscr D}\bigl({\rm e}^{-\pi \HFock^{(\alpha)}}\bigr)$, $\alpha \in [ \, 0, 2 \pi )$, and for 
all $\alpha \in [ \, 0, 2\pi )$ the vector
	\begin{equation}
		\label{Araki}
		\frac{{\rm e}^{-\pi \HFock^{(\alpha)}}\Omega_\circ }{ 
		\|{\rm e}^{-\pi \HFock^{(\alpha)}}\Omega_\circ  \| } = \widehat{\Omega} 
	\end{equation}
is equal to the interacting vacuum vector $\widehat{\Omega}$ 
defined in \eqref{intvacuum};
\item[$ ii.)$] the vector $\widehat{\Omega}$ belongs to the natural 
positive cone $\mathscr{P} \bigr(\mathcal{R}(I_+),\Omega_\circ \bigl)$;
\item[$ iii.)$] the vector $\widehat{\Omega}$  is cyclic and 
separating for the algebras ${\mathcal R}  (I_{\alpha} )$, $\alpha \in [0, 2 \pi)$; 
\item[$ iv.)$] the vector $\widehat{\Omega}$ satisfies the Peierls-Bogoliubov  and the Golden-Thompson 
inequalities:
	\[
		{\rm e}^{- {\pi} \langle \Omega_\circ , V^{(\alpha)} \Omega_\circ \rangle   } 
		\le  \| {\rm e}^{-\pi \HFock^{(\alpha)}}\Omega_\circ  \|  
		\le  \|{\rm e}^{- {\pi} V^{(\alpha)}   } \Omega_\circ \| \; .
	\]
\end{itemize}
\end{theorem}

\goodbreak
\begin{proof}
\quad
\begin{itemize}
\item [$ i.)$] The expression on the l.h.s.~of (\ref{Araki}) is a formula, which is well know from 
the perturbation theory of KMS states (see \cite{DJP}\cite{KL1}).
The identification (\ref{Araki}) follows from  
\begin{equation}
	\label{Vhalbkugel} 
	{\rm e}^{- \pi \HFock^{(\alpha)}} \Omega_\circ 
	= E_0 \, {\rm e}^{- Q  (S_+ )} \Omega_\circ \; .
\end{equation}
This equality follows from the reconstruction theorem, but using $i.)$ it can 
also be verified directly 
using the Trotter product\index{Trotter product formula} formula:
\begin{align*}
		\qquad \qquad E_0 \, {\rm e}^{- \int_0^\theta  {\rm d} \theta' \;  
		V^{(\alpha)} (\theta')  } \Omega_\circ  
		& = \lim_{n \to \infty} E_0 \, {\rm e}^{- \frac{\theta}{n} 
				\sum_{k=1}^n V^{(\alpha)} (k \theta / n )  }  
				\Omega_\circ 
		\\
		& =  \lim_{n \to \infty} E_0 \, \underbrace{ \mathbb{\alpha}_{R_1
		(\frac{n \theta}{n})}^\circ 
				\bigl({\rm e}^{- \frac{\theta}{n} 
				V^{(\alpha)}  } \bigr) 
				\cdots  \mathbb{\alpha}_{R_1(\frac{\theta}{n})}^\circ 
				\bigl( {\rm e}^{- \frac{\theta}{n} 
				V^{(\alpha)}  } }_{n \; \rm{terms}} \bigr) \Omega_\circ \\
		& = s\mbox{-}\lim_{n \to \infty} 
				\Bigl(  \underbrace{ {\rm e}^{- \frac{\theta}{n} 
				\LFock^{(\alpha)}_\circ  } 
				{\rm e}^{- \frac{\theta}{n} 
				V^{(\alpha)}  } \cdots 
				{\rm e}^{- \frac{\theta}{n} 
				\LFock^{(\alpha)}_\circ  }
				{\rm e}^{- \frac{\theta}{n} 
				V^{(\alpha)}  }  }_{n \; \rm{terms}} \Bigr) \Omega_\circ \\
		& = {\rm e}^{- \theta \HFock^{(\alpha)}} \Omega_\circ \; , 
		\qquad 
		0\le \theta \le \pi \; .
\end{align*}
Note that the r.h.s.~in \eqref{Vhalbkugel} is independent of $\alpha$.
\item [$ii.)$] If we approximate the interaction $V^{(\alpha)}$ by a sequence of bounded 
interactions $V_n^{(\alpha)}$, then Araki's perturbation theory of KMS states ensures that the vectors 
	\[
		\widehat{\Omega}_n \doteq \frac{{\rm e}^{-\pi ( \LFock_\circ^{(\alpha)}+V_n^{(\alpha)}) }\Omega_\circ }{ 
		\|{\rm e}^{-\pi (\LFock_\circ^{(\alpha)}+V_n^{(\alpha)})}\Omega_\circ  \| } \; ,  
		\qquad   \alpha \in [ \, 0, 2\pi ) \; ,  
	\]
are all in in the cone $\mathscr{P} \bigr(\mathcal{R}(I_+),\Omega_\circ \bigl)$. The latter is strongly closed, 
so the statement follows from the fact that $\widehat{\Omega}_n$ is 
strongly converging to $\widehat{\Omega}$.
\item [$iii.)$]
Recall from \eqref{E-int-vec} that  
	\[
		{\rm e}^{- \int_0^\pi  {\rm d} \theta \;  V^{(\alpha)} (\theta)  } \Omega_\circ \in \bbH_+ \; .  
	\]
Hence, for
$0\le \theta_1 \le \ldots \le  \theta_n \le  \pi$ and $A_1, \ldots, A_n \in \mathscr{U}(I_\alpha)$, the vectors 
	\[
		\qquad \quad \mathbb{U}^{(\alpha)}(\theta_n) A_n 
		\mathbb{U}^{(\alpha)}(\theta_{n-1}- \theta_n) A_{n-1}
			\cdots  \mathbb{U}^{(\alpha)} (\theta_1-\theta_2)  
			A_1  {\rm e}^{- \int_0^{\pi - \theta_1} {\rm d} \theta \;  V^{(\alpha)} (\theta)  } \Omega_\circ   
	\]
are in $\bbH_+$. Because of \eqref{V>0}, they form a total set in $\bbH_+$. Therefore, 
the vectors 
	\begin{align*}
		\qquad &  
		E_0 \mathbb{U}^{(\alpha)}(\theta_n) A_n 
		\mathbb{U}^{(\alpha)}(\theta_{n-1}- \theta_n) 
			\cdots   \mathbb{U}^{(\alpha)} (\theta_1-\theta_2)  
			A_1 {\rm e}^{- \int_0^{\pi - \theta_1} {\rm d} \theta \;  
				V^{(\alpha)} (\theta)  } \Omega_\circ
		 \nonumber \\ 
		& \quad = {\rm e}^{-\theta_{n} \HFock^{(\alpha)} } A_n 
		E_0 \mathbb{U}^{(\alpha)}(\theta_{n-1}- \theta_n) 
			\cdots   \mathbb{U}^{(\alpha)} (\theta_1-\theta_2)  
			A_1 {\rm e}^{- \int_0^{\pi - \theta_1} {\rm d} \theta \;  
				V^{(\alpha)} (\theta)  } \Omega_\circ
		 \nonumber \\ 
		& 
		\quad = {\rm e}^{-\theta_{n} \HFock^{(\alpha)} } A_n 
		{\rm e}^{- (\theta_{n-1}- \theta_n) \HFock^{(\alpha)} } E_0
			\cdots   \mathbb{U}^{(\alpha)} (\theta_1-\theta_2)  
			A_1  {\rm e}^{- \int_0^{\pi - \theta_1} {\rm d} \theta \;  
				V^{(\alpha)} (\theta)  } \Omega_\circ
		 \nonumber \\ 
		& \quad = {\rm e}^{-\theta_{n} \HFock^{(\alpha)} } A_n 
		{\rm e}^{- (\theta_{n-1}- \theta_n) \HFock^{(\alpha)} } 
		\cdots 
		{\rm e}^{-(\theta_{1} - \theta_2) \HFock^{(\alpha)} } A_1 
		{\rm e}^{-(\pi - \theta_{1}) \HFock^{(\alpha)} }\Omega_\circ    
		 \nonumber \\ 
		& \quad = {\rm e}^{-\theta_{n} \LFock^{(\alpha)} } A_n 
		{\rm e}^{- (\theta_{n-1}- \theta_n) \LFock^{(\alpha)} } 
		\cdots 
		{\rm e}^{-(\theta_{1} - \theta_2) \LFock^{(\alpha)} } A_1 
		{\rm e}^{\theta_{1} \LFock^{(\alpha)} } {\rm e}^{- \pi \HFock^{(\alpha)} }
		\Omega_\circ  
	\end{align*}
form a total set in $\widehat{\mathcal H} (S^1)$. Thus multi-time analyticity \cite{A5} 
and \eqref{Araki} imply that the vectors 
	\[
	A_n (t_n)  
		\cdots 
		A_1 (t_1)
		\widehat{\Omega}  
	\]
with $A_i (t) = {\rm e}^{ i t \HFock^{(\alpha)} } A_i {\rm e}^{- i t \HFock^{(\alpha)} } $, 
$t \in {\mathbb R}$, form a total set in $\widehat{\mathcal H} (S^1)$ too. 

\item [$iv.)$] The  Peierls-Bogoliubov  and the Golden-Thompson 
inequalities (see \cite{A4}) were generalised to the present case in~\cite[Theorem 5.5]{DJP}.
\end{itemize}
\end{proof}

\subsection{Perturbation theory for modular automorphisms}
Next recall Araki's perturbation theory for modular automorphisms~\cite{A2}\cite{A3}, 
which has been generalised to unbounded perturbations by Derezinski, Jaksic and Pillet~\cite{DJP}.

\goodbreak

\begin{theorem} 
\label{H-neu}
The operator $\pi L^{(\alpha)}$ is the generator of the modular group 
	\[
		t \mapsto \Delta_{W^{(\alpha)}}^{i t} \; , 
		\qquad t \in \mathbb{R}  \; , 
	\]
for the pair $\bigr(\mathcal{R}(I_\alpha), \widehat{\Omega} \bigl)$.
\end{theorem}

\begin{proof}  Combining 
Theorem~\ref{9.2.4} and Corollary~\ref{coll:9.3.3}, the result follows immediately.  
\end{proof} 

\begin{theorem} 
\label{H}
The relative modular operator for the triple $\bigr(\mathcal{R}(I_\alpha),
\Omega_\circ, \widehat{\Omega}\bigl)$ is
	\begin{equation}
	\label{rel-mod-op}
		\Delta_{\widehat{\Omega},\Omega_\circ}
		= \frac{ {\rm e}^{-2\pi \HFock^{(0)} }  }{ 
		\|{\rm e}^{-\pi \HFock^{(\alpha)}}\Omega_\circ  \|^2 } \, ; 
	\end{equation}
the corresponding relative modular conjugation $J_{\widehat{\Omega},\Omega_\circ}$
coincides with the (free) modular conjugation $J_\circ^{(0)}$ introduced in \eqref{s1-reflections}. 
\end{theorem} 

\begin{remark} In the terminology introduced by Araki in \cite{A2}, 
$V^{(\alpha)}$ is the \emph{relative Hamiltonian} for the triple 
$\bigr(\mathcal{R}(I_\alpha),\Omega_\circ, \widehat{\Omega}\bigl)$. 
\end{remark}

\begin{proof} 
Let $A\in \mathcal{R} (I_+)$. It follows from \eqref{hosaut} that
	\begin{align*}
		J_\circ^{(0)} A^*\widehat{\Omega} 
		=  
		\widehat{\alpha}_{\Lambda_1 (t)} (A) 
		\widehat{\Omega} \Bigl. \Bigr|_{t= i \pi}
		= \frac{ \widehat{\alpha}_{\Lambda_1 (t)} (A) 
		{\rm e}^{-\pi \HFock^{(0)} }  \Omega_\circ }{ 
		\|{\rm e}^{-\pi  \HFock^{(0)} }\Omega_\circ  \| }  \Bigl. \Bigr|_{t= i \pi}
			= \frac{ {\rm e}^{-\pi \HFock^{(0)} } A \Omega_\circ }{ 
		\|{\rm e}^{-\pi  \HFock^{(0)}  }\Omega_\circ  \| }   \, . 
	\end{align*}
Since $J_\circ^{(0)} J_\circ^{(0)} = \mathbb{1}$, this verifies that 
	\[
		 J_\circ^{(0)} \frac{ {\rm e}^{-\pi \HFock^{(\alpha)}}  A \Omega_\circ }{ 
		\|{\rm e}^{-\pi \HFock^{(\alpha)}}\Omega_\circ  \| }  
 		=
		A^* \Omega \; , \qquad A\in \mathcal{R} (I_+)  \; . 
	\]
Hence the relative modular operator $\Delta_{\widehat{\Omega},\Omega_\circ}$ 
is given by \eqref{rel-mod-op}
and $J_\circ^{(0)}$ is the relative modular conjugation 
$J_{ \widehat{\Omega} ,\Omega_\circ}$. 
\end{proof}

\begin{corollary}
The one-parameter family  (see \eqref{s1-reflections} and \eqref{rel-mod-op}
for more explicit expressions)
	\begin{equation}
	\label{b3}
		u_t 
		\doteq \Delta_{\widehat{\Omega}, \Omega_\circ}^{it} {\Delta^{(0)}_{\circ}}^{-it} \; , \qquad
		t \in \mathbb{R} \; , 
	\end{equation}
of unitaries in $\mathcal{R} (I_+)$ is continuous and satisfies the \emph{cocycle condition} 
	\begin{equation}
	\label{b4}
		u_t \widehat{\alpha}_{\Lambda_1 (t)} (u_s) = u_{t+s} \; , 
		\qquad t, s \in \mathbb{R} \; . 
	\end{equation}
Moreover, $u_t$  \emph{intertwines} the interacting  
$\widehat{\alpha}_{\Lambda_1 (t)}$ and the free 
boosts $\widehat{\alpha}_{\Lambda_1 (t)}^\circ$, \emph{i.e.}, 
	\begin{equation}
	\label{sigma-neu}
		\widehat{\alpha}_{\Lambda_1 (t)} (A) 
		= u_t \widehat{\alpha}_{\Lambda_1 (t)}^\circ (A) u_t^* \; , 
		\qquad A\in \mathcal{R} (I_+) \; 
		\quad t \in \mathbb{R} \; .   
	\end{equation}
\end{corollary}

\begin{theorem}[Uniqueness of the interacting de Sitter vacuum state]
\label{keyresult3}
For each $\alpha \in [ \, 0, 2 \pi ) $, the restricted state 
	\[
		\widehat{\omega}_{\upharpoonright \mathcal{R} (I_\alpha) } (A) =  
		\langle \widehat{\Omega} , A \widehat{\Omega} \rangle \; , 
		\qquad A \in \mathcal{R} (I_\alpha) \; , 
	\]
is the unique $\widehat{\alpha}_{\Lambda^{(\alpha)}}$-KMS state 
on~$\mathcal{R} (I_\alpha)$ and (therefore)
$\widehat{\omega}$ is the unique de Sitter vacuum state 
for the $W^*$-dynamical system $\bigl( \mathcal{R} (S^1), 
\widehat{\alpha}_{\Lambda} \bigr)$.
\end{theorem}

\begin{proof}
For the free field the $\widehat{\alpha}^{\circ}_{\Lambda^{(\alpha)}}$-KMS state 
on~$\mathcal{R} (I_\alpha)$ is unique, thus 
$\mathcal{R} (I_\alpha)$ is a factor, and uniqueness of the 
interacting state now is 
a direct consequence of \cite[Proposition~5.3.29]{BR}, as was 
kindly pointed out to us by Jan Derezinski. 
\end{proof}
 
\chapter{Local Algebras for the Interacting Field}
\label{ch:10}

Before we will discuss the  
covariant net of local algebras for the interacting 
quantum field, we will quickly verify that the newly constructed 
representation of $SO(1,2)$ constructed in the previous section 
respects finite speed of propagation. Hence, it is in the class 
of representations of the Lorentz group, which are called \emph{causal}.

\section{Finite speed of propagation for the ${\mathscr P}(\varphi)_2$ model}
\label{sec:fsol3}

The set	(see Proposition \ref{ialpha} for an explicit formula)
	\[
			I (\alpha , t) 
		= S^1 \cap \Gamma \bigl( \Lambda^{(\alpha)} (t) I \bigr)
	\]
describes the localisation region for the Cauchy data, which 
can influence space-time points in the set $\Lambda^{(\alpha)} (t) I$, $t \in \mathbb{R}$, fixed. 
Here $\Gamma (X) = \cup_{x \in X} \Gamma^- (x)\cup  \Gamma^+ (x)$ denotes the union of the 
future and the past of the points in some subset $X$ of $dS$. 

\begin{theorem}
\label{fst-2-theorem}
\emph{(Finite speed of propagation)}.
Let $I \subset S^1$ 
be an open interval. Then 
	\begin{equation}
		\widehat 
		\alpha_{\Lambda^{(\alpha)} (t)}
 		\colon  
		{\mathcal R} (I)\hookrightarrow 
		{\mathcal R} \bigl( I (\alpha , t)  \bigr) \; .
		\label{e6.1ef}
	\end{equation}
\end{theorem}

\begin{proof}
The following argument  is similar to the one given in the proof of \cite[Theorem 4.1.2]{GJcp}. 
We have seen in Theorem~\ref{fst-theorem}  that 
	\begin{equation}
		\label{e6.1d}
		\widehat  \alpha^{\, \circ}_{\Lambda^{(\alpha)} (t)} \colon 
		{\mathcal R} (I) \hookrightarrow 
		{\mathcal R} \bigl( I (\alpha , t) \bigr)
		\; .
	\end{equation}
We can now exploit the fact that according to 
Theorem \ref{th5.2} iii.) the automorphism 
$\widehat \alpha_{\Lambda^{(\alpha)} (t)}$ is unitarily implemented 
by~${\rm e}^{i t \LFock^{(\alpha)}}$, where
$\LFock^{(\alpha)}=\overline{\LFock^{(\alpha)}_\circ + V (\mathbb{cos}_\alpha) }$ with
	\[
		 V (\mathbb{cos}_\alpha)=\int_{S^1} r {\rm d} \psi \, \cos( \psi - \alpha)  \;   
		 {:}{\mathscr P}( \mathbb{\Phi} (0,\psi)){:}  \; .
	\]
Trotter's product formula \cite[Theorem VIII.31]{RS} yields 
	\[
	{\rm e}^{i t \LFock^{(\alpha)} }= \operatorname*{s-lim}\limits_{n\to\infty} 
	\left({\rm e}^{i t\LFock^{(\alpha)}_\circ /n}{\rm e}^{i tV (\mathbb{cos}_\alpha)/n}
	\right)^{n} \; . 
	\]
Hence
	\begin{equation}
	 \widehat \alpha_{\Lambda^{(\alpha)} (t)} (A)
	 = \operatorname*{s-lim}\limits_{n\to\infty} \left( \widehat\alpha^{\circ}_{\Lambda^{(\alpha)} (t/n)}\circ
	\widehat {\gamma}^{(\alpha)}_{t/n} \right)^{n}(A) \; , 
	\qquad A\in {\mathcal R} ( S^1 ) \; ,
	\label{e6.1e}
	\end{equation} 
with 
	\[ 
		\widehat {\gamma}^{(\alpha)}_t  (A) 
		= {\rm e}^{i  t V (\mathbb{cos}_\alpha)  }A
		{\rm e}^{-i t V (\mathbb{cos}_\alpha) } \; . 
	\]
Note that $\widehat {\gamma}^{(\alpha)}$ has zero propagation 
speed \cite{GJcp}, which means that for every open interval ${J} \subset  S^1 $ 
there exists $V_{\rm loc}^{(\alpha)}$ affiliated\footnote{Let ${\mathcal R}$ be a von 
Neumann algebra acting on a Hilbert space ${\mathcal H}$. A closed and densely defined 
operator $A$ is said to be affiliated with ${\mathcal R}$
if $A$ commutes with every unitary operator $U$ in the commutant of ${\mathcal R}$. }
to ${\mathcal U} ({J})$ such that,  for all $ t \in {\mathbb R}  $,
	\[
		{\rm e}^{it  V (\mathbb{cos}_\alpha) } A {\rm e}^{-it V (\mathbb{cos}_\alpha) }
		= {\rm e}^{it V_{\rm loc}^{(\alpha)}} A {\rm e}^{-it V_{\rm loc}^{(\alpha)}} \; , 
		\qquad   A \in {\mathcal R}  ({J})   \; . 
	\]
Here 
	\[
		 V_{\rm loc}^{(\alpha)}=\int_{J} r {\rm d} \psi \, \cos( \psi - \alpha)  \;   
		 {:}{\mathscr P}( \mathbb{\Phi} (0,\psi)){:}  \; .
	\]
Consequently,  
	\begin{equation}
	\label{zeroprop}
		\widehat {\gamma}^{(\alpha)}_t ({\mathcal R} (J)) 
		= {\mathcal R} (J)  
	\qquad  \forall t  \in {\mathbb R}  \; . 
	\end{equation}
Now (\ref{e6.1ef}) follows from \eqref{e6.1e} and  \eqref{e6.1d}.
\end{proof} 

\begin{theorem} \label{p6.1b}
\emph{(Time slice axiom).}
For $I \subset S^1 $, let ${\mathcal B}^{(\alpha)}_{\epsilon}(I)$  
denote the
von Neumann algebra generated by 
	\[
		\left\{ \widehat \alpha_{\Lambda^{(\alpha)} (t )} (A) 
		\mid A\in {\mathcal U} (I), \; |t |< \epsilon \right\}.
	\]
Then
	\begin{equation}
		\label{intersectBa}
		\bigcap_{\epsilon >0} {\mathcal B}^{(\alpha)}_{\epsilon}(I)
		= {\mathcal R}(I) \; , \qquad I  \subset S^1 \; .
	\end{equation}
Both sides in \eqref{intersectBa} are independent of $\alpha$.
\end{theorem}

\begin{proof} 
Let us first prove that $\bigcap_{r>0} 
{\mathcal B}^{(\alpha)}_{\epsilon}(I)\subset {\mathcal R}(I)$. 
Using \eqref{e6.1ef} 
and ${\mathcal U}(I)\subset {\mathcal R}(I)$, we see that 
	\[
	 {\mathcal B}^{(\alpha)}_{\epsilon}(I)\subset 
	 {\mathcal R} \bigl( I (\alpha, \epsilon ) \bigr)
	 \qquad \epsilon  >0 \; .
	 \]
According to Proposition \ref{l6.1}, the local time-zero algebras are regular from 
the outside. This
implies~$\bigcap_{\epsilon >0} 
{\mathcal B}^{(\alpha)}_{\epsilon}(I)\subset {\mathcal R}(I)$.

Let us now prove that ${\mathcal R}(I)\subset 
\bigcap_{r>0} {\mathcal B}^{(\alpha)}_{\epsilon}(I)$. 
Using again Proposition \ref{l6.1} (this time using that the local 
time-zero algebras are regular from 
the inside),  it suffices to show that  
for each open interval $J$ with~$\overline{J}\subset I$  
	\begin{equation}
		\label{e6.02b}
		{\mathcal R}({J})\subset {\mathcal B}^{(\alpha)}_{\epsilon}(I) 
		\qquad  
		\forall r>0  \; . 
	\end{equation}
We note that in the proof of Theorem \ref{p6.1} we have already shown that 
	\[
		{\mathcal R}({J})\subset \bigcap_{\epsilon >0}
		\bigl\{ \widehat {\alpha}^{\circ}_{\Lambda^{(\alpha)} (t )} (A) 
		\mid A\in {\mathcal U} (I), \; |t |< \epsilon \bigr\}.
	\]
Thus, it remains to show that this statement remains 
true, if $\widehat {\alpha}^{\circ}$ is replaced 
by $\widehat {\alpha}$. In other words, we have to show 
that for $A\in {\mathcal U} (I)$ and 
some $\epsilon >0$, we have
	\begin{equation}
	\label{l-alpha-A}
		{\rm e}^{i t\LFock^{(\alpha)}_\circ }
		A{\rm e}^{-i t\LFock^{(\alpha)}_\circ } \in 
		{\mathcal B}^{(\alpha)}_{\epsilon}(I) \; ,  
		\qquad |t|<  \epsilon \; .
	\end{equation}
For $|t|\leq \delta$ the unitary group ${\rm e}^{i t \LFock^{(\alpha)} }$ 
with 
	\[
		\LFock^{(\alpha)}=\overline{\LFock^{(\alpha)}_\circ 
		+  V (\mathbb{cos}_\alpha) }
	\]
induces the dynamics $\widehat \alpha_{\Lambda^{(\alpha)} (t)}$ 
on ${\mathcal U} (I) \subset {\mathcal R} ({I})$.  Apply \cite[Proposition 2.5]{GeJII} 
to obtain 
	\[
		{\rm e}^{i t \LFock^{(\alpha)}_\circ }
		=\operatorname*{s-lim}\limits_{n\to\infty}{\rm e}^{i t \LFock^{(\alpha)}_n } \; ,  
		\qquad \: t\in {\mathbb R} \; ,
	\]
for $\LFock^{(\alpha)}_n 
=\overline{\LFock^{(\alpha)}_\circ + V (\mathbb{cos}_\alpha) 
- V^{(\alpha)}_{n}}$, where $V^{(\alpha)}_n =
V (\mathbb{cos}_\alpha) \mathbb{1}_{\{| V (\mathbb{cos}_\alpha)  |\leq n\}}$. 
Since $V^{(\alpha)}_n$ is bounded,
	\[
	\LFock^{(\alpha)}_n
	= \overline{\LFock^{(\alpha)}_\circ + V (\mathbb{cos}_\alpha) - V^{(\alpha)}_n}
	= \LFock^{(\alpha)} - V^{(\alpha)}_n \;  
	\]
and Trotter's formula yields
	\[
 		{\rm e}^{i t \LFock^{(\alpha)}_n }
		= \operatorname*{s-lim}\limits_{p\to \infty} 
		\left({\rm e}^{i t \LFock^{(\alpha)}/p}{\rm e}^{-i t
		V^{(\alpha)}_n /p} \right)^{p} \; .
	\]
Hence, for $A\in  {\mathcal U} (I) \subset {\mathcal R} ({I})$ 
and $|t|< \epsilon$,
	\begin{align*}
	& {\rm e}^{i t \LFock^{(\alpha)}_\circ  }A
	{\rm e}^{-i t \LFock^{(\alpha)}_\circ  } \\
	& \quad = \operatorname*{s-lim}\limits_{n\to\infty}\; 
	\operatorname*{s-lim}\limits_{p\to \infty} \; 
	\bigl({\rm e}^{i t \LFock^{(\alpha)}  /p}
	{\rm e}^{-i t V^{(\alpha)}_n/p} \bigr)^{p}A \bigl({\rm e}^{i t
	V^{(\alpha)}_n/p}{\rm e}^{-i t \LFock^{(\alpha)}  /p}\bigr)^{p} \; .
	\end{align*}
But for $p \in {\mathbb N}$
	\begin{multline*}
		\left({\rm e}^{i t  \LFock^{(\alpha)}  /p}
		{\rm e}^{-i t V^{(\alpha)}_n /p} \right)^{p}
		A 
		\left({\rm e}^{i t V^{(\alpha)}_n /p}
		{\rm e}^{-i t \LFock^{(\alpha)}  /p}\right)^{p}  
		 \\
		= \left( \widehat \alpha_{\Lambda^{(\alpha)} (t/p)} 
		\circ \widehat {\gamma}^{(\alpha)}_n (-t/p) \right)^{p}(A) \; ,
	\end{multline*}
where $\widehat {\gamma}^{(\alpha)}_n$ is the dynamics implemented by the unitary
group $t \mapsto {\rm e}^{-i t V^{(\alpha)}_n }$. 
Just like $ \widehat {\gamma}$, the automorphisms $\widehat {\gamma}_n$ have zero 
propagation speed; see \eqref{zeroprop}. This implies 
	\[
	\bigl( \widehat \alpha_{\Lambda^{(\alpha)} (t/p)} \circ 
	\widehat {\gamma}^{(\alpha)}_n (t/p) \bigr)^{p}(A)
	\in {\mathcal B}^{(\alpha)}_{ \epsilon}(I) \; ,  
	\qquad |t|< \epsilon  \; , \quad p \in {\mathbb N} \; . 
	\]
Taking the limit $n \to \infty$ (and recalling that  
that $ \widehat {\gamma}
= \lim_{n \to \infty} \widehat {\gamma}_n$ has zero 
propagation speed too), \eqref{l-alpha-A} follows. 
\end{proof} 

\begin{remark} Thus, 
if we use the left hand side of \eqref{intersectBa} to define ${\mathcal R}_{\rm int} (I)$, then 
	\[
		{\mathcal R}_{\rm int} (I)= {\mathcal R} (I) \; \qquad 
		\forall I \subset S^1 \; . 
	\]
We have seen earlier that ${\mathcal R}_{\rm int} (I_\alpha)= {\mathcal R}  (I_\alpha)$  for all 
half-circles $I_\alpha$. If $I$ is contained in some half-circle, then it follows that 
	\[
		{\mathcal R}_{\rm int} (I) = \bigcap_{I \subset I_\alpha} {\mathcal R}_{\rm int}  (I_\alpha)
	\]
can be identified with the intersection 
of all algebras ${\mathcal R}_{\rm int}  (I_\alpha)$ associated to the half-circles $I_\alpha$, which contain $I$.
\end{remark}

\section{The Haag--Kastler Axioms}
\label{sec:10.2-new}

Let us recall the definition \eqref{free-loc-cov-alg} of the algebra for the free scalar boson field 
associated to the wedge $W_1$. It acts on the covariant Fock space ${\mathcal H} (dS) 
= \Gamma ({\mathfrak h}(dS))$. Using the unitary operator ${\mathfrak U}$
introduced in Proposition~\ref{Prop5.7} (and further specified in 
Proposition \ref{prop:4.10.5}, in particular in~\eqref{s1-to-dS}), we can map 
the interacting dynamics and the interacting vacuum vector to the covariant Fock 
space ${\mathcal H} (dS)$:
\label{covintvacuumpage}	
	\[
		U(\Lambda) \doteq \Gamma ({\mathfrak U}) \widehat{U}(\Lambda)
		\Gamma ({\mathfrak U}^{-1}) \; , \qquad \Lambda \in O(1,2) \; , 
		\qquad
		\Omega \doteq  \Gamma ({\mathfrak U}) \widehat{\Omega} \; . 		
	\]
The group of (anti-)unitary operators $U(\Lambda)$, $\Lambda \in O(1,2)$, gives rise 
to a group of automorphisms: 
	\[
		\alpha_\Lambda (A) = U(\Lambda) A U(\Lambda)^{-1} \; , 
		\qquad \Lambda \in O(1,2) \; , \qquad A \in \mathscr{A}_\circ(dS) \; . 
	\]

\begin{definition}
\label{locoinfield}
We define the following von Neumann algebras:  
\begin{itemize}
\item[$ i.)$] For an arbitrary wedge $W= \Lambda W_1$, $\Lambda \in
  SO_0(1, 2)$, we set 
	 \begin{equation}
	 \label{a-w-U}
	 	{\mathscr A} ( W ) 
		\doteq \alpha_\Lambda \bigl( \mathscr{A}_\circ (W_1) \bigr) \;  . 	
	\end{equation}
\item[$ ii.)$] 
For an arbitrary bounded, causally complete, convex region (these are the de Sitter analogs of 
the double cones) ${\mathcal O} \subset dS$, we set   
	\begin{equation} 
		\label{localfield}
		 {\mathscr A} ({\mathcal O}) 
		\doteq \bigcap_{W \supset {\mathcal O} } {\mathscr A}  \bigl( W \bigr) \; . 
	\end{equation}
\end{itemize}
The inclusion preserving map 
	\[
		{\mathcal O} \mapsto {\mathscr A} ({\mathcal O}) 
	\]
is called, in a slight abuse\footnote{Technically speaking, this family of algebras is not a ``net'', since 
in de Sitter space not every pair of double cones is contained in a double cone.} of the term,
the net of local von Neumann algebras for the interacting bosonic field on the de Sitter space $dS$. 
\end{definition}

Finite speed of propagation as expressed in \eqref{e6.1ef} 
implies that for any wedge $W$, the algebra $\mathscr{A} (W) $ is contained 
in the time-zero Weyl algebra
\begin{equation}
		\left\{ W_F(h)  \mid  h \in \mathfrak{h} (J'')  , \; J \doteq \Gamma (W) \cap S^1 \; 
 		\right\} '' \, .
		\label{e6.1ef-2}
\end{equation}
As a consequence, the time-zero algebras for the free and the interacting theory coincide: 

\begin{theorem}[J\"akel \& Mund \cite{MJ-2}] 
\label{AA0}
The net of local algebras satisfies the following property: for any interval $I \subset S^1$ on the Cauchy surface
we have
	\[
		\mathscr{A} ({\mathcal O}_I) = \mathscr{A}_\circ ({\mathcal O}_I) \; , 
		\qquad I \subset S^1 \; . 
	\]
\end{theorem}

\begin{proof} The proof follows ideas exposed in the proof of Proposition~\ref{Prop-ii.4}~$ii.)$. 
The key step is to show that for any wedge $W$ which contains ${\mathcal O}_I$, we have 
	\[
		\mathscr{A}  (W')  \subset \left\{ W_F (h) \mid h \in \mathfrak{h} \bigl( ( I^c )'' \bigr) \right\} '' ,
	\]
where $I^c \doteq S^1 \setminus \overline{I} $ and $( I^c )''$ denotes it causal completion.
As the edges of $W$ are necessarily space- or light-like 
to~${\mathcal O}_I$, this inclusion follows from \eqref{e6.1ef}. By duality, 
	\[
		\mathscr{A}  (W)  \supset 
		 \mathscr{A}_\circ ( {\mathcal O}_I)  \, , 
	\]
whenever $W$ includes ${\mathcal O}_I$. 
\end{proof}

\goodbreak

\begin{remark}
The circle $S^1$, which we use to identify the free field and the interacting field, could be replaced by 
any space-like geodesic $\Lambda S^1$, $\Lambda \in SO_0(1,2)$. The Fock space simply carries two (in fact, 
infinitely many if one just varies the coupling constants) nets of local algebras, namely 
$\mathcal{O} \mapsto \mathscr{A}_\circ (\mathcal{O})$ and $\mathcal{O} \mapsto 
\mathscr{A} (\mathcal{O})$, 
and one may identify them on any of the space-like geodesic $\Lambda S^1$, $\Lambda \in SO_0(1,2)$.
\end{remark}

\goodbreak 
We can now state the following important results: 

\begin{theorem}[J\"akel \& Mund \cite{MJ-2}]
\label{th:6}
The net $ {\mathcal O} \mapsto \mathscr{A} ({\mathcal O}) $ representing the interacting 
quantum theory satisfies the following Haag--Kastler axioms:
\begin{enumerate}
\item [$i.)$]  {\em (Isotony).} The local algebras satisfy 
	\[
		\mathscr{A} ({\mathcal O}_1) \subset \mathscr{A} ({\mathcal O}_2)  
		\quad \hbox{if} \quad {\mathcal O}_1 \subset {\mathcal O}_2  \; . 
	\]
Here ${\mathcal O}_1$ and ${\mathcal O}_2$ are either double cones or wedges (but the result extends to
arbitrary regions once \eqref{Additivity} has been established). 
\item [$ii.)$]  {\em (Locality).} The local algebras satisfy 
	\[
		\mathscr{A} ({\mathcal O}_1) \subset \mathscr{A} ({\mathcal O}_2)' 
		\quad \hbox{if} \quad {\mathcal O}_1 \subset {\mathcal O}_2 ' \; . 
	\]
Here ${\mathcal O}'$ denotes the space-like complement of 
${\mathcal O}$ in $dS$ and $\mathscr{A} ({\mathcal O})'$ is the commutant 
of $\mathscr{A} ({\mathcal O})$ in $\mathscr{B}\bigl({\mathcal H} (dS)\bigr)$. 
\item [$iii.)$] {\em (Covariance).}
The representation $U \colon \Lambda \mapsto U ( \Lambda) $ acts geometrically, \emph{i.e.}, 
	\[
		U ( \Lambda)  \mathscr{A} ({\mathcal O}) U ( \Lambda)^{-1} 
		= \mathscr{A} (\Lambda {\mathcal O} )  \; , \qquad \Lambda  \in SO_0 (1,2) \; .  
	\]
\item [$iv.)$] {\em (Existence and Uniqueness of the Vacuum \cite{BoB}).} 
There exists a unique (up to a phase\footnote{The phase is uniquely fixed, if one insists 
that the vector $\Omega$ lies in the natural positive 
cone~$\mathcal{P}^\natural (\mathscr{A}_\circ (W_1), \Omega_\circ)$.}) 
unit vector in ${\mathcal H} (dS)$, namely $\Omega$, which 
\begin{enumerate}
\item [a.)] is  invariant under the action of $U  (SO_0(1,2))$; 
\item [b.)] satisfies the geodesic KMS condition: for every wedge $W = \Lambda W_1$, 
$\Lambda \in SO_0(1,2)$, the partial state
	\[ 
		\qquad
		\qquad 
		\omega_{\upharpoonright \mathscr{A}(W)} (A) \doteq \langle \Omega, A  \Omega \rangle \; , 
		\quad A \in  \mathscr{A} (W) \; ,
	\]
satisfies the KMS-condition at inverse temperature~$\beta = 2 \pi r $ with respect to the one-parameter 
group $t \mapsto U (\Lambda_{W}( t/r)) $,  $t \in {\mathbb R} $.
\end{enumerate}
\item[$v.)$]  {\em (Additivity).}  
For $X$ a double cone or a wedge, there holds
	\begin{equation} 
		\label{Additivity} 
		\mathscr{A}(X) = \bigvee_{{\mathcal O\subset X}} \mathscr{A}({\mathcal O}) \; .
	\end{equation}
The right hand side denotes the von Neumann algebra generated by the 
local algebras associated to double cones ${\mathcal O}$ contained in $X$.
(It thus makes sense to define~$\mathscr{A}(X)$ for arbitrary regions
$X$ by Eq.~\eqref{Additivity}.) 
\item [$v'.)$] {\em (Weak additivity).}
For each double cone ${\mathcal O} \subset dS$ there holds 
	\[  
 	 	\bigvee_{\Lambda \in SO_0 (1,2) } \mathscr{A}  (\Lambda
                {\mathcal O}) = \mathscr{A}_\circ  (dS)  
                  \quad \bigl( = \mathscr{B}\bigl( {\mathcal H} (dS) \bigr) \bigr) 
                  \; . 
	\] 
\item [$vi.)$] {\em(Time-slice axiom \cite{CF}).}  
Let $I$ be an interval on a geodesic Cauchy surface and let $I''$ be its causal completion.
Let $\Xi \subset I'' $ be a neighbourhood of $I$. Then 
	\[
		\mathscr{A}( \Xi ) = \mathscr{A}(I'') \; ,  
	\] 
where both algebras are defined via Eq.~\eqref{Additivity}. In particular, the algebra of observables located within an 
arbitrary small time--slice coincides  with the algebra of all observables.
\end{enumerate}
\end{theorem}

\begin{proof} 
Property $i.)$, \emph{isotony}, follows directly from the definition; see \eqref{localfield}.
Next, let us establish property $ii.)$, \emph{locality}.
If $\mathcal{O}_1$ and~$\mathcal{O}_2$ are two
space-like separated causally complete, open and bounded regions,
then there exists a wedge  $W = \Lambda W_1$ such that
	\[
		\mathcal{O}_1 \subset W 	\quad \text{and} \quad \mathcal{O}_2 \subset W' \, . 	
	\]
Now the interacting net inherits wedge duality $\mathscr{A} (W')=\mathscr{A}(W)'$
from the identity $\mathscr{A}_\circ (W_1')=\mathscr{A}_\circ (W_1)'$
using \eqref{a-w-U}.   
These facts imply locality. 

Now, let us prove property $iii.)$, \emph{covariance}. 
Let $\Lambda \in SO_0(1,2)$ be fixed. By construction, the set of all wedges equals
$\{ \Lambda W_1 \mid \Lambda \in SO_0(1, 2) \}$. Thus, 
	\begin{align*}
	{\mathscr A} (\Lambda {\mathcal O} ) 
	& =  \bigcap_{\Lambda {\mathcal O} \subset  \Lambda W} 
	{\mathscr A} ( \Lambda W)
	 =   \bigcap_{ {\mathcal O} \subset W} 
	U ( \Lambda) {\mathscr A} ( W) U ( \Lambda)^{-1} 
	\\
	& =  U ( \Lambda)
	\Bigl( \; \bigcap_{ {\mathcal O} \subset W}  {\mathscr A} (W) \Bigr) U ( \Lambda)^{-1}
	=  U ( \Lambda)  {\mathscr A} ( {\mathcal O} )   U ( \Lambda)^{-1} \; ,
	\end{align*}
proving covariance. 

Property  $iv.)$ is established next: \emph{existence} of the de Sitter vacuum is guaranteed by 
construction, as the state induced by the vector $\Omega$ is a thermal state for the Hawking 
temperature $(2\pi r)^{-1}$ with respect to modular group for the pair $\bigl(\mathscr{A}_\circ (W_1), 
\Omega\bigr)$. By covariance, this property extends to arbitrary wedges. \emph{Uniqueness of the 
de Sitter vacuum state}, was established in Theorem~\ref{keyresult3}. 

Next, we will establish property $v.)$, \emph{additivity}.
The inclusion 
	\[ 
		\mathscr{A}(X) \supset \bigvee_{{\mathcal
    			O\subset X}} \mathscr{A}({\mathcal O}) 
	\]
is a consequence of isotony. Moreover, if $X$ is a double cone, then 
$X$ itself is among the double cones on the right hand side, so the inclusion $\subset$  
automatically holds.  It remains to prove the inclusion $\subset$ if $X$ is a wedge.
If $X=W_1$, then $\mathscr{A}(X)$ coincides with $\mathscr{A}_\circ(X)$, for which 
\eqref{Additivity} implies
	\[
		\mathscr{A}_\circ(X) =  \bigvee_{R_0 (\alpha) I\subset I_+}\mathscr{A}_\circ \bigl(R_0 (\alpha) 
		{\mathcal O}_I  \bigr) \; , 
		\qquad \alpha \in [0, 2\pi) \; , 
	\]
where $I$ is an (arbitrarily small) open interval contained in $I_+\doteq
W_1\cap S^1$, whose causal completion ${\mathcal O}_I=I''$.  Thus,
	\[
		\mathscr{A} \bigl(W_1 \bigr) 
		= \bigvee_{R_0 (\alpha) I \subset I_+}\mathscr{A} \bigl(R_0 (\alpha) {\mathcal O}_I \bigr)
			\subset \bigvee_{{\mathcal O}\subset X}\mathscr{A}({\mathcal O}) \; .        
	\]
Thus, for $X=W_1$ the inclusion $\subset$ holds. By covariance, it
also holds if $X$ is any other wedge. 

Property $v'.)$, \emph{weak additivity}, follows from a similar argument: for each double cone
${\mathcal O} \subset dS$ there exists a Lorentz transformation
$\Lambda_0 \in SO_0(1,2)$ such that ${\mathcal O}= \Lambda_0  \mathcal{O}_I$ 
for some open interval $I \subset S^1$. Now
	\begin{align*}
 	 	\bigvee_{\Lambda \in SO_0(2,1)} \mathscr{A}  \bigl( \Lambda {\mathcal O} \bigr) 
                & = \bigvee_{\Lambda \in SO_0(2,1)} \mathscr{A}  \bigl( \Lambda \Lambda_0 {\mathcal O}_I \bigr) \\
		& = \bigvee_{\Lambda \in SO_0(2,1)} \mathscr{A}  \bigl(\Lambda  {\mathcal O}_I \bigr)
		\supset \bigvee_{ \alpha \in [0, 2\pi) } \mathscr{A}  \bigl(R_0 (\alpha) {\mathcal  O}_I \bigr) \\
		&= \bigvee_{\alpha \in [0, 2\pi) } \mathscr{A}_\circ  \bigl(R_0 (\alpha) {\mathcal  O}_I \bigr) 
		= \mathscr{B}\bigl( {\mathcal H} (dS) \bigr)  \; . 
	\end{align*} 
Again, the last equality relies on the additivity property for the one-particle space. Hence, property $v'.)$, 
\emph{weak additivity}, is established. 

Let us prove property $vi.)$, the \emph{time-slice axiom}. As the geodesic Cauchy surfaces are precisely 
the Lorentz transforms  
of intervals on the equator $S^1$, it 
is sufficient to consider an interval $I \subset S^1$ and a neighbourhood $\Xi \subset I''$
of $I$. If the length of $I$ is less than $\pi r$, then $I''$ is a double cone and ${\mathscr A}(I'')={\mathscr A}_\circ(I'')$. 
Pick any double cone $\mathcal{O}\subset \Xi$ with base on $S^1$. Then, the additivity property  
of the free net implies that 
	\[ 
		{\mathscr A}_\circ(I'') = 
		\bigvee_{R_0 (\alpha) \mathcal{O}\subset \Xi }
   		{\mathscr A}_\circ \bigl(R_0 (\alpha) \mathcal{O} \bigr) \; , \qquad \alpha \in [0, 2\pi) \; .  
	\]
Now ${\mathscr A}_\circ \bigl( R_0 (\alpha) \mathcal{O} \bigr)$ coincides with 
${\mathscr A}\bigl( R_0 (\alpha) \mathcal{O} \bigr)$, and hence the above identity implies 
${\mathscr A}(I'')\subset {\mathscr A}( \Xi )$. The other inclusion follows from isotony. 

\goodbreak
If the length of $I$ is at least $\pi r$, then additivity implies that 
${\mathscr A}(I'')$ is generated by the wedge algebras ${\mathscr A}(R_0(\alpha) W_1)
= {\mathscr A}_\circ(R_0(\alpha) W_1)$, $R_0(\alpha) W_1 \subset I''$. 
Hence, as before, ${\mathscr A}(I'')={\mathscr A}_\circ(I'')$.
As before, the additivity property  of the free net implies that 
	\[ 
		{\mathscr A}_\circ(I'') = 
		\bigvee_{R_0 (\alpha) \mathcal{O}\subset \Xi }
   		{\mathscr A}_\circ(R_0 (\alpha) \mathcal{O}) \; , \qquad \alpha \in [0, 2\pi) \; .  
	\]
Now, just as before,  ${\mathscr A}_\circ 
\bigl( R_0 (\alpha) \mathcal{O} \bigr)$ coincides with 
${\mathscr A}\bigl( R_0 (\alpha) \mathcal{O} \bigr)$, 
and hence the above identity implies 
${\mathscr A}(I'')\subset {\mathscr A}( \Xi )$. The other inclusion follows from isotony. 
\end{proof}

\chapter{The Equations of Motion and the Stress-Energy Tensor}
\label{ch:11}

We conclude our construction of the interacting quantum theory by discussing 
the equations of motion and quantum analogs of the classical 
conversation laws discussed in Section \ref{SET},  
presenting expressions closer to those found in the physics literature.
The notation in this section is somewhat symbolic, but it  
can be given a rigorous meaning: while the 
point-like field and the point-like canonical momentum both do not exist 
as operators on Fock space, they can be understood as operator-valued 
distributions or as quadratic forms. 
Integrals over quadratic forms, when bounded from below, give rise to self-adjoint 
operators, which are (by an abuse of notation) denote by the same symbol; 
see Lemma~\ref{lm:10.2.1} (and its proof) below for a first illustration of what is meant.
These operators could also be 
interpreted as the limits of smeared out quantities, as in Theorem~\ref{uvtheo},
Theorem~\ref{wickooo} and equation \eqref{halberkreis} in Theorem~\ref{th5.2}, 
but we will not dwell on this point.  

\section{The stress-energy tensor}
\label{stress-energy-tensor}

One may introduce canonical time-zero fields $\varphi$ and 
canonical momenta $\pi$: they can be defined in 
terms of the Fock fields $\Phi_F $ on $\Gamma \bigl( \widehat{\mathfrak h}(S^1)\bigr) $ 
(see, \emph{e.g.}, \cite{RS}): 
	\[
		\widetilde \varphi (h) \doteq \Phi_F (h) \;, 
		\qquad 
		\widetilde \pi (g) \doteq  \Phi_F (i \omega g)\; , 
		\qquad h, g \in \widehat{\mathfrak h}(S^1) \quad 
		\text{real valued} \; . 
	\]
Thus $\widetilde \varphi (h) = \mathbb{\Phi}  (0,h) $ and  
$ \widetilde \pi (g) = -i \bigl[ \Gamma (\omega ), \mathbb{\Phi} (0,g) \bigr]$.  
They satisfy the {\em canonical commutation relations}
	\begin{align*}
		[ \widetilde \varphi(\psi), \widetilde \pi (\psi') ] 
			&=    \tfrac{i}{r}   \delta (\psi - \psi')\;  , 
			\\ 
		[ \widetilde \varphi(\psi), \widetilde \varphi(\psi') ] 
		&=   [ \widetilde \pi(\psi), \widetilde \pi (\psi') ] = 0 \; , 
	\end{align*}
in the sense of quadratic forms on $\Gamma \bigl( \widehat{\mathfrak
  h} (S^1)\bigr) $.

However, it is now more convenient to work on the Fock space 
over $L^2 (S^1, r {\rm d} \psi)$, using the map
	\begin{equation}
	\label{h-hat-L2}
		\widehat{\mathfrak h} (S^1) \ni 
		f \mapsto \tfrac{1}{\sqrt{2 \omega}} f \in L^2 (S^1, r {\rm d} \psi)   
	\end{equation}
to identify the two realisations of Fock space. The canonical fields 
and the canonical momenta then take the form 
	\begin{align*}
		\varphi(\psi) & =  \frac{1}{\sqrt{2}}
						\Big(\big(\omega^{-\frac{1}{2}} a\big)(\psi)^* 
						+ \big(\omega^{-\frac{1}{2}} a\big)(\psi)\Big)  \;,   
						\\
		\pi(\psi) & =  \frac{i}{\sqrt{2}}  
		 \Big(\big(  \omega^{\frac{1}{2}}  a\big)(\psi)^* - \big(  \omega^{\frac{1}{2}}  
		a\big)(\psi)\Big)
	\end{align*}
with
	\begin{equation}
	\label{C-A-C} 
		a(\psi) \doteq \sum_{k\in\mathbb{Z}} \frac{{\rm e}^{-ik\psi}}{\sqrt{2\pi r}} a_k
			\quad \text{and} \quad
		a(\psi)^* \doteq \sum_{k\in\mathbb{Z}} \frac{{\rm e}^{ik\psi}}{\sqrt{2\pi r}} a_k^* \; . 
	\end{equation}
Thus
	\[
		\big(\omega^{\pm \frac{1}{2}} a\big)(\psi) 
		=\sum_{k\in\mathbb{Z}} \frac{ \widetilde \omega(k)^{\pm \frac{1}{2}} 
		{\rm e}^{-ik\psi}}{\sqrt{2\pi r}}  a_k
			\quad \text{and} \quad
		\big(\omega^{\pm \frac{1}{2}} a\big)(\psi)^* 
		= \sum_{k\in\mathbb{Z}} \frac{ \widetilde \omega(k)^{\pm \frac{1}{2}}  
		{\rm e}^{ik\psi}}{\sqrt{2\pi r}} a_k^*  \; . 
	\]
Note that  $\big[\pi(\psi'), \; \varphi(\psi)\big]= - \frac{i}{r}  \delta(\psi-\psi')$ still holds.
Using
	\begin{align*}
		a(\psi) & =  \frac{1}{\sqrt{2}} \Big[ \big(\omega^{\frac{1}{2}}\varphi\big)
		(\psi) -i \big(\omega^{-\frac{1}{2}}\pi\big)(\psi)  \Big]  \; ,   \\
		a(\psi)^* & =  \frac{1}{\sqrt{2}} \Big[ \big(\omega^{\frac{1}{2}}\varphi\big)
		(\psi) +i \big(\omega^{-\frac{1}{2}}\pi\big)(\psi) \Big]  \; ,
	\end{align*}
one verifies that 
	\[
		\big[a(\psi')^*, \; a(\psi)\big] =  \tfrac{i}{2}  \left\{  \big[\pi(\psi'), \; \varphi(\psi)\big]
								+  \big[\pi(\psi), \; \varphi(\psi')\big]  \right\}
							=  \tfrac{1}{r} \delta(\psi-\psi')   
	\]
and $\big[a(\psi')^*, \; a(\psi)^*\big]=\big[a(\psi'), \; a(\psi)\big]=0$.

\bigskip
Up till now, we have constructed the generator of the free boost using group theoretic methods. Now we will show that 
the generator can also be expressed explicitly in terms of canonical fields. 

\label{LLSpage}

\begin{lemma} 
\label{lm:10.2.1}
Consider the Fock space over $L^2 (S^1, r \, {\rm d} \psi)$
and set\footnote{We note that normal ordering is not needed at this point.} 
	\begin{align*}
		\LLS^{(\alpha)}_\circ &= {\rm d}\Gamma 
		\bigl( \sqrt{ \omega } \, r \cos_{\psi + \alpha} \sqrt{\omega }  \bigr)  \\
		& =
		 \frac{1}{2} 
			\int_{S^1} r^2  \cos (\psi+ \alpha) \, {\rm d} \psi \;   
			\Bigl( \pi^2 (\psi) + \tfrac{1}{r^2}
			\bigl(\tfrac{ \partial \varphi }{\partial \psi}\bigr)^2 (\psi) 
			+ \mu^2 \varphi^2 (\psi)  \Bigr) \; . 
	\end{align*}
Then $\LLS^{(\alpha)}_\circ $
is the generator of the free boost 
$t \mapsto \widetilde{U}_\circ (\Lambda^{(\alpha)} (t))$  
and, in particular, 
$\LLS_1^\circ \equiv \LLS^{(0)}_\circ$
and 
$\LLS_1^\circ = \LLS^{(\pi/2)}_\circ$
are  the generators of the free boost
introduced\footnote{Note that \eqref{widehat-L} refers to the original Fock space
$\Gamma \bigl(\widehat{\mathfrak h} (S^1)\bigr)$.}  in \eqref{widehat-L}.  
\end{lemma}

\begin{proof}
We write, using the fact that $\widetilde \omega(k)=\widetilde \omega(-k)$,
	\begin{align*}
		\varphi(\psi) & =  \frac{1}{\sqrt{4\pi r }} \sum_{k\in\mathbb{Z}} 
						\widetilde \omega(k)^{-\frac{1}{2}}\Big( {\rm e}^{ik\psi}a_k^* + {\rm e}^{-ik\psi}a_k \Big) \; , \\
		\tfrac{\partial \varphi}{\partial\psi}(\psi) & = \frac{i}{\sqrt{4\pi r }} \sum_{k\in\mathbb{Z}} 
						\widetilde \omega(k)^{-\frac{1}{2 }}k\Big( {\rm e}^{ik\psi}a_k^* - {\rm e}^{-ik\psi}a_k \Big) \; , \\
		\pi(\psi) & =  \frac{i}{\sqrt{4\pi r }} \sum_{k\in\mathbb{Z}} 
						  \widetilde \omega(k)^{\frac{1}{2}} 
						 \Big( {\rm e}^{ik\psi}a_k^* - {\rm e}^{-ik\psi}a_k \Big) \; . 
	\end{align*}
One has
	\begin{align*}
		\mu^2 \varphi(\psi)^2 & =  \frac{\mu^2}{4\pi r } \sum_{k\in\mathbb{Z}} \sum_{l\in\mathbb{Z}} 
						\widetilde \omega(k)^{-\frac{1}{2}}\widetilde \omega(l)^{-\frac{1}{2}} \\
					& \quad \times \Big( {\rm e}^{i(k+l)\psi}a_k^*a_l^* + {\rm e}^{i(k-l)\psi} a_k^* a_l 
							+ {\rm e}^{-i(k-l)\psi} a_k a_l^* + {\rm e}^{-i(k+l)\psi} a_k a_l \Big) \, , \\
	\tfrac{1}{r^2} \bigl(\tfrac{ \partial \varphi }{\partial \psi}\bigr)^2 (\psi) & = 
			- \frac{1}{4\pi r^3} \sum_{k\in\mathbb{Z}}  \sum_{l\in\mathbb{Z}} 
			\widetilde \omega(k)^{-\frac{1}{2}}\widetilde \omega(l)^{-\frac{1}{2}}\, kl \\
					& \quad \times 
				\Big( {\rm e}^{i(k+l)\psi}a_k^*a_l^* - {\rm e}^{i(k-l)\psi} a_k^* a_l - {\rm e}^{-i(k-l)\psi} a_k a_l^* 
						+ {\rm e}^{-i(k+l)\psi} a_k a_l \Big)  \, , \\
		\pi(\psi)^2 & = - \frac{1}{4\pi r } \sum_{k\in\mathbb{Z}}  \sum_{l\in\mathbb{Z}} 
			\widetilde \omega(k)^{\frac{1}{2}}\widetilde \omega(l)^{\frac{1}{2}} \\
					& \quad \times 
				\Big( {\rm e}^{i(k+l)\psi}a_k^*a_l^* - {\rm e}^{i(k-l)\psi} a_k^* a_l - {\rm e}^{-i(k-l)\psi} a_k a_l^* 
						+ {\rm e}^{-i(k+l)\psi} a_k a_l \Big) \, . 
	\end{align*}
Next, define, for $j\in \mathbb{Z}$,
	\[ 
		S_j \doteq \int_{S^1}{\rm d} \psi \; \cos\psi \; {\rm e}^{ij\psi} 
		= \frac{1}{2}\int_{S^1}{\rm d}  \psi \; {\rm e}^{i(j+1)\psi}
				+ \frac{1}{2}\int_{S^1}{\rm d} \psi \; {\rm e}^{i(j-1)\psi} = \pi\big(\delta_{j,-1}+\delta_{j,1}\big) \;.
	\]
It is clear that $S_j=S_{-j}$ for all $j\in \mathbb{Z}$. Hence, we may write
	\begin{align*}
		\frac{1}{2} \int_{S^1} & r \,  {\rm d} \psi \; r \cos\psi \; \pi(\psi)^2  
			= - \frac{r}{8\pi  } \sum_{k \in \mathbb{Z}}  \sum_{l\in\mathbb{Z}} 
				\widetilde \omega(k)^{\frac{1}{2}}\widetilde \omega(l)^{\frac{1}{2}} \\
			& \qquad \qquad \qquad \qquad \times
				\Big( S_{k+l}a_k^*a_l^* - S_{k-l} a_k^* a_l - S_{k-l} a_k a_l^* + S_{k+l} a_k a_l  \Big)  \\
			& = \frac{r}{8  } \sum_{k\in\mathbb{Z}}  \widetilde \omega(k)^{\frac{1}{2}}
				\Big[ -\widetilde \omega(-k+1)^{\frac{1}{2}} a_k^*a_{-k+1}^* 
					- \widetilde \omega(-k-1)^{\frac{1}{2}} a_k^*a_{-k-1}^* \\
			&  \qquad \qquad \qquad  \qquad + \widetilde \omega(k+1)^{\frac{1}{2}} a_k^*a_{k+1} 
			+ \widetilde \omega(k-1)^{\frac{1}{2}} a_k^*a_{k-1} \\
			&  \qquad \qquad \qquad  \qquad + \widetilde \omega(k+1)^{\frac{1}{2}} a_ka_{k+1}^* 
			+ \widetilde \omega(k-1)^{\frac{1}{2}} a_ka_{k-1}^* \\
			& \qquad \qquad \qquad  \qquad - \widetilde \omega(-k+1)^{\frac{1}{2}} a_ka_{-k+1} 
			- \widetilde \omega(-k-1)^{\frac{1}{2}} a_ka_{-k-1} \Big] \; . 
			\\
			& \\
		\frac{1}{2} \int_{S^1} r \, & {\rm d} \psi \; r \cos\psi  \tfrac{1}{r^2} \bigl(\tfrac{ \partial \varphi }{\partial \psi}\bigr)^2 (\psi)  
			= - \frac{1}{8\pi r} \sum_{k\in\mathbb{Z}}  \sum_{l\in\mathbb{Z}} 
					\widetilde \omega(k)^{-\frac{1}{2}}\widetilde \omega(l)^{-\frac{1}{2}} \; kl \\
			& \qquad \qquad \qquad \qquad \times
						\Big(S_{k+l}a_k^*a_l^* - S_{k-l} a_k^* a_l - S_{k-l} a_k a_l^* + S_{k+l} a_k a_l \Big) \\
			& = \frac{1}{8r} \sum_{k\in\mathbb{Z}}  \widetilde \omega(k)^{-\frac{1}{2}}\, k \\
			& \qquad \times
				\Big[ -\widetilde \omega(-k+1)^{-\frac{1}{2}}(-k+1) a_k^*a_{-k+1}^* 
					- \widetilde \omega(-k-1)^{-\frac{1}{2}}(-k-1) a_k^*a_{-k-1}^* \\
			& 	\qquad \qquad 	+ \widetilde \omega(k+1)^{-\frac{1}{2}}(k+1) a_k^*a_{k+1} 
					+ \widetilde \omega(k-1)^{-\frac{1}{2}}(k-1) a_k^*a_{k-1}  \\
			& 	\qquad \qquad  + \widetilde \omega(k+1)^{-\frac{1}{2}}(k+1) a_ka_{k+1}^* 
				+ \widetilde \omega(k-1)^{-\frac{1}{2}}(k-1) a_ka_{k-1}^* \\
			& 	\qquad \qquad  - \widetilde \omega(-k+1)^{-\frac{1}{2}}(-k+1) a_ka_{-k+1} 
				- \widetilde \omega(-k-1)^{-\frac{1}{2}}(-k-1) a_ka_{-k-1} \Big] \; . 
	\end{align*}
Thus,
	\begin{align*}
		\frac{\mu^2}{2} \int_{S^1} r \, &  {\rm d} \psi \; r \cos\psi  \big(\varphi(\psi) \big)^2 \\
			& = \frac{\mu^2 r}{8\pi} \sum_{k\in\mathbb{Z}}  \sum_{l\in\mathbb{Z}} 
					\widetilde \omega(k)^{-\frac{1}{2}}\widetilde \omega(l)^{-\frac{1}{2}} \\
			& \qquad \qquad \qquad \times
					\Big( S_{k+l}a_k^*a_l^* + S_{k-l} a_k^* a_l + S_{k-l} a_k a_l^* + S_{k+l} a_k a_l \Big) \;.
	\end{align*}
Hence, 
	\begin{align*}
		\frac{\mu^2}{2} \int_{S^1} r \, &  {\rm d} \psi \; r \cos\psi  \big(\varphi(\psi) \big)^2 \\
			& = \frac{r}{8} \sum_{k\in\mathbb{Z}}  \mu^2\widetilde \omega(k)^{-\frac{1}{2}}
			\\
			& \qquad \times
					\Big[ \widetilde \omega(-k+1)^{-\frac{1}{2}} a_k^*a_{-k+1}^* 
						+ \widetilde \omega(-k-1)^{-\frac{1}{2}} a_k^*a_{-k-1}^* \\
			& \qquad \qquad  + \widetilde \omega(k+1)^{-\frac{1}{2}} a_k^*a_{k+1} 
				+ \widetilde \omega(k-1)^{-\frac{1}{2}} a_k^*a_{k-1} \\
			& \qquad \qquad  + \widetilde \omega(k+1)^{-\frac{1}{2}} a_ka_{k+1}^* 
				+ \widetilde \omega(k-1)^{-\frac{1}{2}} a_ka_{k-1}^* \\
			& \qquad \qquad  + \widetilde \omega(-k+1)^{-\frac{1}{2}} a_ka_{-k+1} 
				+ \widetilde \omega(-k-1)^{-\frac{1}{2}} a_ka_{-k-1} \Big] \; . 
	\end{align*}
Rearranging the terms in order to join terms having factors involving
the operators $a$ in common and using the fact that $\widetilde \omega(-k)=\widetilde \omega(k)$ for all $k\in\mathbb{Z}$, 
we get
	\begin{align}
		\label{eq:almost-there}
	\LLS_\circ^{(0)} & = \frac{r}{8 } \sum_{k\in\mathbb{Z}}  
			\Bigg[
					-\widetilde \omega(k)^{-\frac{1}{2}}\widetilde \omega(k-1)^{-\frac{1}{2}}
						\underbrace{ \Big( \widetilde \omega(k)\widetilde \omega(k-1) + {r}^{-2}
						(-k+1)k -  \mu^2  \Big)}_{A} 
						a_k^*a_{-k+1}^* 
				\nonumber\\
			&    \qquad \qquad  
					-\widetilde \omega(k)^{-\frac{1}{2}}\widetilde \omega(k+1)^{-\frac{1}{2}}
						\underbrace{ \Big( \widetilde \omega(k)\widetilde \omega(k+1) + {r}^{-2}
						(-k-1)k  - \mu^2  \Big) }_{B} 
						a_k^*a_{-k-1}^* 
				\nonumber\\
			&    \qquad \qquad  
					+\widetilde \omega(k)^{-\frac{1}{2}}\widetilde \omega(k+1)^{-\frac{1}{2}}
						\underbrace{ \Big(  \widetilde \omega(k)\widetilde \omega(k+1) + {r}^{-2}
						(k+1)k  + \mu^2  \Big) 
									}_{-B + 2\widetilde \omega(k)\widetilde \omega(k+1)} 
						a_k^*a_{k+1} 
				\nonumber\\
			&    \qquad \qquad  
					+\widetilde \omega(k)^{-\frac{1}{2}}\widetilde \omega(k-1)^{-\frac{1}{2}}
						\underbrace{ \Big(  \widetilde \omega(k)\widetilde \omega(k-1) + {r}^{-2}
						(k-1)k  + \mu^2  \Big) 
									}_{-A + 2\widetilde \omega(k)\widetilde \omega(k-1)} 
						a_k^*a_{k-1} 
				\nonumber\\
			&    \qquad \qquad  
					+\widetilde \omega(k)^{-\frac{1}{2}}\widetilde \omega(k+1)^{-\frac{1}{2}}
						\underbrace{ \Big(  \widetilde \omega(k)\widetilde \omega(k+1) + {r}^{-2}
						(k+1)k  + \mu^2  \Big)
									}_{-B + 2\widetilde \omega(k)\widetilde \omega(k+1)}  
						a_ka_{k+1}^* 
				\nonumber 
			\\
						&    \qquad \qquad  
					+\widetilde \omega(k)^{-\frac{1}{2}}\widetilde \omega(k-1)^{-\frac{1}{2}}
						\underbrace{ \Big(  \widetilde \omega(k)\widetilde \omega(k-1) + {r}^{-2}
						(k-1)k  + \mu^2  \Big) 
									}_{-A + 2\widetilde \omega(k)\widetilde \omega(k-1)} 
						a_ka_{k-1}^* 
				\nonumber 
			\\
				&    \qquad \qquad  
					-\widetilde \omega(k)^{-\frac{1}{2}}\widetilde \omega(k-1)^{-\frac{1}{2}}
						\underbrace{ \Big( \widetilde \omega(k)\widetilde \omega(k-1) + {r}^{-2}
						(-k+1)k  - \mu^2  \Big) }_{A} 
						a_ka_{-k+1} 
				\nonumber 
					\end{align}
				\begin{align}
			&    \qquad \qquad 
			-\widetilde \omega(k)^{-\frac{1}{2}}\widetilde \omega(k+1)^{-\frac{1}{2}}
			\underbrace{ \Big( \widetilde \omega(k)\widetilde \omega(k+1)  + {r}^{-2}
						(-k-1)k  - \mu^2  \Big) }_{B} 
						a_ka_{-k-1}
			\Bigg] . 
			\nonumber \\
	\end{align}
Due to \eqref{eq:useful-1} and \eqref{eq:useful-2} both $A$, $B$ vanish.

Returning with these informations to (\ref{eq:almost-there}), we get
	\begin{align}
		\label{eq:almost-there-2}
		\LLS_\circ^{(0)} & =  \frac{r}{4 } \sum_{k\in\mathbb{Z}}  
				\Bigg[
		\widetilde \omega(k)^{\frac{1}{2}}
		\widetilde \omega(k+1)^{\frac{1}{2}} a_k^*a_{k+1} 
		+\widetilde \omega(k)^{\frac{1}{2}}
		\widetilde \omega(k-1)^{\frac{1}{2}} a_k^*a_{k-1} 
				\nonumber \\
			& \qquad \qquad
			+\widetilde \omega(k)^{\frac{1}{2}}
			\widetilde \omega(k+1)^{\frac{1}{2}} a_ka_{k+1}^* 
			+\widetilde \omega(k)^{\frac{1}{2}}
			\widetilde \omega(k-1)^{\frac{1}{2}} a_ka_{k-1}^* 
				\Bigg] \; . 
		\end{align}
Now, we have
	\begin{align}
		\label{eq:upIUYiuy-1}
			\sum_{k\in\mathbb{Z}} \widetilde \omega(k)^{\frac{1}{2}}\widetilde \omega(k\pm1)^{\frac{1}{2}} a_k & a_{k\pm1}^* 
			  =   \sum_{k\in\mathbb{Z}} \widetilde \omega(k)^{\frac{1}{2}}\widetilde \omega(k\pm1)^{\frac{1}{2}} a_{k\pm1}^* a_k
			\nonumber \\
			& \quad \stackrel{k\to k\mp 1}{=}   \sum_{k\in\mathbb{Z}} \widetilde \omega(k)^{\frac{1}{2}} 
				\widetilde \omega(k\mp1)^{\frac{1}{2}} a_k^* a_{k\mp1} \; .
	\end{align}
In the first equality in (\ref{eq:upIUYiuy-1}) we have  used the fact that $a_k$ and $a_{k\pm1}^*$ commute.

Inserting (\ref{eq:upIUYiuy-1}) 
into (\ref{eq:almost-there-2}), we get, finally,
	\begin{equation}
		\label{eq:there}
	\LLS_\circ^{(0)} = \frac{r}{2 } \sum_{k\in\mathbb{Z}}  
		\bigg[
			\widetilde \omega(k)^{\frac{1}{2}}\widetilde \omega(k+1)^{\frac{1}{2}} a_k^*a_{k+1} 
			+\widetilde \omega(k)^{\frac{1}{2}}\widetilde \omega(k-1)^{\frac{1}{2}} a_k^*a_{k-1} 
		\bigg] \;.
	\end{equation}
Just like \eqref{ladderoperators}, equation \eqref{eq:there} expresses the generator of the 
boosts $t \mapsto \Lambda_1(t)$ in terms of ladder operators. The difference between the two 
formulas is due to the change of the scalar product, see \eqref{h-hat-L2}. On the other hand, 
one has
	\begin{align*}
		\int_{S^1} r \,  {\rm d} \psi\; & \big(\omega^{1/2} a\big)(\psi)^*\operatorname{\mathbb{cos}} 
		\big(\omega^{1/2} a\big)(\psi)
		\\
		& =  \frac{r}{2}\sum_{k\in\mathbb{Z}}\sum_{l\in\mathbb{Z}}
			\left(  \frac{1}{2\pi}\int_{S^1} {\rm d} \psi\; 
				{\rm e}^{i(k-l+1)\psi} + {\rm e}^{i(k-l-1)\psi} \right) 
				\widetilde \omega(k)^{1/2}\widetilde \omega(l)^{1/2} a_k^* a_l
		\\
		& =  
			\frac{r}{2}\sum_{k\in\mathbb{Z}}\sum_{l\in\mathbb{Z}}
				\Big(\delta_{l,\, k+1}+\delta_{l,\,k-1}\Big) 
				\widetilde \omega(k)^{1/2}\widetilde \omega(l)^{1/2} a_k^* a_l
		\\
		& =  
			\frac{r}{2}\sum_{k\in\mathbb{Z}}
				\Big( \widetilde \omega(k)^{1/2}\widetilde \omega(k+1)^{1/2} a_k^* a_{k+1}
				+ \widetilde \omega(k)^{1/2}\widetilde \omega(k-1)^{1/2} a_k^* a_{k-1} \Big)\;.
	\end{align*}
Therefore,
	\begin{align*}
		\LLS_\circ^{(0)} & = \frac{1}{2} \int_{S^1}  r \, {\rm d} \psi r \cos\psi
				\left( \pi(\psi)^2 + \tfrac{1}{r^2} \bigl( \tfrac{\partial \varphi}{\partial\psi}(\psi) \bigr)^2 
				+ \mu^2  \big(\varphi(\psi) \big)^2 \right) \\
		& 
		= \int_{S^1} r \, {\rm d}  \psi\; \big(  \omega^{1/2}  a\big)(\psi)^*
		r \operatorname{\mathbb{cos}} \big( \omega^{1/2}  a\big)(\psi) 
		\\
		& = {\rm d} \Gamma ( \sqrt{\omega} \,  r \operatorname{\mathbb{cos}} \sqrt{\omega} ) 
		\;.
	\end{align*}
In the last line we have used the normalisation introduced in \eqref{C-A-C}. We also note that 
according to \eqref{eq:omega-1} the coefficients $\omega(k)$ is proportional to $r^{-1}$. 
\end{proof}

\begin{remark}
\label{rm:11.1.2} 
We note that ${\rm d}\Gamma 
		\bigl( \sqrt{ \omega } \, (r \cos_{\psi + \alpha} \, \circ \, \chi_{I_+})
		\sqrt{\omega }  \bigr) $, with $\chi_{I_+}$ the characteristic function of the 
half-circle, is a well-defined operator; see Remark~\ref{rm:C6}.  
But it can not equal\footnote{
We would like to thank Rainer Verch for helpful discussions on 
this issue.}
	\begin{equation}
	\label{quantum-energy-inequality}
		 \frac{1}{2} 
			\int_{I_+} r^2  \cos (\psi+ \alpha) \, {\rm d} \psi \;   
			 \Bigl( {:} \, \pi^2 (\psi) + \tfrac{1}{r^2}
			\bigl(  \tfrac{ \partial \varphi }{\partial \psi}\bigr)^2 (\psi) 
			+ \mu^2 \varphi^2 (\psi) \, {:}  \Bigr)  \; , 
	\end{equation}
as the latter expression (which should be viewed as a quadratic form) would 
\emph{not} preserve\footnote{
The cancelations of the terms not preserving the particle number
in the proof of Lemma~\ref{lm:10.2.1} were based on the identity
	\[
		\int_{S^1}{\rm d} \psi \; \cos\psi \; {\rm e}^{ij\psi} 
		= \pi\big(\delta_{j,-1}+	\delta_{j,1}\big) \; .
	\]
Such a simple identity does not hold, if the integral is restricted to a half-circle.}
the particle number.  
Moreover, in case \eqref{quantum-energy-inequality} gives rise to 
an unbounded operator, it would be affiliated to the von Neumann
algebra ${\mathcal R} (I_+)$ introduced in~\eqref{vN-RI}, while the 
operator ${\rm d}\Gamma 
		\bigl( \sqrt{ \omega } \, (r \cos_{\psi + \alpha} \, \circ \, \chi_{I_+})
		\sqrt{\omega }  \bigr) $
is \emph{not} affiliated to the local algebra $\mathcal{R}(I_+)$; see also 
Remark~\ref{rm:6.5.4}.
\end{remark}

The {\em energy density\/} ${{\rm T}^{0}}_{0}(\psi)$ is the restriction of the energy density in 
the time-zero plane (in the ambient Minkowski space) to the Cauchy 
surface $S^1$, \emph{i.e.}, for $\psi \in S^1$, 
	\begin{align}
	\label{energymomentumdensity2}
		{{\rm T}^{0}}_{0}  (\psi) &=  \frac{1}{2}  \left( {:} \,  \pi^2 (\psi)
			+ \tfrac{1}{r^2} \bigl( \tfrac{ \partial \varphi }{\partial \psi} (\psi) \bigr)^2 
			+ \mu^2 \varphi (\psi)^2  +  \mathscr{P} (\varphi(\psi)) \, {:} \right)  \; .  
	\end{align}
We note that due to \eqref{eq:there} normal ordering does 
not affect the first three terms in the bracket.
Recall that, according to \eqref{e1.4}, the time-zero field $\varphi (h)$ equals 
$\mathbb{\Phi}(\delta \otimes h)$, and that $\delta \otimes h \in \mathbb{H}^{-1}(S^2)$ 
for $h \in  \widehat{\mathfrak{h}} (S^1)$. Hence,
normal ordering in the last term on the r.h.s.~is with respect to the covariance 
$C( f, g) = \langle \overline{f} , g \rangle_{\mathbb{H}^{-1}(S^2)}$, as see \eqref{wick}. 

The following formulas should be compared with the classical expressions derived in Section \ref{SET}.

\label{KLSpage}

\begin{theorem}
\label{Q-conserved-quantities}
Let $\KLS_0 $ denote the generator of the rotations which 
keep $S^1$ invariant. The following operator identities
hold on the Hilbert space $\Gamma  \bigl( L^2( S^1, r {\rm d} \psi )\bigr) $:
	\begin{align*} 
			\LLS^{(\alpha)} &=  \int_{S^1} \,  r^2 
			  \cos  (\psi + \alpha)  \, {\rm d} \psi  \;    
			  {\rm T}_{00}(\psi)  \; , \qquad
		\KLS_0  = \int_{S^1} \,  r  \; {\rm d} \psi \;   
		{\rm T}_{10}(\psi)   
		\; ,
	\end{align*}
with $ {T^{1}}_{0} = r^{-2} \pi \, (\partial_\psi \varphi) $; see \eqref{T01} for the 
corresponding classical quantity.  The r.h.s.~in both equations has to be 
interpreted as in Theorem~\ref{wickooo}. 
\end{theorem}

\begin{proof} Recall that $\LLS^{(\alpha)}_\circ  
= {\rm d} \Gamma \bigl(\sqrt{ \omega} \, r \, 
\cos_{\psi + \alpha} \sqrt{ \omega} \bigr)$.
Moreover, according to Theorem \ref{keyresult1},
	\[
	 \LLS^{(\alpha)}  = {\rm d} \Gamma ( \omega^{-1/2} ) \; 
	 \overline{{\rm d} \Gamma \bigl( \omega r \cos_{\psi 
	 + \alpha} \bigr)+ V(\cos_{\psi + \alpha}) } 
	 \; {\rm d} \Gamma ( \omega^{1/2}  )
	\] 
with $V(\cos_{\psi + \alpha}) =  \int_{S^1} r^2 \,{\rm d} \psi \,  \cos (\psi 
+ \alpha) \; {:} \mathscr{P} (\mathbb{\Phi} 
(0,\psi)) {:} $ acting on $\Gamma (\widehat{\mathfrak h} (S^1))$. 
Taking \eqref{h-hat-L2} into account, we find that
	\[
		{\rm d} \Gamma ( \omega^{-1/2} ) V^{(\alpha)} 
		{\rm d} \Gamma ( \omega^{1/2}  )= 
		  \frac{1}{2} \int_{S^1} \,  r^2  \cos  (\psi 
		  + \alpha)  \, {\rm d} \psi  \;   {:} \mathscr{P} (\varphi(\psi)) {:}   \; . 
	\]
Thus the expression  given for $\LLS^{(\alpha)}$ follows.
Next consider the angular momentum operator:
	\begin{align*} 
		\KLS_0 & = \int_{S^1} \,  r \; {\rm d} \psi \;  
		{:} \, \pi \,  (\partial_\psi \varphi) {:}  \\
		& = \frac{i}{4 \pi}  \sum_{k\in\mathbb{Z}} \sum_{j\in\mathbb{Z}}
		\int_{S^1}  \,  {\rm d} \psi \;   
						{:} \; \widetilde \omega(k)^{\frac{1}{2}} 
						\Big( {\rm e}^{ik\psi}a_k^* - {\rm e}^{-ik\psi}a_k \Big) 
						\\
		& \qquad \qquad \qquad \qquad \qquad \qquad \times  \partial_\psi   
						\widetilde \omega(j)^{-\frac{1}{2}}
						\Big( {\rm e}^{ij\psi}a_j^* + {\rm e}^{-ij\psi}a_j \Big) \; {:} 
		\\
		& = \frac{i}{4 \pi}  \sum_{k\in\mathbb{Z}} \sum_{j\in\mathbb{Z}}
		\frac{ \widetilde \omega(k)^{\frac{1}{2}} } {\widetilde \omega(j)^{-\frac{1}{2}}} 
		\Bigl( a_k^* a_j  \int_{S^1} \,  {\rm d} \psi \;   
		{\rm e}^{ik\psi}  \partial_\psi  {\rm e}^{-ij\psi} + a_j^* a_k  
		\int_{S^1} \,  {\rm d} \psi \;   {\rm e}^{-ik\psi}  \partial_\psi  {\rm e}^{ij\psi} \Bigr) \\
		& =    \sum_{k\in\mathbb{Z}} k a_k^* a_k  
		= {\rm d} \Gamma ( -i \partial _\psi) \; . 
	\end{align*}
We note that  the spectrum of $\KLS_0 $ is $\mathbb{Z}$, 
independent of $r$. 
\end{proof}

\goodbreak

In connection with Remark \ref{classical-energy} it is worth while noting the following result:

\begin{proposition}[$\varphi$-bounds] 
\label{phi-bound}
For $c$ sufficiently large 
and $g\in \widehat{\mathfrak h}(S^1) $,  there exists some $C>0$ 
such that 
	\begin{equation}
		\label{e3.1b}
		\left\|\varphi (g) \left( \int_{S^1} r \, {\rm d} \psi \;  
		 {{\rm T}}_{00} (\psi) + c \cdot \mathbb{1} \right)^{-\frac{1}{2}} \right\|
		\leq C\|g\|_{\widehat{\mathfrak h}(S^1)} 
	\end{equation}
and
	\begin{equation}
		\label{e3.1c}
		\pm \varphi(g)\leq C \|g\|_{\widehat{\mathfrak h}(S^1)}
		\left( \int_{S^1} r \, {\rm d} \psi \;  {\rm T}_{00} (\psi) +c \cdot \mathbb{1} \right)^{\frac{1}{2}} 
	\end{equation}
In particular, $ \int_{S^1} r \, {\rm d} \psi \; {\rm T}_{00} (\psi) 
$ is bounded from below.
\label{3.1}
\end{proposition}

\begin{proof}
One easily obtains (see, \emph{e.g.},~\cite[Theorem~V.20]{S} 
or \cite[Theorem~6.4 (ii)]{DG}) that
for $c\gg 1$ there exists a constant $C>0$ such that
	\begin{equation}
		\label{e3.1bb}
		\bigl({\rm d} \Gamma ( \omega )
		+ \mathbb{1} \bigr)\leq C \; \left( \int_{S^1} r \, {\rm d} \psi \; {\rm T}_{00} (\psi) 
		+c \cdot \mathbb{1} \right)  \:  .
	\end{equation}
Since $ {\rm d} \Gamma  ( \omega )$ 
is bounded from below 
and has compact resolvent on 
$\Gamma \bigl(\widehat{\mathfrak h}(S^1) \bigr)$,
it follows that 
	\begin{equation}
	\label{energy}
		 \int_{S^1} r \, {\rm d} \psi \;  {\rm T}_{00} (\psi) 
	\end{equation}
is bounded from below with a compact resolvent and hence has a
ground state. (As a remark we add that 
the uniqueness of this ground state follows from a Perron-Frobenius 
argument (see e.g.~\cite[Theorem~V.17]{S}).) Since 
	\[
		\omega  \geq m^\circ >0 
	\]
for some $m^\circ >0 $, we  see  that it suffices to check (\ref{e3.1b}) with
\eqref{energy} replaced by the number operator $N$, which is immediate.
To prove \eqref{e3.1c} we use   \eqref{e3.1bb} and the well known 
bound (see, e.g., \cite[Appendix]{Ge})
	\[
		\pm \varphi (g)\leq \| 
		g\|_{\widehat{\mathfrak h}(S^1)}
		\bigl({\rm d}\Gamma (\omega  )+ \mathbb{1} \bigr) \; .
	\]
\end{proof}

An important aspect of Theorem~\ref{wickooo} is additivity of 
the interaction term  on the right 
hand side: for any two disjoint intervals $I$ and $J$ of $S^1$, we 
have 
	\begin{equation}
	\label{sum-t00}
		\int_{I \cup J } r \, {\rm d} \psi \;   {:}  \mathscr{P} (\varphi(\psi)) {:}  = 
		\int_{I  } r \, {\rm d} \psi \;  {:}  \mathscr{P} (\varphi(\psi)) {:}  + 
		\int_{J } r \, {\rm d} \psi \; {:}  \mathscr{P} (\varphi(\psi)) {:}  \; . 
	\end{equation}
Note that in \eqref{sum-t00}  
the meaning of integrands is as in 
Theorem~\ref{wickooo}, \emph{i.e.}, 
the integrated quantities have a rigorous meaning as 
unbounded operators. A similar argument holds for  
the integrals over the 
mass term  $  {:}  \mu^2 \varphi (\psi)^2 {:}  \, $.

On the other hand, 
a formal integral over some interval 
$J$ of the first two terms on the right hand side 
in \eqref{energymomentumdensity2} may fail to yield 
an operator and only give a quadratic form, due to the ultra-violett difficulties 
which may appear by using a characteristic function $\chi_J$ 
to cut-off the densities.
And even if such an integral would define an unbounded operator, 
it would not yield a second quantised one-particle operator
(unless $J$ is the empty set or equals $S^1$) as it would not 
preserve the particle number; see Remark~\ref{rm:11.1.2}. 
 
\section{The equations of motion}
\label{sec:11.2}

Equations of motion for interacting quantum fields on Minkowski space were 
first derived by Glimm and  Jaffe \cite{GJ2}, Schrader \cite{Sch} and, in $2+1$ space-time 
dimensions, by Feldman and~Raczka \cite{FR}. Formulas similar to the ones presented in 
this section were given in~\cite{FHN}.

For arbitrary $x \in dS$, we define 
	\begin{equation}
	\label{covinteractingfield}
		\Phi_{int} (x) \doteq U(\Lambda) \varphi (0) U^{-1} (\Lambda) \; , \quad 
		x = \Lambda \left( \begin{smallmatrix} 0\\0\\r \end{smallmatrix} \right)  
		= \Lambda \left( \begin{smallmatrix} 0\\ r \sin 0 \\r \cos 0 \end{smallmatrix} \right)\; . 
	\end{equation}
In order to ensures that $\Phi_{int} (x)$ is (in the sense of quadratic forms) well-defined, one has to show that  
(in the sense of quadratic forms)
	\[
		U (\Lambda_2 (t)) \varphi (0) U^{-1} (\Lambda_2(t)) = \varphi (0)   \quad 
		\forall t \in \mathbb{R} \; , 
	\]
The latter follows from the Trotter product formula and 
Lemma \ref{lm:L2-invariance}, which implies that 
	\[
		\lim_{n \to \infty} U_\circ (\Lambda_2 (t)) \varphi (\delta_n) U^{-1} (\Lambda_2(t)) 
		= 
		\lim_{n \to \infty}  \varphi ({\rm e}^{it \LLS_2} \delta_n  )  
		= \lim_{n \to \infty}  \varphi (\delta_n  ) \; , \quad 
		\forall t \in \mathbb{R} \; ,  
	\]
on the (dense) form domain  (see \eqref{gamma-circ})
$\Gamma^\circ \bigl( \mathfrak{h}_\circ (S^1)\bigr)$
with $\mathfrak{h}_\circ (S^1)$ consisting of vectors of the form~\eqref{ggg}.

\begin{theorem}
\label{th:11.2.1} 
The interacting quantum field ${\Phi}_{\rm int} ( x ) $, $x \in dS$,   satisfies the 
{\em covariant} equation of motion: 
	\begin{equation}
	\label{eq-mo}
		\Bigl( \square_{dS}+\mu^2 \Bigr) {\Phi}_{\rm int} ( x ) 
		=   - {:} {\mathscr P}' ( {\Phi}_{\rm int} ( x )){:}_{\, \mathfrak{h}(dS)} \; ,
	\end{equation}
where ${\mathscr P}$ is a polynomial, 
	\[
		{:} {\Phi}^n_{\rm int} ( x (t, \psi)){:}_{\mathfrak{h}(dS)} 
		= \lim_{k \to \infty} {:} {\Phi}^n_{\rm int}  ( \delta (\, . \, - t ) \otimes 
		\delta_k (\, . \, - \psi )){:}_{\, \mathfrak{h}(dS)}
	\]
and 
	\[
		{:} {\Phi}_{\rm int}^{n} ( \delta \otimes \delta_k  ){:}_{\, \mathfrak{h}(dS)}
						\doteq \sum_{m=0}^{[n/2]}\frac{n!}{m!(n-2m)!}
					{\Phi}_{\rm int}^{n} ( \delta_t \otimes \delta_k)^{n-2m} 
					\Bigl(-\tfrac{1}{2}  \| \delta_k \|^2_{
					\widehat{\mathfrak{h}}(S^1)}  \Bigr)^{m} \; ;  
	\]
see Theorem~\ref{wickooo}. Note that here $\delta_t \otimes \delta_k \in \mathfrak{h}(dS)$ 
and not in $\mathbb{H}^{-1}(S^2)$.
\end{theorem}

\goodbreak

\begin{proof} 
Without restriction of generality, we may assume that $x =x (t, \psi)$ is a point which lies in the 
double-wedge $\mathbb{W}_1$.  Recalling from \eqref{varepsilon} that 
	\[
			\square_{\mathbb{W}_1}+\mu^2
				=  \frac{1}{r^2 \cos^2 \psi}\,(\partial_t^2+  \varepsilon^2) \; , 
	\]
with $\varepsilon^2  \doteq  - (\cos (\psi)  \, \partial_\psi)^2 + (\cos \psi )^2 \, \mu^2 r^2$, 
the {\em equations of motion} in their covariant form \eqref{eq-mo}
are verified once we have shown that  
	\begin{align*}
		\bigl( \partial^2_t + \varepsilon^2 \bigr) \Phi_{\rm int}(x) & 
		= \bigl( \partial^2_t  - (\cos (\psi)  \, \partial_\psi)^2 
		+ \cos^2 ( \psi ) \, \mu^2 r^2 \bigr) \Phi_{\rm int}(x) \\
		& 
		= - r^2 \cos^2 (\psi ) \,  {:} {\mathscr P}' ( \Phi_{\rm int}(x)){:}_{\, \mathfrak{h}(dS)} \; . 
	\end{align*} 
Using definition (\ref{covinteractingfield}), we find
	\[
		\frac{\partial^2}{\partial t^2} \Phi_{\rm int} (x)  
		= -  r^2 \; [ \LLS^{(0)}, [ \LLS^{(0)},  \Phi_{\rm int} (x) ]]  \; . 
	\] 
Following \cite[p.~224]{RS}, we compute (using $[V^{(0)}, \varphi (\psi) ] = 0$)
	\[ 
		[ \LLS^{(0)}, \varphi (\psi) ]  = [ \LLS_\circ^{(0)}, \varphi (\psi) ]  
		 = 
		 - i  \,\cos (\psi) \, \pi (\psi) \; 
	\]
and	
	\[
		[ \LLS^{(0)}, [ \LLS^{(0)}, \varphi (\psi) ]]  
		= [ \LLS_\circ^{(0)} , [ \LLS^{(0)} , \varphi (\psi) ]] 
			+ [ V^{(0)}, [ \LLS^{(0)} , \varphi (\psi) ]] \; . 
	\]
The first term on the right hand side yields 
	\begin{align}
			[ \LLS_\circ^{(0)}, [ \LLS^{(0)}, \varphi (\psi) ]]
			&= - i  \, \cos (\psi) \, [ \LLS_\circ^{(0)},  \pi (\psi) ]   \nonumber \\  
			&
			=  -
			 \tfrac{1}{r^2} \bigl(  \cos (\psi)  \partial_\psi \bigr)^2 \varphi (\psi) 
			 +   \cos (\psi) \, \mu^2  \varphi (\psi) \; .
	\end{align}
The second equality follows from partial integration (see \eqref{energymomentumdensity2}), \emph{i.e.}, 
	\begin{align*}
		\frac{1}{2} \int {\rm d} \psi' \; r \cos (\psi' ) \, & 
				\left[  \tfrac{1}{r^2} \bigl( \partial_{\psi'} \varphi(\psi') \bigr)^2 ,  \pi (\psi) \right] \\
				 & = - \tfrac{1}{r^2}
				\int {\rm d} \psi' \;  \partial_{\psi'} r \cos (\psi' ) \partial_{\psi'} \varphi(\psi')   
				\underbrace{ \left[  \varphi(\psi') ,  \pi (\psi) \right] }_{= - \frac{i}{r} \delta(\psi' - \psi) }  \; .
	\end{align*}
The second term yields
	\begin{align*} 
		[ V^{(\alpha)}, [ \LLS^{(0)}, \varphi (\psi) ] ] 
			&= 
			- i \, \cos (\psi) \, [ V^{(\alpha)},  \pi (\psi) ]  \nonumber \\ 
			&=  
			- i 
			\cos (\psi) \int {\rm d} \psi' \; r \cos (\psi' ) \, 
				\left[   \,{:}  {\mathscr P} (\varphi(\psi'))  {:}  \; ,  \pi (\psi) \right]  \nonumber \\  
			& = \cos^2 (\psi) \;  {:} \, {\mathscr P}' (\varphi (\psi)) \, {:}   \; . 
	\end{align*}
In the last equality we have used
 	\begin{align*} 
		\Bigl[ 	\, 	{:} \, {\varphi}^{n} \bigl( \delta_k ( \, . \, - \psi' ) \bigr) \, {:} \; , 
		 \, \pi (\psi) \Bigr] 
		 & = \sum_{m=0}^{[n/2]}\frac{n!}{m!(n-2m)!}
					\Bigl(-\tfrac{1}{2}  \| \delta_k \|^2_{
					\widehat{\mathfrak{h}}(S^1)}  \Bigr)^{m} \; 
			\\
		& \qquad \qquad \times \bigl[ \varphi^{n-2m} 
		\bigl( \delta_k ( \, . \, - \psi' ) \bigr) , \pi(\psi) \bigr]  
		\\
		 & = n \sum_{m=0}^{[n-1/2]}\frac{(n-1)!}{m!(n-1-2m)!}
					\Bigl(-\tfrac{1}{2}  \| \delta_k \|^2_{
					\widehat{\mathfrak{h}}(S^1)}  \Bigr)^{m}  
								\\
		& \qquad \qquad \times \varphi^{n-1-2m} \bigl( \delta_k ( \, . \, - \psi ) \bigr)  
		\\
			&=    n \; 	{:} \, {\varphi}^{n-1} \bigl( \delta_k ( \, . \, - \psi )\bigr)   \, 	{:} \; .  
	\end{align*}
In summary, we find
	\begin{align*}
		\frac{\partial^2}{\partial t^2} \Phi_{\rm int} (x)  
		& = 
		\underbrace{\Bigl( \bigl(\cos (\psi) \partial_\psi \bigr)^2 
		- r^2 \cos (\psi) \, \mu^2 \Bigr)}_{= - \varepsilon^2}
		{\rm e}^{i \LLS^{(0)} t}  \varphi(\psi) {\rm e}^{- i \LLS^{(0)} t}
			\nonumber
		\\ 
		& 	\qquad \qquad \qquad \qquad 
			 -  r^2 \cos^2 (\psi )\; {:} {\mathscr P}' (  {\rm e}^{i \LLS^{(0)} t}  
			 \varphi(\psi) {\rm e}^{- i \LLS^{(0)} t}) {:}_{\, \mathfrak{h}(dS)}  \; ,   
	\end{align*}
which verifies the claim. 
\end{proof}

\chapter{Summary}

\section{The conceptional structure}
\label{sec:12.1}

The time-like geodesics on de Sitter space, corresponding to freely
falling observers, are integral curves of Killing vector  
fields. These vector fields are generated by one-parameter groups of Lorentz 
boosts. In the visible and accessible part (and only there) of the
observer's universe, which is a (Rindler) wedge $W$, this vector field is
future directed (Chapter~1). The Lorentz boosts for the various
time-like geodesics generate the Lorentz group $SO_0(1,2)$; see
Chapter~2. We recall the representation theory of 
$SO_0(1,2)$ and provide an outline of harmonic analysis on the de
Sitter space (Chapters~3 and 4). The dynamics for an observer is now 
given by the one-parameter group of unitary operators implementing the
Lorentz boosts associated to the observer's time-like geodesic.
In Chapter 5 we present a first connection to (classical) field
theory: various classical   
dynamical systems are associated to the homogeneous Klein-Gordon
equation, with emphasis  
on the causal structure expressed by the support properties of the
solutions of the  Klein-Gordon  
equation. Unfortunately, the analysis of the \emph{non-linear} (classical) Klein-Gordon equation is 
beyond the scope of this work. 

In Chapter 6, we connect the results of Chapter 4 with those of Chapter 5, exploiting  
Kay's notion of \emph{one-particle quantum structures}.  We also discuss (modular)
localisation on the one-particle level, both in the covariant and the canonical setting. 
In Chapter~\ref{2Q}, we define the Haag--Kastler 
$\mathcal{O}\to \mathcal{A}_\circ(\mathcal{O})$ for the free
massive scalar field on the de Sitter space. The Weyl operators, as well 
as the field operators, 
arise by the second quantisation from the one-particle structures. 
If one requires de Sitter invariance and stability of the free falling motion against small adiabatic 
perturbations, the de Sitter vacuum state $\omega_\circ$ is unique. 
It is induced by the Fock vacuum vector $\Omega_\circ$. For each wedge $W$, 
$\omega_\circ$ is a $2\pi r$-KMS state for the algebra $\mathcal{A}_\circ(W)$ and the Lorentz boost 
$t \mapsto \Lambda_W(t)$, $t \in \mathbb{R}$. This is the 
\emph{geodesic KMS condition} of Borchers and Buchholz.  
The central piece of this work is Chapter 8. The basic idea is to construct an interacting 
vacuum vector $\widehat{\Omega}$ within the canonical Fock space, with
the help of (an extension of) 
Araki's perturbation theory (initially applied to each wedge
individually) for KMS states and the associated   
modular automorphisms. Since no wedge is physically distinguished, one
may expect that the 
various vector valued analytic functions, which arise by choosing different wedges, have common 
boundary values both on the de Sitter space and on the Euclidean sphere\footnote{A specific Euclidean  
sphere is fixed by distinguishing a space-like geodesic $S^1$ in de Sitter space. All other space-like geodesic
arise by applying a Lorentz transformation $\Lambda$ to $S^1$. As
the Lorentz transformed Euclidean upper hemispheres $\Lambda S_+$  
all lie in the tuboid ${\mathcal T}_-$, our arguments do not depend on a particular choice for 
the Euclidean sphere $S^2$.}, and that 
$\widehat{\Omega}$ can be recovered from a rotation invariant, Euclidean, 
interacting vacuum vector $\mathbb{\Omega}$,  which defines a reflection 
positive\footnote{Our formulation of reflection positivity 
differs from the one used in the pioneering work by Figari, H\o egh-Krohn  and Nappi \cite{FHN}, where
two antipodal points were taken out of the sphere and reflection positivity was formulated with 
respect to a half-circle connecting these two points.} state on the sphere,
by applying the Osterwalder--Schrader projection. We show that for polynomial interactions, 
this procedure is viable, as the interaction is given by an integral over point-like fields, 
which allows foliations; see \eqref{as-in-V} below. In particular, 
in Chapter 9 we show that rotation invariance of 
the Euclidean interacting vacuum vector implies the existence of a virtual representation 
of SO(3) on the canonical Fock space, which can be analytically continued to a 
new, interacting representation of the Lorentz group.
As expected, these results can be rephrased\footnote{That the operator 
sum \eqref{operator-sum-summary} exists, does not follow
directly from Araki's perturbation theory. 
It is justified in our work by Euclidean arguments, using the method of local symmetric semi-groups.
However, one could try to avoid Euclidean techniques and instead use ideas
similar to those given in the proof of Theorem~\ref{thm:11.1.4} to justify \eqref{operator-sum-summary}.}
in terms of Araki's perturbation theory of modular automorphisms.
In Chapter 10, the net of local algebras for the interacting field theory is 
constructed in a standard manner: 
the interacting algebra $\mathcal{A} (W_1)$ for the wedge $W_1$ is identified
with the free algebra $\mathcal{A}_\circ(W_1)$ for the wedge $W_1$, and the interacting
algebras associated with arbitrary wedges are defined by covariance
under the new representation. The algebras for smaller regions are
defined by intersections\footnote{Our proof that the intersections 
of wedge-algebras are non-trivial  
relies on finite speed of propagation. In the present case, this important 
property is another consequence of the fact that the interaction is given by 
an integral over point-like fields.}. The interacting net of algebras satisfies the 
Haag--Kastler axioms proposed by Borchers and Buchholz. The local time-zero 
algebras for the interacting and the free model coincide. In Chapter~\ref{ch:11}, 
the generators of the boosts and the rotations are expressed in terms of integrals over 
the density for the quantized stress-energy tensor. The covariant equation of motion  
for the interacting quantum fields is verified. 

One may argue, that Euclidean techniques play a central role in our work.
While some may actually favour starting from a Euclidean setting, others may argue 
that a description based entirely on the physical space-time provides a better understanding of 
time-dependent phenomena. While it may be feasible to avoid Euclidean 
techniques\footnote{To some extent, 
we have already done so by avoiding the 
usage of path integrals or stochastic processes. Instead, we have introduced 
a Euclidean Fock space, which appears more natural from a group theoretical
perspective.}, this somehow (at least in our opinion) misses 
a physically relevant aspect: Stability of the de Sitter vacuum
implies the existence of certain analytic functions, and the latter allow extensions into the tuboids, 
which contain the Euclidean sphere.  In the Wightman framework on Minkowski space, 
similar analyticity properties, ensuring positivity of the energy and thereby stability, 
are absolutely essential. On the de Sitter space, the Euclidean 
aspects just underline the fundamental  role  Tomita-Takesaki modular theory plays 
in the representation theory of the Lorentz group. 

\section{Wightman function, particle content and scattering theory}

The interacting ${\mathscr P}(\varphi)_2$ model on the de Sitter space is in
many ways the simplest model, which satisfies all the basic expectations such as 
finite speed of propagation, 
particle production\footnote{Clearly, the interacting dynamics mixes the $n$-particle sectors 
of Fock space. But, as argued below, the notion of \emph{particles} has to be taken 
with a grain of salt.}, causality, and so on. 
One interesting aspect of our construction is that it allows to describe a new type of interacting 
scalar particles, which have no analog in Minkowski space. From a group theoretical perspective, 
they emerge from representations of $SO(1,2)$ belonging to the \emph{complementary series}. 
These particles are sensitive to the curvature of space-time 
as their de Broglie wave-length is of the 
same order (or larger) than the spatial extension of the de Sitter universe. 
One may expect that they share some 
properties with infra-particles on Minkowski space; 
their absence in the decomposition presented in Theorem~\ref{spectheo} 
can be attributed to their poor localisation properties.
Little is known about their existence or significance. 
But what we can say from the present work is that they allow 
us to add an interaction, and that the
resulting interacting quantum theory on the two-dimensional 
de Sitter space is well-defined. 

It should be emphasized that from a physicists viewpoint 
the present work  can only be a starting point for further work. 
Let us mention three questions, which one might pose 
for the ${\mathscr P}(\varphi)_2$ model on de Sitter space:  
\begin{itemize}
\item [$i.)$]
A legitimate question to ask is whether the interacting \emph{Wightman functions} are 
distributions, which are the boundary values of analytic functions, and what exactly are the 
domains of analyticity of the latter. A proposal for these domains can be found in~\cite{BEM}. 

\item [$ii.)$]
As the thermalisation effects mentioned in Sect.~\ref{sec:12.1} play a significant role 
for the long time behaviour, there is no asymptotically free movement. 
A \emph{scattering theory}, 
which takes these aspects into account, has not yet been formulated,
thus is not available either.  
In short, the \emph{particle content} of ${\mathscr P}(\varphi)_2$
model on the de Sitter space  
has yet to be revealed. 

\item [$iii.)$]
Based on the arguments given in the proof of Proposition~\ref{phi-bound},
one might hope that there is an isolated, discrete contribution 
at the effective mass of the ``dressed particle'' to the K\"allen-Lehmann measure \cite{Bros}
of the Wightman two-point function, 
below the beginning of the support of the continuous contribution to this measure. 
But at the moment, this is merely a speculation. 

\end{itemize}

These questions are related to a (still missing\footnote{In fact, in the past colleagues have 
questioned whether there \emph{exist} stable, physical particles in an interacting theory on de 
Sitter space (see, \emph{e.g.}, \cite{BEM-2}).}) `particle interpretation'  for the interacting  
theory on de Sitter space. The free (or \emph{bare}) $n$-particle subspaces are still present, 
when we describe the interacting theory in Fock space. But they are not invariant under the 
interacting dynamics, and hence their physical significance is unclear. 
From a conceptual perspective, it looks more natural to consider the 
GNS representation for the interacting vacuum state, 
to view the vacuum state induced by the 
GNS vector as a no-particle state and to search for a manifestation (see \cite{Bu1}) 
of  physical (``dressed'') particles\footnote{As there is not 
even an energy operator, arguments based  on a discrete contribution 
to the joint energy-momentum spectrum can not be used to identify (asymptotically free) particles.}. 
Such a search for particle should be based on particle detectors; see \cite{Enss} for important work 
in Minkowski space.
Hence, much remains to be done. 
But, as our work demonstrates,  the dynamics, the vacuum state and the stress-energy tensor 
are well defined and actually given by simple, explicit formulas.  
Therefore, we are optimistic that many physically interesting 
aspects of this model will be revealed in the not too distant future.   

\section{A detailed summary}

We introduce the interacting theory at a late stage in this work. This was done partially to emphasize the 
important role the representation theory of the Lorentz group\footnote{Our approach, 
which starts from the space-time symmetries, differs radically from the more 
common scheme to ``quantize''  a classical system, described
by a partial differential equation.}
plays in our work, but 
mostly to make the material accessible to mathematicians. Many notions and concepts 
are introduced early on, and are thereby somewhat separated (for good or for worse) 
from their quantum field theoretic context. Hence, it may be worth while trying 
to connect a few dots in the detailed account of our work presented now. 

Let us now describe our work in some more detail.  
In Chapter 1, we describe the two-dimensional de Sitter space
	\begin{equation*}
		dS \doteq \left\{  
		x \equiv (x_0, x_1, x_2) \in \mathbb{R}^{1+2} 
		\mid 
		x_{0}^{2} - x_{1}^{2} - x_{2}^{2} = - r^2 \right\} \; , 
		\quad r >0 \; ,
	\end{equation*}
as a Lorentzian manifold. In fact, de Sitter space 
emerges as a solution of the Einstein equations (Sect.~\ref{sec:1.1}). 
The embedding $dS \hookrightarrow \mathbb{R}^{1+2}$
is compatible with the metric and the causal structures
(Sect.~\ref{sec:1.2}, Lemma~\ref{lm:1.2.1}). Using 
geographical coordinates, we provide a conformal map from 
$dS$ to the cylinder~$ (- \tfrac{\pi}{2} , \tfrac{\pi}{2} ) \times S^1$. 
The Laplace-Beltrami operator $\square_{dS}$ takes a particular 
simple form in these coordinates. 
We describe the future and the past, the space-like complement 
and the causal completion of an arbitrary region in de Sitter space. 
We discuss the existence of geodesics and geodesic distances. 
In particular, we emphasize the role of space-like geodesics as Cauchy surfaces; 
by a coordinate transformation, any space-like geodesic can be identified 
with the circle~$S^1 = \{ x \in dS \mid x_0 = 0 \}$.
In Sect.~\ref{subsec:1.3}, the Lorentz group $SO_0(1,2)$ is introduced. The causal 
structure of $dS$ is discussed in Sect.~\ref{sec:1.4}. 
The intersection of the past (or future) of a region 
in $dS$ with the Cauchy surface $S^1$ is computed explicitly (Proposition~\ref{ialpha}).  
Causally complete regions, 
in particular wedges and double cones, are introduced in Sect.~\ref{sec:1.5}. 
The coordinates 
	\begin{equation*} 
		 x (t, \psi) \doteq \Lambda_{1} \left( t \right) \, R_0 \left( - \psi \right) 
			\begin{pmatrix}
					0 \\
					0  \\
					r  
			\end{pmatrix} \; , \qquad \, t \in \mathbb{R} \; , 
		\quad  {\textstyle  -\frac{\pi}{2}< \psi < \frac{ \pi}{2}}   \; ,
	\end{equation*} 
for the wedge  $W _1\doteq \bigl\{  x \in dS \mid x_2 > |x_0 | \bigr\} $	
yield a Laplace-Beltrami operator $\square_{\mathbb{W}_1}$, 
which is convenient for the description of the one-parameter group $t \mapsto \Lambda_{1} \left( t \right)$
of boosts leaving the wedge $W _1$
invariant, as
 	\begin{equation*}
			\square_{\mathbb{W}_1}+\mu^2
				=  \frac{1}{r^2 \cos^2 \psi}\,(\partial_t^2+  \varepsilon^2) \; , 
	\qquad
		\varepsilon^2  \doteq  - (\cos \psi  \, \partial_\psi)^2 
		+ (\cos \psi )^2 \, \mu^2 r^2 \; . 
	\]
A key aspect in our work is that the various one-parameter 
subgroups of the Lorentz group 
$SO_0(1,2)$ allow analytic continuations.  For the convenience 
of the reader, we recall the relevant 
domains of analyticity (called tuboids) in complexified de Sitter 
space $dS_{\mathbb{C}}$, first described 
by Bros and Moschella (Sect.~\ref{tuboidsds}).  The existence of 
these tuboids establishes the 
connection to Euclidean field theory on the sphere, as the 
Euclidean hemispheres 
	\begin{equation*} 	
		S_\mp = \bigl\{ ( i\lambda_0,x_1, x_2) \in  (i\mathbb{R}) 
		\times \mathbb{R}^2 \mid \lambda_0^2 +  x_1^2 +  x_2^2 
		= r^2, \mp \lambda_0 > 0 \bigr\}
	\end{equation*} 
are contained in the \emph{tuboids}
	\begin{equation*} 
 		{\mathcal T}_\pm = \bigl( \mathbb{R}^{1+2} \mp i V^+ \bigr) 
		\cap dS_{\mathbb{C}} \; , 
		\qquad
		V^+ \doteq \bigl\{   y \in \mathbb{R}^{1+2} 
		\mid  y \cdot  y >0,  y_0 > 0 \bigr\} \, .  
	\end{equation*} 
The tuboids ${\mathcal T}_\pm$ are described in some more 
detail in Lemma~\ref{lemma1.2} (due to the 
authors) and  Lemma~\ref{tds} (due to Bros and Moschella). 
As one may expect, various analytic continuations in one 
parameter into the tuboid coincide, if they coincide at the 
boundary (Theorem~\ref{flattube}). 
Geographical and path-space coordinates for the Euclidean sphere are 
introduced in Sect.~\ref{sec:1.7}. 

Chapter~\ref{isometrygroup} is devoted to a discussion 
of the group of space-time symmetries, namely,  the Lorentz
group~$SO_0(1,2)$. The orbits of $SO_0(1,2)$ in the ambient Min\-kowski space 
are described in Lemma~\ref{lm:2.1.1}. The action of the Lorentz group
on the light cone is described in some detail, as it gives rise to 
the unitary irreducible representations of $SO_0(1,2)$, which we  
present in Sect.~\ref{RRLC} and Sect.~\ref{UIRc}. As we apply 
the method of induced representations, it is useful to study the stability group 
of an arbitrary point on the light cone $\partial V^+$.  
The stability group of  a specific point in $\partial V^+$ is the 
group of horospheric translations $q \mapsto D(q)$;  see
Lemma~\ref{stabil}. Of course, the stability groups of other points on
the light cone are all isomorphic to  this nilpotent group. Various decompositions 
of the Lorentz group are presented in the sequel: the Cartan 
decomposition in Sect.~\ref{sec:2.3}, the Iwasawa decomposition 
in Sect.~\ref{sec:2.4} and the Hannabus decomposition in 
Sect.~\ref{sec:2.5}. The forward light-cone, the mass hyperboloid, the de Sitter 
space and the circle $\gamma_\circ$ on the forward light-cone can all be viewed as cosets, emerging 
from the various decompositions of the Lorentz group; see Sect.~\ref{circle-mass-shell}. 
The complex Lorentz group, presented in Sect.~\ref{sec:2.7}, contains both 
the Lorentz group,  which keeps de Sitter space invariant, and the group of rotations $SO(3)$, 
which keeps the Euclidean sphere invariant, as subgroups. The Lie algebras 
$\mathfrak{so}(1,2)$ and $\mathfrak{so}(3)$ are dual symmetric Lie algebras
in the sense of Cartan. In Chapter~\ref{interactingdesitter} we will exploit this fact 
to (re-)construct representations of $SO_0(1,2)$
from the representations of $SO(3)$ by the method of virtual representations, due to Fröhlich, 
Osterwalder and Seiler. However, some elementary aspects of this rather involved 
construction are already discussed in Sect.~\ref{sec:2.7}.
In the sequel, analytic continuations 
of the boosts are described, as well as their relations to reflections. These relations will
give rise to the Bisognano-Wichmann theorem. The latter plays a crucial role in Chapter~9. 

In Chapter 3, we apply Mackey's theory of induced representations to the special case of the 
Lorentz group. While Bargmann's pioneering work on the Lorentz group
used Schur's multiplier representations, the authors find it more convenient to follow 
the geometrical ideas underlying Mackey's approach.
And although there are many accounts in the literature
of the theory of integration on homogeneous spaces, 
the authors believe that it is convenient that we recall the relevant notions in Sect.~\ref{sec:3.1}. 
For example, the existence of a strongly quasi-invariant measure is stated in Theorem~\ref{th:3.1.12}. 
The method of induced representations is presented in Sect.~\ref{sec:3.2};
see in particular Definition~\ref{Fmu}.  The  Wigner representation 
is given in Proposition~\ref{prop:3.2.9}.
The reducible representation provided by the push-forward on the forward
light-cone is decomposed w.r.t.~the eigenvalue of the Casimir operator (Sect.~\ref{RRLC}, 
Theorem~\ref{spectheo}), using the Mellin transform introduced in Proposition~\ref{prop:3.3.4}. 
We note that in this 
decomposition only the unitary irreducible representations belonging to the principal 
series appear.
Bargmann's unitary irreducible representations $\widetilde{u}\equiv \widetilde{u}_\nu$
of $SO_0(1,2)$, which include the representations in the principal and the complementary 
series, are recovered in Sect.~\ref{UIRc}. In particular, in Proposition~\ref{Prop:2.1} the one-particle Hilbert space
$\widetilde{\mathfrak h}\equiv \widetilde{\mathfrak h}_\nu$ is introduced, 
and in Theorem~\ref{TH-irr} the irreducibility of the
representation $\widetilde{u}_\nu$ is established. We note that the
functions in $\widetilde{\mathfrak h}$ are  
homogeneous functions on the forward light cone~$\partial V^+$ 
an thus are specified by the values they take on a circle on~$\partial V^+$.   
A direct proof of a result by Sally, concerning 
intertwiners, is given next (Proposition~\ref{Prop:2.1.0}). As~$\partial V^+$ is not invariant 
under time-reflections, these intertwiners are needed to extend 
the representation of $SO_0(1,2)$ to a representation of $O(1,2)$;  
see Sect.~\ref{sec:time reflection}. 
In particular, in Proposition~\ref{prop:3.6.1} we present an anti-unitary implementer of the 
time-reflection~$T$ and a unitary implementer of the (partial) parity 
transformation $P_2$.  
In the sequel, a description of the unitary irreducible representations of the Lorentz group on two 
mass-shells is given (Sect.~\ref{UIRm}, Theorem~\ref{th:3.7.1}). 
Explicit formulas, due to Bros and Moschella, are provided in 
Theorem~\ref{th:3.7.2}. These formulas are of interest, as they relate the induced 
representations on the forward light cone~$\partial V^+$ to 
the momentum space representation of the Poincar\'e group in the Minkowski space case.  

Chapter~\ref{Harmanaly} is devoted to harmonic analysis on the hyperboloid. 
The Harish-Chandra plane waves 
	\begin{equation*}
		 x \mapsto   (  x_\pm \cdot  p )^{s} \; , 
		 \qquad x \in dS \; , \quad p \in \partial V^+ \; , 
	\end{equation*}
are presented in Sect.~\ref{planwaves}.  These distributions 
are solutions of the homogeneous Klein--Gordon,  
which should be viewed as the boundary values of functions 
analytic in the tuboids introduced in 
Section~\ref{tuboidsds}: 
	\[
			(\square_{\partial V^+} + \mu^2 ) 
			\left(  x_\pm  \cdot  p \right)^s = 0 \; , \qquad -s(s+1) = 
		\mu^2 r^2  \; . 
	\]
The symbol $x_\pm$, which is used instead of $x \in dS$,   
indicates this fact. The plane waves are important, as they 
enter in the definition of the Fourier--Helgason 
transformation ${\mathcal F}_{\pm}$ (Definition \ref{def:4.2.1}),
	\begin{equation*}
 	 {\mathcal D} (dS)  \ni f \mapsto	\widetilde {f}_\pm (  p, s) 
		= \int_{dS} {\rm d} \mu_{dS} ( x ) \; f( x ) \; (  x_\pm \cdot  p )^s \; , 
	\end{equation*}
presented in Sect.~\ref{Fourier--HelgasonTransformation}. 
Analyticity properties of the Fourier--Helgason transformation with 
respect to the parameter $\nu$, 
	\[
		\nu = \begin{cases} \sqrt{\mu^2 r^2 - \tfrac{1}{4}} = m r \; ,  
		& \text{for} \quad \mu^2  = \tfrac{1}{4r^2}+ m^2 \ge \tfrac{1}{4r^2} \; , 
		\\
		 i \sqrt{\tfrac{1}{4} - \mu^2 r^2} = i mr  \; , 
		& \text{for} \quad  -\tfrac{1}{4r^2} < m^2 \le 0 \; ,
		\end{cases}
	\]
labelling the unitary irreducible representations~$\widetilde{u}_\nu$, 
are stated in Lemma~\ref{nu-analyicity}. 
A Hardy space decomposition 
(Theorem~\ref{hardy}), due to Bros and Moschella, and 
the Plancherel theorem (Theorem~\ref{plancherel}), due to Molchanov, 
are stated (without proofs) for the sake of completeness, as we consider them important 
structural results. The remainder of Chaper 4 deals with two more realisations of the unitary 
irreducible representations of $SO_0(1,2)$; anticipating their physical meaning we call them 
\emph{covariant} and \emph{canonical} representations. We start by 
defining a map ${\mathcal F}_{+ \upharpoonright \nu}$ 
(in analogy to Minkwoski space, we call it the restriction of the Fourier transform to the 
upper mass-shell)
from test-functions $f \in  {\mathcal D}_{\mathbb{R}} (dS)$ to elements 
	\[
		{\widetilde f}_\nu \doteq  \sqrt{ \frac{c_\nu {\rm e}^{-  \pi \nu}  r   }{\pi}} 	\; 		
				 	 \widetilde {f}_+  (  \, . \, , s^+)  \; , 
		\qquad 
		c_\nu = -\frac{1}{2\sin  (\pi s^+ )}   \; , 
		\quad
			s^+= -\frac{1}{2}  - i \nu \; , 
	\]
of the Hilbert space $\widetilde {\mathfrak h}_\nu$, which carries the induced representation 
$\widetilde{u}_\nu$
of the Lorentz group. This allows us to define a (complex valued) semi-definite quadratic form 
	\begin{equation}
	\label{s-nuclear}
	{\mathcal D}_{\mathbb{R}} (dS) \ni f, g \mapsto 
	\langle \widetilde f_\nu , \widetilde g_\nu \rangle_{ \widetilde {\mathfrak h}_\nu }
	\end{equation}
on the test-functions. On the equivalence classes $[f] \in  \mathcal{D}_{\mathbb{R}}(dS) / 
\ker {\mathcal F}_{+ \upharpoonright \nu} \,$, the quadratic form \eqref{s-nuclear} defines
a (complex valued) scalar product 
	\begin{equation}
		 \label{s-ophs2a}
		\langle [f] , [g] \rangle_{{\mathfrak h} (dS)} \doteq 
		\langle \, \widetilde f_\nu , \widetilde g_\nu  \,  \rangle_{ \widetilde {\mathfrak h}_\nu } \;   .   
	\end{equation}
The completion of $\mathcal{D}_{\mathbb{R}}(dS) / \ker {\mathcal F}_{+ \upharpoonright \nu} $ 
with respect to this scalar product is the covariant one-particle Hilbert space $\mathfrak{h}(dS)$.
The push-forward\footnote{We suppress the dependence on 
$\nu$ whenever possible.} 
	\begin{equation*} 
		u (\Lambda) [f] =  [ \Lambda_*f ]  \; , \qquad \Lambda \in O(1,2) \; , 
	\end{equation*}
on the test-functions extends to a (anti-) unitary representation of the Lorentz group $O(1,2)$
on $\mathfrak{h}(dS)$ (Proposition~\ref{prop:4.4.5}). 
By construction, this representation is unitarily equivalent to the irreducible 
representation $\widetilde{u}$ on $\widetilde{\mathfrak{h}}$
of $O(1,2)$ constructed in Section~\ref{UIRc}.

The kernel of the scalar product \eqref{s-ophs2a} specifies the Wightman 
two-point function ${\mathcal W}^{(2)} (  x_1 ,  x_2 )$
on the de Sitter space, 
	\begin{equation*}
	\langle [f]  ,   [g] \, \rangle_{{\mathfrak h} (dS)} = \int_{dS \times dS} {\rm d} \mu_{dS} 
	( x_1) {\rm d} \mu_{dS} (  x_2)   { f ( x_1 ) } 
		{\mathcal W}^{(2)} (  x_1 ,  x_2 ) g ( x_2) \; . 
	 \end{equation*}
This distribution ${\mathcal W}^{(2)} (  x_1 ,  x_2 )$
has been studied previously in great detail by Bros and Moschella. In particular, 
these authors have shown that ${\mathcal W}^{(2)} (  x_1 ,  x_2 )$
can be expressed in terms of Legendre functions: for $( z_1,   z_2) 
\in {\mathcal T}_ + \times {\mathcal T}_-$,
	\begin{equation}
		\label{s-tpf-2}
	{\mathcal W}^{(2)} ( z_1 ,  z_2) 
	=  c_\nu \, P_{s^+} \left(  \tfrac{z_1 \cdot  z_2}{r^2} \right) 	\; , \qquad m> 0 \; . 	 
	\end{equation}
The boundary values ${\mathcal W}^{(2)} (  x_1 ,  x_2 )$ of \eqref{s-tpf-2} are taken as 
$\Im z_1 \nearrow 0$ and $\Im z_2 \searrow 0$. 
In Sect.~\ref{sec:4.5}, we analytically  continue the Wightman two-point function 
from the circle $S^1$ (where they equal $c_\nu \, P_{- \frac{1}{2} - i \nu}  
		 \bigl( - \tfrac{  \vec{{\tt x}}\cdot \vec{{\tt y}} }{r^2} \bigr)$ for 
		 $\vec{{\tt x}} , \vec{{\tt y}} \in S^1 \subset dS$) to the 
Euclidean sphere: for $f, g \in C^\infty_{\mathbb{R}}(S^2)$, 
we define the \emph{covariance} $C$ by setting
	\begin{equation*}
		  C(f,g) \doteq \frac{r^2}{ 2}  \int_{S^2}  {\rm d} \Omega (\vec{{\tt x}})  
		  \int_{S^2}  {\rm d} \Omega (\vec{{\tt y}}) \; 
		{ f(\vec{{\tt x}}) } \, c_\nu \, P_{- \frac{1}{2} - i \nu}  
		 \bigl( - \tfrac{  \vec{{\tt x}} \cdot \vec{{\tt y}} }{r^2} \bigr) \,   g(\vec{{\tt y}})  \; ,  
	\end{equation*}
where $\vec{{\tt x}} \cdot \vec{{\tt y}} $ now denotes the Euclidean scalar product of the 
vectors $\vec{{\tt x}}, \vec{{\tt y}} \in S^2 \subset \mathbb{R}^3$. It is interesting to note 
that (up to a constant) $P_{-\frac{1}{2} - i \nu}  
\bigl( - \tfrac{  \vec{{\tt x}} \cdot \vec{{\tt y}} }{r^2} \bigr)$ is just the kernel
of the operator $(- \Delta_{S^2} + \mu^2)^{-1}$,  
\emph{i.e.}, 
	\[
		C(f,g) = \langle  f , g \rangle_{ \mathbb{H}^{-1} (S^2) } 
		\; , \qquad f, g \in C^{\infty}(S^2) \; . 
	\] 
Hence the Euclidean one-particle Hilbert space is the \emph{Sobolev space} $\mathbb{H}^{-1}(S^2)$.
This Hilbert space allows a decomposition into closed subspaces.
For a compact subset $K \subset S^2$, 
we define a subspace $\mathbb{H}^{-1}_{\upharpoonright K} (S^2)$ of 
the $\mathbb{H}^{-1}(S^2)$ by setting
	\begin{equation*}
		\mathbb{H}^{-1}_{\upharpoonright K} (S^2) = 
		\bigl\{ f \in \mathbb{H}^{-1}(S^2) \mid {\rm supp\,} f \subset K \bigr\}   \; . 
	\end{equation*}
This definition gives rise to Dimock's pre-Markov property (Lemma~\ref{dlemma}): 
let $e_0$ and $e_\pm$ denote the orthogonal projections from 
$\mathbb{H}^{-1}(S^2)$ onto $\mathbb{H}^{-1}_{\upharpoonright S^1} (S^2)$
and $\mathbb{H}^{-1}_{\upharpoonright \overline{S_\mp}} (S^2)$, respectively. 
Then 
	\[
			e_\mp   e_\pm  = e_0
			\qquad \hbox{on \; $\mathbb{H}^{-1} ( S^2 )$.}
	\]
Thus $\mathbb{H}^{-1}_{\upharpoonright S^1} (S^2) =   \mathbb{H}^{-1}_{\upharpoonright \overline{S_+}} (S^2) 
\cap   \mathbb{H}^{-1}_{\upharpoonright \overline{S_-}} (S^2) $.

By construction, the de Sitter space and the Euclidean sphere share the circle~$S^1$. Hence, 
in order to connect the 
de Sitter space (quantum) theory with the Euclidean (quantum) theory, we will use 
the canonical formulation of the one-particle theory given in Sect.~\ref{Sect: canon-HS}. 
In Definition~\ref{zeit-null-Hilbert} the canonical one-particle Hilbert 
space 
	\[
		\widehat{\mathfrak h} ( S^1) \cong \mathbb{H}^{-1}_{\upharpoonright S^1} (S^2)
	\]
is specified, using the Wightman two-point function given in \eqref{s-tpf-2}. 
The Fourier coefficients of the Legendre function can be expressed
in terms of $\Gamma$-functions (Proposition~\ref{thhm}), setting
	\[
		\langle h , h' \rangle_{\widehat{{\mathfrak h}} (S^1)}  = 
		\bigl\langle h  , \tfrac{1}{2\omega} 
		h' \bigr\rangle_{L^2( S^1, r {\rm d} \psi)} \; ,  
	\]
where $\omega $ is a strictly positive self-adjoint operator 
on $L^2( S^1, r {\rm d} \psi)$ with Fourier coefficients 
	\begin{equation*}
		\widetilde {\omega}(k) 
		=  \widetilde {\omega}(- k) =  \frac{k+s^+}{r} \, 
					 \frac{\Gamma \left( \frac{k+s^+}{2} \right)}
					 { \Gamma \left( \frac{k-s^+}{2} \right)}
			\frac{ \Gamma \left( \frac{k+1-s^+}{2} \right) }
			{ \Gamma \left( \frac{k+1+s^+}{2} \right)} >0  \; ,
			\quad k  \in \mathbb{Z} \; . 
	\end{equation*}
In Theorem~\ref{UIR-S1} we provide a simple and explicit description of the unitary irreducible 
representation $\widehat{u}(\Lambda)$, $\Lambda \in O(1,2)$,  on the canonical 
Hilbert space~$\widehat{\mathfrak h} ( S^1)$. 
The Casimir operator is represented by a multiple 
of the identity on~$\widehat{\mathfrak h} ( S^1)$; see Corollary~\ref{cor:4.6.7}. 

In Sect.~\ref{sec:4.7} we present a first glimpse at the Osterwalder-Schrader reconstruction
theorem, at the level of one-particle spaces:   
we discuss how a unitary representation of the rotation group 
on the Sobolev space $\mathbb{H}^{-1} ( S^2 )$ 
over the Euclidean sphere $S^2$ gives rise to a unitary representation of the Lorentz group on the 
de Sitter space. In particular, Lemma~\ref{lm:4.8.1} states that the Osterwalder-Schrader projection 
$e_0$ from the Euclidean one-particle space $\mathbb{H}^{-1} ( S^2 )$ to the canonical 
one-particle space $\widehat{\mathfrak h} ( S^1)$
gives rise to a virtual representation of $SO(3)$ on $\widehat{\mathfrak h} ( S^1)$.  
This allows for a first application of the main result (Theorem~\ref{FOS})
in the theory of \emph{virtual representations}, due to Fr\"ohlich, Osterwalder, and Seiler:
the virtual representation of $SO(3)$ can be analytically continued to a 
unitary representation of $SO_0(1,2)$. The latter is the identical to the canonical representation 
$\widehat{u}(\Lambda)$, $\Lambda \in O(1,2)$, constructed in Sect.~\ref{Sect: canon-HS}. 
Thus, for the sake of illustration, we have gone a full circle. However, the Euclidean formulation will 
actually be useful, when we perturb one-parameter groups of boosts in Chapter~9. There, 
Euclidean techniques are not only used to resolve domain questions, but also to show that 
the perturbed one-parameter groups generate a perturbed representation of the entire Lorentz group.

In order to arrive at a covariant description of the interacting Haag--Kastler net 
in Chapter 10, we investigate how the canonical 
one-particle space is related to the covariant one-particle space. 
In Sect.~\ref{sec:4.8}, we establish the following result (Theorem~\ref{st-kappa}): 
each test function $f \in \mathcal{D}_{\mathbb{R}}(dS)$ on de Sitter space can be 
decomposed into a time-reflection symmetric 
and a time-reflection anti-symmetric part, and for both there exists a representer  who has its support 
on the Cauchy surface (as the Fourier--Helgason transform extends to distributions supported on $S^1$); 
see \eqref{s1-to-dS} below for further details.   

Given a one-particle representation of the Lorentz group, free quantum fields are 
easily defined, by the method of second quantisation: 
in Sect.~\ref{Fockspace}, we define creation operators ${ a}^*({f})$ 
and annihiliation operators ${ a}({f})$, field operators 
	\[ 
		\Phi_F (f) \doteq \frac{1}{ \sqrt{2}}\bigl ( { a}^*({f}) + { a}({f}) \bigr)^- \;  , 
		\qquad f  \in {\mathfrak h}  \; , 
	\]
and Weyl operators ${ W}_F ({f}) \doteq {\rm e}^{ i  \Phi_F (f) }$, $f  \in {\mathfrak h}$, 
all acting on the bosonic Fock space
	\[
			\Gamma ( {\mathfrak h} )  \doteq\oplus_{n= 0}^{\infty}  \Gamma^{(n)} ( {\mathfrak h} ) \;  , 
			\qquad
			\Gamma^{(n)} ( {\mathfrak h} )
			= \underbrace{{\mathfrak h} \otimes_s \cdots \otimes_s {\mathfrak h} }_{n-times}\; . 
	\]
Well-known theorems by Araki (Theorem~\ref{araki}), Leyland, Roberts and 
Testard (Theorem~\ref{th:4.9.2}) and Eckmann and Osterwalder (Theorem~\ref{EckOsw}) 
describe properties the net of local von Neumann algebras for the free massive 
scalar quantum field inherits from the one-particle description. These results are used in 
Chapter~7.

In Chapter~\ref{ch:5} we provide a brief account of classical field theory on de Sitter space. 
In particular, in Sect.~\ref{sec:5.1}
we derive the classical equations of motion from a classical Lagrange density 
	\begin{equation*}
		{\mathcal L}(\mathbb{\Phi}) 
		=  \frac{1}{2}  d \mathbb{\Phi} \wedge * d \mathbb{\Phi}  
		- \frac{\mu^2}{2} \mathbb{\Phi} * \mathbb{\Phi}  - P( \mathbb{\Phi})  * 1  \; . 
	\end{equation*}
In Sect.~\ref{SET} the Noether theorem (Theorem~\ref{noether}) is presented. The latter provides  
explicit expressions for the conserved quantities associated to the boosts and the rotations. 
These classical quantities are expressed in terms of integrals over components 
of the classical energy-stress tensor ${T^\mu}_{\nu}$:
	\[
		L^{\scriptscriptstyle Class.}_1  = \int_{S^1} \, r^2 \, \cos\psi \; {\rm d} \psi \;    {T}_{00}  
		\qquad \text{and} \qquad 
		K^{\scriptscriptstyle Class.}_0 = \int_{S^1} \,   r \; {\rm d} \psi \; {T}_{10} \; ,
	\]
with 
	\begin{align*}
		{T^0}_{0}  &= \frac{1}{2} \left( \mathbb{\pi}^2 +  r^{-2} \bigl( \partial_\psi  \mathbb{\Phi} \bigr)^2 + 
					\mu^2 \mathbb{\Phi}^2 \right)  + P  (\mathbb{\Phi} ) \; , 
		\\
		{T^{1}}_{0} & 
		= r^{-2} (\partial_\psi \mathbb{\Phi} )\, \mathbb{\pi}  \; ,  \qquad  \mathbb{\pi} 
		= \tfrac{\partial}{\partial x_0} \mathbb{\Phi}	 \; . 	 
	\end{align*}
We will provide nearly identical formulas for the corresponding quantum quantities in Chapter 11. 

The covariant classical dynamical 
system is presented in Sect.~\ref{CCDS}, taking advantage of the existence of retarded and advanced 
solutions $\mathbb{E}^{\pm}f$
of the inhomogeneous Klein-Gordon equation (Theorem~\ref{fundamental}): 
	\begin{equation*}
			 (\square_{dS}+\mu^2)  \mathbb{\Phi}   
		 =  f \; , \qquad f \in{\mathcal D}_\mathbb{R} (dS) \; . 
	\end{equation*}
The difference between the retarded and the advanced solution of 
the inhomogeneous equation, namely 
	\begin{equation*}
	\mathbb{\Phi}=\mathbb{E} f \; , \qquad \text{with} 
	\quad  \mathbb{E} = \mathbb{E}^{+} - \mathbb{E}^{-} \; , 
	\end{equation*}
is a solution of the homogenous Klein--Gordon 
equation $(\square_{dS}+\mu^2)  \mathbb{\Phi} = 0$.
Next, we recall a result of Bär, Ginoux and Pf\"affle (Theorem~\ref{solutions}) describing 
the smooth real valued solutions of the homogeneous Klein-Gordon equation.  
Support properties of the solutions associated to test-functions $f \in{\mathcal D}_\mathbb{R} (dS)$ 
with compact support are 
discussed in Lemma~\ref{Lm3.8}. In Proposition~\ref{Prop4-10} we state that the covariant 
symplectic space 
	\begin{equation*}
	{\mathfrak k}(dS) \doteq  
	{\mathcal D}_{\mathbb{R}} (dS) / 
		(\square_{dS}+\mu^2) {\mathcal D}_{\mathbb{R}} (dS)   
	\end{equation*}
carries a representation  
	\[
		\Lambda \mapsto {\mathfrak z} (\Lambda)\; ,  \qquad \Lambda \in O(1,2) \; , 
	\]
of the Lorentz group, defined by  ${\mathfrak z} (\Lambda) [f] = [ \Lambda_* f]$, 
and thus gives 
rise to the \emph{covariant} classical dynamical system (Definition~\ref{Def4-11}). 
The restriction of this dynamical system to a wedge, presented in Sect.~\ref{sec:5.4} provides
explicit expression for the propagator using formulas from Sect.~\ref{sec:1.5}; 
see Lemma~\ref{PropWedge}. The description 
of the \emph{canonical} classical dynamical system, given in Sect.~\ref{CaCDS}, takes 
advantage of the solution of the Cauchy problem for the Klein--Gordon equation, 
provided by Dimock (Theorem~\ref{cauchyproblem}):
let~$(\mathbb{\phi}, \mathbb{\pi}) \in  
C^\infty (S^1) \times C^\infty (S^1) $. 
Then there exists a unique $\mathbb{\Phi} \in C^\infty (dS)$ satisfying the 
homogeneous Klein--Gordon equation  with Cauchy data
	\begin{equation}
		\label{s-3.26} 
		\mathbb{\Phi}_{\upharpoonright {\mathcal C}} = \mathbb{\phi} \; , 
		\quad (n \mathbb{\Phi})_{\upharpoonright {\mathcal C}} = \mathbb{\pi} \; . 
	\end{equation}
The space of Cauchy data $\widehat {\mathfrak k} (S^1) 
		\doteq C^\infty_\mathbb{R}(S^1)\times C^\infty_\mathbb{R}(S^1) $,
together with the {\em canonical symplectic form} 
	\begin{equation*}
		\widehat \sigma\big((\mathbb{\phi}_1,\mathbb{\pi}_1),
		(\mathbb{\phi}_2,\mathbb{\pi}_2)\big)
		\doteq \langle \mathbb{\phi}_1,\mathbb{\pi}_2 
		\rangle_{L^2(S^1, \, r\,  {\rm d} \psi ) }-  
		\langle\mathbb{\pi}_1,\mathbb{\phi}_2 \rangle_{L^2(S^1, \, r \, {\rm d} \psi)} \; , 
	\end{equation*}
forms a symplectic space $ \widehat {\mathfrak k} (S^1)$, which carries a 
representation $\Lambda  \mapsto \widehat  {\mathfrak z} ({\Lambda}) $,
$\Lambda \in O(1,2) $, defined  by  
	\begin{equation*}
		\widehat {\mathfrak z} (\Lambda) (\mathbb{\phi},\mathbb{\pi})  
		\doteq\big(  ( \Lambda_* \mathbb{\Phi} )_{ \upharpoonright S^1} \; ,
		( n  \Lambda_* \mathbb{\Phi})_{\upharpoonright S^1}\big) \; .
	\end{equation*}
Here $\mathbb{\Phi}$ is the unique $C^\infty$-solution of the homogeneous Klein--Gordon 
equation with Cauchy data given by \eqref{s-3.26}.
The canonical classical dynamical system $( \widehat {\mathfrak k} (S^1), \widehat {\mathfrak z})$
is introduced in Proposition~\ref{nocheinlabel}. As one may expect, a symplectic map
connects the canonical to the covariant dynamical system (Proposition~\ref{cauchy-symplectic}). 
In Proposition~\ref{fsol}, we state how the localisation properties of the solutions of the Klein-Gordon equation 
manifest themselves on the space $\widehat {\mathfrak k} (S^1)$.
An explicit description of the canonical classical dynamical system
is given in Proposition~\ref{porp4.13}, taking advantage of the results of Sect.~\ref{sec:5.4}.
The restriction of the canonical classical dynamical system
to two half-circles gives rise to a \emph{double classical linear system} in the sense 
of Kay (Proposition~\ref{ssdd}). 

In Chapter 6,  we relate (following Kay) different realisations of the one-particle 
Hilbert space and the one-particle dynamics with the corresponding classical dynamical systems 
described in Chapter 5: given a classical dynamical system for the Klein--Gordon equation on the 
de Sitter space (in either the covariant or the canonical formulation) there is a \emph{unique one-particle 
quantum system} associated to it, characterised by the one-particle \emph{geodesic KMS condition}.  
The covariant de Sitter one-particle structure is given by the identity map
	\begin{align}
	\label{s-K1PStrucHe}
		K  \colon  \qquad {\mathfrak k}(dS) 
		& \rightarrow  {\mathfrak h} (dS) \nonumber \\
		  [f] 
		  & \mapsto   [f]\; , \qquad f  \in {\mathcal D}_{\mathbb{R}} (dS) \; , 
	\end{align}
which intertwines the (anti-) unitary representation $u$ of $O(1,2)$ on ${\mathfrak h} (dS)$
with symplectic representation ${\mathfrak z} $ of $O(1,2)$ on ${\mathfrak k}(dS)$:
	\begin{equation} 
		\label{s-eqUHe}
		 u (\Lambda)\circ K = K \circ  {\mathfrak z} (\Lambda)  \;, 
		 \qquad \Lambda \in O(1,2)\; .   
	\end{equation}
As a consequence of the geodesic KMS condition, the
\emph{pre-Bisogano Wichmann property} holds: for any wedge $W$,
	\begin{equation*}
		u (\Lambda_W ( i \pi)) [g] 
		= u (\Theta_{W}) [g] \; , 
		\qquad [g] \in K {\mathfrak k}(W) \; , 
	\end{equation*}
where $\Theta_{W}$ is the reflection at the edge of the wedge $W$ (Theorem~\ref{one-p-BW}). 
Hence, the one-particle Tomita operator $s_{\scriptscriptstyle W}$  has the polar decomposition 
	\begin{equation*} 
			s_{\scriptscriptstyle W} 
			= u  \bigl(\Theta_{W}\bigr) \,  
			u\bigl(\Lambda_{\scriptscriptstyle W}( i\pi) \bigr) \; . 
	\end{equation*}

On our way towards a canonical one-particle structure, we start with a detour, which also 
serves to relate our work to earlier work \cite{FHN}. 
In Sect.~\ref{KMSops}, we describe one-particle KMS structures with respect to the 
boosts $t \mapsto \Lambda_1(t)$, taking advantage of explicit formulas derived in 
Sect.~\ref{sec:5.4}. 
We define an $\mathbb{R}$-linear map $\widehat{K}_\beta \colon\widehat {\mathfrak k}(S^1)
\to\widehat {\mathfrak  d} (S^1) $, $\beta >0$,  by setting
	\begin{equation*}
		\widehat{K}_\beta (\mathbb{\phi},{\mathbb \pi}) \doteq
			\big((1+ \rho_\beta )^{{\frac{1}{2}}}+\rho_\beta^{\frac{1}{2}}\, j \big) \;
 			\widehat{K}_\infty (\mathbb{\phi},{\mathbb \pi}) \; , 
		\qquad (1+\rho_\beta) = (1  -  {\rm e}^{-\beta|\varepsilon|} )^{-1} \; . 
	\end{equation*}
It was recognised by Borchers and Buchholz that the 
proper, ortho\-chron\-ous Lorentz group $SO_0(1,2)$ 
can be unitarily implemented if and only if~$\beta = 2 \pi r$. In fact, for 
$\beta = 2 \pi r$, the unitary map 
	\begin{align*} 
		\mathbb{v} \colon \widehat {\mathfrak  d} (S^1) & \to \widehat{{\mathfrak h}}  (S^1) \\
		h & \mapsto 
		\tfrac{1}{\sqrt{r}}  \; 
		|\,  \mathbb{cos}_\psi |^{-1}\, \big( \rho_{2 \pi}^{\frac{1}{2}} (P_1)_* - (1+\rho_{2 \pi})^{\frac{1}{2}}  \big) h \; , 
	\end{align*}
allows us to implement the rotations $R_0 (\alpha)$, $\alpha \in [0, 2 \pi)$, in the double $(2 \pi r)$-KMS 
one-particle structure. According to Proposition~\ref{prop:6.4.3}, the map 
	\begin{align*}
	\label{K1PStrucHe}
		 \widehat{K} \colon  \qquad \widehat{\mathfrak k}(S^1) 
		& \rightarrow  \widehat{{\mathfrak h}}  (S^1) \nonumber \\
		 (\mathbb{\phi},\mathbb{\pi})  		  & \mapsto   \tfrac{1}{\sqrt{r}}  ( - \mathbb{\pi} +  i \,\omega
		 r \, \mathbb{\phi}) \; .
	\end{align*}
yields a double $2\pi$-KMS one-particle structure 
	\[
		\bigl(\widehat{K} , \widehat{{\mathfrak h}}  (S^1), {\rm e}^{i t \, 
		r  \omega \,  \mathbb{cos}_\psi } , 
		C (P_1)_* \bigr) 
	\]
for the classical double dynamical system 
$\bigl( \widehat {\mathfrak k}  \left(S^1 \setminus \{ - \frac{\pi}{2} , 
\frac{\pi}{2} \} \right) , \widehat \sigma, \widehat {\mathfrak z} (\Lambda_1( \frac{t}{r}) ),  
\widehat {\mathfrak z} (P_1T )  \bigr)$ in the sense
of~\ref{dbops}, unitarily equivalent to the canonical de Sitter one-particle structure
$\bigl(\, \widehat{K}_{2 \pi} , \widehat {\mathfrak  d} (S^1), {\rm e}^{i \frac{t}{r} \varepsilon},  j \bigr) $, 
in agreement with Theorem~\ref{ThB1}. The main result of
Chapter 6 can be summarised as follows: there exists a unitary map ${\mathfrak U}$  
from the canonical one-particle Hilbert space $\widehat{{\mathfrak h}}  (S^1)$ to 
the covariant one-particle Hilbert space ${\mathfrak h} (dS)$, 
which  intertwines  the representations $\widehat{u} (\Lambda)$ and 
$ u (\Lambda)$, $ \Lambda \in O(1,2)$, and the one-particle 
structures (Proposition~\ref{Prop5.7}).
In other words, the following diagram commutes:

\begin{picture}(250,140)

\put(60,80){$\nearrow$}
\put(0,70){${\mathcal D}_{\mathbb{R}} (dS)$}
\put(220,70){${\mathfrak U} $}
\put(130,70){${\mathbb T} $}
\put(50,95){$\widehat  {\mathbb P}$}
\put(50,45){$ {{\mathbb P}}$}
\put(60,55){$\searrow$}

\put(90,110){$ (\widehat {\mathfrak k} (S^1 ), \widehat {\mathfrak z})$}
\put(160,110){$\longrightarrow$}
\put(190,110){$(\widehat{{\mathfrak h}}  (S^1), \widehat{u})$}

\put(90,30){$({\mathfrak k} (dS), {\mathfrak z})$}
\put(160,30){$\longrightarrow$}
\put(190,30){$({\mathfrak h} (dS), u)\; .$}

\put(165,40){$K$}
\put(165,120){$ \widehat{K}$}

\put(120,85){\vector(0,-3){20}}
\put(210,85){\vector(0,-3){20}}

\end{picture}

\vskip -.8cm
\noindent
Moreover, according to Proposition~\ref{prop:4.10.5}, 
the unitary map $\mathfrak{U} \colon \widehat{{\mathfrak h}}  (S^1) \to {\mathfrak h}(dS)$ 
is  the linear extension of the map
	\begin{equation}
	\label{s1-to-dS}
		h_1 + i \omega r h_2 \mapsto [\delta \otimes h_1] + 
		[\delta' \otimes h_2] \; .
	\end{equation}
This map respects the local structure: the restrictions $\mathfrak{U}_{\upharpoonright I}$
of the map $\mathfrak{U}$ to some interval  $I \subset S^1$
maps  
	\begin{equation*}
		\widehat{{\mathfrak h}}  ( I) 
		\doteq
		\overline{ \bigl\{ h \in \widehat{{\mathfrak h}}  (S^1)   \mid  
		{\rm supp\,}  \Re h \subset I \, ,    \, {\rm supp\,}  \omega^{-1}\Im h  \subset I \bigr\} }
	\end{equation*}
to ${\mathfrak h}({\mathcal O}_I)$, 
with ${\mathcal O}_I= I''$ the causal completion of $I$. 

The final two statements in Chapter 6 concern the causal structure. 
They will ensure that the interacting covariant Haag--Kastler net 
is non-trivial. The unitary operator 
	\[
		{\rm e}^{i t \omega r \,  \mathbb{cos}_\psi } \colon 
		\widehat{{\mathfrak h}}  (I)
		\hookrightarrow
		\widehat{{\mathfrak h}}  \bigl(  \Gamma(\Lambda_1(t)I) \cap S^1 \bigr) \; , 
		\qquad t \in \mathbb{R} \;  , 
	        \]
with $\Gamma \bigl( \Lambda_1 (t) I \bigr)$ the causal dependence region of $\Lambda_1(t) I$.
                In particular, the unitary group $t \mapsto {\rm e}^{it (\omega r \,  
\mathbb{cos}_\psi )_{\upharpoonright I_\pm}}$ leaves $\widehat{{\mathfrak h}}  (I_\pm)$ invariant 
(Proposition~\ref{halpha}).

In Chapter~\ref{2Q}, we describe the net of local von Neumann algebras 
	\begin{equation*}
		{\mathscr A}_\circ ({\mathcal O}) \doteq  \pi^\circ \bigl({\mathfrak W}
		({\mathcal O}) \bigr)'' \; ,  \qquad {\mathcal O} \subset dS \; , 
	\end{equation*} 
for the free massive (\emph{i.e.}, $\mu^2 >0$) 
scalar field on the de Sitter space. Here 
${\mathfrak W}({\mathcal O}) \equiv {\mathfrak W}({\mathfrak k}({\mathcal O}), \sigma)$
is a local Weyl algebra for the covariant dynamical system $({\mathfrak k}(dS), \sigma)$
and $\pi^\circ$ denotes the Fock representation. The maps 
\eqref{s-K1PStrucHe} and \eqref{s-eqUHe} are used to 
specify the \emph{covariant} quantum dynamical system (Definition~\ref{def:7.1.1}).
The de Sitter vacuum state is characterised  by the geodesic KMS condition 
(Definition~\ref{vs}); 
Theorem~\ref{th2.5} states that for the free dynamics such a state is unique. 
Moreover, the local von Neumann algebra ${\mathscr A}_\circ ({\mathcal O}) $ 
is equal to the von Neumann algebra generated 
by ${\rm W}_{F} (f )$, $f \in {\mathfrak h}({\mathcal O})$. 
Proposition~\ref{prop:7.1.6} ensures that the state induced by the Fock  vacuum vector
satisfies the geodesic KMS condition; hence, for the free field, the state induced by the 
Fock  vacuum vector is the de Sitter vacuum state. The Haag--Kastler axioms for the free 
field are summarized in Theorem~\ref{th:7.1.5}. 
The Reeh-Schlieder property (Theorem~\ref{th:7.1.8}) 
follows from the an argument due to Bros and Moschella.  
The \emph{canonical} net of local von Neumann algebras on the circle $S^1$, 
	\begin{equation*}
			{\mathcal R} (I) \doteq  \pi^\circ \bigl( \,
		\widehat{\mathfrak W} (\widehat {\mathfrak k}(I) )\bigr)'' \; ,  \qquad I \subset S^1 \; ,  
	\end{equation*}
is presented in Sect.~\ref{sec:7.2}. The free de Sitter vacuum state for the canonical 
quantum dynamical system
is described in Theorem~\ref{th:7.2.2}. The local von Neumann algebras associated to open
intervals on the circle $S^1$ are hyperfinite type III$_1$
factors (Theorem~\ref{th:7.2.4}). The proof of this result is due to Figliolini and Guido; 
it uses the fact that the net of local von Neumann algebras is regular from the inside 
and from the outside (Proposition~\ref{l6.1}).  The canonical net of local von Neumann algebras 
satisfies finite speed of propagation (Theorem~\ref{fst-theorem}), 
\begin{equation*}
		\widehat 
		\alpha^{\circ}_{\Lambda_1 (t)}
 		\colon  
		{\mathcal R} (I)\hookrightarrow 
		{\mathcal R} \bigl( I (t)  \bigr) \; , 
		\qquad I (t) 
		= S^1 \cap \Gamma \bigl( \Lambda_1 (t) I \bigr) \; , 
	\end{equation*} 
and the time-slice axiom (Theorem~\ref{p6.1}). 
Euclidean fields on the sphere are described in Sect.~\ref{sec:7.3}. 
Euclidean Haag--Kastler axioms for the free field are stated in 
Theorem~\ref{th:7.3.1}. 
The Markov property (Theorem~\ref{pre-martheo}) implies reflection 
positivity (Theorem~\ref{martheo}). Reflection positive state are 
introduced in Definition~\ref{def:7.3.7}. 
The Euclidean Fock vacuum state is reflection positive (Corollary~\ref{cor:7.3.8}).

Chapter~\ref{ch:8} is dedicated to a construction of the 
\emph{interacting vacuum state}. In order to make the polynomial 
interactions well defined, an ultra-violet renormalisation is
necessary. In two space-time dimension, this can be achieved by normal ordering 
monomials, setting
	\begin{equation*}
			{:} \, \mathbb{\Phi}(f)^{n}\, {:}  
				\doteq \sum_{m=0}^{[n/2]}\frac{n!}{m!(n-2m)!}
					\mathbb{\Phi}(f)^{n-2m} 
					\Bigl(-\tfrac{1}{2}  \| f \|^2_{\mathbb{H}^{-1}(S^2)}  \Bigr)^{m} \; , 
					\quad f \in  \mathbb{H}^{-1} (S^2) \; , 
	\end{equation*}
where $[ \, . \, ]$ denotes the integer part. We note that $f$ can \emph{not} be replaced 
by a two-dimensional $\delta$-function on the sphere, as the $\mathbb{H}^{-1}$-norm
would not be finite. To gain a better understanding of this ultra-violet problem, we  
describe the short-distance behaviour of 
Euclidean covariance in Theorem~\ref{c-log}, following ideas of 
De Angelis, de Falco and Di Genova.
In Sect.~\ref{sec:8.2} we use Tomita-Takesaki modular theory to
define non-abelian $L^p$-spaces, only to restrict to the abelian case 
in Sect.~\ref{sec:8.2.1}. In Sect.~\ref{sec:8.2.2}, we provide
$L^p$ estimates for Eucildean field operators. In Sect.~\ref{sec2.2}, 
the ultra-violet renormalization of the interaction on the Euclidean 
sphere (Theorem~\ref{uvtheo}) and the Cauchy surface (Theorem~\ref{wickooo})
is carried through: we show that for $n \in {\mathbb N}$ and $f\in L^{2}(S^2, {\rm d} \Omega)$
the following  limit exists in $\kern -.2cm \bigcap \limits_{1\leq p<\infty} \kern -.2cm 
L^{p}( \mathscr{U}(S^2) , \Omega_\circ )$:
	\[
		\lim_{m \to \infty}\int_0^{2 \pi} r \, {\rm d} \theta \int_{-\pi/2}^{\pi/2} r \cos \psi \, 
		{\rm d} \psi \;  f(\theta,\psi) \; 
		{:}\, \mathbb{\Phi} \bigl(  \delta^{(2)}_{m}(\, .-\theta, \, .- \psi)  \bigr)^n \, {:}  \; .
	\]
It is denoted by $\int_{S^2} {\rm d} \Omega \; f(\theta,\psi)  \; {:} \, \mathbb{\Phi}( \theta,\psi)^{n} \, {:} \; $. 
Similarly, the limit 
	\begin{equation}
	\label{as-in-V}
		\lim_{k\to \infty}\int_{S^1} r {\rm d} \psi \;  h (\psi)\,  {:}\mathbb{\Phi}(0, 
		\delta_{k}(.-\psi))^{n}  {:}  \; , 
		\qquad h \in L^2 ( S^1, {\rm d} \psi) \; .  
 	\end{equation}
exists  
in $\bigcap_{1\leq p<\infty}L^{p}( \mathscr{U}(S^1) , \Omega_\circ )$.
It is denoted by $\int_{S^1}  r {\rm d} \psi \;  h (\psi)\,  {:}\mathbb{\Phi}(0, \psi)^{n} {:} \; $.
Before treating the interaction itself, we show that the Euclidean field allows a foliation; 
this is the content of Lemma~\ref{1.0b}. 
A foliation of the interaction is provided in Theorem~\ref{2.2}:
for $h\in L^{2} \bigl( I_+ \, , r {\rm d} \psi \bigr)$ and $g \in C^\infty (S^{1})$,
	\begin{align*}  
		\int_{S^{1}} r \, {\rm d} \theta \; g(\theta) \; \mathbb{U}_\circ^{(0)} (\theta) & 
		V ( \operatorname{\mathbb{cos}} h)  
		\mathbb{U}_\circ^{(0)} (-\theta)  
				= \int_{S^2} {\rm d} \Omega \;   
				g(\theta)  h(\psi) \; {:}{\mathscr P}(\mathbb{\Phi}(\theta,\psi)) {:} 
	\end{align*}
as unbounded operators acting on the Euclidean Fock space $ \bbH$. 
Here the unitary operator $\mathbb{U}_\circ^{(0)} ( \theta )$ 
implements the rotation $R_1(\theta)$ on $ \bbH$ and
	\begin{equation}
		\label{s-bbv-interaction}
		V (h) \doteq \int_{S^1}  r {\rm d} \psi \;  h(\psi) \, {:} {\mathscr P}(\mathbb{\Phi}(0,\psi)) {:} \; ,
		 \qquad h \in L^2 (S^1, r {\rm d} \psi)\; .
	\end{equation}
In particular, 
	\[ 
		Q(S^2)= 
		\int_{S^2} {\rm d} \Omega \; {:}{\mathscr P}(\mathbb{\Phi}(\theta,\psi)) {:} 
				= \int_{S^{1}} r \, {\rm d} \theta \; V^{(0)} ( \theta) \, ,  
	\]
where $V^{(0)} ( \theta) \doteq \mathbb{U}_\circ^{(0)} ( \theta ) 
V (   \chi_{ I_+  } \, \mathbb{cos} ) \mathbb{U}_\circ^{(0)} ( - \theta )$. 

The existence of the vacuum vector in Fock space is guaranteed in Sect.~\ref{sec:8.4}, 
by Theorem~\ref{th:8.4.1}, due to Glimm:
if the real-valued polynomial ${\mathscr P}$ is bounded from below, 
	\begin{equation*}
		{\rm e}^{- Q (S^2)}  \in L^{1}( \mathscr{U}(S^2) , \Omega_\circ ) \; . 
	 \end{equation*}
A similar result for the time-zero interaction on a half-circle is stated in Theorem~\ref{thm:8.4.3}:
	\begin{equation*}
		{\rm e}^{- t V (\chi_I \, \mathbb{cos})}  \in L^{1}( \mathscr{U}(S^1) , \Omega_\circ ) \; , 
		\qquad \forall t >0 \;.  
	 \end{equation*}
Hence, the Euclidean interacting vacuum vector  
	\begin{equation*}	 
		\mathbb{\Omega} \doteq \frac{  {\rm e}^{-  Q \left(S_+ \right)} \Omega_\circ } 
		{ \|  {\rm e}^{-  Q \left(S_+ \right)} \Omega_\circ \| } \in \bbH_+  \, .  
	\end{equation*} 
The canonical interacting de Sitter vacuum vector 
	\begin{equation*}
		\widehat{\Omega} \doteq \frac{ E_0 \mathbb{\Omega} }
		{ \left\|  E_0  \mathbb{\Omega} \right\| }
		 \in \widehat{\mathcal H} (S^1)  
	\end{equation*}
is cyclic and separating for the abelian algebra generated by the time-zero 
fields (Theorem~\ref{th:8.4.2}). At this point we state 
a general result, which ensures semi-boundedness of an operator 
sum (Theorem~\ref{thm:8.4.4}). It is applied in Chapter~\ref{ch:11}.

Chapter~\ref{interactingdesitter} deals with the interacting representation of $SO(1,2)$. 
The key idea is to use the Osterwalder-Schrader projection $E_0$ to 
define a virtual representation of $SO(3)$ on the canonical Fock space $\widehat{\mathcal H}(S^1)$, 
by setting 
	\begin{equation} 
		\label{s-eqDefPa}
		\mathcal{P} (R) E_0 A \mathbb{\Omega}  
		\doteq E_0\, \mathbb{\alpha}^\circ_{R }(A) \mathbb{\Omega} \; , 
		\qquad A , \mathbb{\alpha}^\circ_{R }(A)  \in \mathscr{U}(S_+)  \;  , 
		\quad R \in SO(3) \; .  
	\end{equation} 
As the rotations (in general) do not preserve the upper hemisphere, 
domain questions arise from \eqref{s-eqDefPa}.
In Sect.~\ref{sec:9.1}, 
the generator of the unitary group implementing the interacting boosts $t \mapsto \Lambda_1(t)$
is reconstructed, starting from the unitary group implementing the rotations $\theta \mapsto R_1(\theta)$
acting on the Euclidean sphere. Note that the maps $t \mapsto \Lambda_1(t)$ and $\theta \mapsto R_1(\theta)$
have common fixed points, lying on the circle $S^1$.
In Proposition~\ref{SemiGroupL}
we state that \eqref{s-eqDefPa} gives rise to a local symmetric semi-group, and 
using the main result on local symmetric semi-groups (Theorem~\ref{klein-l-f}), 
due to Fr\"ohlich and (independently) Klein \& Landau, this yields a unitary 
representation of the interacting boosts on the canonical Fock space.   
These boosts and the rotations leaving the Cauchy surface invariant are 
combined in Sect.~\ref{sec5.8}. In Theorem~\ref{uo} we present the 
resulting interacting unitary representation of the Lorentz group. 
In Definition~\ref{def:6}, we define the corresponding automorphisms on the 
canonical Fock space $\widehat{\mathcal H}(S^1)$. In the sequel (Theorem~\ref{9.2.4}), 
we express the Tomita operator for the pair $\bigr( \mathcal{R}(I_+),
\widehat{\Omega} \bigl)$ in terms of the generator of the boost and the implementer
of the reflection at the edge of the wedge. We note that the abstract construction 
in Sect.~\ref{sec5.8} does not yield explicit expressions for the interacting generator 
of the boosts $t \mapsto \Lambda_1(t)$.
In Sect.~\ref{arpermodaut}, we overcome this issue, as we establish 
the connection of our work with Araki's
perturbation theory of modular automorphisms. One of our main 
result is the justification of the operator sum 
	\begin{equation}
	 	\LFock^{(0)} = \overline{\LFock_\circ^{(0)}+V (\mathbb{cos}) }   \; , 
		\label{operator-sum-summary}
	\end{equation} 
where $V (\mathbb{cos})$ is given by \eqref{s-bbv-interaction}, 
with ${\mathscr P}$ a real valued polynomial, bounded from below. Note that both operators, 
$\LFock_\circ^{(0)}$ and $V (\mathbb{cos})$
are unbounded from both above and below. The formula~\eqref{s-bbv-interaction} for the second operator 
is merely an abbreviation, as the proper definition of this operator requires an 
ultra-violett renormalisation; see \eqref{as-in-V}. 
The sum \eqref{operator-sum-summary}
provides the crucial link between the free and the interacting quantum field 
theory, as $\widehat{L}^{(0)}$ is the generator of the one-parameter 
unitary group implementing the boost which keeps the wedge $W_1$ invariant. 
Various expressions 
for the canonical 
interacting vacuum vector $\widehat{\Omega}$ are given in Theorem~\ref{th5.2}: 
for example, 
	\begin{equation*}
		\widehat{\Omega} = \frac{ {\rm e}^{-\pi \widehat{H}^{(0)} } \Omega_\circ }
		{ \|{\rm e}^{-\pi \widehat{H}^{(0)}  } \Omega_\circ  \| } \; , 
	\end{equation*}
with
	\[
	 	\widehat{H}^{(0)}  := \widehat{L}_\circ^{(0)} + \int_{0}^{\pi}  r {\rm d} \psi \;  r
		\cos (\psi) \, {:} {\mathscr P}(\mathbb{\Phi}(0,\psi)) {:}   \; . 
	\] 
The \emph{relative modular operator}	
	\begin{equation*}
		\Delta_{ \widehat{\Omega},\Omega_\circ}
		= \frac{ {\rm e}^{-2\pi \widehat{H}^{(0)} }  }{ 
		\|{\rm e}^{-\pi \widehat{H}^{(0)}}\Omega_\circ  \|^2 } \, ; 
	\end{equation*}
for the triple $\bigr(\mathcal{R}(I_+),
\Omega_\circ, \widehat{\Omega} \bigl)$ computed in Theorem~\ref{H}. 
The uniqueness of the interacting de Sitter vacuum state is established in 
Theorem~\ref{keyresult3}.

In Chapter~\ref{ch:10}, we first define the canonical net of local algebras associated to the interacting model. 
This net has finite speed of propagation (Theorem~\ref{fst-2-theorem}), \emph{i.e.}, 
  \begin{equation*}
		\widehat 
		\alpha_{\Lambda_1(t)}
 		\colon  
		{\mathcal R} (I)\hookrightarrow 
		{\mathcal R} \bigl( I (t)  \bigr) \; , 
		\qquad I (t) 
		= S^1 \cap \Gamma \bigl( \Lambda_1 (t) I \bigr) \; . 		
 \end{equation*}
It also satisfies the time-slice axiom (Theorem~\ref{p6.1b}). 
The covariant net of local von Neumann algebras is given by 
the intersection of wedge algebras (Definition~\ref{locoinfield}): 
\begin{itemize}
\item[$ i.)$] for an arbitrary wedge $W= \Lambda W_1$, $\Lambda \in
  SO_0(1, 2)$, we set 
	 \begin{equation*}
	 	{\mathscr A} ( W ) 
		\doteq \alpha_\Lambda \bigl( \mathscr{A}_\circ (W_1) \bigr) \;  . 	
	\end{equation*}
\item[$ ii.)$] 
for an arbitrary bounded, causally complete, convex region (these are the de Sitter analogs of 
the double cones) ${\mathcal O} \subset dS$, we set   
	\begin{equation*} 
		 {\mathscr A} ({\mathcal O}) 
		\doteq \bigcap_{W \supset {\mathcal O} } {\mathscr A}  \bigl( W \bigr) \; . 
	\end{equation*}
\end{itemize} 
The local time-zero algebras for the interacting and the free 
model coincide (Theorem~\ref{AA0}), \emph{i.e.}, 
	\[
		\mathscr{A} ({\mathcal O}_I) = \mathscr{A}_\circ ({\mathcal O}_I) \; , 
		\qquad I \subset S^1 \; . 
	\]
This implies that the von Neumann algebras $\mathscr{A} ({\mathcal O})$, $ {\mathcal O} \subset dS$, 
for bounded, causally complete regions ${\mathcal O}$ are hyperfinite type III$_1$ factors, just as in the 
non-interacting case.
The Haag--Kastler axioms for the net of local von Neumann algebras 
	\[
		{\mathcal O} \mapsto \mathscr{A} ({\mathcal O}) \; , 
		\qquad {\mathcal O} \subset dS \; , 
	\]
associated to the interacting model are stated 
in Theorem~\ref{th:6}. 

In the last chapter, Chapter~\ref{ch:11}, we discuss the importance of the quantized stress-energy 
tensor, whose components are 
	\begin{align*}
		{{\rm T}^{0}}_{0}  (\psi) &=  \tfrac{1}{2}  \bigl( \pi^2 (\psi)
			+ \tfrac{1}{r^2} \bigl( \tfrac{ \partial \varphi }{\partial \psi} (\psi) \bigr)^2 
			+ \mu^2 \varphi (\psi)^2  + {:}  \mathscr{P} (\varphi(\psi)) {:} \bigr)  \; , 
				\\
			{T^{1}}_{0} (\psi) & = r^{-2}  \pi (\psi) \, 
			(\partial_\psi \varphi) (\psi)  \; . 
	\end{align*}
The generators of the boosts and the rotations are expressed in terms of 
integrals over the quantized
stress-energy tensor density (Theorem~\ref{Q-conserved-quantities}): the 
following operator  identities
hold on the canonical Fock space:
	\begin{align*} 
			\widehat{L}^{(0)} &=  \int_{S^1} \,  r^2 
			  \cos  (\psi)  \, {\rm d} \psi  \;    {\rm T}_{00}(\psi)  \; , \qquad
		\widehat{K}_0  = \int_{S^1} \,  r  \; {\rm d} \psi \;   {\rm T}_{10}(\psi)   
		\; .
	\end{align*}
Note that the notation on the r.h.s.~in both equations is symbolic; it has to be interpreted as 
in~\eqref{as-in-V}.  Glimm-Jaffe type $\phi$-bounds are 
given in Proposition~\ref{phi-bound}:
for $c\gg 1$ and $g\in \widehat{\mathfrak h}(S^1) $,
	\begin{equation*}
		\pm \varphi(g)\leq C \|g\|_{\widehat{\mathfrak h}(S^1)}
		\left( \int_{S^1} r \, {\rm d} \psi \;  {\rm T}_{00} (\psi) 
		+c \cdot \mathbb{1} \right)^{\frac{1}{2}} 
		\; . 
	\end{equation*}
In particular, $ \int_{S^1} r \, {\rm d} \psi \; {\rm T}_{00} (\psi) $ is bounded from below.
The {\em covariant} equation of motion  
	\begin{equation*}
		\bigl( \square_{dS}+\mu^2 \bigr) {\Phi}_{\rm int} ( x ) 
		=   - \, {:} \, {\mathscr P}' ( {\Phi}_{\rm int} ( x ))\, {:}_{\, \mathfrak{h}(dS)} \; ,
	\end{equation*}
for the interacting quantum fields
	\begin{equation*}
		\Phi_{int} (x) \doteq U(\Lambda) \varphi (0) U^{-1} (\Lambda) \; , \qquad 
		x = \Lambda \left( \begin{smallmatrix} 0\\0\\r \end{smallmatrix} \right) \; . 
	\end{equation*}
is established in Sect.~\ref{sec:11.2};
see Theorem~\ref{th:11.2.1}.

\newpage
\appendix

\renewcommand{\theequation}{\Alph{chapter}.\arabic{equation}}
\renewcommand{\thetheorem}{\Alph{chapter}.\arabic{theorem}}

\chapter{A local flat tube theorem}

The principle of uniqueness of the analytic continuation is at the heart of 
complex analysis in several variables. The \emph{Edge of the Wedge Theorem},
and the \emph{Jost-Lehmann-Dyson Representation} are crucial in deriving 
basic results in algebraic quantum field theory.  
The following theorem \cite[Theorem 14.2]{RS} is know as the 
\emph{degenerate} or \emph{flat tube theorem}. 

\begin{theorem}
Let $T_1(x_1 , z_2)$ and $T_2(z_1 , x_2)$ be tempered distributions in 
the $x$ variables $(x \in \mathbb{R})$ and polynomially bounded 
analytic functions in the variables in 
	\[
		\bigl\{ z \in \mathbb{C} \mid | \Im z | < 1 \bigr\} \; ,
	\] 
\emph{i.e.}, for each $z_2$ with $| \Im z_2 | < 1$, 
$T_1(x_1 , z_2)$ is a distribution in $x_1$ and for each
$g \in S(\mathbb{R})$, the map
	\[
		z_2 \mapsto 
		\int {\rm d} x_1 \; g(x_1) T_1(x_1 , z_2)
	\]
is analytic. Suppose that 
$T_1(x_1 , x_2 + i 0) = T_2 (x_1 + i 0 , x_2 )$
as distributions in the two variables. Then there exists a function 
$f(z_1, z_2)$ analytic in
	\[
		\mathcal{T}_{\widetilde{S}}
		= \bigl\{ x + i y 
		\mid x \in  \mathbb{R}^2 , 
		y \in \widetilde{S} \bigr\} \; , 
	\]
with 
	\[
		\widetilde{S} = \bigl\{ ( y_1 , y_2 )  \in \mathbb{R}^2 
		\mid |y_1| + |y_2| < 1 \bigr\} \; , 
	\]
so that $T_1(x_1 + i 0, z_2) = f ( x_1 + i 0 , z_2)$ 
and $T_2(z_1 , x_2 + i 0 ) = f ( z_1 , x_2 + i 0 )$.
\end{theorem}

This is a \emph{global} statement. For our applications,  
we will need a \emph{local} statement. Before we proceed, we present 
a variation of this result (stated here for analytic functions with continuous 
boundary values only) that is better suited for our applications. 
Let
	\[
		\mathbb{S}_\beta \doteq 
		\bigl\{ z \in \mathbb{C} \mid | \Im z | < | \beta | \, , \, 
		\beta \cdot (\Im z) > 0 \bigr\} \; , 
		\qquad \beta \in [ - \infty, \infty] \; , 
	\] 
denote a \emph{strip} in the complex plane $\mathbb{C}$.
Recall that a function $f \colon D \to \mathbb{C}$
is called \emph{regular} whenever it is continuous and its restriction to the interior 
of $D \subset \mathbb{C}$ is analytic.

\begin{theorem}[D'Antoni \& Zsid\'o, Theorem~2.5 \cite{DA-Z-2001a}]
\label{A-th}
Let $0 \ne \beta_1 , \beta_2 \in [ - \infty , \infty ]$
and $F \colon \mathbb{R}^2 \to \mathbb{C}$ be a bounded
map. Assume that $F$ has a bounded, continuous extension
	\[
		F
		\colon
		\mathcal{T}_{ \{ (y_1, y_2) \in \mathbb{R}^2 \mid
		y_1 \cdot \beta_1 \ge 0 , y_2 \cdot \beta_2 \ge 0 ,
		| y_1 | < | \beta_1 | , | y_2 | < | \beta_2 | , y_1 \cdot y_2 = 0 \} } 
		\to \mathbb{C}
	\]
such that the functions 
\begin{align*}
z_1 & \mapsto F (z_1,s) 
\in \mathbb{C} \; , \qquad z_1 \in \mathbb{S}_{\beta_1} \cup \mathbb{R} \; ,   
\\
z_2 & \mapsto 
F (t, z_2) \in \mathbb{C} \; , \qquad z_2 \in \mathbb{S}_{\beta_2}  \cup \mathbb{R} \; , 
\end{align*}
are regular for every $s, t \in \mathbb{R}$ fixed. 
Then there exists a unique continuous extension $ \widetilde{F} \colon
\mathcal{T}_{ \widetilde{S} }  \to \mathbb{C}$ of $F$, with 
	\[
		\widetilde{S}
		= \bigl\{ (y_1, y_2) \in \mathbb{R}^2 \mid
		y_1 \cdot \beta_1 \ge 0 , y_1 \cdot \beta_1 \ge 0 ,
		\tfrac{y_1}{\beta_1} + \tfrac{y_2}{\beta_2} < 1 \bigr\} \; , 
	\]
which is analytic in the interior. Moreover,
$\widetilde{F}$ is bounded and the maps 
$\mathbb{S}_{\beta_1} \ni \zeta \mapsto 
\widetilde{F} (\zeta,s) $ and $ \mathbb{S}_{\beta_2} \ni 
\zeta  \mapsto \widetilde{F} (t,\zeta) $,  
are analytic for every fixed $s, t \in \mathbb{R}$, respectively. 
\end{theorem}

\begin{remark}
We note that the original formulation of this theorem (due to D'Antoni and 
Zsid\'o) addresses \emph{vector valued functions}.
\end{remark}

A \emph{local flat tube theorem} (for vector valued functions)
has been provided by D'Antoni and 
Zsid\'o \cite{DA-Z, DA-Z-2001a}. In fact, they have proven  
a vectorial flat tube theorem for truncated tubes, claiming essentially 
that for every $\beta_1, \beta_2 , r_1 , r_2 >0$ there are 
	\[
		0 < \beta_i' < \beta_i \; , 
		\qquad 
		0 < r_i' < r_i \; , 
		\qquad
		i = 1,2  \; ; 
	\]
with $\beta_i' \to \beta_i $, $r_i' \to +\infty$,  when $r_i \to +\infty$, 
$i = 1,2$, such that extendibility to the degenerated tube 
	\[ 
		\bigl\{ (z_1, z_2) \in  \mathcal{T}_{ \{ (y_1, y_2) \in \mathbb{R}^2 
		\mid \, 0 \le \frac{y_1}{\beta_1} , \frac{y_2}{\beta_2} 
		\le 1 , \, y_1 \cdot y_2 = 0 \} }
		\mid | \Re z_1 | < r_1 \; , \; | \Re z_2 | < r_2 
		\bigr\}
	\]
of $F$ implies that $F_{| \{ (t, s) \in \mathbb{R}^2 \mid |t| \le r_1' , |s| \le r_2' \}}$
has a continuous extension $\overline{F}$ on 
	\begin{equation}
	\label{local-tube} 
		\bigl\{ (z_1, z_2) \in  \mathcal{T}_{ \overline{T_{\beta_1', \beta_2'}}  }
		\mid | \Re z_1 | < r_1' \; , \; | \Re z_2 | < r_2' 
		\bigr\}
	\end{equation}
which is analytic in the interior. Here
	\[
		T_{\beta_1, \beta_2} \doteq 
		\left\{ y = (y_1, y_2) \in \mathbb{R}^2 \mid
		y_1 \cdot \beta_1 > 0 , y_2 \cdot \beta_2 > 0 , 
		\tfrac{y_1}{\beta_1} + \tfrac{y_2}{\beta_2} < 1 \right\} \; . 
	\]
A set of the form \eqref{local-tube} is called a \emph{local tube}. 

\bigskip
We now present the basic geometric idea. 
Consider the \emph{disk segment} $\mathbb{G}_{\beta, r}$
bounded by $[ - r, r]$ 
and by a circular arc through the three points $- r, i \beta, r$
in the complex plane. For $r'$ and $\beta'$ sufficiently small, these disk 
segments contain the square $[ - r' , r'] + i [0, \beta']$  
\cite[Equ.~(3.1)]{DA-Z}:  
	\begin{equation}
	\label{appropriate-size}
		 [ - r' , r'] + i [0, \beta'] \subset \mathbb{G}_{\beta, r}  
		 \quad 
		 \text{if}
		 \; \;  r' \le \tfrac{r}{2}  
		 \; \;   
		\text{and} 
		 \; \;  \beta' \le \tfrac{\beta}{2} \; . 
	\end{equation}
Next, consider the \emph{sector} 
$\mathbb{A}_{\theta_\circ} \doteq \bigl\{ \rho \, {\rm e}^{i \theta} 
		\mid \rho > 0 \, , \,
		0 < \theta < \theta_\circ \bigr\} $   
in the complex plane. The conformal map  
	\begin{align*}
		\qquad \qquad
		\qquad \qquad \quad
		\mathbb{G}_{\beta, r} & \to 
		\mathbb{A}_{\theta( \beta, r)}
		\\
		z &\mapsto \frac{r+z}{r-z} \; , 
		\qquad \quad \tan \tfrac{\theta(\beta, r)}{2}
		= \tfrac{\beta}{r} \; , \quad 0 < \theta(\beta, r) < \pi \; , 
	\end{align*}
maps the circular arcs with endpoints $-r$ and $r$ in the open upper 
half-plane in the rays 
	\[
		(0, + \infty) \cdot {\rm e}^{i \theta} \subset \mathbb{A}_{\theta( \beta, r)} \; . 
	\]
Now, the \emph{principal branch} of the logarithm 
maps the sector $\mathbb{A}_{\theta( \beta, r)}$
to the strip $\mathbb{S}_1$. Therefore, the conformal map
	\begin{align}
		\Phi_{\beta, r} \colon \mathbb{G}_{\beta, r} \cup (-r, r) 
		& \to \mathbb{S}_1 \cup \mathbb{R} \; , 
		\nonumber \\
		z 
		& \mapsto \frac{1}{\theta(\beta, r)} \ln \frac{r+z}{r-z}  \; , 
	\label{DAZ-phi}
	\end{align}
maps any circular arc in $\mathbb{G}_{\beta, r}$ with endpoints 
in $-r$ and $r$ in a straight line in $\mathbb{S}_1$ parallel to $\mathbb{R}$.
Note that we have added a part of the boundary (the open interval connecting 
the points $-r$ and $r$) to the domain of definition of $\Phi_{\beta, r}$. We 
note that 
	\[
		\bigl( \Phi_1^{-1} + \Phi_2^{-1}\bigr) \mathcal{T}_{ \{ (\eta_1, \eta_2) 
		\in \mathbb{R}^2 
		\mid \, \eta_1, \eta_2 \ge 0 ,  \eta_1 + \eta_2 < 1 \} }
	\]
contains 
	\[
		\bigl\{ (z_1, z_2) \in \mathcal{T}_{ \overline{T_{\beta_1', \beta_2'}}  }
		\mid | \Re z_1 | < r_1' \; , \; | \Re z_2 | < r_2' 
		\bigr\}
	\]
for $	\beta_1', \beta_2', r_1', r_2'$ of appropriate size; 
see \eqref{appropriate-size}. 

\bigskip
Combining the properties of the map $\Phi_{\beta, r}$ with Theorem~\ref{A-th}, 
we arrive at the following local tube theorem 
(again, we present only a simplified 
version of the original theorem by D'Antoni \& Zsid\'o which addresses 
vector valued functions): 

\begin{theorem}
[D'Antoni \& Zsid\'o, Theorem~3.3 
\cite{DA-Z-2001a} ]
\label{local-flat-tube}
Let $0 \ne \beta_1, \beta_2 \in \mathbb{R}$, $r_1, r_2 > 0$ and let
	\[
		F \colon (-r_1, r_1) \times (-r_2, r_2) \to \mathbb{C} 
	\]
be a bounded map. Assume that the functions 
	\begin{align*}
		z_1  & \mapsto F 
		( z_1 , s) \; , 
		\qquad z_1   
		\in \mathbb{G}_{\beta_1, r_1} \cup (-r_1, r_1) \; , 
		\\
		z_2  & \mapsto F  
		(t, z_2 ) \; , \qquad 
		z_2  \in \mathbb{G}_{\beta_2, r_2} \cup (-r_2, r_2)  \; ,  
	\end{align*}
are regular, for every $t \in (- r_1, r_1)$ and for every $s \in (- r_2, r_2)$, respectively. 
If $\beta_1', \beta_2' \in \mathbb{R} $ and $r_1', r_2' > 0$ 		
satisfy 
	\[ 
		\beta_j' \cdot \beta_j > 0 \; , \qquad \left( \tfrac{r_j'}{r_j} \right)^2 +  
		\tfrac{\beta_j'}{\beta_j}
		\sqrt{ 1 + \bigl( \tfrac{\beta_j}{r_j} \bigr)^2} 
		< 1 \; , \qquad j= 1, 2 \; ,
	\]
then $F_{ | \{ (t, s) \in \mathbb{R}^2 \mid | t | \le r_1', |s| \le r_2' \}} $ has a unique
continuous extension $\overline{F}_{ \beta'_1, \beta'_2, 
r_1', r_2'}$ on 
	\[
		\bigl\{ (z_1, z_2) \in \mathcal{T}_{\overline{T_{\beta_1', \beta_2'}}}
		\mid | \Re z_1 | \le r_1' , \; | \Re z_2 | \le r_2 '\bigr\}
	\]
which is analytic in the interior. Moreover, the maps 
\begin{align*}
z_1  & \mapsto 
\overline{F}_{ \beta_1', \beta_2', r_1', r_2'} (z_1, s) \; , 
\qquad z_1 \in \bigl\{ z \in \mathbb{S}_{\beta_1'} \mid
| \Re z | < r_1' \bigr\} \; , 	\\
z_2  & \mapsto 
\overline{F}_{ \beta_1', \beta_2', r_1', r_2'} (t, z_2) \; , 
\qquad z_2 \in \bigl\{ z \in \mathbb{S}_{\beta_2'} \mid
| \Re z | < r_2' \bigr\} \; , 	
\end{align*}
are analytic for every $s \in (-r_2, r_2)$ and $ t \in (-r_1, r_1)$, respectively.
\end{theorem}

\begin{remark} We can now take the union over all the 
$\beta_1', \beta_2' \in \mathbb{R} $ and $r_1', r_2' > 0$ which satisfy the relevant 
inequalities. Denoting 
	\begin{align*}
	 & D_{\beta_1, \beta_2, r_1', r_2'} = \bigcup_{\beta_1' \cdot \beta_1 >0, 
	 \beta_2' \cdot \beta_2 >0, r_1'> 0, r_2'>0
	 \atop \left( \frac{r_j'}{r_j} \right)^2 +  
		\frac{\beta_j'}{\beta_j}
		\sqrt{ 1 + \frac{\beta_j}{r_j} } < 1 \; , \; \;  j= 1, 2 } 
	 \bigl\{ (z_1, z_2) \in  \mathcal{T}_{\overline{T_{\beta_1', \beta_2'}}}
		\mid | \Re z_1 | \le r_1' , \; | \Re z_2 | \le r_2 ' \bigr\} 
	 \\
	 & \quad = \bigl\{ 
	 (z_1, z_2) \in \mathbb{C}^2 \mid \Re z_i < r_i ,  
	 \tfrac{\Im z_i}{\beta_i} \ge 0, i= 1,2, 
	 \tfrac{\Im z_1}{\beta_1}
	 \tfrac{r_1 \sqrt{r_1^2 + \beta_1^2}}{r_1^2 - (\Re z_1)^2}
	 +
	 \tfrac{\Im z_2}{\beta_2}
	 \tfrac{r_2 \sqrt{r_2^2 + \beta_2^2}}{r_2^2 - (\Re z_2)^2} < 1
	 \bigr\} \; ,
	\end{align*}  
$F$ has a unique continuous extension 
$\overline{F} \colon D_{\beta_1, \beta_2, r_1', r_2'} \to X $, which 
is analytic in the interior. Moreover, $\overline{F}$ is bounded and  
the complex-valued maps
	\begin{align*}
	z_1 & \mapsto \overline{F}(z_1, s)  \; , 
	\qquad 
	z_1 \in \bigl\{ \zeta \in \mathbb{C} \mid | \Re \zeta | < r_1 \, , 
	0 < \tfrac{\Im \zeta}{\beta_1} 
	\tfrac{r_1 \sqrt{r_1^2 + \beta_1^2}}{r_1^2 - (\Re \zeta)^2} < 1 \bigr\}
	\\
	z_2 & \mapsto \overline{F}(t,z_2)  \; , 
	\qquad 
	z_2 \in \bigl\{ \zeta \in \mathbb{C} \mid | \Re \zeta | < r_2 \, , 
	0 < \tfrac{\Im \zeta}{\beta_2} 
	\tfrac{r_2 \sqrt{r_2^2 + \beta_2^2}}{r_2^2 - (\Re \zeta)^2} < 1 \bigr\}
	\end{align*}
are analytic for every $s \in (-r_2, r_2)$ and $t \in (- r_1, r_1)$, respectively.
\end{remark} 	

We may also consider the case that the restrictions 
apply only to one variable:

\begin{corollary}
Let $T > 0$ and let $F \colon \mathbb{R} \times (-T, T) \to \mathbb{C}$
be a bounded map. Assume that 
the functions 
\begin{align*}
		z_1 & \mapsto F(z_1, s) \; , 
		\qquad z_1 \in \bigl\{ z \in \mathbb{C} \mid 0 \le \Im z < \pi \bigr\}  \; , 
		\\
		z_2 & \mapsto F (t, z_2) \; , 
		\qquad z_2 \in \mathbb{G}_{\pi, T} \cup ( - T, T ) \; , 
\end{align*}
are regular for every $s \in (- T, T)$ and for every $t \in \mathbb{R}$, respectively.  
If $\beta >0$ and $T_\circ > 0 $ satisfy 
	\[ 
		\left( \tfrac{T_\circ}{T} \right)^2 +  
		\tfrac{\beta}{\pi}
		\sqrt{ 1 + \bigl( \tfrac{\pi}{T} \bigr)^2} 
		< 1 \; ,
	\]
then $F_{ | \{ (t, s) \in \mathbb{R}^2 \mid  |s| \le T_\circ \}} $ has a unique
continuous 
extension $\overline{F}_{\beta , T_\circ }$ on 
	\[
		\bigl\{ (z_1, z_2) \in \mathcal{T}_{\overline{T_{\pi, \beta}}}
		\mid  | \Re z_2 | \le T_\circ \bigr\}
	\]
which is analytic in the interior. Moreover,
the maps 
	\begin{align*}
	 z_1 & \mapsto \overline{F}_{ \beta , T_\circ} (z_1, s) \; , 
	\qquad 
	z_1 \in \mathbb{S}_{\pi}   \; , 
	\\
	z_2 & \mapsto 
	\overline{F}_{ \beta , T_\circ} (t, z_2) \; , 
	\qquad 
	z_2 \in \bigl\{ \zeta \in \mathbb{S}_{\beta} \mid
	| \Re \zeta | < T_\circ \bigr\} \; , 	
\end{align*}
are analytic for every $s \in  (- T, T)$ and $ t \in \mathbb{R}$, respectively.
Consequently, denoting 
	\begin{align*}
	 D_{\beta, T_\circ} & \doteq \bigcup_{\beta >0, 
	 T_\circ >0
	 \atop \left( \frac{T_\circ}{T} \right)^2 +  
		\frac{\beta}{\pi}
		\sqrt{ 1 + \frac{\pi}{T} } < 1 } 
	 \bigl\{ (z_1, z_2) \in  \mathcal{T}_{\overline{T_{ \pi , \beta }}}
		\mid  | \Re z_2 | \le T_\circ \bigr\} 
	 \\
	 & = \bigl\{ 
	 (z_1, z_2) \in \mathbb{C}^2 \mid | \Re z_2 | < T_\circ , 
	 \Im z_1  \ge 0,  \Im z_2  \ge 0, 
	 \tfrac{\Im z_1 }{\pi} 
	 +
	 \tfrac{\Im z_2}{\pi}
	 \tfrac{T \sqrt{T^2 + \pi^2}}{T^2 - (\Re z_2)^2} < 1
	 \bigr\} \; , 
	\end{align*}
$F$ has a unique continuous extension 
$\overline{F} \colon D_{\beta, T_\circ} \to \mathbb{C} $, which 
is analytic in the interior. Moreover, $\overline{F}$ is bounded and  
the maps
	\begin{align*}
	z_1 & \mapsto \overline{F}(z_1, s) \in \mathbb{C} \; , 
	\qquad
	z_1 \in \bigl\{ \zeta \in \mathbb{C} \mid  
	0 < \tfrac{\Im \zeta}{\pi} < 1 \bigr\} \; , \\
	z_2 & \mapsto \overline{F}(t,z_2) \in \mathbb{C} \; , 
	\qquad 
	z_2 \in \bigl\{ \zeta \in \mathbb{C} \mid | \Re \zeta | < r_2 \, , 
	0 < \tfrac{\Im \zeta}{\pi} 
	\tfrac{T \sqrt{T^2 + \pi^2}}{T^2 - (\Re \zeta)^2} < 1 \bigr\} \; , 
	\end{align*}
are analytic for every $s \in (-T, T)$ and $t \in \mathbb{R}$, respectively. 
\end{corollary}

\chapter{One particle structures}
\label{AKay}

Let $G$ be a group. A (classical) \emph{linear dynamical system}\index{dynamical system} 
$({\mathfrak k},\sigma, \{ T_g\}_{g \in G})$ is a real symplectic vector space\index{symplectic 
space} $({\mathfrak k},\sigma)$ together with a  group of \emph{symplectic 
transformations}\index{symplectic transformation} $\{ T_g\}_{g \in G}$. If~${\mathfrak h}$ is a 
complex Hilbert space with scalar product $\langle \, .\, , \, .\, \rangle $, then $({\mathfrak h}, 
2 \Im \langle \, .\, , \, .\, \rangle)$ is a symplectic space. If, in addition, a unitary representation 
$\{ u (g)  \}_{g \in G}$ of $G$ is given, then $( {\mathfrak h}, 2 \Im \langle \, .\, , \, .\, \rangle , 
\{ u (g)  \}_{g \in G}  )$ is a linear dynamical system.

\begin{definition}
\label{def:A.1}
Given a  linear dynamical system $({\mathfrak k},\sigma, \{ T_g\}_{g \in G})$, a symplectic 
transformation $K \colon {\mathfrak k}\to {\mathfrak h}$  defines  a \emph{one-particle 
quantum structure}\index{one-particle quantum structure} on a Hilbert space~${\mathfrak h}$, 
if there exists  a group of unitary operators such that the following diagram commutes

\bigskip
\label{page48}
\vskip -.8cm

\begin{picture}(180,140)

\put(120,100){$\longrightarrow$}
\put(60,100){$({\mathfrak k}, \sigma )$}
\put(125,110){$K$}
\put(55,70){$ {T_g}$}
\put(200,70){$u(g)$}
\put(125,50){$K$}

\put(170,100){$ ({\mathfrak h}, 2\Im \langle \, .\, , \, .\, \rangle)$}

\put(60,40){$({\mathfrak k}, \sigma )$}

\put(170,40){$({\mathfrak h}, 2\Im \langle \, .\, , \, .\, \rangle)$ \; \; . }

\put(120,40){$\longrightarrow$}

\put(75,85){\vector(0,-3){20}}
\put(190,85){\vector(0,-3){20}}

\end{picture}

\vskip -1cm
\end{definition}

\noindent
By definition,  $K$ is injective.  Kay \cite{Kay0, Kay1, Kay2} has shown 
that one can associate several essentially unique one-particle quantum 
structures to a given classical dynamical system.

\begin{definition}
Given a \emph{linear dynamical system} \index{linear dynamical system}
$({\mathfrak k},\sigma, \{ T_t\}_{t \in \mathbb{R}})$, the symplectic transformation $K$  specifies
\begin {itemize}
\item [---] a \emph{one-particle structure with positive energy}\index{one-particle structure with 
positive energy}, if 
\begin{itemize}
\item[$ i.)$] $t \mapsto u(t)$ is strongly continuous and its generator $\varepsilon\ge 0$ is positive;   
\item[$ ii.)$]  $K{\mathfrak k}$ is dense in ${\mathfrak h}$.   
\end{itemize}
\goodbreak
\item [---] a \emph{one-particle $\beta$-KMS structure}\index{one-particle $\beta$-KMS structure}, if 
\begin{itemize}
\item[$ iii.)$] the map $t\mapsto \langle K {\mathfrak f},u(t)K {\mathfrak g} \rangle$, ${\mathfrak f}, 
{\mathfrak g} \in {\mathfrak k}$, is analytic in the strip $\{ t \in \mathbb{C} \mid 0< \Im t<\beta \} $, 
continuous at the boundary, and satisfies the one-particle $\beta$-KMS condition \index{one-particle 
$\beta$-KMS condition}
	\begin{equation} 
		\label{o-p-kms-condition}
			\quad \qquad \langle K{\mathfrak f},u(t+i\beta)K{\mathfrak g} \rangle 
			= \langle u(t) K{\mathfrak g},K{\mathfrak f} \rangle \; , \; \;   t\in\mathbb{R}  \; ,  
			\; {\mathfrak f}, {\mathfrak g} \in {\mathfrak k} \; ;
	\end{equation}
\item[$ iv.)$] $K{\mathfrak k}+iK{\mathfrak k}$ is dense in ${\mathfrak h}$.
\end{itemize}
\end{itemize}
Note that the Hilbert space ${\mathfrak h}$ and the one-parameter group $t \mapsto u(t)$ acting on it, 
although denoted by the same letters in $ i.)$--$ ii.)$ 
{\em and} $ iii.)$--$ iv.)$, are necessarily different in the two distinct cases.  
\end{definition}

\begin{proposition}
[Kay \cite{Kay1}, Theorems 1a \& 1b] 
\label{Kay Th}
There exists a unique  (up to unitary equivalence) one-particle structure with positive energy 
for which  zero is not an eigenvalue of the generator of $t \mapsto u(t)$. Moreover, for each 
$\beta >0$ there exists a unique  (up to unitary equivalence) one-particle $\beta$-KMS structure 
for which  zero is not an eigenvalue of the generator of $t \mapsto u(t)$.
\end{proposition} 

\paragraph{\it Notation.} If ${\mathfrak h}$ is a complex vector space, then the \emph{conjugate 
vector space}\index{conjugate vector space} $\overline {\mathfrak h}$ is the real vector space 
${\mathfrak h}$ equipped with the complex structure $-i$. We denote by 
	\[ 
		{\mathfrak h} \ni h\mapsto  \overline{h} \in \overline{\mathfrak h} 
	\] 
the $\mathbb{C}$-linear identity operator. If ${\mathfrak h}$ is a Hilbert space, then the conjugate 
Hilbert space~$\overline{\mathfrak h}$ is equipped with the scalar product~$\langle \overline{h_{1}}, 
\overline{h_{2}} \rangle \doteq \langle h_{2}, h_{1} \rangle$. If $a\in {\mathcal  L}({\mathfrak h})$, 
then we denote by $\overline{a}\in {\mathcal  L}(\overline{\mathfrak h})$ the {\em linear} operator
$\overline{a}\overline{h}\doteq \overline{ah}$. 

\bigskip 
Given a one-particle structure with positive energy there exists an associated  one-particle 
$\beta$-KMS structure:

\begin{proposition}  \label{Kbeta} Let $(K, {\mathfrak h},\{ u(t)\}_{t \in \mathbb{R}})$ be a one-particle  
structure with positive energy for a classical dynamical system $({\mathfrak k},\sigma, 
\{ T_t \}_{t \in \mathbb{R}})$. If $K{\mathfrak k} \in {\mathcal  D}(\varepsilon^{-1/2})$, then  
\label{page49}
	\begin{align*}
		K_{\scriptscriptstyle \rm AW}  \mathfrak{f} 
		&\doteq (1+\varrho)^{\frac{1}{2}} K\mathfrak{f} \oplus  \varrho^{\frac{1}{2}} K\mathfrak{f}  \; , 
				\qquad \varrho\doteq ({\rm e}^{\beta\varepsilon}-1)^{-1} \; ,   \\  
		{\mathfrak h}_{\scriptscriptstyle \rm AW}  
		&\doteq  {\mathfrak h}\oplus \overline{{\mathfrak h}} 
		\; ,   \\  
		u_{\scriptscriptstyle \rm AW} (t) &\doteq  u(t)\oplus \overline{u(t)} \; ,  
	\end{align*}
defines a one particle $\beta$-KMS structure for  $({\mathfrak k},\sigma, \{ T_t \}_{t \in \mathbb{R}})$. 
\end{proposition}

\begin{remarks} \quad
\begin{enumerate}
\item [$ i.)$] 
The subscripts used in $K_{\scriptscriptstyle \rm AW}$, ${\mathfrak h}_{\scriptscriptstyle \rm AW}$ and 
$u_{\scriptscriptstyle \rm AW} (t)$ pay tribute to the fundamental work of Araki and Woods~\cite{AW}.
\item [$ ii.)$] 
$({\mathfrak h}_{\scriptscriptstyle \rm AW},\{ u_{\scriptscriptstyle \rm AW} (t)\}_{t \in \mathbb{R}})$~is a 
one-particle $\beta$-KMS structure  for the dynamical system 
$({\mathfrak h}, \Im \langle \, . \,  ,  \, . \, \rangle ,\{ u(t)\}_{t \in \mathbb{R}})$, specified by  
${\mathcal K}_{\scriptscriptstyle \rm AW} \colon {\mathfrak h} \to {\mathfrak h}_{\scriptscriptstyle \rm AW}$, 
	\[ 
		h \mapsto (1+\varrho)^{\frac{1}{2}} h \oplus  \varrho^{\frac{1}{2}} h  \; . 
	\]
\item [$ iii.)$] 
$\overline{u(t)}=\overline{{\rm e}^{it\varepsilon}}={\rm e}^{-it 
\varepsilon}$, hence the generator of the one-parameter group
	\[
		t \mapsto \overline{u(t)} 
	\]
has  {\em negative}  spectrum. 
\item [$ iv.)$] The space
${\mathfrak h}^{\scriptscriptstyle \rm L}\doteq \{ K_{\scriptscriptstyle \rm AW} \mathfrak{f}\mid 
\mathfrak{f} \in  {\mathfrak k} \}$ is a real subspace in ${\mathfrak h}_{\scriptscriptstyle \rm AW}$. 
Moreover, ${\mathfrak h}^{\scriptscriptstyle \rm L} + i {\mathfrak h}^{\scriptscriptstyle \rm L}$ is 
dense in ${\mathfrak h}_{\scriptscriptstyle \rm AW}$
and ${\mathfrak h}^{\scriptscriptstyle \rm L}  \cap i {\mathfrak h}^{\scriptscriptstyle \rm L}  = \{0\}$. 
Thus one can define, following Eckmann and Osterwalder \cite{EO} (see also \cite{LRT}), a 
closeable operator 
	\begin{equation} 
		\begin{matrix}
			s: & {\mathfrak h}^{\scriptscriptstyle \rm L} & 
				+ & i {\mathfrak h}^{\scriptscriptstyle \rm L}  
				& \to&  {\mathfrak h}^{\scriptscriptstyle \rm L}  & 
				+ & i {\mathfrak h}^{\scriptscriptstyle \rm L}  \\
				& f & + & i g & \mapsto & f & - & i g
						\end{matrix} \; \; . 
	\end{equation}
The polar decomposition of its closure $\overline {s} = j \delta^{1/2}$
provides 
\begin{itemize}
\item[---] an anti-unitary involution (\emph{i.e.}, a \emph{conjugation}\index{conjugation})
	\begin{equation} 
		\label{eckmanj}
		\begin{matrix}
		j: & {\mathfrak h}\oplus \overline{{\mathfrak h}}  & \to&  {\mathfrak h}\oplus \overline{{\mathfrak h}}  \\
			& f  \oplus   g & \mapsto &   \overline{g}  \oplus  \overline{ f }
		\end{matrix}  \; \;  ; 
	\end{equation}
\item[---] a $\mathbb{C}$-linear, positive operator $\delta^{1/2} $, such that
	\begin{equation}
		\label{eckmand}
		\delta^{it}=  u_{\scriptscriptstyle \rm AW} (- t \beta)\; ,  \qquad t \in \mathbb{R} \; . 
	\end{equation}
\end{itemize}
\eqref{eckmanj} implies $ j {\mathfrak h}^{\scriptscriptstyle \rm L}=   {\mathfrak h}^{\scriptscriptstyle \rm R}$
and  \eqref{eckmand} implies that
$\{ \delta^{it}\}_{t \in \mathbb{R}}$  leaves the subspaces~${\mathfrak h}^{\scriptscriptstyle \rm L}$ 
and~${\mathfrak h}^{\scriptscriptstyle \rm R}$ invariant. 
\item [$ v.)$] 
Sometimes we denote $K_{\scriptscriptstyle \rm AW}$ by
$K_{\scriptscriptstyle \rm AW}^{\scriptscriptstyle \rm L}$. This is useful as one encounters as well
the map 
$K^{\scriptscriptstyle \rm R}_{\scriptscriptstyle \rm AW} 
\colon {\mathfrak k} \to {\mathfrak h}_{\scriptscriptstyle \rm AW}$, 
	\begin{equation}
	\label{eqUbeta2}
	 	K^{\scriptscriptstyle \rm R}_{\scriptscriptstyle \rm AW} \mathfrak{g} 
		\doteq   \varrho^{\frac{1}{2}} K\mathfrak{g} \oplus  (1+\varrho)^{\frac{1}{2}} K\mathfrak{g} \;  ,
	 \end{equation} 
which maps ${\mathfrak k}$ to the symplectic complement 
${\mathfrak h}^{\scriptscriptstyle \rm R} \subset {\mathfrak h}_{\scriptscriptstyle \rm AW}$ 
of  ${\mathfrak h}^{\scriptscriptstyle \rm L}$.
\item [$ vi.)$] \label {sechs} 
The triple  $(K^{\scriptscriptstyle \rm R}_{\scriptscriptstyle \rm AW}, {\mathfrak h}_{\scriptscriptstyle \rm AW},
\{ u_{\scriptscriptstyle \rm AW} (t) \}_{t \in \mathbb{R} } )$ provides a $(-\beta)$-KMS structure for
the linear dynamical system $({\mathfrak k},\sigma, \{T_t\}_{t \in \mathbb{R} } )$.
\end{enumerate}
\end{remarks}

\bigskip
The existence of  $vi.)$ motivated Kay \cite{Kay1, Kay2}  to investigate the possibility of doubling the 
classical dynamical system as well:

\begin{definition}
\label{dcldsy}
Let $\underline { {\mathfrak k}} ={\mathfrak k}_{\scriptscriptstyle \rm R} 
\oplus {\mathfrak k}_{\scriptscriptstyle \rm L}  $ be the direct sum of  two symplectic subspaces 
${\mathfrak k}_{\scriptscriptstyle \rm R} $ and~$  {\mathfrak k}_{\scriptscriptstyle \rm L}  $ 
such that 
	\[
		\underline { \sigma} (  \mathfrak{f},   \mathfrak{g}) 
			= 0 \quad \hbox{if}  \quad \mathfrak{f} \in {\mathfrak k}_{\scriptscriptstyle \rm L}
			\quad \hbox{and}  \quad  \mathfrak{g} \in 
			{\mathfrak k}_{\scriptscriptstyle \rm R} \; . 
	\]
Let $\{ \underline { T}_{\, t} \}_{t \in \mathbb{R}}$ be a one-parameter group of symplectic maps,  
which leaves~${\mathfrak k}_{\scriptscriptstyle \rm L} $ and ${\mathfrak k}_{\scriptscriptstyle \rm R}$ invariant.
Furthermore, let $ \underline{\imath}$ be an anti-symplectic involution such that 
	\[
		[ \, \underline  {T}_{\, t} , \underline {\imath}\, ]=0 \quad \hbox{and} \quad
		\underline {\imath} \, {\mathfrak k}_{\scriptscriptstyle \rm L} 
		= {\mathfrak k}_{\scriptscriptstyle \rm R} \; .
	\]
The quadruple $(\, \underline{{\mathfrak k}}, \underline {\sigma}, \{ \underline { T}_{\, t}\}_{t \in \mathbb{R} } , 
\underline {\imath}\,)$ is called a \index{double classical linear dynamical system} {\em double} (classical) 
linear dynamical system. 
\end{definition}

\goodbreak
It follows that  
$\underline {\imath}\, {\mathfrak k}_{\scriptscriptstyle \rm R} = {\mathfrak k}_{\scriptscriptstyle \rm L}$.
In other words, the following diagram commutes:
\vskip -.5cm
\begin{picture}(200,140)

\put(120,100){$\longrightarrow$}
\put(60,100){$({\mathfrak k}_{\scriptscriptstyle \rm L}, \underline{ \sigma})$}
\put(125,110){$\underline{\imath}$}
\put(55,70){$ {\underline{T}_{\, t}}$}
\put(200,70){$\underline{T}_{\, t}$}
\put(125,50){$\underline{\imath}$}

\put(170,100){$ ({\mathfrak k}_{\scriptscriptstyle \rm R}, \underline{\sigma})$}

\put(60,40){$({\mathfrak k}_{\scriptscriptstyle \rm L}, \underline{\sigma})$}

\put(170,40){$({\mathfrak k}_{\scriptscriptstyle \rm R}, \underline{\sigma})$ \; \; . }

\put(120,40){$\longleftarrow$}

\put(75,85){\vector(0,-3){20}}
\put(190,85){\vector(0,-3){20}}

\end{picture}

\vskip -.8cm
\goodbreak
\begin{definition} 
\label{dbops} (Kay \cite{Kay1}, Def.~3).
A \emph{double $\beta$-KMS one-particle structure}\index{double $\beta$-KMS one-particle structure}, 
\emph{i.e.}, a quadruple  $(\underline{K},  {\mathfrak h} ,  
\{ \underline{\delta}^{-i t / \beta}\}_{t \in \mathbb{R}}  , \underline{j} )$,
associated to a double  linear classical dynamical system 
$(\underline{{\mathfrak k}}, \underline{\sigma}, \{ \underline{T}_{\, t} \}_{t \in \mathbb{R}}, 
\underline{\imath}\, )$ consists of 
\begin{itemize}
\item [$ i.)$] a complex Hilbert space ${\mathfrak h}$; 
\item [$ ii.)$] a $\mathbb{R}$-linear symplectic map $\underline{K} \colon \underline{ {\mathfrak k} }\to  {\mathfrak h} $ 
such that $\underline{ K}{\mathfrak k}_{\scriptscriptstyle \rm L} 
+ i \underline{ K }{\mathfrak k}_{\scriptscriptstyle \rm L} $ is dense in ${\mathfrak h}$; 
\item [$ iii.)$] a strongly continuous unitary group 
$t \mapsto \underline{\delta}^{-i t / \beta} $ such that 
\begin{itemize}
\item[---] $\underline{\delta}^{-i t / \beta} \circ \underline{ K } = \underline{ K }\circ  \underline{ T}_{\, t}  $ 
for all $t \in \mathbb{R} $; 
\item[---] $\underline{K }{\mathfrak k}_{\scriptscriptstyle \rm L} 
+ i \underline{ K }{\mathfrak k}_{\scriptscriptstyle \rm L} 
\subset {\mathscr D} \bigl( \underline{\delta}^{1 /2} \bigr)$;
\end{itemize}
\item [$ iv.)$] an anti-unitary operator $\underline{j}$ such that $ \underline{j} \circ \underline{ K }
= \underline{ K} \circ \underline \imath$ on $\underline{ {\mathfrak k}}$ and 
	\[ 
		\underline{j} \underline{\delta}^{1 /2} f 
		=   f \qquad  \forall f \in \underline{ K }{\mathfrak k}_{\scriptscriptstyle \rm L} \; . 
	\]
\end{itemize}
\end{definition}

\bigskip
\noindent
The operator $\underline{\delta}$ is positive,  
$\underline {K} {\mathfrak k}_{\scriptscriptstyle \rm R} +  i \underline {K} {\mathfrak k}_{\scriptscriptstyle \rm R} $ 
is dense in~$ {\mathfrak h}$,  
	\[ 
		\underline { K} {\mathfrak k}_{\scriptscriptstyle \rm R} + i \underline {K} 
		{\mathfrak k}_{\scriptscriptstyle \rm R}  \subset {\mathscr D} \bigl( \underline{\delta}^{-1 /2}\bigr) 
	\]
and $\underline{j} \, \underline{\delta}^{-1/2}  g =  g $ for all 
$g \in \underline{ K }{\mathfrak k}_{\scriptscriptstyle \rm R} $. 

\goodbreak
\begin{theorem}[Kay \cite{Kay1}, Theorem 2] 
\label{ThB1}
There exists a unique, up to unitary equivalence, double $\beta$-KMS-structure for which
the generator~$\underline{\varepsilon}$ of 
the one parameter group 
	\[
\underline{\delta}^{i t } ={\rm e}^{-i t \beta \underline{\varepsilon}}, \qquad \beta >0 \; , 
	\] 
has no zero eigenvalue. 
\end{theorem}

\chapter{Sobolev spaces on the circle and on the sphere}
\label{apC}
If $h \in L^2 (S^1, {\rm d} \psi)$, then $h$ has a \emph{Fourier series}\index{Fourier series} 
	\begin{equation}
	\label{sobolev}
	h (\psi) = \sum_{k\in {\mathbb Z}} \widetilde{h}(k) \, 
	{\rm e}^{ik \psi} \; , 
	\qquad \widetilde{h}(k) 
	= \frac{1}{2 \pi} \int_{S^1} {\rm d} \psi \; h (\psi) {\rm e}^{-ik  \psi} \;  . 
	\end{equation}
The infinite sum on the r.h.s.~ converges  in $L^2 (S^1, {\rm d} \psi)$. In fact,  
the infinite sum $ \sum_{k\in {\mathbb Z}} 
\widetilde{h}(k) {\rm e}^{ik \psi} $ exists, 
iff $| \widetilde{h}(k) |= o(k^{-N})$ for all $N \in \mathbb{N}$.

By the \emph{Weierstra\ss\ approximation theorem} 
the polynomials  
	\[
	\sum_{k=-N}^N \widetilde{h}(k) \, 
	{\rm e}^{ik \psi} \; , \qquad N \in {\mathbb N} \; , 
	\]
are dense in the sup norm in $C(S^1)$. \emph{Parseval's identity} 
\index{Parseval's identity} states that 
	\begin{equation}
	\label{parseval}
	 \sum_{k\in {\mathbb Z}} | \widetilde{h}(k) |^2 
	 = \frac{1}{2 \pi} \int_{S^1} {\rm d} \psi \; | h (\psi) |^2 \; . 
	 \end{equation}
In case $h \in C^1 (S^1)$, 
the Fourier series converges uniformly and absolutely.

\begin{definition}
\label{def:c1}
 Let $0 \le p \le \infty$.  
The \emph{Sobolev space} \index{Sobolev space} of order $p$ is given by
	\[
		{\mathbb H}^p (S^1)  \doteq \Bigl\{ h \in L^2 (S^1) \mid 
		\sum_{k\in {\mathbb Z}}(1+k^2)^p 
		| \widetilde{h}(k) |^2 < \infty \Bigr\} \; , 
	\]
where the $\bigl\{ \widetilde{h}(k) \bigr\}$ are the Fourier 
coefficients of $h$, see \eqref{sobolev}. 
\end{definition}	

${\mathbb H}^p (S^1)$ is a Hilbert space with the inner product 
	\[
	\Bigl\langle \, \sum_{j\in {\mathbb Z}} \widetilde{h}(j) 
	\, {\rm e}^{i j \psi}  \; , 
	\; \sum_{k\in {\mathbb Z}} \widetilde{g}(k) \, {\rm e}^{ik \psi} 
	\Bigr\rangle_{{\mathbb H}^p (S^1)} 
	= \sum_{k\in {\mathbb Z}}(1+k^2)^p \; 
	\overline { \widetilde{h}(k) } \widetilde{g}(k) 
	\]
for $h, g \in {\mathbb H}^p (S^1)$ with Fourier 
coefficients $\bigl\{ \widetilde{h}(j) \bigr\}$, 
$\bigl\{ \widetilde{g}(k) \bigr\}$, respectively. The norm is 
given by 
	\[
	\| h \|_{{\mathbb H}^p (S^1)}  = \Bigl( \, \sum_{k\in {\mathbb Z}} (1+ k^2)^p \; 
	| \widetilde{h}(k) |^2 \Bigr)^{1/2} \; . 
	\]
The trigonometric polynomials are dense in ${\mathbb H}^p (S^1)$.

\begin{definition} 
\label{sob-circ}
For $0 < p < \infty$, we denote by ${\mathbb H}^{-p}(S^1)$ the dual 
space of ${\mathbb H}^p (S^1)$, 
\emph{i.e.}, the space of bounded linear functionals on ${\mathbb H}^p (S^1)$. 
\end{definition}	

For $\xi \in {\mathbb H}^{-p}(S^1)$ we have 
	\[
	\| \xi \|_{{\mathbb H}^{-p} (S^1)}  
	= \Bigl( \, \sum_{k\in {\mathbb Z}} (1+ k^2)^{-p} \; |b_k|^2 \Bigr)^{1/2} \; , 
	\qquad
	\text{where $b_k = \xi ({\rm e}^{i k \psi})$} \; . 
	\]
\goodbreak
Furthermore, for each sequence $\{ b_k \}$ satisfying 
	\[
	 \sum_{k\in {\mathbb Z}} (1+ k^2)^{-p} \; |b_k|^2 < \infty \; ,   
	\]
there exists a bounded linear functional $\xi \in {\mathbb H}^{-p} (S^1)$ 
with $ b_k = \xi ({\rm e}^{ik \psi})$. 

\goodbreak
\begin{proposition}
The elements in ${\mathbb H}^p  (S^1) $ 
share the following properties:
\begin{itemize}
\item[$ i.)$] if $p <0$, then the elements in 
${\mathbb H}^p  (S^1) $ are generalised functions;
\item[$ ii.)$] if $p = 0$, then ${\mathbb H}^0 (S^1)  = L^2(S^1, {\rm d} \psi)$; 
\item[$ iii.)$] if $p < 1/2 $, then the characteristic function $\chi_I$ 
of an interval $I \subset S^1$
is an element of ${\mathbb H}^p  (S^1)$, but not for $1/2 \le p $;  
\item[$ iv.)$] if $1/2< p $, then the functions 
$ f \in {\mathbb H}^p  (S^1) $ are continuous;
\item[$ v.)$] if $1 \le p $, then the functions 
$ f \in {\mathbb H}^p  (S^1) $ 
are bounded and 
differentiable almost everywhere.
\end{itemize}
\end{proposition}

The following standard result on (fractional) Sobolev spaces
will be helpful. 

\begin{proposition}
\label{prop:C4}
The norm 
on ${\mathbb H}^{1/2}(S^1)$ specified in Definition~\ref{def:c1}
and the norm $||| \, . \, |||$ specified by
	\begin{equation}
	\label{Gagliardo}
		||| h |||
		= 
		\left( \| h \|^2_{L^2(S^1, {\rm d} \psi)} + 
		 \underbrace{ \int_{S^1} {\rm d} \psi \int_{S^1}
		 {\rm d} \psi' \; \frac{|h(\psi)-h(\psi')|^2}{|\psi-\psi'|^{2}}}_{= A(h)^2} 
		\right)^{1/2}  \; ,
	\end{equation}
where the square root of the second term in the bracket is called the Gagliardo (semi-) norm 
$A(h)$ of $h$, are equivalent.  
\end{proposition}

\begin{proof}
By changing of variable, choosing $\alpha=\psi-\psi'$, we get
for $A(h)^2$: 
	\begin{align*}
		\int_{S^1} {\rm d} \psi' \,
		\left(\int_{S^1} {\rm d} \psi \,
		\frac{|h(\psi)-h(\psi')|^2}{|\psi-\psi'|^{2}} \right)\, 
%		& = \int_{S^1} {\rm d} \alpha \int_{S^1} {\rm d} \psi' \; 
%		\frac{|h(\alpha+\psi')-h(\psi')|^2}{|\alpha|^{2}}    
%		\\
%		&= \int_{S^1} {\rm d} \alpha
%		\left(\int_{S^1} {\rm d} \psi' \; \left| 
%		\frac{h(\alpha+\psi')-h(\psi')}{|\alpha| }\right|^2 \right)  
%		\\
		&= \int_{S^1} {\rm d} \alpha \, 
		\left\| \frac{h(\alpha+\cdot)-h(\cdot)}{|\alpha| } 
		\right\|^2_{L^2(S^1, {\rm d} \psi)}   
		\\
		&= \int_{S^1} {\rm d} \alpha 
			\, \left( 2 \pi \sum_{k\in {\mathbb Z}} 
			\left| \widetilde{\frac{h(\alpha+\cdot)-h(\cdot)}{|\alpha| }}(k) \right|^2 \right)   \; , 
	\end{align*}
where Parseval's identity~\eqref{parseval} has been used. Now, 
	\begin{align*}
		\int_{S^1} {\rm d}\alpha
		\, \left(   
		\sum_{k\in {\mathbb Z}} 
			\left| \widetilde{\frac{h(\alpha+\cdot)
			-h(\cdot)}{|\alpha| }}(k) \right|^2 \right)  & 
			= \int_{S^1} {\rm d}\alpha
			\left( \sum_{k\in {\mathbb Z}} \frac{| {\rm e}^{i \alpha k }-1|^2}
			{|\alpha|^{2}}\, 
			| \widetilde{h}(k) |^2  \right)
		\\
%		&  = 2 \int_{S^1} {\rm d}\alpha 
%			\left( \sum_{k\in {\mathbb Z}} \frac{(1-\cos \alpha k)}{|\alpha|^{2}}\, 
%			| \widetilde{h}(k)|^2  \right) 
%		\\
		&  = 2  \sum_{k\in {\mathbb Z}} | \widetilde{h}(k)|^2  
		\int_{0}^{2 \pi} {\rm d}\alpha \, \frac{(1-\cos \alpha k)}{|\alpha|^{2}} \, .  
	\end{align*}
We have used that $| {\rm e}^{i \alpha k} -1 |^2 =
		2 \bigl( 1 - \cos (\alpha k) \bigr) $. 
Now, 
	\begin{align*}
		\int_{0}^{2 \pi} {\rm d}\alpha \, \frac{(1-\cos \alpha k)}{|\alpha|^{2}}
		&  =  |k| \; c_k \; , 
	\end{align*}
with $c_k$ uniformly bounded: since $ 1- \cos \alpha' \ge 0$ for all $\alpha'$, we have
	\begin{align*}
		c_k \doteq 
		\int_{0}^{2 \pi |k|} {\rm d}\alpha' \, \frac{(1-\cos \alpha' )}{|\alpha'|^{2}}\, 
%		& \le 		
%		\int_{0}^{\infty} {\rm d}\alpha' \, \frac{(1-\cos \alpha' )}{|\alpha'|^{2}}\, 
%\\
		& < \infty \; . 
%		\le  
%		\underbrace{ \int_{0}^{1} {\rm d}\alpha' \, 
%		\frac{(1-\cos \alpha' )}{|\alpha'|^{2}}}_{< \infty} 
%		+ 2 \underbrace{\int_{0}^{\infty} {\rm d}\alpha' \, \frac{1}{1 + |\alpha'|^{2}}
%		}_{< \infty}\,  \; .
	\end{align*}
It follows that the norm given by \eqref{Gagliardo}, namely
	\[
		||| h |||
		= 
		\sqrt{ \sum_{k\in {\mathbb Z}}  \; 
		 | \widetilde{h}(k) |^2   + 
		 4 \pi  \sum_{k\in {\mathbb Z}} c_k \, |k| \, | \widetilde{h}(k)|^2   }  \; ,
	\]
is equivalent to the norm  
$\| h \|_{{\mathbb H}^{1/2}(S^1)}^2$ specified in Definition~\ref{def:c1}, namely 
	\[
		\| h \|_{{\mathbb H}^{1/2} (S^1)}  = \sqrt{ \sum_{k\in {\mathbb Z}} \sqrt{1+ k^2} \; 
		| \widetilde{h}(k) |^2 } \; . 
	\]
\end{proof}

\begin{corollary}
Multiplication by a Lipschitz function defines a bounded operator 
in ${\mathbb H}^{1/2}(S^1)$. 
\end{corollary}

\begin{proof} 
By Proposition~\ref{prop:C4}, we may use the 
norm $||| \cdot |||$ on  ${\mathbb H}^{1/2}(S^1)$. 
Thus assume that $f$ is Lipschitz with the Lipschitz constant $L$. It follows that
	\begin{align*}
		| f (\psi)h(\psi)-f(\psi')h(\psi')|
		& \le | f(\psi)| \cdot |h(\psi)-h(\psi')|+|h(\psi')| \cdot |f(\psi)-f(\psi')|
		\\
		& \le \| f \|_{\infty} \cdot |h(\psi)-h(\psi')|
		+ L \; | h (\psi')| \cdot 
		|\psi-\psi'| \; .
 	\end{align*}
Since $\| f h \|_{L^2(S^1)} \le \| f \|_\infty \| h \|_{L^2(S^1)} $ and 
$\int_{S^1} {\rm d} \psi = 2 \pi$,	
we have the estimate
	\begin{align*}
		A(fh) & = \left( \int_{S^1} {\rm d} \psi \int_{S^1} {\rm d} \psi' \; 
		\frac{ |f(\psi)h(\psi)-f(\psi')h(\psi')|^2}{ |\psi-\psi'|^{2} }  
		\right)^{1/2}
		\\
		& 
		\le  \| f \|_\infty \, A (h)   + 2 \pi L
		\underbrace{ \left(  \int {\rm d} \psi'  
		 \; |h(\psi')|^2 \right)^{1/2}}_{=  \| h \|_{L^2(S^1)}}
		\; .
	\end{align*} 
A Lipschitz function is continuous and bounded on $S^1$ and 
${\mathbb H}^{1/2}(S^1) \subset L^2(S^1)$, hence the statement follows.
\end{proof}

\begin{remark}
\label{rm:C6}
Clearly, the function $ \psi \mapsto \cos \psi \cdot \chi_{I_+} (\psi)$, 
$\psi \in S^1$, is Lipschitz continuous and therefore defines a bounded
multiplication operator on ${\mathbb H}^{1/2}(S^1)$. 
\end{remark}

\goodbreak

Next we consider the sphere. The surface element is 
	\[
		{\rm d} \Omega = \cos \psi {\rm d} \psi  {\rm d} \theta \; . 
	\]
We denote by $L^2 (S^2,  {\rm d} \Omega )$ the set of measurable functions
$f$ on the sphere $S^2$ for which
	\[
		\| f \|_{L^2 (S^2,  {\rm d} \Omega )}^2  \doteq 
		\int_{S^2} {\rm d} \Omega \; | f (\theta, \psi) |^2 < \infty \; . 
	\]
A function $f \in L^2 (S^2, \cos \psi {\rm d} \psi  {\rm d} \theta )$ can 
be expanded, in the $L^2$-sense, 
into its Fourier (Laplace) series (with respect to spherical harmonics)
where 
	\begin{equation}
	\label{F-S2}
		\widetilde f_{\ell, k} \doteq \int_{S^2} {\rm d} \Omega \; 
		f (\theta, \psi) \, \overline{Y_{\ell, k} (\theta, \psi)} \; . 
	\end{equation}

\begin{definition}
\label{sobolev-S2}
The Sobolev $\mathbb{H}^p (S^2)$, $p \ge 0$, 
is the closure of the set of $C^\infty (S^2)$ functions with respect to the norm
	\[
		\| f \|_{\mathbb{H}^p (S^2)} \doteq \Bigl( \sum_{\ell = 0}^\infty \sum_{k= - \ell}^{\ell} 
		\left( \ell + \tfrac{1}{2} \right)^{2p} | \widetilde f_{\ell, k} |^2 \Bigr)^{1/2} \; . 
	\]
\end{definition}

The space $\mathbb{H}^p (S^2)$ is a Hilbert space with inner product 
	\[
		\langle f, g \rangle_{\mathbb{H}^p (S^2)} \doteq 
		\sum_{\ell = 0}^\infty \sum_{k= - \ell}^{\ell } 
		\left( \ell + \tfrac{1}{2} \right)^{2p} \; 
		\overline{ \widetilde f_{\ell, k} } \, \widetilde g_{\ell, k} \; , 
		\qquad f, g \in \mathbb{H}^p (S^2) \; .
	\]
By construction, $\mathbb{H}^0 (S^2)= L^2 (S^2,  {\rm d} \Omega )$. 	

\begin{definition} For $0 < p < \infty$, we denote by ${\mathbb H}^{-p}(S^2)$ 
the dual space of ${\mathbb H}^p (S^2)$, 
\emph{i.e.}, the space of bounded linear functionals on ${\mathbb H}^p (S^2)$. 
\end{definition}	

For $\xi \in {\mathbb H}^{-p}(S^2)$ we have 
	\begin{equation}
	\label{H-p}
		\| \xi \|_{{\mathbb H}^{-p} (S^2)}  = \Bigl( \, 
		\sum_{\ell = 0}^\infty \sum_{k= - \ell}^{ \ell } 
	 	\left( \ell + \tfrac{1}{2} \right)^{-2p} \; | b_{\ell, k} |^2 \Bigr)^{1/2} \; , 
	\end{equation}
where $b_{\ell, k} = \xi (Y_{\ell, k})$. Furthermore, for each 
sequence $\{ b_{\ell, k} \}$ satisfying 
	\[
	 	\sum_{\ell = 0}^\infty \sum_{k= - \ell}^{ \ell } 
	 	\left( \ell + \tfrac{1}{2} \right)^{-2p} \; | b_{\ell, k} |^2 < \infty \; ,   
	\]
there exists a bounded linear functional $\xi \in {\mathbb H}^{-p} (S^2)$ 
with $b_{\ell, k} = \xi (Y_{\ell, k})$. 

\chapter{Some identities involving Legendre functions}

In the sequel, we will use the following well-known properties 
of the \index{Gamma function} Gamma function:
	\begin{align}
			\Gamma(z+1) & =  z  \Gamma(z) \; ,
				\label{eq:gamma-1} \\
			\Gamma(z)\Gamma(1-z) & =  \frac{\pi}{\sin(\pi z)} \; ,
				\label{eq:gamma-2} \\
			\Gamma(z) \Gamma(-z) & =  -\frac{\pi}{z\sin(\pi z)}\; ,
				\label{eq:gamma-3} \\
			\Gamma(2z)	& = \frac{2^{2z-1}}{\sqrt{\pi}}\; \Gamma(z)\Gamma\left(z+\tfrac{1}{2}\right) \; ,
				\label{eq:gamma-4} \\
			\Gamma\big(\overline{z}\big) & = \overline{\Gamma (z)} \; .
				\label{eq:gamma-5}
	\end{align}
They are valid except when the arguments are non-positive integers.

The \emph{Legendre function}\index{Legendre function} $P_s$ solves \cite[8.820]{Grad} the differential equation 
	\[
		\frac{{\rm d}}{{\rm d} z} (1- z^2)\frac{{\rm d}}{{\rm d} z} P_{s}(z) + s(s+1) P_{s}(z) =0 \; . 
	\] 
$P_s$ is analytic in $ z \in \mathbb{C} \setminus (-\infty, \;
-1)$, that means, it has a cut  on the negative real axis.  

\begin{remark}
Setting $z= - \cos \psi$ we find
	\[ \frac{1}{\sin_\psi } 
						\frac{\partial}{\partial \psi}   \sin_\psi \frac{\partial}{\partial \psi} 	
						P_{s}( - \cos \psi) + s(s+1) P_{s}( - \cos \psi) =0 \; . 
	\] 
The \emph{associated Legendre functions} \index{associated Legendre functions}
	\begin{align*}
		P_s^k (\cos \psi) & \doteq (-1)^k (\sin \psi)^k \tfrac{ {\rm d}^k }{ {\rm d} (\cos \psi)^k} \bigl( P_s (\cos \psi) \bigr) \\
		P_s^{-k} & \doteq (-1)^k \tfrac{(s-k) !}{(s+k)! } P_s^{k} \; ,  
		\qquad k= 0, 1,2, \ldots \; ,
	\end{align*}
are analytic in $\mathbb{C}\setminus (-\infty, \; +1)$. 
\end{remark}
\goodbreak

\begin{lemma}
The function 
	\[
		S(z)\doteq \sqrt{z^2-1}
	\]
is analytic in $\mathbb{C}\setminus (-\infty, \; 1)$ and one has 
	\begin{equation}
	\label{c-root}
		 \lim_{\epsilon \downarrow 0} S\big(\epsilon(1\pm i)\big)= {\rm e}^{\pm i\pi/2} \; .
	\end{equation}
\end{lemma}

\begin{lemma} The Fourier series of the Legendre function is given by
	\begin{equation}
		P_s(-\cos\psi) \; = \; 
							p(0)+ 2\sum_{k=1}^\infty p(k) \cos(k\psi)
		\;,
		\label{eq:Fourier-Pmuk-1}
	\end{equation}
where, for $k\in \mathbb{N}_0$, 
	\begin{equation}
		p(k) \doteq (-1)^k \frac{\Gamma(s -k + 1)}{\Gamma(s+k+1)} 
				\left( \lim_{\epsilon\to 0_+}P_s^k \big(\epsilon(1 + i)\big)\right)
				\left(\lim_{\epsilon\to 0_+} P_s^k \big(\epsilon(1 - i)\big)\right) 	\;.
	\label{eq:pk-1}
\end{equation}
\end{lemma}

\begin{proof}
For $| \arg (z-1)| <\pi$ and $| \arg (w-1)| <\pi$ and $\Re z >0$ and $\Re w >0$, one has \cite[page 202]{Lebedev}
	\begin{align}
		P_s\Big(zw& -\sqrt{z^2-1} \sqrt{w^2-1} \cos\psi\Big)
		\label{eq:addition-formula}
		 \\
		&	= P_s(z)P_s(w) + 2 \sum_{k=1}^\infty (-1)^k \frac{\Gamma(s -k + 1)}{\Gamma(s+k+1)} 
 				P_s^k (z) P_s^k (w) \cos(k\psi)\;.
		\nonumber
	\end{align}
(This relation is also found in \cite[page 78]{snow}.) Hence, setting $z=\epsilon(1+i)$, $w=\epsilon(1-i)$, and taking
the limit $\epsilon \downarrow 0$, we have for the l.h.s.\ of
(\ref{eq:addition-formula}),
	\[ 
		\lim_{\epsilon \downarrow 0}  P_s\Big(2\epsilon^2 -S\big(\epsilon(1+i)\big)S\big(\epsilon(1+i)\big) \cos\psi\Big)
			\; = \; 
		P_s\left( - \cos\psi\right) \; .
	\]
Setting $z=i\epsilon$,  $w=-i\epsilon$ and taking the limit
$\epsilon \searrow 0$ on the  r.h.s.\ of (\ref{eq:addition-formula}), the lemma follows.
\end{proof}

\begin{lemma}
	\begin{equation}
		\lim_{\epsilon \downarrow 0} P_s^k \big(\epsilon(1 \pm i)\big)
			=  
			\frac{ {\rm e}^{\pm ik\pi/2}\sqrt{\pi} }{2^k}\,
			\frac{\Gamma(s + k + 1)}{  \Gamma(s -k + 1)} 
			\frac{1}{\Gamma\left(\frac{k-s+1}{2}\right)\Gamma\left(\frac{k+s}{2}+1\right)}
			\;.
		\label{eq:Jinboiui-2}
	\end{equation}
\end{lemma}

\begin{proof}
According to \cite[Eq.\ 7.12.27, page 198]{Lebedev}, one has,
for $k\in \mathbb{N}_0$, $|z-1|<2$ and $\mbox{arg}(z-1)<\pi$,
	\begin{align*}
		P_s^k (z)
				& =  
				\frac{ \big(z^2-1\big)^{k/2} \Gamma(s + k + 1)}{2^k \Gamma(k + 1) \Gamma(s -k + 1)} \;
				F\left( k-s, \; k+s+1, \; k+1; \; \frac{1-z}{2}\right)
				\\
				& =  
				\frac{S(z)^k  \Gamma(s + k + 1)}{2^k \Gamma(k + 1) \Gamma(s -k + 1)}
				F\left( k-s, \; k+s+1, \; k+1; \; \frac{1-z}{2}\right)
				\;,
	\end{align*}
where $F$ is the hypergeometric function \index{hypergeometric function}
	\begin{align}
		F(\alpha , \; \beta, \; \gamma,\; z )
			& \doteq
			1 + \sum_{n=1}^\infty \frac{(\alpha)_{n} (\beta)_{n} }{n! (\gamma)_{n}} \; z^n 
			\nonumber \\
			& = 
				\frac{\Gamma(\gamma)}{\Gamma(\alpha)\Gamma(\beta)}
				\sum_{n=0}^\infty
				\frac{\Gamma(\alpha+n)\Gamma(\beta+n)}{\Gamma(\gamma+n)}
				\; \frac{z^n}{n!} \; ,
		\label{eq:def-hipergeometrica}
	\end{align}
valid for $|z|<1$. Here $(q)_n$ is the \index{Pochhammer symbol} Pochhammer symbol, which yields  
	\[
			(q)_n \doteq 
				\begin{cases} 1 & \text{if \ $n=0$} \; ; \\
							q(q+1) \cdots (q+ n-1) & \text{if \ $n>0$}\; . 
				\end{cases} 
	\]
$F$  is analytic in the whole open unit disk $|z|<1$. Therefore,
	\begin{align}
		\lim_{\epsilon\to 0_+}P_s^k \big(\epsilon(1 \pm i)\big)
			&=  
			\frac{{\rm e}^{\pm ik\pi/2}}{2^k}
			\frac{\Gamma(s + k + 1)}{ \Gamma(k + 1) \Gamma(s -k + 1)} 
			\nonumber \\
			&\qquad \qquad \times F\left( k-s, \; k+s+1, \; k+1; \; \frac{1}{2}\right)
			\;.
		\label{eq:Jinboiui}
	\end{align}

The value of $F(\alpha, \; \beta, \; \gamma; \; z)$ at the point
$z=1/2$ cannot be easily computed from the power series definition
\eqref{eq:def-hipergeometrica}. However, the hypergeometric function satisfies the following
relation (see \cite[Eq.\ 9.6.11, page 253]{Lebedev}):
	\begin{align}
		\label{eq:smart-identity}
			& F\left( 2\alpha , \; 2\beta, \; \alpha + \beta + \tfrac{1}{2};\; \tfrac{1-z}{2} \right)
				\\
			&\qquad  \qquad =  
			\frac{\Gamma\left(\alpha+\beta+\frac{1}{2}\right)\Gamma\left(\frac{1}{2}\right)  }{
			\Gamma\left(\alpha+\frac{1}{2}\right)\Gamma\left(\beta+\frac{1}{2}\right)}
			\, F\left(\alpha, \; \beta, \; \tfrac{1}{2}; \; z^2\right)
			\nonumber \\
			& \qquad \qquad  \qquad \qquad +
			z \frac{\Gamma\left(\alpha+\beta+\frac{1}{2}\right)\Gamma\left(-\frac{1}{2}\right) }{
			\Gamma\left(\alpha\right)\Gamma\left(\beta\right)}
			\, F\left(\alpha+\tfrac{1}{2}, \; \beta+\tfrac{1}{2}, \; \tfrac{3}{2}; \; z^2\right)
			\;,
			\nonumber
	\end{align}
valid for all $z\in\mathbb{C}\setminus\big( (-\infty, \; -1)\cup(1, \;
\infty)\big)$ and for all $\alpha+\beta+\frac{1}{2}\not\in -\mathbb{N}_0$
(\emph{i.e.}, for all $\alpha+\beta+\frac{1}{2}\neq 0, \; -1, \; -2,
\ldots$).  Taking $z=0$ in (\ref{eq:smart-identity}), one finds  
	\begin{equation}
		F\left(2\alpha , \; 2\beta, \; \alpha + \beta + \frac{1}{2};\; \frac{1}{2} \right)
		\; = \; 
		\frac{\Gamma\left(\alpha+\beta+\frac{1}{2}\right)\Gamma\left(\frac{1}{2}\right)
		  }{
		\Gamma\left(\alpha+\frac{1}{2}\right)\Gamma\left(\beta+\frac{1}{2}\right)}
			\;,
	\label{eq:oijosidcn}
	\end{equation}
since $F\left(\alpha, \; \beta, \; \frac{1}{2}; \; 0\right)=1$ 
(see (\ref{eq:def-hipergeometrica})). By choosing 
	\[
		\alpha \; = \; \frac{k-s}{2}
		\quad \mbox{ and } \quad
		\beta \; = \; \frac{k+s+1}{2}
	\]
one has 
$\alpha+\beta+\frac{1}{2}  =  k+1$ (which is non-zero for $k\in\mathbb{N}_0$)
and it follows from (\ref{eq:oijosidcn}) that
	\begin{equation}
	\label{d020}
		F\left( k-s, \; k+s+1, \;   k+1; \; \frac{1}{2} \right)
		\; = \; 
		\frac{\sqrt{\pi}\, \Gamma(k+1)}{\Gamma\left(\frac{k-s+1}{2}\right)\Gamma\left(\frac{k+s}{2}+1\right)}
		\;,
	\end{equation}
as $\Gamma\left(\frac{1}{2}\right)=\sqrt{\pi}$.
Inserting \eqref{d020} into (\ref{eq:Jinboiui}), one gets \eqref{eq:Jinboiui-2}.
\end{proof}

\begin{proposition} 
\label{w-coefficients}
The Fourier series of the Legendre function 
\index{Fourier series of the Legendre function}
is given by
	\begin{equation}
		P_s(-\cos\psi) = p(0)+ 2\sum_{k=1}^\infty p(k) \cos(k\psi) \;,
	\end{equation}
where, for $k\in \mathbb{N}_0$, 
	\begin{equation}
	\label{eq:pk-6}
		p(k)  = -\frac{\sin\big(\pi s \big)}{\pi} \frac{1}{ (k +s)} \; 
				\frac{\Gamma\left( \frac{k-s}{2} \right)}
					{\Gamma\left(\frac{k+s}{2}\right)}
				\frac{\Gamma\left( \frac{k+s+1}{2} \right)}{\Gamma\left(\frac{k-s+1}{2}\right)} \; .
	\end{equation}
\end{proposition}

\begin{proof}
Inserting (\ref{eq:Jinboiui-2}) into 
(\ref{eq:pk-1}), one gets
	\begin{equation}
		p(k) = 
			(-1)^k \frac{\pi}{2^{2k}}\,
			\frac{\Gamma(s + k + 1)}{  \Gamma(s -k + 1)} \; 
			\frac{1}{\Gamma\left(\frac{k-s+1}{2}\right)^2
			\Gamma\left(\frac{k+s}{2}+1\right)^2} \; ,
			\quad k\in \mathbb{N}_0 \;.
		\label{eq:pk-2}
	\end{equation}
Now, using the well-known properties
(\ref{eq:gamma-1})--(\ref{eq:gamma-5}) of the Gamma function, we start
a series of manipulations, in order to write $p(k)$ in a more
convenient fashion.

In (\ref{eq:pk-2}) we consider the factor
	\[
		\frac{\Gamma(s + k + 1)}{\Gamma\left(\frac{k+s}{2}+1\right)}
		= 
		\frac{\Gamma(s + k + 1)}{\Gamma\left(\frac{k+s+1}{2}+\frac{1}{2}\right)}
		=  
		\frac{\Gamma(2z)}{\Gamma\left(z+\frac{1}{2}\right)} \;,
	\]
by taking $z=\frac{k+s+1}{2}$. From (\ref{eq:gamma-4}), one has
$\frac{\Gamma(2z)}{\Gamma\left(z+\frac{1}{2}\right)}=
\frac{2^{2z-1}}{\sqrt{\pi}}\Gamma(z)$. Hence,
	\[
		\frac{\Gamma(s + k + 1)}{\Gamma\left(\frac{k+s}{2}+1\right)}	
			= 
		\frac{2^{2z-1}}{\sqrt{\pi}}\Gamma(z)
			=  
		\frac{2^{k+s}}{\sqrt{\pi}}\Gamma\left( \frac{k+s+1}{2} \right) \;.
	\]
Inserting this into (\ref{eq:pk-2}), we get
\begin{align}
	p(k)  = \;  \; &(-1)^k \sqrt{\pi}\, 2^{s-k} \; 
			\frac{\Gamma\left( \frac{k+s+1}{2} \right)}{\Gamma(s -k + 1)\Gamma\left(\frac{k-s+1}{2}\right)^2
			\Gamma\left(\frac{k+s}{2}+1\right)}
		\nonumber \\
		 \stackrel{(\ref{eq:gamma-1})}{=}  &(-1)^k 
			\frac{\sqrt{\pi}\, 2^{s-k+1}}{s^2 - k^2} \; 
			\frac{\Gamma\left( \frac{k+s+1}{2} \right)}{\Gamma(s -k)\Gamma\left(\frac{k-s+1}{2}\right)^2
			\Gamma\left(\frac{k+s}{2}\right)} \;.
\label{eq:pk-3}
\end{align}
Now, we write $\Gamma\left( \frac{k-s+1}{2} \right) = \Gamma\left( z+\frac{1}{2} \right)$ with 
$z=\frac{k-s}{2}$ and, using (\ref{eq:gamma-4}), we deduce that 
	\begin{align*}
		\Gamma\left( \frac{k-s+1}{2} \right)^2 
		& = 
		\Gamma\left( z+\frac{1}{2} \right)^2
		\stackrel{(\ref{eq:gamma-4})}{=} 
		\left( 
		     \frac{\Gamma(2z)}{\Gamma(z)} \frac{\sqrt{\pi}}{2^{2z-1}}
		\right)^2
		\\
		&= 
		\left( 
		\frac{\sqrt{\pi}}{2^{k-s-1}}
		\frac{\Gamma(k-s)}{\Gamma\left( \frac{k-s}{2} \right)}
		\right)^2
		=  
		\frac{\pi}{2^{2k-2s-2}}
		\frac{\Gamma(k-s)^2}{\Gamma\left( \frac{k-s}{2} \right)^2}
		\;.
	\end{align*}
Returning to (\ref{eq:pk-3}), we find 
	\begin{equation}
		p(k) \; = \; 
			(-1)^k  \frac{2^{ k-s-1} }{\sqrt{\pi}\big(s^2 - k^2\big)} \; 
			\frac{\Gamma\left( \frac{k+s+1}{2} \right)
		      \Gamma\left( \frac{k-s}{2} \right)^2 }{
			\Gamma(s -k) \Gamma(k-s)^2\,
			\Gamma\left(\frac{k+s}{2}\right) } \;.
			\label{eq:pk-4}
	\end{equation}
We now write
	\[ 
		\Gamma(s -k) \Gamma(k-s)
			\; \stackrel{(\ref{eq:gamma-3})}{=} \;
				-\frac{\pi}{(s -k)\sin\big(\pi (s -k)\big)}
			\; = \; 
				\frac{(-1)^{k}\pi}{(k -s)\sin\big(\pi s \big)} \; ,
	\]
and, inserting this identity into (\ref{eq:pk-4}), we get
	\begin{equation}
		p(k) \; = \; 
			-\sin\big(\pi s \big)
			\frac{2^{ k-s-1} }{\pi^{3/2} (k +s)} \; 
			\frac{
			\Gamma\left( \frac{k+s+1}{2} \right)
		      \Gamma\left( \frac{k-s}{2} \right)^2 }{
			\Gamma(k-s)\,
			\Gamma\left(\frac{k+s}{2}\right) }
				\;.
		\label{eq:pk-5}
	\end{equation}
Taking $z=\frac{k-s}{2}$ yields 
	\[ 
		\frac{\Gamma\left( \frac{k-s}{2} \right)}{\Gamma(k-s)}
		\; = \; 
		\frac{\Gamma(z)}{\Gamma(2z)}
		\; \stackrel{(\ref{eq:gamma-4})}{=} \; 
		\frac{\sqrt{\pi}}{2^{2z-1}}\;
		\frac{1}{\Gamma\left(z+\frac{1}{2}\right)} 
		\; = \; 
		\frac{\sqrt{\pi}}{2^{k-s-1}}\;
		\frac{1}{\Gamma\left(\frac{k-s+1}{2}\right)}
		\; .
	\]
Returning with this result to (\ref{eq:pk-5}), we find \eqref{eq:pk-6}.
\end{proof}

\begin{remark}
Comparing (\ref{eq:Fourier-Pmuk-1}) with (\ref{eq:Fourrier-Pmu}) we see that
$p_k=\sqrt{2\pi r} \, p\big(|k|\big)$, for all $k\in\mathbb{Z}$. Thus,
from the definition (\ref{eq:definition-omegak-0}) we get
\begin{equation}
\label{eq:symmetry-of-tildeomega}
\widetilde{\omega}(k)=\widetilde{\omega}(-k)
\end{equation}
for all $k\in\mathbb{Z}$.  Actually, we can directly establish that
$p(k)=p(-k)$ for all $k \in \mathbb{Z}$. This is the content of the next lemma.
\end{remark}

\begin{lemma}
For all $k\in\mathbb{Z}$, we have $p(k)=p(-k)$.
\end{lemma}

\begin{proof}
Until now we considered $k\in\mathbb{N}_0$ but, for $s\not\in\mathbb{Z}$,
(\ref{eq:pk-6}) is well-defined for all $k\in\mathbb{Z}$ and we will now show
that $p(k)=p(-k)$ for all $k\in\mathbb{Z}$. Let 
	\[ 
			{\mathscr F}(k) \doteq \frac{1}{ (k +s)} \; \frac{ \Gamma\left( \frac{k-s}{2} \right) }{
											\Gamma\left(\frac{k+s}{2}\right) }
										\frac{\Gamma\left( \frac{k+s+1}{2} \right)}
											{\Gamma\left(\frac{k-s+1}{2}\right)} \;.
	\]
Then,
	\[
			{\mathscr F}(-k) = \frac{1}{ (-k +s)} \; \frac{ \Gamma\left( \frac{-k-s}{2} \right) }{
											\Gamma\left(\frac{-k+s}{2}\right) }
					\frac{\Gamma\left( \frac{-k+s+1}{2} \right)}{\Gamma\left(\frac{-k-s+1}{2}\right)} \;.
	\]
Now,
	\[
		\frac{ \Gamma\left( \frac{-k-s}{2} \right) }{
			\Gamma\left(\frac{-k+s}{2}\right)}
		= \frac{ \left( -k+s \right) }{\left( -k-s \right)}
			\frac{  \sin\left(\pi\frac{-k+s}{2}\right) }{
      				\sin\left( \pi\frac{-k-s}{2} \right) }
			\frac{\Gamma\left(\frac{k-s}{2}\right)}
				{ \Gamma\left( \frac{k+s}{2} \right)}
	\]
and
	\[
		\frac{ \Gamma\left( \frac{-k+s+1}{2} \right) }{
				\Gamma\left(\frac{-k-s+1}{2}\right)}	
		= \frac{ \Gamma\left( 1- \frac{k-s+1}{2} \right) }{
				\Gamma\left( 1-\frac{k+s+1}{2}\right)}
		= \frac{\sin\left( \pi\frac{k+s+1}{2}\right)}{\sin\left( \pi\frac{k-s+1}{2} \right)}
			\frac{\Gamma\left( \frac{k+s+1}{2}\right)}{
				\Gamma\left( \frac{k-s+1}{2} \right)} \;.
	\]
Therefore,
	\begin{align}
		\label{eq:yubuuytFytf}
		{\mathscr F}(-k) & = \frac{1}{ (-k +s)} \; 
					\frac{ \left( -k+s \right) }{
 						     \left( -k-s \right)}
				\left( \frac{ \sin\left(\pi\frac{-k+s}{2}\right)}{
    						  \sin\left( \pi\frac{-k-s}{2} \right)}
					\frac{\sin\left( \pi\frac{k+s+1}{2}\right)}{
						\sin\left(\pi \frac{k-s+1}{2} \right)}
				\right)
								 \\
			& \qquad \qquad \qquad \qquad \qquad \qquad \qquad \qquad	\times\left[	\frac{\Gamma\left(\frac{k-s}{2}\right)}{
						    \Gamma\left( \frac{k+s}{2} \right)}
					\frac{\Gamma\left( \frac{k+s+1}{2}\right)}{
							\Gamma\left( \frac{k-s+1}{2} \right)}
				\right] \nonumber \\
			& = \frac{1}{ -k -s} \;  \left( \frac{ \sin\left(\pi\frac{-k+s}{2}\right)}{
      										\sin\left( \pi\frac{-k-s}{2} \right)}
									\frac{\sin\left( \pi\frac{k+s+1}{2}\right)}{
										\sin\left(\pi \frac{k-s+1}{2} \right)}
								\right)
				 \left[\frac{\Gamma\left(\frac{k-s}{2}\right)}{	
						\Gamma\left( \frac{k+s}{2} \right)}
					\frac{\Gamma\left( \frac{k+s+1}{2}\right)}{
						\Gamma\left( \frac{k-s+1}{2} \right)}
				\right] \;. \nonumber
	\end{align}
Using 
$\sin \alpha \sin \beta = \frac{\cos(\alpha-\beta)-\cos(\alpha+\beta)}{2}$, 
we get
	\begin{align*} 
	\frac{ \sin\left(\pi\frac{-k+s}{2}\right)}{ \sin\left( \pi\frac{-k-s}{2} \right)}
	\frac{\sin\left( \pi\frac{k+s+1}{2}\right)}{\sin\left( \pi\frac{k-s+1}{2} \right)}
	&= \frac{\cos \left( -\pi k -\frac{\pi}{2} \right) - \cos\left( \pi s +\frac{\pi}{2} \right) }
		{\cos\left( -\pi k -\frac{\pi}{2}\right) - \cos\left( -\pi s +\frac{\pi}{2}\right)}
	\\
	&=  \frac{\cos\left( \pi s +\frac{\pi}{2}\right)}{\cos\left( \pi s -\frac{\pi}{2}\right)}
	=  -1 \;.
	\end{align*}
Hence, returning to (\ref{eq:yubuuytFytf}),
	\[
		{\mathscr F}(-k) =  \frac{1}{ k +s} \; 
				\frac{\Gamma\left(\frac{k-s}{2}\right)}
					{\Gamma\left( \frac{k+s}{2} \right)}
				\frac{\Gamma\left( \frac{k+s+1}{2}\right)}{
					\Gamma\left( \frac{k-s+1}{2} \right)}
			=  {\mathscr F}(k) \;.
	\]
This establishes that $p(k)=p(-k)$ for all $k\in\mathbb{Z}$.
\end{proof}

\begin{proposition} Let $\widehat{{\mathfrak h}}  (S^1)$ be   
as defined in \eqref{eq:def-scalar-product-2}, and let 
$f, g \in {\mathscr D}(\omega)$. It follows that  
\label{prop:E5}
	\begin{equation}
		\langle \omega \, f , \omega \, g \rangle_{\widehat{{\mathfrak h}}  (S^1)} = -\frac{1}{2 \sin(\pi s^+)}
		\int_{S^1 \times S^1} \kern-.3cm {\rm d} \psi\, {\rm d} \psi' \; \overline{f(\psi')} \, 
		P_{s}'\big( - \cos(\psi' - \psi) \big) \, g(\psi) \;.
		\label{C-32}
	\end{equation}
\end{proposition}

\begin{proof} In what follows we will denote the coefficients $p(k)$, $p_k$ and
$\omega(k)$ by $p_s(k)$, $p_{k, \, s}$ and $\omega_s(k)$, respectively.

For $s\in \mathbb{C}\setminus\mathbb{Z}$, define
\begin{align*}
			\langle \kern -.2cm \langle \; f ,\, \; g  \; \rangle \kern -.2cm \rangle
			& \doteq   c_\nu \int_{S^1} r\, d\psi \int_{S^1} r \, d\psi' \; \overline{f(\psi')} \, g(\psi)
				\, P_{s}' \big( - \cos(\psi' - \psi) \big) 
				\\
			& = -\frac{1}{  2 
			\sin(\pi s^+)}  \int_{S^1} r\, d\psi \int_{S^1} r \, d\psi' \; \overline{f(\psi')} \, g(\psi)
							\, P_{s}'\big( - \cos(\psi' - \psi) \big)   \;.
\end{align*}
We write
	\begin{equation*}
		P_{s}'\big( - \cos \varphi \big) 
		=  \sum_{k\in\mathbb{Z}} p_{k, \,s}^1 \; \frac{{\rm e}^{ik\varphi}}{\sqrt{2\pi r}} \; ,
	\end{equation*}
and, as in (\ref{eq:Jhoitrytfsd}), we get
	\begin{equation}
			\langle \kern -.2cm \langle \; f ,\, \; g  \; 
			\rangle \kern -.2cm \rangle
		 =  -\frac{\sqrt{2\pi r}}{  2   \sin(\pi s^+)} 
		 \sum_{k\in\mathbb{Z}} p_{k, \,s}^1 \, \overline{f_k} \, g_k \;,
		\label{C-33}
	\end{equation}
where $f_k$ and $g_k$ are the Fourier coefficients of $f$ and $g$, respectively, 
\emph{i.e.},
	\[
		f_k \doteq \int_{S^1} r \, d\psi' \;  f(\psi') \frac{{\rm e}^{-ik\psi'}}{\sqrt{2\pi r}}
			\qquad \mbox{ and } \qquad
		g_k \doteq \int_{S^1} r \, d\psi \;  g(\psi) \frac{{\rm e}^{-ik\psi}}{\sqrt{2\pi r}} \; .
	\]
Taking the mixed derivatives $\partial_z\partial_w$ of both sides in
(\ref{eq:addition-formula}), we find 
\begin{align}
	P_s'' \bigl( zw-& \sqrt{z^2-1} \sqrt{w^2-1} \cos\psi \bigr)  
			\left(  w- \tfrac{z \sqrt{w^2-1}}{\sqrt{z^2-1}}  \cos\psi  \right)
			\left(  z- \tfrac{w\sqrt{z^2-1} }{\sqrt{w^2-1}}  \cos\psi  \right)
		\nonumber \\
		&  + 	P_s' \bigl( zw-\sqrt{z^2-1} \sqrt{w^2-1} \cos\psi \bigr)
					\left(  1- \frac{zw}{\sqrt{z^2-1}\sqrt{w^2-1}} \cos\psi  \right)
		\nonumber \\
		& \qquad = P_s'(z)P_s'(w) + 2 \sum_{k=1}^\infty (-1)^k
				\frac{\Gamma(s -k + 1)}{\Gamma(s+k+1)} 
				{P_s^k }'(z) {P_s^k} '(w) \cos(k\psi)\;.
\label{C-34}
\end{align}
Writing  $z=\epsilon(1+i)$, $w=\epsilon(1-i)$ (with $\epsilon>0$) and taking
the limit $\epsilon \downarrow 0$, we get from \eqref{C-34} that 
\begin{align*}
	P_s' ( - \cos\psi ) & =  \bigl(P_s'(0)\bigr)^2 + 2 \sum_{k=1}^\infty 
								(-1)^k \frac{\Gamma(s -k + 1)}{\Gamma(s+k+1)} 
								\left( \lim_{\epsilon_\to 0_+} {P_s^k }'\big( \epsilon(1+i) \big) \right) 
						\\
						& \qquad \qquad \qquad \qquad \qquad \qquad \times 
								\left( \lim_{\epsilon_\to 0_+} {P_s^k} '\big( \epsilon(1-i) \big) \right) 
								\cos(k\psi)
						\\
						& =  p^1_s(0) + 2 \sum_{k=1}^\infty p^1_s(k)\cos(k\psi) \;,
\end{align*}
with
	\[
	p^1_s(k) \doteq (-1)^k \frac{\Gamma(s -k + 1)}{\Gamma(s+k+1)} 
						\left( \lim_{\epsilon_\to 0_+} {P_s^k }'\big( \epsilon(1+i) \big) \right) 
						\left( \lim_{\epsilon_\to 0_+} {P_s^k} '\big( \epsilon(1-i) \big) \right) \;.
	\]
Now, according to \cite[Eq.\ (7.12.16), page 195]{Lebedev}, 
	\[
		\bigl( z^2 - 1 \bigr) {P_s^k }'\big( z \big) = s z P_s^k ( z )
		- (s+k)P_{s-1}^k ( z ) \qquad 
		\forall k
		\in \mathbb{N}_0 \; . 
	\]
Hence,
	\[
		\lim_{\epsilon \downarrow 0} {P_s^k }'\bigl( \epsilon(1\pm i) \bigr) 
			= (s+k)\lim_{\epsilon_\to 0_+} P_{s-1}^k \bigr(  \epsilon(1\pm i) \bigr) \; ,
	\]
and from this we infer that 
\begin{align*}
		p^1_s(k) & = (-1)^k (s+k)^2 \frac{\Gamma(s -k + 1)}{\Gamma(s+k+1)} 
					\left(  \lim_{\epsilon \downarrow 0} P_{s-1}^k \bigl(  \epsilon(1 + i) \bigr) \right)
					\\
				& \qquad \qquad \qquad \qquad \qquad \qquad \qquad \times 
					\left( \lim_{\epsilon \downarrow 0} P_{s-1}^k  \bigl(  \epsilon(1 - i) \bigr) \right)
					\\
				& = (-1)^k (s+k)(s-k)  \frac{\Gamma(s -k )}{\Gamma(s+k)} 
					\left( \lim_{\epsilon \downarrow 0} P_{s-1}^k \bigl(  \epsilon(1 + i) \bigr) \right)
					\\
				& \qquad \qquad \qquad \qquad \qquad \qquad \qquad \times 
					\left( \lim_{\epsilon \downarrow 0} P_{s-1}^k \bigl(  \epsilon(1 - i) \bigr) \right)
					\\
				& \stackrel{(\ref{eq:pk-1})}{=} (s+k)(s-k) p_{s-1}(k) \; .
\end{align*}
Since $p_{s-1}(k)=p_{s-1}(-k)$, $k\in\mathbb{Z}$, it follows from the last equality that
$p^1_s(k)=p^1_s(-k)$,  $k\in\mathbb{Z}$, and 
\[
	P_s' ( - \cos\psi ) = \sum_{k\in\mathbb{Z}} p_{k, \, s}^1 \frac{ {\rm e}^{ik\psi} }{\sqrt{2\pi r}}
\]
with
\[
 	p_{k, \, s}^1 = \sqrt{2\pi \, r} \; p^1_s(k) = \sqrt{2\pi \, r} \; (s+k)(s-k) p_{s-1}(k) \; ,
	\qquad k\in\mathbb{Z} \; . 
\]
Now, 
\begin{align}
		(s+k)(s-k) p_{s-1}(k) & \stackrel{(\ref{eq:pk-6})}{=} 
						 \frac{-\sin\big(\pi s -\pi\big)}{\pi} \frac{(s+k)(s-k)}{ (k +s-1)} \; 
						\label{eq:IUbuytytryxsga}
						\\
						& \qquad \qquad \qquad \qquad \times \frac{ \Gamma\left( \frac{k-s+1}{2} \right)}
										{\Gamma\left(\frac{k+s-1}{2}\right)}
									\frac{\Gamma\left( \frac{k+s}{2} \right)}
										{\Gamma\left(\frac{k-s+2}{2}\right)} \; . 
										\nonumber
\end{align}
Since $\sin (\pi s -\pi )=-\sin (\pi s )$, 
$\Gamma\left(\frac{k-s+2}{2}\right)=\Gamma\left(\frac{k-s}{2}+1\right)
= \frac{k-s}{2} \Gamma\left(\frac{k-s}{2}\right)$ 
and
\[
(k +s-1)\Gamma\left(\tfrac{k+s-1}{2}\right)=2\Gamma\left(\tfrac{k+s-1}{2}+1\right)
=
2\Gamma\left(\tfrac{k+s+1}{2}\right)
\; , \]
relation (\ref{eq:IUbuytytryxsga}) becomes
	\[
		(s+k)(s-k) p_{s-1}(k) = -(s+k) 
					\frac{\sin\big(\pi s\big)}{\pi}
					\frac{\Gamma\left( \frac{k-s+1}{2} \right)}
						{\Gamma\left(\frac{k+s+1}{2}\right)}
					\frac{\Gamma\left( \frac{k+s}{2} \right)}{\Gamma\left(\frac{k-s}{2}\right)} \;.
	\]
Comparing with 
(\ref{eq:omega-1}) yields 
	\[
		(s+k)(s-k) p_{s-1}(k) =  - \frac{\sin (\pi s )}{\pi} \; \widetilde \omega_s(k) \, r \; .
	\]
Here $\omega_s \equiv \omega$, with the index indicating the 
dependence of $\omega$ on $s$. The latter  had been suppressed in the main text. 
Hence,
	\[  
		p_{k, \, s^+}^1 = -\sqrt{\tfrac{2 r }{\pi}}\sin (\pi s )\widetilde \omega_s(k) \, r
	\]
Returning to (\ref{C-33}), we find 
	\begin{align*}
			\langle \kern -.2cm \langle \; f ,\, \; g  \; \rangle \kern -.2cm \rangle \,  & 
			 =  -\frac{\sqrt{2\pi r}}{ 2  \sin(\pi s^+)} \sum_{k\in\mathbb{Z}} p_{k, \,s}^1 \, \overline{f_k} \, g_k  \\
		&	= \sum_{k\in\mathbb{Z}} r^2 \omega(k) \, \overline{f_k} \, g_k
		\\
		&	= \sum_{k\in\mathbb{Z}} r^2 \omega(k) \, 
		\overline{\int_{S^1} r \, d\psi' \;  f(\psi') \frac{{\rm e}^{-ik\psi'}}{\sqrt{2\pi r}}} \, 
		\int_{S^1} r \, d\psi \;  g(\psi) \frac{{\rm e}^{-ik\psi}}{\sqrt{2\pi r}}
		\\
		&=  r^3   \, 
		\int_{S^1\times S^1} d\psi d\psi' \;  \overline{f(\psi')} 
		\frac{\sum_{k\in\mathbb{Z}}{\rm e}^{-ik(\psi-\psi')}}{2\pi } \, 
		(\omega g)(\psi) 
		\\
		&	=  r^2 \left\langle  f, \; \omega g \right\rangle_{L^2(S^1, \, \; r d\psi)} 
		= 	 r^2  \left\langle \omega f, \; \omega g \right\rangle_{\widehat{{\mathfrak h}}  (S^1)} \;.
	\end{align*}
\end{proof}
	
\backmatter

\bibliographystyle{amsalpha}

\printindex

\end{document}